# ABSTRACT BOOK
# VITM 2024

Duke University, April 22-24, 2024, Durham, NC

# Foundations and case examples

Monday, April 22, 2024

9:00 am - 9:40 am

Chaired by Ioannis Sechopoulos & Andrew Maidment

# A Road to VITM: Previous workshops and symposia on virtual trials in breast imaging

Predrag R. Bakic,[§,1] Stephen J. Glick,[2] Hilde Bosmans,[3] Kyle J. Myers,[4] Andrew D.A. Maidment[5]

[1]Lund University, Malmö, Sweden; [2]Food and Drug Administration, Silver Spring, MD; [3]Katholieke Universiteit Leuven, Belgium; [4]Puente Solutions LLC, Phoenix, AZ; [5]University of Pennsylvania, Philadelphia, PA

## BACKGROUND AND PURPOSE

Computer simulation of the anatomy, imaging, and interpretation has been used for a long time for assessing performance of breast imaging technology. A combined approach was initiated around 2009-10, when the phrase virtual clinical trials (VCTs) emerged. Several VCT workshops and symposia were organized by the imaging community. We briefly review presentations and discussion points from the VCT meetings at International Workshops on Breast Imaging (IWBI), Annual Meetings of the American Association of Physicists in Medicine (AAPM), and SPIE Medicinal Imaging conferences.

## REVIEW OF VCT MEETINGS ON MEDICAL IMAGING CONFERENCES

### VCT Symposium at AAPM 2013, Indianapolis, IN
*(Chairs: P.R. Bakic, K.J. Myers)*

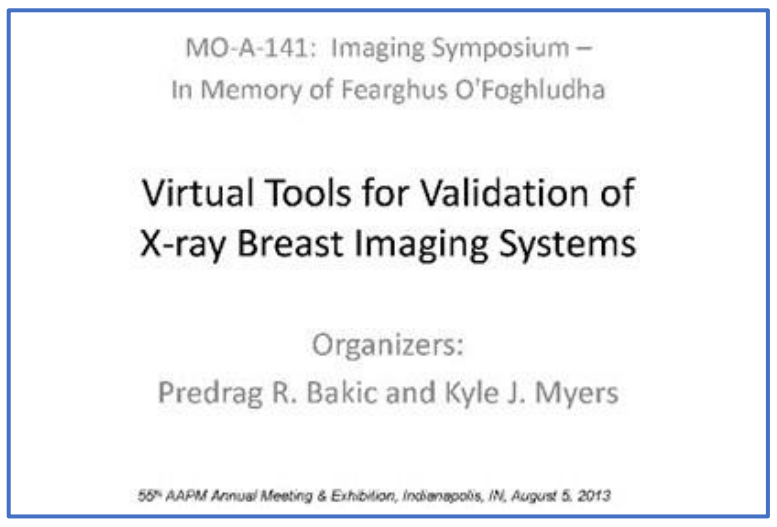

A two-hour symposium included four invited presentations on the real time and small-scale anatomy simulation, computational phantom generation from clinical data, and in silico task-based assessment of imaging systems. The presenters included: Predrag R. Bakic, Kyle J. Myers, Ingrid Reiser, Nooshin Kiarashi, and Rongping Zeng.[1]

### VCT Symposium at AAPM 2014, Austin, TX
*(Chairs: K.J. Myers, P.R. Bakic)*

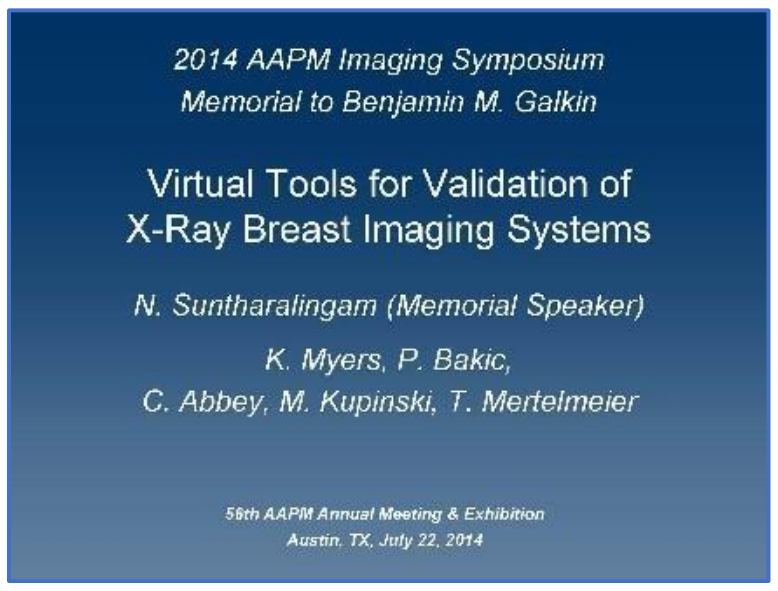

A two-hour symposium included three invited presentations on the statistical approach to evaluate phantom realism, observer models and realism testing, and simulation methods from an industry perspective. Presenters included: Kyle J. Myers, Predrag R. Bakic, Craig K. Abbey, Matthew A. Kupinski, and Thomas Mertelmier.[2]

---

§ Predrag.Bakic@med.lu.se

## VCT Symposium at AAPM 2016, Washington, D.C.

*(Chairs: P.R. Bakic, K.J. Myers)*

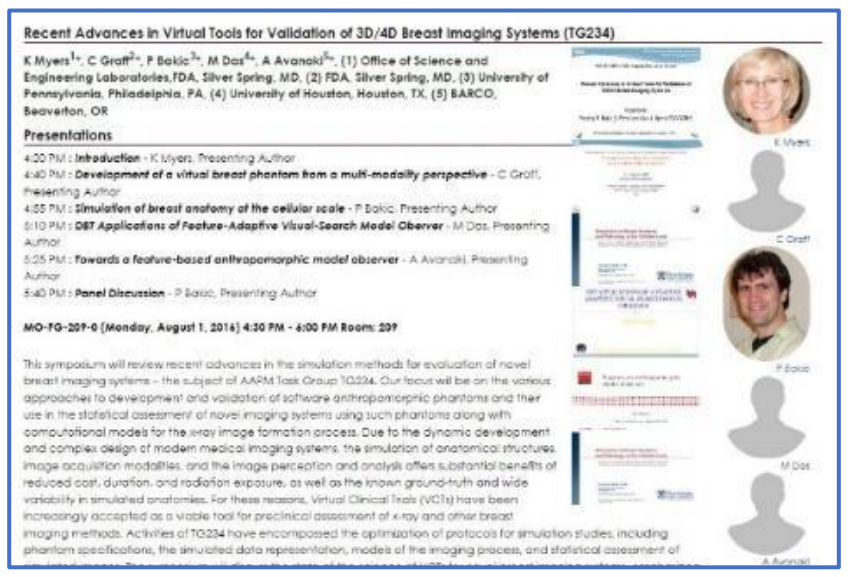

A two-hour symposium included four invited presentations on the multi-modality virtual phantoms, cellular scale simulations, and visual search and anthropomorphic model observers. Presenters included: Kyle J. Myers, Christian Graff, Predrag R. Bakic, Mini Das, and Ali N. Avanaki.[3]

## VCT Workshop at IWDM 2016, Malmö, Sweden

*(Chairs: P.R. Bakic, A.D.A. Maidment)*

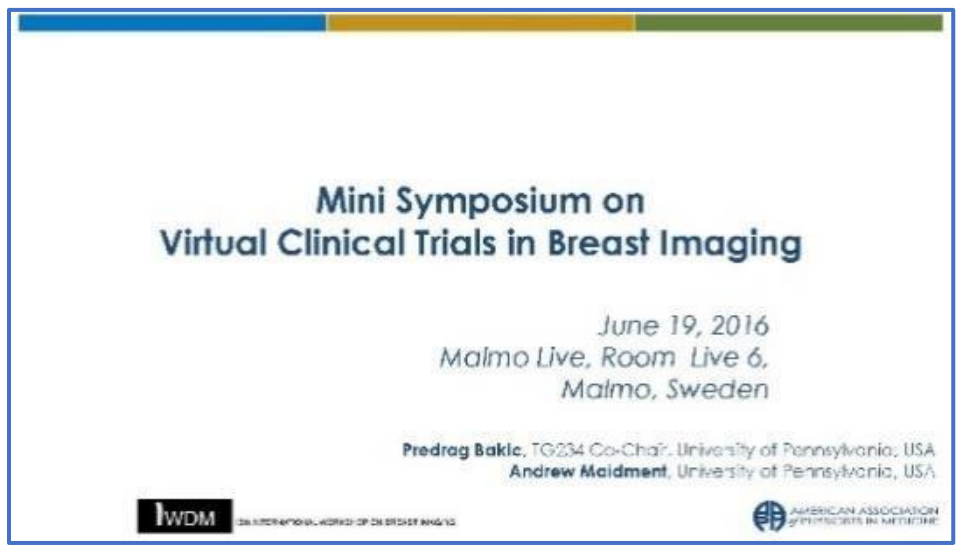

A two-hour workshop included five invited updates from academia and regulatory agencies. The panel discussion suggested future directions: define diagnostically relevant tasks, share VCT modules, develop standards, track clinical outcomes, extend to multimodality imaging, simulate inter-operator variability, and validate simulations. Presents included: Predrag R. Bakic, Andrew D.A. Maidment Stephen J. Gick, Hilde Bosmans, and Alistair Mackenzie.[4]

## VCT Workshop at IWBI 2018, Atlanta, GA

*(Chairs: P.R. Bakic, S.J. Glick)*

A four-hour workshop included four invited updates from academia, industry, and regulatory agencies. Presented were also five proffered short talks. The panel discussion focused on VCT realism, inter-operability, need for data standards, and support to multimodality and multiscale simulations. Presenters included: Predrag R. Bakic, Stephen J. Glick, Rongping Zeng, Ann-Kathrine Carton, Andrew D.A. Maidment, Alistair Mackenzie, Thomas Mertelmeier, Christiana Balta, Janne Vignero, Bruno Barufaldi, and Oliver Diaz.[5]

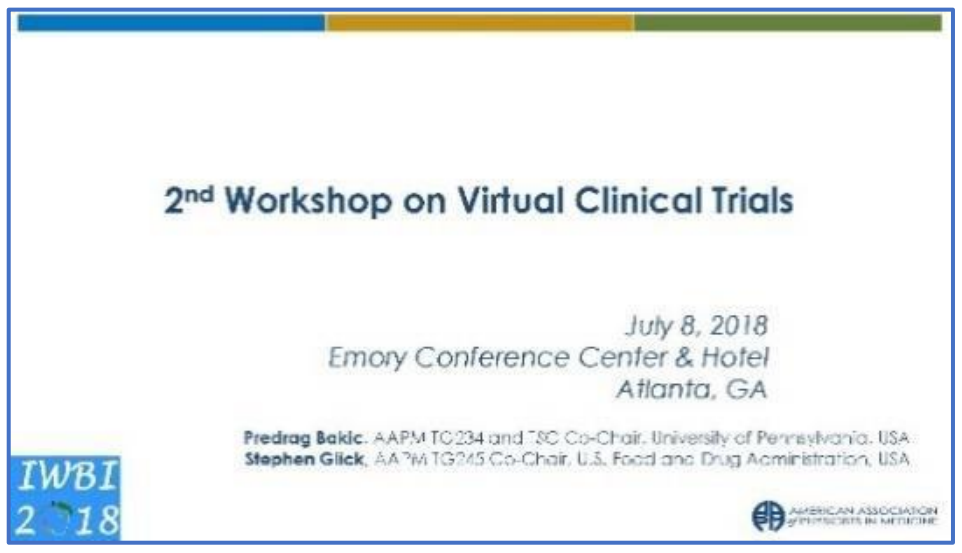

## VCT Workshop at SPIE 2021, Digital Forum

*(Chairs: H. Bosmans, F.W. Samuelson)*

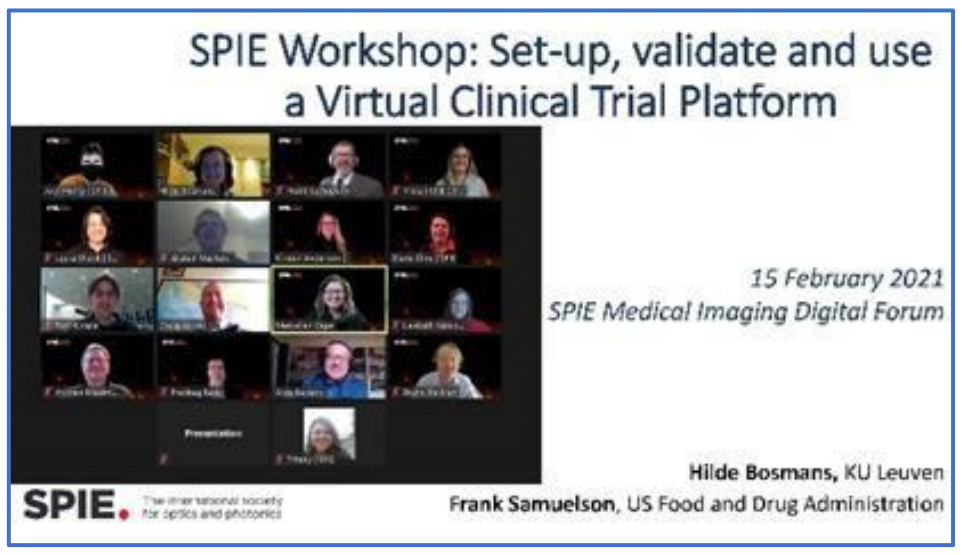

A two-hour workshop included eight presentations from academia, industry, and regulatory agencies on the set-up, validation, and use of VCT platforms, with a live demonstration of a VCT platform. Presenters included: Hilde Bosmans, Frank W. Samuelson, Ehsan Samei, Predrag R. Bakic, Andrew D.A., Maidment, Aldo Badano, Alistair Mackenzie, Tom Kimpe, Bruno De Man, and Maryellen Giger

## VCT Workshop at IWBI 2022, Leuven, Belgium

*(Chairs: P.R. Bakic, S.J. Glick)*

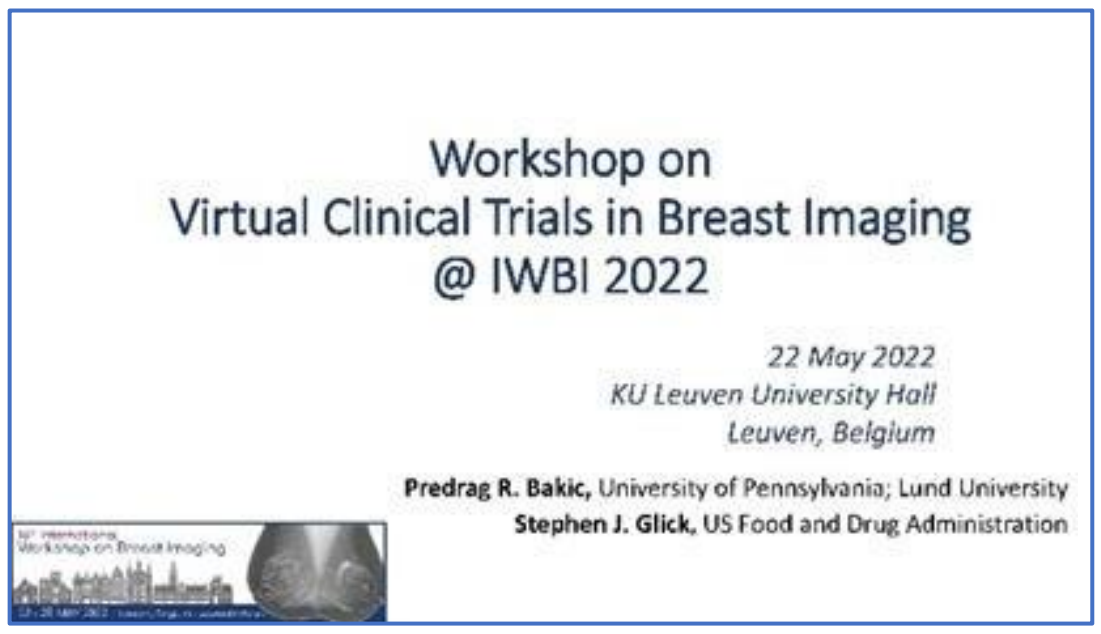

A four-hour workshop with two invited lectures from academia, and eight proffered short talks. The panel discussion focused on collaborative studies, exchanging simulation software, cross-comparison, risk of bias, and community building – *which led to the International Summit on Virtual Trials in Medicine – VITM!*[6] Presenters included: Predrag R. Bakic, Stephen J. Glick, Raymond J. Acciavatti, Ann-Katherine Carton, Marcelo A.C. Vieira, Lisbeth Vancoillie, Antonio Sarno, Lucas R. Borges, Richad Osuala, Juhun Lee, Kristina Bliznakova, Ehsan Samei, Hilde Bosmans, Andrew D.A., Maidment, Paul Kinahan, and Robert M. Nishikawa.[7]

# DISCUSSION

## Common research and discussion topics

Previous scientific meetings on virtual trials highlighted several recurring discussion topics, e.g., evaluation of virtual trials realism for various simulation tasks and developing data and simulation standards.

## Major discussions & developments

Early meetings (AAPM 2013-16) emphasized mainly the activities and discussions within the AAPM Task Group on Virtual Tools for the evaluation of New 3D/4D Breast Imaging Systems (TG234).

With VCT workshops at IWBI 2016-2022, the wider medical imaging community has become more and more familiar with the VCT approaches and benefits. Thus, the discussions reflected on the topics of interest at the time, e.g., sharing the simulations and developing appropriate data standards (IWBI 2016); multimodality/multiscale simulations (IWBI 2018); and collaborative studies, cross-comparison, and further community building (IWBI 2022) – which led to the initial planning of VITM 2024!

## Future directions

VITM 2024 program reflects the current topics of interest and open questions. Future directions in virtual trials research are summarized within the VITM 2024 Discussion sessions on: Bridging real and virtual trials, Intersection of Physics and Biology simulations; Overcoming barriers to VCT wide accessibility and implementation; and Enhancing diversity in digital patients.

# CONCLUSION

VITM reflects the development of the virtual trial community formed over many years, and recurring needs for tools to communicate, evaluate, expand, and maximize the benefits of computer modeling in medical imaging.

## ACKNOWLEDGEMENTS AND CONFLICT OF INTEREST

Previous VCT seminars and workshops have been supported by AAPM/TG234, IWBI, and SPIE funding, as well as the funding sources of participating presenters.

# Streamlined large-scale construction of virtual twins of aortic root complexes for in-silico trials of cardiac valve implants

Matheson, B.A[12], Dou, H.[12], Teleanu, C.[12], Hyde-Linaker, G.[3], Bryan, R.[3], Palmer, M.[4], Horner, M.[4], Allocco, D.[5], Flynn, D.M.[5], Blackman, D.[6], Ravikumar, N.[12], Frangi, A.F.[178], Taylor, Z.A.[12]

1 Centre for Computational Imaging and Simulation Technologies in Biomedicine (CISTIB); 2 University of Leeds; 3 Synopsys, Inc.; 4 Ansys, Inc.; 5 Boston Scientific Corporation; 6 Leeds Teaching Hospitals NHS Foundation Trust; 7 Christabel Pankhurst Institute; 8 University of Manchester

## BACKGROUND AND PURPOSE

In-silico trials (ISTs) hold transformative potential in healthcare. However, creating large-scale, virtual patient cohorts remains difficult, especially given the challenges of automating high-quality image segmentation and meshing processes for complex anatomies. We introduce a fully automated methodology for generating aortic root virtual twins (VTs). The work is motivated by development of ISTs of Transcatheter Aortic Valve Implants (TAVIs).

## METHODS

Our VT workflow comprises the following steps (figure 1): 1) segmentation of aorta and left ventricle outflow tract in cardiac computed tomography angiograms; 2) identification of aorta centreline and aortic root anatomical landmarks; 3) landmark-based warping of a template leaflet model; 4) segmentation of calcifications from a landmark-based region of interest around the valve plane; 5) computational mesh generation; 6) device-anatomy registration; and 7) boundary condition assignment. Steps 1-5 were performed using Synopsys Simpleware software, and 6-7 using Ansys LS-DYNA augmented with custom Python scripts.

The workflow was tested on a study population of 212 patients from the REPRISE III clinical trial. The resulting mesh quality was assessed using standard metrics (Aspect Ratio/Jacobian). To confirm the models' utility, balloon aortic valvuloplasty was simulated on a subset of the cohort.

## RESULTS

A sample of the resulting VTs is visualised in figure 2. Histograms of mesh quality metrics, showing the high quality of models, are presented in figure 3. Sample simulation results, highlighting predicted strain in a key tissue region associated with device-induced conduction disturbances (1), are depicted in figure 4.

## CONCLUSION

Our approach enables automated construction of high-quality VTs for complex aortic root anatomies. Given the labour-intensiveness of manual processes, automated workflows are paramount for IST scalability. Automated approaches promote traceability, verification, and reproducibility, all of which are essential for regulatory compliance. Future work will focus on using the VTs to predict incidence of conduction disturbances in TAVI cohorts.

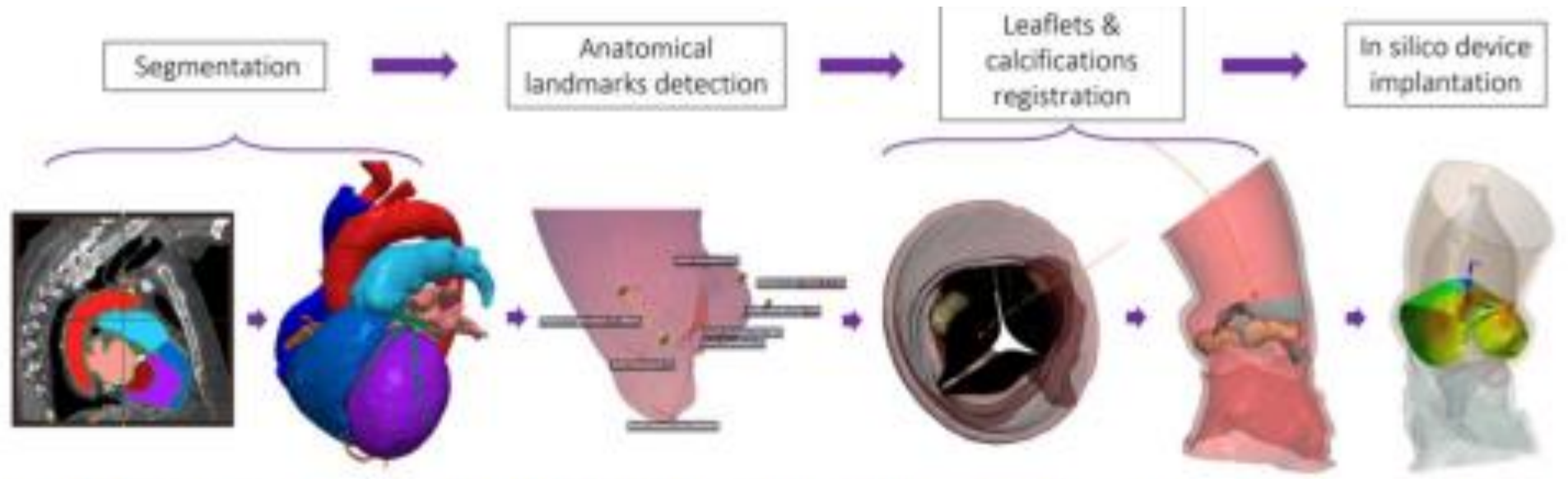

Figure 1: Schematic representation of the VT workflow, illustrating the sequence of processes from segmentation to boundary condition assignment. This workflow encompasses: a) segmentation of aorta and left ventricle outflow tract, b) anatomical landmarks detection, c) leaflets and calcifications registration, and d) in silico device implantation.

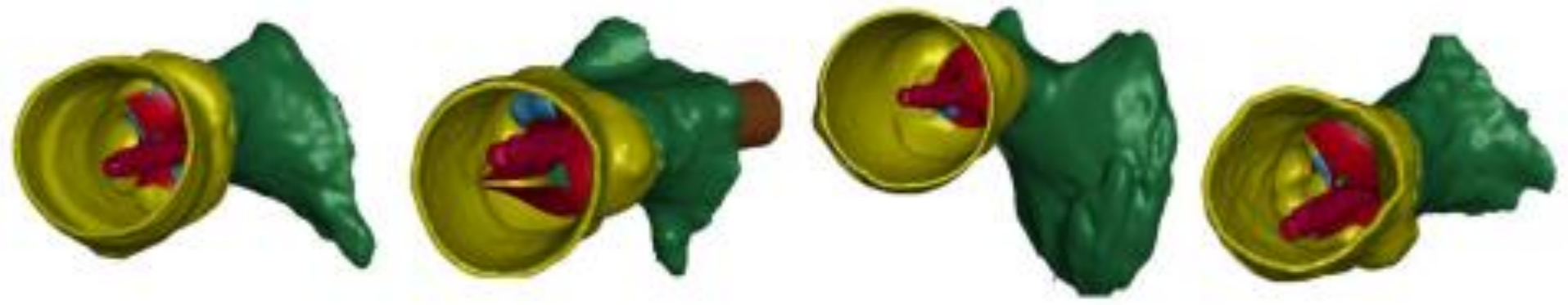

Figure 2: A selection of aortic root virtual twins Simulation Input Files rendered in LS-Dyna. The figure showcases a variety of Balloon Aortic Valvuloplasty models.

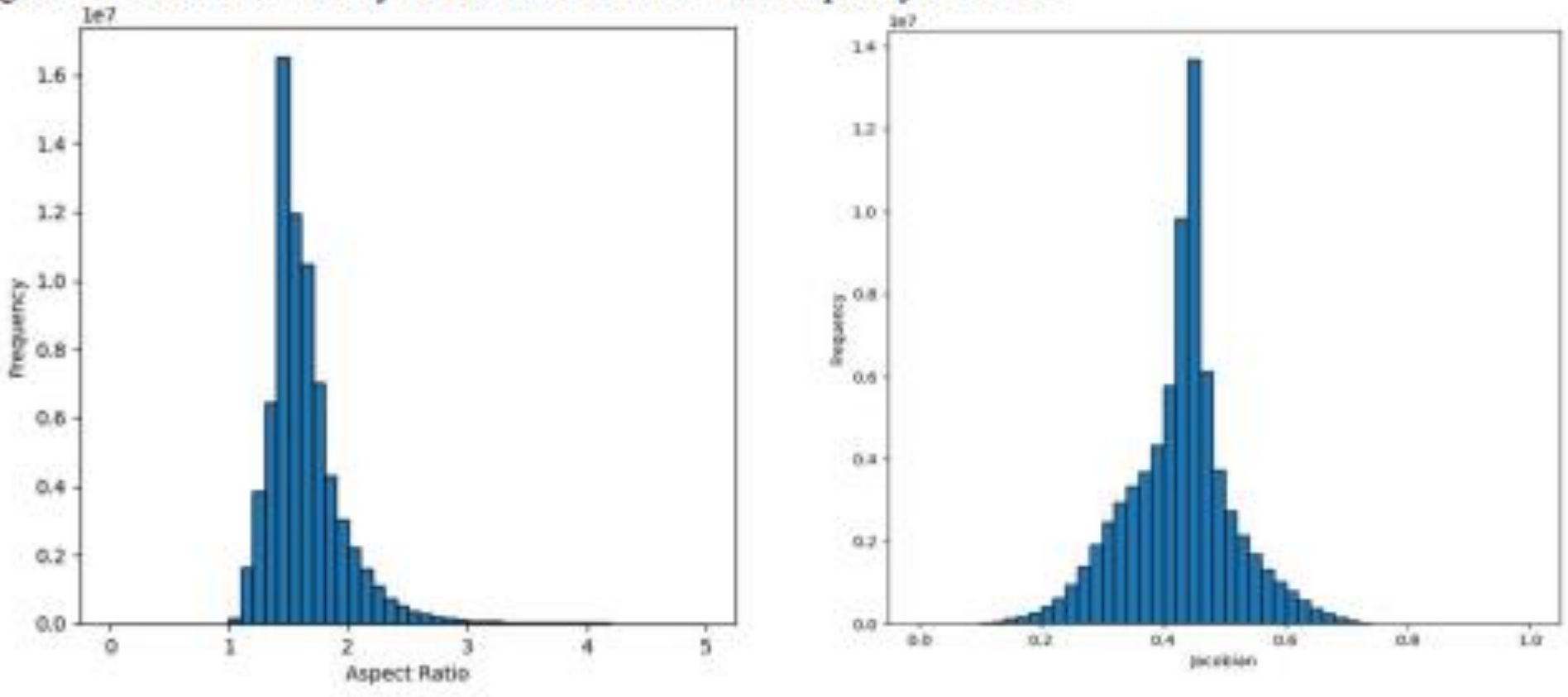

Figure 3: Histograms illustrating the distribution of mesh quality metrics, specifically Aspect Ratio and Jacobian.

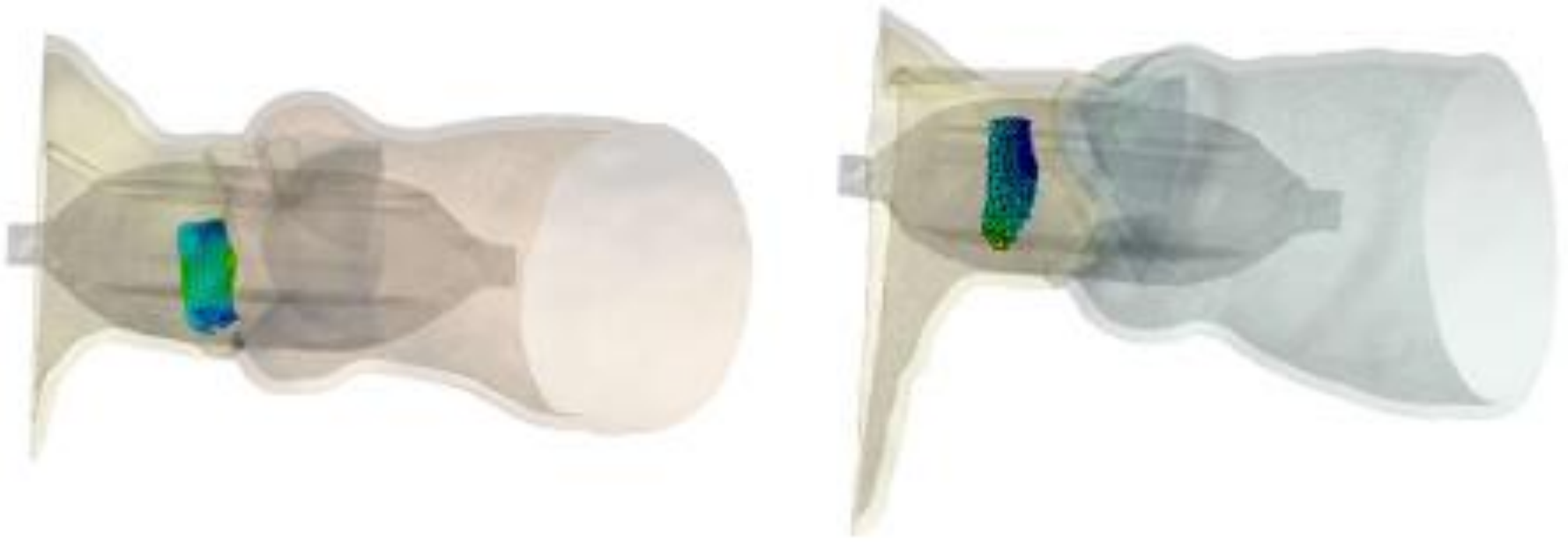

Figure 4: Visualisation of strain within the isolated conduction regions, specifically where the membranous interventricular septum interfaces with the left ventricle (2).

# Assessment of the Relative Value of Different Body Size Measurands to Inform the Needed Radiation Dose in CT Imaging: A Virtual Trial Study

Njood Alsaihati, Francesco Ria, Justin Solomon, Jered Wells, Ehsan Samei

## BACKGROUND AND PURPOSE

In recent years, the utilization of computed tomography (CT) has increased, driven by advancements in technology and its versatile applications.[1] Alongside this expansion, concerns have also increased regarding the potential risks associated with radiation exposure.[2,3] In this scenario, the effective management of radiation dose is crucial to minimize radiation exposure. However, similar to other X-ray based medical imaging modalities, the level of radiation dose used during CT image acquisition directly impacts the resulting image quality.

Specifically, when X-ray beams pass through the patient body, they undergo attenuation, scattering , or transmission. Moreover, as the body size increases, the probability of X-ray attenuation increases, resulting in higher energy deposition and a reduction in transmitted beam intensity. Consequently, fewer X-ray photons reach the detector for a given exposure leading to increased relative quantum noise and degrading the final quality of the image. As a result, achieving consistent image quality across different patient body sizes necessitates the administration of appropriate levels of radiation dose. [4,5] However, determining the appropriate radiation dose tailored to the individual is complicated by the range of body habitus and size among patients encountered in clinical operations.

This scenario is further complicated by the fact that patient body size is not characterized by a singular metric; it can be described by a multitude of measurements. Common clinical metrics like body weight and body mass index (BMI) are frequently used to describe patient size, but patients with identical body weights or BMIs may be associated with different body habitus thus limiting their utility in CT. A more useful metric for CT imaging was introduced in the American Association of Physicists in Medicine (AAPM) Report 220.[6] The concept of water-equivalent diameter ($D_w$) accounts for both geometric size and tissue attenuation variations within the patient. Nonetheless, the appropriateness of each metric in dictating the CT scanner radiation output to achieve optimal image quality has not yet been determined.

The purpose of the study was to assess the utility of different measurands of patient size in terms of their reliability to dictate the needed radiation dose to achieve a targeted image noise magnitude. To do so, we exploited virtual imaging trials (VITs) framework which stands as a crucial tool because it benefits from the ground truth representation of patient sizes and shapes.

# METHODS

## 1. Image Acquisition

In this study, six virtual patients (XCAT phantoms, Duke University) were used to represent a diverse spectrum of body habiti.[7] These virtual patients underwent simulated CT imaging using a scanner-specific simulator known as DukeSim developed at Duke University.[8,9] The imaging process involved two protocols. First, chest CT scans simulated at seven fixed mA levels (40, 66, 132, 158, 198, 264, and 528 mA), both with and without noise. Second, CT scans covering the head, neck, and shoulder regions simulated across 10 fixed mA levels (50, 70, 90, 120, 170, 220, 300, 400, 550, 750, mA), both with and without noise.

## 2. Data Analysis

Following image acquisition, a comprehensive analysis was conducted as follows. $D_w$ was calculated from each image slice ($D_{w_slice}$), at the center slice ($D_{w_center}$), averaged across all slices ($D_{w_average}$), and as the median across all slices ($D_{w_median}$). Additionally, patient weight and BMI were investigated as patient size candidates.

Average noise magnitude values for each slice ($\sigma_{slice}$) were determined by subtracting each noise-free image from its noisy counterpart. Using the Beer-Lambert Law and the relationship between noise and mA, a functional model was developed to establish the relationship between noise, size, and mA, as represented by the equation:

$$\sigma_{slice} = c_1[c_2\, mA\, e^{-(c_3 size)}]^{-c_4} \tag{1}$$

In equation 1, parameters c1, c2, c3, and c4 represent the fitting parameters. The model was fit to the data using curve fitting function. In this analysis, $D_{w_slice}$ was assumed to be the gold standard size metric for determining the required mA to achieving a target noise level. The other size metrics were evaluated in relation to $D_{w_slice}$. Two figures of merit were used to evaluate the efficacy of each size metric in predicating the needed mA and noise magnitudes accurately. In particular, the root mean square error (RMSE), and the mean percent error (MPE) in mA and noise were computed using the formula:

$$MPE = \left|\frac{\alpha_{predicted_i} - \alpha_{predicted_{ref}}}{\alpha_{predicted_{ref}}}\right| x\ 100 \tag{2}$$

where $\alpha_{predicted_i}$ refers to the predicted values of mA or noise for the evaluated size metric (i.e. $D_{w_average}$, $D_{w_median}$, $D_{w_center}$, weight, and BMI). $\alpha_{predicted_{ref}}$ is the predicted values of mA or noise for the gold standard metric (i.e. $D_{w_slice}$).

# RESULTS

The results in Table 1 and Table 2 present the mean percent error of the various size metrics in predicating the required tube current and achieving the targeted noise level across different body regions: chest region and head, neck and shoulder region. These metrics are compared relative to the gold standard metric $D_{w_slice}$.

In chest region, weight and BMI exhibit the highest percentage error in predicating the required tube current and targeted noise level. The RMSE values for the noise model corresponding to $D_{w_slice}$ , $D_{w_average,}$ $D_{w_median,}$ $D_{w_center,}$ weight, and BMI are 3.12, 7.00, 6.92, 6.95, 8.33, and 7.28, respectively. Additionally, Figure 1 show surface fits illustrating the relationship between mA, size metrics, and $\sigma_{slice}$ for the chest protocol data.

Conversely, in the head, neck, and shoulder regions, all size metrics, including $D_w$ metrics, weight, and BMI, demonstrate significantly higher percentage errors. Furthermore, the RMSE values for the noise model associated with $D_{w_slice}$ , $D_{w_average}$, $D_{w_median}$, $D_{w_center}$, weight, and BMI are 0.87, 5.65, 5.73, 5.66, and 5.65, respectively.

*Table 1 Mean percent error in image noise (σ) and tube current (mA) for various size metrics in the chest region*

| Size metric | Mean % error in $\sigma$ | Mean % error in mA |
|---|---|---|
| $D_{w_average}$ | 12.5 | 20.9 |
| $D_{w_median}$ | 12.5 | 20.3 |
| $D_{w_center}$ | 12.5 | 20.3 |
| Weight | 18.2 | 32.9 |
| BMI | 14.4 | 23.5 |

*Table 2 Mean percent error in image noise (σ) and tube current (mA) for various size metrics in the head, neck and shoulder regions*

| Size metric | Mean % error in $\sigma$ | Mean % error in mA |
|---|---|---|
| $D_{w_average}$ | 32.7 | 56.1 |
| $D_{w_median}$ | 32.8 | 56.3 |
| $D_{w_center}$ | 33.3 | 56.5 |
| Weight | 32.9 | 56.4 |
| BMI | 33.1 | 56.1 |

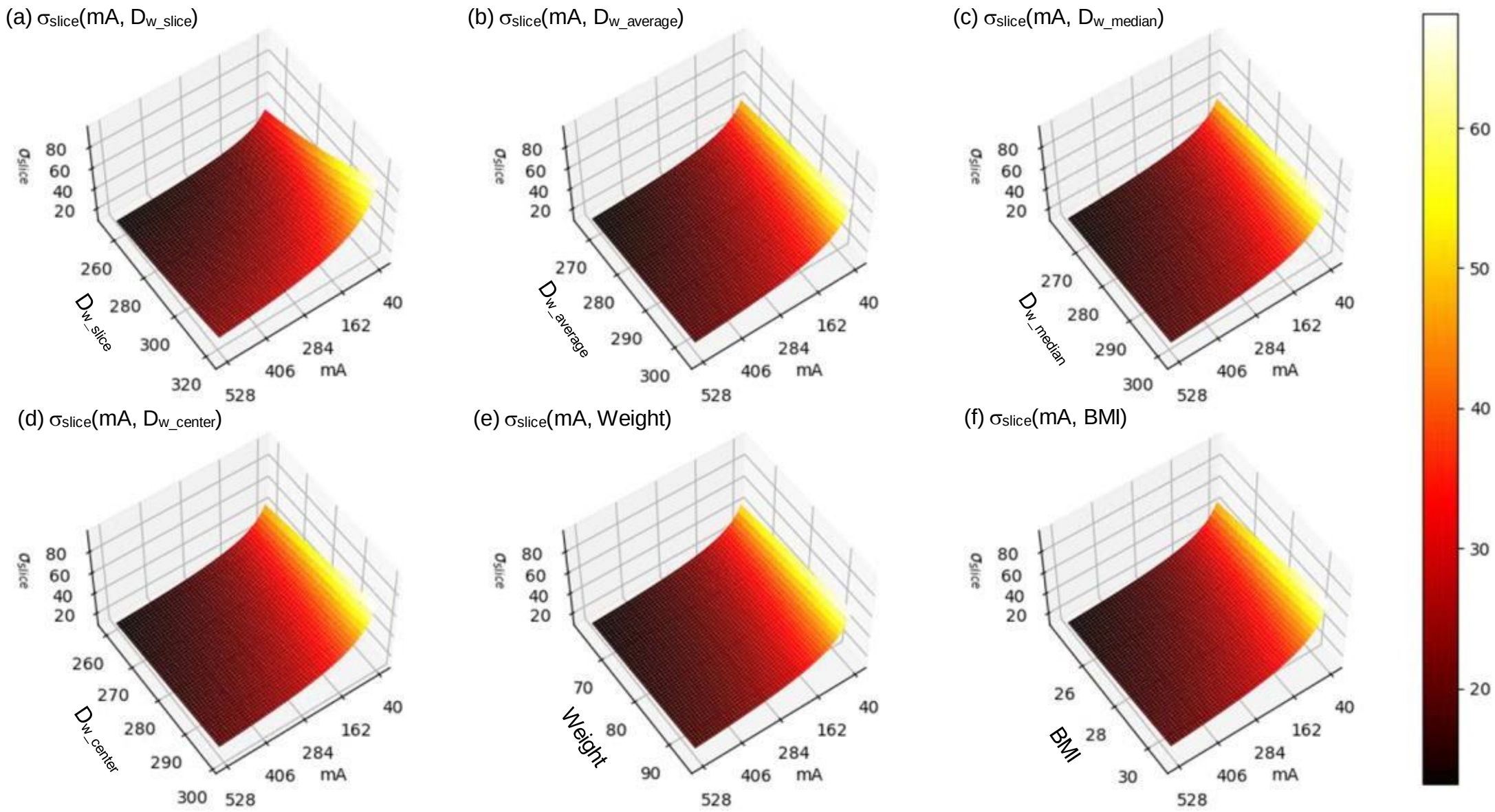

*Figure 1 3D surface fits showing the relationship between mA, size metrics, and average noise magnitude σslice for chest protocol data.*

# DISCUSSION

In this study we assessed how different patient size surrogates can predict the required radiation output to obtain a specific image noise magnitude in CT studies. Our analysis of the chest data shows weight and BMI do not accurately capture variations in size and attenuation resulting in the highest mean percentage error in mA and noise. On the other hand, the head, neck, and shoulder data results highlight the importance of considering anatomical variations when assessing size metrics for predicating the required radiation dose to target a specific noise magnitude. The substantial variations in size and attenuation along the z-axis requiring dynamic estimates of patient size. Metrics such as $D_{w_average}$ and $D_{w_median}$ are region-dependent. For example, in the chest region, these metrics values may closely align, whereas differences are observed when analyzing the head, neck, and shoulder regions together.

Achieving optimal image quality with minimal radiation exposure is crucial in medical imaging. However, traditional assessments using patient images are limited due to the lack of a definitive ground truth and incomplete patient data, such as weight and BMI records, which may sometimes not be recorded or updated on the day of imaging the patient. Moreover, concerns about repeated patient imaging due to radiation exposure further complicate these assessments. VITs offer a solution by simulating diverse patient types and imaging conditions. Leveraging data from virtual patients represented by XCAT phantoms and DukeSim scanner, such assessments become feasible, providing access to never-before-available information for refining CT imaging protocols to account for variations in patient body habitus and composition.

## CONCLUSION

Not all size estimates carry the same level of relevance to inform dose for CT imaging. Care should be exercised when predicting scanner performance using metrics non-representative of patient body habitus. The findings are dependent on the body part being imaged. Furthermore, the study highlights the significance of VITs in addressing these challenges. VITs offer a more reliable representation of patient sizes and shapes, providing valuable insights in assessing the utility of different size metrics in determining the necessary radiation dose to achieve targeted image quality.

## ACKNOWLEDGEMENTS AND CONFLICT OF INTEREST

This work was supported by the National Institute of Biomedical Imaging and Bioengineering of the National Institutes of Health under award number P41-EB028744 and R44EB031658.

# Applications of virtual trials

Monday, April 22, 2024

9:40 am - 10:20 am

Chaired by Ioannis Sechopoulos & Andrew Maidment

# In silico clinical trial to assess the influence of orientation on the primary stability of a cementless acetabular cup

Mark Taylor (Flinders University)

## Background

Achieving primary fixation of cementless acetabular components is critical for ensuring longevity of the device. The initial mechanical environment, particularly the bone strains immediately adjacent to the cup and the micromotions at the bone implant interface are dependent on a number of factors including cup placement and orientation. The Lewinnek safe zone (LSZ) of 40 degrees of inclination and 15 degrees of version plus/minus 10 degrees has been widely adopted as the target for acetabular cup placement. However, studies have reported that actual cup alignment can vary from 20 to 60 degrees of inclination and 0 to 40 degrees of version. The influence of alignment, both within and outside the LSZ, on primary stability are poorly understood. The aim of this study is to assess the influence of acetabular component alignment within an in silico clinical trial framework.

## Methods

Finite element models of 57 implanted hemi-pelvii were generated from CT scans of a cohort of end stage osteoarthritis patients using a custom scripted ISCT work flow (Simpleware, Synopsys Inc.) which included: auto-segmentation, defining an anatomical reference frame; sizing and positioning of the cementless acetabular component; calibration of CT scan to define bone material properties; mesh generation; definition of contact surfaces; defining loading and boundary conditions to simulate level gait and post-processing. Each pelvis was implanted with the following cup orientations (inclination:version angles in degrees): 40:15 (reference); 30:5; 30:25; 50:5; 50:25; 20:-5; 20:35; 60:-5 and 60:35. Bone strain immediately adjacent to the implant and mircomotions (gap and shear micromotions) were used to assess primary stability. Analysis was performed at the levels of the individual subjects and the entire cohort.

## Results and Discussion

There was significant variation in the predicted strains and micromotions for the reference cup alignment (40:15) and both were moderately correlated to the average bone modulus of the pelvii. The mean and 95th percentile equivalent strains varied from approx. 950 to 5300 microstrain and 3900 to 18500 microstrain respectively. Similar trends were observed for the predicted micromotions. Within the LSZ, the influence of cup positioning only had a minor effect on the bone strains, with an average variation of 5%. However, with upto 20% variation was predicted for an individual. There was greater variation in the predicted micromotions, with an average variation across the cohort of approx. 25% and individuals experiencing variations of up to 55%. Outside of the LSZ, the mean strain variation across the cohort increased to 12% with peak of upto 39% reported for an individual. Similar trends were observed for the predicted micromotions.

In conclusion, at both the individual subject and cohort levels, micromotion was more sensitive to changes in the acetabular cup alignment than the peri-prosthestic bone strains.

# Illustration of potential Virtual Imaging Trial applications in dental CBCT imaging

Merken K[1], Marshall NW[1], Nuyts J[2], Massera RT[1], Jacobs R[3,4], Bosmans H[1]

[1]KU Leuven, Department of Imaging and Pathology, Division of Medical Physics & Quality Assessment, Leuven, Belgium
[2]KU Leuven, Department of Imaging and pathology, Division of Nuclear Medicine & Molecular Imaging, Leuven, Belgium
[3]KU Leuven, Department of Imaging and Pathology, Division of Oral and Maxillofacial Surgery, Leuven, Belgium
[4]Karolinska Institute, Department of Dental Medicine, Stockholm, Sweden

## BACKGROUND AND PURPOSE

A literature search in the field of dentomaxillofacial radiology revealed several ex-vivo/in-vitro studies and few in-vivo studies in which technical imaging components have been optimized towards better performance in dental practice. In present study, the feasibility of a Virtual Imaging Trial (VIT) platform to potentially set-up similar studies in a virtual setting is illustrated.

## METHODS

An in-house developed framework and the necessary digital phantoms were prepared for the following potential studies: i) the impact of intra-canal post material type (Ni-Cr alloy, fiberglass, gutta-percha) and acquisition settings (mA, kV) on root fracture (RF) visibility; ii) the effect of a metal artefact reduction algorithm on RF visibility in a tooth treated endodontically and restored with a metal post; iii) image artefact levels from candidate new restorative materials, like graphene; iv) effect of patient rigid motion on image artefacts. In addition, features not available on the real system were modeled, i.e., automatic exposure control and extended mA and kV ranges, were added to potentially study the full extent of the impact of these parameters. Patient dose levels were also quantified.

## RESULTS

The generated images showed the influence of different restorative materials, dose levels, rigid motion, and post-processing on the quality of the final images. Tendencies retrieved from illustrating several conditions were consistent with findings in literature. Images were considered sufficiently realistic. Furthermore, the platform was able to simulate scenarios that are difficult or impossible to replicate physically in a controlled and repeatable way.

## CONCLUSION

The dental CBCT platform has considerable flexibility, and can model a range of tasks in VITs, from RF visibility, through to an exploration of image artefacts. Optimization studies are possible through simultaneous modelling of task and patient dose. The platform also has application as teaching tool.

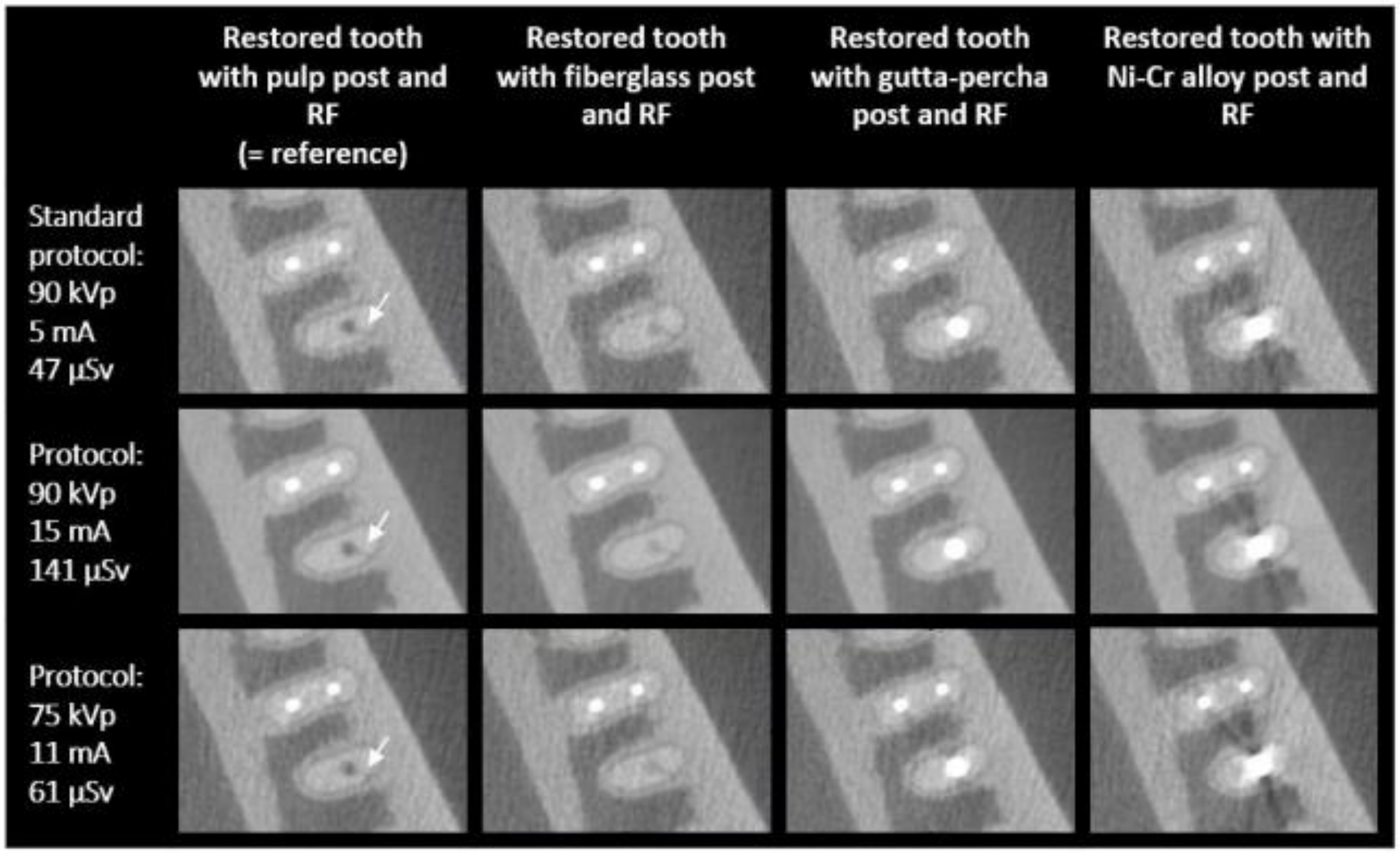

*Figure 1: Illustration of the impact of intra-canal post material type and acquisition settings on RF visibility. A small FOV of 4x4 cm² with a reconstructed isotropic voxel size of 0.08 mm was selected, similar as in clinical practice.*

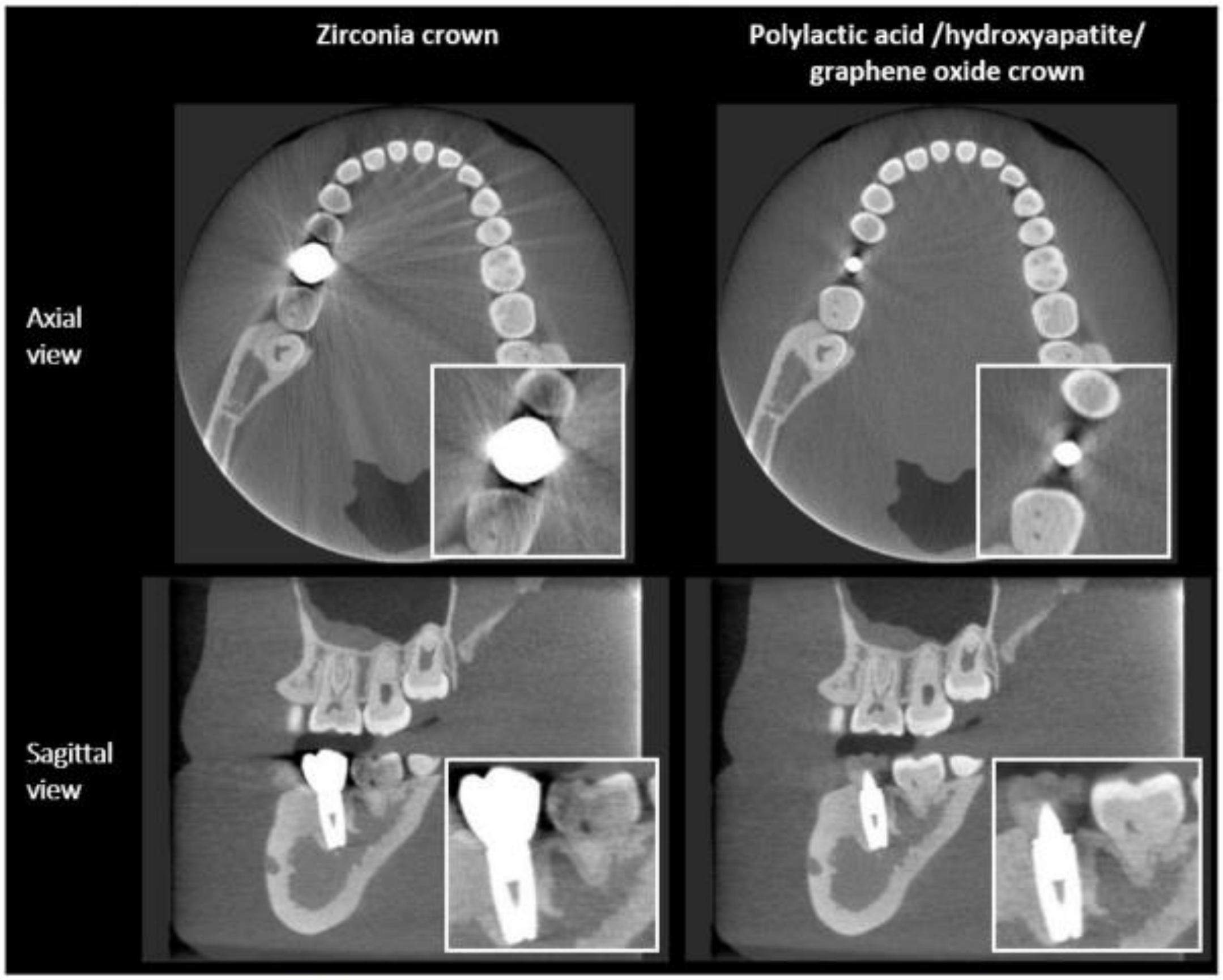

*Figure 2: Illustration of the image artefact levels using a graphene based material instead of the nowadays commonly used Zirconia as crown material. For the implant and abutment, Titanium was used. The images were simulated at the standard clinical setting of 90 kVp and 5 mA, and a FOV size of 11x8 cm² with a reconstructed isotropic voxel size of 0.2 mm. The effective dose for the chosen set-up was 196 µSv.*

# The use of phantoms in virtual imaging trials for microcalcification detection optimization in digital breast tomosynthesis

Cockmartin L, Houbrechts K, Marinov S, Marshall N W, Vancoillie L, Verhoeven H, Sanchez de la Rosa R, Carton A-K*, Bosmans H*

*Contributed equally as senior authors

## ABSTRACT

When designing virtual imaging trials (VIT), it is essential to determine the required level of realism of the task and background. This work aims to validate two simulated breast backgrounds:physical L1 phantom (acrylic spheres in water) and virtual anthropomorphic phantom (Stochastic Solid Breast Texture (SSBT)) against patients for microcalcification detection in digital breast tomosynthesis (DBT) using a hybrid VIT. Patient and L1 images were acquired on a GE Senographe Pristina™ DBT system. Realistic DBT projections of the SSBT phantom were simulated using the Computer Assisted Tomography SIMulator (CatSim) software (GE HC), including quantum and electronic noise, detector Modulation Transfer Function and x-ray scatter at an energy of 23keV. Voxel models of individually segmented calcifications were created from a micro-CT dataset of biopsy specimens. Calcifications of similar sizes were grouped into clusters based on the distribution of real clusters in DBT. Projections of microcalcifications were created using CatSim and inserted into background images. The same cluster templates were used for all three backgrounds to allow pairwise comparison. Reconstructed volumes of 40x40x40mm$^3$ were included in an observer study with 120 signal present and 120 signal absent volumes for each background type. Three readers localized and rated the presence of a cluster on a scale from 1-4. Area under the alternative free-response receiver operating characteristic (AFROC) curve was used as figure of merit. The reader-averaged areas under the curve for patients, SSBT and L1 were 0.70±0.04, 0.74±0.04 and 0.76±0.03 respectively. Microcalcification detectability was similar in patient and SSBT backgrounds($p=0.09$), yet small differences were found between patients and L1($p=0.02$). No difference was seen between L1 and SSBT($p=0.44$).
The SSBT phantom is a validated alternative for patients in VIT when studying microcalcification detection in DBT, while L1 offers a close approximation.

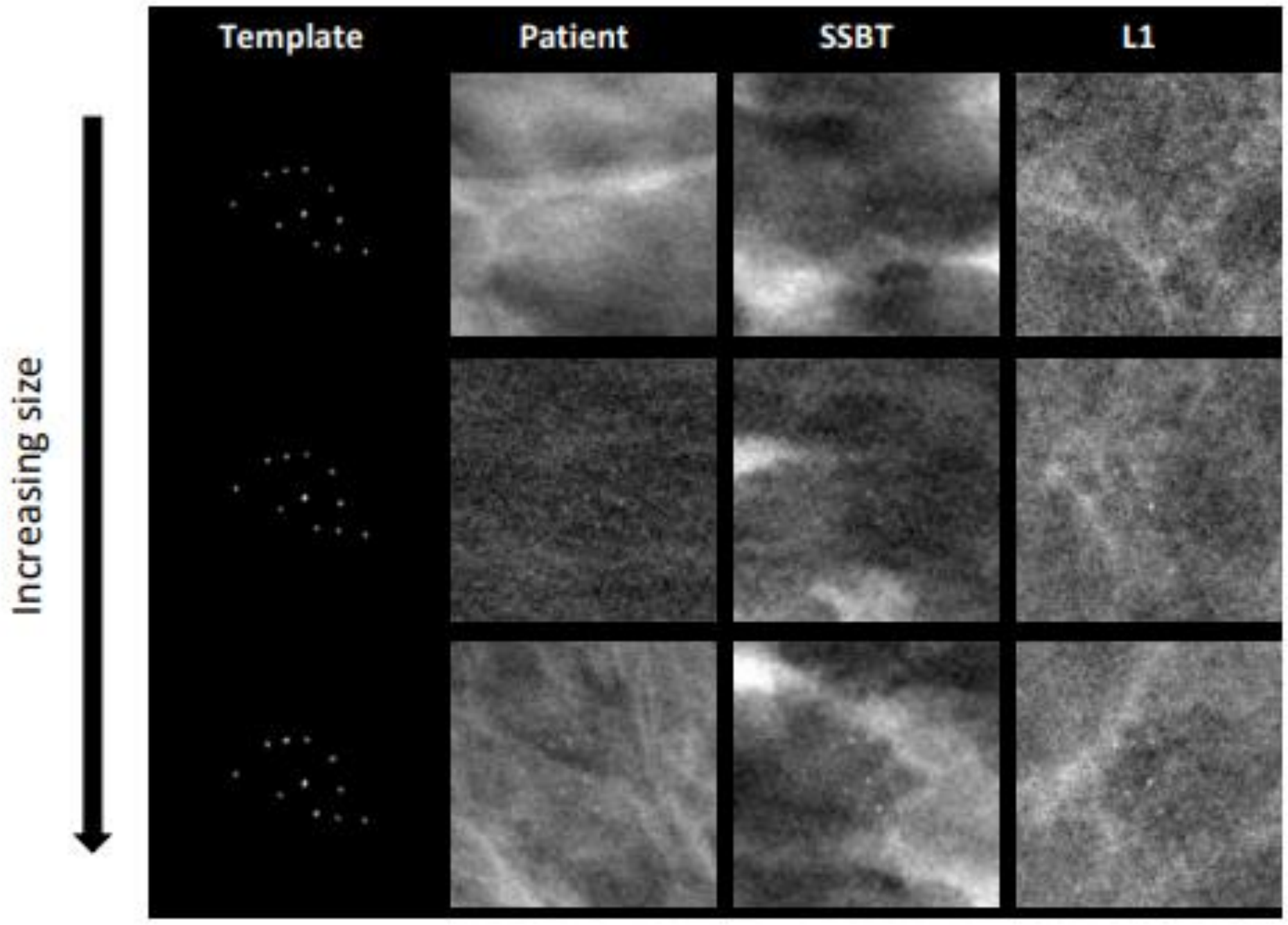

**Figure 1:** A cluster is formed by combining individual calcifications of similar size. After generating the cluster templates using a VIT framework, they are inserted into three different backgrounds: patient, virtual SSBT phantom and physical L1 phantom DBT images. The figure shows the in-plane DBT reconstructed image for all three backgrounds for three clusters with increasing size from top to bottom row.

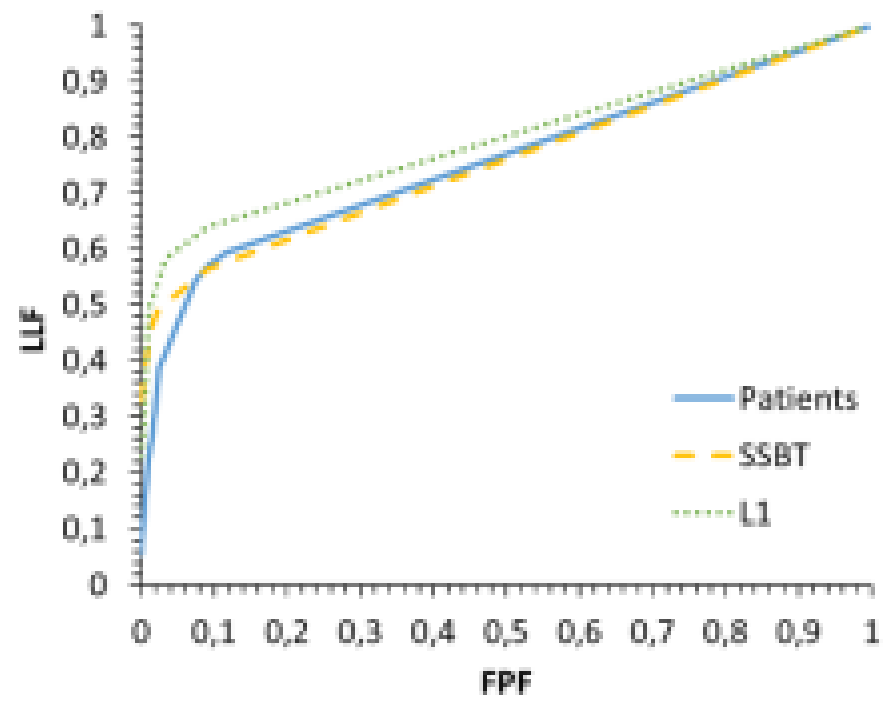

| | Patients | SSBT | Physical L1 |
|---|---|---|---|
| **Reader 1** | | | |
| **TP - TN - FP - FN** | 59 - 88 - 22 - 48 | 65 - 89 - 15 - 48 | 71 - 87 - 22 - 37 |
| **AUC** | 0.72 | 0.74 | 0.77 |
| **Reader 2** | | | |
| **TP - TN - FP - FN** | 51 - 90 - 25 - 51 | 57 - 94 - 5 - 61 | 65 - 94 - 10 - 48 |
| **AUC** | 0.68 | 0.73 | 0.76 |
| **Reader 3** | | | |
| **TP - TN - FP - FN** | 58 - 87 - 21 - 51 | 63 - 95 - 3 - 56 | 63 - 96 - 6 - 52 |
| **AUC** | 0.70 | 0.76 | 0.76 |
| **Mean AUC** | **0.70** | **0.74** | **0.76** |
| **CI 95%** | (0.66, 0.74) | (0.71, 0.78) | (0.73, 0.79) |
| **Mean precision** | **0.71** | **0.90** | **0.85** |

**Figure 2:** The reader-averaged alternative free-response receiver operating characteristic (AFROC) curve of the three different background types. The true positive (TP), true negative (TN), false positive (FP) and false negative (FN) results of the individual readers together with their area under the AFROC curve and precision can be found in the table.

# Computational patient specific modeling

Monday, April 22, 2024

10:50 am - 11:30 am

Chaired by Ann-Katherine Carton & Paul Kinahan

# Finite element modeling of patient- and disease-specific lung respiration for the lung function assessment

Amar Kavuri, Fong Chi Ho, Jerry Worthy, Zhuohan Zhou, Ehsan Samei, W. Paul Segars, Ehsan Abadi

Center for virtual imaging trials, Duke University

## BACKGROUND AND PURPOSE

The validity of studies of lung disease through virtual trials depends on accurate mechanical modeling of lung respiration.[1] Current respiration models are generic and lack the necessary variability in motion, especially considering the presence of lung diseases of varying severity [2,3]. The aim of this work was to develop a respiration model that incorporates disease-specific mechanics and respiratory forces.

## METHODS

Five parametric respirations were generated accounting for Chronic Obstructive Pulmonary Disease (COPD) abnormalities. The overall framework of this study is shown in Figure 1. The anatomy of the lungs and disease abnormalities (emphysema and air-trapping) were modeled by segmenting CT scans of patients of different sex (2M, 3F), age (43.3±10.9 years), and BMI (25.2±6.12), replicating the method used for the generation of the state-of-the-art human models (e.g., XCAT) [4]. The diseased lung models were converted into tetrahedral meshes using finite element modeling software (ABAQUS). Tissue specific mechanical properties (Young's modulus and Poisson's ratio) were assigned to each tissue type (Table 1) [5]. Patient-specific forces and boundary conditions were applied to both lungs' surfaces. The magnitudes of forces were adjusted iteratively until a target volume was achieved for each time frame. The resultant motion vectors and lung models were utilized to generate voxelized phantoms over the respiratory cycle.

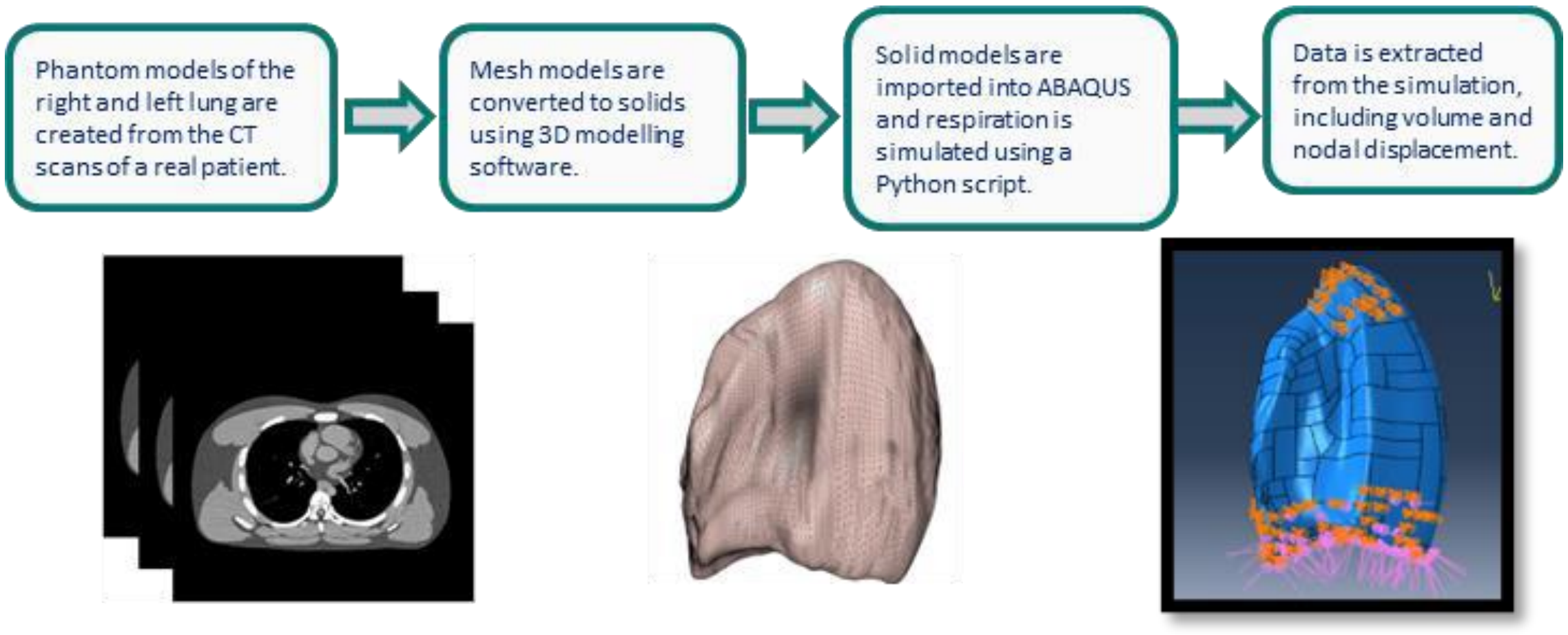

*Figure 8: Computational framework used to model respiration.*

*Table 3:Mechanical properties of lung tissues used for simulation: Young's modulus (E), Poisson's Ratio (ν), and Volume Density (ρ) [5].*

| Tissues | Young's modulus, E (MPa) | Poisson's Ratio, ν | Density, ρ (g/ml) |
|---|---|---|---|
| Healthy Lung | $3.74 \times 10^{-3}$ | 0.3 | 0.26 |
| Emphysema | 49 | 0.4 | 0.045 |
| Air-trapping | 33 | 0.33 | 0.078 |

## RESULTS

Figure 2 shows an example lung model simulated at full inspiration and at an intermediate respiration level. The left lung was set to include multiple areas of emphysema while the right lung was defined as healthy. The results show the bottom of the lungs displace more as expected compared to that of the top due to diaphragm pressure. The diseased lung is also seen to compress less compared to that of the healthy lung.

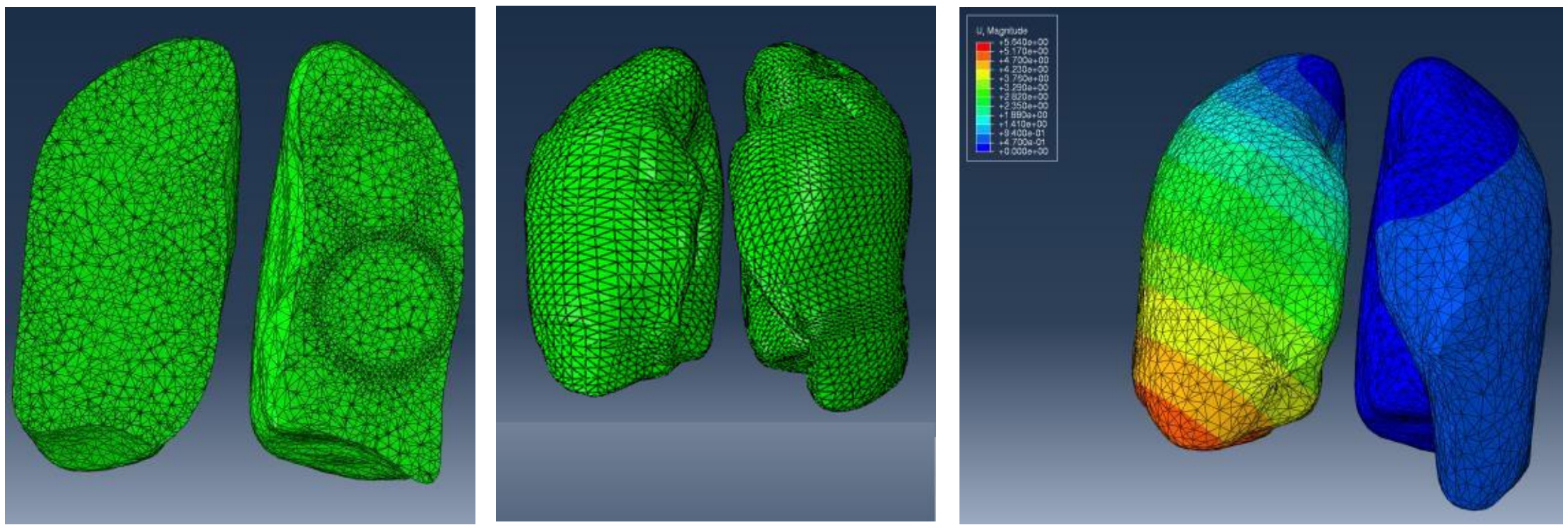

*Figure 9: Simulated lung models at full inspiration (left) and an intermediate respiration level (middle) with left lung includes disease areas. The magnitude of displacement is represented by color(right).*

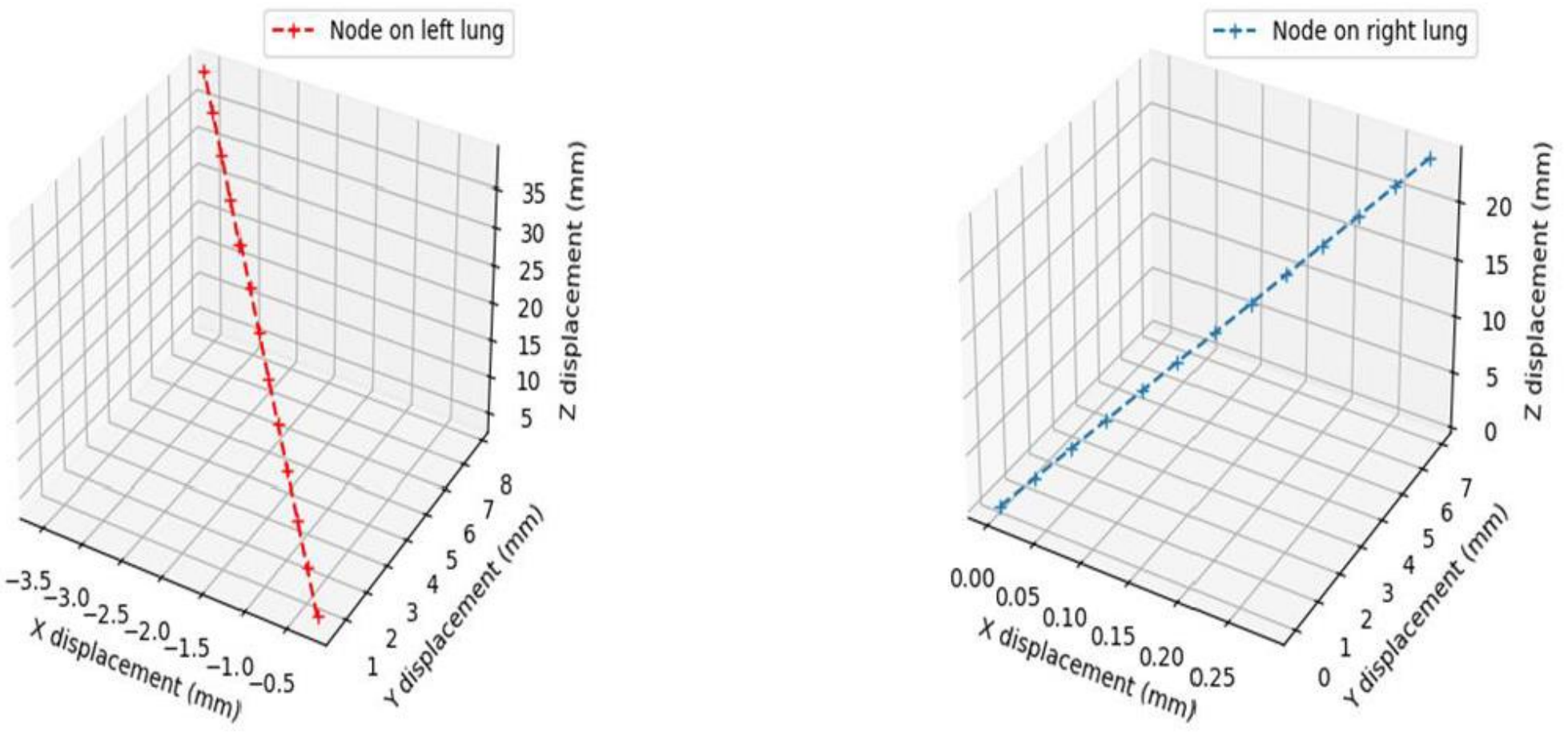

*Figure 10: 3D trajectory of sample nodes calculated from the biomechanical finite element model to demonstrate the tracking of specific nodes on the left and right lungs during the respiration process.*

Figure 3 demonstrates the 3D trajectory of sample nodes during the respiration simulation process.

## DISCUSSION

The generated lung models demonstrate distinct volumetric alterations in diseased regions compared to healthy parenchyma and differences in respiration motion attributed to the presence or absence of disease. Although the simulation possesses certain capabilities, it is not without its drawbacks. Currently, each tissue that is simulated is characterized by a uniform composition. In our subsequent research, we plan to explore methods to represent the variability within each organ's tissue.

## CONCLUSION

This work developed an accurate biomechanical finite element model for lung respiration using patient- and disease-specific mechanics. The model will enable more reliable and accurate virtual trials such as studying the relevance of quantitative CT in assessing lung function in disease.

## ACKNOWLEDGEMENTS AND CONFLICT OF INTEREST

This study was supported in part by grants from the National Institutes of Health (R01EB001838, R01HL155293, and P41EB028744).

# Generation of digital breast phantoms with patient-based internal structures from breast CT images

Martina Nassi[1], Koen Michielsen[1], Ioannis Sechopoulos[1,2,3] and Juan J. Pautasso[1]

1. Department of Medical Imaging, Radboud University Medical Center, Nijmegen, the Netherlands
2. Dutch Expert Centre for Screening (LRCB), Nijmegen, The Netherlands
3. Technical Medical Centre, University of Twente, Enschede, The Netherlands

## BACKGROUND AND PURPOSE

To generate digital breast phantoms from patient breast CT (bCT) scans, incorporating random realistic shapes and patient-based internal structures.

## METHODS

Principal Component Analysis (PCA) was applied to the breast surface of 488 patient bCT scans acquired from multiple sites, representing shape variance with linearly-independent vectors. These vectors were used to generate 400 breast models with varying shapes, voxelized through mesh reconstruction. Complete phantoms were created by integrating these models with fibro-glandular tissue distributions retrieved from 119 pre-segmented bCT scans from one site, using non-rigid image registration. Tissue distribution was characterized through glandular density computation along the axial, coronal, and sagittal directions. To compare the anatomical texture of synthesized and real breasts, 50 coronal 2D regions of interest (ROIs) measuring 43.7 mm on a side were extracted from each breast, and a Hann window was applied to reduce edge effects. The 2D noise power spectrum (NPS) of each ROI was calculated, averaged across ROIs, and radially averaged to yield a 1D NPS. Power-law functions were fitted to these radial profiles and the slopes compared to prior publications on segmented bCT images.

## RESULTS

The fourteen-component PCA model, explaining 99.99% of data variance, yielded a median mean absolute error of 0.78 mm (IQR=[0.67, 0.96] mm) compared to the original shapes. Fibroglandular tissue concentrated caudally and anteriorly, exhibiting a symmetrical lateromedial distribution. Synthesized breasts, compared to real ones, demonstrated a percent relative error in median peak magnitude of 41%, 31%, and 88% along the axial, sagittal and coronal directions, respectively, indicating major challenges in tissue alignment near the nipple. In the spatial frequency range of 0.07 lp/mm to 1 lp/mm, the average fitted power-law slope was found to be 2.7 (SD=0.2) for synthesized breasts and 2.3 (SD=0.3) for real breasts (both median $R^2 > 0.99$), differing from Metheany et al's findings (2.1, SD=0.3). The statistically significant difference in these average values likely reflects glandular density discrepancies.

## CONCLUSION

The developed model can simulate digital breast phantoms with different shapes and patient-based internal structures. This tool holds promise for conducting virtual clinical trials, providing a valuable resource for 3D breast imaging optimization and evaluation.

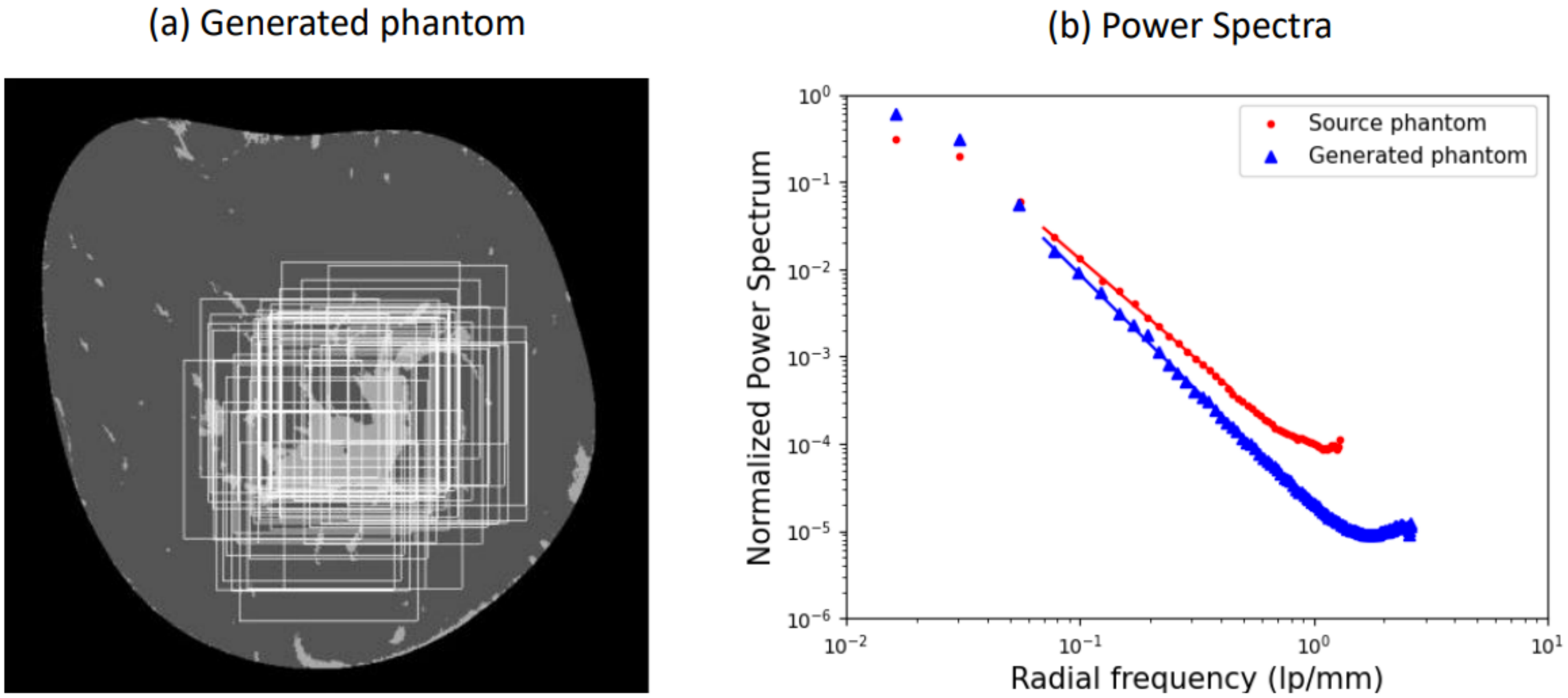

Power spectra for a generated and a source phantom. A slice from the generated phantom (a) is shown with sample patches (squares) from multiple elevations overlaid. Log-log plots of normalized noise power spectra (b) are well fit by power-law functions over the range of frequencies dominated by anatomical variability.

# Deep Learning-Driven Development of Patient-Specific Digital Twins with Automatic Quality Control

Lavsen Dahal, Mobina Ghojoghnejad, Dhrubajyoti Ghosh, Yubraj Bhandari, Fong Chi Ho, David Kim, Fakrul Islam Tushar, Ehsan Abadi, Ehsan Samei, Joseph Lo+, William Paul Segars+

Center For Virtual Imaging Trials
Duke University

## BACKGROUND AND PURPOSE

Patient-specific phantoms improve virtual imaging trials by accurately representing the population at large. We present an automated phantom generation framework comprising 3 modules: (1) deep- learning segmentation of up to 140 anatomical structures, (2) volumetric- and statistical-based quality assessment to reject poor segmentation masks, and (3) generating 3D polygon mesh representations. External validation yielded DSC for selected structures as follows: bladder 0.87 clavicle 0.92, femur 0.94, gallbladder 0.80, heart 0.85, kidneys 0.92, liver 0.97, lungs 0.97, pancreas 0.83, pelvis 0.95, ribs 0.81, scapula 0.89, spleen 0.97, sternum 0.88, and stomach 0.89. The segmented masks were used to define patient-specific phantoms represented as 3D polygon meshes. We have designed 2500+ new anthropomorphic phantoms.

## INTRODUCTION

Virtual Imaging Trials (VIT) offer cost-effectiveness, speed and scalability in the evaluation and optimization of existing medical imaging technologies compared to traditional clinical trials [1-4]. VIT requires a virtual patient population that mimics the anatomy and physiology of the patient.

Computational anthropomorphic phantoms are used to create the virtual patient population. These simulated phantoms are expected to be realistic to be indicative of true population regarding organ volume, boundary, tissue properties, and blood flow.

There have been several attempts in building successful anthropomorphic phantoms for this purpose. Full body phantoms were build based on the Visible Male and Female datasets [6] for adult male and female [5] . Computational phantoms of the male and female newborn patients were defined using Nonuniform rational B-spline (NURBS) surfaces [7]. Sixty adult and ninety pediatric computational phantoms of varying age, height, and body mass percentiles representative of large population have been released in last decade [8, 9]. However, the current phantom library sample size is inadequate for use in virtual imaging studies. Several studies have shown that sample size in thousands is required to improve the reliability of the virtual imaging trials [10-12]. However, the generation of computational phantoms has faced significant challenges due to the painstakingly slow and resource-intensive process of manually segmenting organ boundaries. This bottleneck has impeded the expansion of phantom libraries that can effectively represent the population at large.

In this work, we propose an automatic phantom generation framework that can generate a large number of phantoms with improved quality and realism and publicly release 1000 patient-specific digital twins of diverse age, race, sex, and body habitus. We use the deep learning-based algorithm to segment up to 140 structures from full-body Patient CTs. Automatic quality control module is proposed that can flag cases of segmentation failures using volume, anatomical knowledge, and statistical Gaussian Mixture Model. An interactive web application is publicly released for displaying phantoms. The phantoms are simulated using an in-house developed virtual scanner for underweight, healthy, and obese category patients and the efficacy of these phantoms for virtual imaging trials is demonstrated.

## METHODS

We used public and private models to train, validate and test segmentation models. The private dataset from Duke University Health System is used for creating anthropomorphic phantoms. The nn-UNet [13] model is subjected to quality control for selecting favorable cases, which are subsequently employed in constructing phantoms which can be simulated via scanners to facilitate virtual imaging trials. Four training datasets with a total of 5527 images were included in our study. The public dataset TotalSegmentator [14] and a private dataset from Duke University Medical Center (DukeCT) was used for model development and validation, and AMOS [15] and XCAT [7] were used for the evaluation of segmentation model.

A deep learning-based algorithm [13] is trained for the purpose of segmenting using ground-truth labels from public models and pseudo-labels obtained from other public and private models, allowing for more efficient utilization of available labeled data. We train a total of four models, each specializing in segmenting different sets of structures. The first model focuses exclusively on bone segmentation, encompassing a set of 62 classes. The second and third models are designed to segment 61 and 17 classes, respectively, covering a diverse range of anatomical structures. To create accurate ground truth labels for the training set, we employ a combination of strategies. We leverage the labels from the publicly available dataset called TotalSegmentator. Additionally, we utilize a combination of private models and public models to generate automated labels. This approach allows us to expand the training data and incorporate a wider range of anatomical variations and complexities. Consequently, our training methodology is supervised learning using pseudo-labels.

## RESULTS

The segmentation model was validated on one public dataset (AMOS) and a private dataset (XCAT). A comparison was made between our model, a publicly available model called TotalSegmentator, and a commercial model. The results are shown in box plots (Figure 1), indicating that our model performs similarly to the public models. However, the commercial model sometimes failed to produce segmentation results when the structure’s field of view was partial, or the commercial model lacked confidence based on quality control measures. To ensure a fair comparison and respect the commercial model’s design choice, cases with a Dice similarity coefficient (DSC) of 0 were removed for the private model in the box plot analysis. The box plot represents the reference standard mask available for the evaluated structures. It should be noted that not all models could segment all structures; hence, not all structure DSC values could be reported.

In our study, we implemented an anatomical knowledge-based quality control method for bone segmentation, leveraging bone symmetry as a criterion. This approach, albeit efficient in identifying poor segmentation instances, may introduce a systematic bias by potentially rejecting cases where bones are more than 50% dissimilar in volume. Despite this limitation, the method proved effective for preliminary screening: out of 3648 initial cases, 3581 images or 98% met the symmetry criterion. Images failing to meet the threshold were flagged for subsequent re-evaluation, pending improvements in the bone segmentation model.
We obtained a dataset of 3581 images from the first anatomical knowledge-based Quality Control set. This initial phase was followed by performing basic volumetric evaluations of different organs. Our additional analysis criteria called for the removal of any CT scan in which over 25% of the anatomical structures were missing due to improper or insufficient scanning. Utilizing a volume threshold technique—that is, finding zero volume instances in the organs under examination—made this determination easier. A refined data set of 3435 CT images was produced because of this filtration process.
Following that, a thorough statistical Quality Control (QC) phase was applied to these images, focusing on key organs such as the liver, lungs, stomach, spleen, pancreas, heart, gluteus, and esophagus. This QC step reduced the dataset to 2561 images by locating and eliminating outliers within these structures.
We present the demographic composition of phantoms built based on data from a single institution, reflecting the patient demographics of that health system. The population is primarily White, followed by Black, with a smaller representation of Asian individuals and other races categorized as 'Other'. Additionally, the sampling from PET/CT scans introduces a systematic bias, with a predominance of male subjects, likely attributable to the higher incidence of prostate cancer imaging.

## CONCLUSION

We build computational phantoms using the fusion of deep learning techniques stringent quality control and announce their public release. To offer the world a glimpse into the realm of these digital marvels, we have designed an interactive web application with the power to categorize and display phantoms by age, sex, and race. The applicability of these phantoms for conducting virtual imaging trials is demonstrated, highlighting their versatility across a broad spectrum of applications. This versatility underscores the potential of our phantoms to revolutionize virtual trials and beyond, offering a wide range of possibilities for research and development in the field of medical imaging.

## ACKNOWLEDGEMENTS

This work was funded by the Center for Virtual Imaging Trials, NIH/NIBIB P41-EB028744, and NIH/NCI

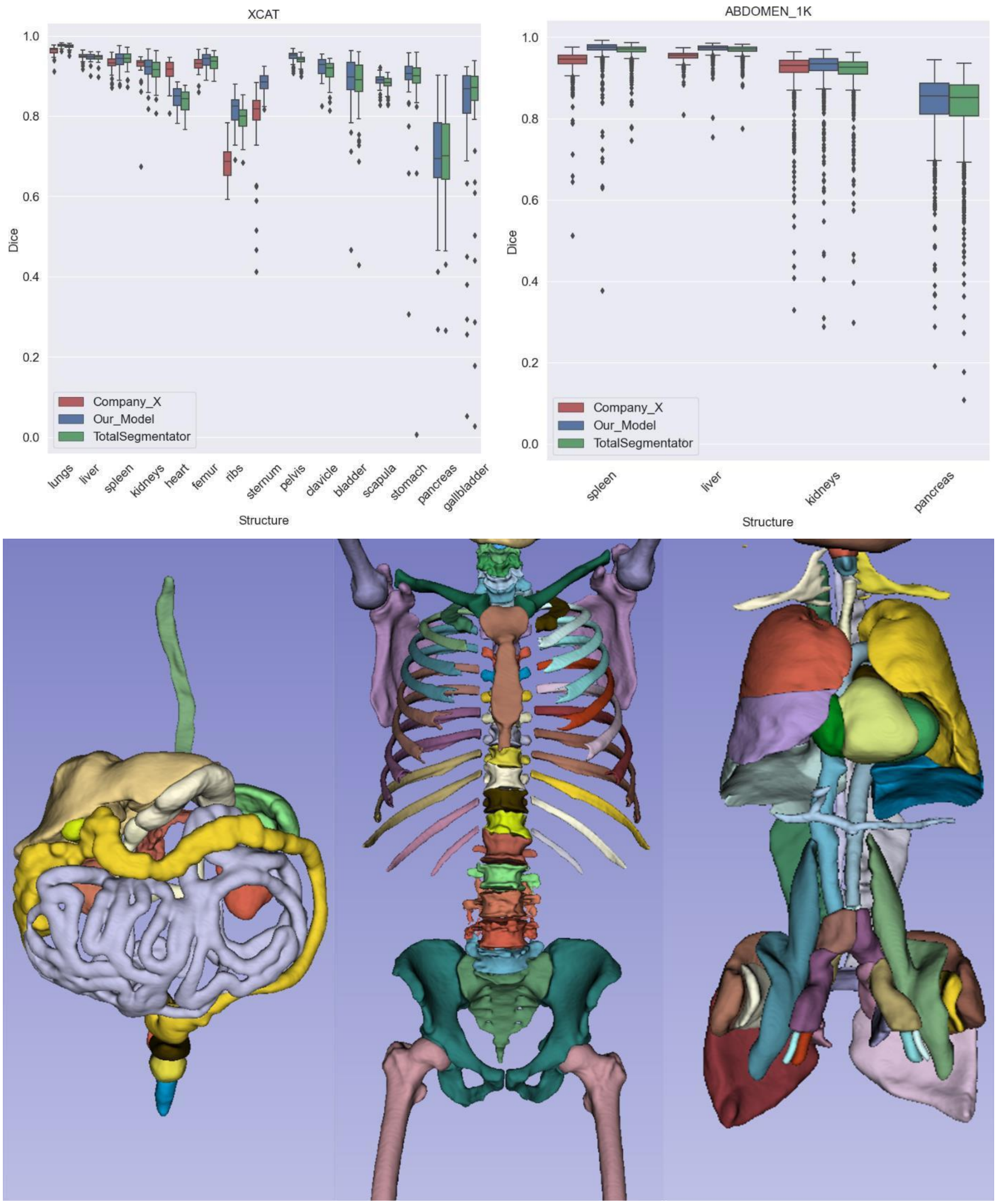

*Figure 11 Box plots for DSC comparison for a private and a public dataset for multiple structures (top) and 3D rendered segmentation masks obtained from our model (bottom).*

# Computational organ modeling

Monday, April 22, 2024

11:30 am - 12:10 pm

Chaired by Ann-Katherine Carton & Paul Kinahan

# Library of Realistic, Image-based 4D Beating Heart Models

Ethan Malin, Can Ceritoglu, Tilak Ratnanather, Ehsan Samei, and Paul Segars

## BACKGROUND AND PURPOSE

To create a population of detailed, anatomically variable 4D beating heart models for medical imaging research.

## METHODS

We obtained several sets of 4D CT data from healthy patients selected from the PROMISE national clinical trial. Each patient dataset was electrocardiogram-gated and contained 10-12 time frames over the cardiac cycle. The 4D data for each patient was imported into the Simpleware ScanIP software (licensed by Synopsis). The AS Cardio tool automatically segmented the four chambers of the heart, the left ventricular myocardium, the great vessels, and the coronary arteries from each time frame of the data as well as output 17 landmarks corresponding to major features of the heart. The segmentations from frame 1 of each dataset were used to create the initial instance of the heart with each structure defined as a polygon mesh via a marching cubes algorithm. Given the landmarks and segmentations for each frame, the MC-LDDMM mapping algorithm was used to calculate the frame-to-frame motion of the heart. The 3D motion fields were applied to the cardiac model from frame 1 to make it beat, creating a time-changing mesh model. Cubic spline curves were fit to the time changing locations of each vertex point of the meshes creating a 4D continuous model from which any number of time points can be sampled over the cardiac cycle. Example heart models were imported into whole-body XCAT computational phantoms and imaged with the DukeSim CT simulator, under various imaging parameters, to demonstrate the utility of the models

## RESULTS

The image-based cardiac models accurately mimic the twisting, contracting motion of the heart for various anatomically variable subjects. When combined with DukeSim, realistic virtual cardiac imaging data can be produced for research.

## DISCUSSION

The library of realistic cardiac models can provide a vital tool for 4D cardiac imaging studies.

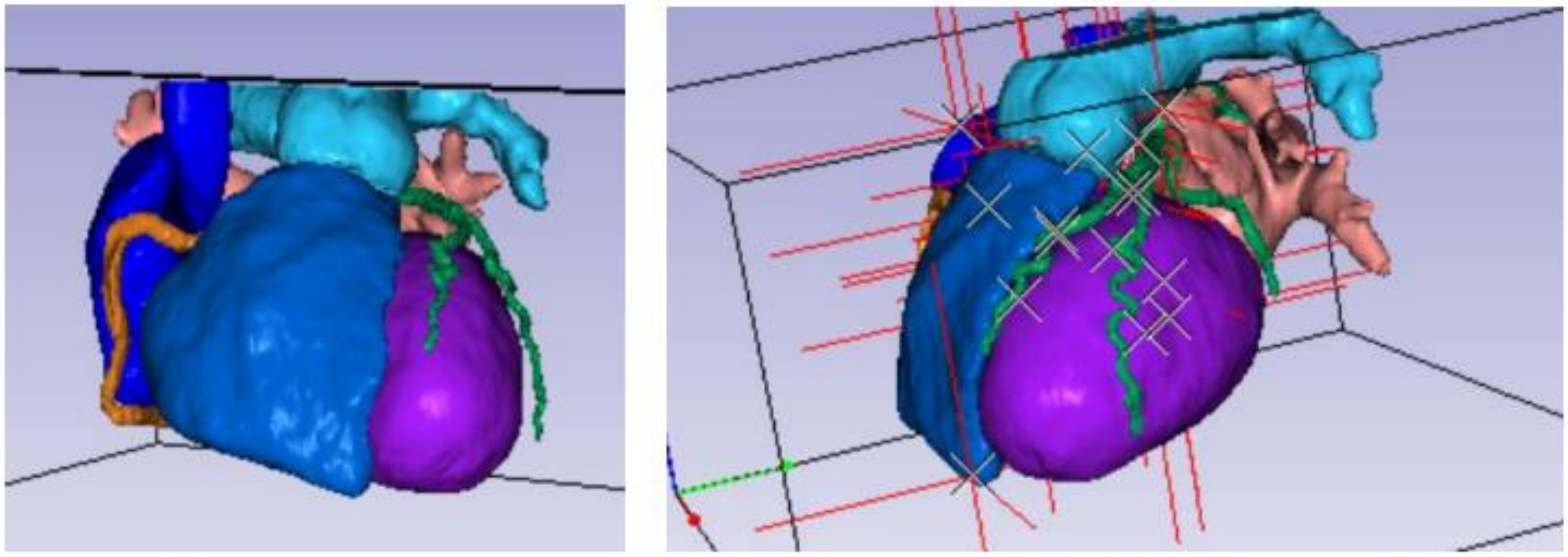

*Fig. 1. (Left) Sample view of the 3D volume segmented by the Simpleware AS Cardio software from the first phase of a patient 4D dataset. (Right) Rotated view of the same phase with the addition of the 17 landmarks that Simpleware outputs.*

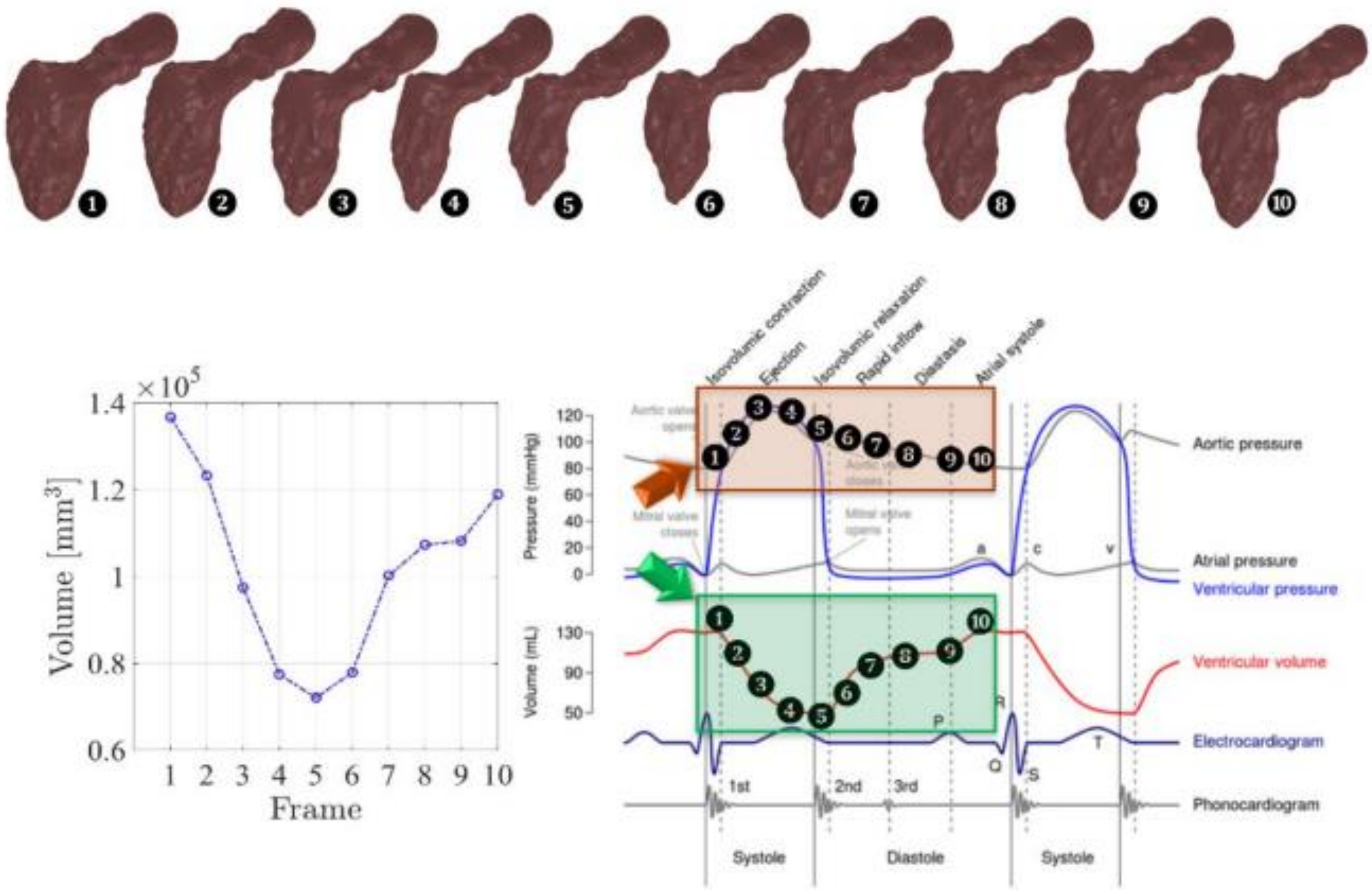

*Fig. 2. Motion of the left ventricle over the cardiac cycle. Time curves can be fit to the time-changing locations of the mesh vertices to define a continuous 4D model. This was done for each segmented structure from each patient to create a library of 4D anatomically variable beating heart models.*

# New features to improve realism in Perlin Noise-based digital breast phantoms

Magnus Dustler[1,2], Hanna Tomic[1], Anna Bjerkén[2], Predrag Bakic[1]

1. Diagnostic Radiology, Department of Translational Medicine, Lund University
2. Medical Radiation Physics, Department of Translational Medicine, Lund University

## BACKGROUND AND PURPOSE

Many tasks in medical imaging optimization are difficult or impossible to carry out with the use of clinical data, either because they would require multiple repeated image acquisitions of the same human subject, or because they require control of parameters which are unavailable to the researcher, such as the size and location of a tumour. Realistic software phantoms can be an option for such tasks. Software phantoms allow otherwise unavailable imaging or anatomy related parameters to be varied and their effect on image quality estimated, enabling optimisations that cannot be done using clinical data alone.

Digital phantoms created through a variety of means are currently in use in breast imaging research[1]. Digital breast phantoms based on the Perlin Noise algorithm – a form of gradient noise – has the potential to realistically represent both the appearance and variability of real breast tissue. The algorithm properly known as *Noise*, was first described by Ken Perlin in 1985[2]. It and its derivatives have since become ubiquitous in computer graphics and special effects. It is a flexible algorithm which generates smoothly and continuously varying structures of a set spatial frequency, in any dimension. We have previously described the basic method of using Perlin noise to generate breast tissue[3-6]. The realism of the resulting phantoms have been verified by expert breast radiologists[7]. This study describes new features that we are currently implementing in order to improve the appearance and realism of phantoms generated with the method.

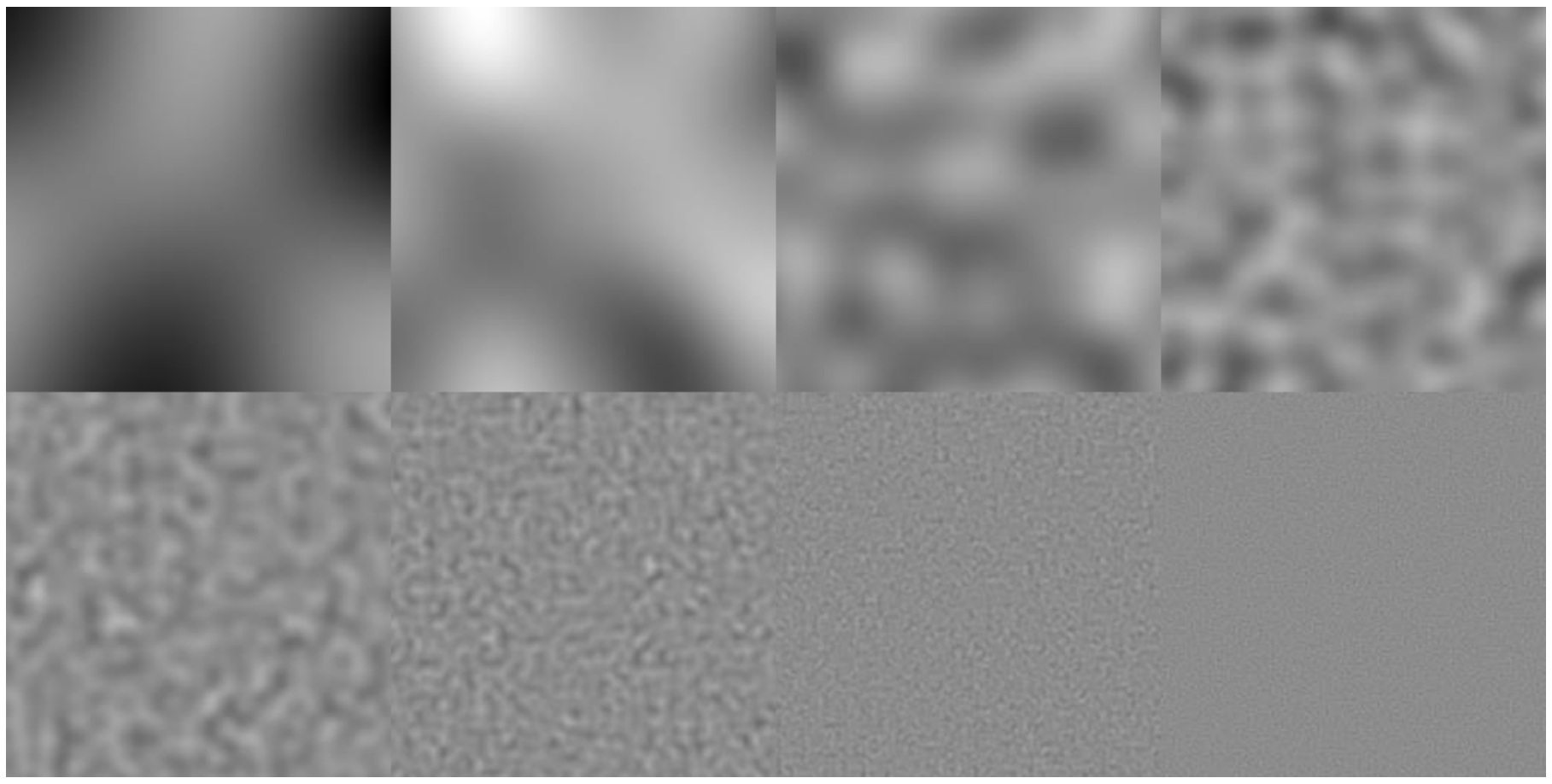

*Figure 12 Perlin Noise building blocks of different frequencies*

# METHODS

The basic Perlin Noise algorithm generates smoothly varying random structures of a set frequency or scale, controlled by a set of randomly selected gradients (Figure 1). Our in-house developed method uses a fractal 3D-implementation of the Noise algorithm. It combines a set of noise volumes with increasing frequency, providing structures with features of a range of different scales. By applying perturbation functions to the combined volume, a ridged appearance approximating fibrogladular breast tissue is achieved. Basically, this takes one minus the absolute value of the noise (which normally varies from -1 to 1) and increases the slope of the noise function to create sharper transitions. Figure 2 shows these steps for a sample volume.

From this state, a threshold value between 0 and 1 is applied, binarizing the noise distribution and creating strand-like features of a specific thickness. To capture the variation inherent in breast tissue, the complete phantom volume is built up by combining several volumes of Perlin Noise with different thresholds in order to get a range of differently sized features. Separate thresholds and other parameters is used to generate different types of tissue, e.g. dense and adipose. Different tissues can then be mapped to different attenuation properties for image generation. The finished Perlin volumes are then mapped to realistic breast shapes and global distributions of dense and adipose tissue.

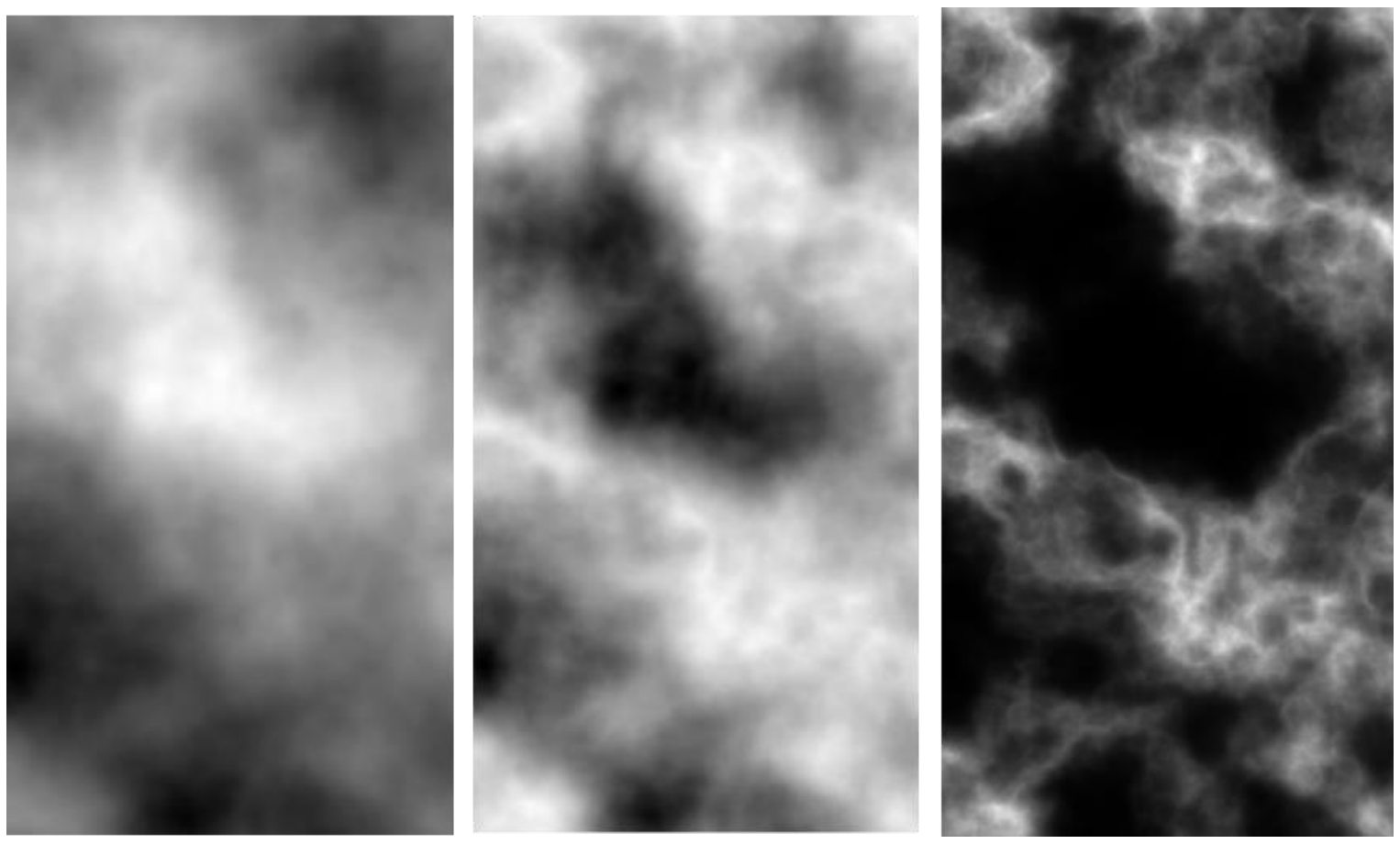

*Figure 2 From left to right: 1. A combined Perlin volume of several (16) constituent noise volumes of different frequencies. 2. The absolute value of this volume, showing a ridged appearance of the noise, and 3. The previous noise volume raised to the power of ten to create sharper slopes/ridges.*

## Blending dense and adipose tissue

By partly reusing the same set of random gradients, different tissue can be more realistically blended together. This could include tumours, that could be made to correlate with the appearance of the surrounding tissue.

To blend more adipose and more dense sections of the breast together, Perlin Noise was generated as described above, with different combinations of parameters for dense and adipose tissue. By using the same basic set of random gradients for both adipose and dense tissue, structures will continue between dense and adipose compartments and not stop abruptly at compartment borders. However, by also varying the frequency content by cutting lower frequencies from the adipose tissue, we can subtly change the appearance of the Perlin structures so that they appear to more naturally blend from dense fibrous strands into less dense structures in the adipose tissue. Dense and adipose regions of the breast were assigned by generating a simple fractal Perlin volume of the same size as the phantom volume, thresholding it into two regions and assigning one of them as dense and one as adipose.

## Varying density over time

Breast tissue changes in appearance over time, and the most typical age related change is that density decreases, both locally and globally. Simulating this is important in order to realistically simulating e.g. repeat screening rounds of the same population. Our tissue simulations are well suited to this as they can easily be modified by changing threshold values and distributions while maintaining the overall shape and appearance of the simulated tissue. By gradually increasing the thresholds that control strand thicknesses and dense area extent, a gradual change in density occurs. Local density changes by varying strand thickness, global density changes by varying the extent of the dense area.

# RESULTS

Figure 3 simultaneously illustrates the simulation of realistic blending of dense and adipose tissue and density changes due to age. Note that for clarity a simple 3D block of tissue is used, with no realistic breast shape applied.

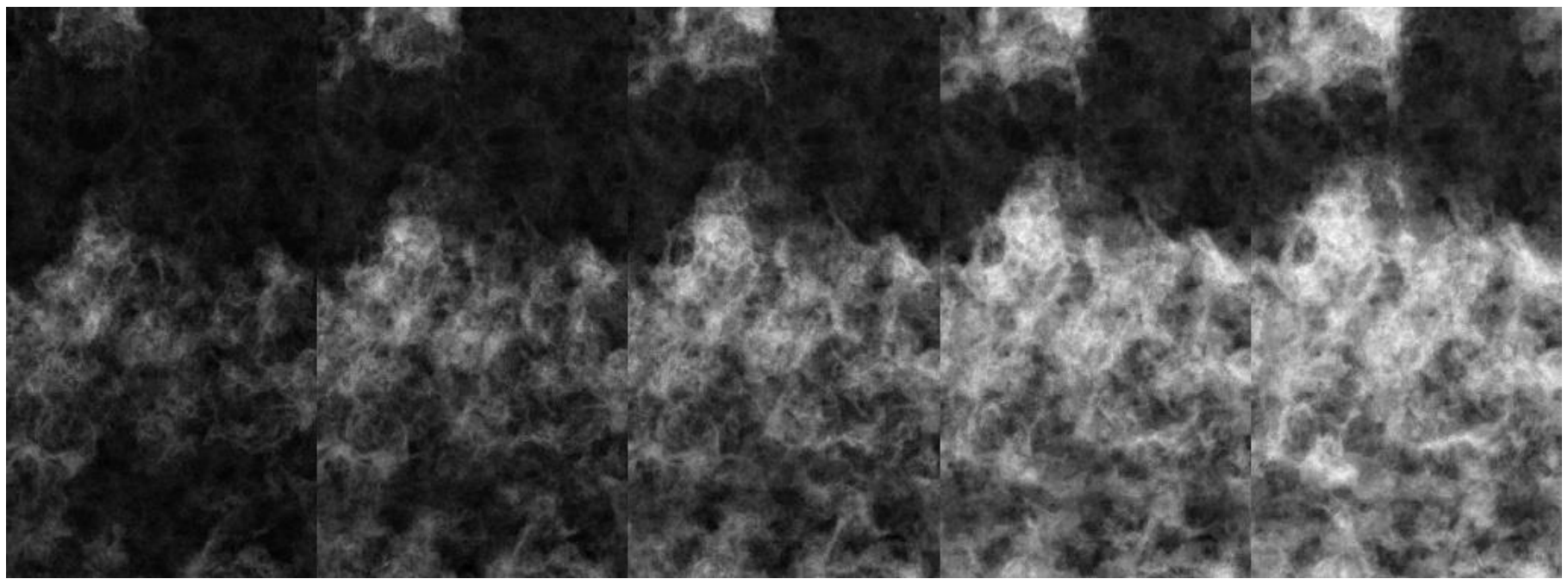

*Figure 3 Blended dense and adipose tissue, going from more dense (right) to less dense (left). Note that both the dense and adipose tissue changes, and that both the local appearance and global distribution of the dense tissue alters over time. Also note the way in which fibrous structures continue across dense/adipose tissue borders and naturally blend together.*

# DISCUSSION

Previous studies have shown that breast phantoms generated using our method can be made to realistic enough for radiologists to sometimes mistake them for real breasts. The added possibility to gradually change the appearance of the breast tissue so as to approximate age-related changes is important especially for virtual clinical trials of screening programs, as the appearance of a breast cannot be expected to realistically remain consistent over time.

Validation of the new imaging features through human observer studies are needed and are currently in the pipeline. A limitation of the current simulation model is that it lacks directionality. In order to make structures look as if they are emanating from an imagined nipple like how the ducts of a real breast would, in future work we plan to control the gradients of the low frequency noise.

## CONCLUSION

Breast phantoms generated using our Perlin Noise-based method are flexible and easily adaptable. A more realistic blending of dense and adipose tissue and gradual density changes over time as demonstrated here can potentially add to that realism.

# The Virtual Kidney

David Cox, Paul Segars, Ehsan Samei

## BACKGROUND AND PURPOSE

Computational phantoms can serve as virtual patients in medical imaging simulations, or virtual imaging trials, to conduct studies where it is difficult to impossible to use real patients. They allow repeated imaging of the same subjects under various conditions to evaluate and optimize imaging devices and techniques. We develop the XCAT phantoms that are widely used for this purpose. These phantoms, consisting of adults and pediatrics at different ages, heights, and weights, were developed using low-resolution PET-CT imaging data. High resolution details were added to each phantom based on anatomical templates derived from the Visible Human imaging data from the National Library of Medicine (NLM). The Visible Human anatomical images have an in-plane resolution of 0.33mm. With the higher resolution, it is possible to create more detailed vessels and inner structures for the organs that can be mapped into each different XCAT anatomy. Figure 1 shows the current detail, derived from the NLM data, present in the kidneys of an example adult male XCAT phantom.

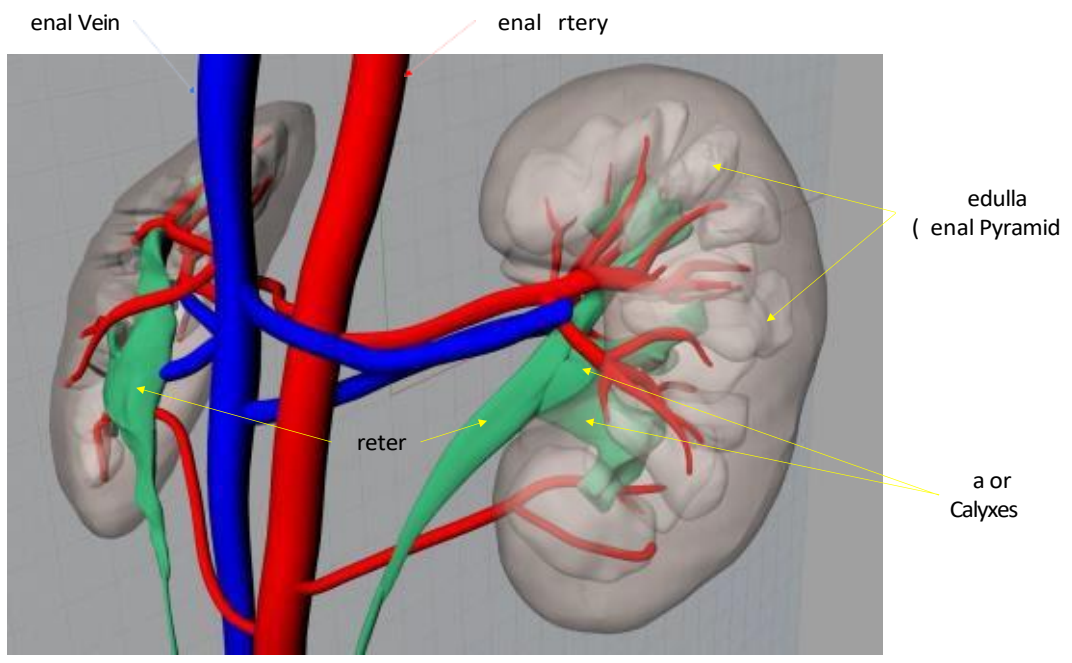

*Figure 1. A model of human kidneys for an adult male based on 0.33mm resolution images.*

Even higher resolution data is now coming available via the Human Organ Atlas project, https://human-organ-atlas.esrf.eu/. For this, hierarchical phase-contrast tomography (HiP-CT) is being used to image excised whole organs at a resolution of 25 microns and selected regions of organs at a level of a few microns. The images from the Human Organ Atlas offer unprecedented detail down to the cellular level for whole organs such as kidneys as well as the brain, liver, lungs, etc. Our goal is to use these images to construct a new generation of phantoms with ultra high-resolution detail for use in simulated imaging trials. The specific focus of this work is the kidney, analyzing the HiP-CT data for the first time to see what structures can be segmented via thresholding and the use of the marching cubes algorithm to build 3D models.

## METHODS

The marching cubes algorithm (Lorensen and Cline) produces isosurfaces representing the boundaries of objects. In our case, we want to visualize the various structures within a kidney. The specific structures that can be visualized from HiP-CT images depend on the tissue types and their x-ray attenuation. Marching cubes finds an isosurface that delineate objects within a threshold range. By analyzing HiP-CT images with marching cubes for different ranges of intensity values, we produce isosurfaces that delineate various structures. Thus, marching cubes allows us to model the different kidney structures.

We implemented software in Python to construct isosurfaces using the scikit marching cubes implementation. We utilized the LADAF-2020-31 image set of a complete kidney available from the Human Organ Atlas (Tafforeau). The scikit implementation produces isosurfaces that are represented as 3D triangular mesh models. Different models were obtained by specifying different intensity values (referred to as the level of interest parameter in the scikit documentation) as an input parameter to the marching cubes algorithm.

The image set of the kidney consisted of 4282 JPEG2000 images, each 3287 × 2215 pixels in the X and Y dimensions. The dataset covered the whole organ with an isotropic resolution of 25 microns. To visualize the anatomy as 3D models, the images were stacked in the Z direction. Stacking all 4282 images produced a 3D model of the entire kidney. To see inside the kidney, fewer images were stacked. For example, stacking 2141 images allowed us to produce a 3D model of half a kidney allowing us to more easily see interior structures. In some cases, we created slices through the kidney by stacking smaller numbers of images. For example, a slice would be produced from a stack of 100 images. (Note, when the images are stacked, each “pixel” acquires a Z-value. The term “voxel” is used to refer to these three-dimensional values).

The resulting 3D meshes are false colored to help visualize structures. In the typical case a triangle bisects a cube formed by four voxels. The colors of the triangles in a mesh depended on the actual intensity values of the nearest voxels. In a common case a triangle bisects a cube and its vertices are interpolated to be located between voxels. The final color of the triangle is then calculated by assigning colors to the voxels and interpolating those values.

In some cases, the images were pre-processed using contrast stretching. Contrast stretching doesn’t alter the final 3D mesh models. However, it does allow a different false coloring of the triangles in the mesh making. When intensity values fall within a narrow range, the triangles tend to have one color. Through contrast stretching, the triangle can be represented with a wider range of colors to assist in visualization.

ltering the “level of interest” parameter to the marching cubes algorithm produces different isosurfaces and, therefore, different 3D models. The different 3D models allowed us to test segment the kidney into different tissue types and structures.

## RESULTS

The images are provided as JPEG200 files and intensity values are represented at 16-bit gray levels. Thus, gray levels vary from 0 to 65,535. These are normalized to values from 0.0 to 1.0 with 0.0 representing pure black and 1.0 representing pure white. The distribution of gray levels is depicted in Figure 2. Figure 2 consists of three distinct peaks representing at least three distinct populations of intensity values.

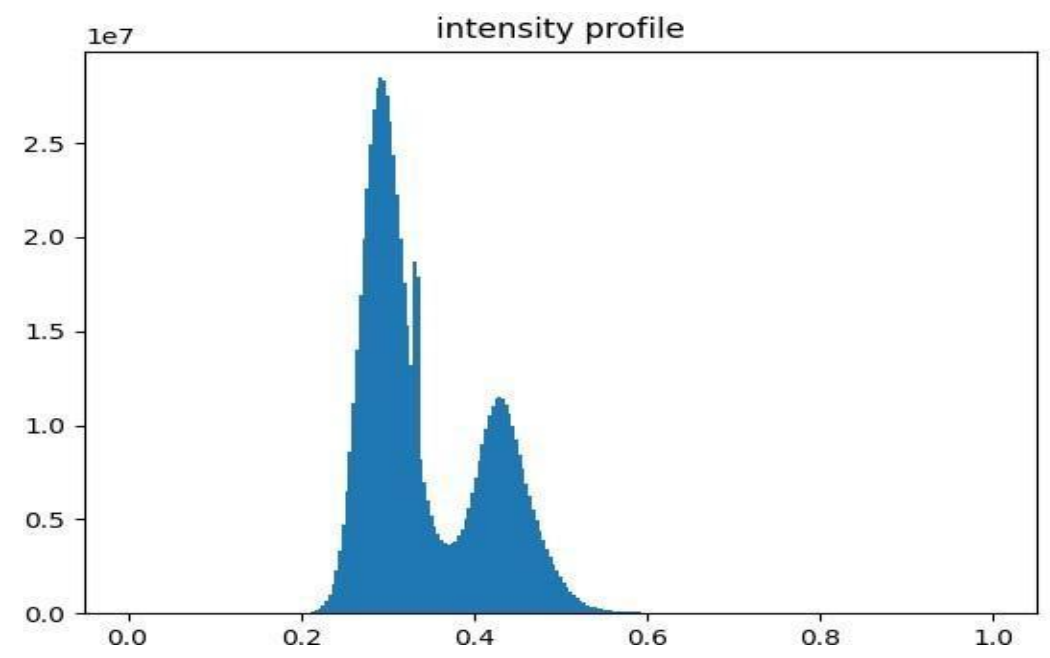

*Figure 2. Intensity profile for HiP-CT JPEG2000 images with intensities normalized to 0.0 to 1.0.*

Figure 3 illustrates a 3D mesh model of a slice of the kidney consisting of a stack of 101 images (specifically, images 1000 to 1100). The image on the top represents the entire slice. The orange structures correspond to blood vessels in typical anatomical diagrams of the kidney. The blue-white regions represent either tissues that have finer structures or, possibly, noise from the imaging process.

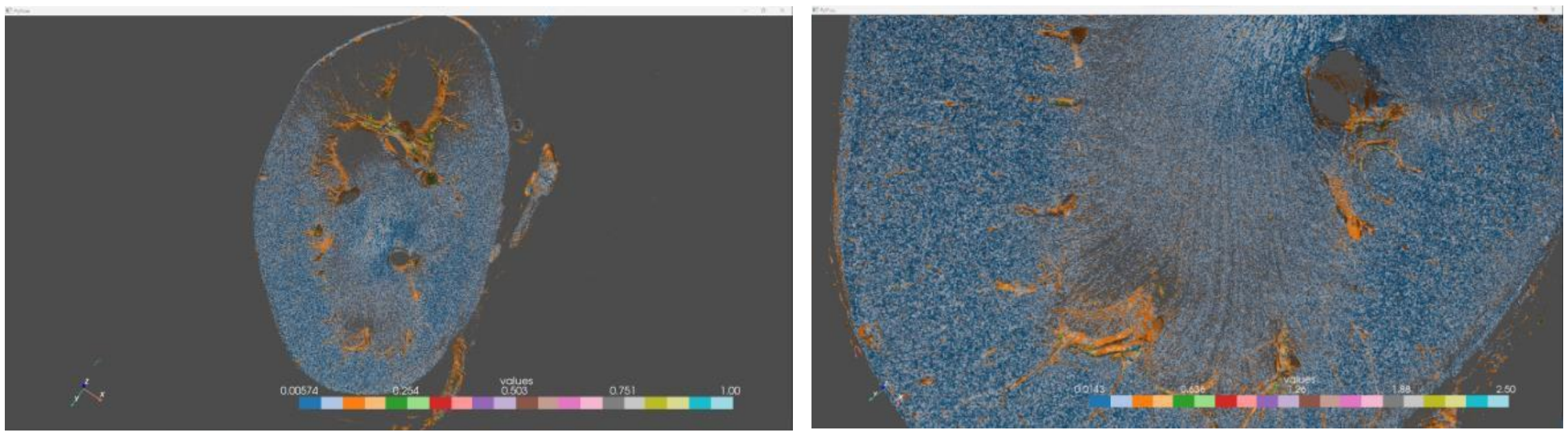

*Figure 3. Kidney mesh model for a "slice" of images 1000 to 1100 and an isolevel (or "level of interest") equal to 0.4. Top is the entire kidney and bottom is a close-up view revealing finer structural details.*

The kidney has arteries and veins that transport blood into and out of the organ. It also has renal vessels that direct urine away from the kidney and tissue consisting of a million nephrons and glomeruli that filter urine from the blood into the renal vessels. Figure 4 illustrates a 3D mesh model of the network of blood vessels segmented from the images with the isolevel set to 0.51.

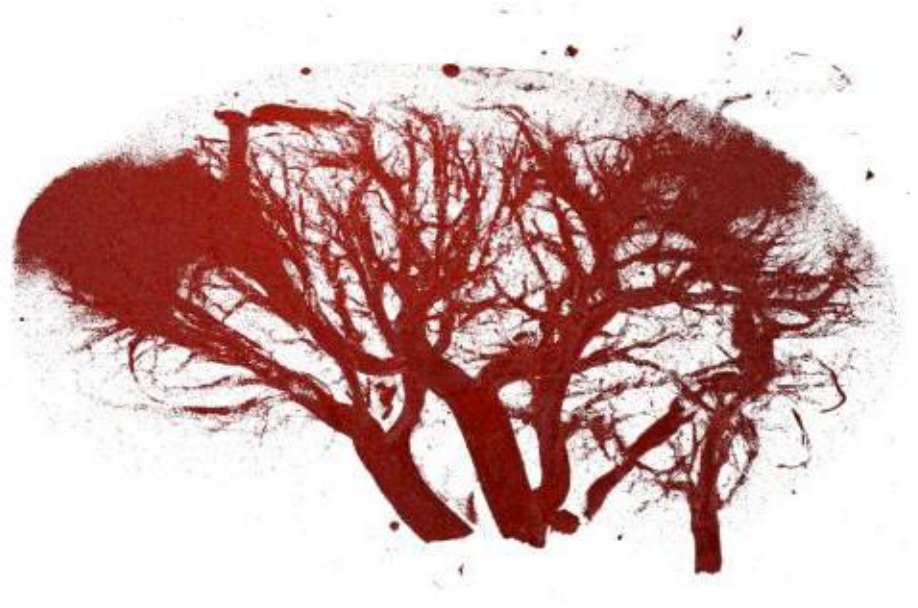

*Figure 4. Network of blood vessels visualized using the marching cubes algorithm with an isolevel of 0.51.*

Segmenting the renal tubules and urine collection system is problematic using thresholding. Figure 5 is an example of the isosurfaces created for an isolevel of 0.34. In the center of the kidney extending to the right are structures that correspond to renal tubules and ureters found in typical anatomical renderings of the kidney. However, a considerable amount of noise is also present. Separating the signal from the noise in these images remains an ongoing task.

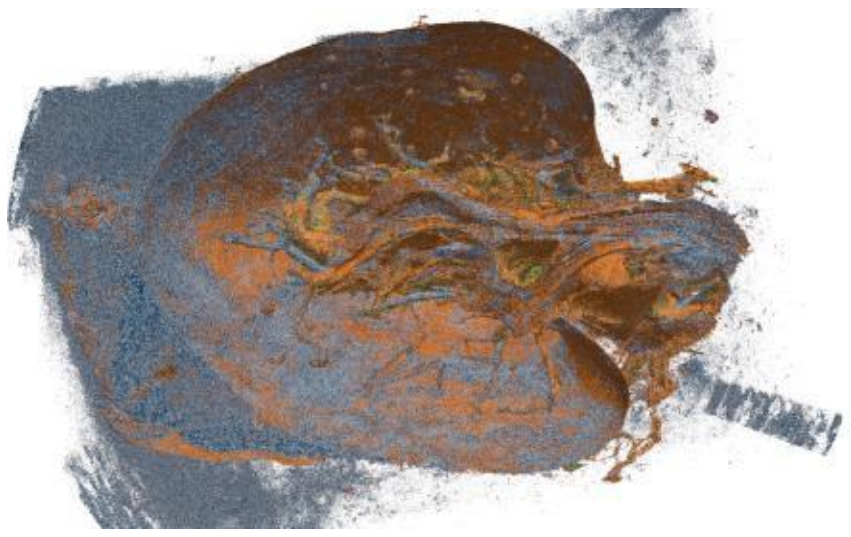

*Figure 5. Example of using thresholding to segment renal tubules and the urine collection system of the kidney.*

Figure 6 illustrates an example of segmenting the filtration system of the kidney – the one million nephrons and glomeruli in the kidney. While the isosurfaces generated in Figure 6 correspond to what one might expect from anatomical renderings of the kidney, it is hard to conclude that the isosurfaces in Figure 6 correspond to real anatomy or are the result of noise in the images.

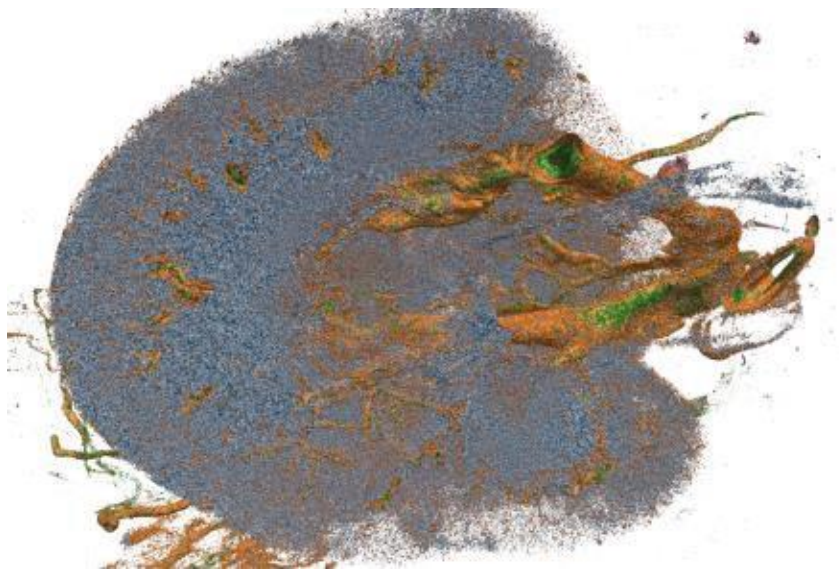

*Figure 6. Example of using thresholding to segment the filtration system of the kidney. The regions colored blue are possible sites of nephrons/glomeruli. However, more work remains to ensure that these isosurfaces are not imaging noise.*

## DISCUSSION

High resolution imaging such as HiP-CT allows us to resolve the whole organ at 25 micron resolution – roughly one fifth to one tenth the size of a glomerulus in the kidney. Our experiments illustrate the ability to analyze these images using algorithms such as marching cubes to create 3D models segmented into different highly detailed structures. Marching cubes fits surfaces to intensity bands within the image through thresholding. It gave us an initial view to see what is possible to define in the high-resolution images by separating intensity values. More work will investigate other segmentation algorithms to better define the interior structures of the kidney to very small detail.

## CONCLUSION

Technology now allows CT imaging of excised whole organs at the cellular and near-cellular level. This study represents our initial analysis of the data to see what structures the intensity values in the images define. Thresholding can be seen to bring out noisy representations of the vessels and structures within the kidney. nfortunately, there isn’t a clear separation of all the ob ects. Our next step is to establish a segmentation pipeline using a combination of manual, semi-automatic, or automatic techniques to better delineate structures to convert to 3D models. The 3D models will then be used to greatly enhance the anatomical detail in our series of XCAT phantoms for use in virtual imaging trials.

## ACKNOWLEDGEMENTS AND CONFLICT OF INTEREST

This work was supported in part by grants from the NIH (P41EB028744 and RO1EB001838).

# Computational models of diseases

Monday, April 22, 2024

2:45 pm - 3:25 pm

Chaired by Kristina Bliznakova & Joseph Lo

# Subject-specific airway modeling of patients with chronic bronchitis and small airway disease

Fong Chi Ho, W. Paul Segars, Ehsan Samei, Ehsan Abadi

## BACKGROUND AND PURPOSE

The geometry of the human bronchial tree plays a pivotal role in gas-exchange efficiency and dissipation, impacting factors such as airway resistance, ventilation, and gas flux. These geometric attributes, including the size and shape of airways, influence the airflow dynamics, leading to variations in gas exchange efficiency across different individuals. Advanced imaging techniques, such as high-resolution computed tomography (CT), have facilitated the quantitative analysis of these geometric factors, offering insights into their correlation with pulmonary function and disease states[1].

A geometrical airway attribute of clinical relevance is the homothety ratio, defined as the ratio of the child-to-parent branch diameter or length in the bronchial tree [2]. This ratio reflects the tapering and branching patterns of the airways, indicating the airway remodeling in pathological conditions. Previous measurements have shown a homothety ratio of 0.80 to 0.85 for normal human lungs, and about 0.77 in patients associated with the presence of chronic bronchitis (CB) and small airways disease (SAD), both of which are phenotypes of chronic obstructive pulmonary disease (COPD)[3]. A hallmark of COPD is the geometric changes in the airways that restrict airflow. Therefore, analyzing airway structure can improve the diagnosis and understanding of COPD. One example is a study that showed correlations between airway geometry (distribution of minimal inner cross-sectional area of airways, expressed by a fractal index) and airflow limitation and future body mass index reduction in COPD patients, as predicted by Hess– urray's law.[4, 5]

Structural changes in the bronchial tree as well as the physiological impairment of airflow have correlations with COPD GOLD level, i.e., *[6] a post-bronchodilator $FEV_1$/FVC (*forced expired volume in one second/ forced vital capacity*). A limitation of the $FEV_1$/FVC is predicted* from the population without COPD at the similar age, gender, and body composition as the patient. Thus, there is a need for quantifying the impact of airway changes on pulmonary function in a patient-specific manner which may be achieved through patient-specific airway modeling to [7, 8]. Despite the recognized importance of accurate COPD airway modeling, a notable gap has existed in developing models that incorporate physical airway-related measurements derived from patients to reflect the individual's specific airway geometry and pathological conditions.

This study aimed to develop a subject-specific airway anatomy model for individuals with diagnosed COPD, incorporating phenotype characteristics and $FEV_1$ measurements. The model is further applicable to virtual imaging trials towards personalized diagnostics and evaluation of treatment strategies.

## METHODS

### 1. Computational model

Ten patient-specific human models were generated based on the clinical cases from the COPDGene Phase 1with various COPD severities [9-11]. The homothety ratios in the airway diameter ($h_{d,\,COPD}$) and length ($h_{l,\,COPD}$) were measured per generation from the airways that were segmented from corresponding CT images. The segmented airway tree was grown asymmetrically based on the patient's measured $FEV_1$ and homothety ratios, and further informed based on the characteristics of CB and SAD from literature [12-14]. For the modeling SAD, mucus occlusions were inserted based on parametric response mapping (PRM) maps of non-emphysematous region [15, 16].

## 2. Airway modeling

Bronchial tree follows Poiseuille's law in an optimal airflow under resting conditions for the 6th generation and beyond. This infers a fractal-shaped tree structure with minimized energy loss within a confined volume[17]. The structure with a fractal dimension of 3 and a homothety ratio of 0.7937 aligns with the Hess–Murray law [5]. The air resistance across generations is proportional to the bronchiole length homothety ratio and inversely proportional to the fourth power of the corresponding diameter homothety ratio ($R \propto \frac{1}{h} \propto \frac{h_l}{h_d^4}$) [2]. *The total air resistance ($R_N$) from generation 1 to generation N is calculated using a homothety ratio $h_p$ per generation p and the initial air resistance $R_0$ in generation 0 as follows:*

$$R_N = R_0 + \sum_{p=1}^{N} \frac{1}{2^p} \frac{R_0}{(h_1 \times h_2 \times \ldots \times h_p)^3} \tag{1}$$

The airway geometry on ventilation follows $\Delta P_N = R_N \Phi$, *with* $\Delta P_N$ being the total air pressure drop in the airways from generation 0 to N, $R_N$ the air resistance from generation 0 to N, and *$\Phi$ the air flux in liter per second. The air resistance is inversely proportional to the air flux.* In airway diseases like SAD and CB in COPD, the variation in bronchiole length compared to non-diseased individuals is assumed to be neglectable compared to the variation of inner diameters of the bronchioles. By using $FEV_1$ and predicted $FEV_1$ as the air flux *$\Phi$ for COPD patients and healthy patients accordingly*, a ratio of *$\Phi_{COPD}$ to $\Phi_{healthy}$* can be calculated as follows:

$$\frac{\Phi_{COPD}}{\Phi_{healthy}} = \frac{FEV_{1,COPD}}{FEV_{1,predicted}} = \frac{R_{N,healthy}}{R_{N,COPD}} = \frac{R_0\left[1+\sum_{p=1}^{N}\frac{1}{2^p}\left(\frac{h_{l,healthy}}{h_{d,healthy}^4}\right)^p\right]}{R_0\left[1+\sum_{p=1}^{N}\frac{1}{2^p}\left(\frac{h_{l,COPD}}{h_{d,COPD}^4}\right)^p\right]} \tag{2}$$

*After simplification, the diseased airways can be modeled asymmetrically using different homotheties in length ($h_{l,1,COPD}$, $h_{l,2,COPD}$) and diameter ($h_{d,1,COPD}$, $h_{d,2,COPD}$) on each of the bifurcated bronchioles. Since* minor differences among the homothety ratios of individuals can induce a significant change in respiratory performance[2], *the airway-length homothety ratio, in the presence of COPD with CB and SAD, maintained a small variability (5%) as the non-diseased airways. The airway-diameter homothety ratio is deduced based on the following equation with clinically measured $FEV_1$. The homothety ratios in the first multiple generations ($N_{seg}$) were measured from the initial segmentation[13].*

$$\frac{FEV_{1,COPD}}{FEV_{1,predicted}} = \frac{\sum_{p=1}^{N_{seg}}\frac{1}{2^p}\left(\frac{h_{l,healthy}}{h_{d,healthy}^4}\right)^p + \sum_{p=N_{seg}}^{N}\frac{1}{2^p}\left(\frac{h_{l,healthy}}{h_{d,healthy}^4}\right)^p}{\sum_{p=1}^{N_{seg}}\left(\frac{h_{d,1,COPD}^4}{h_{l,1,COPD}}+\frac{h_{d,2,COPD}^4}{h_{l,2,COPD}}\right)^{-p} + \sum_{p=N_{seg}}^{N}\left(\frac{h_{d,1,COPD}^4}{h_{l,1,COPD}}+\frac{h_{d,2,COPD}^4}{h_{l,2,COPD}}\right)^{-p}} \tag{3}$$

The mucus occlusion in airways was included in the airway pruning and air resistance calculation to indicate the loss of gas-exchange functionality, realistically reflecting the $FEV_1$ measurement.

## 3. Quantitative Assessment

The percentage of airway pruning was quantified and compared between diseased and healthy airways. The modeled airways were validated by comparing the averaged number of terminal bronchioles per mL of lung *to in* vivo measurements of 28 explanted lungs scanned by high resolution micro-CT[18]. To confirm that the proposed models aligns with patient-specific spirometry measurements, the accuracy and variability of the values on the right-hand side of Equation (3) were evaluated by comparing them to the ground truth values of clinical spirometry measurements of $FEV_1/FEV_{1,\ predicted}$, as expressed on the left-hand side of the same equation, using the mean absolute error and standard deviation.

# RESULTS

**Figure 1** shows an example of the proposed airway modeling. **Figure 1a** demonstrates the CT airway segmentation of a patient diagnosed with 'mild' COPD (GOLD level 2 [6]. **Figure 1b** shows the *asymmetrical airways grown using the proposed method without airway-related disease and with variability of 5% in airway-diameter homothety ratios across generations of bronchioles ($h_l$ = 0.85±0.03 and $h_d$ = 0.84±0.04). **Figure 1c** illustrates the asymmetrical airways grown from the same initial airways with COPD characterization of reduced diameter by inflamed bronchus and bronchioles ($h_l$ = 0.84±0.04 and $h_d$ = 0.79±0.03).*

**Figure 2a** demonstrates the averaged percentage change in airway pruning across ten airway modeling using spirometry-induced homothety ratios. The averaged percentage of pruning gradually increases with the airway generation from the 9th generation, and prominently located in the peripheral lung region from the 16th to the 18th generations. This result coheres the pathological change of reduced airway diameter across generations due to the inflammation by CB and SAD[3]. **Figure 2b** shows the percentages of airway pruning across the lobes of the patients by comparing the modelling of diseased airways to the healthy airways.

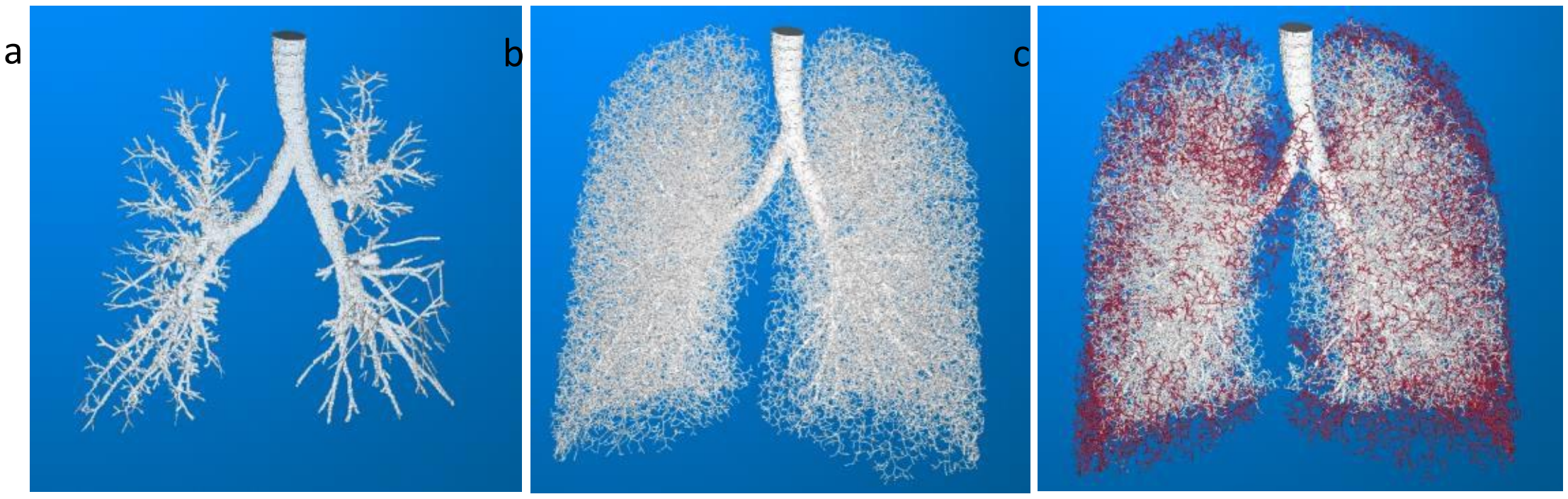

*Figure 1. (a) An example of initial airway segmentation from the patient COPDGene CT data with seven airway generations with CB and SAD. (b) The asymmetrical airways grown without airway-related disease. (c) The asymmetrical airways grown with chronic bronchitis characterization of reduced airway-diameter homothety ratio, resulting in airway pruning (red), including the gas-trapping region calculated from PRM.*

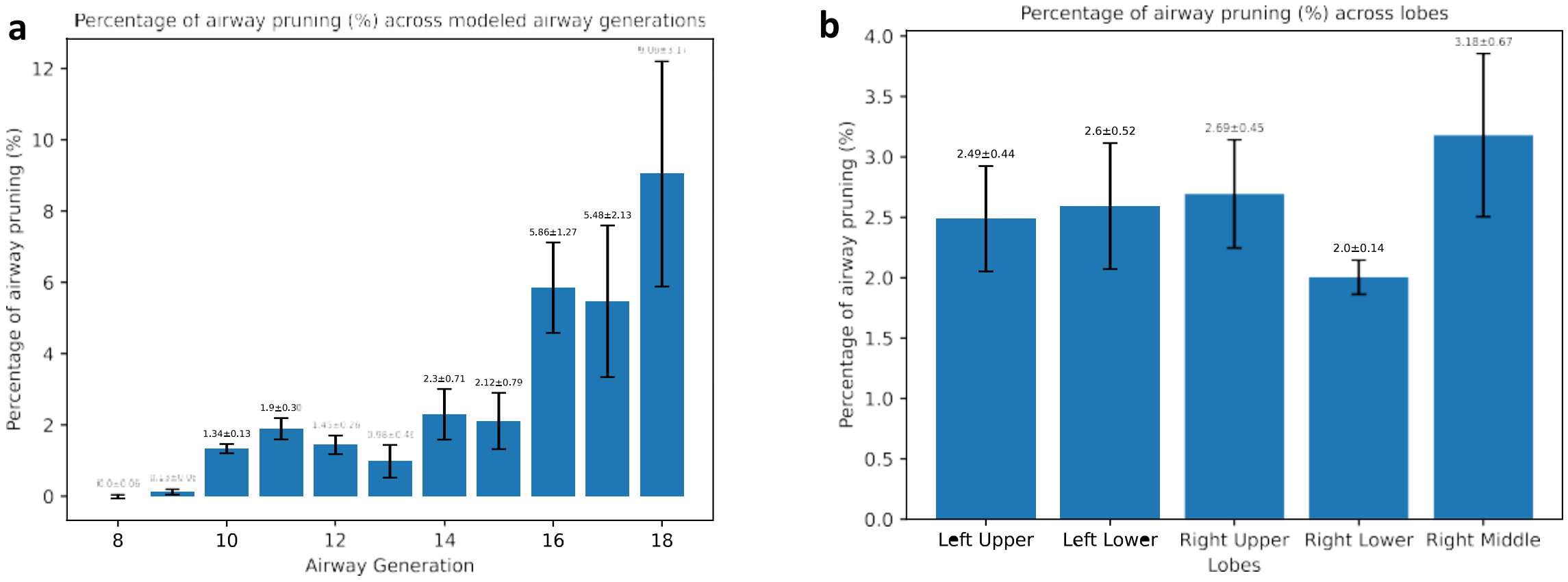

*Figure 2. (a)The percentage of airway pruning across modeled airway generations by comparing the number of airways per generation between a healthy individual and COPD patient. (b) The percentage of airway pruning across lobes by comparing the number of airways per lobe between the healthy individual and COPD patient.*

The average number of terminal bronchioles per mL of lung across ten patients was 3.50±0.15 which cohered to the quantity measured in micro-CT[18]. The ratio of modeled air resistance between COPD patients and healthy individuals with 5% variability was compared with the ratio of clinical spirometry measurement of $FEV_1/FEV_{1,predicted}$, resulting in a difference of 16.32%±6.77%.

## DISCUSSION

This study developed subject-specific COPD airway models that incorporate structural information from CT scans and homothety ratios from the physical airway-related measurements.

## CONCLUSION

This study moves beyond generic airway modeling and bridges the gap between one-size-fits-all modeling and patient-specific needs by developing realistic airway modeling for future virtual imaging trials. The novel methodology can provide sets of realistic virtual patients modeling different severities of COPD to investigate techniques for more accurate COPD diagnostics and management and personalized treatment strategies.

## ACKNOWLEDGEMENTS AND CONFLICT OF INTEREST

This work was supported in part by grants from the NIH (R01HL155293, P41EB028744, and RO1EB001838).

# In silico modeling of growing spiculated breast mass lesions

Miguel A. Lago, Vanday Bundu, Aldo Badano

Division of Imaging, Diagnostics and Software Reliability, OSEL, CDRH, U.S. Food and Drug Administration, Silver Spring, MD, USA

## INTRODUCTION

In silico models of medical imaging have the potential to study scenarios that are unattainable in normal circumstances and allow for less expensive and less time-consuming trials. Here, we focus on the development of models of growing spiculated breast mass lesions. Previously, our group developed a fully in silico clinical trial to study breast cancer detection [1] that replicated the results of a real clinical trial. This study was later expanded as a longitudinal clinical trial with a growing lesion [2-4] in which we characterized the visibility of breast masses at different growing stages. However, the question of how to model the malignancy of lesions still remains. It is known that spiculations is one of the most understood biomarkers for malignancy. In this research, we introduce a spiculation model for the previously reported lesion growing algorithm with varying parameters to control length, thickness, and density. To validate the malignancy appearance of these spiculations, we conducted a reader study with expert radiologists on breast images.

## METHODS

### 1. In silico breast imaging

We generated in silico breast mammograms using the VICTRE pipeline [5, 6]. An anthropomorphic model of the breast is created, including internal tissue composition (glandular, adipose, ligaments, etc.), and compressed between two plates. This breast phantom includes a list of candidate lesion locations based on the terminal duct lobular units. A Monte Carlo X-ray simulation based on MC-GPU [7] was performed, resulting on a Digital Mammography (DM) image for the corresponding compressed breast phantom. Figure 1 shows the steps of this in silico imaging pipeline.

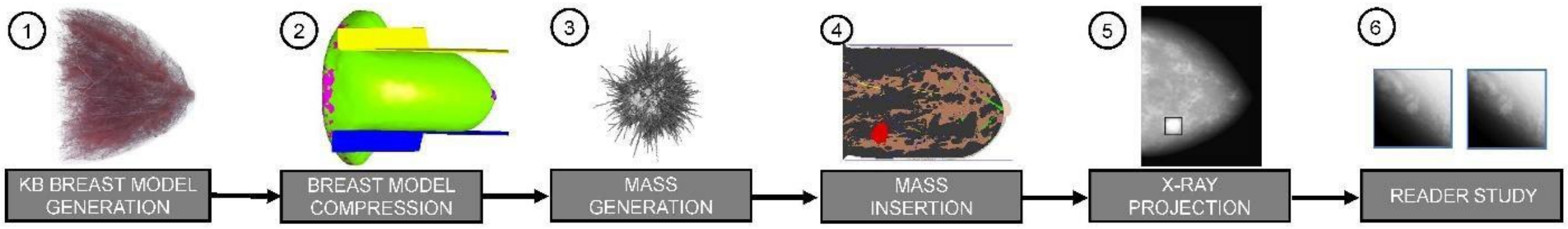

*Figure 1. Pipeline for the in silico clinical trial for breast imaging*

### 2. Lesion growing

The lesion growth model was presented by Sengupta et al. [2, 3] and simulates the internal growth of a breast mass tumor. The growth of these mass models is driven by the interstitial tumor pressure, the concentration of metabolites, and the stiffness of the local tissues. The initial location of the tumor growth is selected from candidate locations during the breast phantom generation. This location defines its final shape and size resulting in unique morphologies for each mass.

A mass lesion core was grown to an average size of 3-mm radius. For each lesion, we set spiculation growing parameters to generate three different levels based on that growing stage. The spiculation generation was controlled by parameters directing the number of spiculations, their initial thickness, and their maximum length. Spiculations grow from the perimeter of the given 3D mass core on an initial perpendicular direction as branches. Each growing step, these spiculations move on a similar path with randomized small direction changes, always moving away from the core. On each iteration, the spiculation branch has a small probability of splitting in two branches. Additionally, as the growing increases, the branch width decreases. A branch is considered dead when its thickness is 0 or the maximum number of iterations is reached. We created spiculations at three different conspicuity levels (L1, L2, and L3), by varying their thickness, length, and number of branches. Figure 2 shows a projection of the 3D model for the same mass lesion with spiculations at different growing stages on the largest spiculation settings (L3). Figure 3 shows an example of the DM projection of the same mass with the three spiculation conspicuity levels (L1, L2, and L3).

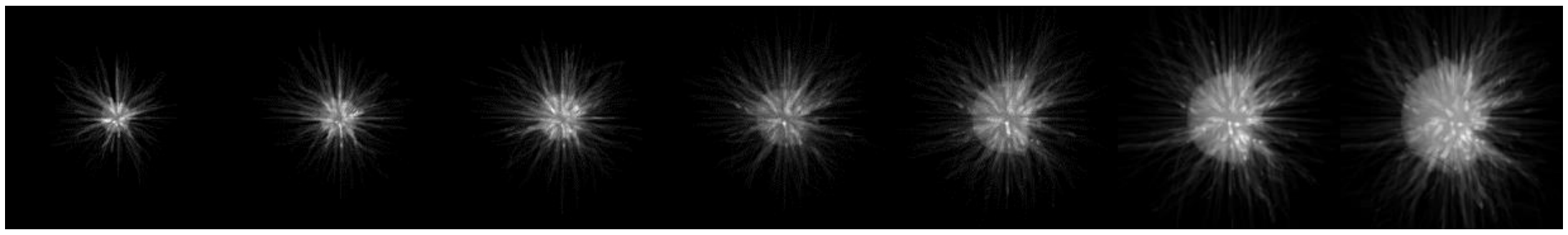

*Figure 2: Example of a projection of the 3D model of a single mass at different growing stages with corresponding growing spiculations at the most conspicuous level (L3). This model is later inserted on the phantom before the projection simulation.*

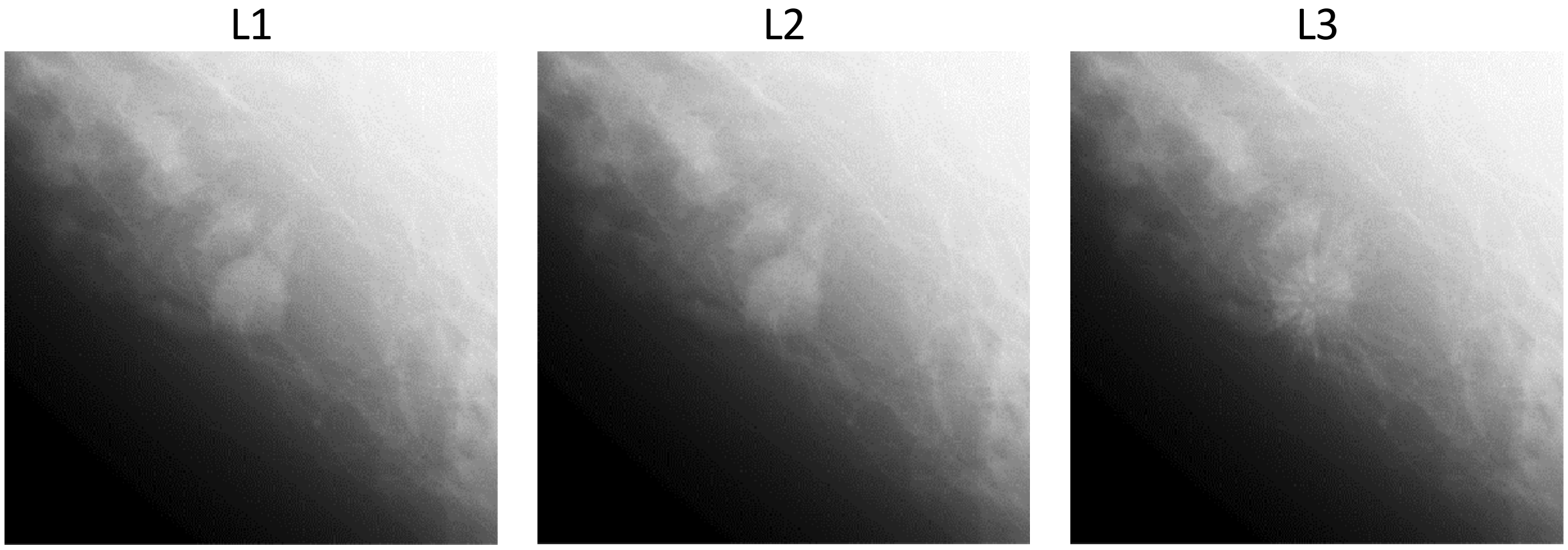

*Figure 3: Example of a single mass at the same growing stage generated with three degrees of spiculations. From least conspicuous spiculations (L1), to most conspicuous spiculations (L3).*

## 3. Reader study design

Ten different mass cores (realizations) were generated, each of them at the three spiculation levels for a total of 30 different images. A perceptual 2-alternative-forced-choice (2AFC) experiment was designed in which participants were shown two mass images next to each other and had to click on the “most malignant mass” from two images.

We conducted this experiment as part of the medical imaging perception lab during the European Society of Breast Imaging conference (EUSOBI) in 2023. Six radiologists, all of them experts on breast imaging, were recruited to participate in this study. The total number of pair-wise comparisons was 300.

## RESULTS

Figure 4 shows the results of the percentage of times the three spiculation levels were chosen over the other image. Radiologists selected the higher spiculation level (L3) as more malignant over the middle level (L2) and the middle level (L2) over the lower level (L1) with significant differences.

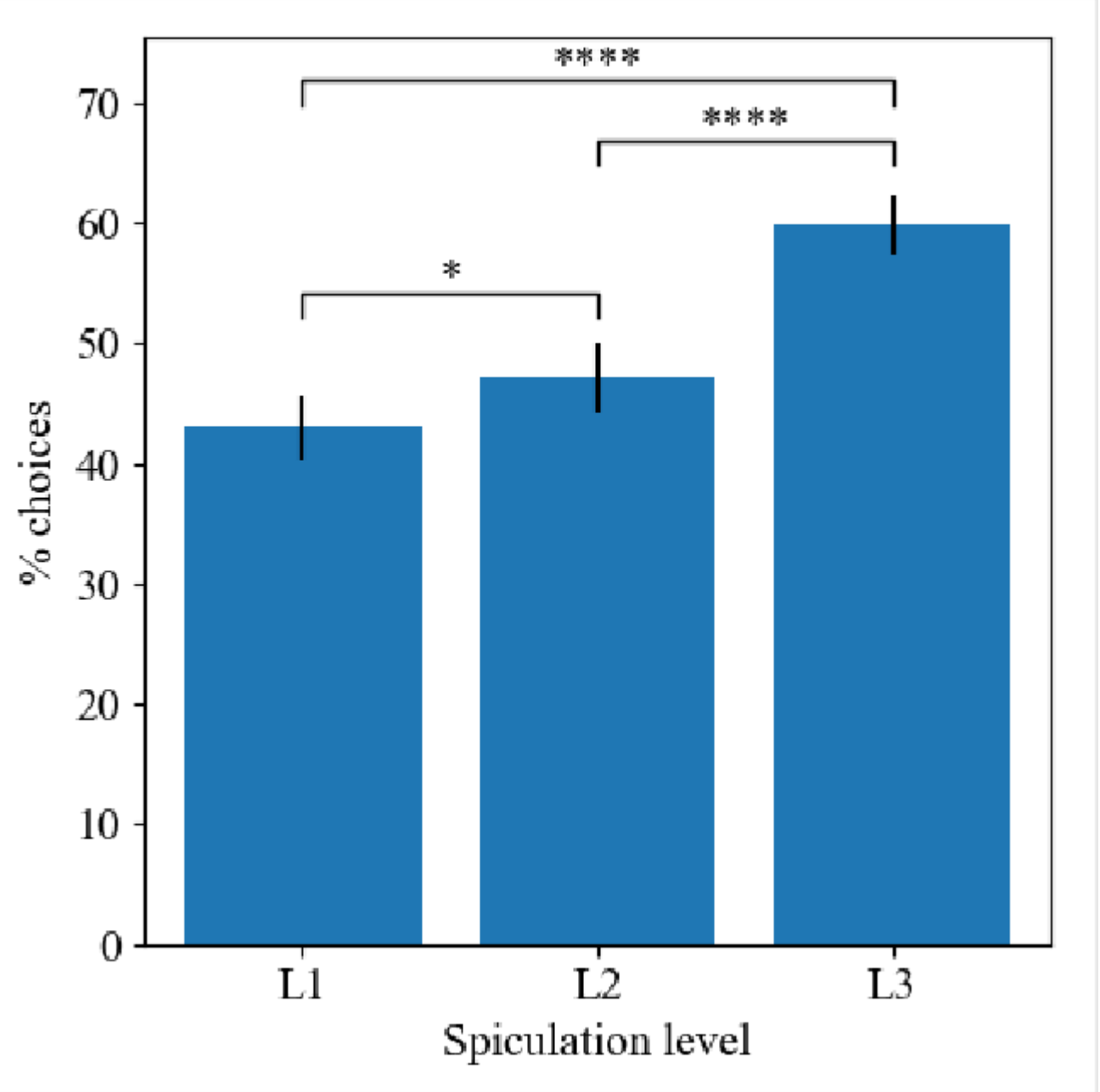

*Figure 4: Forced-choice experiment results comparing radiologists choosing the “more malignant lesion” between three different levels of spiculations. All participants saw the same pairs of images. Error bars are the standard deviation of the mean over 6 participants for 200 image pairs for each level (* p<0.05, **** p< 0.0001).*

## DISCUSSION

The study presented aims to validate the inclusion of spiculations to model, to a certain point, the degree of malignancy of in silico breast masses. A radiologist typically scores a screening case using image features that make them suspicious enough to recall the case. Spiculations is one of many, which is unlikely to happen by itself without any other co-occurring feature. However, by isolating the effect of spiculations, we can understand their significance.

Finally, although the definition of malignancy cannot be ascertained from only imaging, it is important to know what biomarkers or features we can include on our models that can account for it. This may lead to more complex in silico clinical trials that include spiculations (an indicator for malignancy) as one of the features to be considered.

# CONCLUSION

Understanding what makes a lesion malignant and being able to tune our lesion simulation parameters for this will help improve the realism and quality of our in silico imaging trial pipeline. Adding spiculations to lesions allows for the use of the VICTRE pipeline in more complex tasks such as lesion malignancy classification for in silico studies closer to clinical scenarios.

# ACKNOWLEDGEMENTS AND CONFLICT OF INTEREST

Vanday Bundu was supported by an appointment to the Research Participation Program at the Center for Devices and Radiological Health administered by the Oak Ridge Institute for Science and Education through an interagency agreement between the U.S. Department of Energy and the U.S. Food and Drug Administration.

The mention of commercial products, their resources, or their use in connection with material reported herein is not to be construed as either an actual or implied endorsement of such products by the Department of Health and Human Services. This is a contribution of the U.S. Food and Drug Administration and not subject to copyright.

# Biologically informed *in-silico* models of tumor growth dynamics

J.B. Stevens, W.P. Segars, J.G. Liu, and K.J. Lafata

## BACKGROUND AND PURPOSE

To create computational tumor models and improve realism in virtual imaging trials by simulating growth dynamics of avascular tumors via assimilation of imaging data and mathematical biology models.

## METHODS

We developed a tumor growth model that is spatially constrained by boundary conditions derived from PET/CT images. We implemented a stochastic PDE model to capture the dynamic evolution of tumor and immune cell populations. We solved the system of equations numerically using a finite difference scheme. Boundary conditions were derived from gross-tumor-volume contours and were used to create a surface mesh. To simulate growth, we developed a Monte Carlo scheme whereby tumor cells are distributed around concentric spheres according to their local density defined by the solution to the PDE. Immune cell interactions were modelled based on interaction distance (i.e., between immune cells and tumor cells) and interaction probabilities. The resulting cellular point-cloud was voxelised based on cell densities using a recursive, histogram-based approach and Gaussian smoothing. Spatial differences in nutrient density were parameterized via PET habitat analysis and encoded into the model. Cellular entropy and tumor volume were used to quantify changes in texture and morphology.

## RESULTS

Based on simulated tumor cell dynamics, we observed qualitative and quantitative differences in tumor volume, texture, and necrotic core formation. Tumor volume increased with time to a stable value that was greater for less necrotic tumors. Final tumor volume was 18.2% lower for necrotic tumors compared to less necrotic tumors. Mean absolute gradient was used to quantify entropy fluctuation and was equal to 0.4801 and 0.8360 for the regular and necrotic tumors respectively.

## CONCLUSION

We developed a methodology to grow 3D tumors with biological interpretability and characterized the growth across varying levels of necrosis. This technique can inform the realism of in-silico tumor models and enhance the quality of virtual imaging trials.

## Biologically informed *in-silico* models of tumor growth dynamics

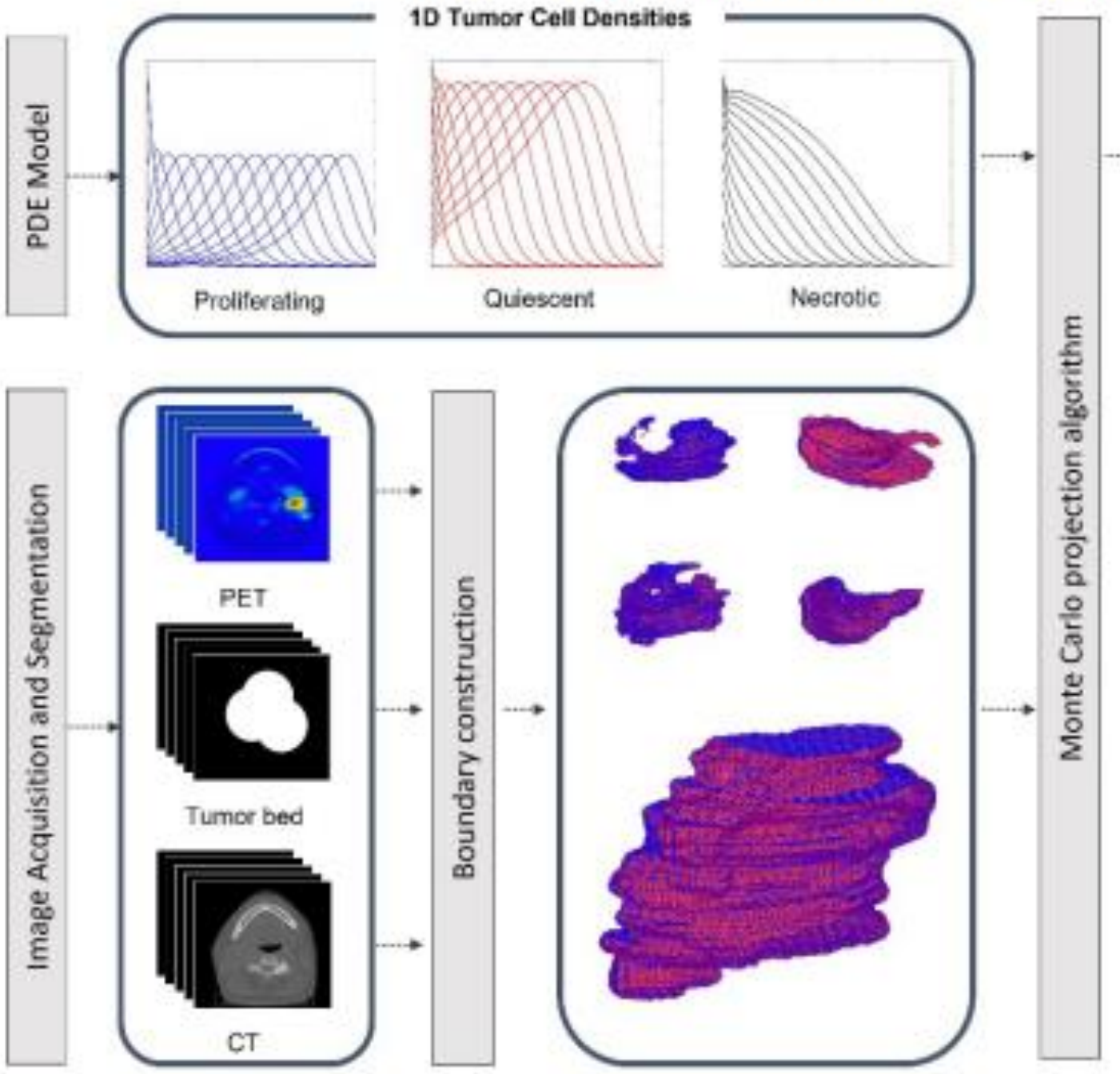

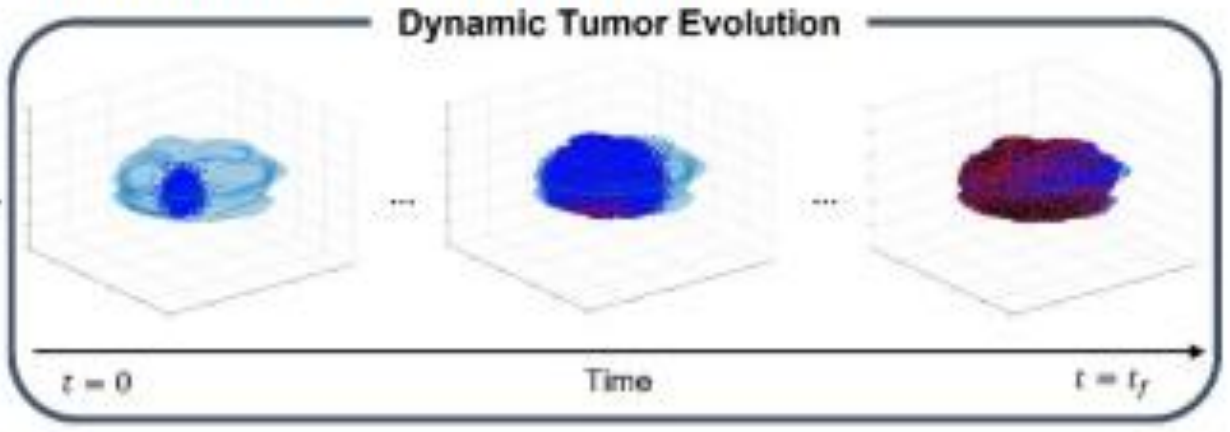

***Figure 1***: *(left) Methodology workflow. Imaging data is used to create boundary surfaces using a mesh grid algorithm. Cell densities are generated and extended to 3D via a Monte Carlo algorithm.*

***Figure 2:*** *(below) A system of coupled PDEs that describe evolution of tumor cells and enforce contact inhibition of migration, cell motility/death and cell division.*

$$\frac{\partial p}{\partial t} = \frac{\partial}{\partial x}\left[\frac{p}{p+q}\frac{\partial}{\partial x}(p+q)\right] + g(c)p(1-p-q-n) - f(c)p$$

$$\frac{\partial q}{\partial t} = \frac{\partial}{\partial x}\left[\frac{q}{p+q}\frac{\partial}{\partial x}(p+q)\right] + f(c)p - h(c)q$$

$$\frac{\partial n}{\partial t} = h(c)q$$

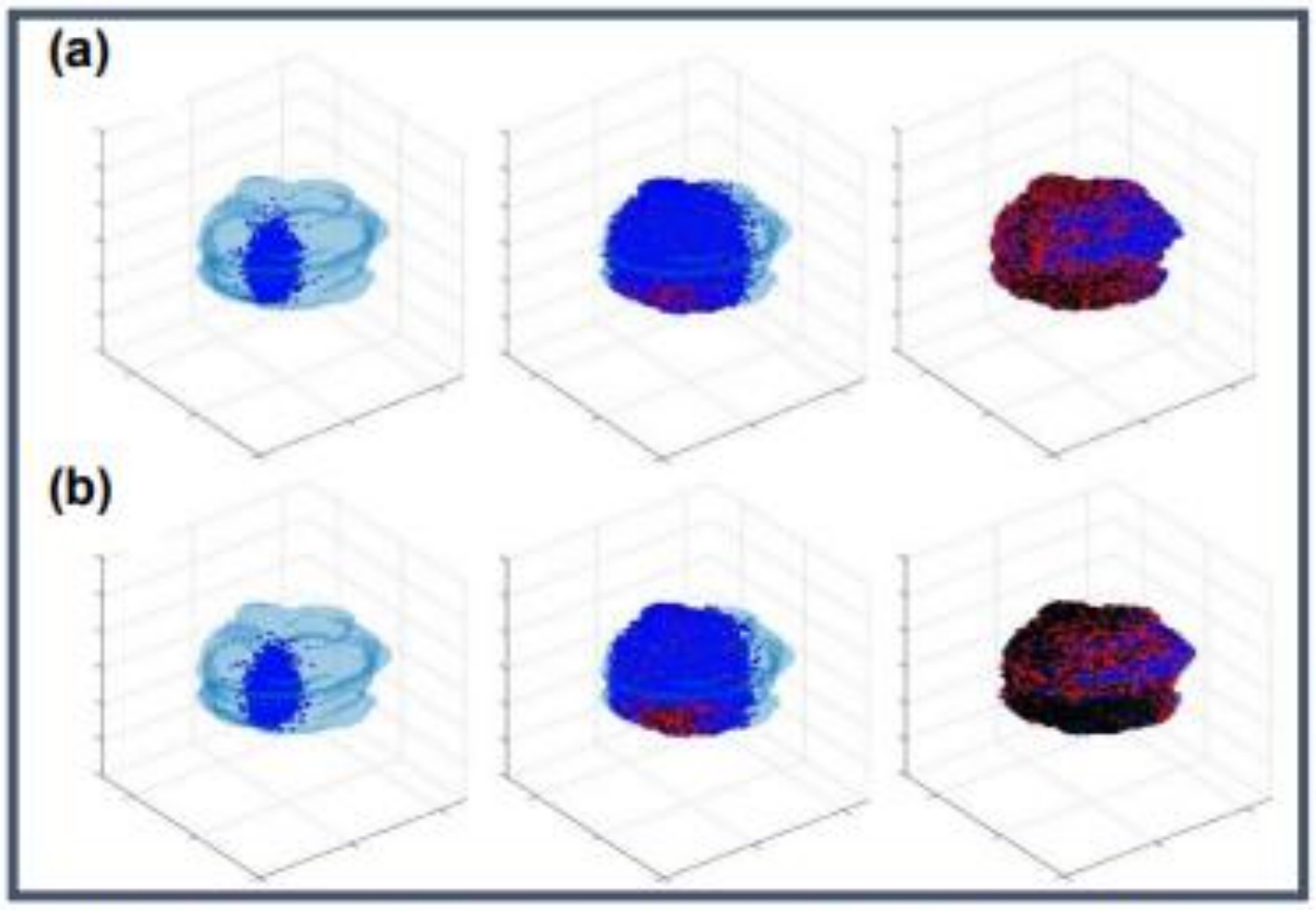

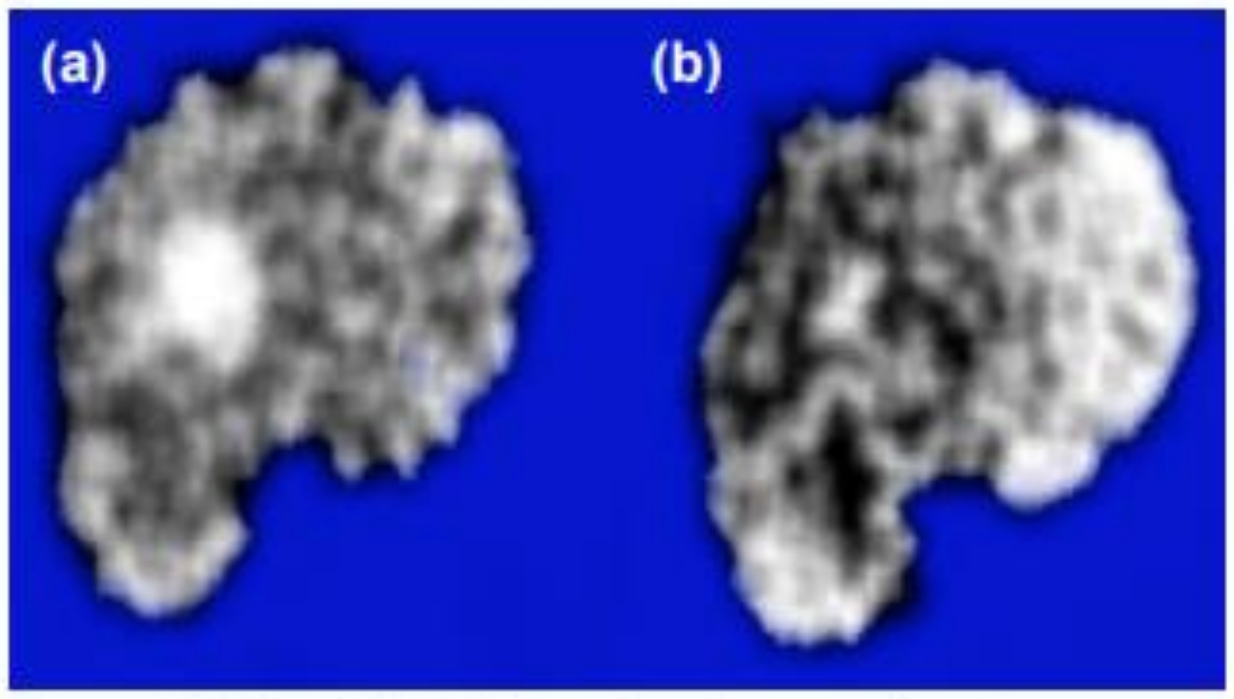

***Figure 4***: *(above) Axial slices from two voxelised tumors. (a) A tumor with greater nutrient access, hence less necrotic and (b) a tumor with lower nutrient access, hence more necrotic.*

***Figure 3***: *A series of snapshots in time showing the spatially bounded growth of two types of tumors. (a) A tumor with less necrosis, which shows fewer necrotic (black) cells at final snapshot and (b) a tumor with more necrosis, which shows more necrotic cells and quicker conversion between live (proliferating and quiescent) and dead cells.*

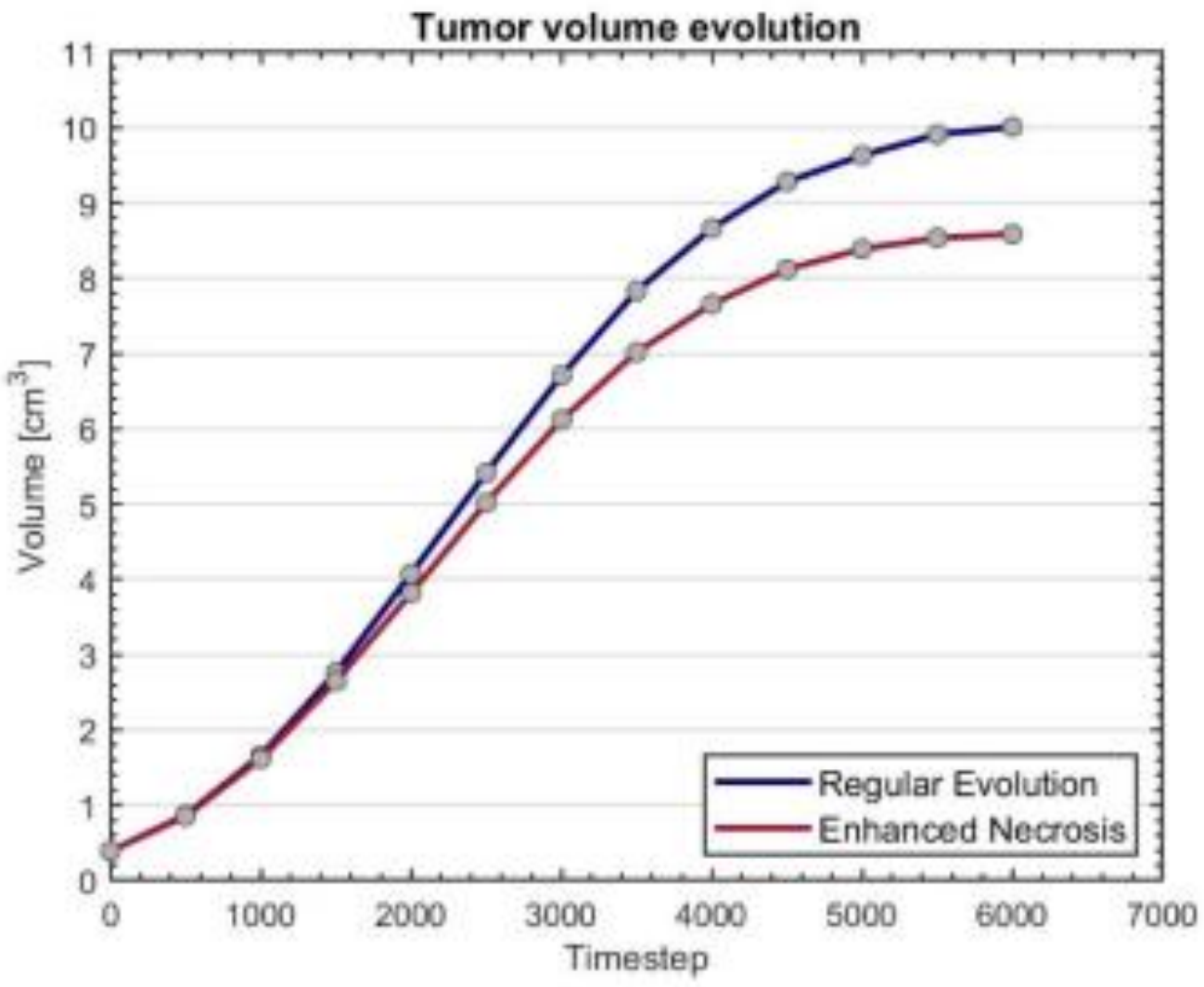

***Figure 5:*** *Tumors that exhibit less necrosis tend to grow more and reach a larger final volume than more necrotic tumors.*

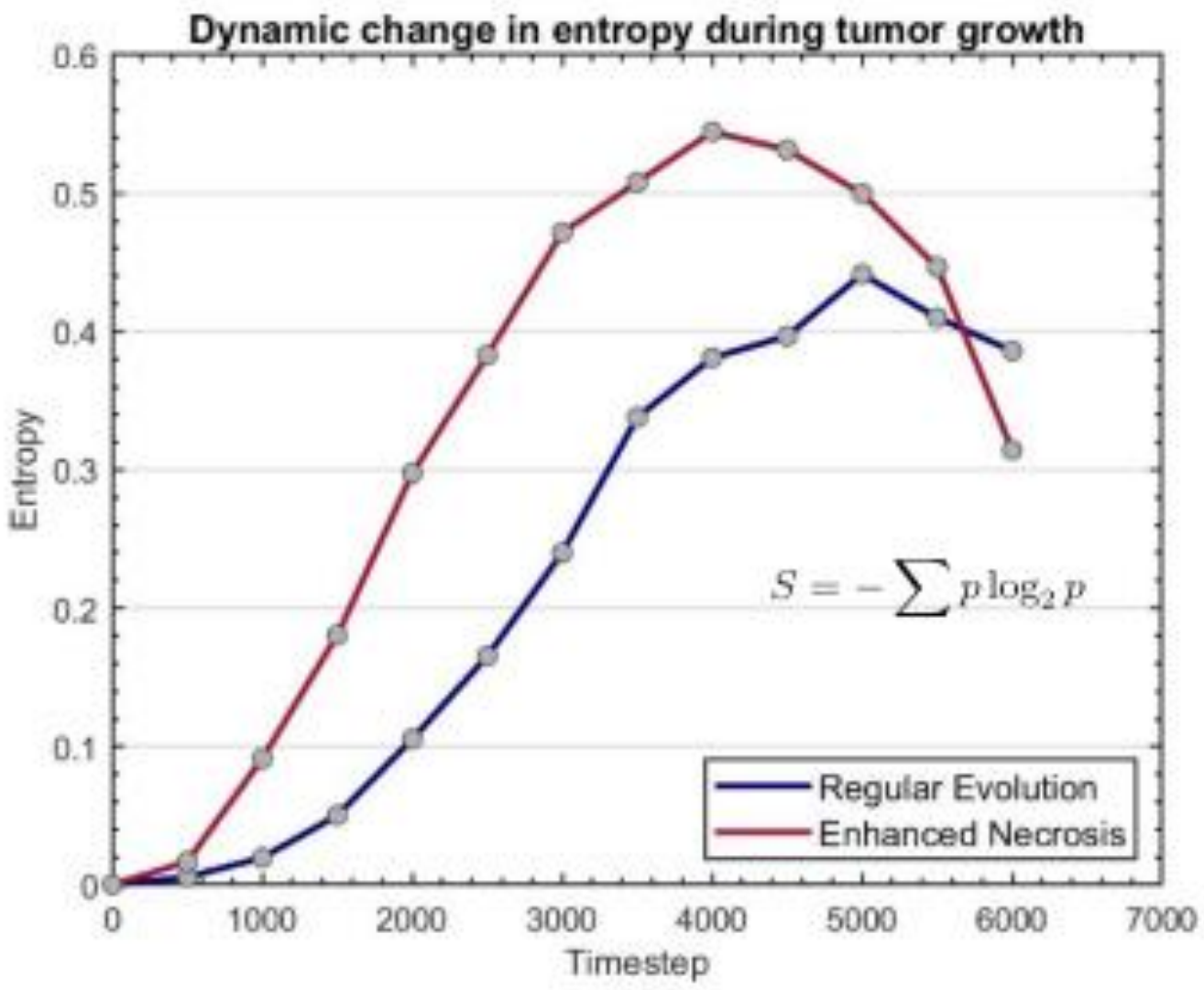

***Figure 6:*** *Entropy behaves differently for tumors with varying nutrient parameters. This is "cellular entropy" i.e., entropy derived from cell density directly, rather than from imaging.*

# Modeling of current imaging systems

Tuesday, April 23, 2024

9:00 am - 9:40 am

Chaired by Paul Kinahan & Ehsan Abadi

# Effects of misalignment between mechanical and x-ray breast images in simultaneous digital breast tomosynthesis and mechanical imaging

Predrag R. Bakic,[§] Rebecca Axelsson, Hanna Tomic, Sophia Zackrisson, Anders Tingberg, Magnus Dustler
Lund University, Department of Translational Medicine, Skåne University Hospital, Malmö, Sweden;

## BACKGROUND AND PURPOSE

Simultaneous digital breast tomosynthesis (DBT) and mechanical imaging (MI) – DBTMI – is a novel breast cancer screening approach. DBT acquires low-dose projections to reconstruct breast slices, which reduces anatomical noise. MI uses a pressure sensor to measure stress on the compressed breast surface. DBTMI can identify false positive detections to improve screening specificity. A preliminary clinical study of 52 women suggested that DBTMI could resolve 33% false positives, which were not resolved by DBT+ultrasound.[1]

An important step in DBTMI is the alignment between MI and radiographic images. The MI clinical descriptor is the Relative Mean Pressure over the lesion Area (RMPA), calculated as the average stress over the 3x3 sensor elements and normalized for the overall average stress on the breast.[2] Potential DBT-MI misalignment may cause errors in RMPA and affect clinical decisions. This work analyzes the effects of misalignment using simulated DBTMI data.

## METHODS

Previously, we simulated DBTMI using computational breast phantoms, ray tracing x-ray acquisition, and finite-element model of the compressed breast and the MI sensor with 10 mm isotropic spatial resolution.[3,4] The phantoms simulated various volumetric density with a spherical lesion. In that previous study, RMPA was calculated assuming an ideal DBT-MI alignment.

In this study, we simulated the misalignment by shifting DBT relative to MI images in orthogonal directions parallel to the MI sensor. Simulation was repeated for different lesion locations, and using four phantoms with different volumetric breast density, VBD (*see Fig.1*). We have identified maximum RMPA changes, relative to the ideal alignment. In addition, the analysis was repeated assuming another MI sensor with 5 mm isotropic resolution.

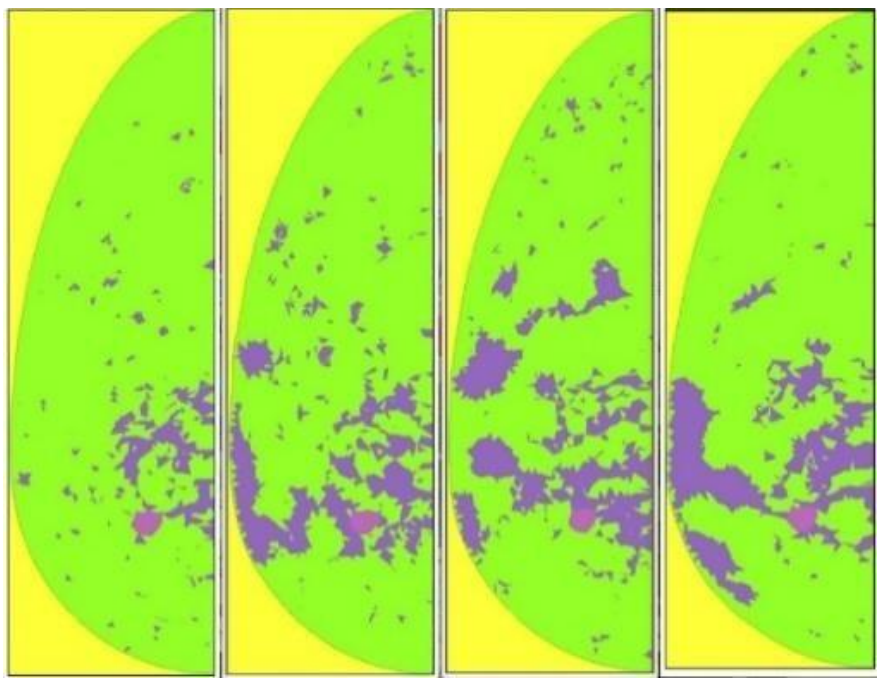

(a)

§ Predrag.Bakic@med.lu.se

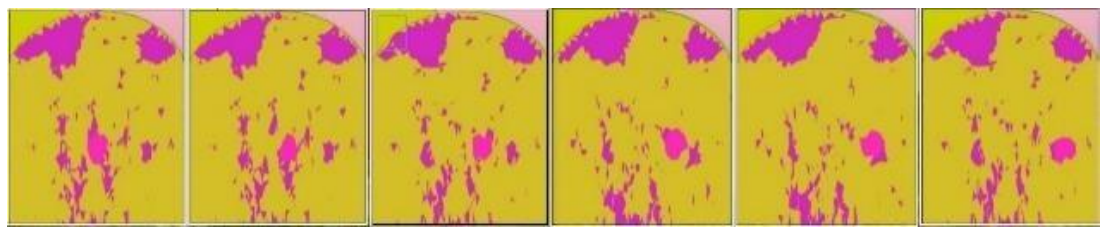

(b)

*Figure 1 Simulated anatomy in our study. (a) ML sections of phantoms with volumetric breast density, VBD, of (L-R): 4%, 7%, 9%, 12%. (b) CC sections of phantoms with the lesion (red) located at (L-R): 5, 4.4, 3.8, 3.2, 2.6, 2.0 cm from the breast surface.*

# RESULTS

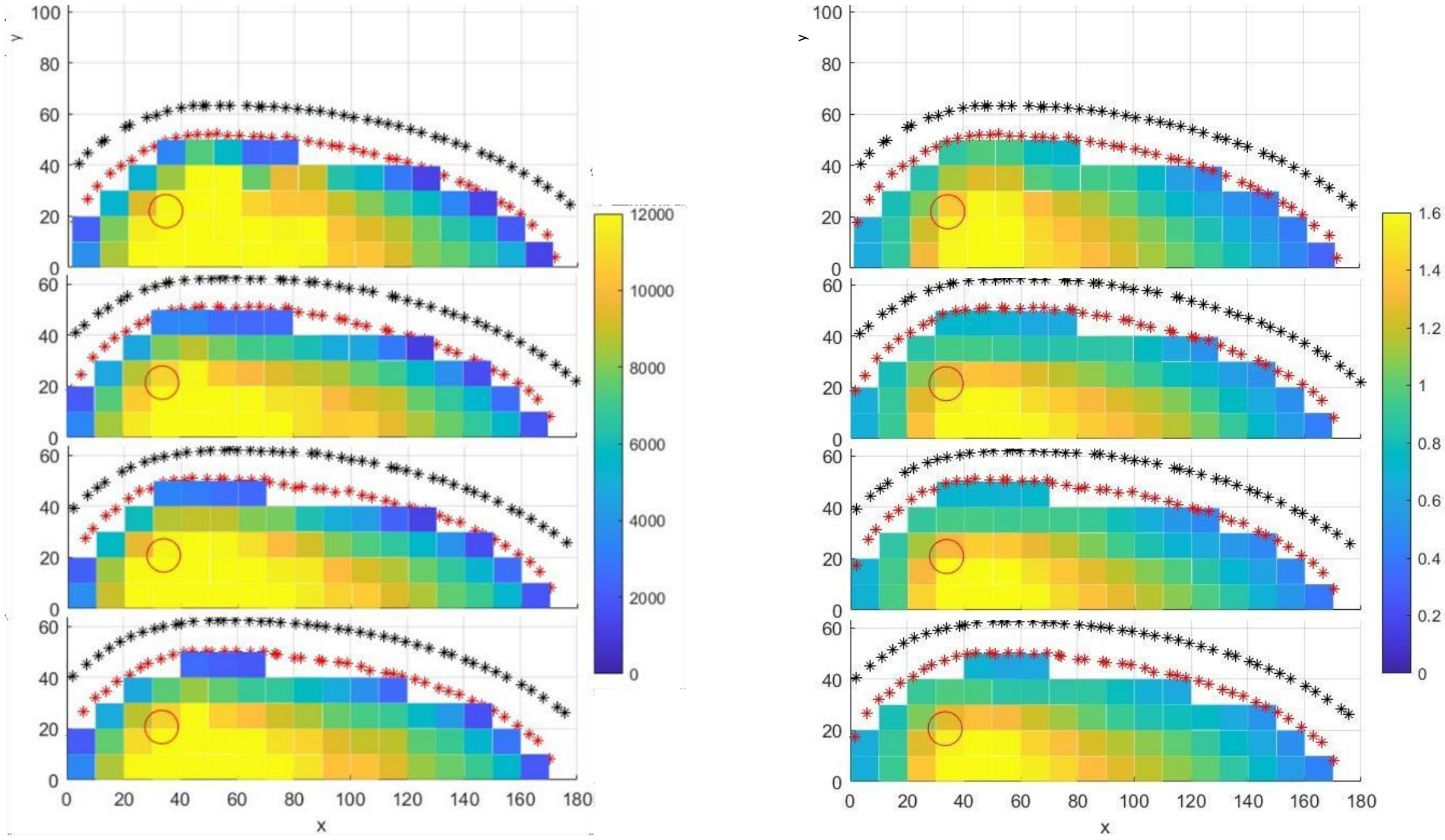

(a) (b)

*Figure 2 Distribution of (a) the stress (kPa) and (b) Relative Mean Pressure over the lesion Area (RMPA), on the surface of simulated compressed breasts with VBD of (top-bottom): 12%, 9%, 7%, and 4%, and MI sensor with the 10-mm resolution.*

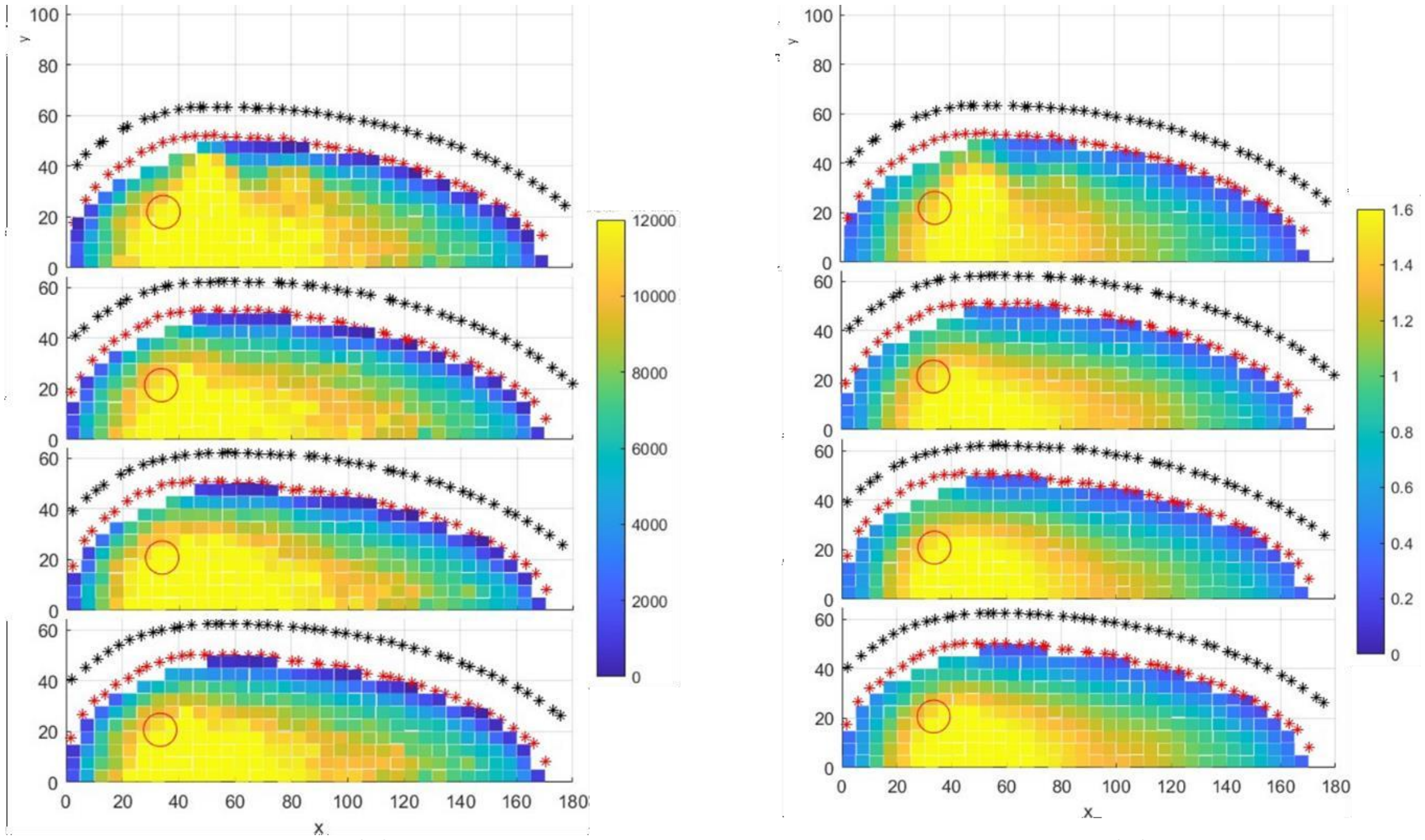

(a) (b)

*Figure 3 Distribution of (a) stress (kPa) and (b) Relative Mean Pressure over the lesion Area (RMPA), on the surface of simulated compressed breasts with VBD of (top-bottom): 12%, 9%, 7%, and 4%, and MI sensor with the 5-mm resolution.*

Figs. 2 and 3 show distributions of the stress and RMPA over the breast phantom surface, for the MI sensor resolution of 10 mm and 5 mm, respectively, as calculated by the FE analysis. The simulated lesion was centered at 33.6 mm and 21.4 mm, in caudal-cranial (*x*) and chest-nipple (*y*) directions, respectively.

Tables 1 and 2 list the stress and RMPA, respectively, at the lesion location (i.e., for the perfect DBT-MI alignment) as well as the relative difference in the stress and RMPA, respectively, for assumed misalignments in x and y directions, using the simulated MI sensor with the spatial resolution of 10 mm. Listed are the values averaged over six lesion depths, shown in Fig. 1(b). Similarly, Tables 3 and 4 list the stress and RMPA at the lesion, respectively, as well as their relative differences for various misalignments, using the simulated MI sensor with the spatial resolution of 5 mm.

*Table 1: The effect on the stress of misalignments in caudal-cranial (x) and chest-nipple (y) directions, for a 10-mm MI sensor*

| **VBD (%)** | **Stress (kPa) @ Lesion** | **Relative Differences in Stress (%)** | | | | *Average over Misalignments* |
|---|---|---|---|---|---|---|
| | | **x-10 mm** | **x+10 mm** | **y-10 mm** | **y+10 mm** | |
| 12 | 12.4±0.5 | 21.5±1.3 | 58.6±5.7 | 16.1±1.4 | 32.2±3.8 | *32.1±7.1* |
| 9 | 11.8±0.7 | 20.9±2.3 | 5.5±3.2 | 13.5±2.6 | 31.9±5.4 | *18.0±7.2* |
| 7 | 12.5±0.8 | 21.6±3.8 | 6.1±2.8 | 16.7±1.9 | 30.4±5.6 | *18.7±7.5* |
| 4 | 11.4±0.5 | 20.5±2.1 | 8.8±5.4 | 22.5±2.7 | 33.9±4.0 | *21.4±7.6* |
| *Average over VBDs* | | *21.1±5.0* | *19.8±9.0* | *17.2±4.4* | *32.1±9.6* | ***22.5±14.7*** |

*Table 2: The effect on the RMPA of misalignments in caudal-cranial (x) and chest-nipple (y) directions, for a 10-mm MI sensor*

| VBD (%) | RMPA @ Lesion | Relative Differences in RMPA (%) | | | | *Average over Misalignments* |
|---|---|---|---|---|---|---|
| | | **x-10 mm** | **x+10 mm** | **y-10 mm** | **y+10 mm** | |
| 12 | 1.42±0.02 | 23.5±0.3 | 12.9±0.7 | 13.0±0.2 | 18.8±0.5 | *17.1±0.9* |
| 9 | 1.29±0.02 | 14.7±0.6 | 6.2±0.8 | 20.3±0.4 | 22.0±3.3 | *15.8±3.5* |
| 7 | 1.34±0.02 | 13.8±0.2 | 6.8±0.7 | 21.8±0.4 | 26.6±2.6 | *17.2±2.8* |
| 4 | 1.31±0.03 | 15.2±0.2 | 6.8±0.8 | 23.6±0.1 | 25.5±3.4 | *17.8±3.5* |
| *Average over VBDs* | | *16.8±0.8* | *8.2±1.4* | *19.7±0.6* | *23.2±5.5* | ***17.0±5.7*** |

*Table 3: The effect on the stress of misalignments in caudal-cranial (x) and chest-nipple (y) directions, for a 5-mm MI sensor*

| VBD (%) | Stress (kPa) @ Lesion | Relative Differences in Stress (%) | | | | *Average over Misalignments* |
|---|---|---|---|---|---|---|
| | | **x-5 mm** | **x+5 mm** | **y-5 mm** | **y+5 mm** | |
| 20 | 12.4±0.9 | 9.8±1.0 | 12.7±2.7 | 8.5±2.9 | 13.8±3.2 | *11.2±5.2* |
| 15 | 12.2±1.5 | 11.4±2.9 | 7.5±6.1 | 7.5±3.9 | 10.7±7.8 | *9.3±11.0* |
| 10 | 13.1±1.5 | 14.0±4.8 | 12.4±4.6 | 15.1±4.9 | 14.7±6.1 | *14.1±10.2* |
| 5 | 11.9±1.0 | 11.1±2.1 | 10.8±2.6 | 11.5±4.5 | 16.1±5.2 | *12.4±7.6* |
| *Average over VBDs* | | *11.6±6.0* | *10.9±8.5* | *10.7±8.3* | *13.8±11.6* | ***11.7±17.6*** |

*Table 4 The effect on the RMPA of misalignments in caudal-cranial (x) and chest-nipple (y) directions, for a 5-mm MI sensor*

| VBD (%) | RMPA @ Lesion | Relative Differences in RMPA (%) | | | | *Average over Misalignments* |
|---|---|---|---|---|---|---|
| | | **x-5 mm** | **x+5 mm** | **y-5 mm** | **y+5 mm** | |
| 20 | 1.48±0.06 | 13.0±0.7 | 16.3±1.5 | 5.5±0.4 | 13.2±1.2 | *12.0±2.1* |
| 15 | 1.51±0.10 | 11.0±1.3 | 6.4±1.4 | 5.4±0.9 | 12.3±2.5 | *8.8±3.3* |
| 10 | 1.55±0.09 | 10.9±1.5 | 7.3±0.7 | 7.2±0.6 | 13.1±1.6 | *9.6±2.4* |
| 5 | 1.51±0.07 | 11.5±0.9 | 8.6±1.4 | 9.7±1.2 | 13.9±1.1 | *10.9±2.4* |
| *Average over VBDs* | | *11.6±2.3* | *9.6±2.6* | *7.0±1.7* | *13.1±3.4* | ***10.3±5.2*** |

## DISCUSSION

The background tissue distribution has substantial effect both on the MI stress at the lesion location, and on the relative difference in stress caused by DBT-MI misalignment, as seen in Table 1. On the other hand, the relative difference in RMPA (Table 2) is lower as compared to the relative difference in the MI stress (Table 1): 17.0 vs. 22.5%, as averaged over all the misalignments and VBDs. This result is expected, since the RMPA calculation includes local averaging and normalization of the stress, which reduces the variation.

When simulating the MI sensor with a higher spatial resolution (i.e., with the 5-mm sensor element size), the observed relative difference in the stress (Table 3) and RMPA (Table 4) are lower as compared to using a 10-mm sensor: 11.7% vs. 22.5% for the stress, and 10.3% vs. 17.0% for the RMPA. This suggests that the 5-mm sensor may reduce variations in RMPA due to potential DBT-MI misalignment. We are working with the MI sensor manufacturer to prototype and test a 5-mm MI sensor. In addition, we have been developing DBTMI fiducial markers to improve the accuracy of the DBT-MI alignment.

We have observed that with a 5-mm sensor, the calculated RMPA values at the lesion tend to be higher than with a 10-mm sensor, e.g., 1.51 vs. 1.34 (as averaged over all VBDs, Tables 4 and 2). This may be caused by the size of the 3x3 ROI for local averaging of the stress when calculating RMPA. With the 5-mm sensor, the size of this ROI is $(15\ \text{mm})^2$ , while with a 10-mm sensor, the ROI size is $(30\ \text{mm})^2$. The ROI is centered at the lesion location, and the 3x3 averaging over a smaller physical ROI would likely yield a higher value. This is of importance when discriminating between malignant and benign lesions based upon the RMPA values; the discriminating criteria may differ for MI sensors with different resolution.

Our preliminary simulation analysis of the effect of misalignment shows an agreement with the estimated variation in RMPA from clinical DBTMI. Our preliminary analysis of 31 clinical DBTMI datasets suggested an average relative difference in RMPA of 14.3%±12.2%,[5] which is comparable to the relative difference of 17.0%±5.7% from our simulation. Analysis of a larger number of clinical datasets is ongoing.

## CONCLUSION

Preliminary simulation suggests that misalignment between MI and x-ray images can affect clinical decisions. The effect depends on the lesion position, the background distribution of dense tissue, and the spatial resolution of the MI sensor. Further tests and the sensor optimization to ensure correct alignment are ongoing.

## ACKNOWLEDGEMENTS AND CONFLICT OF INTEREST

This work was supported by grants from the European Commission H2020 Marie Skłodowska-Curie Actions Fellowship, the grants from Cancerfonden, Bröstcancerförbundet, and Stiftelsen för Cancerforskning vid Onkologiska kliniken vid Universitetssjukhuset MAS.

# System-specific simulation of mammography *for processing* images

Franziska Mauter[1,2], Marta Pinto[2], Marcel Reginatto[1], Ruben van Engen[4], Mathias Anton[1], Ioannis Sechopoulos[2,3,4]

[1]Divison of Ionising Radiation, Physikalisch-Technische Bundesanstalt, Braunschweig, Germany
[2]Department of Medical Imaging, Radboud University Medical Center, Nijmegen, Netherlands
[3]Dutch Expert Centre for Screening (LRCB), Nijmegen, Netherlands
[4]Technical Medicine Centre, University of Twente, Enschede, The Netherlands

## BACKGROUND AND PURPOSE

To develop a fast Python-based mammography simulator that generates system-specific for processing images using deep learning to estimate the X-ray scatter components.

## METHODS

Thickness maps per tissue type are generated from a digital 3D phantom via a ray-tracing projector. Using an X-ray-spectrum model and system-specific measurements of tube outputs, air-kerma-to-mean-signal relationships, modulation transfer functions, noise power spectra, and signal-mean-to-variance relationships, a raw projection image that includes system performance limitations is generated. A U-net trained on Monte Carlo (MC) simulations of homogeneous breast phantoms is used to predict the scatter signal of the final image (s. Fig.). Simulated images were visually inspected by an expert in the field of mammography image quality. Computation times were averaged for the generation of mammograms derived from 20 virtually compressed patient breast CT data with compressed breast thicknesses between 3.9 cm and 6.8 cm. Average scatter prediction time of the U-net and performance measures based on mean relative difference (MRD) and mean absolute relative difference (MARD) inside the breast area were derived from an independent MC test set of 100 simulations.

## RESULTS

Visual inspection of the simulated images by an expert indicates reasonable results. A complete simulation takes 198 s, resulting in a minimum time saving of approx. 90 % compared to pure MC simulations. Scatter prediction by the U-net takes, on average, 140 ms with a median MRD of -0.1 % (Q1: -0.2 %, Q3: 0.0 %) and a median MARD of 2.5 % (Q1: 1.8 %, Q3: 3.2 %).

## CONCLUSION

Using a U-net to estimate the scatter signal of a mammogram gives good results and significantly reduces the computation time for the entire simulation.
A comprehensive objective validation of the simulation results against real acquisitions will be performed to ensure the authenticity of the generated data.

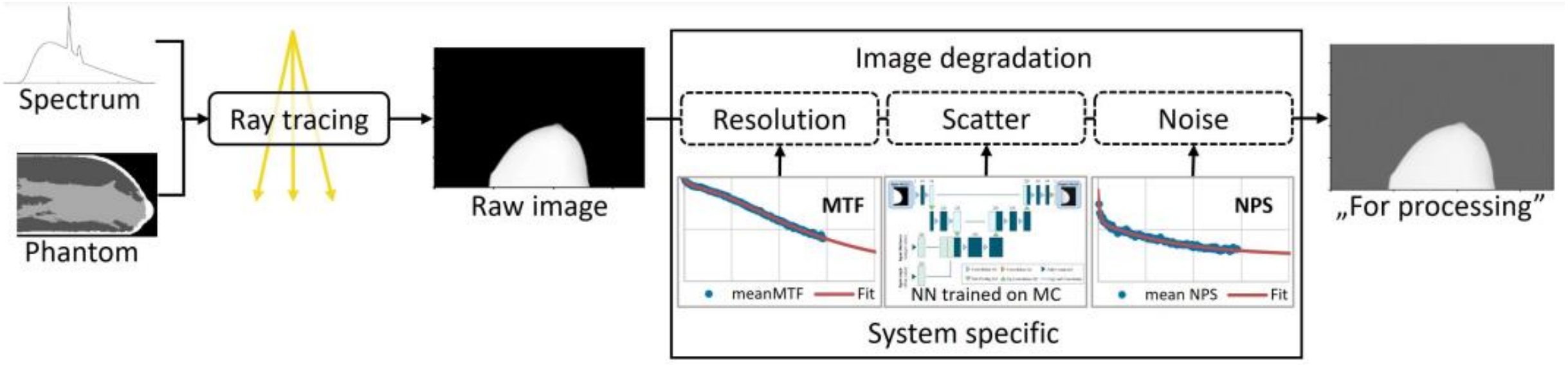

# Implementing detailed antiscatter grid modelling for Virtual Imaging Trials applications

Rodrigo T. Massera[1], Karen Merken[1], Nicholas Marshall[1,2] and Hilde Bosmans[1,2]

1. Medical Imaging Research Centre, Medical Physics and Quality Assessment, KU Leuven, 3000, Leuven, Belgium

2. Competentiecentrum medische stralingsfysica, UZ Gasthuisberg, Herestraat 49, 3000, Leuven, Belgium

## BACKGROUND AND PURPOSE

This work explores the application of detailed grid modeling in Virtual Imaging Trials, and presents a case study for digital mammography (DM). Antiscatter grids are widely used in diagnostic x-ray imaging as a scatter rejection technique. However, assessing the impact of grids on image quality for a specific task using physical measurements can be challenging. In this regard, VITs are an important tool to simulate the impact of antiscatter grids for tasks that are closely related to what is found in clinical applications. To better understand the implications of the antiscatter grids on image quality, the modelling avoided simplifications as much as possible, while maintaining reasonable simulation times. The grid structure is simulated, along with interactions within the grid.

## METHODS

Breast phantoms were generated (section B.1.) with four distinct group sizes and glandular volume. Afterwards, three types of lesions were generated (section B.2.): microcalcifications, spiculated masses and lobular masses. Digital mammography images were simulated (sections B.3. and B.4.) using a combination of Monte Carlo simulations with ray tracing techniques, and the image quality was compared between two types of antiscatter grid and without grid in place.

### B.1. Phantom generations

The breast phantoms used in this study were generated using the BreastPhatom software[1] in combination with the VictrePipeline[2]. The maximum thickness of the compressed breast phantoms was increased to 7 cm to include thicker breasts found in the population undergoing DM examination, for this, the volume of the uncompressed breast was also increased. In total, four different breast sizes were generated based on thickness: (i) 3.5 cm, (ii) 5.0 cm, (iii) 6.0 cm and (iv) 7.0 cm.

### B.2. Tasks

Two categories of lesions were considered in this study, masses and microcalcifications. Two types of masses were generated using the breastMass software[2], lobular (a more regular shape), and spiculated (irregular shape with spicules emerging from the center), The diameters (excluding the spicules) ranged from 2.5 mm to 6.0 mm. The clusters of microcalcifications were obtained from the software described in Van Camp et al[3]. The number of microcalcifications in the cluster varied from 9 to 15, with an average size ranging from 120 micrometers to 180 micrometers. They were randomly distributed in a 0.125 $cm^3$ cube.

### B.3. System geometry and general parameters

The source to detector distance was 65 cm. The x-ray field size irradiated the entire detector (12 cm x 24 cm). The detector was modelled as an amorphous selenium layer of thickness 0.25 mm. X-ray spectra were simulated using a W/Rh anode/filter combination with a tube potential ranging from 26 kV to 32 kV, adjusted by breast size, The spectra were obtained from the SpekPy software[4]. The breast was positioned at 1.5 cm above the detector, and between the compression and support plates (1 mm thickness, PMMA). The linear grid has a height of 1.5 mm, focal distance of 65 cm, a frequency of 31 l/cm and a ratio of 5. The septa were composed of lead, and the interspace material of paper fibre. The cellular grid has a height of 2.4 mm, focal distance of 65 cm, frequency of 15 cells/cm, and a ratio of 3.8. The septa were composed of copper, and the interspace material is air. For both grids, the total cover is 0.4 mm of carbon fibre.

**B.4. Image formation**

First, the necessary reciprocal motion of the antiscatter grids were studied by simulating flood images of the bucky. This motion was modelled as a rectilinear uniform movement around the central position of the grid, perpendicular to the septa alignment of the linear grid. The cellular grid was rotated approximately 48 degrees to minimize artifacts, and the same linear motion was modelled. The required displacement was determined by analyzing the NNPS of the flood images until the curves were uniform at high frequencies (up to 4 $mm^{-1}$), where grid artifacts are usually present. Afterwards, flood images were acquired with other energies (2 keV step) to cover the entire range of the x-ray spectra.

Each breast phantom was ray-traced using the Siddon's algorithm[5] with supersampling to avoid aliasing, and an array was stored containing the projected distance per pixel for each material. The same procedure was performed with the lesions. The position of the lesion was selected by regions of the breast where the compressed breast thickness is close to the maximum (to avoid the borders), and where glandular tissue is present, to avoid extremely fatty regions . The lesions were included in the phantom by subtracting the adipose tissue distances from the array and replacing them with the lesion material. This allowed the quick insertion of lesions without repeating the ray tracing step several times.

The transmission probability per pixel was calculated from the projected distances and the attenuation coefficients for each material. From the initial number of particles, the transmission probabilities were integrated over the desired x-ray spectrum, combined with the flood images to return the average deposited energy per pixel and the respective quantum noise. For the scatter images, the simulations were carried out directly with the phantom in place and x-ray spectrum, however a low-resolution image (downsampled by 20:1) was obtained to speed up simulation times. The primary and scattered images were combined with the same number of photons emitted by the x-ray source, and the MTF and NNPS curves measured on a real system (MAMMOMAT Inspiration, Siemens, Germany) were applied to modulate the signal and noise in the projection images.

All the simulations in this work were carried out with a modified PENELOPE[6] (v. 2018) and penEasy[7] (v. 2022) Monte Carlo code. These modifications include the usage of the Message Passing Interface (MPI) to parallelize the computation in a cluster. The antiscatter grid hybrid implementation is described in previous work,[8] which combines the analytical model proposed by Zhou et al. [9] with the tracking of scattered particles within the grid. In this work, we extended our previous hybrid method to include 2D grids.

The wICE cluster from KU Leuven was used to carry out the simulations. The nodes used consisted of 2 Intel Xeon Platinum 8360Y CPUs@2.4 GHz (36 cores each), with 256 GB of RAM. Each primary and scatter simulation took between 2 to 4 hours. The projection images took approximately 10 minutes to be generated.

# RESULTS

## C.1 Antiscatter grid motion study

Figure 1 shows the impact of the grid reciprocal linear motion for linear and cellular grids (i.e. they oscillate between the central position). For the results, we require that the grid completes one complete cycle during the exposure. In 1(a) a linear grid is stationary, producing large artifacts on the image. In figure 1(b) the same grid performs a reciprocal motion of 1.0 cm, basically removing all artifacts. In 1(c), a cellular grid performs a reciprocal motion of 1.0 cm, which is not sufficient to remove all artifacts. In 1(d), the movement is increased to 2.5 cm, producing a uniform image.

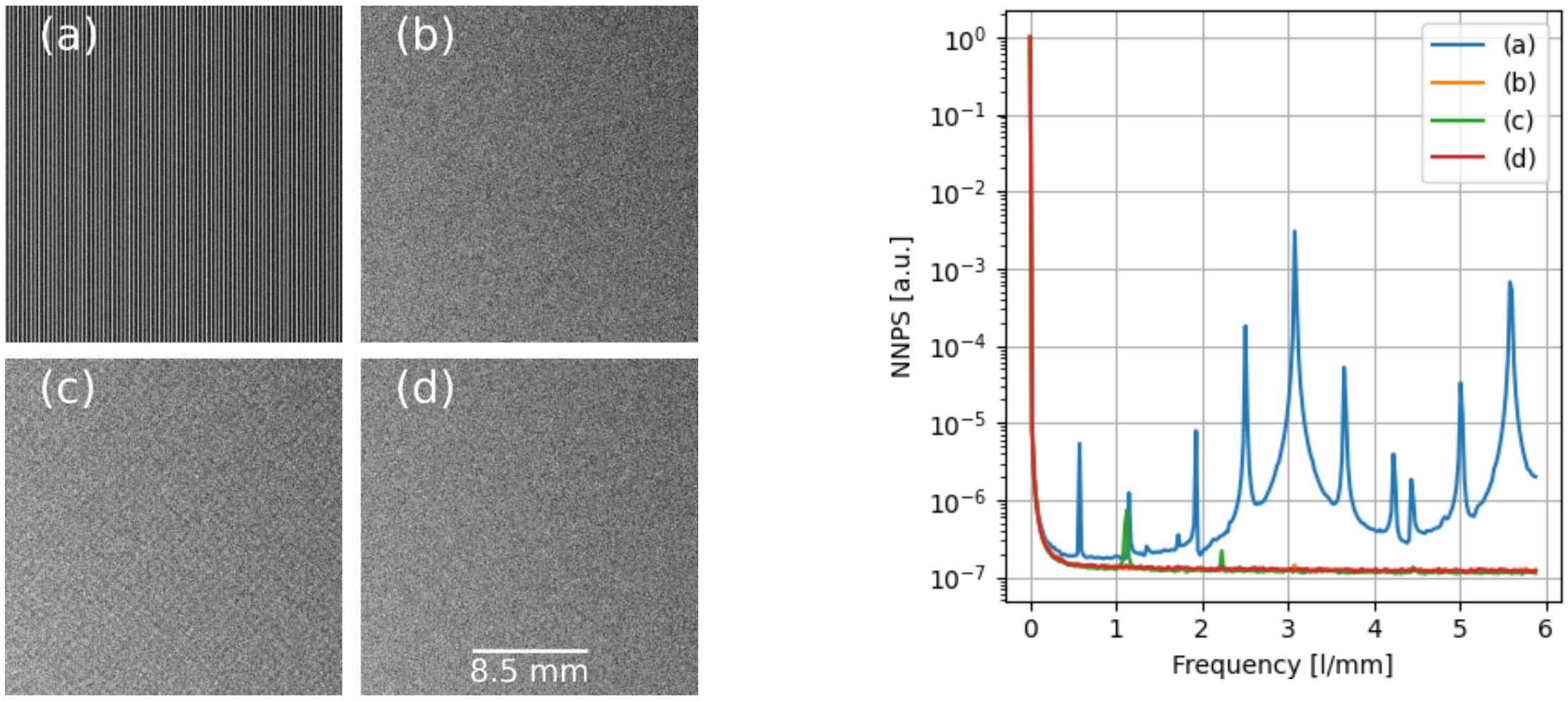

Figure 1: On left: flood images of the antiscatter grids, with and without motion. Right: NNPS calculations of the images showing the peaks where the artifacts are present. Linear grid: (a) without motion and (b) with 1.0 cm maximum displacement. Cellular grid: 1.0 cm and 2.5 cm maximum displacement.

## C.2 Example of generated images

Digital mammography images generated in this work are shown in Figure 2. The acquisitions consisted of the cases without antiscatter grid, with linear grid and with cellular grid. For the cluster of microcalcifications, the breast thickness was 7 cm, and the mean glandular dose (MGD) was fixed at 2 mGy. For the masses, a 5.8 cm breast thick was used, with a constant MGD of 1.5 mGy.

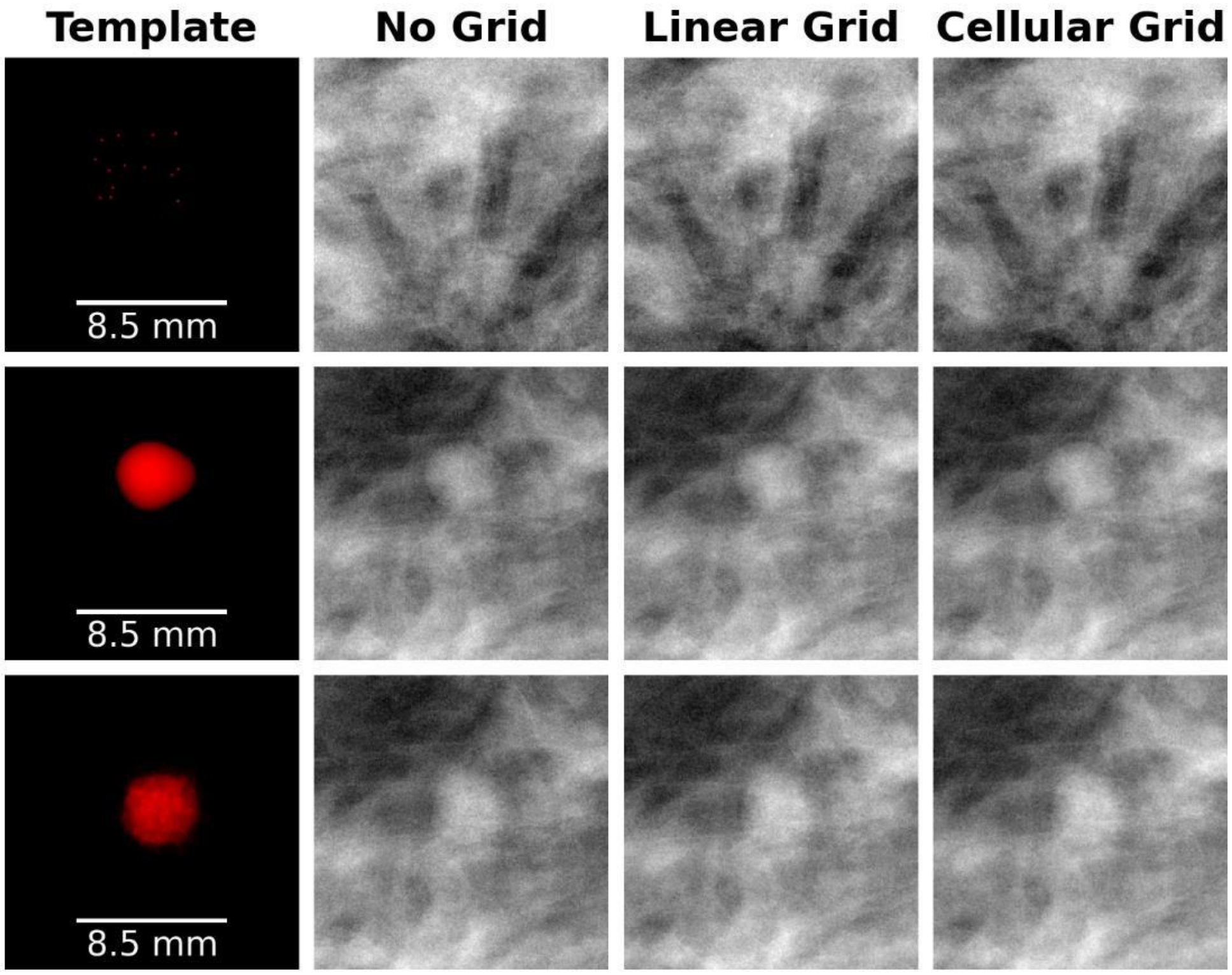

Figure 2: Comparison between the antiscatter grids and without grid case for different lesions. First row: cluster of microcalcifications; second row: lobular mass; third row: spiculated mass.

## DISCUSSION

Simulating the antiscatter grid structure allows the study of grid-related artifacts on the projected images. Based on this information, it is possible to optimize the reciprocal motion of the grid to generate uniform images, avoiding an excessive cutoff effect from primary photons. In addition, this provides the contribution of secondary photons generated from photon-interactions within the grid (scattering from the interspace material or fluorescence photons from the septa material) that can be absorbed by the x-ray detector, potentially lowering the performance of the grid. Although this effect is minimized in mammography due to the low tube potential used, the scattering within the grid can be more significant in modalities that employs higher tube potentials, such as in diagnostic x-rays. For the microcalcification region shown in Figure 2, the linear grid has a primary (Tp) and scattered (Ts) transmission of 0.79 and 0.18, respectively, while the cellular grid presented values of 0.81 and 0.12 for Tp, and Ts, respectively. The grid performance can be assessed by a visual analysis exemplified in Figure 2, where the detection of microcalcifications is compared between the grids, in addition to the discrimination between the mass lesion types.

## CONCLUSION

The application of Virtual Imaging Trial methods to assess the performance of antiscatter grid in x-ray medical imaging provides the possibility of assessing the impact on image quality in scenarios closer to that present in the clinics. Physical assessments are usually made in uniform phantoms under steady state conditions while virtual methods allow grid parameters and scattered radiation to be quantified in specific areas of patient anatomy, where the local scatter conditions can vary. The methodology used in this work can be extended to other imaging modalities, including digital breast tomosynthesis, angiography imaging systems and cone beam CT.

## ACKNOWLEDGEMENTS AND CONFLICT OF INTEREST

This work is part of a research collaboration agreement with Siemens Healthlineers. Part of the resources and services used in this work were provided by the VSC (Flemish Supercomputer Center), funded by the Research Foundation - Flanders (FWO) and the Flemish Government.

# Modeling of emerging imaging systems

Tuesday, April 23, 2024

9:40 am - 10:20 am

Chaired by Paul Kinahan & Ehsan Abadi

# Modeling propagation-based x-ray phase-contrast imaging: validity of the projection approximation

Giavanna Jadick and Patrick La Rivière

## BACKGROUND AND PURPOSE

X-ray phase-contrast imaging (XPCI) shows promise for improving image contrast without necessarily increasing radiation dose by probing a subject's refractive properties in addition to its absorptiveness. XPCI would be highly useful in medical scenarios, which often require rapid, high-contrast imaging of tissues with similar x-ray attenuation coefficients, such as a cancerous tumor embedded in healthy tissue. Prior to clinical implementation, corresponding XPCI forward models must be designed to accurately assess prototyped imaging setups, design phase retrieval algorithms, and optimize system parameters. However, a number of challenges have limited the computational application of developed models to date.

An object's refractive or "phase" properties are not immediately apparent in traditional contact x-ray imaging. Several approaches have been proposed for encoding phase-contrast information in an x-ray wave field's intensity.[1] Propagation-based imaging is one prominent technique with the unique advantage of not requiring specialized optics equipment given an x-ray source with sufficient spatial coherence. By moving the detector further from an object, phase differences in the x-ray wave field magnify and manifest as bright and dark intensity fringes at object edges. Unfortunately, discrete modeling of propagation-based XPCI is computationally expensive due to high x-ray wave field sampling requirements. This difficulty is compounded when attempting to invert the forward model for phase retrieval. Consequently, the projection approximation—the assumption that refractive effects within an object are negligible—is commonly applied to simplify computation relative to more accurate approaches.[2] As x-ray imaging technology rapidly advances, achieving larger detector fields-of-view and smaller pixel sizes, the conditions of the approximation are less satisfied, and the quantitative accuracy of phase images may be compromised.[3,4]

It is essential to understand the validity of approximations under various conditions to make informed decisions in imaging simulations. The purpose of this work is to assess the accuracy of the projection approximation in propagation-based XPCI modeling. We consider different x-ray energies, detector resolutions, and object thicknesses, characterizing scenarios in which more computationally intensive forward modeling is likely necessary.

## METHODS

We utilized the Fresnel number $N_F$ as an initial estimate of projection approximation validity:

$$N_F = \Delta^2(\lambda z)^{-1} \quad (1)$$

where $\Delta$ is spatial resolution, $\lambda$ is x-ray wavelength, and $z_0$ is object thickness.[2] The energy $E$ is related to wavelength as $E = hc/\lambda$, where $h$ is Planck's constant and $c$ is the speed of light in a vacuum. The condition $N_F \gg 1$ roughly indicates that refractive effects are negligible, suggesting the projection approximation is reasonable.

However, it is highly idealized in that it neglects more realistic simulation elements, such as detector noise and blur. We computed $N_F$ at a continuum of parameters representing three conventional x-ray imaging modalities (Table 1).

*Table 2. Parameters used for the Fresnel number calculations.*

| Imaging modality | Energy $E$ [keV] | Resolution $\Delta$ [μm] | Thickness $z_0$ [cm] |
|---|---|---|---|
| Micro CT | 20 | 0.1 – 5.0 | 0.1 – 2.0 |
| Mammography | 20 | 10 – 120 | 1 – 20 |
| Clinical CT | 50 | 50 – 600 | 10 – 50 |

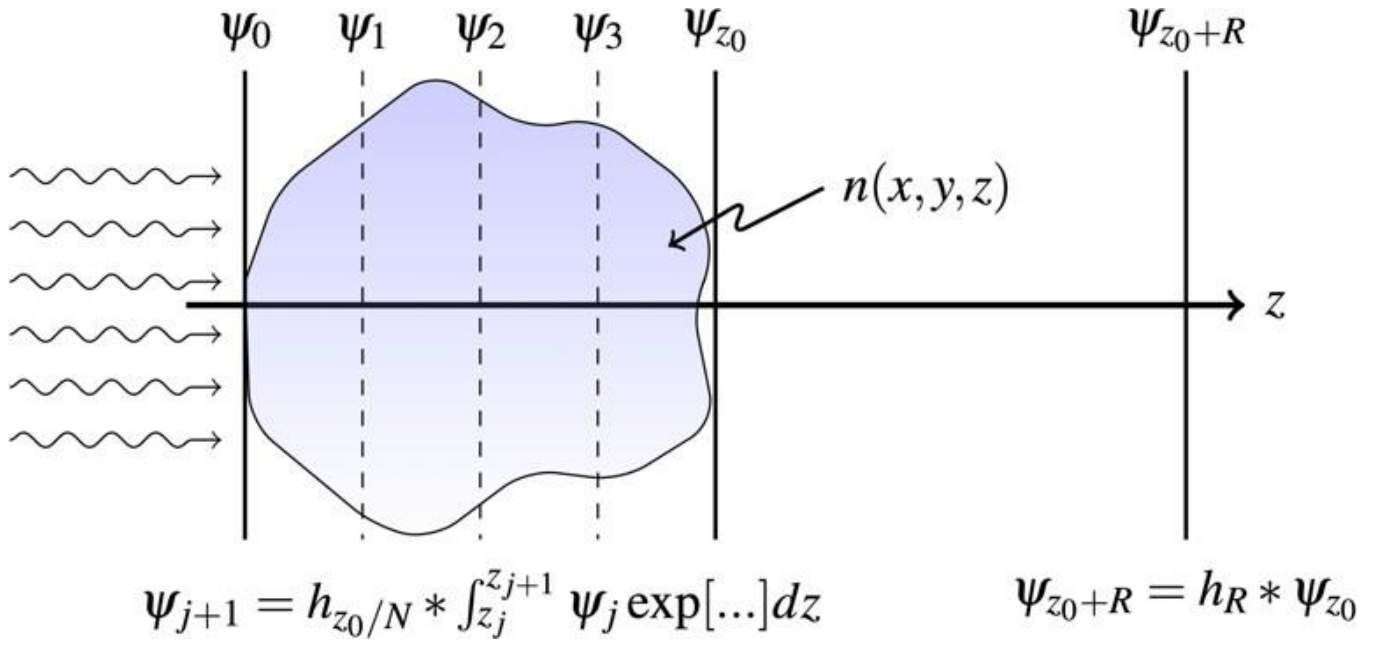

*Figure 3. Schematic of a multislice simulation using N slices, where* exp[…] *indicates line integrals through the object's complex index of refraction (Eq. 2) and $h_z$ is the Fresnel operator for free-space propagation over distance z.*

The Fresnel number findings were used to motivate parameter choice for a synchrotron micro CT-inspired simulation study in which propagation-based XPCI projection images of a computational phantom were generated using either the projection approximation or the more accurate "multislice" forward model.

The incident wave $\psi_0$ was modeled as coherent and monochromatic ($E$ = 20 keV). Each phantom voxel ($x, y, z$) was defined according to its material's complex index of refraction, $n = 1 - \delta + i\beta$, at energy $E$. Under the projection approximation, the wave exiting the phantom is written in terms of the accrued amplitude ($\beta$) and phase ($\delta$) shifts,

$$\psi_{z_0}(x,y) = \psi_0(x,y) \exp\left[-k \int_0^{z_0} dz\, \beta(x,y,z) + i\delta(x,y,z)\right], \tag{2}$$

where $k = 2\pi E/hc$ is the wavenumber, $z$ is the optical axis, and $z_0$ is the location of the object exit plane.[2]

For the multislice forward model (Fig. 1), the object is instead divided into $N$ slices. In principle, each slice $j$ should be sufficiently thin to satisfy the projection approximation. The object exit wave is then computed iteratively, with the wave field at each successive slice $\psi_{j+1}$ found by multiplying the wave incident from the previous slice $\psi_j$ by the line integrals through the object slice (as in Eq. 2) and then applying free-space propagation over the slice width $z_0/N$. We modeled the free-space propagation of a wave as a convolution with the Fresnel operator,

$$\psi_{z_0+R}(x,y) = \Im^{-1}\left\{\Im\{\psi_{z_0}(x,y)\} \cdot \exp\left[-\frac{i\pi hc}{E} R(\nu_x^2 + \nu_y^2)\right]\right\}, \tag{3}$$

where $\Im$ indicates the two-dimensional Fourier transform with respect to $x$ and $y$.[2]

After simulating the incident wave field's modulation through the object with either the projection approximation or multislice model, free-space propagation to the detector was simulated as a convolution with the Fresnel operator over distance *R*. At the detector plane, the wave field was downsampled to match the target resolution and convolved with a normalized, two-dimensional Lorentzian point-spread function with a full-width at half-max equal to 1-μm.[5] The final image *I* was then set to the magnitude of the final complex wave field $\psi_{z0+R}$.

We conducted simulations of two phantoms. First, we used a simple 1-μm PMMA sphere in a 5-mm slab of water to quantify image accuracy when varying the number of multislices from *N* = 0 (projection approximation) to 50. The 50-slice image was used as a reference for calculating divergence *D* relative to the most accurate simulation:

$$D_N = \left( \frac{1}{n_{\text{pixels}}} \sum_{x,y} [I_N(x,y) - I_{\text{ref}}(x,y)]^2 \right)^{0.5}. \quad (4)$$

Two pixel sizes ($\Delta$ = 0.05, 0.5 μm) and three object-to-detector distances (*R* = 0, 50, 100 mm) were simulated.

In the second simulation, we modeled a 5-mm thick, metal-stained zebrafish sample and a detector placed *R* = 50 mm from the object with sampling $\Delta$ = 0.5 μm. Wave modulation through the object was simulated using either the projection approximation or the multislice approach with 2, 4, and 16 slices.

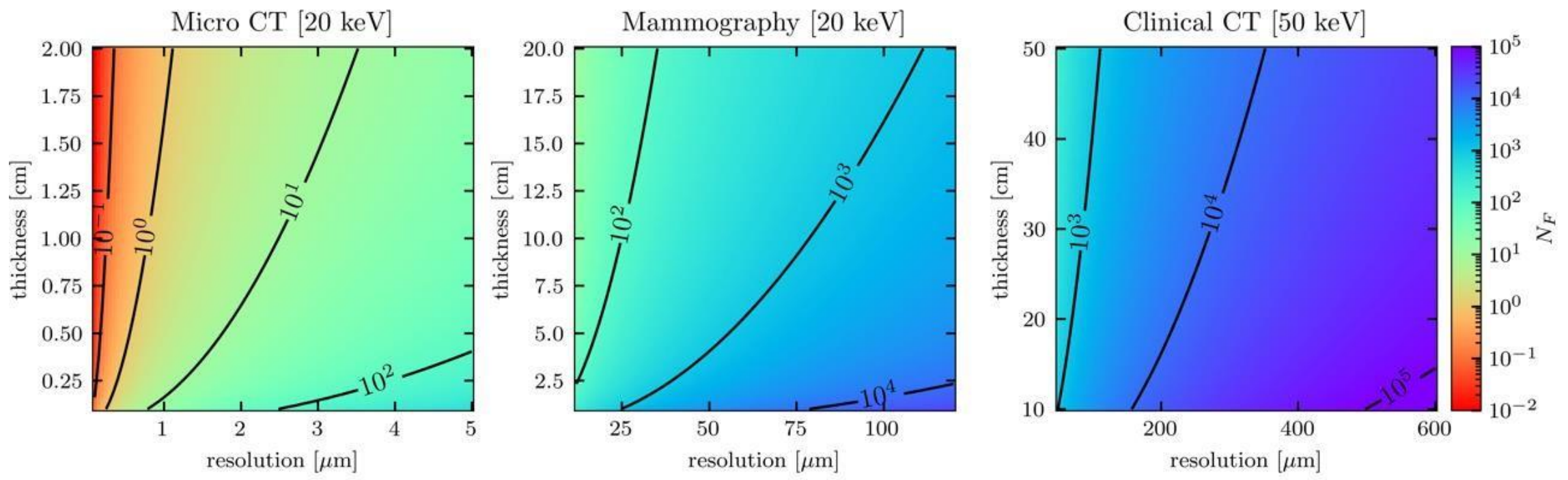

*Figure 4. Heatmaps of the Fresnel number evaluated at parameters typical for three x-ray imaging modalities.*

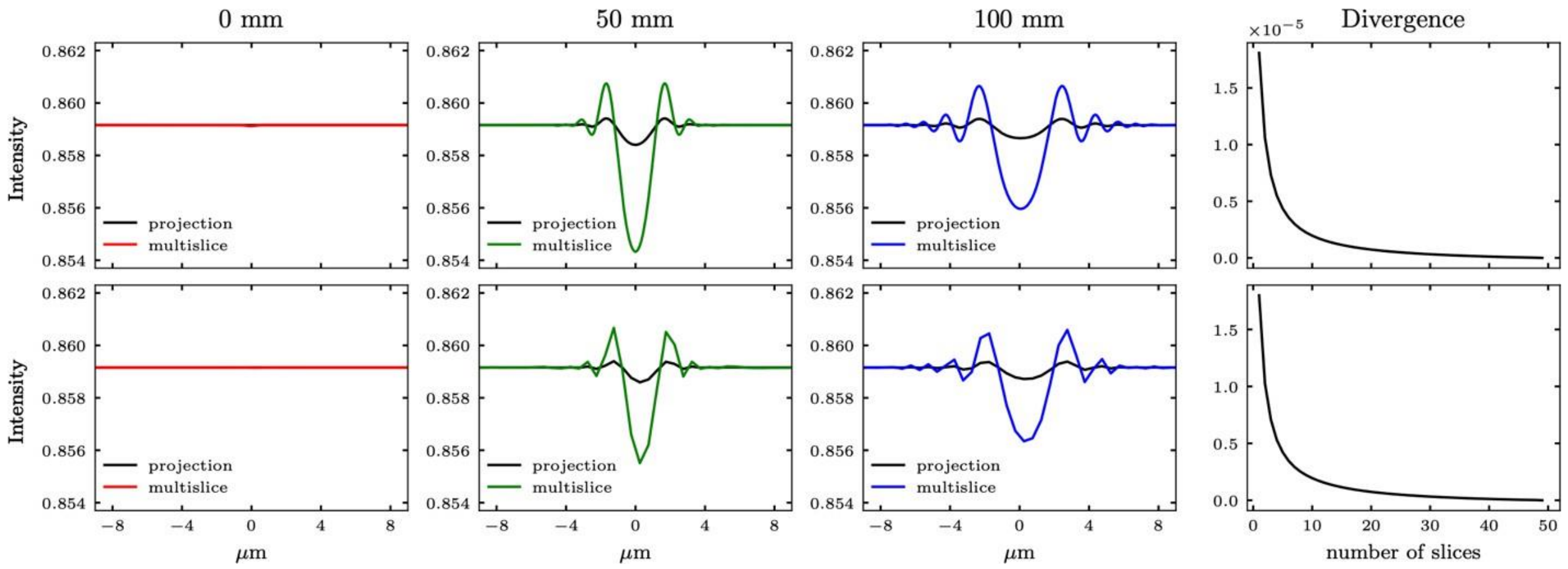

*Figure 5. Projection approximation vs. multislice (N=50) line profiles through the 1-μm PMMA sphere with (top) 0.05-μm and (bottom) 0.5-μm detector sampling. The right column shows exit wave divergence as a function of number of slices N.*

# RESULTS

Figure 2 shows heatmaps of the Fresnel number for ranges of parameters corresponding to micro CT, mammography, and clinical CT imaging scales. The condition $N_F \gg 1$ is well satisfied for the mammography and clinical CT regimes but not at high resolutions of interest for micro CT. It is possible that with thicker samples or lower energies, high-resolution mammography may benefit from multislice modeling.

Figure 3 shows line profiles through the simulated images of the 1-μm PMMA sphere in water. This imaging task has poor x-ray absorption contrast, indicating great potential for contrast improvements using XPCI. The projection approximation always underestimates the amplitude of edge fringes, especially in the high-resolution images where fringe width is on the same order as the sampling distance. This effect worsens as the free-space propagation distance to the detector is increased, which increases the visibility of refractive effects. This is apparent in the zebrafish images, where accuracy noticeably improves as the number of slices increases (Fig. 4). The qualitative improvement approaches a plateau at slice widths thinner than approximately 1 mm, with a number of slices in the range from 4 ($N_F$ = 3.2) to 16 ($N_F$ = 12.9) depending on the desired simulation accuracy. This is consistent with expectations based on the Fresnel number heatmaps in Fig. 2.

# DISCUSSION

We present an assessment of the validity of the projection approximation in forward modeling of propagation-based XPCI. This approximation is a hallmark of widely implemented phase retrieval techniques such as Paganin's method for a single defocused image.[6] While such approaches may yield qualitatively pleasing images, their quantitative accuracy can be compromised when approximation conditions are not well satisfied. This is of particular interest in the context of emerging micron- and sub-micron-resolution x-ray detectors.[3,4] By characterizing the scenarios in which more accurate forward modeling is needed, one can additionally determine when the more convenient approximate methods are sufficient, saving computational expense.

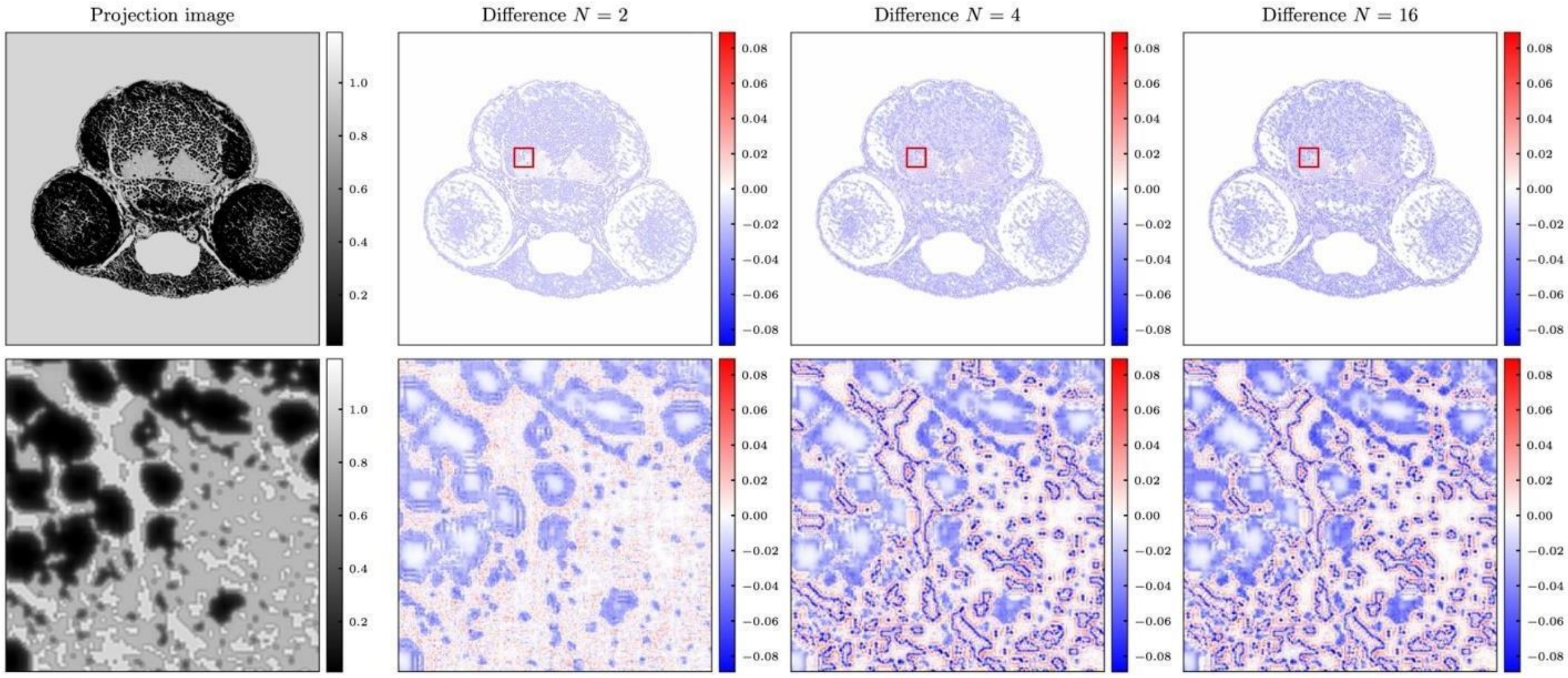

*Figure 6. The zebrafish images simulated using the projection approximation (left) and its difference with multislice (right columns) over a 2.5-mm field-of-view (top) and a 0.15-mm region-of-interest (bottom) as indicated by the red squares. For 1, 2, 4, and 16 slices, the Fresnel number is 0.8, 1.6, 3.2, and 12.9,*

For emerging micro CT systems, quantitative XPCI will likely require phase retrieval algorithms based on a more accurate forward model for wave field propagation through an object, such as the multislice approach. The precise conditions depend on the imaging scenario at hand—object thickness, detector characteristics, and object-to-detector distance. Using our simple 1-μm sphere simulations, one can define a maximum acceptable error threshold to determine the minimum number of multislices needed in the forward model. For our zebrafish example, the qualitatively visible accuracy approaches a plateau with slices thinner than 1 mm. The framework can be extended to other propagation-based XPCI setups.

## CONCLUSION

The projection approximation is computationally convenient for discrete propagation-based XPCI modeling. Based on an analytical estimate, we expect the approximation is sufficiently accurate for XPCI simulations at clinical CT scales. For high-resolution systems such as micro CT, multislice modeling is likely necessary. Using a simulation study, we quantify accuracy divergence as a function of the number of multislices used. This work motivates the development of propagation-based phase retrieval algorithms based on this more accurate forward model.

## ACKNOWLEDGEMENTS AND CONFLICT OF INTEREST

The authors declare no conflicts of interest. This work is supported by the United States National Science Foundation Graduate Research Fellowship Program under Grant No. 2140001.

# Simulation Tools to Optimize the Scanning Motion for Next-Generation Tomosynthesis

Raymond J. Acciavatti, Priyash Singh, Chloe J. Choi, Andrew D. A. Maidment
University of Pennsylvania, Department of Radiology, 3400 Spruce Street, Philadelphia PA 19104

## BACKGROUND AND PURPOSE

Digital breast tomosynthesis (DBT) is now in widespread use for breast cancer screening. In current clinical systems, the x-ray source moves only in the left-right (LR) direction (Fig. 1), yielding a cone-beam artifact in the perpendicular direction. This artifact can be visualized with a Defrise phantom.[1] The phantom consists of multiple plastic plates at low-frequency spacing. The phantom is well visualized in the LR direction but not in the posteroanterior (PA) direction, since rays must traverse multiple plates in the PA direction before reaching the detector.

Our research lab has constructed a prototype next-generation tomosynthesis (NGT) system to analyze how image quality in DBT can be improved. The NGT system is capable of source motion with an additional degree of freedom in the PA direction, unlike a clinical DBT system. Our previous work demonstrated that the cone-beam artifact is suppressed with a T-shaped scanning motion.[1] The purpose of this study is to investigate a new motion following a convex path (forming a loop). We use theoretical modeling and breast-shape analysis to investigate whether this new design improves three measures of image quality relative to the T scan; namely, the cone-beam artifact, out-of-plane blurring, and breast-outline overestimation effects.

## METHODS

In addition to a conventional scan, T and convex scans (Fig. 2) are simulated with 191.7 mm of PA motion. All motions have 13 projections and a LR range of 191.7 mm. We model the AXS-2430 detector (Analogic Canada Corporation, Montreal, Quebec) used in the NGT system. This detector has 0.085 mm pixelation and an active area of 304.64 mm × 239.36 mm. The detector is –24.46 mm below the breast support. The source motion is constrained by a spherical arc (radius 744.05 mm) centered on the centroid of the breast support.

The cone-beam artifact is analyzed with a Defrise phantom (frequency 0.20 $mm^{-1}$, thickness 50.0 mm). The attenuation coefficient is treated as sinusoidal (as opposed to a square wave), as we have developed a closed-form solution for the projection image of a sinusoidal test object. Modulation is calculated in the LR and PA directions in the central slice of the phantom by normalizing signal to the corresponding signal in a uniform phantom with equivalent dimensions.

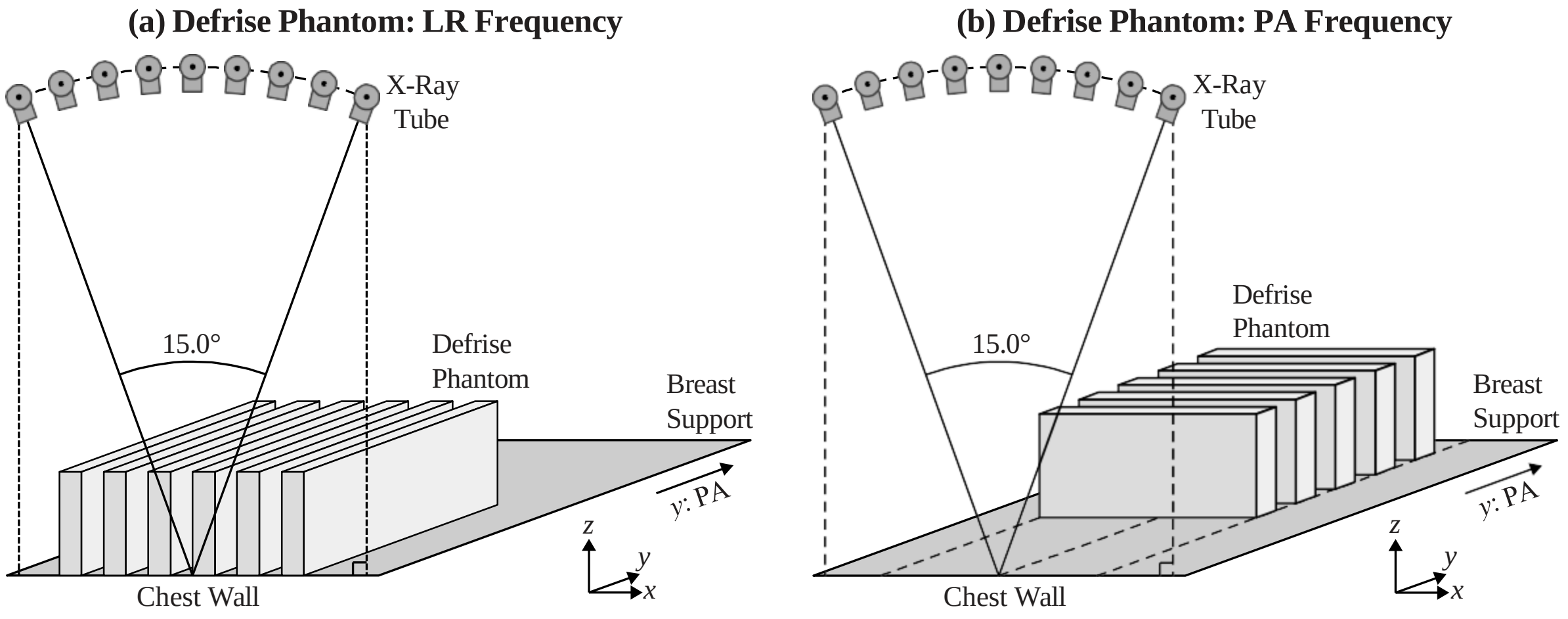

*Figure 1. Defrise phantoms are simulated with two frequency orientations: left-right (LR) and posteroanterior (PA).*

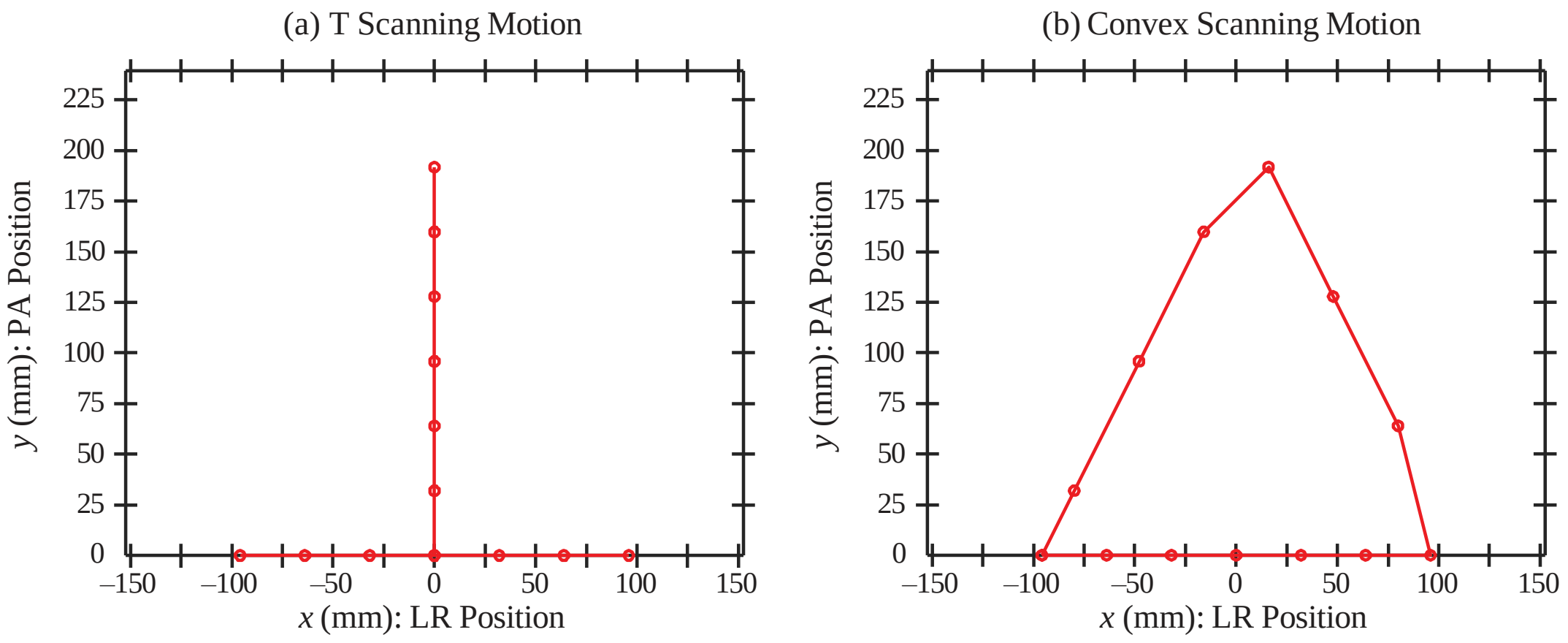

*Figure 2. The NGT system supports PA motion unlike current clinical DBT systems. Two simulated motions are shown.*

To measure out-of-plane blurring, the point spread function (PSF) is calculated in the *z* direction. Three metrics are derived from the PSF; *i.e.*, full width at quarter maximum (FWQM), full width at half maximum (FWHM), and full width at three-quarters maximum (FW3QM). These metrics are illustrated in Fig. 3 at the point (0, 54.0, 25.0) (in mm) under the convex motion. Our previous work[2] measured the spatial distribution of each metric under 1,000 random samplings of points in a volume with dimensions 200.0 × 100.0 × 50.0 (in mm). Cumulative histograms of each metric are plotted with 95% confidence intervals using bootstrapped resampling (200 resamples of 1,000 random points).

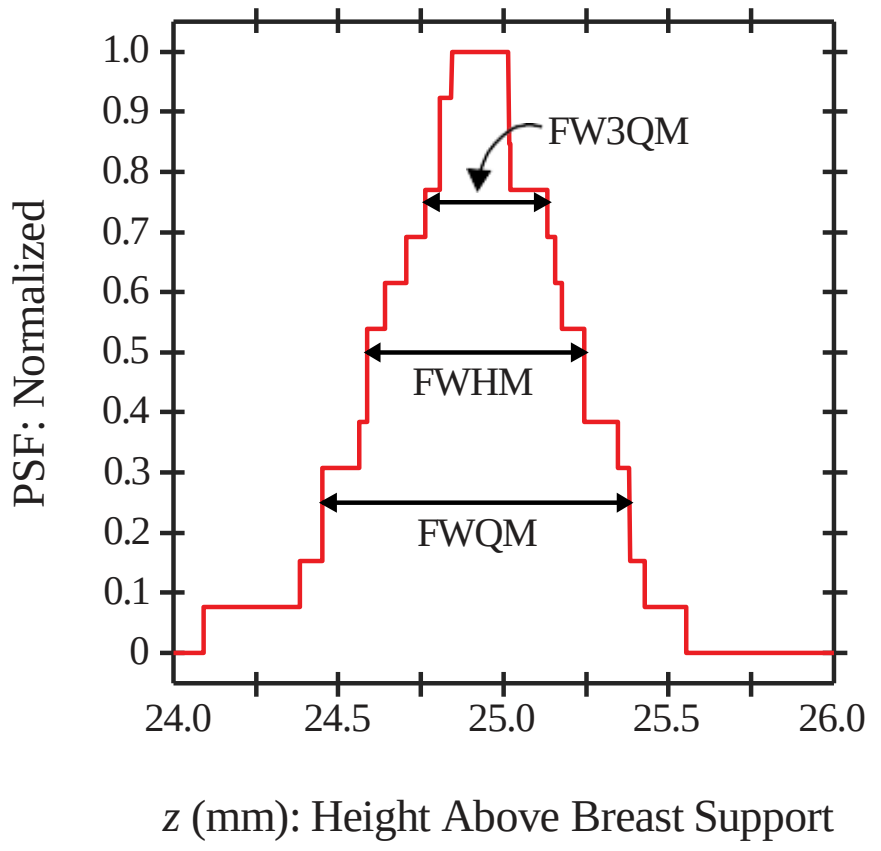

*Figure 3. PSF calculated in z*

For calculation of the 3D breast outline, we use the model of the compressed breast developed by Rodríguez-Ruiz *et al.*[3]; specifically, the average breast with advanced 3D curvature (chest wall-to-nipple distance of 97 mm and 1.0 $mm^3$ voxels). To summarize this calculation, which was described in our earlier work[4], detector pixels corresponding to air signal in each projection are identified. Rays are then backprojected from these pixels to the source, identifying voxels corresponding to air. This information is combined across projections to generate the breast outline. Simulations of the breast-outline segmentation were performed with 2 × 2 detector element binning.

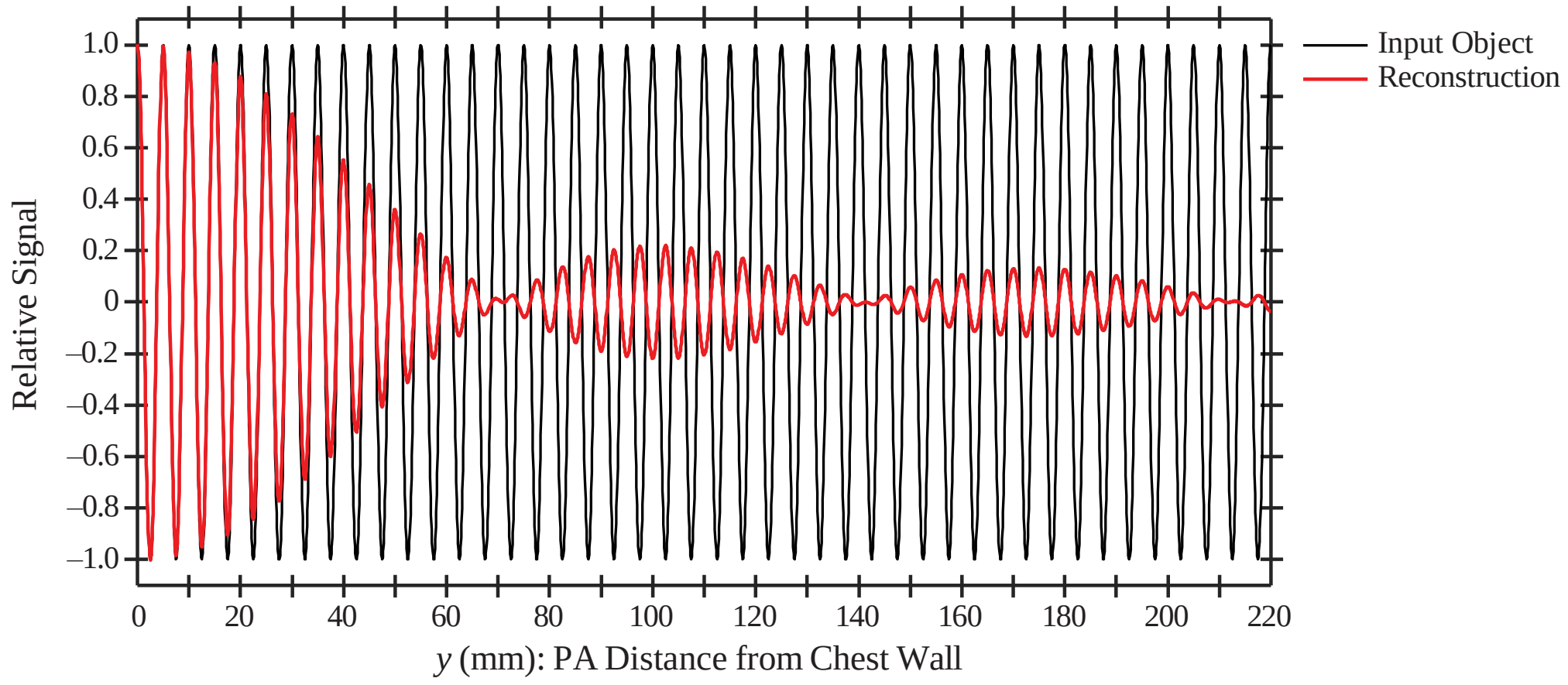

*Figure 4. Defrise phantom reconstruction in the PA direction illustrating the cone-beam artifact in the conventional geometry.*

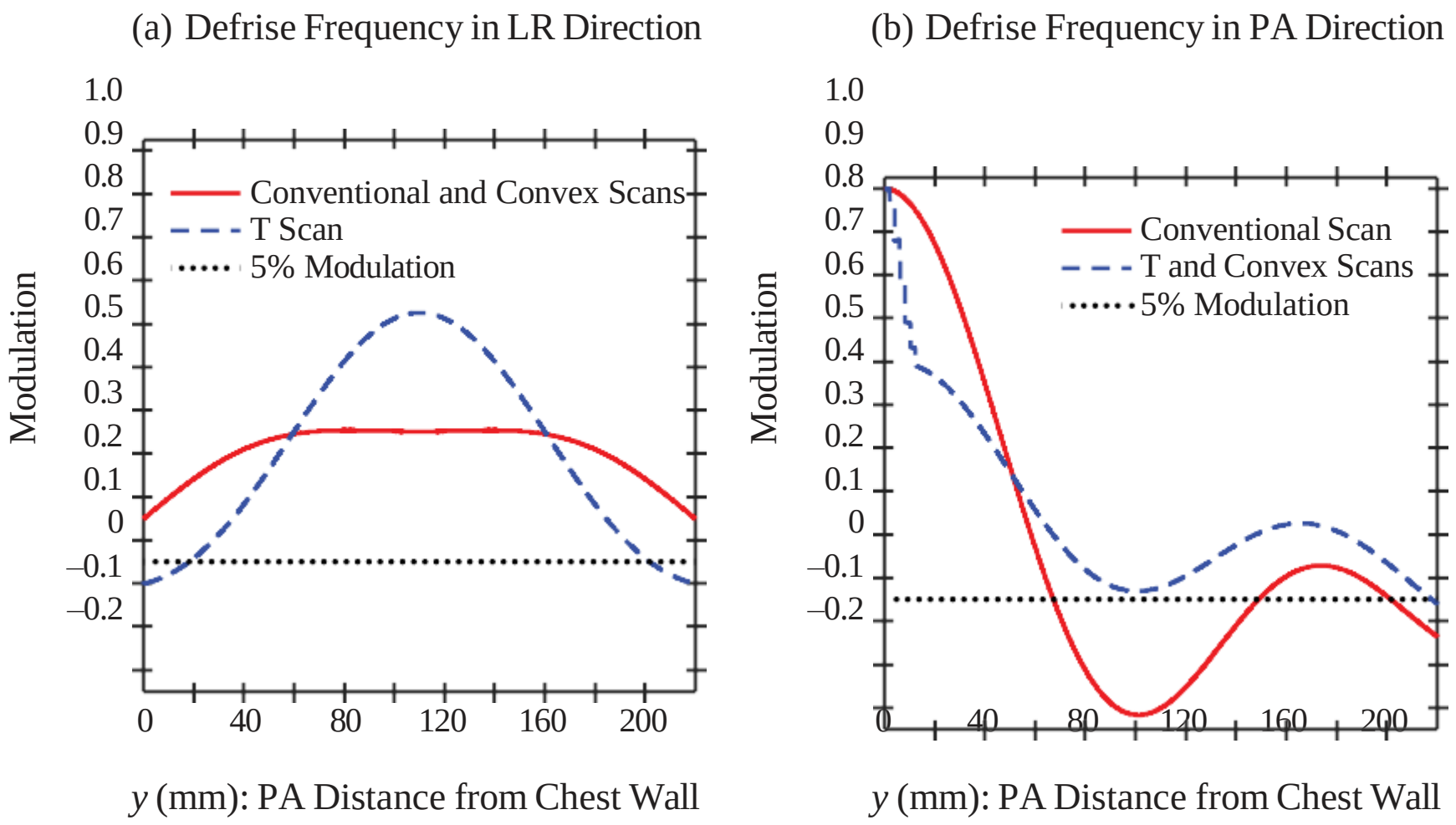

*Figure 5. The threshold of 5% is set for detectable modulation in Defrise phantom plot profiles under various scanning motions.*

# RESULTS

In the conventional scan, there is a cone-beam artifact in the PA direction (Fig. 4), as shown by the drop in signal distal to the chest wall. The signal amplitude falls to zero at the position 70.7 mm distal to the chest wall. There is a 180° phase shift between 70.7 and 141.4 mm distal to the chest wall. In the LR direction, Defrise modulation is kept above 5% under both conventional and convex motions but not under the T motion [Fig. 5(a)]. In the PA direction, both T and convex motions offer an advantage over the conventional motion; modulation is kept above 5% [Fig. 5(b)].

In Fig. 6, PSF width metrics are analyzed with cumulative histograms. Shaded areas denote 95% confidence intervals determined from bootstrapped resampling of 1,000 random points in the volume. The three PSF width metrics are minimized most noticeably under a convex scanning motion, benefiting image quality.

The breast-volume overestimate (difference relative to input phantom, 625.9 mL) is 33.3, 27.8, and 27.3 mL under conventional, T, and convex motions, respectively. Thus, the volume overestimate is reduced by 16.5% and 18.0% under T and convex motions, respectively, relative to the conventional motion. Three slices in the reconstruction are illustrated (Fig. 7). The T and convex motions offer an advantage over the conventional motion in the inferior slice (*z* = 0.5 mm); the breast outline in the PA direction more closely matches the input phantom. All three scans yield similar breast outlines in the mid-slice (*z* = 24.5 mm), and these three outlines all match the input phantom well. The breast outlines of the three scans are essentially identical in the superior slice (*z* = 50.5 mm), but overestimate the breast size in the PA direction relative to the input phantom.

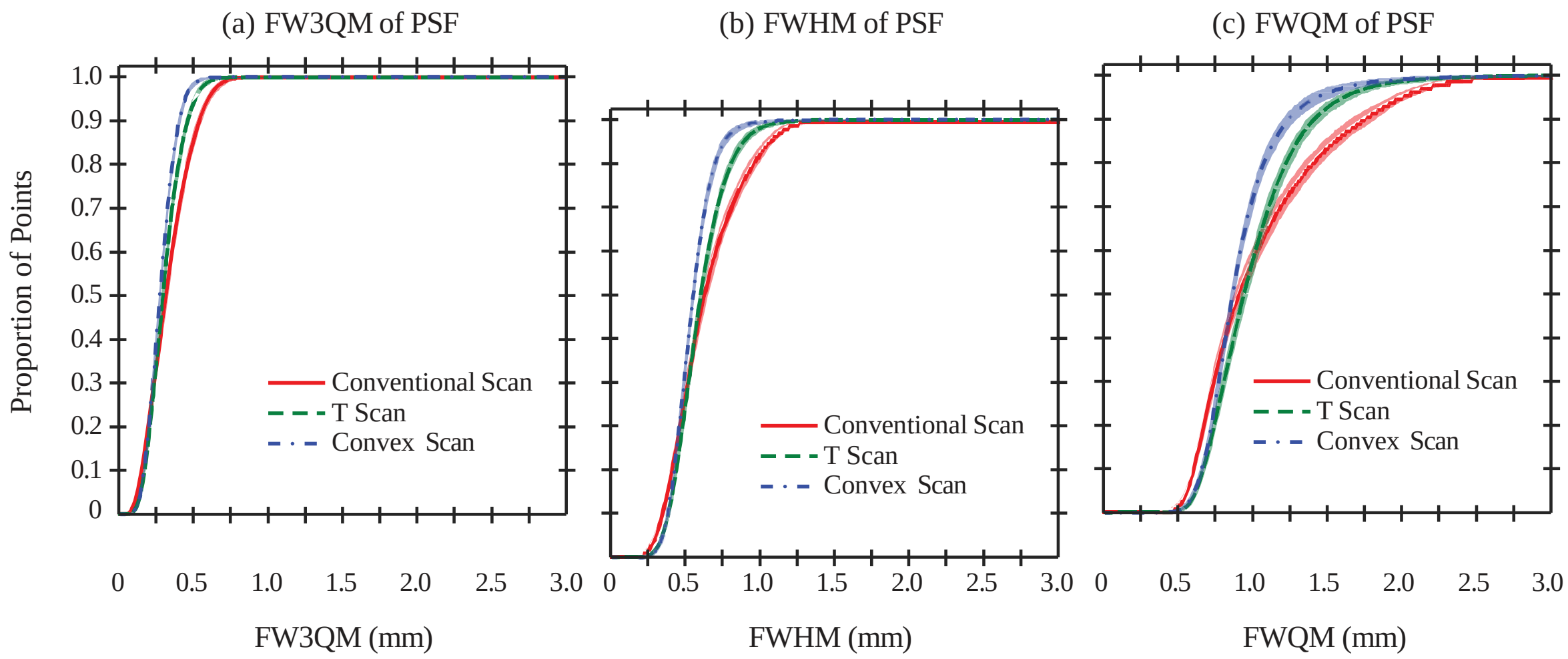

*Figure 6. Blurring in the z direction is mitigated by the convex scanning motion as illustrated by three PSF width measures.*

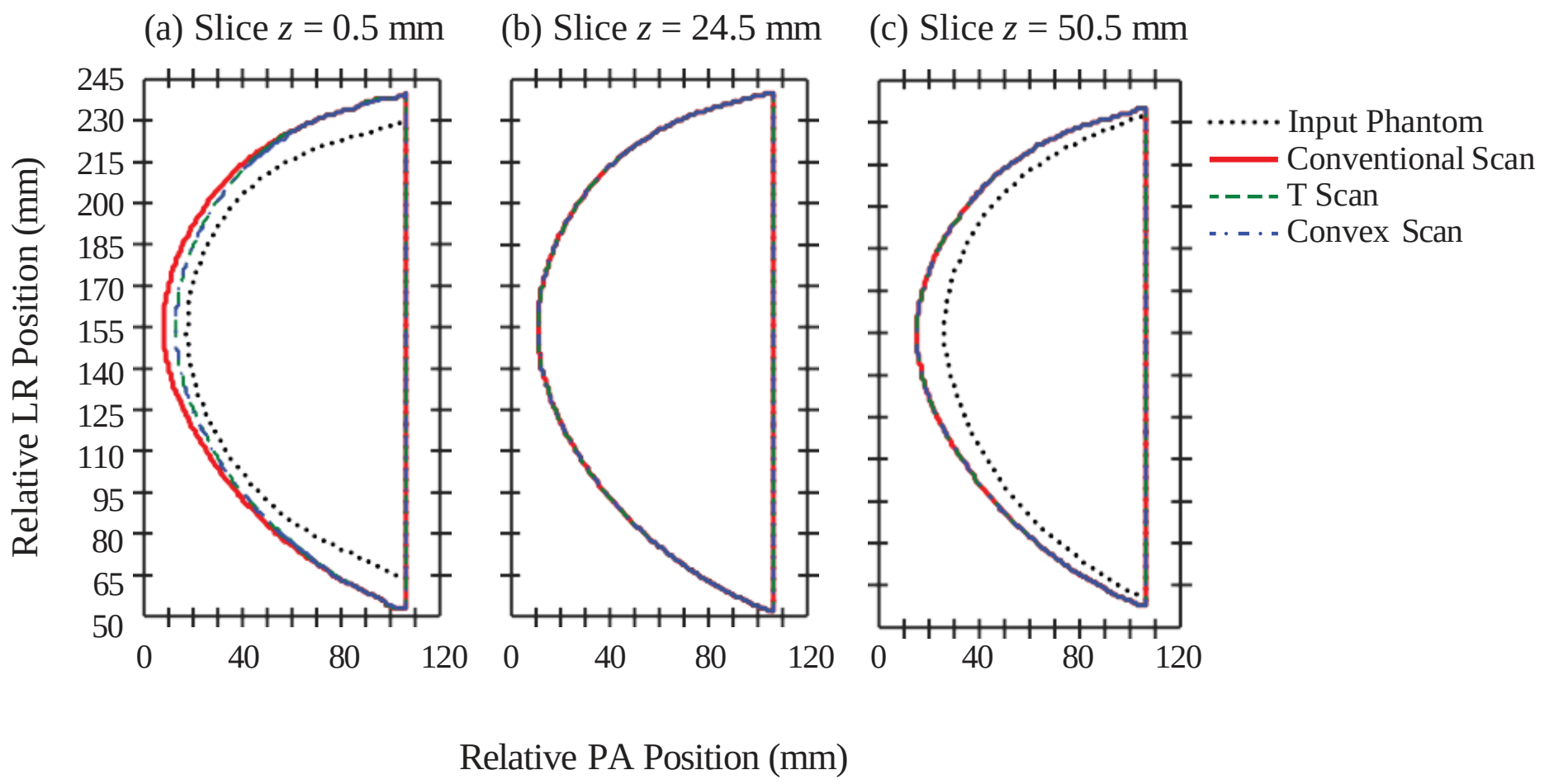

*Figure 7. Breast-outline segmentations in three slices of the reconstruction under conventional, T, and convex scanning motions.*

# DISCUSSION

Our previous work demonstrated advantages of the T scan over the conventional scan; specifically, in improving the PA Defrise modulation in physical experiments[1], minimizing out-of-focus breast-density artifacts in the *z* direction in virtual phantoms[5], and improving visualization of the PA extent of the breast outline in a physical phantom[4]. The convex scan proposed in this paper continues to demonstrate advantages over the conventional scan, but also offers two potential benefits over the T scan; specifically, in terms of LR Defrise modulation and PSF width metrics in the *z* direction.

## CONCLUSION

This paper describes simulation tools that can rapidly identify a promising new scanning motion. Our next step is to validate these simulations with physical experiments using the NGT system. Future work should also investigate the scanning motions in terms of lesion detectability in breast phantoms.

## ACKNOWLEDGEMENTS AND CONFLICT OF INTEREST

Andrew D. A. Maidment and Raymond J. Acciavatti are inventors on patents and patent applications related to next-generation tomosynthesis (NGT) technology. Andrew D. A. Maidment has received research support from Hologic Inc., Barco NV, and Analogic Corporation; is the spouse to an employee and shareholder of Real Time Tomography (RTT), LLC; is a member of the scientific advisory board of RTT; and is an owner of Daimroc Imaging, LLC. Support was provided by Breast Cancer Alliance, Inc. (2022 Young Investigator Grant) and the National Institutes of Health (T32EB009384, R01CA196528, R37CA273959, and P30CA016520). The content is solely the responsibility of the authors and does not necessarily represent the official views of the funding agencies.

# Installing photon-counting CT in DukeSim – A validated and computationally efficient simulation framework accounting for scanner-specific physics and acquisition geometry

Cindy McCabe, Brian Harrawood, Ehsan Samei, Ehsan Abadi

## ABSTRACT

The purpose of this study was to develop a computationally efficient CT simulator that realistically accounts for conditions of clinical photon-counting CT (PCCT) systems. This simulator was built upon a state-of-the-art platform (DukeSim) which simulate CT projection images of computational phantoms given the geometry and physics of the scanner and acquisition settings. To model the photon-counting detection process, we utilized a Monte Carlo-based detector model with the known properties of the detectors to inform the signal and variance for a 3-by-3-pixel array at each preset energy threshold, as well as the spatio-energetic covariance between the thresholds for the 3-by-3-pixel arrays. DukeSim was augmented to account for the geometry and physics of PCCT systems, including correlated noise between thresholds, and low-dose physics for photon starved regions. We validated the simulation platform against experimental measurements using a physical phantom and a clinical PCCT scanner (NAOETOM Alpha, Siemens). The images were acquired at four dose levels (CTDIvol of 1.5, 3.0, 6.0, and 12.0 mGy) using an ACR phantom – a cylindrical phantom with four inserts (bone, air, acrylic, and polyethylene). Each acquisition was reconstructed with three kernels (Br40, Br48 and Br56) and two slice thicknesses (0.4mm, and 1.0mm) using a 512-matrix size. The experimental acquisitions were replicated using our developed simulation platform. The real and simulated images were quantitatively compared in the matter of image quality metrics (HU values, noise magnitude, noise power spectrum, and modulation transfer function). The discrepancy between the real and simulated data was on average 1.25 HU in terms of noise magnitude and 0.0026 mm-1 in terms of the frequency at 50% MTF. We successfully developed a validated virtual imaging platform for the emerging CT technology, enabling the comparison of various CT systems, as well as the optimization of imaging parameters (including dose optimization) for specific clinical tasks.

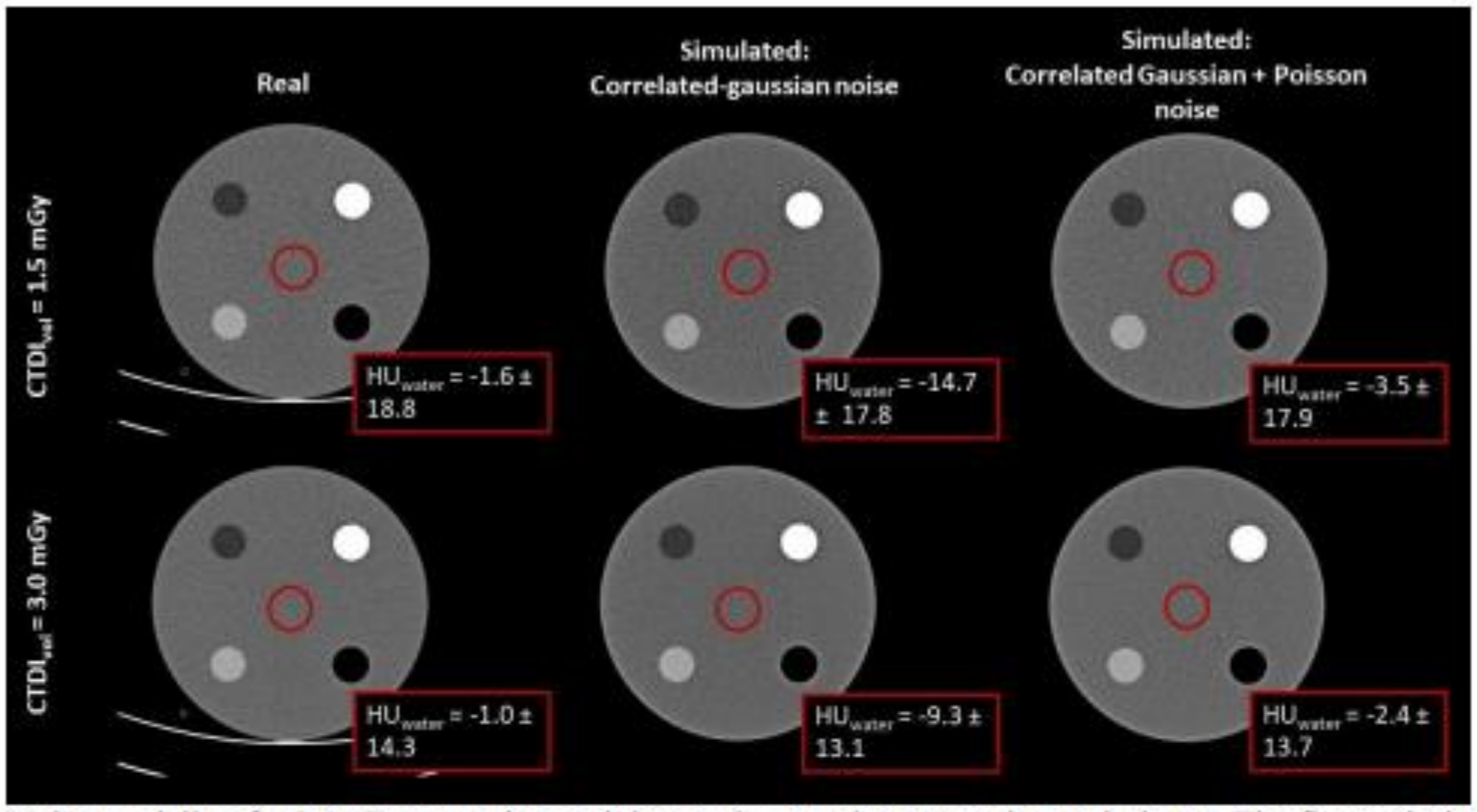

**Figure 1**: Noise modeling for PCCT comparing real data and two noise generation techniques. The first technique utilizes correlated gaussian noise—however, this leads to HU inaccuracy especially at low doses as seen in the middle column. The second noise model uses a correlated gaussian noise model when there are enough photons detected, and it will switch to a Poisson noise generation technique in a photon starved regions.

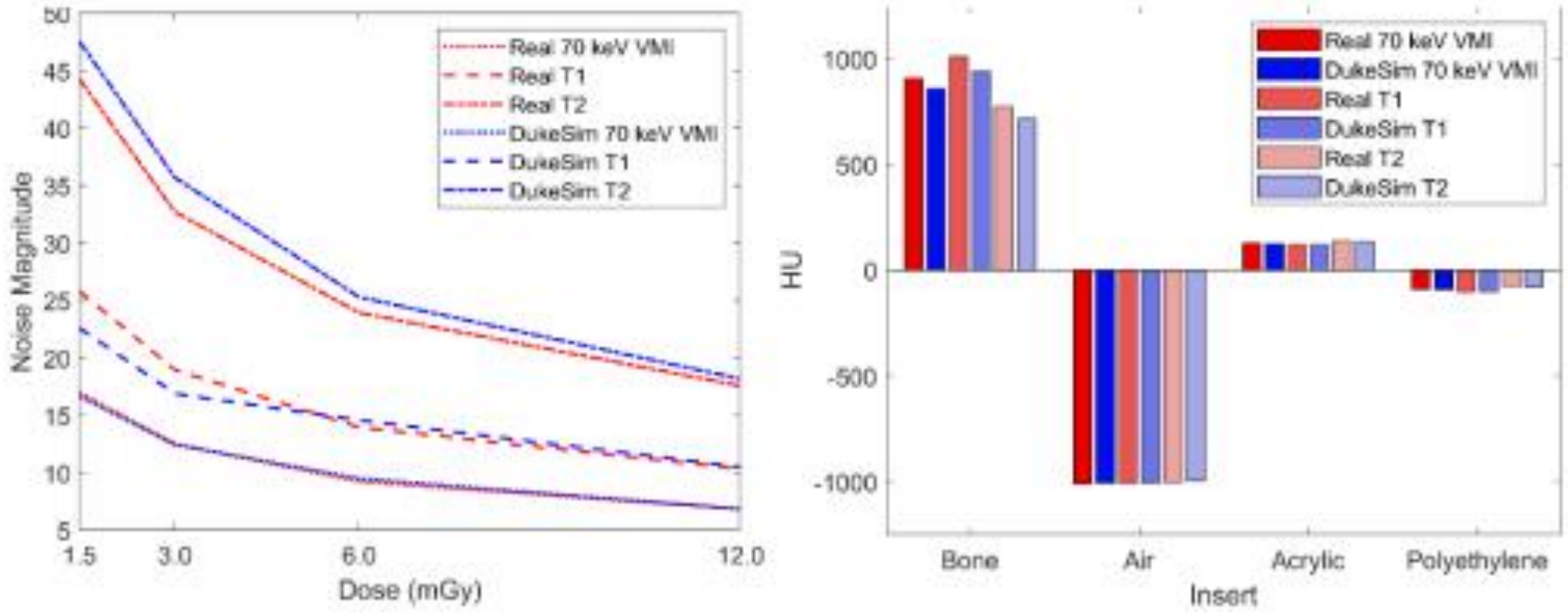

**Figure 2**: Noise magnitude plot (left) for four dose levels (1.5, 3.0, 6.0, 12.0 mGy) using the Br40 kernel. Real data is represented in red and simulated data is represented in blue. HU values (right) for four inserts for each image series at 1.5mGy using the Br40 kernel. Simulated data is represented by a lighter shade of the bar color compared to experimental data.

# Late-breaking demonstrations

Tuesday, April 23, 2024

10:50 am - 11:30 am

Chaired by Kristina Bliznakova & Stephen Glick

# CT Performance in Quantifying Small Airway Disease using Paired Inspiratory-expiratory Scans: A Virtual Imaging Study

Fong Chi Ho, Amar Kavuri, Saman Sotoudeh-Paima, Mobina Ghojogh Nejad, Ehsan Samei, W. Paul Segars, Ehsan Abadi

## BACKGROUND AND PURPOSE

**Chronic obstructive pulmonary disease** (COPD) is characterized by multiple underlying conditions, including emphysema and chronic airway inflammation[1]. The pathological changes of the lungs in COPD have been demonstrated using quantitative computed tomography (CT) for chronic bronchitis and small airways disease (SAD) respectively, by measuring the inflamed bronchial wall thickness with bronchiole internal diameter larger than 2 mm[2], and density measurements of the nonemphysematous gas-trapping region with bronchiole internal diameter less than 2 mm by parametric response mapping (PRM)[3].

The PRM registers CT images in both end-respiration states to measure the density differences in emphysema and SAD region simultaneously, which has shown a progression from SAD to emphysema with increasing COPD severity [4]. Thus, an accurate quantification of functional Small Airway Disease (fSAD) as gas trapping in airway anatomy is vital to reflect the severity and the early diagnosis of both emphysema and SAD. However, the density-based fSAD measurement does not fully reflect the major abnormality in SAD being the complete airway occlusion by the chronic mucus hypersecretion and concentration. Therefore, mucus score is utilized as a clinically related biomarker to quantify the occurrence of complete airway occlusion by mucus plugs using a bronchopulmonary scoring system[5]. Previous studies have demonstrated correlations between mucus score of COPD patients without mucus-related symptoms and the associated reduction in clinical disease measurement accuracy[6]. Hence, the evaluation of airway occlusion on CT images complements the COPD patient assessment.

However, quantifying airway occlusion is limited by the CT image quality. CT image quality has been recently improved by the introduction of photon-counting CT (PCCT) that offers enhanced spatial resolution and reduced image noise compared to conventional energy-integrating CT (EICT) systems. This advancement holds the potential to more accurately quantify mucus plugging and the anatomy of functional Small Airway Disease. To assess such potential, virtual imaging trials can be utilized since, providing a ground truth known framework for a systematical evaluation and the parameter optimization investigation. In this study, we developed and utilized the first suite of virtual COPD patients with mucus modeling at both end-respiration states to evaluate the PCCT against EICT in mucus quantification by a novel imaging biomarker, PercOcclusion.

## METHODS

### Section 1. Computational Models with COPD

Five anthropomorphic, computational human models (XCAT phantoms) were developed at both end-inspiration and end-expiration states ($XCAT_{INSP}$ and $XCAT_{EXP}$) with varied severity of COPD based on the clinical COPDGene Phase 1 dataset. The organs and major structures were defined for each XCAT phantom [7, 8] by segmenting the clinical CT cases in both end-respiration states using a vendor-based segmentation program (DirectOrgan 2.0, Siemens) and other segmentation methods [9-11].

The emphysema and fSAD masks per end-expiration CT case were segmented by the PRM followed by a median filter of 5x5x5 pixels and a connected component analysis to reduce the image noise and eliminate the smaller blobs segmentation, respectively. The filtered PRM masks were fine-tuned by a radiologist to classify as healthy parenchyma, emphysema, and fSAD regions.

After modeling the XCAT for both end-respiration states using initial organ segmentation, a large deformation diffeomorphic metric mapping (LDDMM) was performed to transform the $XCAT_{EXP}$ to the $XCAT_{INSP}$ by spatially aligning every organ across the states[12]. The same diffeomorphic transformation was applied to obtain a PRM in inspiration ($PRM_{INSP}$) from the PRM masks in expiration ($PRM_{EXP}$) containing the definition of emphysema, fSAD, and overlapped region. Based on the fSAD distribution on $PRM_{INSP}$, mucus of various densities, shapes, locations, angles, and lengths were modeled in the $XCAT_{INSP}$ (Section 2). Mathematically extended models for the airways and vessels were also incorporated in the $XCAT_{INSP}$ implementing bronchitis characteristics following pathological COPD patient statistics[13]. The $XCAT_{INSP}$ was then statistically textured with parenchyma and its secondary pulmonary lobules defined using the techniques of Abadi et al [13]. An inverted LDDMM of the same diffeomorphisms was performed to complete a $XCAT_{EXP}$ with parenchyma, detailed airways and vessels, and one-to-one mucus mapping to the $XCAT_{INSP}$. The complete workflow of phantom generation in both respiration states is illustrated in **Figure 1**.

## Section 2. Mucus modeling

Chronic entrapped mucus was modeled with the distributions of shape (round, oval, lobulating, and complex), location (anterior and posterior), and angle (acute and obtuse) based on contrast-enhanced CT measurements [14]. The mucus was installed in the earliest generation of bronchiole in every connected region of gas trapping indicated by the $PRM_{INSP}$. Mucus size was modeled to fit 10% to 100% of the airway in varied lengths, showing different levels of bronchiole occlusion and airflow limitation. A diverse density of mucus was assigned with resulting 0 HU to 130 HU in CT acquisitions, demonstrating the gradual production of mucus protein and a steady reabsorption of the water content in the chronic entrapped mucus[15].

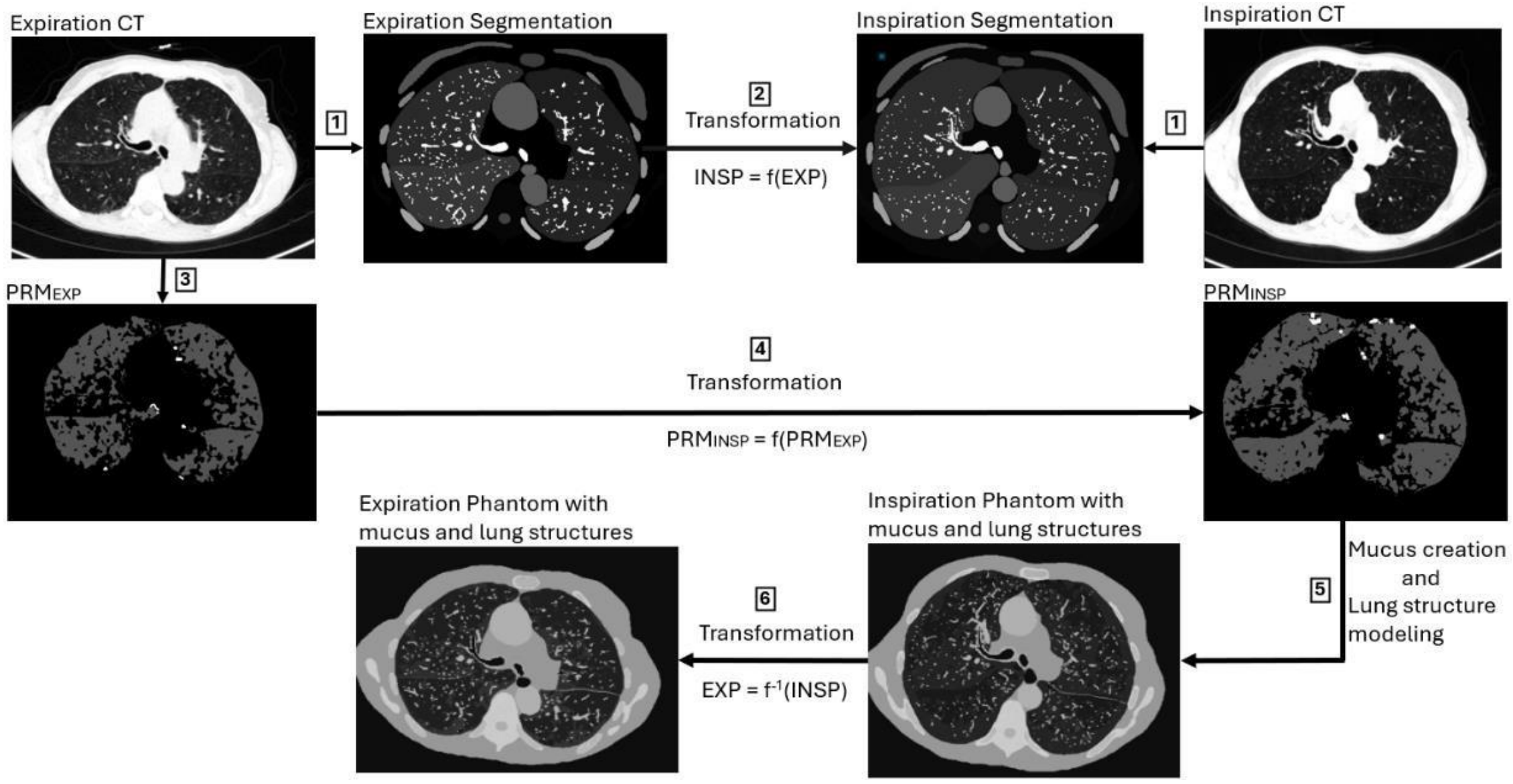

***Figure 1**. The workflow in paired inspiratory-expiratory phantom creation. (1) Initial CT organ segmentations. (2) LDDMM performed diffeomorphic transformation. (3) Fine-tuned $PRM_{EXP}$ masks with classified emphysema and fSAD regions.* ***(4) LDDMM mapping of the***

***$PRM_{EXP}$ to $PRM_{INSP}$. (5)** $XCAT_{INSP}$ modeled with detailed lung structures and mucus as informed by the $PRM_{INSP}$. (6) $XCAT_{EXP}$ with detailed lung structure and one-to-one mapping of mucus from $XCAT_{INSP}$ using an inverted LDDMM.*

## Section 3. Virtual Acquisitions

To simulate CT images from the phantoms, we used the DukeSim CT simulator [16]. In this study, the phantom projections were generated by DukeSim simulating a clinical PCCT and an EICT (NAEOTOM Alpha, SOMATOM Definition Flash, Siemens). The virtual CT acquisitions used 120kV at dose levels of 20, 60, and 100 mAs with typical tube current modulation (TCM)[16]. The projections were reconstructed using weighted filtered back projection (wFBP) and iterative reconstructions techniques (SAFIRE and PNR) in vendor-specific reconstruction software (ReconCT, Siemens), with a field of view of 500 mm in 2 pixel sizes (0.49 and 0.98 mm), two slice thicknesses (0.75 and 1.5 mm for FLASH and 0.4 and 1.5 mm for NAEOTOM Alpha), and two reconstruction kernels (Q40 and Qr61).

## Section 4. Quantitative Assessments

The mucus was quantified using a novel imaging biomarker, 'PercOcclusion', which quantifies the percentage of the airway in the area obstructed by the mucus to distinguish the partial bronchiole occlusion from the complete occlusion.

$$PercOcclusion = \frac{Max(Area_{mucus})}{Area_{airway}} \; x \; 100\%$$

The PercOcclusion was measured and referenced to the ground truth, which was established by generating a mono-energetic representation of the phantoms at the effective energy (70 keV) of the acquisition process. The accuracy and variability of the biomarker measurements for each $XCAT_{INSP}$ across all cases were evaluated against that of the ground truth in terms of the absolute error mean and standard deviation.

## RESULTS

**Figure 2** illustrates the variability of mucus modeled across four shapes, two locations, and two angles. Full occlusion and 30% occlusions are shown. **Figure 3** shows a qualitative comparison of the EICT (FLASH) and PCCT (NAEOTOM Alpha) scanners imaged at 100mAs and reconstructed in 2 different kernels at the smallest slice thickness available (0.75 mm for FLASH and 0.4 mm for NAEOTOM Alpha). In this example, a mucus with a ground truth density of 0 HU and a PercOcclusion of 53 was inserted in the airway based on the $PRM_{INSP}$ measurement. The same mucus was labeled in a magnified area for every scanner condition. The partially occluded mucus is visible in both EICT and PCCT images but is shown as complete occlusion in the EICT images. The improved spatial resolution of the PCCT scanner shows superior details in the mucus silhouette compared to the EICT scanner, distinguishing the partial occlusion, and offering a closer resemblance to the ground truth. The use of the sharp kernel in the PCCT scanner results in a sharper boundary and less dilated mucus, which can lead to a more accurate quantification.

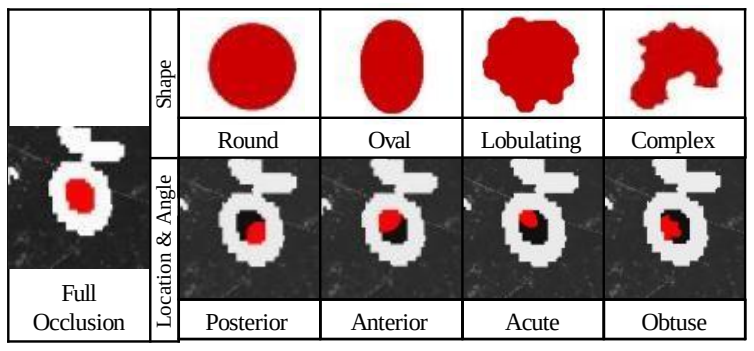

***Figure 2.** Variability of mucus modeled across shape, location, angle and size based on distributions measured from contrast-enhanced CT images.*

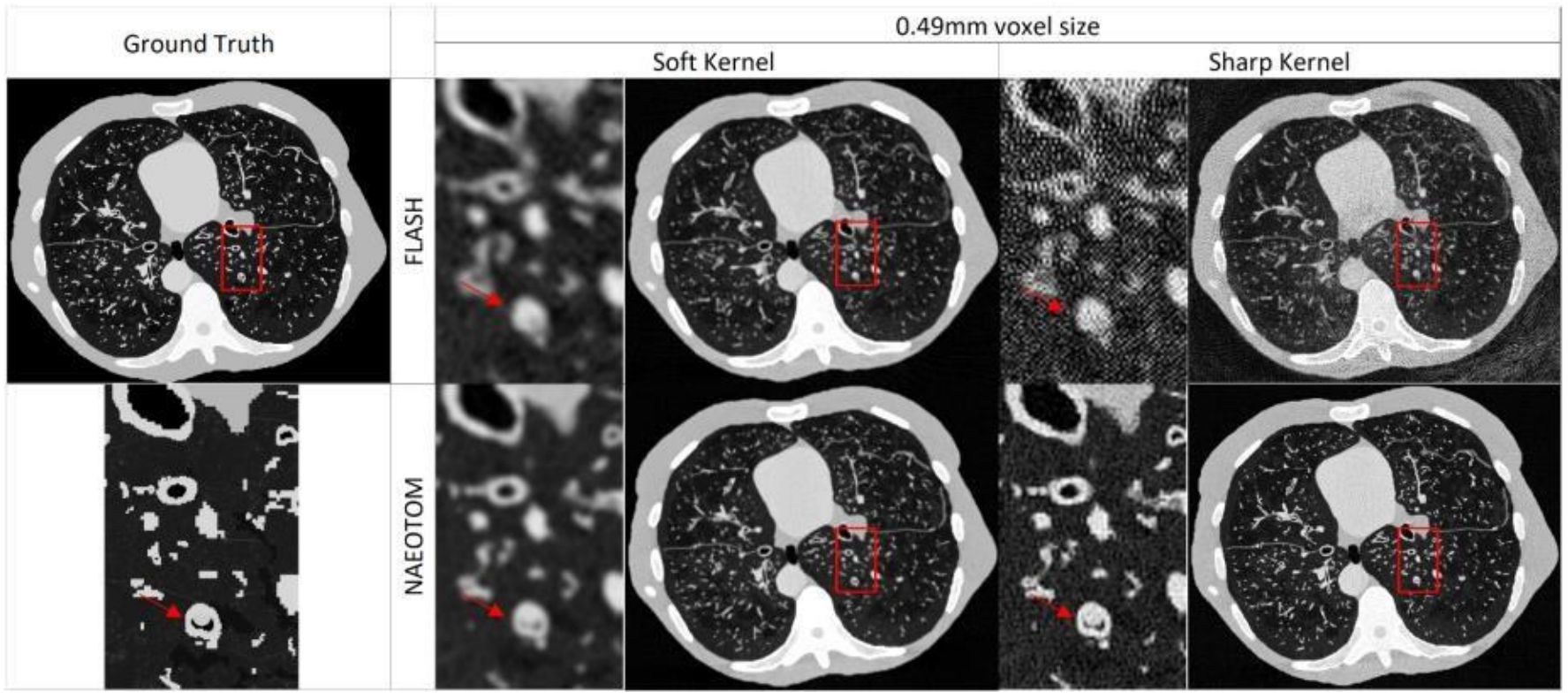

*Figure. 3. A qualitative comparison for EICT (FLASH) and PCCT (NAEOTOM) scanners using 2 kernels against the ground truth rendition of an XCAT phantom with mucus of 0 HU and 53% occlusion. The EICT and PCCT images were reconstructed using wFBP and iterative reconstruction, respectively.*

**Table 1** shows the average and standard deviations of the PercOcclusion absolute error measurements across multiple protocols. The PercOcclusion accuracy was improved by using higher dose levels, smaller voxel sizes, and shaper kernels. Holding other settings constant, switching from the soft to sharp kernel improved the accuracy by 2.74% on EICT and 15.03% on PCCT, respectively. The accuracy improved by 27.04% for PCCT and insignificantly for EICT by increasing the dose level from 60mAs to 100mAs. Using a pixel size of 0.49 mm instead of 0.98mm reduced the average absolute mean error by 12.76% for EICT and 31.54% for PCCT. Comparing the absolute mean errors among the scanners at optimized settings (100mAs, 0.49mm voxel size, sharp kernel), PCCT showed more accurate quantifications by an average of 23.15%, compared to EICT.

***Table 1.*** *The absolute PercOcclusion mean and standard deviation of error measurements across protocols and settings*

| | Energy-integrating CT (wFBP) | | | | Photon-counting CT (PNR) | | | |
|---|---|---|---|---|---|---|---|---|
| | 60mAs | | 100mAs | | 60mAs | | 100mAs | |
| PercOcclusion (%) | 0.49mm | 0.98mm | 0.49mm | 0.98mm | 0.49mm | 0.98mm | 0.49mm | 0.98mm |
| Soft | 30.15±31.39 | 32.51±32.58 | 26.12±28.07 | 30.34±30.73 | 26.12±28.07 | 30.34±30.73 | 18.36±24.33 | 24.89±27.14 |
| Sharp | 29.05±29.95 | 33.79±32.91 | 24.64±26.03 | 28.38±28.87 | 24.64±26.03 | 28.38±28.87 | 11.71±15.63 | 19.99±23.47 |

# DISCUSSION

This study highlights the potential of PCCT, alongside optimized imaging parameters (higher dose levels, smaller voxel sizes, and sharper kernels) in the setting of virtual imaging trial with COPD patient models and a novel imaging biomarker, PercOcclusion, to enhance the mucus occlusion quantification in small airways, a critical aspect of COPD pathology not fully addressed by previous methods.

# CONCLUSIONS

By leveraging PCCT technology and developing a virtual imaging trial with modeled COPD patient mucus at different respiratory states, the study presents a novel approach to overcoming the limitations of direct clinical measurements with enhanced diagnostic accuracy of SAD and COPD, facilitating earlier detection and more effective management.

## ACKNOWLEDGEMENTS AND CONFLICT OF INTEREST

This work was supported in part by grants from the NIH (R01HL155293, P41EB028744, and R01EB001838).

# PyAnsys-heart: a python library for LS-DYNA multi-physics heart simulation

Martijn Hoeijmakers, Wenfeng Ye, Karim El Houari, Clémentine Shao, Michel Rochette, Mark Palmer (Ansys)

## ABSTRACT

Physics-based computer simulations of the heart have huge potential in the medical device industry and clinical practice, for instance to accelerate and improve device designs, assist clinical decision making, or guide treatment planning. The importance of modeling choices with respect to electrophysiology, mechanics and fluid dynamics, and their respective coupling strongly depends on the application of interest. LS-DYNA is a finite element solver that offers the necessary multi-physics capabilities and features for heart modeling. However, setting up these models and obtaining physiological results is still a highly manual process and requires expertise in LS-DYNA usage, heart physiology, and scripting. In this paper we propose a python-based high-level interface to LS-DYNA, that is free-to-use and dedicated to heart modeling. We introduce the relevant heart modeling features that are available and introduce the modular python library to set up and drive these simulations. Two example models are presented: a bi-ventricular mechanical model and a full heart electro-mechanical coupled model.

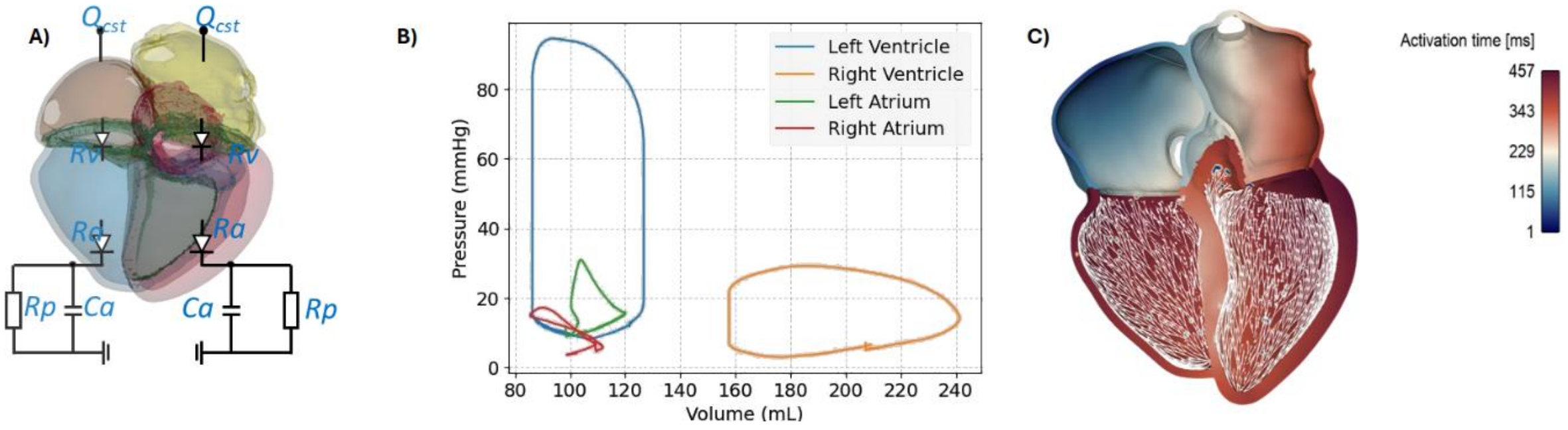

*Figure 1: A) Whole heart geometry and diagram describing the coupling with the 0d circulation model. B) Pressure volume loops for the four cavities obtained from the coupled 3D-0D model. C) Electrical activation time field*

# The Setup of Porous Media Valve Model with Ansys LS-DYNA ICFD using Patient Specific Heart Models from Synopsys Simpleware

Chien-Jung (Peggy) Huang, George Hyde-Linaker, Chris Goddard, Rodrigo Paz, Facundo Del Pin, Christoph Maurath, Rebecca Bryan, Thierry Marchal, Mark Palmer

## BACKGROUND AND PURPOSE

The numerical simulation of cardiovascular flow using patient specific geometries is helpful for understanding diseased hearts and the design of medical devices. To obtain valid and credible numerical results, a good quality mesh is crucial. This means all elements meet certain criteria including aspect ratio, growth rate, no intersections, etc. For patient-specific geometries, it usually requires a large amount of manual effort on transformation from medical images to smooth and high-quality mesh. Building an automated workflow can save a lot of time on model preparation. As to cardiac flow simulation and analysis, the heart valves play an important role. However, it is hard to obtain the detailed leaflet geometry through CT scans. It is labor intensive to reconstruct the valves leaflet out of patient-specific geometries, and it needs more computation time to perform fluid-structure interaction simulation. A lumped model adding resistance to the flow to simulate the close of valves is a more practical approach. This study aims to build and optimize an automated workflow of generating mesh of heart geometries and apply porous media models for valves to simulate the cardiac flow in a beating heart.

## METHODS

### 1. Workflow from Simpleware to LS-DYNA

The Synopsys Simpleware [1] software offers automated segmentation solutions that can convert medical imaging data into high-quality mesh of patient-specific geometries. From CT scans, Simpleware can extract regions of interest, perform segmentation, and generate simulation ready mesh with minimum manual effort (Figure 1). The workflow has been further improved so that the generated mesh is more suitable for flow simulations. Simpleware generates surface mesh of the boundary of blood pool, and automatically assigns different part id numbers to different parts of the heart, inlets, and outlets. This makes it easier to set boundary conditions on different parts. The outcome mesh consists of low skewness triangular elements, flat inlet and outlet surfaces with clean edges (Figure 2). There are no intersections of separate surfaces, and the small structures in ventricles are preserved and smooth. All the mesh is exported into one file. It is then imported into LS-Prepost [2] to be edited and transformed into LS-DYNA keyword format, then it is ready for flow simulation for LS-DYNA ICFD solver [3]. The ICFD solver is able to generate volume mesh from input surface mesh. LS-Prepost can record the commands used, so that the same steps can be easily applied on other heart geometries from Simpleware. The whole process can be completed in less than 20 minutes.

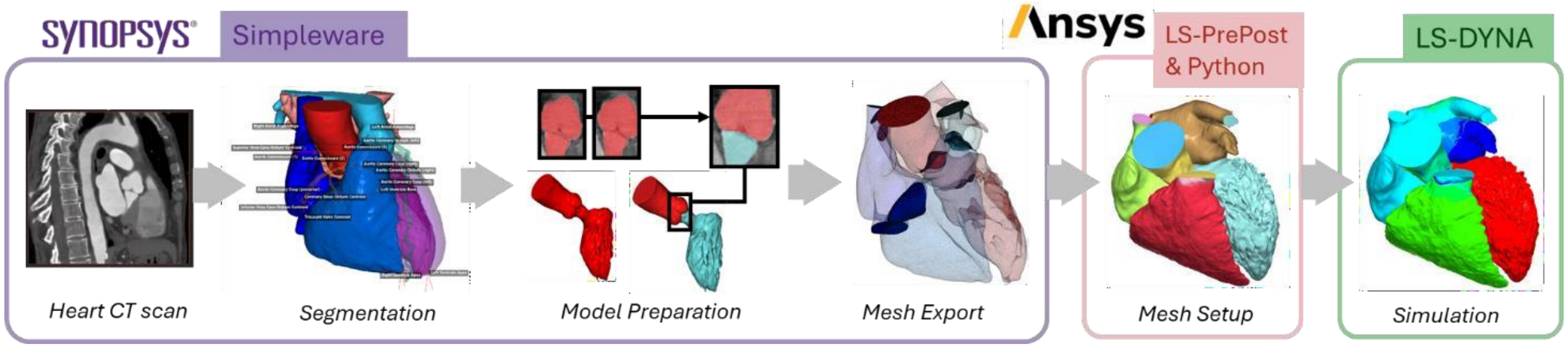

*Figure 7 The automated workflow from CT scans to flow simulations.*

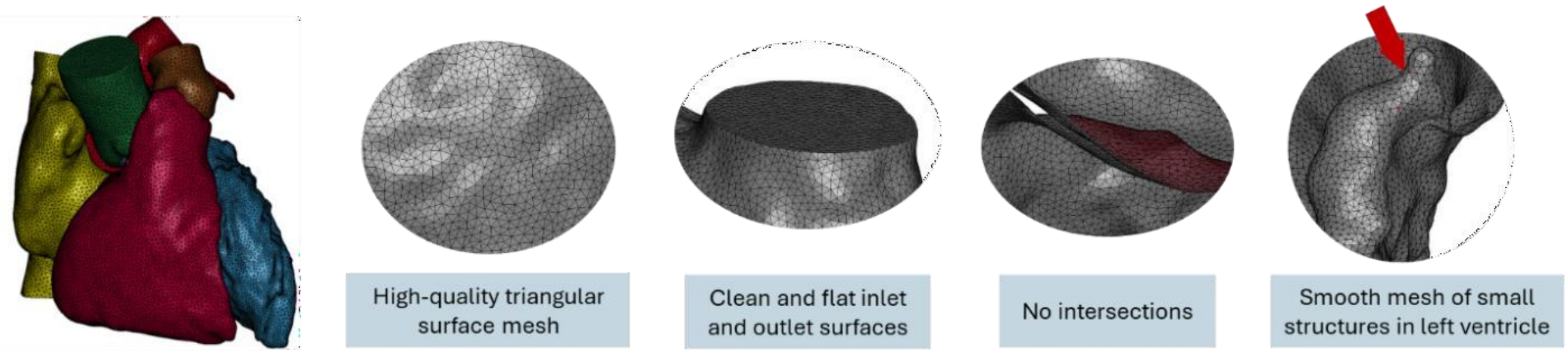

*Figure 8 The surface mesh generated by Simpleware.*

The workflow can also be applied on a series of scans taken at different phases in the cardiac cycle on the same patient. It has been successfully used on three sets of 4D heart scans of different patients, each with 10 to 12 frames. The mesh topologies in all the frames generated by Simpleware are identical, meaning that it is possible to calculate the movement of every node. A python script is developed to convert the mesh of different frames into keywords of prescribed motion on each node. The flow simulation in the beating heart can be set up in less than 30 minutes.

## 2. Porous media valve model

To consider the existence of the valves and their influence on the cardiac flow, a porous media model has been developed in the ICFD solver. The porous media model introduces resistance to the flow to simulate the opening and closing mechanism of valves through the parameter of permeability. With a low value of permeability, the resistance is high representing a closed valve, and vice versa. The permeability can be either a user-defined profile or computed from the pressure difference across the porous region. The porous media model defines a porous media region within a certain shape. The region is defined with parameters of the shape, and it can be set to be moving attached to parts of the geometry. There is no need to make any changes to the established mesh. In this study, cylindrical regions of porous media at valve locations are used as shown in Figure 3.

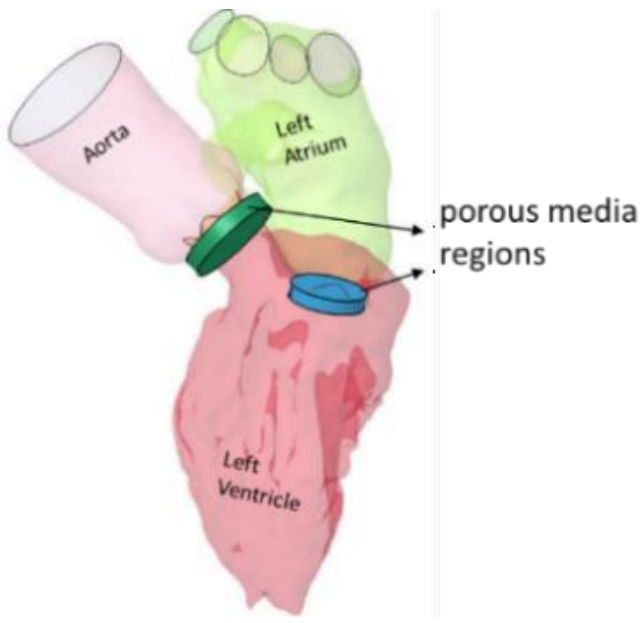

*Figure 9 The set-up of the porous media regions for left heart at the location of mitral and aortic valves.*

# RESULTS

## 1. Flow simulation inside a beating heart

The automated workflow has been successfully applied on three sets of 4D heart scans. The meshes of blood pools of the four chambers, aorta and pulmonary artery have been generated, and their movements have been converted into nodal prescribed motion. Figure 4 shows an example of a blood pool geometry at the beginning and end of the systole phase. Flow simulations inside these three beating hearts with LS-DYNA ICFD solver have been performed without existence of valves, and the obtained pressure fields are shown in Figure 5.

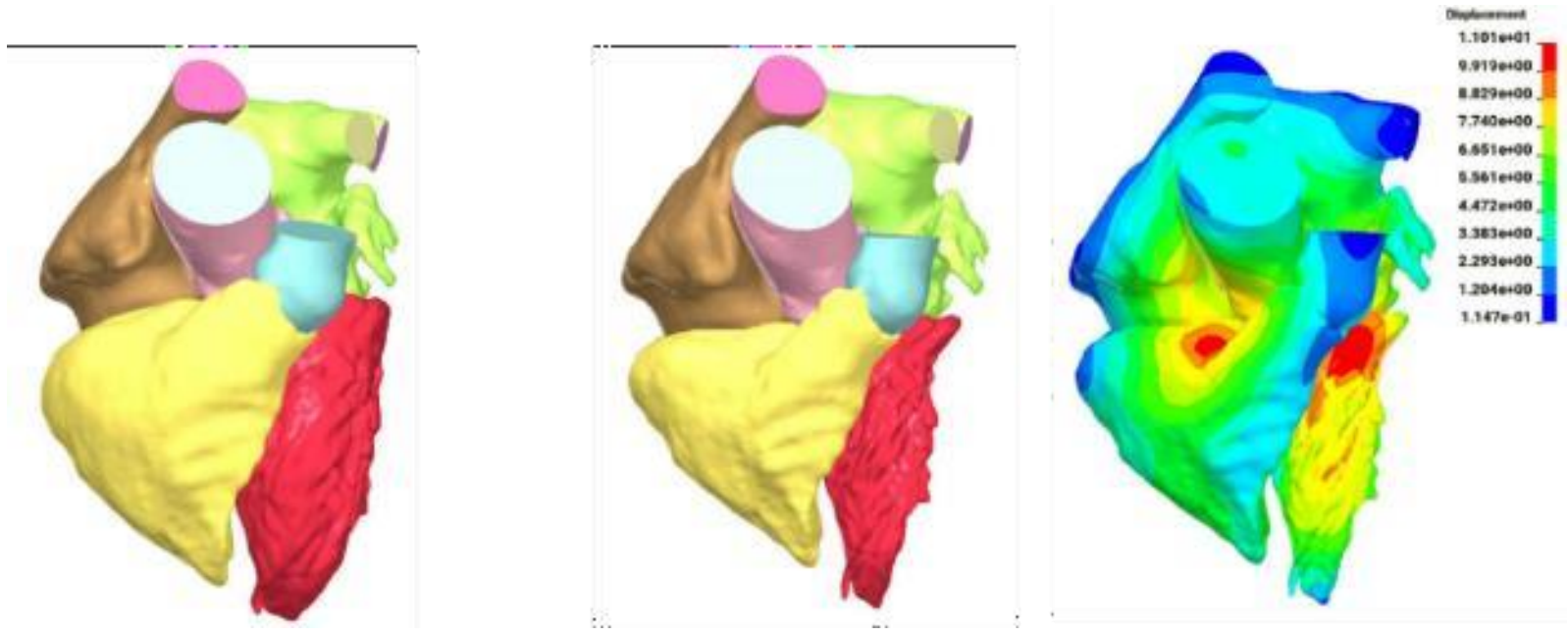

*Figure 10 The geometry at the beginning (left), end (middle) of the contraction phase and the displacement on each node (right).*

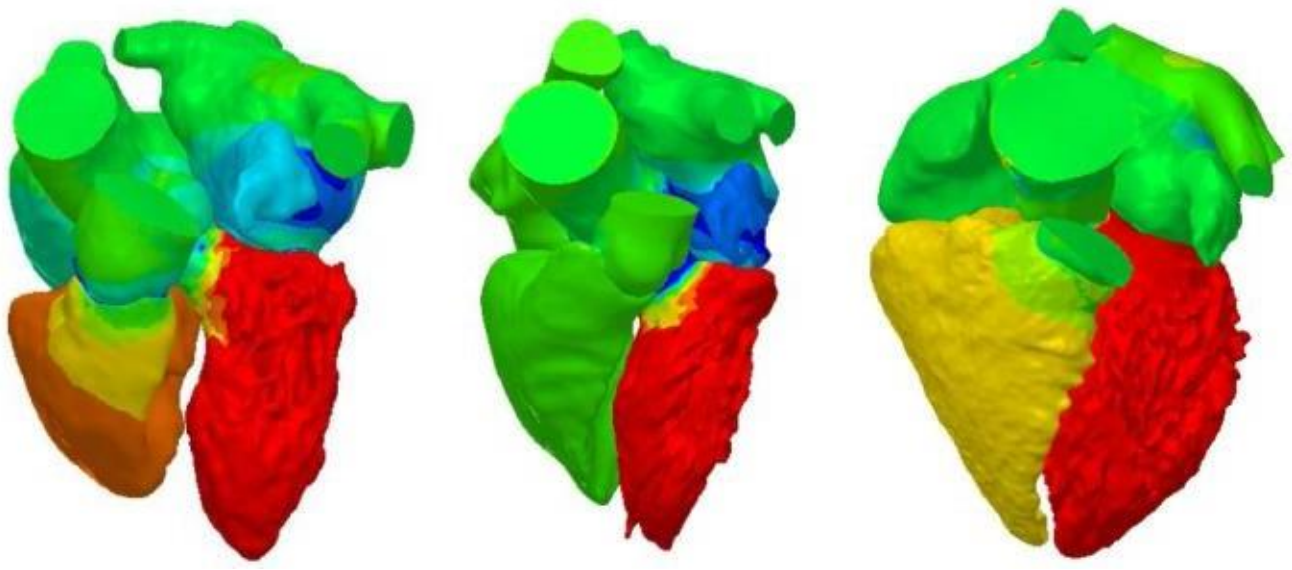

*Figure 11 The obtained pressure field of three different heart models at the beginning of the contraction phase (without valves).*

## 2. Porous model with prescribed permeability

The porous media valve model is added to the left heart in one of the three established heart models. From the scans, the systole and diastole phases can be identified. Therefore, the profiles of permeability for mitral and aortic valves can be designed as shown in Figure 6. The permeabilities remain constant initially, and the permeability profiles increase or decrease during the transition between systole and diastole phase. From the obtained velocity fields at systole and diastole phases in Figure 6, it can be observed that the flow from the left ventricle is blocked at the mitral valve during systole phase and the flow is blocked at the aortic valve during diastole phase.

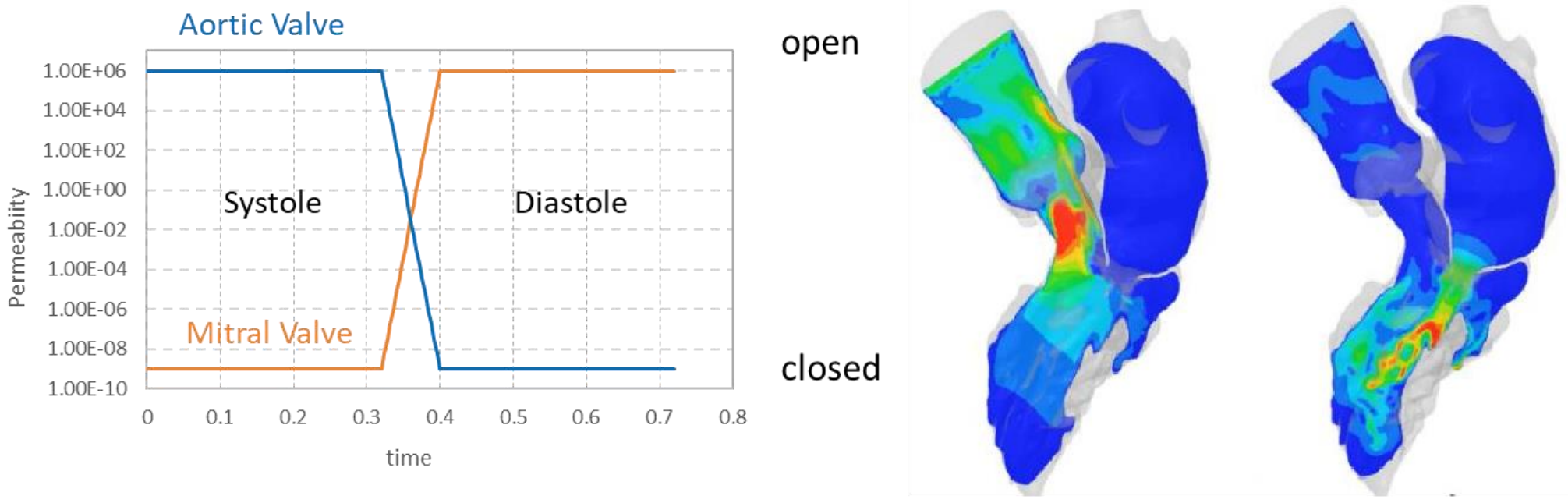

*Figure 12 The prescribed permeability function (left) and the velocity magnitude fields (right) on cross-sectional plane at systole and diastole phases.*

## 3. Porous model with permeability as function of pressure difference

In a more realistic model for valves, it should consider the fact that the valve closes and opens due to its interaction with the nearby flow field. An assumption is made that the heart valves open and close due to pressure difference across them. For example, for aortic valve, if the ventricle has greater pressure compared to aorta, the aortic valve opens. Otherwise, the aortic valve closes. Therefore, in the model, the permeability is set to be a *tanh* function of the pressure difference across valve regions, as stated in Satiago et al. [4]:

$$k = \left(\frac{k_{max} - k_{min}}{2}\right) \cdot \left(1 + \tanh\left(\frac{\Delta p - \Delta p_{ref}}{S}\right)\right) + k_{min}. \quad (1)$$

where $k$ stands for permeability and $k_{max}$ and $k_{min}$ are its maximum and minimum values, $\Delta p$ and $\Delta p_{ref}$ are the pressure difference and a pressure reference value, and parameter $S$ adjusts the slope. The obtained permeability profiles of the aortic and mitral valves are plotted in Figure [7].

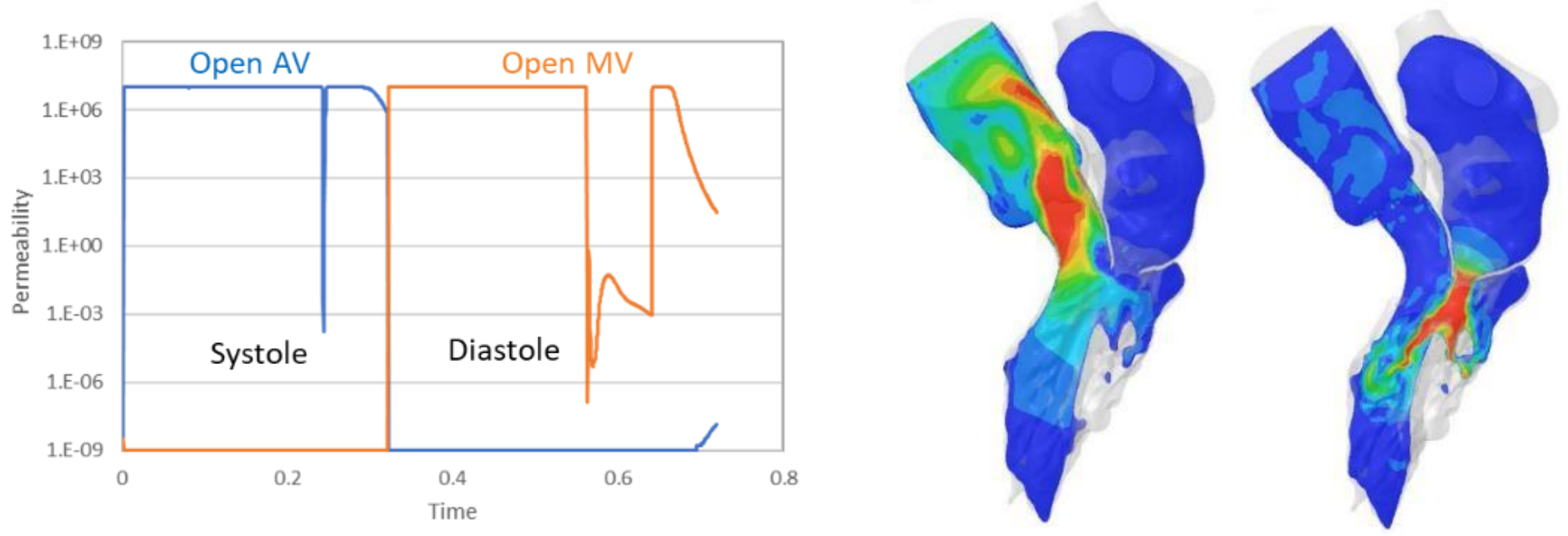

*Figure 13 The obtained permeability profile (left) according to the pressure difference and the velocity magnitude fields (right) on cross-sectional plane at systole and diastole phases.*

# DISCUSSION

The simulation results of flow in three hearts demonstrates that the automated workflow can be applied on various patients with minimal manual effort, and the flow simulation in the beating hearts can be successfully performed with LS-DYNA ICFD solver. Without considering valves, the flow gets pushed to atrium and aorta or pulmonary artery when the ventricle contracts. After adding the porous model on the valve areas in the left heart, the flow gets pumped to the aorta during systole, and fills the ventricle from the atrium during the diastolic phase. For the case that the permeability is determined by pressure difference, the valves open and close as expected. However, at two locations there is sudden drop of the permeability. It is because the movement of the heart is linearly interpolated from 10 frames. Between the transition of frames, the sudden change of the boundary motion introduces a peak in pressure field, affecting the value of permeability. If more frames are used in one cardiac cycle or fluid-structure interaction can be introduced, the boundary motion can be smoother thus this issue can be resolved. Currently, Simpleware can generate an all-on-one model including mesh of blood pool, myocardium, and outer tissues. A multi-physics heart model coupling fluid, structural, and electrophysiology is under development.

## CONCLUSION

This study demonstrates a nearly automated and fast workflow from medical scans to flow simulation in beating hearts. The porous media model is easy to set up and can effectively simulate the mechanism of the valves. This method saves a large amount of time and manual labor as well as provides a route to applications within large scale multi subject analysis.

## ACKNOWLEDGEMENTS AND CONFLICT OF INTEREST

No potential competing interest was reported by the authors.

# Towards a Validated Digital Twin Using Computational Physiology via the Pulse Physiology Engine

Aaron Bray, Emily Veenhuis, M.S., Jeffrey B. Webb, M.S., and Rachel B. Clipp, PhD.

## BACKGROUND AND PURPOSE

The Pulse Physiology Engine (Pulse) is an open source computational physiology engine designed to develop and integrate multiscale models for an individualized whole-body predictive patient physiology model[1]. Pulse is composed of lumped-parameter models that represent the fluid dynamics and thermodynamics of the systems within the body, including the cardiovascular, respiratory, and renal systems. These system models are then coupled with models that represent diffusion, pharmacokinetics, pharmacodynamics, and baroreceptor and chemoreceptor feedback[2,3]. This combination produces a dynamic representation of the whole-body physiology for the healthy state. The model parameters are then scaled to represent the effects of disease, injury, and treatment[4–6]. These same parameters can be modified to create a digital twin of a specific patient. Previously, a “standard” male and a “standard” female patient have been validated[7]; however, a digital twin requires the validation for a range of patients. This study modifies these parameters to validate a range of specified patients to demonstrate the ability of Pulse to act as a digital twin.

## METHODS

The Pulse patient framework was updated to allow for additional patient-specific parameters to be set. This includes the ability to specify blood pressures through a combination of any two of the following: pulse, diastolic, systolic, or mean arterial pressure. The remaining blood pressure can be calculated from known equations. Initial validation for the standard female also demonstrated that the stroke volume and cardiac output were not calculated with a gender-specific equation. The cardiovascular system in Pulse was updated to specific the cardiac output based on gender, then the stroke volume was calculated based on the cardiac output and the patient-specific heart rate. Pulse uses automated spreadsheets for all systems and levels of validation. Previously, the literature (validation) values were statically specified for a male young adult patient. We first linked the Pulse patient specification file to the spreadsheet at comparison time. This then provides access to the gender, weight, age, etc. We then added dependencies to scale the validation parameters based on the literature and patient-specific parameters. For example, the cardiac output and blood flow distributions are gender-dependent, the lung volumes are weight-dependent, and the heart rate and blood pressures are specified in the patient file.

*Table 1 Patient Population Input Parameters for the Pulse Physiology Engine Digital Twin Study*

| | **Male** | **Female** |
|---|---|---|
| **Age (year)** | 18, 45, 65 | 18, 45, 65 |
| **Height (cm)** | 163, 180.34, 190 | 151, 162.56, 175.5 |
| **Body Mass Index** | 16, 23.71, 29 | 16, 22.31, 29 |
| **Body Fat Fraction** | 0.02, 0.21, 0.25 | 0.1, 0.32, 0.2 |
| **Heart Rate (beats/minute)** | 60, 72, 100 | 60, 72, 100 |
| **Respiration Rate (breaths/minute)** | 8, 12, 20 | 8, 12, 20 |
| **Mean Arterial Pressure (mmHg)** | 70, 87, 100 | 70, 87, 100 |
| **Pulse Pressure (mmHg)** | 30, 40.5, 60 | 30, 40.5, 60 |

A patient population was created by varying patient input parameters to represent a diverse population. The patient input parameters are shown in Table 1. A total of 13,122 patient combinations are generated using the patient parameters in Table 1. Each patient combination was tested for validity by testing for certain edge case combinations that can introduce inconsistencies within the model, leading to unrealistic scenarios that defy stabilization criteria. These conflicts are most frequently observed with extreme height and BMI combinations and heart rate and blood pressure combinations. Invalid patient parameter combinations were thrown out, resulting in a total of 4,374 valid patients to be simulated. To simulate a patient, the Pulse tuning algorithms scale the model parameters to produce the specified patient. After the parameters were tuned and the computation reached a stable state, the simulation was further executed for two minutes. A validation comparison was then completed for each simulated patient. Each output parameter from Pulse was compared to the patient-specific validation values produced by the dynamic spreadsheet. The percent error is then calculated and categorized into good (less than 10% error), fair (between 10% and 30% error), and poor (greater than 30% error).

# RESULTS

Each system in Pulse was analyzed separately to assess performance. The parameter output errors were grouped for each system and the percent of parameters than were in the good, fair, and poor categories was calculated. This process was completed for the “standard” male and female patients, and then for the cumulative male and cumulative female patients. All systems except the renal system were found to have 80% of parameters categorized with less than 10% error and over 90% were categorized with less than 30% error. Renal demonstrated a 40-60% good error performance and 60-80% were categorized as fair or good. This further revealed that Pulse validates at a higher rate for the male patients than the female patients. This is particularly true for the cardiovascular, respiratory, and tissue systems.

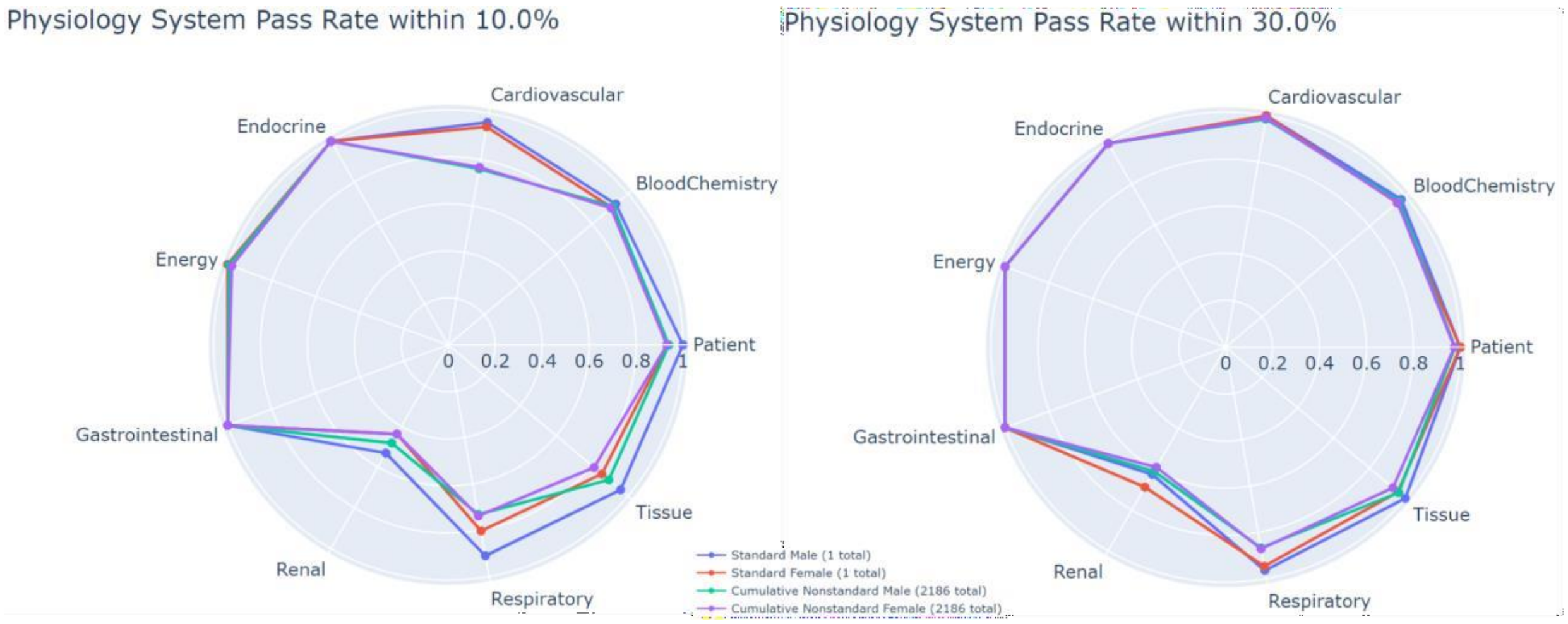

*Figure 1 The radar plots illustrate the overall validation for the "standard" male and female patients and the cumulative male and female populations. All systems except the renal system were found to have 80% of parameters categorized with less than 10% error and over 90% were categorized with less than 30% error.*

The box plots in Figure 2 show the validation spread across the unique patients as each input parameter is varied while holding the remaining values at the standard value. This allows us to study the sensitivity of each input parameter on the validation of the output parameters. This reveals that validation of the cardiovascular system is most sensitive to the heart rate variation followed by the blood pressure. The respiratory system showed a similar sensitivity to the respiration rate. This is unsurprising as the heart rate and respiratory rate are the "drivers" of each of these symptoms. The downstream effects of modifying these parameters in the Pulse system may not be well controlled. Both the validation spreadsheet and the model tuning algorithm may need to be adjusted to stabilize the system with respect to varying these two parameters. The respiratory system further showed sensitivity to the Body Mass Index due to the scaling of the various lung volumes based on size. This scaling may be causing large changes in validation at the extremes of this value. The renal system showed consistent sensitivity to all parameters with poor validation outcomes across all patients, including the "standard" male and female patients.

## DISCUSSION

Overall, our population study of Pulse patients shows positive validation results as we vary patient parameters. This indicates that creating digital twins is achievable with Pulse. While many of our systems perform extremely well, the renal system fails validation for even the standard patients and further fails with patient-specific parameters. The consistent failure across the outputs to inputs demonstrates that the fundamentals of the renal system need to be investigated and improved to reflect validated physiologic function. Further sensitivity analysis conducted on the tuning algorithm itself may reveal improvements to address validation sensitivity to the "driver" components of heart rate and respiration rate. Validating Pulse against a cohort of patients using electronic health record data. We are exploring the use of data from PhysioNet to complete this study; however data reflecting non Intensive Care Unit patients can be challenging to locate.

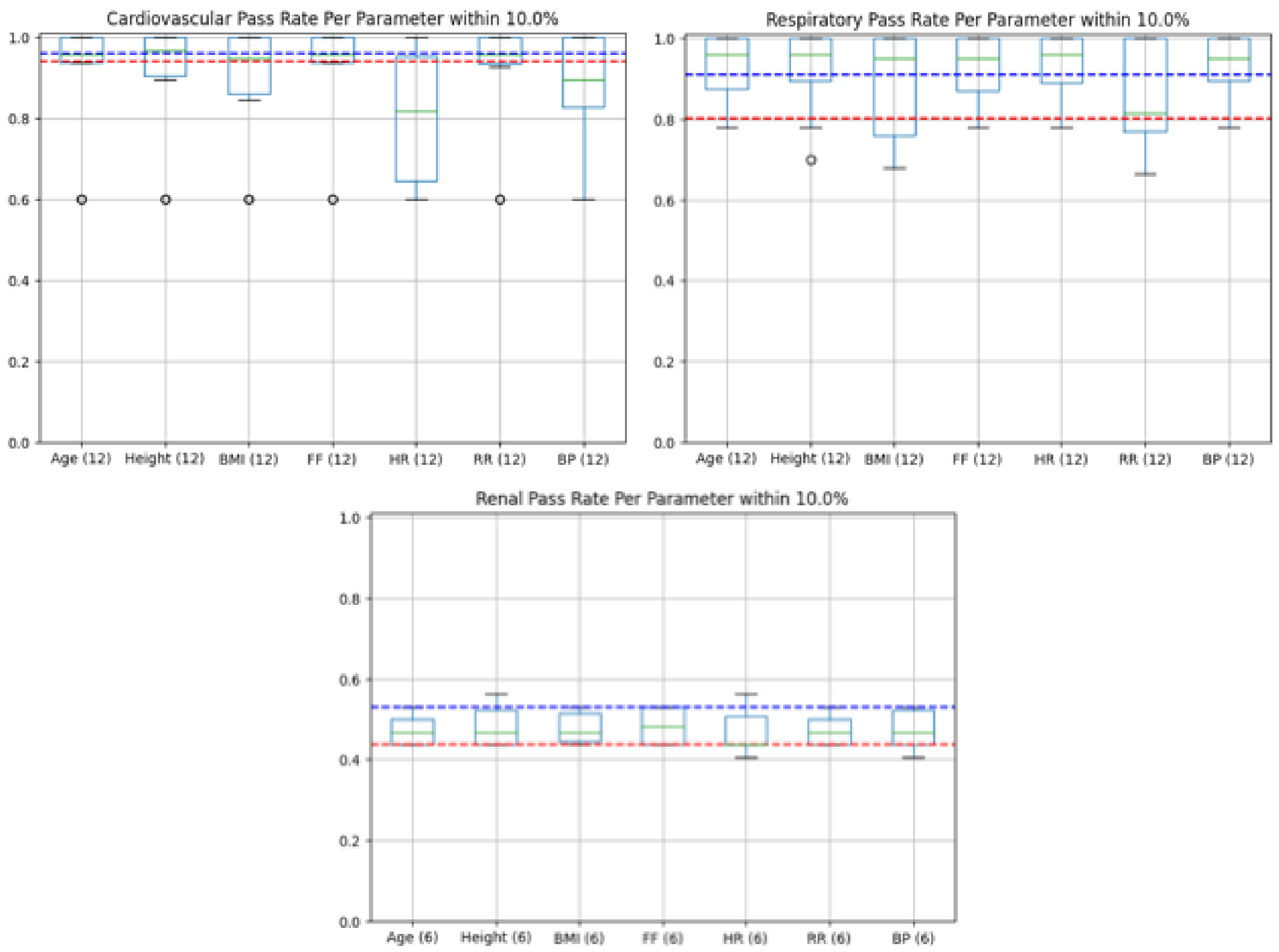

*Figure 2 The box plots illustrate the sensitivity of the validation parameters when varying the input parameters individually. This is shown for the cardiovascular and respiratory systems, where the blue line is "standard" male and the red line is the "standard" female results.*

## CONCLUSION

This study shows strong promise for the use of Pulse as a digital twin, which could provide support for virtual trials related to specific medical conditions and treatments. However, model improvements and further validation could address current deficiencies in Pulse.

## ACKNOWLEDGEMENTS AND CONFLICT OF INTEREST

We would like to acknowledge our funding that supported this work by the National Institute of Biomedical Imaging and Bioengineering of the National Institutes of Health under award number R01EB031808. The authors declare that they have no conflicts of interest.

# FAST (Fast Analytical Simulator of Tracer)-PET: an accurate and efficient PET analytical simulation tool

Suya Li[1], Mahdjoub Hamdi[1], Kaushik Dutta[1], Tyler J. Fraum[1], Richard Laforest[1], Kooresh I. Shoghi[1,2]
1Department of Radiology, Washington University School of Medicine, St Louis MO USA
2Department of Biomedical Engineering, Washington University in St Louis, St Louis MO USA

## BACKGROUND AND PURPOSE

Simulation of Positron Emission Tomography (PET) images is essential in the development and validation of quantitative imaging workflows and advanced image processing pipelines. Existing Monte Carlo or analytical PET simulators often compromise on either efficiency or accuracy. We aim to develop and validate FAST (Fast Analytical Simulator of Tracer)-PET, a novel analytical framework to simulate PET images accurately and efficiently.

## METHODS

An overview of the workflow is shown in Figure 1a. FAST-PET simulates PET images by performing precise forward projection, scatter, and random estimation that match the Siemens Vision-600 PET/CT scanner geometry and statistics. Although the same process should be applicable to other scanner models, we focus on the Biograph Vision-600 here. Calibration and validation of FAST-PET were performed through comparison with an experimental scan of a National Electrical Manufacturers Association (NEMA) Image Quality (IQ) phantom. Further validation was conducted between FAST-PET and Geant4 Application for Tomographic Emission (GATE) in clinical image simulations.

## RESULTS

Figure 1b depicts representative slices of experimental and simulated (FAST-PET and GATE) 5-seconds and 120-seconds scans of the NEMA phantom. The distributions of mean activity shown in Figure 1c are similar among all three methods for long (120s) and short (5s) acquisition times. Figure 1d depicts representative FAST-PET and GATE slices of clinical patient simulations with diagonal and opposite diagonal profiles. Scatter plots and concordance correlation coefficient values in Figure 1e confirm the agreement between GATE and FAST-PET in terms of both mean activity and variability. FAST-PET significantly outperforms GATE in efficiency, simulating a PET image in about 2.5 minutes compared to GATE's 56 hours on a 24-core, 3GHz Intel computer.

## CONCLUSION

FAST-PET has been developed and validated as an analytical simulation tool, designed to produce PET images that mirror those acquired from actual scanners and GATE simulations, while markedly reducing the processing time.

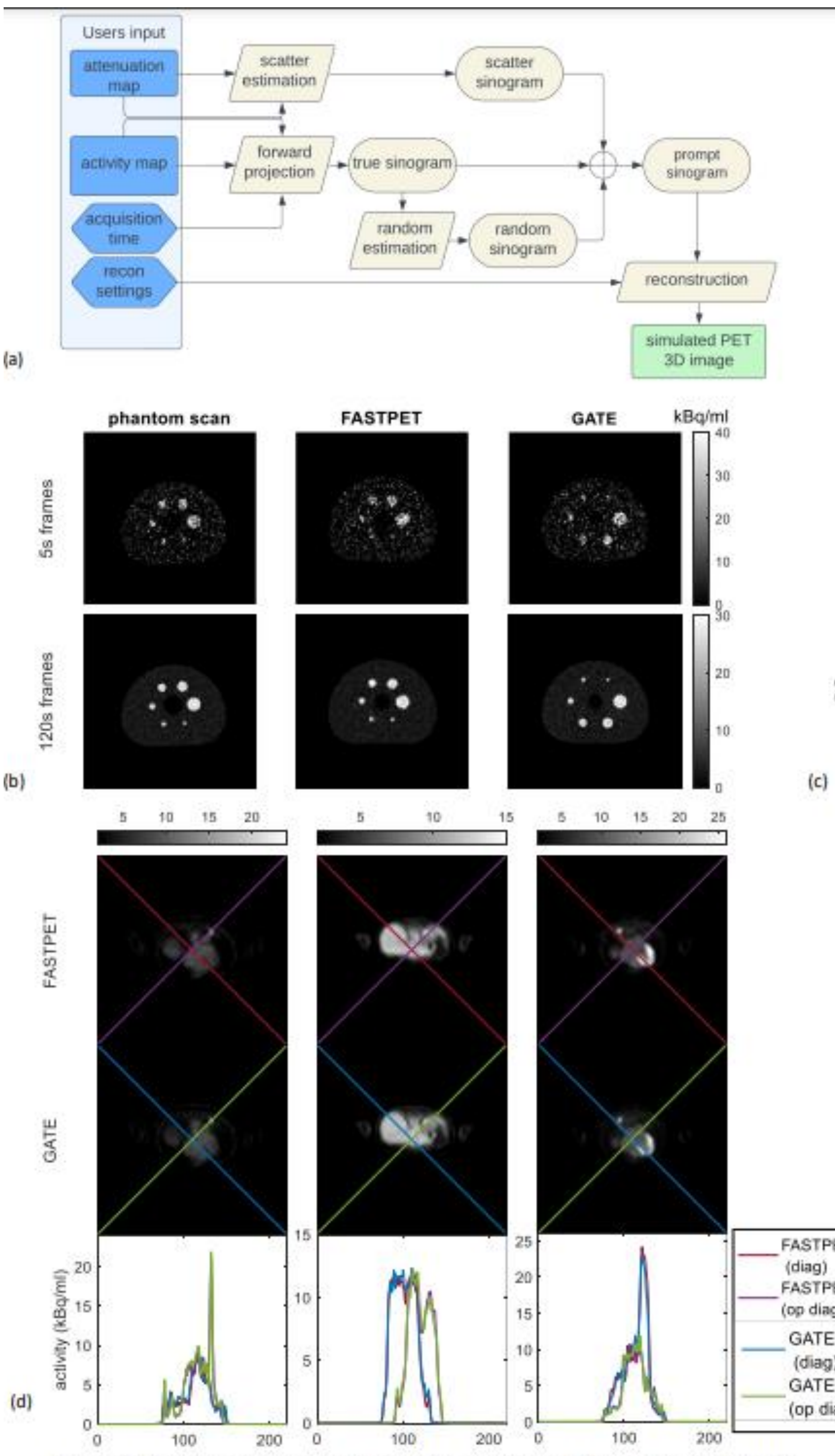

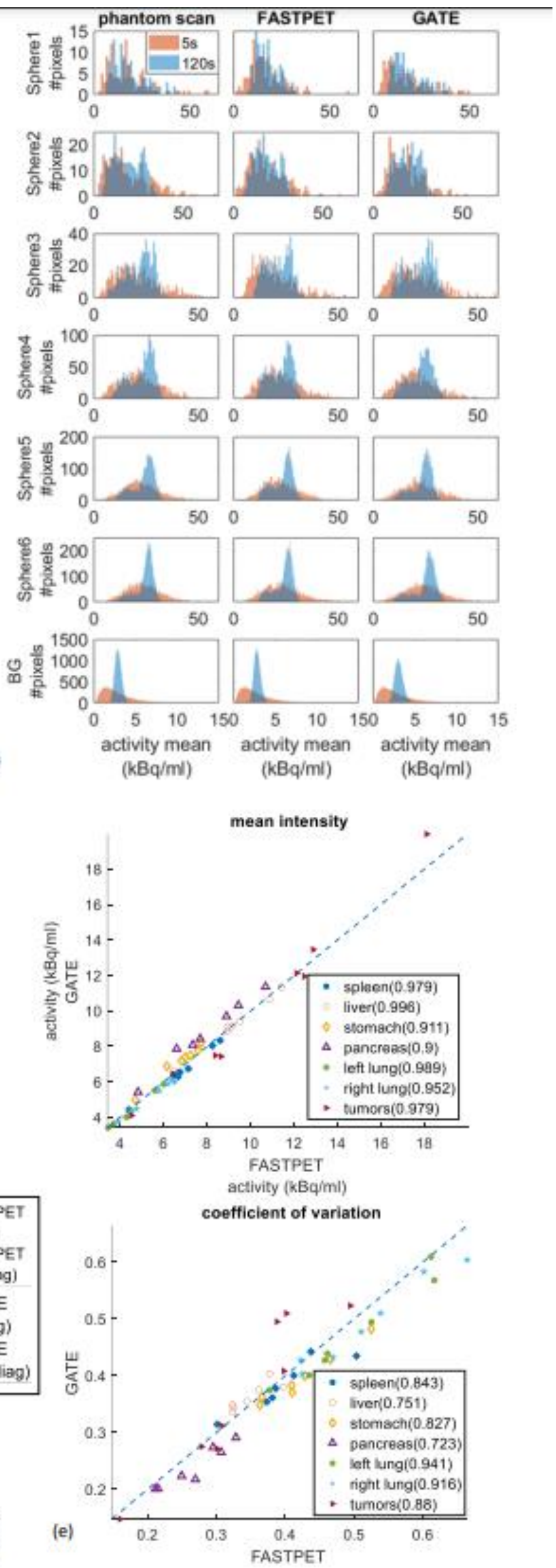

Figure1. (a) Overview of simulation workflow. (b) 5-second and 120-second frame of NEMA IQ phantom generated from physics phantom scan, FASTPET simulation and GATE simulation. (c) Distributions of activity mean within ROIs of spheres (from small to big: sphere 1 to 6) and background (BG) in phantom scan, FASTPET and GATE simulation images. (d) FASTPET and GATE simulated 5-minute clinical PET scans and their diagonal and opposite diagonal profiles. (e) Scatter plots comparison of mean activity and COV within organs and tumors between FASTPET and GATE simulation. Numbers in the parentheses are the CCC values between FASTPET and GATE for each organ and tumor.

# Stochastic modeling

Tuesday, April 23, 2024

11:30 am - 12:10 pm

Chaired by Joseph Lo & Ann-Katherine Carton

# Multi-organ volume estimations using automated segmentation in a large clinical population

Mobina Ghojogh Nejad, Lavsen Dahal, William Segars, Ehsan Samei, Joseph Lo (Duke University)

## Background

There are limited data on internal organ volumes in a large clinical population. Access to organ volume data may alter decision making in clinical practice for opportunistic screening of presymptomatic diseases, risk prediction, or to optimally select matched donors for organ transplantations. For example, imaging-based volumetry of the liver is vital in the preoperative planning for major hepatic resection or liver transplantation as well as in the determination of future remnant liver volume. Despite clinical relevancy of organ volume estimations, there are limited standardized methods to calculate organ volumes in clinical settings. Historically, morphologic assessment of internal organs was limited to two-dimensional linear estimations given the ease of performing such measurements from cross-sectional images. [9] The variability in morphology or orientation of the organ may cause inconsistent measurements leading to overestimation or underestimation of the volumes.[10] Three-dimensional volume measurements traditionally relied on tedious work of manual outlining of organ boundaries in each imaging sections.[11] This limitation precludes development of large-scale dataset-based volume estimations in various clinical entities. Here we first introduce an image analysis framework for automated segmentation of multiple organs and volume estimations and then report the results for a large clinical population.

## Methods

In this retrospective study, we used public and private models to train, validate and test segmentation models. The private dataset was provided by Duke University Health System, through institutional review board (IRB) approved protocols and was employed to create anthropomorphic phantoms. The nn-UNet model underwent the quality control for selecting favorable cases which subsequently were used to construct the phantoms. Four datasets with a total of 5895 images were included in our study. The public dataset of TotalSegmentator and a private dataset from Duke (DukeCT) were used for model development and validation, and other public datasets including CT_ORG, ABDOMEN_1K, AMOS, and XCAT were used for the testing and evaluation of the segmentation model. The labels for DukeCT dataset were generated automatically using public and commercial models. These labels were used as the reference standard to train the segmentation model.

A proportion of images as well as corresponding medical records for subjects with extremely high or low volumes were reviewed to assess validity of the segmentations.

**Results:** The dataset comprises CT scan images obtained from 2265 individuals, including 467 females and 1985 males, with a mean age at the time of imaging recorded as 67±23 years.

The demographic distribution of the dataset revealing 24% of the individuals are of African American descent, 34% are of Hispanic/Latino origin, and 56% belong to the white racial category. Mean of height, weight, and body mass index (BMI) were 1.71±0.12 meter, 83±20 kg, and 28±5.9 $kg/m^2$, respectively. We calculated mean volume of 1673±470 mL for liver. 2 subjects demonstrated liver volume less than 600 mL and liver volume was more than 3000 mL in 36 individuals. Among the low-volume outlier cases, situs inversus totalis led to incomplete segmentation of mispositioned liver in one case, while the other had severe cachexia as a result of widespread systemic metastasis without any error in liver segmentation by the model. The review of imaging data in 3 subjects with abnormally high liver volumes revealed multiple hepatic masses, metastatic lesions or liver infiltration.

For spleen, we measured mean volume of 258±167mL. 14 cases found to have zero spleen volumes. Reviewing image segmentations and medical records confirmed previous complete splenectomy in 10 cases. 17 cases demonstrated calculated volumes more than 1000 mL. We reviewed images for 5 cases and confirmed accurate segmentations. The clinical diagnosis was chronic lymphocytic leukemia in two patients, non-Hodgkin lymphoma in two cases, and post-transplant lymphoproliferative disease in one patient.

## Discussion

We introduced an AI based multiorgan automated segmentation platform and later applied the model in a large population of screening patients. We also investigated clinical correlations of extremely low or high estimated volumes to evaluate validity of the segmentation model predictions. Organ volumes data have been reported previously in smaller datasets for limited number of organs. Our results indicate mean liver volume of 1673±460 mL which is consistent with previous reports for liver volume.

Large imaging dataset and the simultaneous multiorgan segmentation model are the strength of our study. However, there are some limitations in this study. Rather than estimating volumes from healthy individuals, our focus on whole body CT imaging led to including many patients who underwent CT imaging for prostate disease, hence the imbalance in the patient sexes. This may cause internal organ volume alternations as a consequence of disease process or therapeutic interventions. This is a common limitation in organ volume studies as CT images are usually performed in individuals with specific risk factors for a disease or already have diagnosed clinical findings.

## Conclusion

We used a multi-organ auto-segmentation algorithm to estimate organ volumes in a large clinical dataset. Our model could generate accurate segmentations and subsequent volume estimations in the presence of pathologic processes. Our results are generally in accordance with previously reported organ volumes.

**Table 1**. Calculated mean of organ volume using automated segmentation model

| Organ | Volume (mL) |
|---|---|
| Stomach | 262 ±157 |
| Gallbladder | 26.8 ± 23.3 |
| lung | 3082 ± 860 |
| Brain | 753± 542 |
| heart | 589 ± 146 |
| Urinary bladder | 128 ± 95 |
| Femur | 673± 263 |
| Humerus | 2601 ± 72 |
| Vertebrae T1 | 23 ± 5 |
| Vertebrae L1 | 55±13 |
| Vertebrae C1 | 14.6 ± 5.9 |
| Pancreas | 75± 27 |
| Seminal vesicles | 8.0 ± 8.3 |
| Adrenal | 8.0 ± 3.3 |
| Liver | 1673 ± 470 |
| Spleen | 258 ± 167 |
| Prostate | 23 ± 26 |
| Left kidney | 157± 50 |

Figure.1 Liver volume distribution by sex
Figure.2 Spleen volume distribution by sex

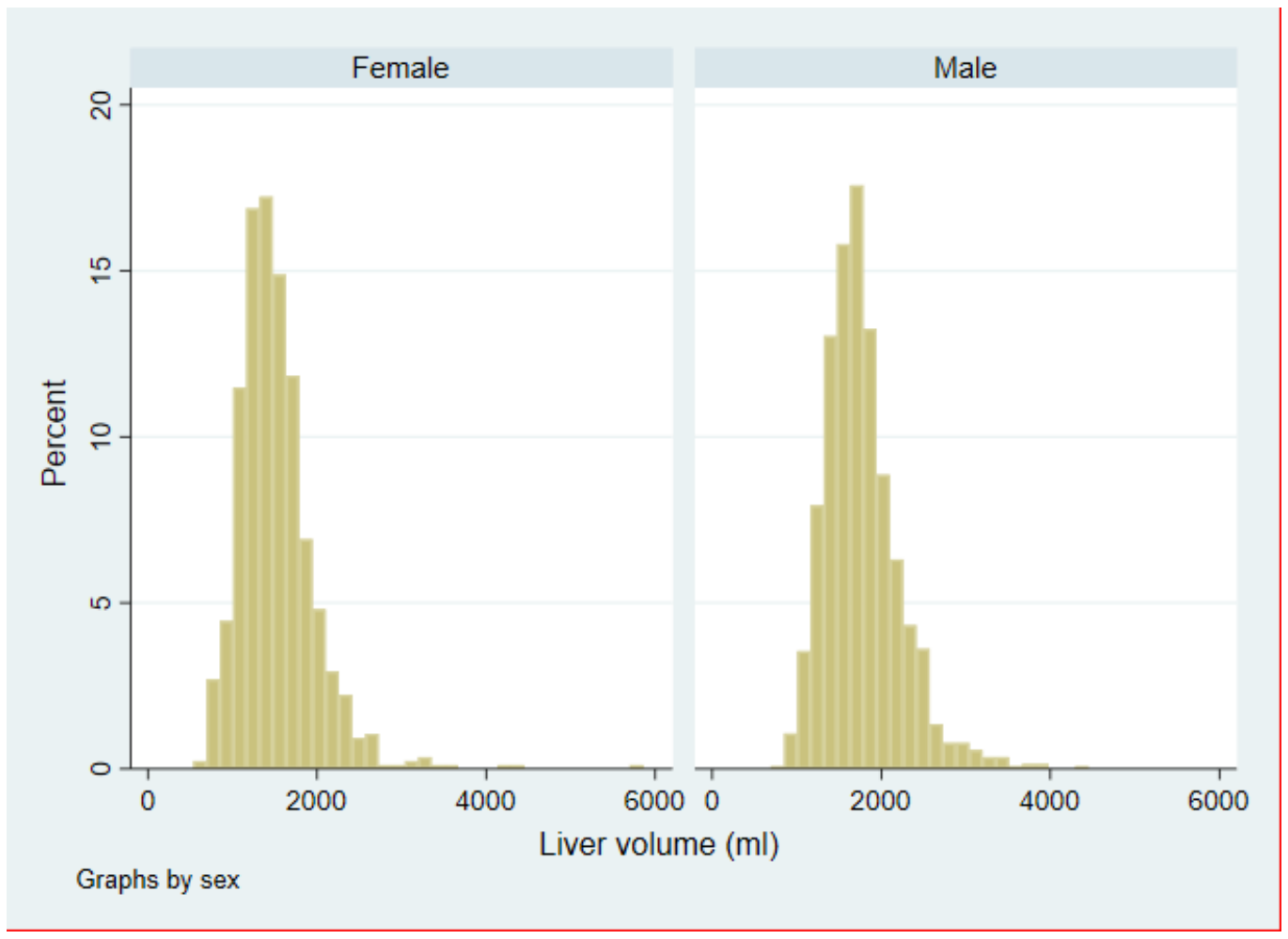

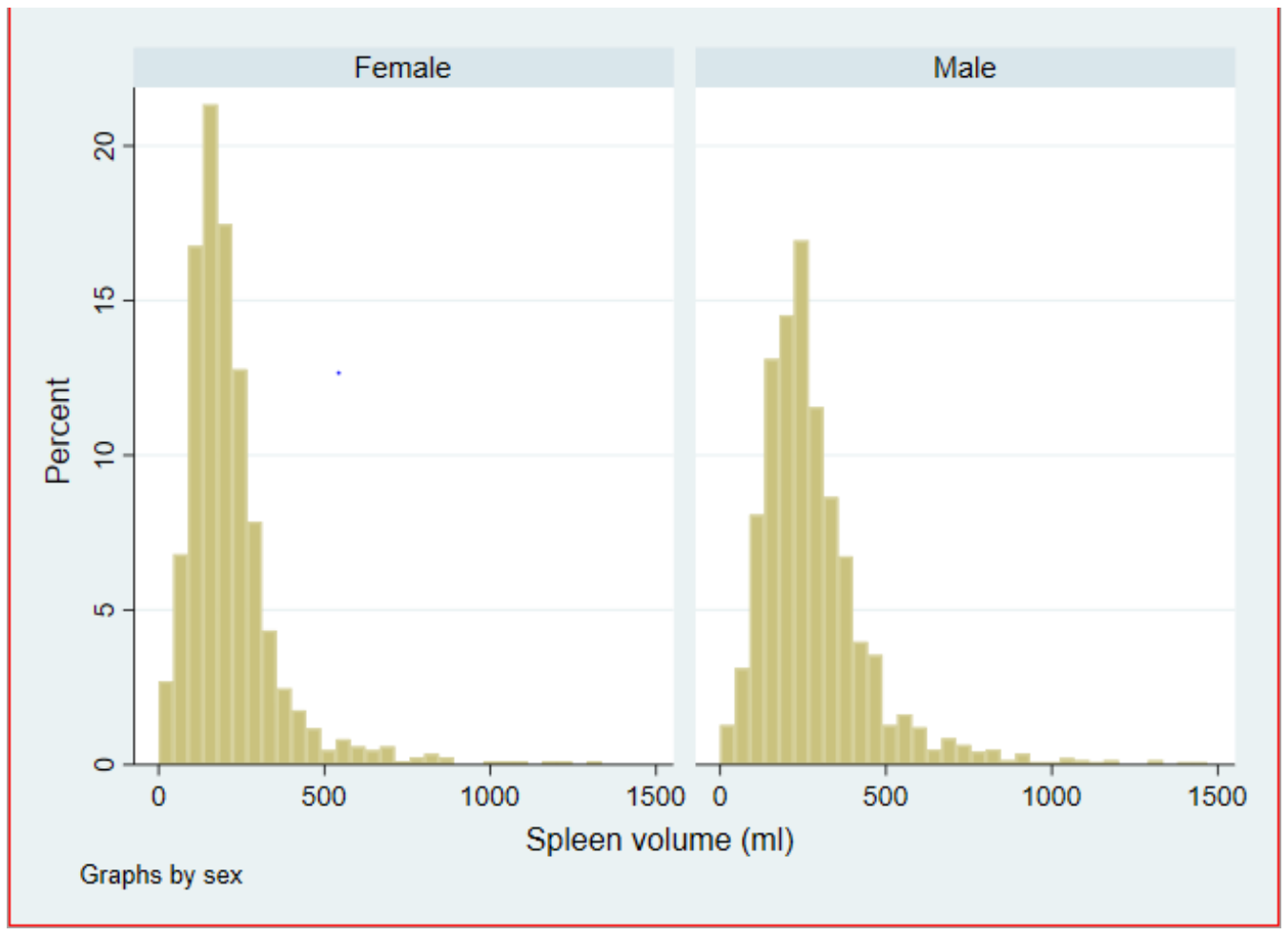

# Stochastic numerical phantoms to enable optoacoustic tomography virtual imaging studies

Umberto Villa (The University of Texas at Austin), Refik Cam ( University of Illinois at Urbana-Champaign ), Panpan Chen ( University of Illinois at Urbana-Champaign ), Hsuan-Kai Huang ( University of Illinois at Urbana-Champaign ), Fu Li ( University of Illinois at Urbana-Champaign ), Luke Lozenski ( Washington University in St Louis ), Evan Scope Crafts (The University of Texas at Austin), Seonyeong Park ( University of Illinois at Urbana-Champaign ), Mark Anastasio (University of Illinois at Urbana-Champaign)

## Purpose

Optoacoustic tomography (OAT) is a non-invasive, ionizing radiation-free, low-cost imaging modality that could help identify new biomarkers for early cancer detection, as well as monitor response to treatment. By combining endogenous hemoglobin optical contrast with ultrasound detection principles, OAT can assess tumor angiogenesis and hypoxia, which result from the increased metabolic activity of aggressively growing malignant tumors. However, widespread application of OAT in a clinical setting requires further investigation. To address this, a computational framework is developed to enable objective assessment of prototype-stage OAT system designs and image reconstruction methods by use of clinically relevant numerical phantoms.

## Methods

The proposed breast and mouse optoacoustic phantoms leverage validated tools (VICTRE and MOBY, respectively) for the generation of realistic anatomical structures. Functional, molecular, optical, and acoustic properties are then assigned to each anatomical structure within physiological ranges. Numerical lesions are modeled to include viable tumor cells, a necrotic core, and/or a peripheral angiogenesis region. A two-compartment pharmacokinetic model is used to define the evolution of contrast agent concentration in dynamic contrast-enhanced studies. These numerical phantoms are then virtually imaged using a GPU-accelerated high-fidelity multiphysics imaging operator coupling photon transport and wave propagation.

## Results

Several case studies will be presented to illustrate the use of the proposed framework to guide OAT system and reconstruction method designs, as well as training and assessing AI-based approaches.

## Conclusion

The proposed framework will enhance the authenticity of virtual OAT studies. It can be widely employed for the investigation and development of advanced image reconstruction and machine learning-based methods, as well as the objective evaluation and optimization of the OAT system designs.

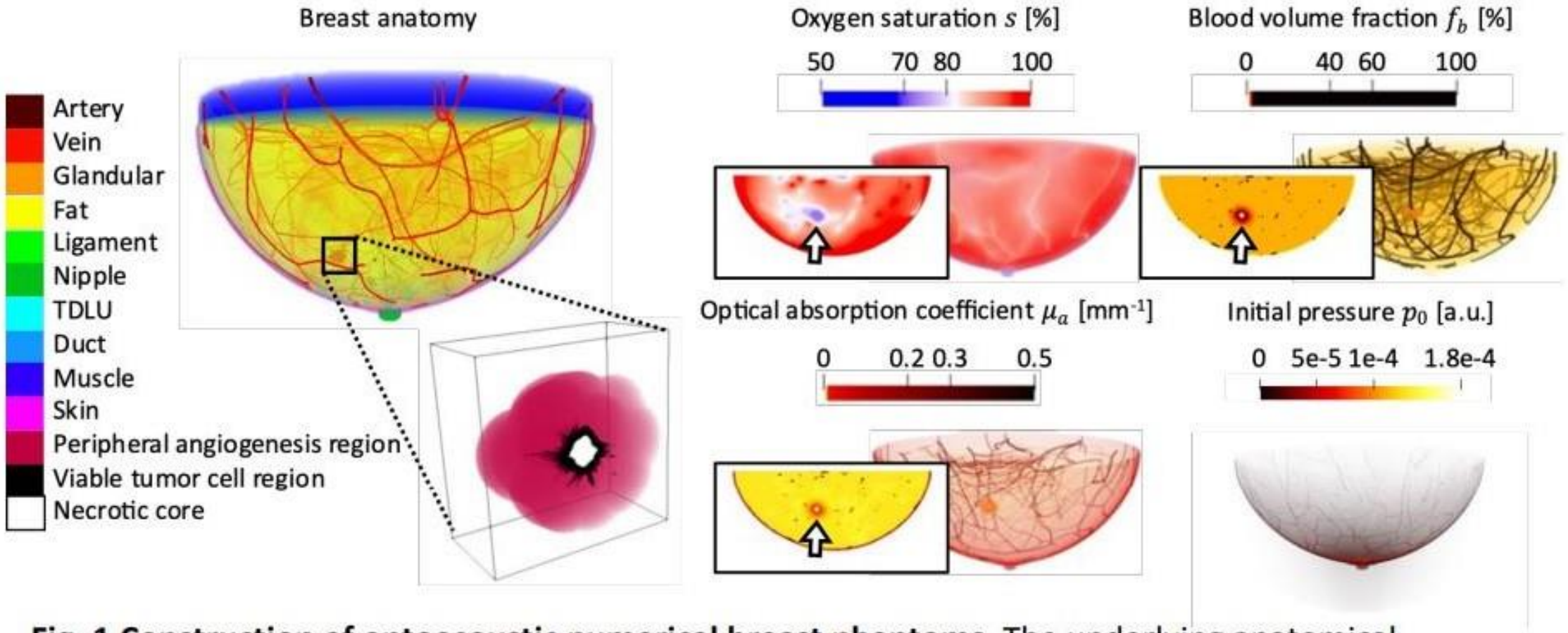

**Fig. 1 Construction of optoacoustic numerical breast phantoms.** The underlying anatomical structures are stochastically generated using tools from the Virtual Imaging Clinical Trial for Regulatory Evaluation (VICTRE) project (Badano et al, 2018). A lesion consisting of a necrotic core, viable tumor cells, and angiogenesis region, is inserted at a physiologically plausible location. Oxygen saturation and hemoglobin concentration in the area surrounding the lesion are prescribed to mimick hypoxia and angiogenesis. Wavelength dependent optical properties are computed based on mixing theory and the otoacoustically induced pressure distribution is estimated by use of Monte Carlo simulation (MCX, Fang and Boas, 2009).

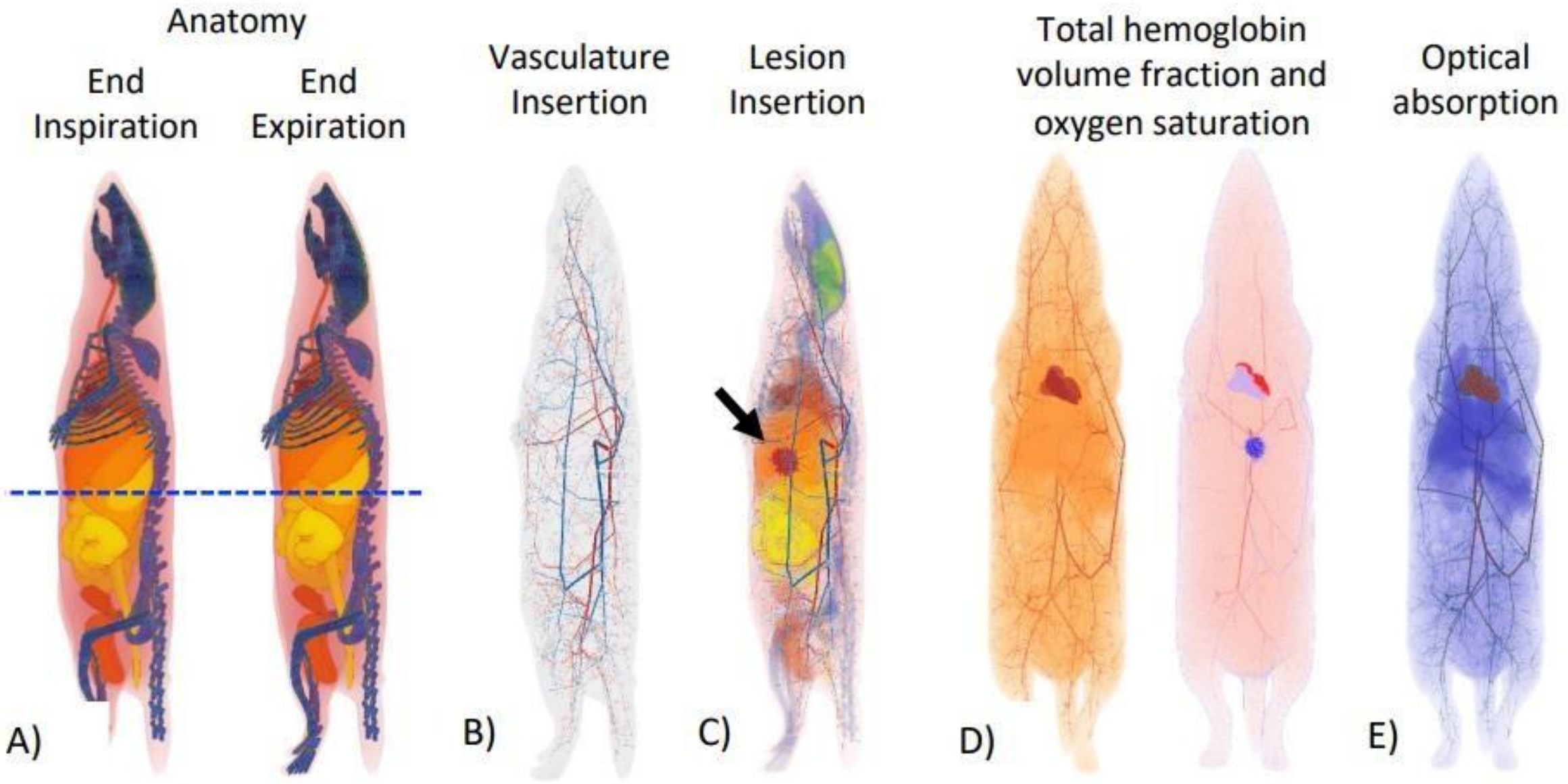

**Fig. 2 Construction of dynamic optoacoustic numerical mouse phantom.** A) Tissue maps and respiratory motion are prescribed using the anatomically realistic MOBY phantom (Segars and Tsui, 2007). B) A vasculature network is generated by use of VItA software (Talou et El, 2021). C) A circumscribed lesion is inserted in the abdomen region. D) Spatiotemporal total hemoglobin, water, lipid volume fractions and oxygen saturation are prescribed based on the dynamic tissue map. D) Optical absorption and scattering coefficient at different wavelengths are computed by use of mixing and Mie theory.

# MC-GPU v2.0: upgraded GPU-accelerated x-ray imaging device simulator

Andreu Badal (FDA)

## Purpose

Introducing MC-GPU v2.0, an upgraded version of the first GPU-accelerated Monte Carlo software for x-ray imaging device simulation.

## Methods

MC-GPU was originally released in 2009 with the vision of leveraging the nascent capability of massively-parallel, general-purpose programmable GPUs to accelerate the CPU-based PENELOPE package for simulating x-ray imaging devices. The combination of faster hardware and optimized software enabled the simulation of applications that required exceedingly long computation times with previous codes, such as cone beam CT scans. Multiple versions of MC-GPU have been developed over the years for specialized applications, such as mammography. In the new major release version of the software, we are bringing together features from different versions of the code, some new capabilities, upgraded physics models, algorithmic optimizations, and multiple bug fixes. A key improvement in the new version is the implementation of a more realistic and flexible detector and anti-scatter grid model that can reproduce multiple detector technologies used in clinical practice or under development, such as photon-counter detectors.

## Results

We will describe the code upgrades and present example simulations that highlight the new capabilities of MC-GPU v2.0. To show the range of modifications, we will focus on the simulation of dedicated breast-CT devices using different detector technologies. An example CT simulation with a direct conversion integrating detector is shown in Figure 1. Validation and verification activities performed during the development of the new code, and a study of the effect of the upgraded physics models, will be presented.

## Conclusion

GPU-accelerated Monte Carlo simulations are a valuable tool for in silico evaluation of x-ray imaging devices, and have been successfully used for device research and development and regulatory evaluation. The improved computational models implemented in MC-GPU v.2.0 extend the class of devices that can be accurately studied with this software, enabling new research opportunities.

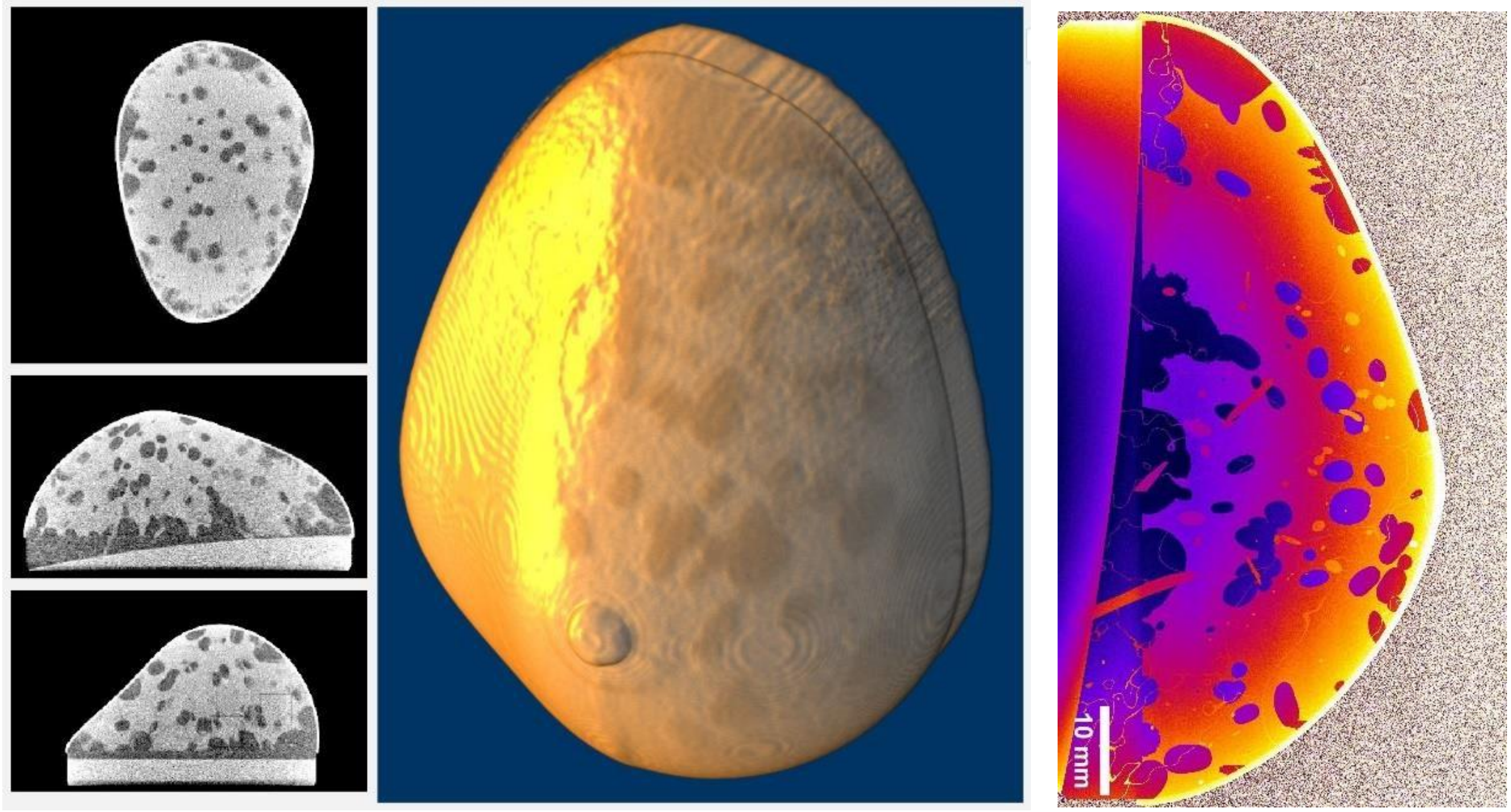

**Figure 1**: Volume rendering of a MC-GPU simulated breast-CT scan reconstructed with a FDK cone-beam CT algorithm (left and center). Voxel dose distribution (with white corresponding to highest dose and blue to lowest) tallied by MC-GPU at the central plane of the VICTRE breast phantom for the complete CT acquisition (right).

# Regulatory science and validation

Tuesday, April 23, 2024

2:45 pm - 3:25 pm

Chaired by Stephen Glick & Predrag Bakic

# Towards Consensus recommendations on reliable development and use of Virtual Imaging Trials

Ehsan Abadi, Bruno Barufaldi, Miguel Lago, Andreu Badal, Claudia Mello-Thoms, Nick Bottenus, Kristen A. Wangerin, Mitchell Goldburgh, Lawrence Tarbox, Alejandro F. Frangi, Andrew Maidment, Paul E. Kinahan, Hilde Bosmans, Ehsan Samei

## ABSTRACT

The concepts and methods underlying Virtual imaging trials (VITs) have been maturing over the last several years. These advances include continued developments of diverse models of human anatomy and diseases, as well as scanner-specific models of various imaging devices. Given the remarkable progress and benefits of VITs across a wide range of imaging modalities and applications, it is essential to establish robust frameworks to facilitate their widespread use within the medical imaging research community. To enable this, certain conditions must be met. Firstly, models of patients and imaging devices need to be credible, i.e., they need to be rigorously verified and validated, ensuring a sufficient level of realism and diversity for the intended imaging trial. Another vital requirement is reproducibility, where virtual resources and documents are designed in a way that enable researchers to seamlessly reproduce data and findings of previous trials. Further, virtual trials should be accessible which necessitates streamlined coding, proper data management schemes, standardized input/output formats with proper headers and identifiers, and user-friendly programs. To achieve these conditions (credibility, reproducibility, and accessibility), consensus guidelines are needed for collaboration, sharing, and interoperability of VIT components, a task that is recently taken on by the AAPM Task Group 387. The purpose of this abstract is to provide a summary of this task group developments with a focus on the key requirements of credibility, reproducibility, and accessibility. We include diverse imaging applications and modalities (e.g., X-ray imaging, nuclear medicine, MRI, Ultrasound), and provide examples for advancements and challenges encountered in each domain. Additionally, we discuss unmet needs and suggest the potential solutions towards positioning virtual trials as a powerful resource for medical imaging research and practice.

# Technical note: computed tomography phantom image dataset for virtual clinical trial validation

Dimitar Petrov[a,]*, Gizem Yegin[b], Rodrigo Trevisan Massera[b], Kwinten Torfs[b], Hilde Bosmans[a,b]

[a]Dept. of radiology LUCMFR, UZ Leuven, Belgium; [b]Medical physics and quality assessment, KU Leuven, Faculty of medicine, Belgium.

*dimitar.petrov@uzleuven.be

## INTRODUCTION

In the medical imaging context, virtual clinical trials (VCT) are an alternative to the classical clinical trials for evaluating and optimizing imaging concepts and technologies[1]. A VCT platform for computed tomography (CT) that simulates images that closely match the images generated on real scanners can serve as a crucial tool for CT quality evaluation and optimization. In this case, all imaging chain links are replaced by virtual counterparts, designed to simulate as close as possible the real patients and imaging system.

Some of the major challenges in VCT are related to the simulation realism and validation of the simulated images. In order to solve these issues, a number of research groups[2], [3], [4] have validated their VCT frameworks against phantom images aiming to match the simulated images to the real acquisitions in terms of CT numbers, sharpness and noise characteristics. Shankar et al. [5] adds the task-based validation of real versus simulated anthropomorphic data. Nevertheless, performing such comparisons requires access to a CT scanner and sometimes an extensive amount of time performing the actual scanning. Also the absence of a publicly available database with calibration images and the corresponding phantom models might hamper the comparison between different VCT frameworks.

The aim of this work is to provide an extensive database of phantom images acquired at 3 CT scanners at different scanning parameters and the corresponding virtual phantom with chemical composition and material density.

## METHODS

### Anthropomorphic phantom

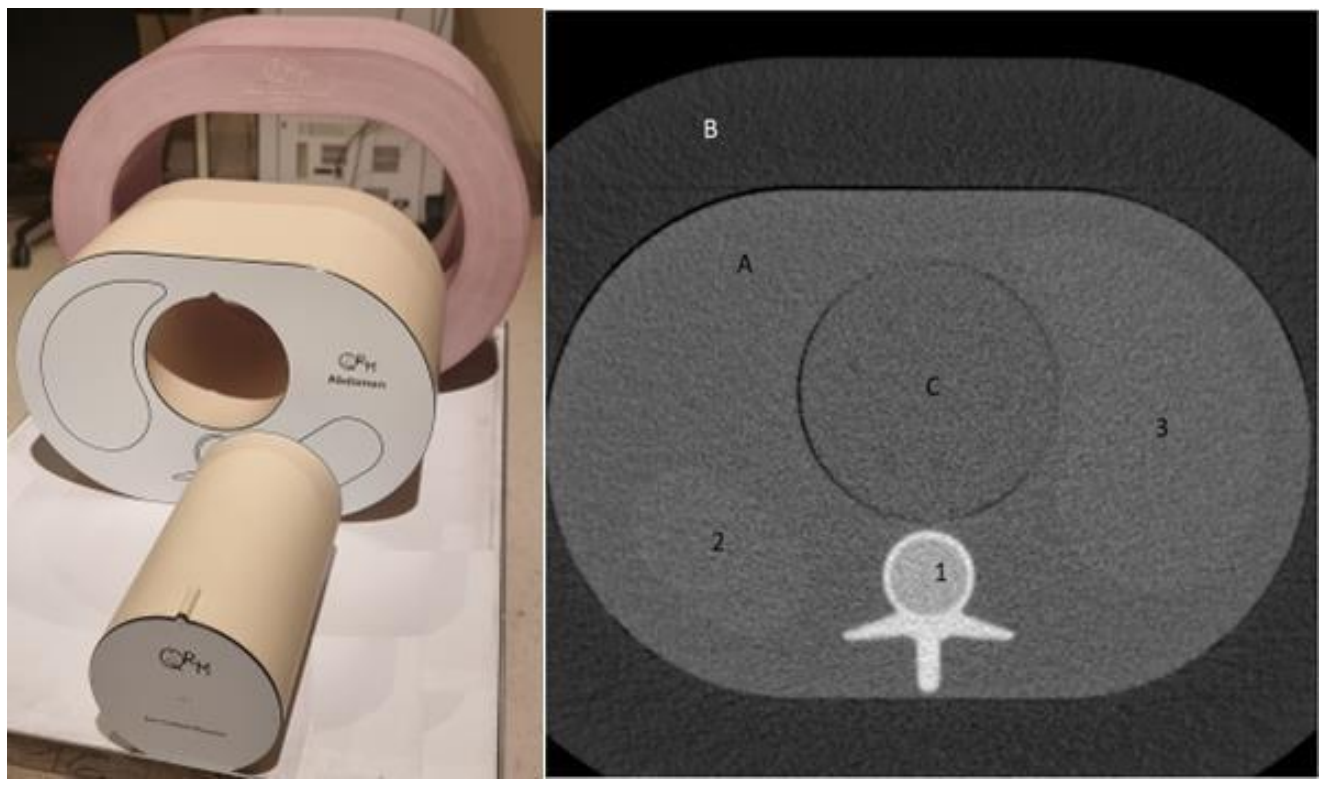

*Figure 14 Left: A photograph of the physical phantom and the three parts. Right: A CT slice of the phantom with the phantom parts and anthropomorphic landmarks denoted with numbers and letters.*

An abdomen phantom (QRM, Mohrendorf, Germany) with anthropomorphic features was used for the CT acquisitions and was later carefully segmented and virtualized. The ellipsoid phantom base (figure 1.A) with axes of 200mm and 300mm is intended to mimic the human tissue with respect to density and X-ray attenuation characteristics of a human abdomen. It comprises a bone-like spine compartment with spongiosa and corticalis mimicking areas with 200HU and 550HU respectively at 120kV (figure 1.1). Liver and spleen compartment (figure 1.2-1.3) provide the appropriate CT values of 60HU at 120kV. All organ simulating compartments are embedded in soft tissue-equivalent material with approximately 30HU at 80-120kV. In order to achieve more clinically relevant patient size, the base was situated within an 400mm x 300mm extension ring (figure 1.B) that simulates a fatty layer with -80HU at 120kV. Within the main phantom body a cylindrical insert with low-contrast targets was situated (figure 1.C and figure 2). The cylindrical insert was produced based on an in-house design dedicated for task-based image quality, i.e. the detection of subtle lesions in a largely homogenous liver. The number of targets was chosen to create enough 'signals' for human or model observer reading. The first half of the cylindrical insert contained a total of 45 spheres with diameters of 4, 6 and 8mm with contrast of -20HU compared to the soft-tissues mimicking background. Each sphere size was 15 times represented within the insert in a uniformly distributed manner. The second half of the cylinder consisted of only background for signal-free image extraction.

All three parts of the phantom were extended to a length of 170mm in order to acquire signal present and signal absent images during the same acquisition, with the aim to save time and avoid potential discrepancy between the two classes. A practical disadvantage is the total weight of the phantom.

## Acquisition data

The physical phantom was scanned on a Siemens SOMATOM Edge energy integrating CT (EI-CT) scanner, a Siemens SOMATOM Force EICT scanner and a Siemens NAEOTOM Alpha photon-counting CT (PC-CT) scanner (Siemens Healthineers, Erlangen, Germany) with a wide range of scanning and reconstruction parameters and their combinations, with a total of 324 acquisitions per EI-CT and 648 acquisition for the PC-CT. Table 1 shows all parameters for each scanner.

*Table 3 This is the caption of a table, should be left aligned above the table*

| Parameters | SOMATOM Edge | SOMATOM Force | NAEOTOM Alpha |
|---|---|---|---|
| Tube voltage [kVp] | 80, 120 | 80, 120 | 90, 120 |
| CTDI [mGy] | 3.6, 7.2, 10.7 | 3.6, 7.2, 10.7 | 3.6, 7.2, 10.7 |
| Reconstruction algorithm | FBP, IR strength 3 | FBP, IR strength 3 | QIR strength 1 and 3 |
| Reconstruction kernel | Br40f, Br54f, Br62f | Br40d, Br54d, Br64d | Br40f, Br56f, Br64f |
| Field of view [$mm^2$] | 250x250, 300x300, 350x350 | 250x250, 300x300, 350x350 | 250x250, 300x300, 350x350 |
| Slice thickness [mm] | 1.5, 3.0, 5.0 | 1.5, 3.0, 5.0 | 1.5, 3.0, 5.0 |
| Matrix size | 512x512 | 512x512 | 512x512, 1024x1024 |

## Virtual phantom

In order to compare the outcome from a VCT framework to the acquired real phantom images, the phantom needed to be virtualized. This was performed via careful segmentation of the compartments mimicking human shaped organs – liver and spine - further smoothed via mathematical functions. The major structures and simpler phantom landmarks were approximated with mathematical functions – phantom body, spleen, cylindrical insert, target spheres and extension ring. The resulting mathematical representation of the phantom was made such that it oversamples 3 times the real phantom size. This was selected as a compromise between phantom accuracy and virtual phantom memory size.

The phantom elemental composition and density were provided by the manufacturer; the 9 material attenuation coefficients were described with certainty up to 5%.

## Database

The image database and phantom characteristics were stored on a cloud server and are available for non-commercial purposes upon request after publication of peer reviewed paper. The image dataset was stored in a structured DICOM format, where the hospital sensitive DICOM tags were removed, and all acquisition parameters were left in the same format as exported from the scanner. The resulting total volume of uncompressed data is 164 GB. The file naming will be given using the following schematic: *(CT scanner)-_(kVp)_(filtration)_(dose level)_(reconstruction kernel)_(reconstruction algorithm)_(Field of view)_(matrix size)_(slice thickness)*

In order to be granted with access to the database, please email: *dimitar.petrov@uzleuven.be*

## Comparison

To illustrate how the virtual phantom and the image database can be used to check the validity of a VCT framework, an example study was conducted. For this purpose the DukeSim framework was used [2] with the standard parameters for EI-CT as configured by the authors. The input parameters were selected to closely match the chosen scanning parameters of the SOMATOM Edge EI-CT scanner (table 2).

*Table 2 This is the caption of a table, should be left aligned above the table*

| Parameters | SOMATOM Edge | DukeSim |
|---|---|---|
| Tube voltage [kVp] | 120 | |
| CTDI [mGy] | 7.2 | ~21 |
| Reconstruction algorithm | FBP | |
| Reconstruction kernel | Br40f | Hann filter; $f_1$=0.4mm$^{-1}$ |
| Field of view [mm$^2$] | 300x300 | |
| Slice thickness [mm] | 3.0 | |
| Matrix size | 512x512 | |

The resulting images were used to conduct a number of tests including CT number accuracy, contrast, sharpness, noise and task-based image quality.

### CT number accuracy

The HU mean measurements were performed using ImageJ [6] with circular ROIs where possible and in the center of the cylindrical insert, liver, extension ring, main phantom body, spongiosa and corticalis compartments of the spine.

### Sharpness

The sharpness was measured in terms of modulation transfer function. The methodology is in-house developed, but inspired by [7]. Due to constrains of the reconstruction algorithm provided with the DukeSim framework, there was insufficient air region reconstructed, so the estimation of the MTF from the air-phantom border was not possible. Thus we adjusted the algorithm to perform the measurement from the border of soft tissue and the corticalis compartment of the spine.

### Noise characteristics

The noise was estimated in two ways 1) using the global noise level (GNL) and 2) via the noise power spectrum both evaluated from the soft tissue region with CT numbers between 0HU and 170HU. The algorithm is also in-house developed, but based on the work of Christianson et al. [8] and Smith et al. [9].

### Task based image quality

The image quality will be evaluated using a recent development of a channelized Hotelling model observer (CHO) reading the low-contrast spheres within the cylindrical insert [10]. The two-layer volumetric CHO design first locates the targets within cropped VOI and then applies a volumetric anthropomorphic CHO tuned to match human results. The results will be represented in a form of Dtr, denoting the smallest barely visible sphere size. Showing higher image quality for lower Dtr values.

## RESULTS OF THE EXAMPLE USECASE

It was possible to create CT images with Dukesim of the virtual phantom and evaluate these images with the same tools as the corresponding real CT scans. The results are shown in the successive graphs.

The HU accuracy test shows a close match between the real and the virtual scan, with a max deviation of 10 HU in the soft tissue region (Figure 4 T-L). Evaluation of sharpness confirms the close correspondence in terms of spatial resolution between both images too (Figure 4 B-L). A similar result is found for the noise power spectrum (Figure 4 B-R). There is a large discrepancy between the GNL measurements with the real images showing higher noise level. This is due to the very different dose levels at which the two methods were compared.

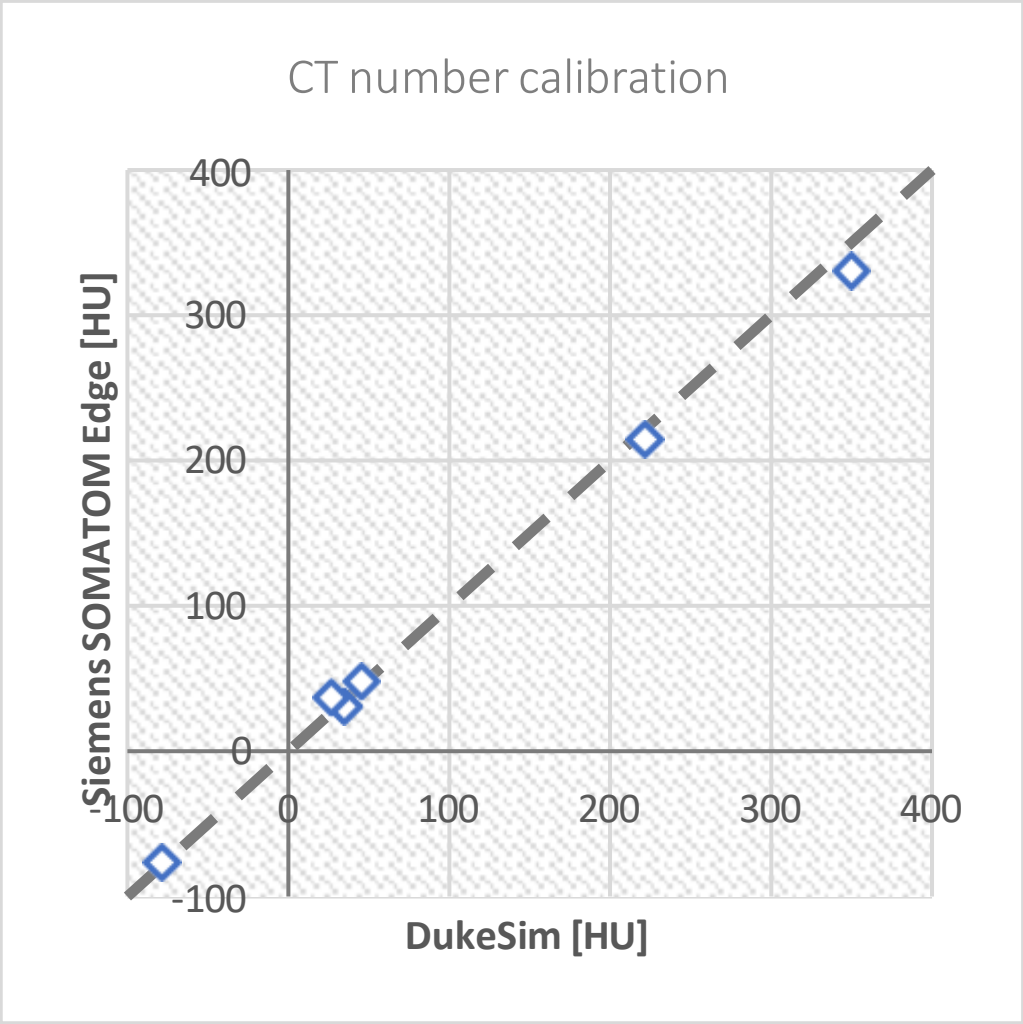

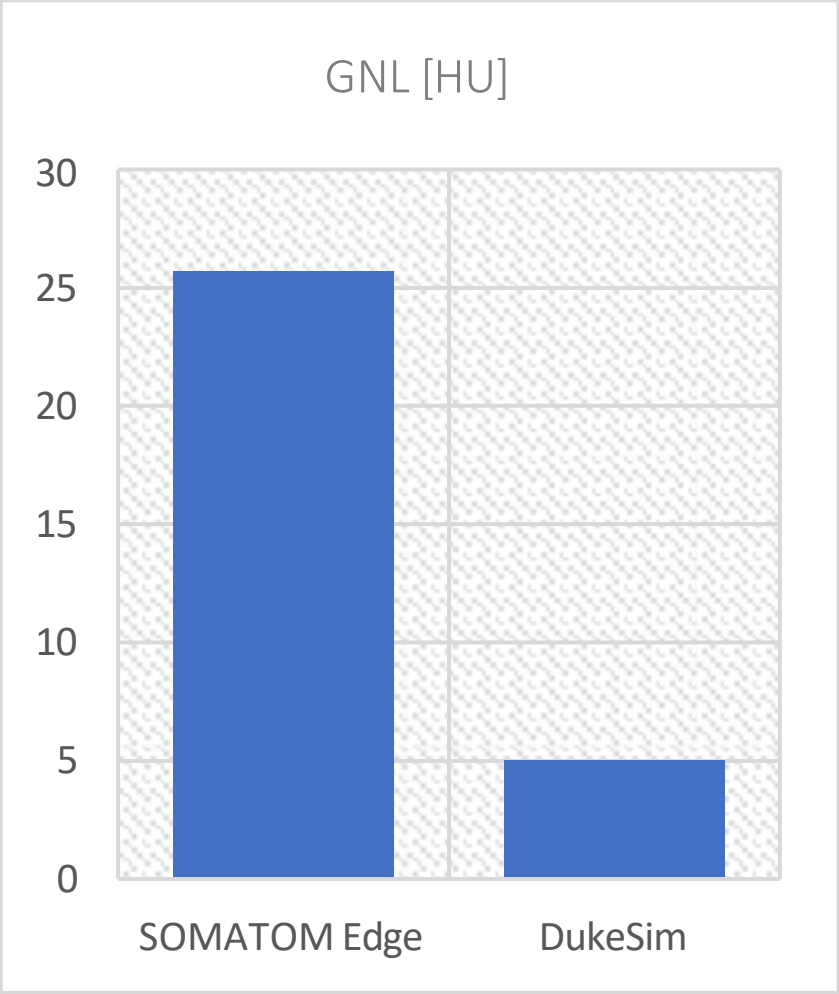

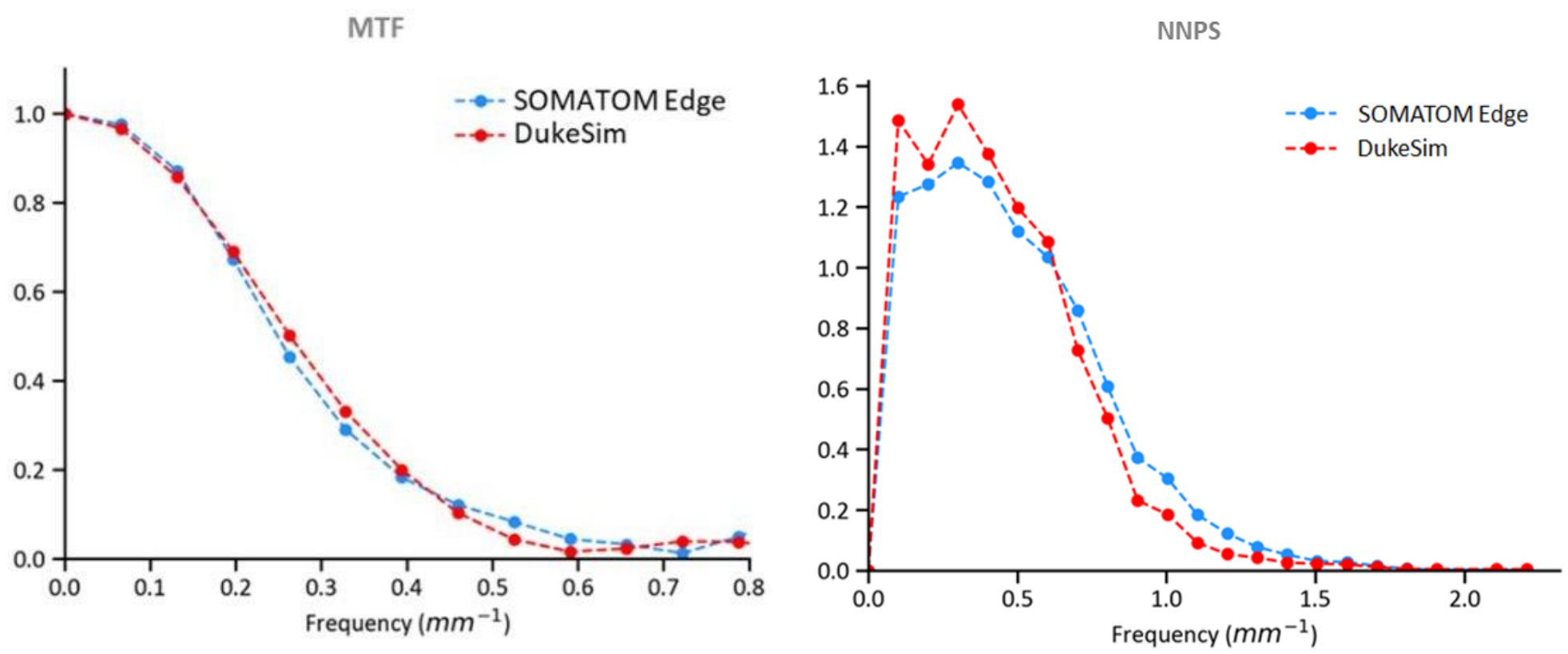

*Figure 4. Top-Left: The CT number accuracy results. The HU measured from the SOMATOM Edge image are plotted against the CT numbers of the images simulated using DukeSim; Top-Right: The global noise level results; Bottom-Left: the modulation transfer function comparison; Bottom-Right: the normalized noise-power spectrum comparison.*

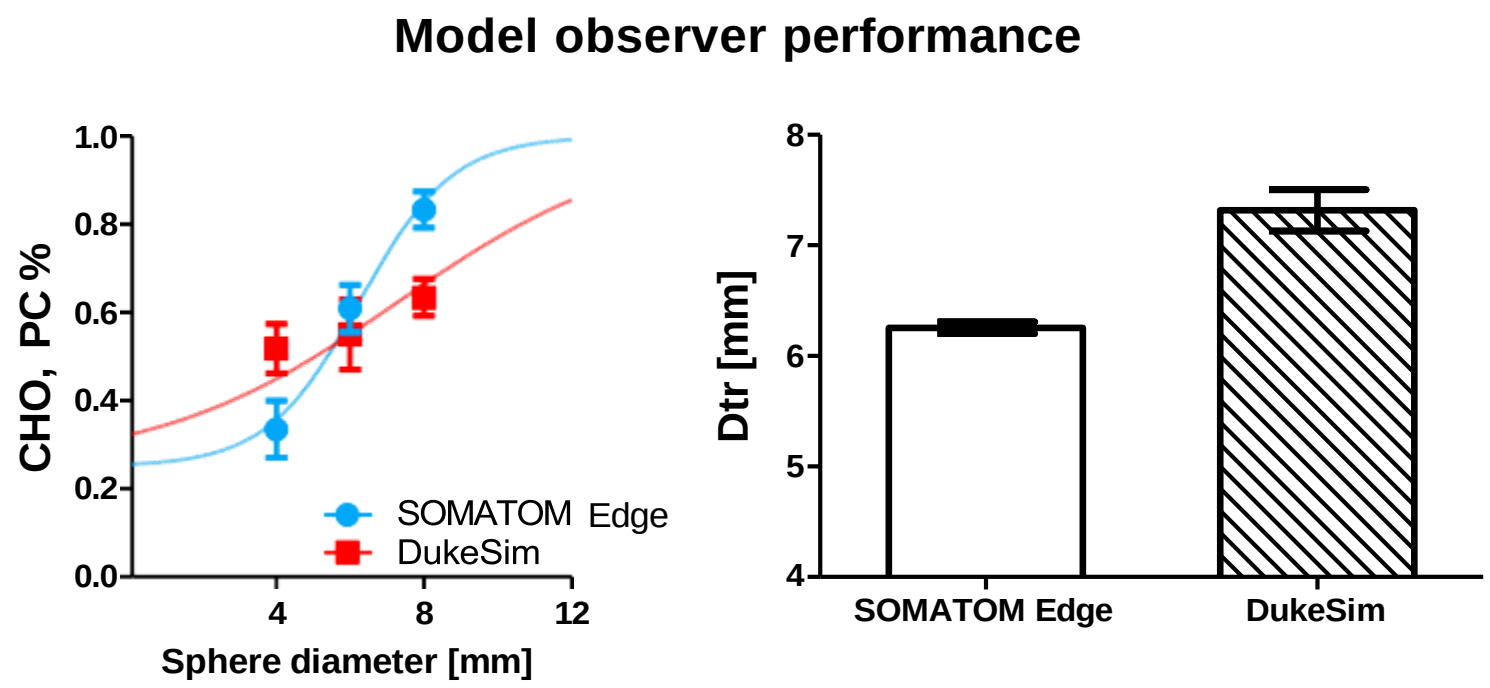

*Figure 5. The task based image quality comparison. The left curve shows the percentage correct results for 4, 6 and 8 mm spheres for both methods with logistic regression curve fitted to the data. The right graph shows the diameter threshold (Dtr) of the smallest visible sphere.*

While MTF and noise power are very similar, the model observer performance results are significantly different (figure 5). This is mainly caused by artifacts seen in the simulated image, due to the limited amount of projections. Deviations as seen in this study are the main reason for the use of a reference data base as proposed in this work.

# CONCLUSION

With this work we provide a dataset of about 1000 CT scans of an anthropomorphic phantom scanned at different scanning parameters and different CT scanners. Along with this dataset we provide a fairly accurate virtual model of the phantom with detailed properties in order to be successfully simulated. The worked example shown in this work summarizes how the dataset and model can be used in practice.

# Comparative Evaluation of Ray-Tracing and Monte Carlo Virtual Clinical Trials Pipelines for Lesion Detection in Digital Breast Tomosynthesis

Chloe J. Choi[1], Bruno Barufaldi[2], João P. V. Teixeira[3], Raymond J. Acciavatti[2], Andrew D. A. Maidment[2]

[1]Department of Bioengineering, University of Pennsylvania, Philadelphia, PA, USA
[2]Department of Radiology, University of Pennsylvania, Philadelphia, PA, USA
[3]Department of Computer Engineering, Federal University of Paraíba, Brazil

## BACKGROUND AND PURPOSE

Ideally, the assessment of new imaging systems should be conducted through clinical trials. However, these studies often face practical limitations, primarily driven by concerns about radiation exposure, as well as significant time and financial constraints. Additionally, optimizing these systems is a formidable challenge, given the multitude of parameters available in modern imaging technology, the inherent diversity in patient populations in terms of anatomy and pathology, and the absence of a definitive ground truth for reference. A promising alternative is the use of Virtual Clinical Trials (VCTs), an emerging technique that assesses medical systems through computer-simulated scenarios involving patients, imaging equipment, and interpretation processes. VCTs have gained substantial popularity in breast imaging research, primarily because of the relatively low occurrence of breast cancer within screening populations[1].

In the realm of VCTs for X-ray imaging systems, Monte Carlo (MC) simulations are regarded as the gold standard. These simulations meticulously model individual photons and their statistical interactions with matter. While highly accurate, MC simulations come with a significant computational burden and time consumption, especially when applied to tomosynthesis or computed tomography (CT), which require the generation of tens or hundreds of projections for a single acquisition[1-4]. Ray-tracing algorithms address these limitations. These algorithms approximate primary x-ray-tissue interactions using the Beer-Lambert law and therefore offer a faster and more efficient means of simulating imaging systems, albeit with a reduced level of realism compared to MC simulations[1,2,5,6].

While the capabilities and applications of x-ray simulators are evident, there is room for improvement in terms of tailoring them to specific scanners and ensuring compatibility between different simulator platforms[1]. This research will primarily focus on the latter aspect. If VCTs are conducted appropriately, they should yield consistent results regardless of the simulation pipeline used. We cross-validated the outcomes obtained from a ray-tracing simulator (OpenVCT) and an MC simulator (MC-GPU)[5,7]. This study aims to address two key questions: 1) Can one VCT pipeline accurately predict the results generated by another VCT pipeline? and 2) Do ray-tracing algorithms offer sufficient realism to potentially replace MC simulators for specific tasks, in this case, lesion detectability in DBT?

## METHODS

The overall experimental design is based on a previous study[8].

# 1. Phantom generation

## i. Simplex-noise phantoms

The texture of the breast parenchyma was modeled using **Simplex noise.**[9] **Simplex noise**, an enhancement over Perlin noise, offers smoother transitions and reduced directional bias through a sophisticated interpolation function and a simplex grid. Ninety-nine plate-shaped phantoms were created using a simplified version of the algorithm from Barufaldi et al[8]. Each phantom measures 308 mm × 240 mm × 45 mm with a uniform voxel size of 200 microns. They were designed to span the full extent of the detector area (304.64 mm × 239.36 mm). No coarse masks were used to ensure an even distribution of the glandular tissue throughout the phantom volume. No refinement using histogram thresholding was performed. The parameters used for Simplex-noise phantom generation are summarized in table 1.[9]

Table 1. Noise parameters used for Simplex-noise phantom generation.

| Noise parameters | Values |
|---|---|
| Range of set levels {[min, min], [max, max]} | {[0.85, 0.95], [0.97, 0.98]} |
| Number of volumes | 3 |
| Range of octaves (min:max in 1 step) | [4, 16] |
| Persistency | 0.73 |
| Lacunarity | 1.5 |

The three-dimensional model of the spherical breast mass, measuring 6 mm in each dimension, was created with an isotropic resolution. We inserted 154 lesions uniformly across the central layer of each Simplex-noise plate phantom, corresponding to the mid-thickness region, by employing the voxel additive technique outlined by Barufaldi et al[9]. This method integrates the lesion into the existing tissue structure by combining their attenuation coefficients, yielding more lifelike imagery. Accordingly, the composite attenuation coefficient is calculated as the cumulative attenuation coefficients of the lesion and the underlying breast tissue. In this study, the weighting factor assigned to the lesion (*wl*) was set at 2.0%[10].

## ii. Conversion of VCTX (OpenVCT) to RAW (MC-GPU)

As MC-GPU does not readily support partial-volume indices, modifications were made to binarize the phantom volume into 100% adipose and 100% glandular tissues. The allocation of each voxel to glandular or adipose tissue was determined stochastically, adhering to the predefined partial volume fractions of each index. This ensured the proportional distribution of tissue types within the phantom aligned with the specified ratios. The density of the inserted lesions was adjusted to achieve an adequate level of detectability. The overall breast density was maintained at a constant level. A subset of 33 phantoms was used for MC-GPU analysis, instead of the larger set of 99 that was used for OpenVCT.

# 2. Image acquisition and reconstruction

OpenVCT and MC-GPU pipelines were used to simulate x-ray projections of the generated phantoms. The image acquisition parameters (Table 1) were simulated to resemble the NGT system with an AXS-2430 detector (Analogic Canada Corporation, Montreal, Quebec). A Docker environment was used to execute MC-GPU as a part of the VICTRE pipeline[11]. No other components of the VICTRE pipeline were used; reconstructions and reader studies were performed outside of the pipeline to ensure consistent image looks between OpenVCT and MC-GPU. The central slice of the reconstruction was generated using commercial software (Briona 9.0.4, Real Time Tomography, Villanova, PA) with simple back projection. Table 2 summarizes the acquisition parameters.

Table 2. Summary of the system parameters.

| X-ray parameters | values |
| --- | --- |
| Angular range | ±7.5° |
| Source-to-Imager distance (mm) | 738 |
| Breast support-to-Imager distance (mm) | 22.5 |
| Detector element size (mm) | 0.085 × 0.085 |
| Detector size (pixels) | 3584 × 2816 |
| Reconstruction size (pixels) | 0.085 × 0.085 |

## 3. Virtual reader studies

An open-source virtual reader studies software called Medical Virtual Image Chain (MeVIC)[12] was used to simulate image displays on a high-resolution monitor and lesion detection studies based on a mathematical virtual reader model. A Channelized Hotelling Observer (CHO) model with 15 Laguerre-Gauss channels and spread 24 was trained and tested.

## 4. Data analysis

To evaluate the spatial dependency in lesion detectability, the reconstructed images were segmented into three distinct rectangular zones (anterior, middle, and posterior), each representing different distances from the chest wall. Independent sets of regions of interest (ROIs) of 201 × 201 pixels were randomly selected for each region to train and test the CHO model. The data comprised [1000, 650] ROIs with lesions (“diseased”) and [1000, 650] without (“healthy”) for each spatial region, for [OpenVCT, MCGPU]. The performance of lesion detection was calculated using regional Receiver Operating Characteristic (ROC) analysis, employing open-source R software tools like *pROC, auctestr*, and *psycho*[13-16], which facilitate computation of the Area Under the ROC Curve (AUC) and the detectability index (d'). The comparison of the ROC curves was conducted using the DeLong test[17].

# RESULTS

When comparing the computational efficiency of the two pipelines, it was observed that one pipeline completed its process within approximately 15 to 20 seconds, whereas the other required significantly more time, ranging from 9 minutes and 15 seconds to 9 minutes and 30 seconds. This indicates a substantial difference in computational speed, with the ray-tracing OpenVCT being markedly faster.

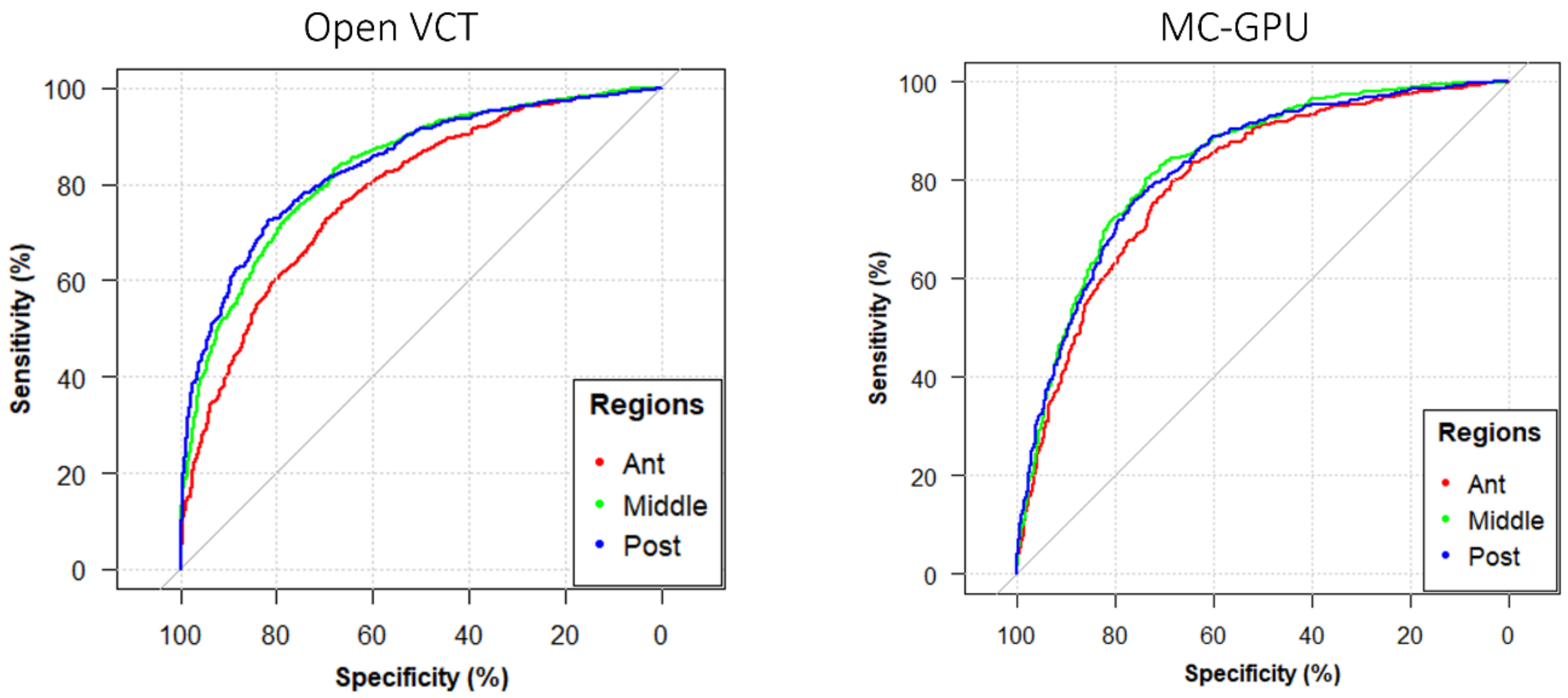

| 1000 ROIs | Ant | Middle | Post |
|---|---|---|---|
| d' | 1.12 | 1.38 | 1.49 |
| AUC | 0.78 (0.760-0.800) | 0.83 (0.815 – 0.850) | 0.84 (0.823 – 0.857) |

| 650 ROIs | Ant | Middle | Post |
|---|---|---|---|
| d' | 1.27 | 1.49 | 1.41 |
| AUC | 0.80 (0.772-0.825) | 0.83 (0.810 – 0.855) | 0.83 (0.804 – 0.848) |

*Figure 1. ROC curves and quantitative metrics from OpenVCT (left) and MC-GPU (right) are shown, with each curve corresponding to a different region.*

Both OpenVCT and MC-GPU experiments demonstrated a consistent trend in lesion detectability. Lesion detectability was significantly lower in the anterior region compared to the middle and posterior regions. Statistical analyses showed significant differences with p-values for anterior versus posterior comparisons being <0.001 for OpenVCT and 0.13 for MC-GPU; and for anterior versus middle, p-values were <0.001 for OpenVCT and 0.06 for MC-GPU. The discrepancies in AUC or detection index (d') between OpenVCT and MC-GPU are attributable to their differing phantom formats. Specifically, OpenVCT accommodates partial volume voxels, whereas MC-GPU does not, necessitating adaptations in the experimental setup.

It is important to note the different numbers of ROIs used in the two experiments: 1000 ROIs for OpenVCT and 650 ROIs for MC-GPU. The reduced number of ROIs in the MC-GPU experiment was due to longer computation times, limiting the analysis to 33 phantoms compared to 99 for OpenVCT.

## DISCUSSION AND CONCLUSION

Monte Carlo simulations are detailed, computationally complex software that serve as the gold standard in simulating x-ray imaging systems because of the high degree of physical realism. Ray-tracing offers a faster, more efficient alternative, albeit with some realism trade-offs. The comparative evaluation of OpenVCT and MC-GPU in DBT simulations revealed comparable results in terms of AUC and d', underscoring the potential utility of ray-tracing VCT pipelines in practical clinical applications.

## ACKNOWLEDGEMENTS AND CONFLICT OF INTEREST

Andrew D. A. Maidment and Raymond J. Acciavatti are inventors on patents and patent applications related to next-generation tomosynthesis technology. Andrew D. A. Maidment has received research support from Hologic Inc., Barco NV, and Analogic Corporation; is the spouse to an employee and shareholder of Real Time Tomography (RTT), LLC; is a member of the scientific advisory board of RTT; and is an owner of Daimroc Imaging, LLC. Support was provided by Breast Cancer Alliance, Inc. (2022 Young Investigator Grant) and the National Institutes of Health (T32EB009384, R01CA196528, R37CA273959, and P30CA016520). The content is solely the responsibility of the authors and does not necessarily represent the official views of the funding agencies.

# Virtual trials and processing methods

Wednesday, April 24, 2024

8:15 am - 8:55 am

Chaired by Joseph Lo & Alex Frangi

# Addressing the multiplicity challenge of spectral CT through physics-regulated estimation of composition in spectral energy CT (PRECISE CT)

Mojtaba Zarei (Duke University), Jayasai Rajagopal ( Duke University ), Nicholas Felice ( Duke University ), Ehsan Samei (Duke University)

## Purpose

An integral component of spectral computed tomography (CT), spectral material decomposition is commonly captured in terms of specific material or energy-dependent renditions. This study aims to utilize a virtual imaging trial (VIT) platform and a deep neural network to decompose spectral information in terms of material primer components, density and atomic number, catheterized as physics-regulated estimation of composition in spectral energy CT (PRECISE CT).

## Method

A VIT platform was employed to generate a comprehensive training dataset and the corresponding ground truth in terms of physical density (ρ) and effective atomic number (Zeff) from 32 computational phantoms. The simulations were done assuming a state-of-the-art photon-counting CT system. This dataset was designed to encompass realistic variations encountered in clinical settings and includes four sets of clinically relevant energy thresholds, two dose levels, four iodinated contrast agent concentrations, and three reconstruction kernels. To enhance the robustness of our model, a novel physical-informed regularization loss was added to a generative adversarial network (GAN) loss. The model was then trained and evaluated using a test set including 16 distinct computational phantoms.

## Results

Density and effective atomic number maps in the test set exhibited a low normalized mean squared error (less than 5% for ρ and less than 1% for Zeff). Additionally, estimated material domains exhibited a peak signal-to-noise ratio of 40 dB for the ρ and 35 dB for the Zeff images. Multiscale-structural-similarity-index-measures for both domains were greater than 0.99.

## Conclusion

The results demonstrate that spectral CT data can decompose into the primary attributes of the material generating the CT contrast. This was enabled by a novel training approach based on VIT that circumvents the limitations of accessing the training dataset and accurate ground truth, coupled with the integration of physical-informed regularization within a GAN. This strategy may address the ongoing challenge of robust rendition of spectral data for efficient interpretation.

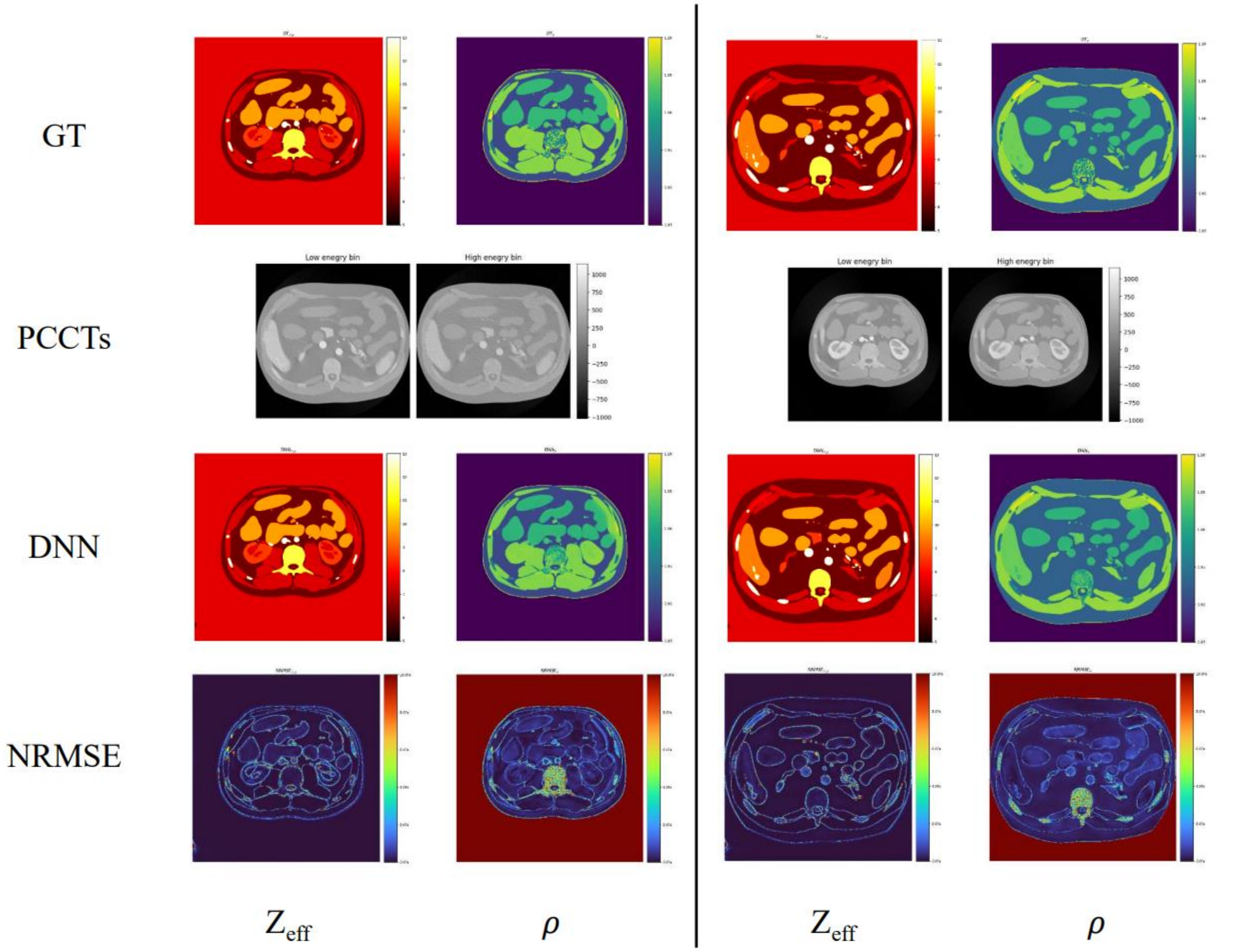

# Development of organ-specific suppression technique for dynamic chest radiography using virtual patients

Futa Goshima[1], Rie Tanaka[1], Ryuichi Nagatani[1], Ryunosuke Goto[1], Noriyuki Ohkura[2], Isao Matsumoto[3]
1. College of Medical, Pharmaceutical & Health Sciences, Kanazawa University, Ishikawa, Japan
2. Department of Respiratory Medicine, Kanazawa University hospital, Ishikawa, Japan
3. Department of Thoracic Surgery, Kanazawa University, Ishikawa, Japan

## BACKGROUND AND PURPOSE

Dynamic chest radiography (DCR) is a flat-panel detector (FPD)-based functional X-ray imaging method. DCR allows the evaluation of lung function, based on changes in lung density during respiration. However, the evaluation is adversely affected by overlapping organ shadows in the lungs. The purpose of this study was to develop a deep-learning-based organ-suppression technique for DCR.

## METHODS

Four-dimensional (4D) extended cardiac-torso (XCAT) phantoms with normal heart rate (60 beats/min), slow breathing (6 breaths/min), and diaphragm displacement (4 cm) were generated. An X-ray simulator was used to create the projection images of the XCAT phantoms. The projection was conducted under the same imaging conditions as in the actual DCR. In addition to the original images, projected images of the organ structures to be suppressed (bone, mediastinum, and diaphragm) were created and used for training pix2pix, to estimate the organ images from original images (Figure 1). The organ suppression images created by subtracting the organ images from the originals, were evaluated based on the structural similarity index measure (SSIM) and peak signal-to-noise ratio (PSNR). The clinical cases were processed using a trained model.

## RESULTS

The proposed method resulted in high values for both metrics in every organ examined (PSNR/SSIM:48.3 dB/0.986 for bone, 35.7 dB/0.826 for mediastinum, and 30.1 dB/0.962 for diaphragm). Visual evaluation confirmed that organ shadows were reasonably suppressed in the resulting images of both the XCAT phantom and real patients (Figure 2, 3 ). The results indicate that the proposed method can effectively reduce organ shadows.

## CONCLUSION

The proposed deep learning-based organ-suppression technique for DCR is expected to improve the evaluation of lung function.

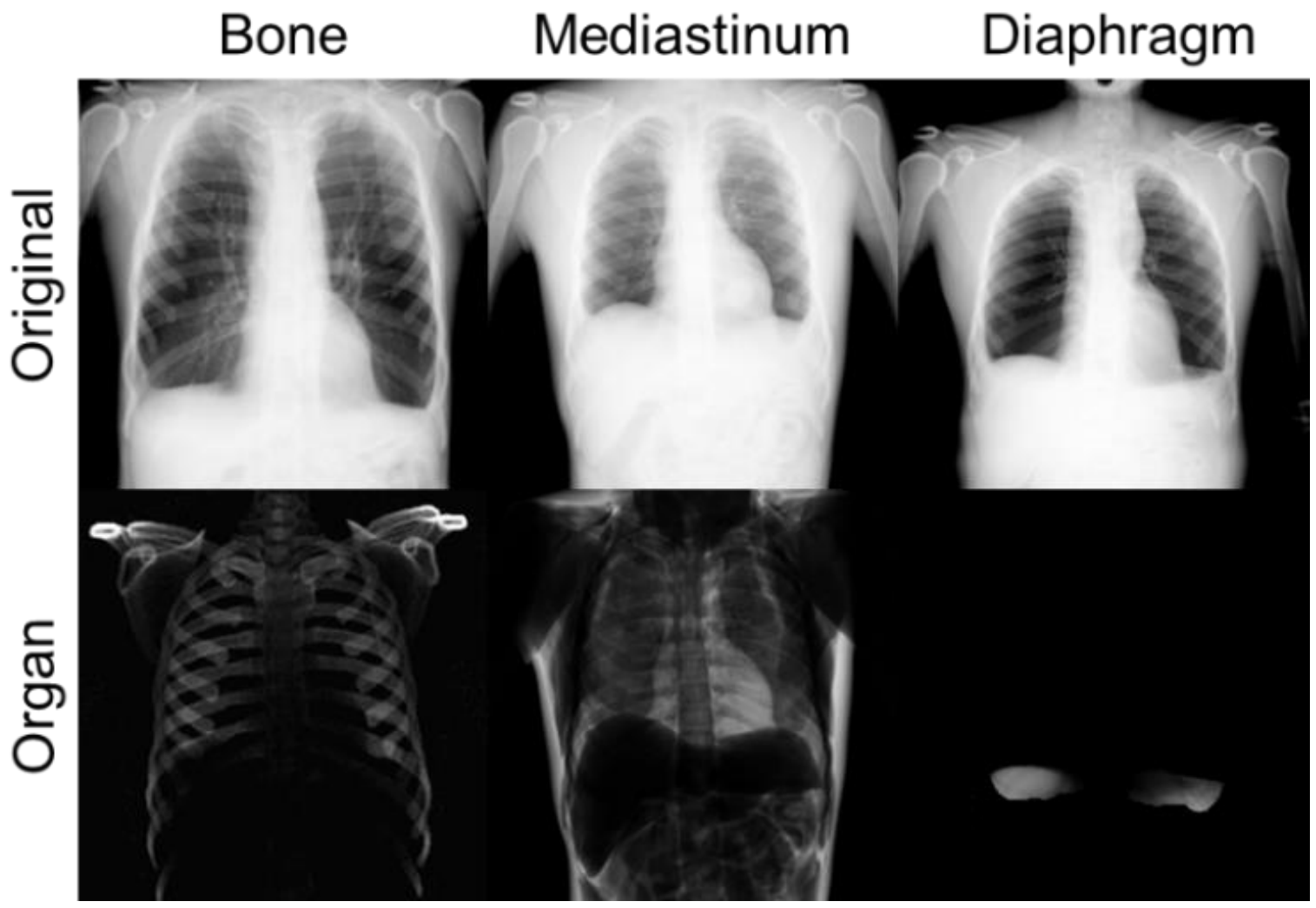

*Figure 1 Projection images of the XCAT phantom with (labelled as original) and without specific organ structures (bone, mediastinum, or diaphragm) (labelled as organ).*

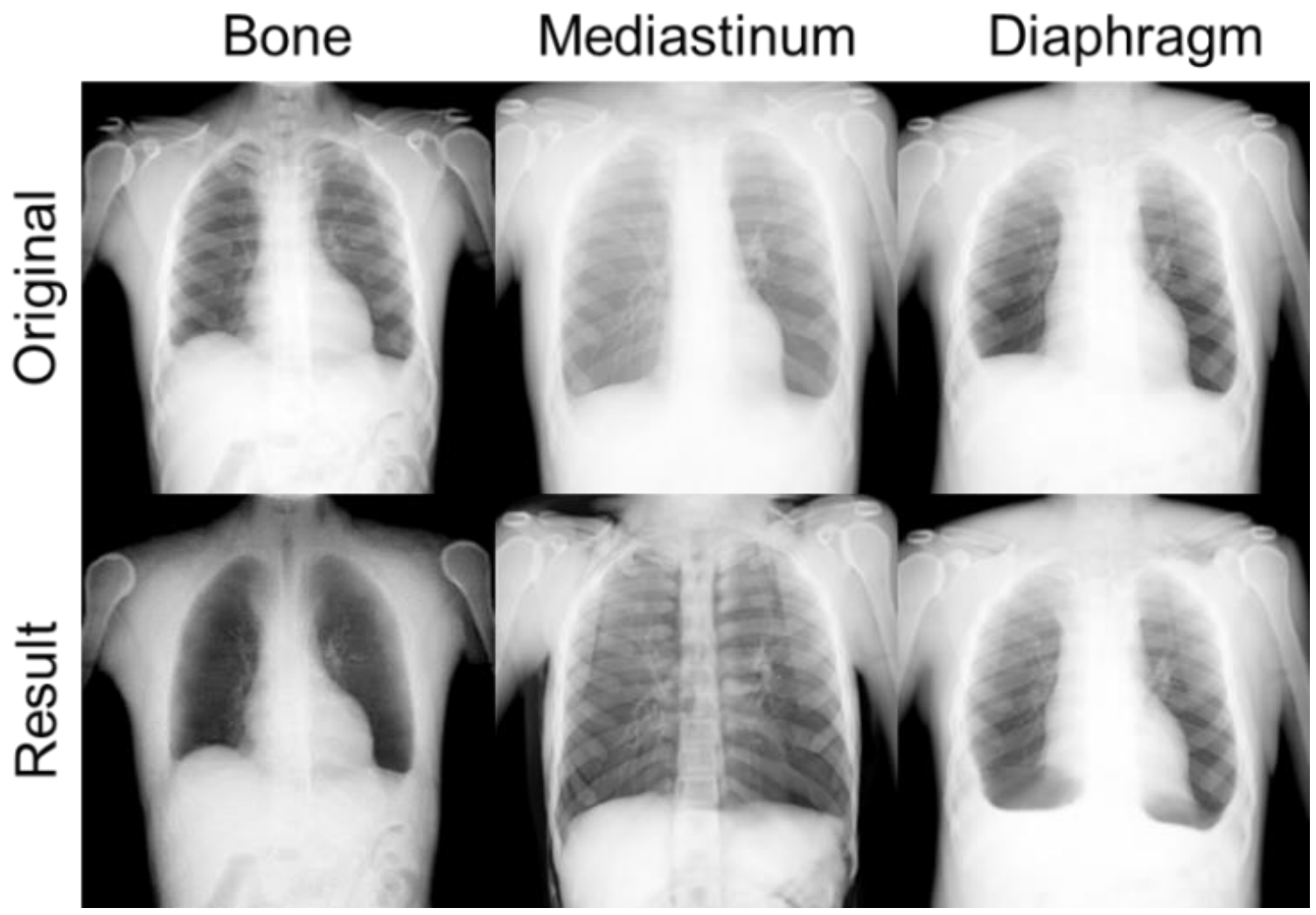

*Figure 2 Examples of organ-specific suppression images in XCAT phantom.*

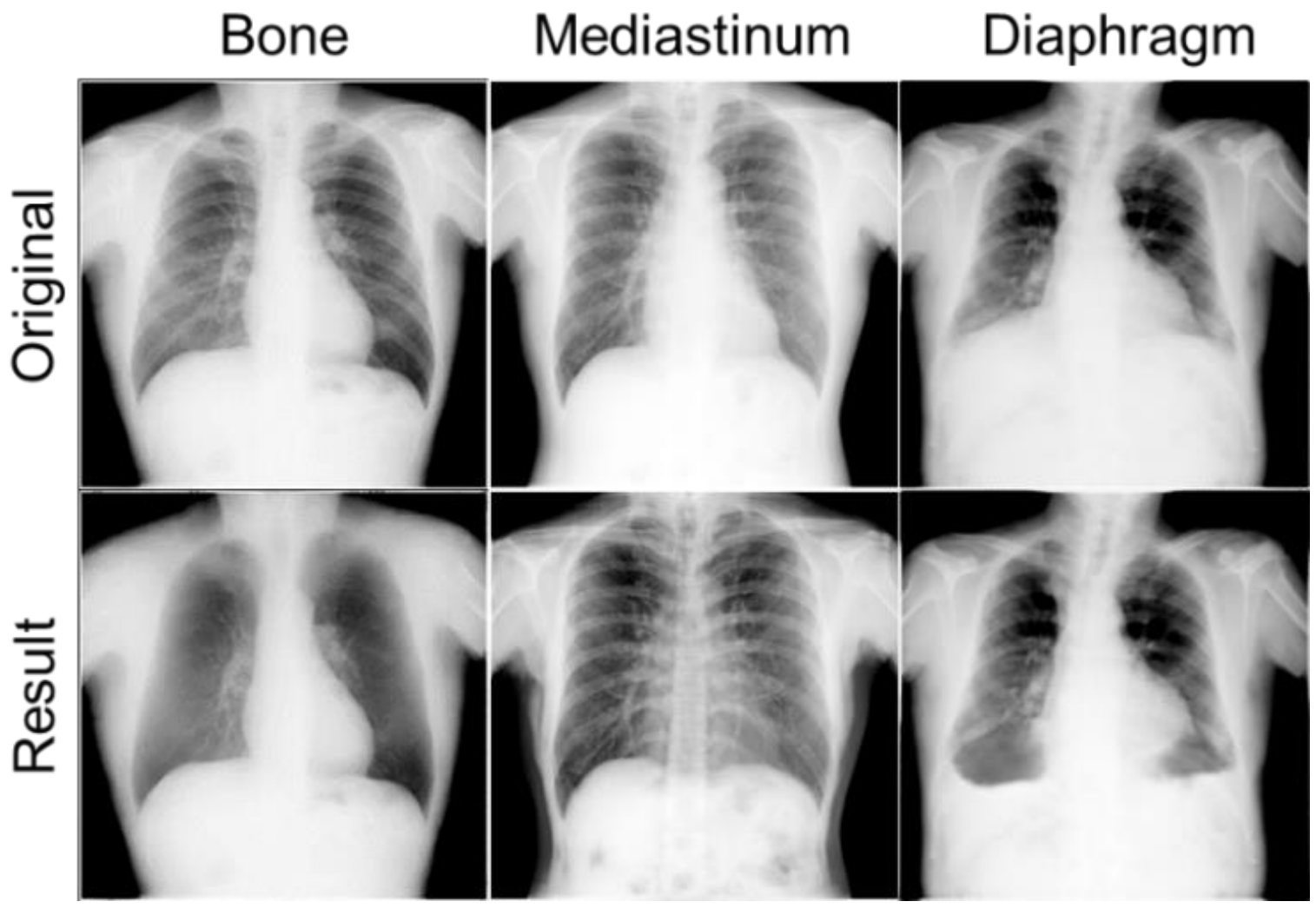

*Figure 3 Examples of organ-specific suppression in images of clinical cases.*

# Deep physics-informed super-resolution of cardiac 4D-flow MRI: a synthetic study in patient-specific geometries

Fergus Shone (CISTIB, University of Leeds), Alejandro Frangi (University of Manchester), Erica Dall'Armellina ( University of Leeds ), Zeike Taylor (University of Leeds), Peter Jimack (University of Leeds)

## Purpose

4D-flow magnetic resonance imaging (MRI) provides non-intrusive blood flow reconstructions in the left ventricle (LV), and has the potential to become a key tool in both research and clinic. However, low spatio-temporal resolution and the presence of significant noise artifacts hamper the accuracy of derived haemodynamic quantities and thus limit the effectiveness of the modality to establish links between haemodynamic abnormalities and pathologies. Further, models that are constrained by boundary conditions are impacted by additional uncertainty which arises due to low spatial resolution of structural cine-MRI.

## Methods

We introduce a physics-informed neural network (PINN) model for cardiac 4D-flow MRI super-resolution. Model training is constrained by low-resolution 4D-flow MRI data, the no-slip boundary condition on the endocardium and the governing physical equations. We utilise a self-adaptive soft-attention scheme (McClenny et al. 2023) with a SIREN network architecture (Sitzmann et al. 2020). Synthetic 4D-flow MRI data was produced in a cohort of 3 LV cases, downsampling by factors of 5 in both space and time, with artificial noise added to achieve a signal-to-noise ratio of 10. To enforce no-slip boundary condition, an approximated endocardium is used, replicating the positional uncertainty present in real 4D-flow MRI studies.

## Results

Our PINN model is able to reconstruct clinically-relevant haemodynamic variables with increased accuracy vs raw 4D-flow MRI data. In Fig. 1 we demonstrate qualitative improvement in the reconstruction of vorticity fields, linked to MI patients (Demirkiran et al. 2022). In Fig. 2 we see our model predicts kinetic energy, also linked to MI (Garg at al. 2018), to a higher degree of accuracy.

## Conclusions

A PINN-based approach is proposed for super-resolution of cardiac 4D-flow MRI. Model performance is demonstrated using a cohort of synthetic cases, where the model is shown to outperform raw 4D-flow MRI data across a range of clinically relevant metrics.

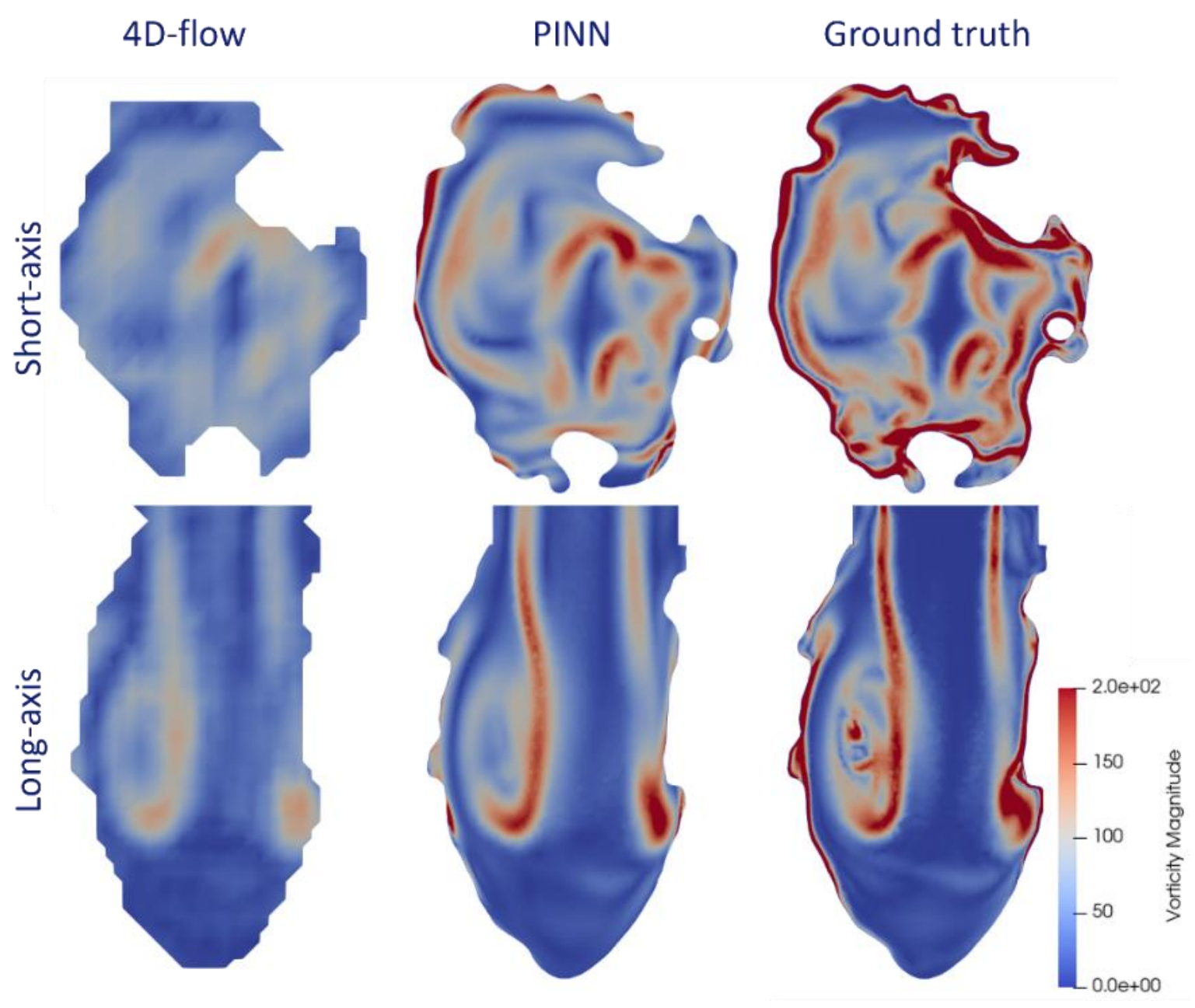

*Figure 2: Contours of vorticity magnitude in late diastole, on a short-axis slice near the base (top row) and long-axis slice (bottom row), with raw 4D-flow MRI data (left column), PINN prediction (middle column) and ground truth (right column).*

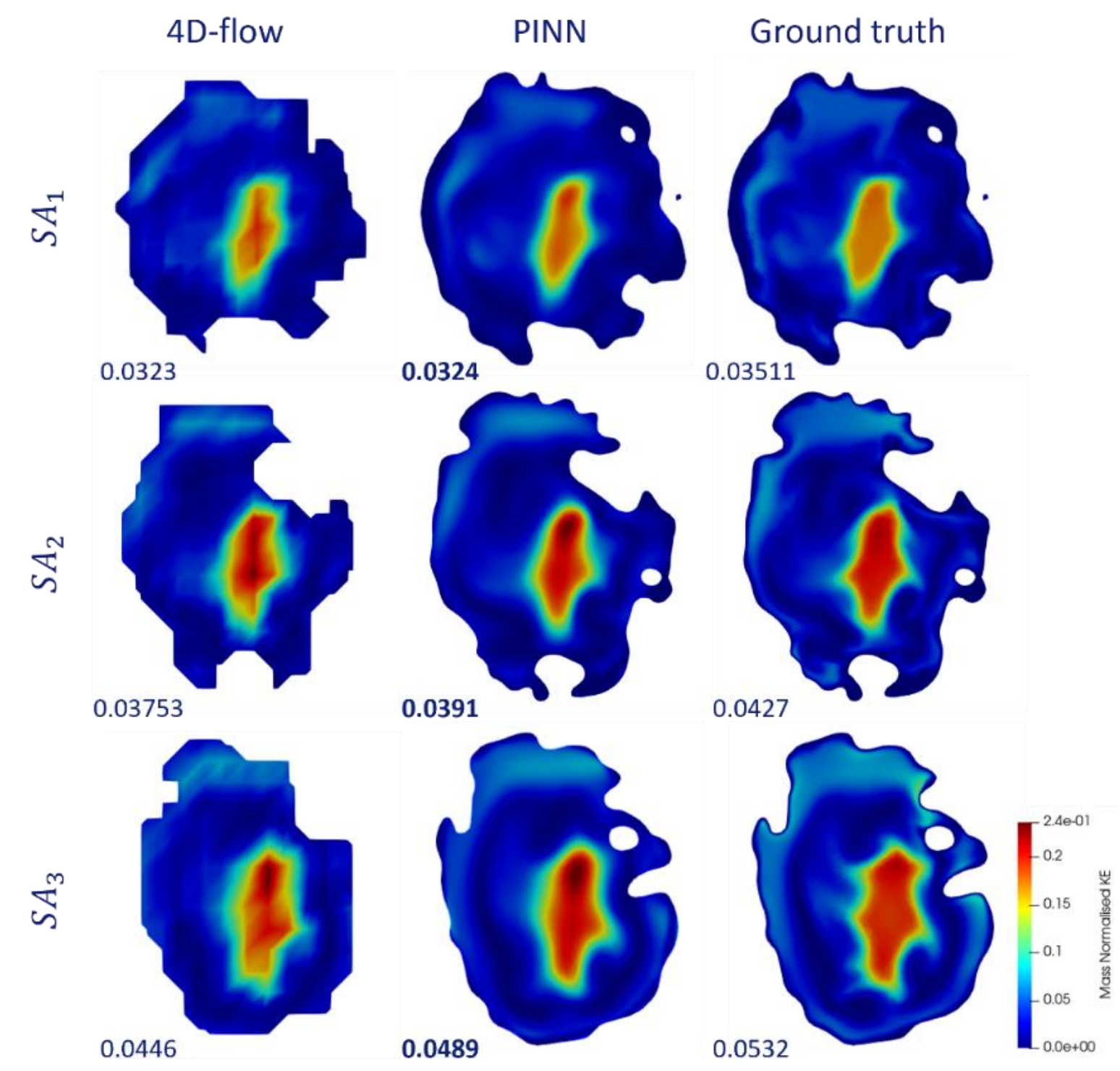

*Figure 3: Contour plots of mass-normalised kinetic energy at three short-axis slices in mid-diastole, with raw 4D-flow MRI data (left column), PINN prediction (centre column) and ground truth data (right column). Reported values are mean in-plane kinetic energy.*

# Virtual trials and artificial intelligence

Wednesday, April 24, 2024

8:55 am - 9:35 am

Chaired by Joseph Lo & Alex Frangi

# Evaluating breast density classification by sparse approximation classifiers and deep networks using simulated digital mammograms

Chelsea E. Harris[a], Predrag R. Bakic[b], and Sokratis Makrogiannis[a]
a. Delaware State University, USA b. Lund University, Sweden

## BACKGROUND AND PURPOSE

Breast density is a significant consideration in assessing the risk of breast cancer. Increased density limits the sensitivity and specificity of mammographic screening, and is correlated with genetic predisposition to cancer. Although deep learning techniques have produced promising results, limited availability of annotated data is a challenge especially as network model sizes increase. We hypothesize that simulated mammograms may help to enrich training datasets and improve classification performance.

## MATERIALS AND METHODS

We aim to evaluate the classification performance of deep networks, sparse approximation, and joint deep and sparse approximation techniques. In our preliminary experiments, we used regions of interests from CBIS-DDSM and INbreast public datasets. We have also simulated digital mammograms of various density levels. Simulated images are generated using open platform OpenVCT and computational breast phantoms, selected to match clinical density distribution, and the mapping between volumetric percent density and BI-RADS categories. We performed classification between the lower two and upper two BI-RADS density categories. We have employed five deep network models: Densenet201, InceptionResNetV2, InceptionV3, ResNet101, and Xception. We have compared performances of sparse representation classification using label-separated dictionary learning (LS-SRC) and deep feature dictionaries (DF-SRC), and CNNs with transfer learning or fine-tuning. The training set is a mixture of simulated and real mammograms. Our test set contains only real mammograms.

## RESULTS

Stratified 10-fold cross validation DF-SRC classification results yielded a breast density classification accuracy of 74.62% and AUC of 0.8139 using only real mammograms for training and testing. While these results are encouraging, the training dataset is limited, and simulated images may address the need for a rich training dataset.

## CONCLUSION

Our preliminary results suggest that deep feature dictionaries may leverage sparse approximations for breast density prediction. We will compare the performance of DF-SRC, LS-SRC and end-to-end CNN classifiers, with and without the use of simulated images.

Table 1: Distribution of each BI-RADS category in our low vs. high density class separation for CBIS-DDSM data.

| | Low density | | High density | | |
|---|---|---|---|---|---|
| BI-RADS | 1 | 2 | 3 | 4 | |
| | 85 | 154 | 93 | 38 | 370 |

Table 2: Breast density classification performance of DF-SRC and LS-SRC methods using benign mass cases from the CBIS-DDSM dataset. The feature dimension provided for LS-SRC experiment results is the resulting feature dimensionality after PCA is applied. (256) and (128) correspond to the undersampling size of the ROIs before dimenstionaity reduction; 256 × 256 and 128 × 128 pixels respectively.

| Method | Feature Dimension | TNR % | TPR % | ACC % | AUC |
|---|---|---|---|---|---|
| Densenet201-SRC | 256 × 1 | 80.00 | 63.08 | 71.54 | 0.7595 |
| InceptionResNetV2-SRC | 512 × 1 | 79.23 | 70.00 | **74.62** | **0.8139** |
| InceptionV3-SRC | 1280 × 1 | 79.23 | 63.08 | 71.15 | 0.7905 |
| ResNet101-SRC | 2048 × 1 | 64.62 | 74.62 | 69.62 | 0.7918 |
| Xception-SRC | 512 × 1 | 76.15 | 64.62 | 70.38 | 0.7575 |
| LS-SRC(256) | 346 × 1 | 47.69 | 70.00 | 58.85 | 0.6201 |
| LS-SRC(128) | 341 × 1 | 55.38 | 62.31 | 58.85 | 0.5725 |

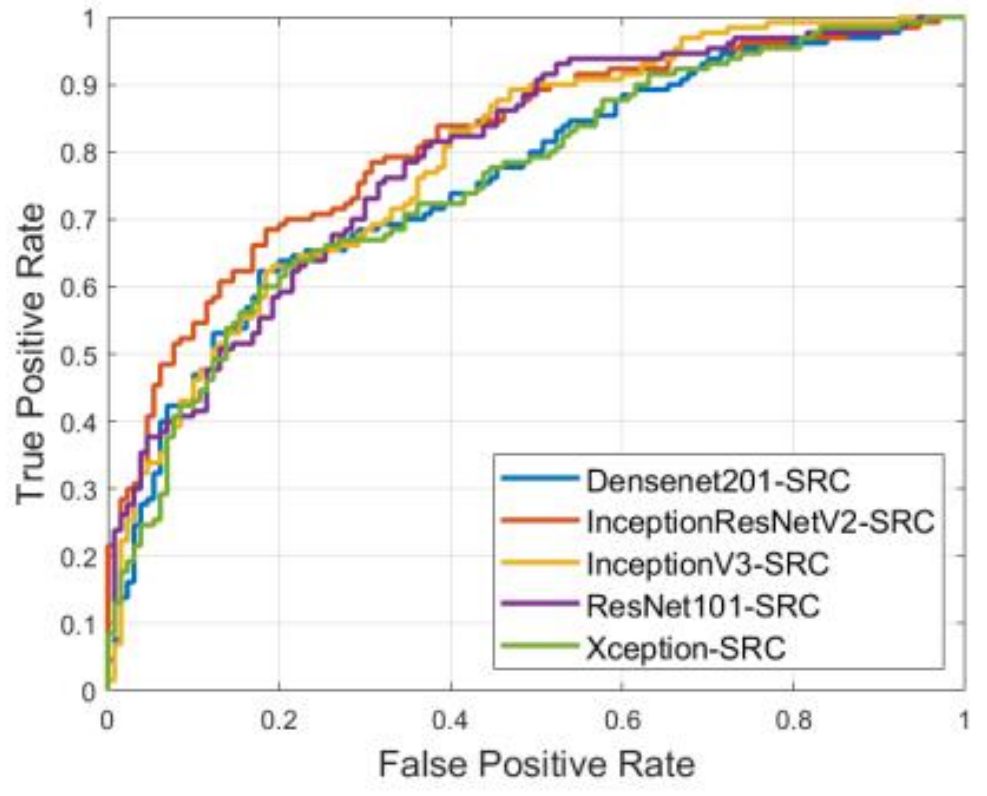

Figure 1: ROC curves for DF-SRC results presented in Table 2. The AUC values are 0.7595, 0.8139, 0.7905, 0.7918, 0.7575 for Densenet201-SRC, InceptionResNetV2-SRC, InceptionV3-SRC, ResNet101-SRC, Xception-SRC methods respectively.

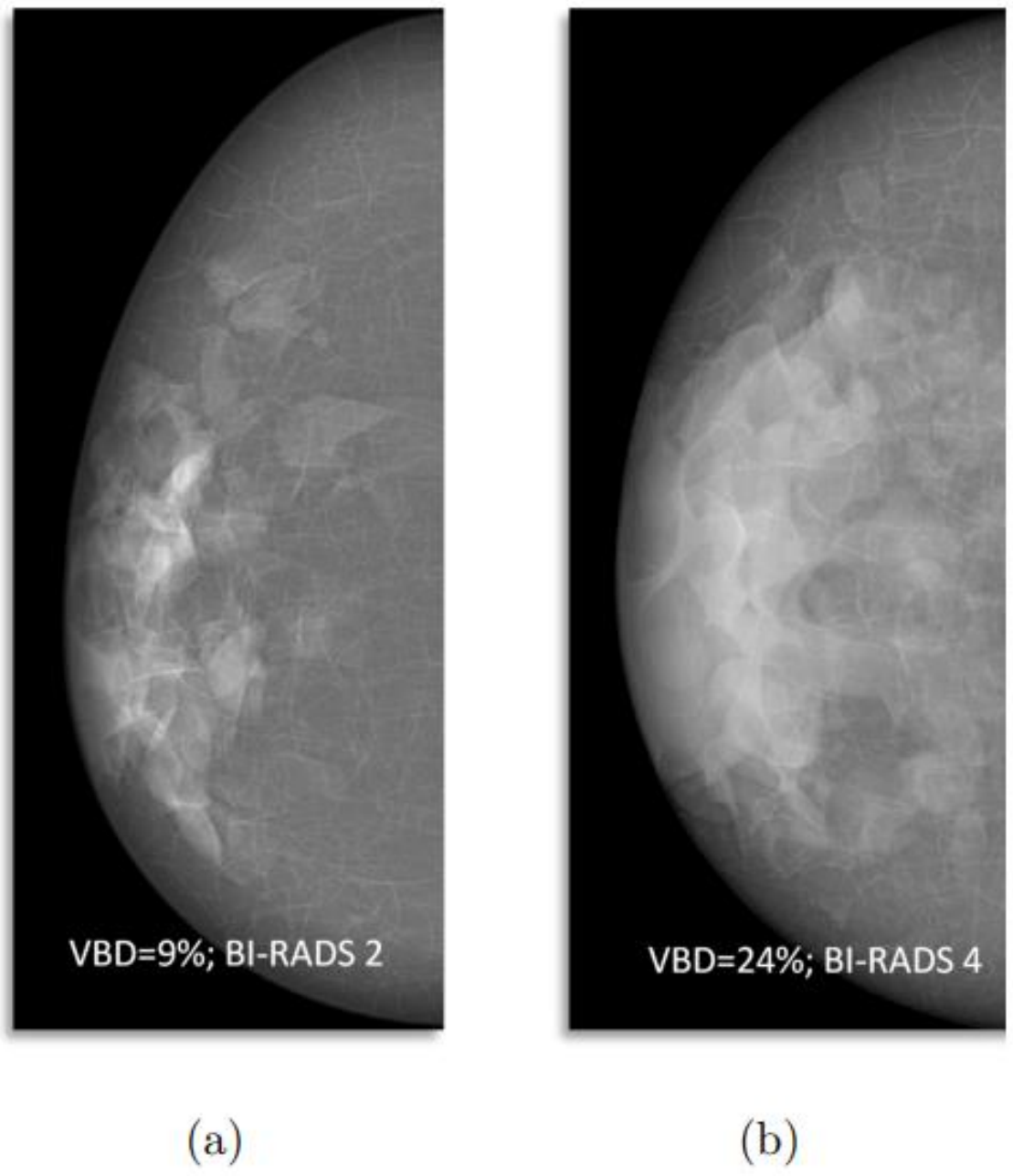

Figure 2: Sample simulated mammograms of 700ml breast phantom, with volumetric breast density of (a) 9%, BI-RADS 2, and (b) 24%, BI-RADS 4.

# Deep learning-based lung volume estimation with dynamic chest radiography using virtual patients

Nozomi Ishihara[1], Rie Tanaka[1], Futa Goshima[1], Haruto Kikuno[1], Noriyuki Ohkura[2], Isao Matsumoto[3]
1. College of Medical, Pharmaceutical & Health Sciences, Kanazawa University, Ishikawa, Japan
2. Department of Respiratory Medicine, Kanazawa University hospital, Ishikawa, Japan
3. Department of Thoracic Surgery, Kanazawa University, Ishikawa, Japan

## BACKGROUND AND PURPOSE

Dynamic chest radiography (DCR) is a recently developed, low-dose functional imaging method. Lung volume estimation with DCR has been investigated; however, the lack of reliable ground truth makes R&D and clinical application difficult. This study aimed to develop a deep learning-based lung volume estimation method with DCR using a four-dimensional extended cardiac-torso (XCAT) phantom with a known lung volume.

## METHODS

Twenty XCAT phantoms were generated and projected under the same imaging conditions as the actual DCR (Figure 1). A convolutional neural network (VGG19) was trained on the projection images and reference lung volumes derived from the log files of the XCAT phantom using a 20-fold cross-validation technique (Figure 2). The trained VGG19 model was also tested using one frame of DCR images of 257 real patients and computed tomography-derived reference lung volumes, and assessed based on the mean absolute error (MAE) and mean absolute percentage error (MAPE). Correlation analysis was also performed on the estimated lung volume using Pearson's correlation coefficient (r), and R-squared of the linear regression ($R^2$).

## RESULTS

The MAE and MAPE for the total lung volume of the virtual and real patients (virtual/real) were 201/691 ml and 3.9/14.5 %, respectively. The trained VGG19 model showed a linear correlation between the reference and estimated lung volumes, with r and $R^2$ of the linear regression for virtual/real patients (virtual/real); 0.99/0.81 and 0.98/0.65 (Figure 3).

## CONCLUSION

A deep learning-based lung volume estimation method for DCR was developed using virtual patients and was effective in clinical cases. Although some issues must be addressed for clinical implementation, the feasibility of DCR-based lung estimation was ascertained in this study.

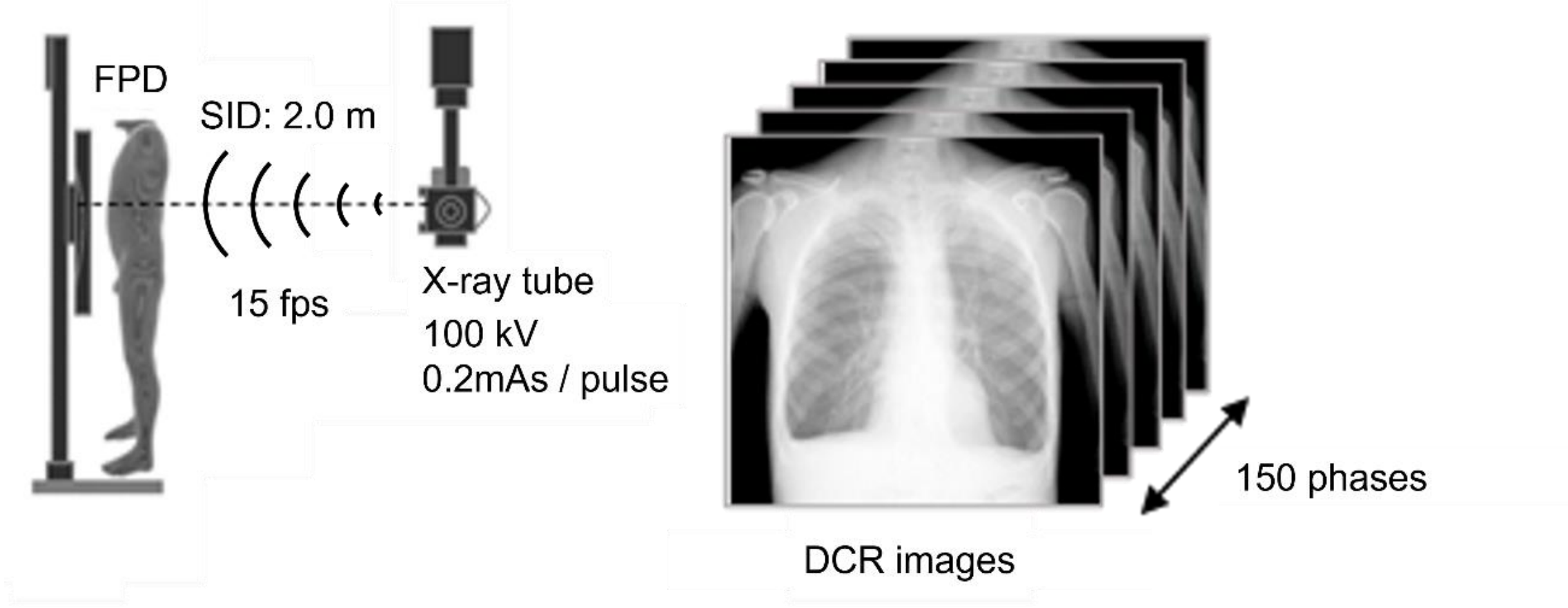

*Figure 1* Projection images of the extended cardiac-torso (XCAT) phantom
SID: Source-to-image distance, FPD: Flat-panel detector, fps: flame per second

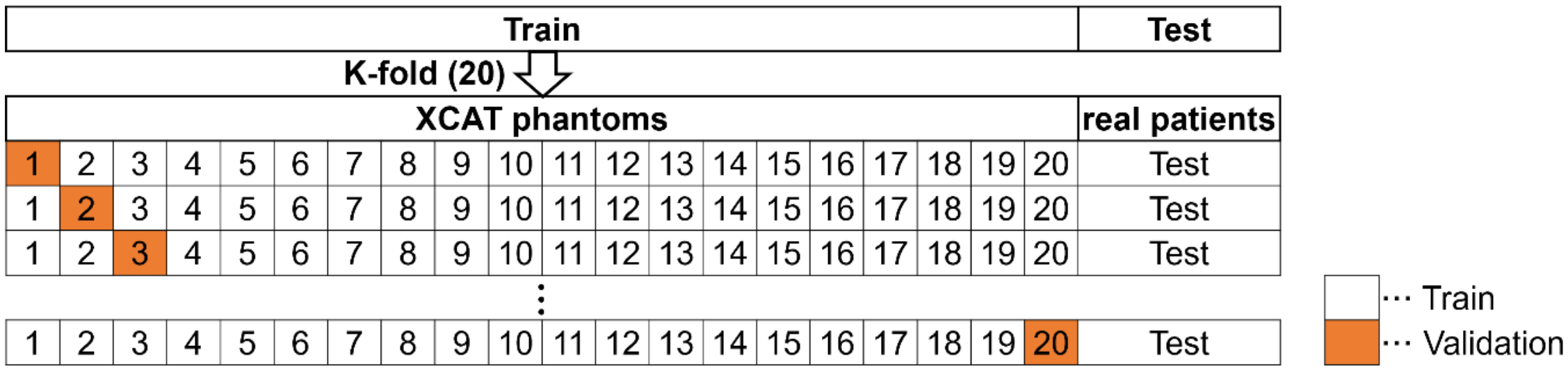

*Figure 2* Datasets for 20-fold cross-validation

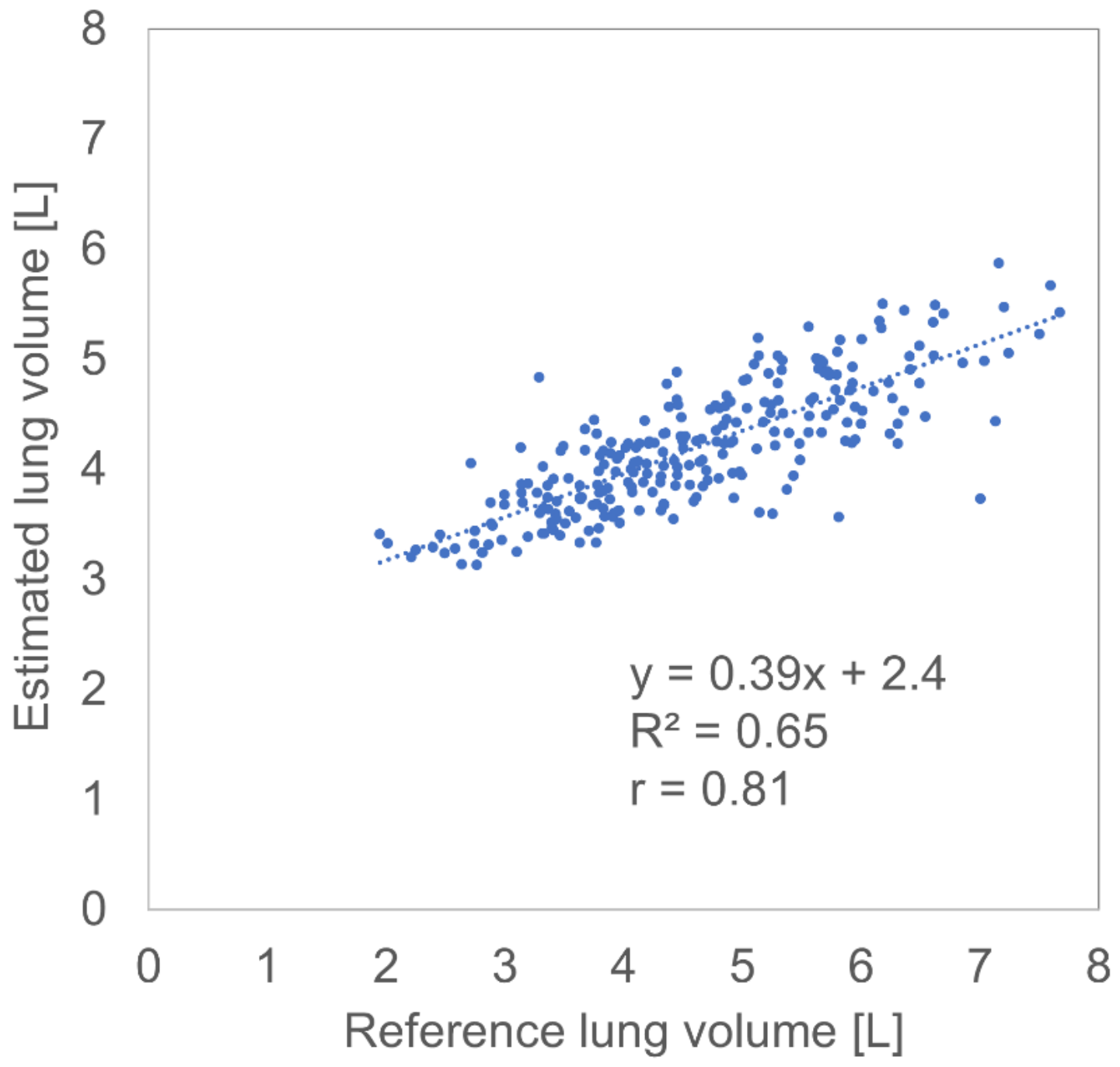

*Figure 3* Relationship between reference and estimated lung volumes of 257 patients by the trained VGG19 model

# Point-cloud Based Segmentation of the Prostate for Adaptive Radiation Therapy Treatment

Jianxin Zhou[1], M. Salvatori[2], K. Fejza[3], G. M. Hermann[3], A. Di Fulvio[1]
1. Department of Nuclear, Plasma, and Radiological Engineering, University of Illinois at Urbana-Champaign, Urbana, IL, USA.
2. IFC-National Research Council, Via Giuseppe Moruzzi, Pisa, Italy.
3. Department of Radiation Oncology, OSF HealthCare, Peoria, IL, USA.

## BACKGROUND AND PURPOSE

Adaptive radiation therapy (ART) is a closed-loop treatment method that adjusts the dose-delivery plan to account for patients' anatomy changes with daily measurements. This adaptive workflow requires fast segmentation methods to contour organs and update the anatomy information. In this work, we propose a fast point-cloud model to segment prostate, bladder, and rectum, which enables the APT for prostate cancer treatment.

## METHODS

The segmentation workflow is shown in Figure 1. We first obtain the gross boundary of the organs with Mask-RCNN and sample the points inside to significantly reduce the data size. Then, we perform segmentation on the point-cloud data to achieve fast contouring. We use 38 patients' CT images from OSF HealthCare to develop the model. Moreover, synthetic CT images are generated through the XCAT digital human phantom to increase the dataset and aid the training. We perform a 7-fold cross-validation to evaluate the model performance.

## RESULTS

Table 1 summarizes the segmentation results. The model achieved an average 0.88±0.05 Dice coefficient for prostate, which outperforms other state-of-the-art segmentation models. As Figure 2 shows, our model doesn't perform well for several OSF images, image 11, image 20, etc. Therefore, we used the XCAT digital human phantom to generate multiple synthetic images with various organ shapes to improve the model performance for these images. Then, the model performances of image 11 & 20 are further improved. The segmentation time of each point-cloud data is 1-3 seconds, using a NVIDIA RTX 3080 GPU.

## CONCLUSION

We developed a point-cloud-based segmentation model to aid the treatment of prostate cancer. The segmentation accuracy of prostate is 0.88±0.05, which can be further improved by adding synthetic images to the training data set. The fast point-cloud-based model can enable the adoption of APT workflow in the treatment of prostate cancer to improve treatment efficiency.

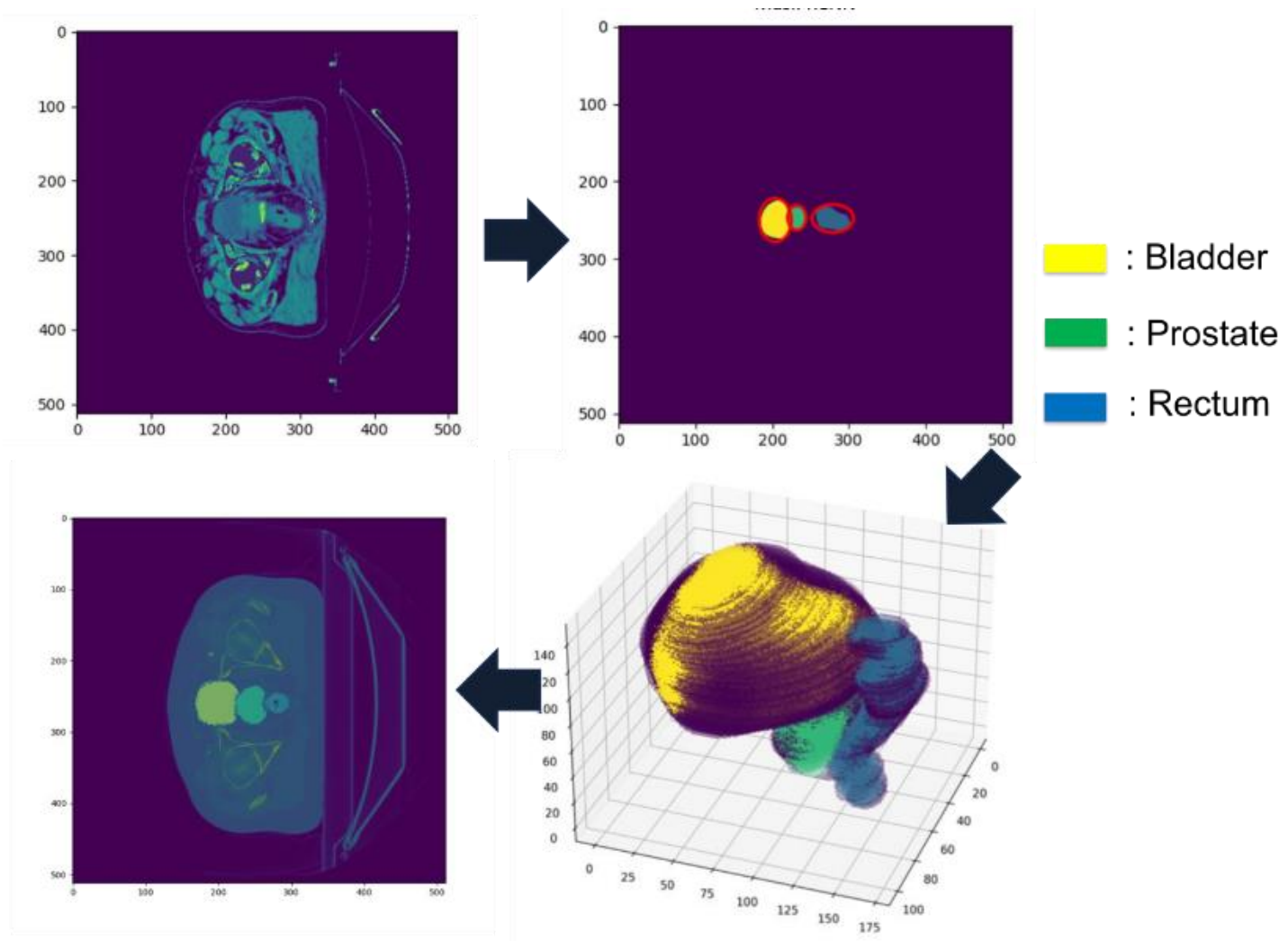

Figure 1: Workflow of the prostate segmentation model with the point-cloud-based neural network.

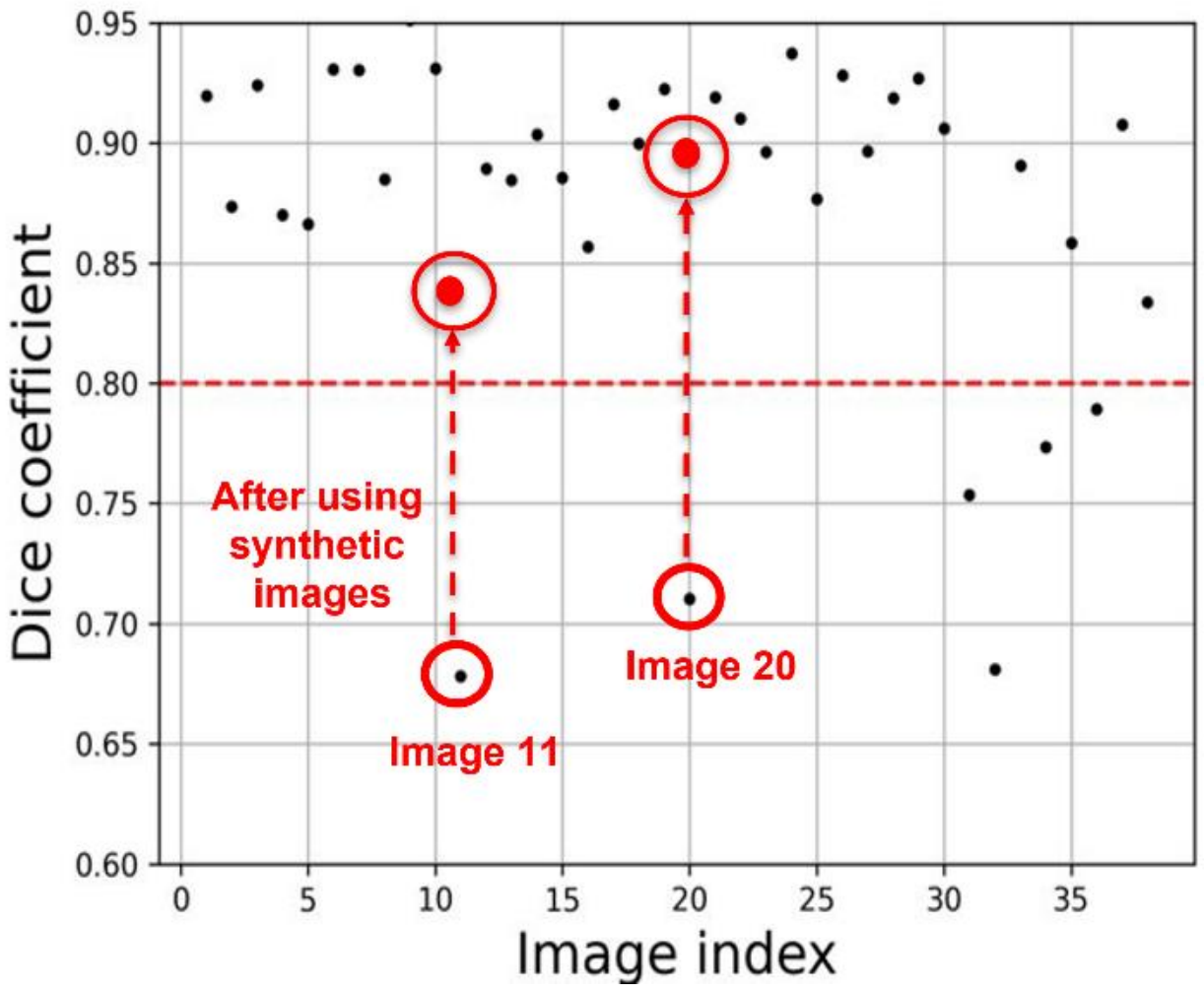

Figure 2: Segmentation result of 38 images and the improved result for image 11 and 20 after using synthetic images.

Table 1: Segmentation result without using synthetic images in the training

| Dice coefficient | Model | Image number | Year | Bladder | Prostate | Rectum |
|---|---|---|---|---|---|---|
| This work | PointMLP | 32 | 2023 | 0.94±0.02 | 0.88±0.05 | 0.83±0.06 |
| N. Tong, et al [1] | MTER-Net | 200 | 2021 | 0.96±0.02 | 0.86±0.06 | 0.86±0.07 |
| M. Kawula, et al [2] | 3D U-Net | 59 | 2022 | 0.97±0.01 | 0.87±0.03 | 0.89±0.04 |
| T. Nemoto, et al [3] | 2D U-Net | 556 | 2020 | 0.94±0.04 | 0.85±0.05 | 0.83±0.09 |
| C. Liu, et al [4] | 3D V-Net | 771 | 2020 | N/A | 0.85±0.06 | N/A |

# Perspective of medical industry on in silico trials

Wednesday, April 24, 2024

9:35 am - 10:15 am

Chaired by Stephen Glick and Liesbeth Vancoillie

# The role of system simulation for product design of deep-silicon photon counting CT

Erik Fredenberg (GE Healthcare)

Photon-counting CT (PCCT) is the next major leap in detector technology for CT with the potential to significantly improve spectral and spatial resolution. GE HealthCare is developing a prototype wide-coverage PCCT system based on edge-on deep-silicon sensors. Computer simulations and virtual clinical trials play an increasingly important role for evaluating imaging systems, not least for the development of new technologies, such as deep-silicon PCCT, where fast turn-around times are crucial. GE HealthCare has therefore, together with academic collaborators, developed a model for the deep-silicon photon-counting detector. The model has been integrated into the CT & X-ray simulation environment CatSim, and it has been used throughout the development process to evaluate system performance and to design the image chain. Our ambition for the near future is to employ the model for understanding the potential of new clinical applications enabled by PPCT and to bring it to use in virtual clinical trials.

# Artificial Intelligence for Healthcare: From Patient Twinning to Precision Therapy

Matt Holbrook (Siemens Healthineers)

The digital twin is a technology-based approach that links the real and digital worlds, turning data into actionable insights. It is made possible by advances in the availability of medical data and power of artificial intelligence. Siemens Healthineers is applying these concepts to workflows to maximize efficiency, better understand patient health, predict changes, and therapy outcomes. This presentation will outline ways in which current system, patient, and organ avatars are combined with disease and predictive models to impact patient health.

# VCT for breast x-ray imaging in GE HealthCare

Ann-Katherine Carton (GE Healthcare)

# Observer evaluation and image quality

Wednesday, April 24, 2024

10:50 am - 11:30 am

Chaired by Francesco Ria & Hilde Bosmans

# Virtual validation of novel 4D noise reduction filter in abdominal dynamic CT perfusion

Authors: Sjoerd A. M. Tunissen[a], Ewoud J. Smit[a], Ioannis Sechopoulos[a, b, c]

a Dept. of Medical Imaging, Radboudumc, Nijmegen, The Netherlands
b Dutch Expert Centre for Screening (LRCB), Nijmegen, The Netherlands
c Technical Medicine Centre, University of Twente, Enschede, The Netherlands

## BACKGROUND AND PURPOSE

To evaluate the performance of a novel noise filtering technique for temporal data, called the 4D similarity filter (4DSF), using virtual imaging methods.

## METHODS

Ten XCAT anthropomorphic digital phantoms, each with an added lesion of 10 mm diameter in the liver, were used. A previously validated computer simulation of a clinical wide-area CT system was used to simulate two protocols; a new dynamic CT perfusion (dCTP) protocol involving 12 phases and a standard clinical protocol involving 4 phases. The tube current and voltage levels were selected based on the 4-phase liver protocol used in our clinic, ensuring that the total dose of the two protocols is equivalent. Tube current was scaled based on patient size, but no axial or angular modulation was used. To validate the performance of the 4DSF, the standard deviation (σ) of the noise in the liver tissues was determined. The detectability of the lesion was determined by means of the Contrast-to-Noise ratio (CNR). Finally, the accuracy of the time intensity curve (TIC) of the lesion was evaluated by means of Mean Absolute Error (MAE) between the TIC before and after 4DSF.

## RESULTS

On average across the 10 XCAT phantoms, the 4DSF reduced the σ of the noise from 217 HU to 30 HU for the 12-phase protocol, and from 97 HU to 25 HU for the 4-phase protocol. The CNR of the lesion improved from 0.5 to 1.8 for the 12-phase protocol, and from 1.0 to 3.1 for the 4-phase protocol. However, the average MAE of the TIC in the 12-phase protocol was 20 HU, and in the 4-phase protocol it was 6 HU.

## CONCLUSION

High radiation dose limits the clinical implementation of dynamic CT perfusion. Therefore, using the 4DSF for dCTP at equal dose as 4-phase imaging, could make dCTP, for liver, feasible in clinical practice. However, the noise reduction comes at the cost of reduced accuracy in the TICs. Therefore, one should be careful when using 4DSF for quantitative analysis instead of detection tasks.

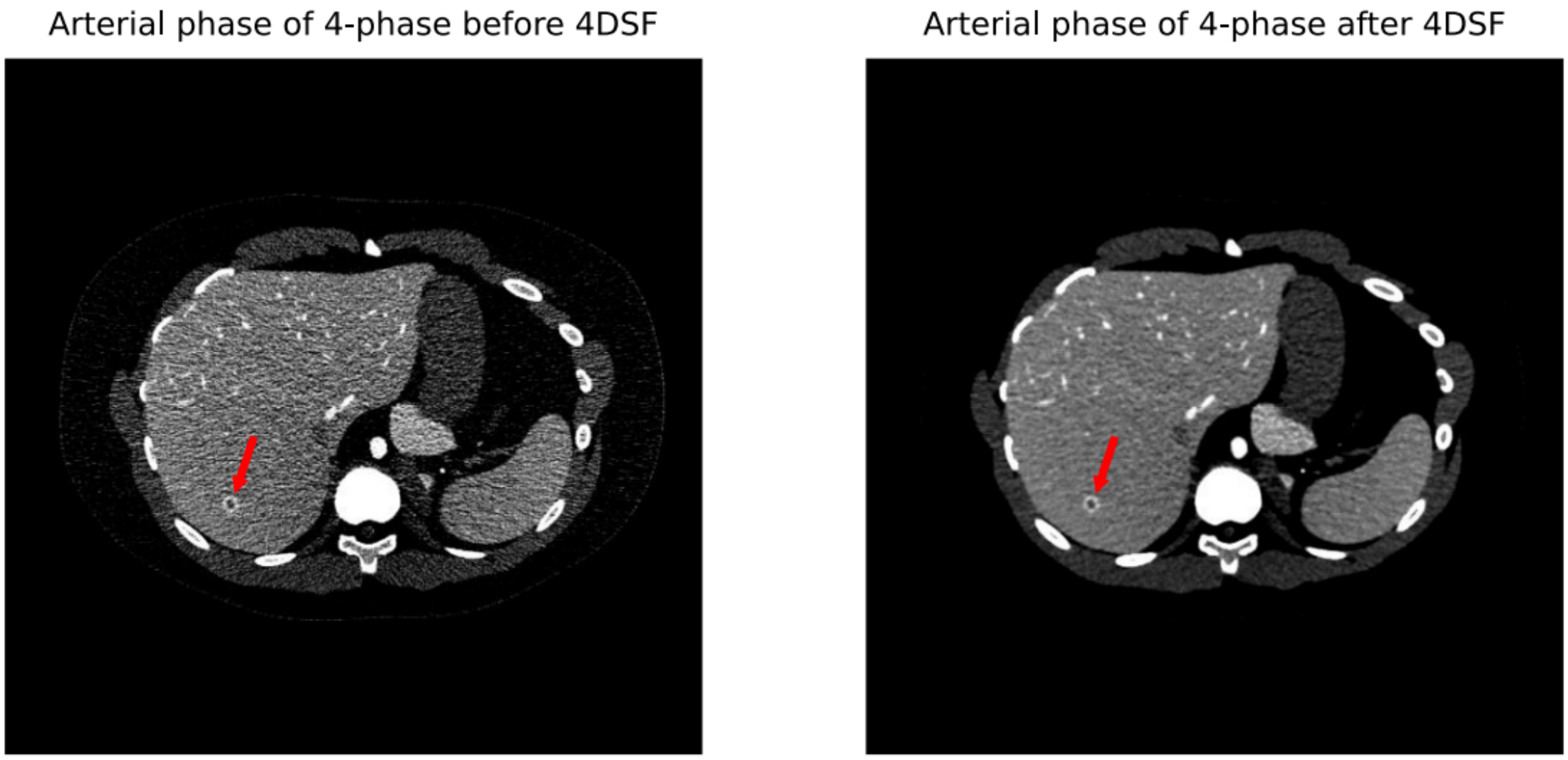

*Figure 1: Arterial phase scan of the 4-phase protocol, before 4DSF (left), and after 4DSF (right). (WW/WL 300/150).*

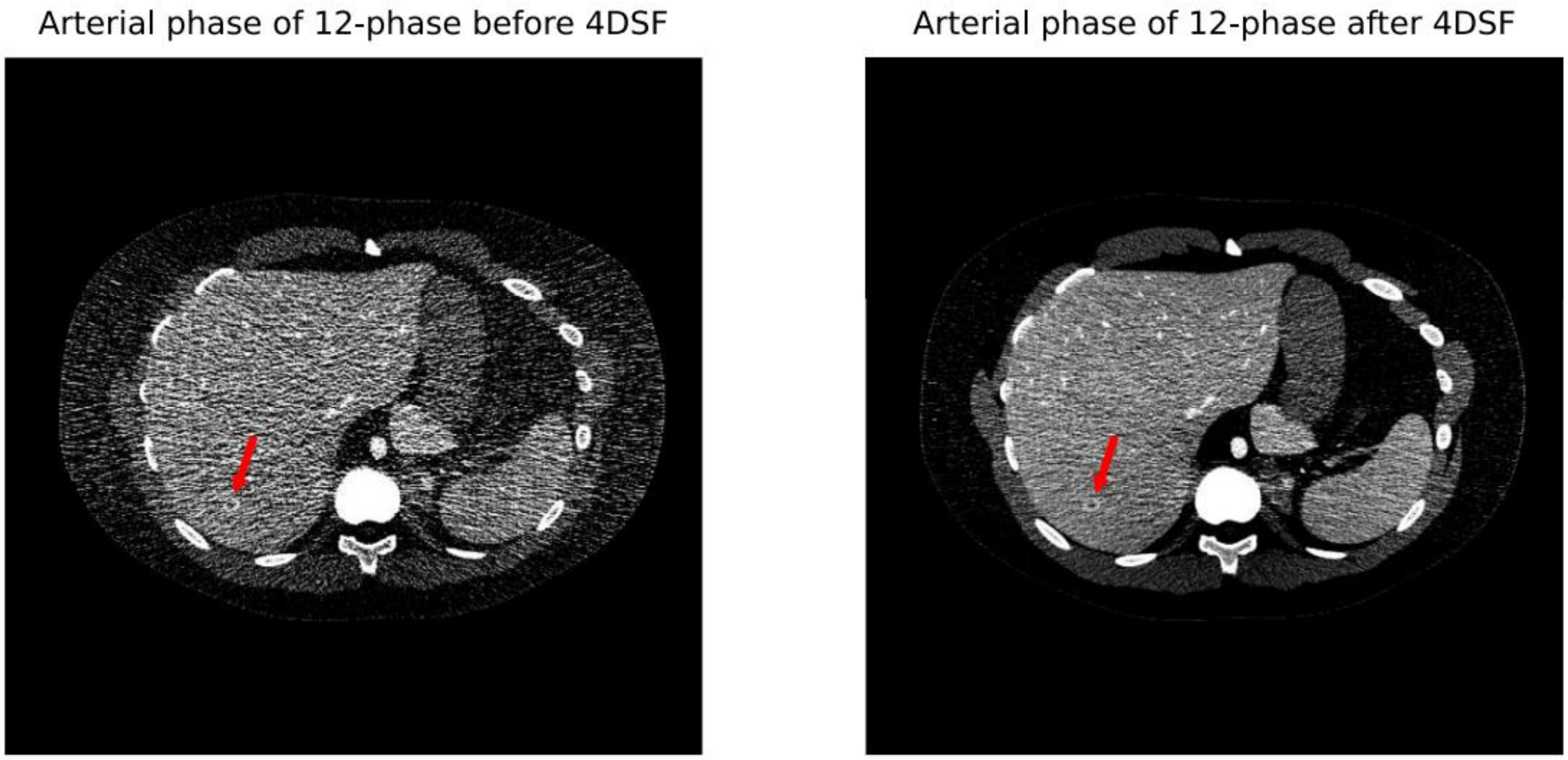

*Figure 2: Arterial phase scan of the 12-phase protocol, before 4DSF (left), and after 4DSF (right). (WW/WL 300/150).*

# Exploring how observer task impacts FFDM VCT performance

Dan Li (FDA CDRH DIDSR), Stephen Glick (FDA), Andrey Makeev ( FDA CDRH DIDSR )

## Purpose

Virtual clinical trials (VCTs) and physical phantom studies have the potential for greatly decreasing the monetary cost and time consumed in the evaluation of new breast imaging technology. In the area of medical imaging, we advocate for a task-based assessment approach when evaluating image quality. Most breast imaging clinical studies reported in the literature involve the observer both detecting whether a suspicious lesion is present, as well as rating how likely it is that the lesion is malignant. In fact, the decision variable commonly used in clinical breast imaging studies to form a receiver operating characteristic (ROC) curve is the probability of malignancy (POM). Designing VCTs and physical phantom studies that accurately model clinical studies is challenging because it is difficult to model malignant and benign lesions. One lesion feature that that is commonly used to discriminate malignant and benign lesions is the border of the lesion, with malignant lesions typically portraying spiculations. To more accurately model clinical breast imaging studies, we explore the use of a classification task instead of a detection task, that evaluates the reader's performance in discriminating between spiculated and lobular (non-spiculated) masses.

## Methods

It is challenging to design VCTs and physical phantom studies that model actual clinical studies due to the difficulty of modeling lesions of different malignancy levels. While there are multiple signals for potential malignancy, spicules of the lesion is what a radiologist typically looks for when predicting likelihood of malignancy. To fill the gap between clinical breast imaging studies and VCT, we explored the use of a classification task instead of a detection task, that evaluates the readers performance in discriminating between spiculated and lobular (non-spiculated) masses. To assess this classification task, we use a convolutional neural network (CNN) model trained to differentiate spiculated and non-spiculated masses using Monte Carlo simulated images. The basis for using CNN observer models is that a previous study showed that the CNN based framework can be trained to approximate the ideal observer. In addition to the mass classification task, we also examine this CNN model observer in a more traditional detection task.

## Results

This study evaluates the effect of dose and resolution on task-based performance and shows that different conclusions might result depending on whether the VCT study evaluates a mass detection task or a mass classification task.

## Conclusion

Depending on the purpose of VCT study, the designed task should be representative enough to model the actual clinical scenario. Evaluation using a mass classification task might provide useful information in assessing breast imaging technology.

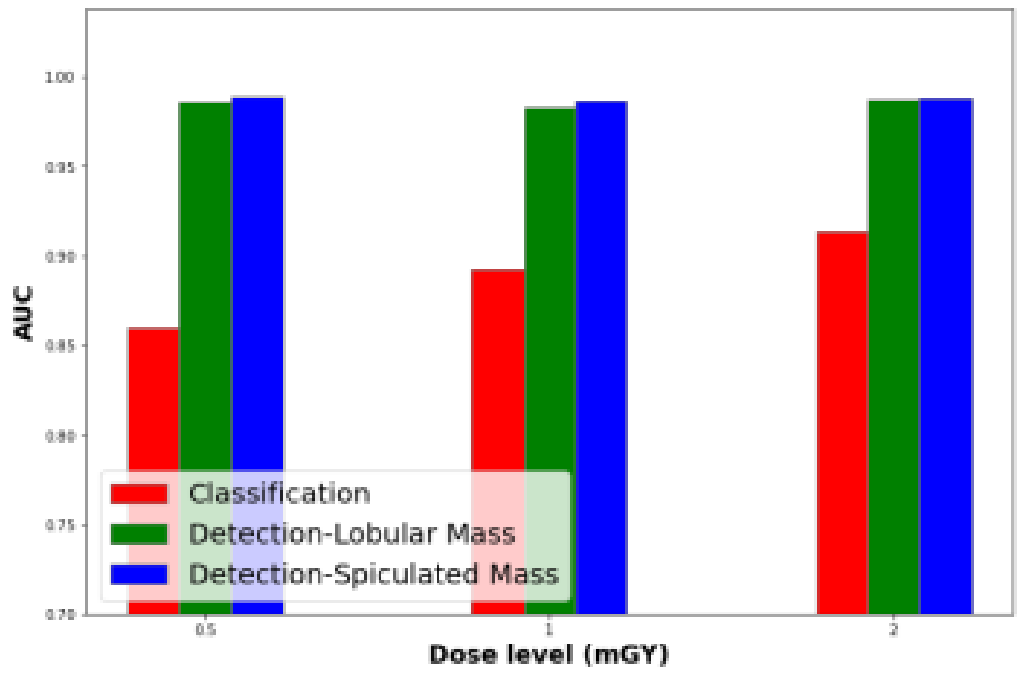

Figure 1. Shown is the Area Under the ROC curve versus dose for three different VCT assessing different tasks; a) detection of a lobular mass, b) detection of a spiculated mass, and c) classification of spiculated and lobular masses. It can be observed that there is very little dependence on dose with the detection tasks, however, there is a small but significant dependence on dose for the classification task.

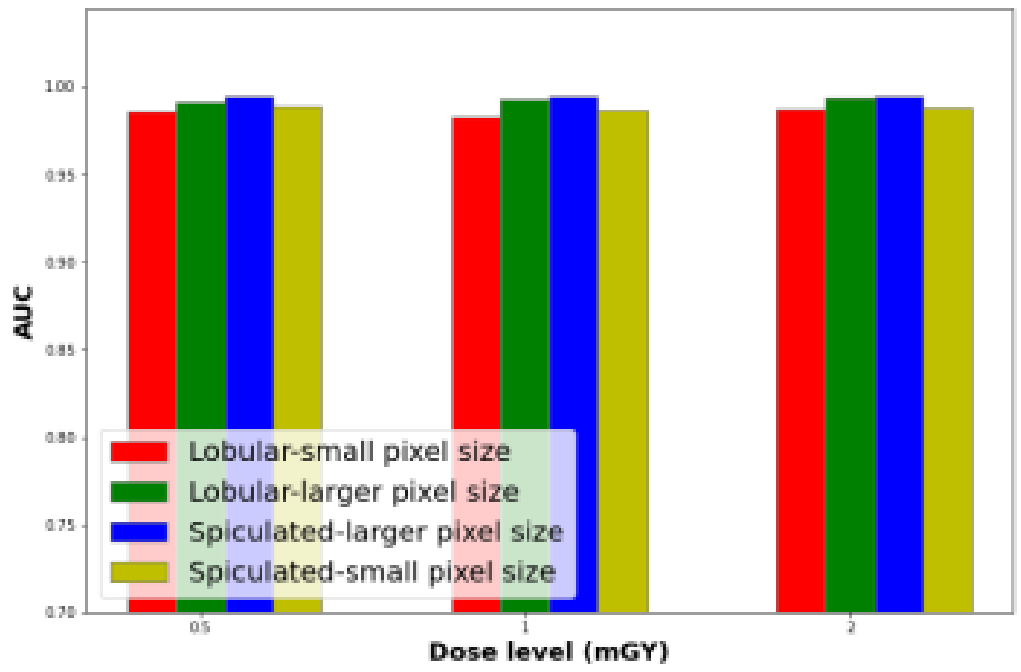

Figure 2. Shown is the AUC versus dose level for VCTs that evaluated a detection task. Results are shown for mammography with a small detector pixel size (85 microns) and for a larger detector pixel size (170 micron pixel). Results suggest that for the mass detection task, there is very little dependence on pixel size or dose.

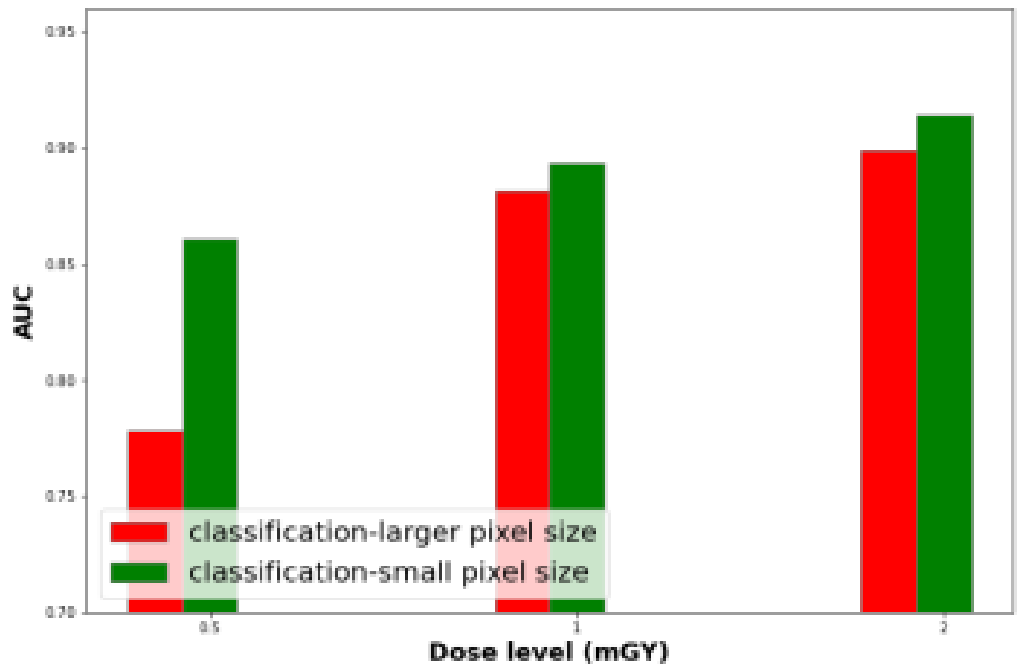

Figure 3. Shown is the AUC versus dose level for VCTs that evaluated a classification task. Results are shown for mammography with a small detector pixel size (85 microns) and for a larger detector pixel size (170 micron pixel). Results suggest that for the mass classification task, there is a small but significant dependence on pixel size and dose.

# Ideal observer approximation in virtual imaging trials

Mark Anastasio (University of Illinois at Urbana-Champaign), Kaiyan Li (University of Illinois Urbana-Champaign), Weimin Zhou ( Global Institute of Future Technology, Shanghai Jiao Tong University ), Umberto Villa ( University of Texas - Austin ), Hua Li ( Washington University in St. Louis )

## Purpose

It is widely accepted that the optimization of medical imaging system performance should be guided by task-based measures of image quality (IQ). Task-based measures of IQ quantify the ability of an observer to perform a specific task, such as detection or estimation of a signal (e.g., a tumor). For signal detection tasks, the Bayesian Ideal Observer (IO) sets an upper limit of observer performance and has been advocated for use in optimizing medical imaging systems and data-acquisition designs. However, except in special cases, the determination of the IO test statistic is analytically intractable. Recently, supervised learning-based methods for approximating the IO performance for medical imaging systems have been developed. These methods hold great potential for virtual imaging trials (ViTs), where large ensembles of simulated but realistic image data can be produced. In this work, we present methods for estimating the IO performance for binary signal detection tasks, detection-localization tasks, and detection-estimation tasks. These methods represent powerful new computational tools that will enhance the effectiveness of ViTs.

## Approach

Convolutional neural networks (CNNs) are employed to approximate the IO for binary signal detection and detection-localization tasks. For detection-estimation tasks, a hybrid approach is developed that combines a multi-task CNN and a Markov-Chain Monte Carlo (MCMC) method to approximate the IO. The use of CNNs to estimate the data space IO performance for computational imaging systems is also investigated.

## Results

For all considered tasks, case studies are presented to demonstrate and validate the IO estimation methods. Validation is accomplished by use of special cases where the IO test statistic is known analytically and/or cases where a MCMC method is available as a reference. For the case of X-ray CT, we demonstrate that a deep learning-based image reconstruction method can result in the loss of more task-specific information than a conventional model-based method.

## Conclusion

This work demonstrates that the proposed CNN-based IO approximation methods can enable the application of IO analyses in general VITs, which may promote advanced evaluations of emerging medical imaging techniques.

# Future prospects and applications

Wednesday, April 24, 2024

11:30 am - 12:10 pm

Chaired by Francesco Ria & Hilde Bosmans

# Virtual Imaging Trial can Unveil the Relationship Between Intravascular Diffusivity and Intravoxel Incoherent Motion in MRI

Mojtaba Lashgari, Zheyi Yang, Miguel O. Bernabeu, Jing-Rebecca Li, Alejandro F Frangi

## BACKGROUND AND PURPOSE

### Introduction

Perfusion is vital for maintaining tissue function by supplying oxygen and nutrients while removing carbon dioxide and waste products. Consequently, perfusion imaging is essential for diagnosing and monitoring microvascular diseases like ischemia, hypercapnia-induced vasoconstriction, and hyperoxygenation-induced vasoconstriction. It also aids in distinguishing between high- and low-grade brain gliomas and serves various other diagnostic purposes, etc. [1 2].

Intravoxel incoherent motion (IVIM) MRI emerges as a promising method for assessing tissue perfusion. A notable strength of IVIM MRI lies in its use of endogenous contrast and offering localized insights into both perfusion and tissue microstructure. These dual capabilities render IVIM MRI an exceptional clinical imaging technique [3].

IVIM involves the transient movement of water molecule spins within an MRI voxel during the measurement period, resulting in a distribution of spins' speeds in orientation and/or amplitude [4]. This distribution originates from the blood flowing within a microvascular network and mimics a pseudo-diffusion process. This distribution arises from blood flow within the microvascular network and resembles a pseudo-diffusion process. The IVIM theory, pioneered by Le Bihan et al. [5], hypothesize that the incoherent blood flow in microvasculature causes spin dephasing in blood when diffusion gradients are applied, as described by the formula:

$$\frac{S}{S_0} = (1-f) \times e^{-b \times D} + f \times e^{-b \times D^*}$$

where S is IVIM signal, $S_0$ IVIM signal at b=0, $f$ is perfusion fraction, $D$ is diffusion coefficient, and $D^*$ is pseudo-diffusion coefficient.

In a standard IVIM imaging experiment, blood self-diffusion appears to attenuate signals distinctively from blood velocity (represented by pseudo-diffusion), yet flow-compensated diffusion gradients challenge this hypothesis [2]. The application of virtual imaging methods offers a promising avenue for resolving this conflict, offering a unique approach to studying states are not physiologically plausible *in vivo*. By simulating complex microcirculatory dynamics within tissues, these methods enable the identification of contributions from various parameters to the signal. Consequently, they facilitate the disentanglement of diffusion and perfusion effects, leading to distinguish effect of blood self-diffusion on IVIM MRI signal.

Thus far, numerical investigations into IVIM imaging have exclusively employed an elementary computational approach. This technique simulates spin displacements solely according to constant blood flow velocities along vascular trajectories $R(t')$, rather than within vascular lumens. Subsequently, the acquired phase $\phi_t$ resulting from these spin displacements is calculated as:

$$\phi_t = \int_0^t \gamma G(t')(R(t'), t')dt'$$

and finally, the approximate IVIM imaging signal is computed using:

$$S = S_0 \times \mathbb{E}\{e^{(i\phi t)}\}$$

where $\mathbb{E}\{\}$ is an expectation of random variable of $i\phi t$ [6].

To investigate the significance of intravascular diffusivity on the simulated IVIM MRI signal, this study uses a new virtual imaging trial to simulate IVIM imaging signals by solving the generalized Bloch-Torrey partial differential equation using a new finite element solver.

# METHODS AND DATA

## Data and material

This study uses a transmission electron image of a retinal vascular plexus network, shown in the center of Figure 1 [7]. The simulations were performed on ARC3, the High-Performance Computing facilities at the University of Leeds. ARC3 consists of 252 nodes with 24 cores (Broadwell E5-2650v4 CPUs, 2.2 GHz) and 128GB of memory each, and an SSD within the node with 100GB of storage.

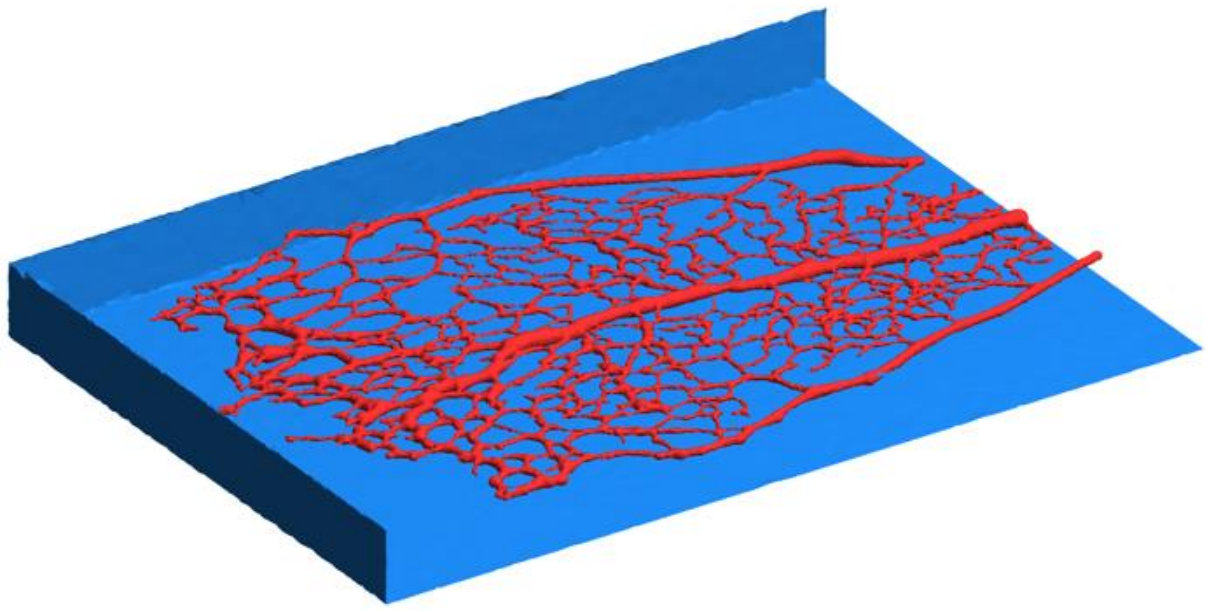

*Figure 1. Sub-set of murine retinal vascular plexus (red) surrounded by extra- vascular space (blue).*

## Method

To simulate IVIM MRI signal, we extend SpinDoctor [8], a diffusion MRI simulation toolbox which solves Bloch-Torrey partial differential equation:

$$\frac{\partial}{\partial t} M_l(\boldsymbol{r}, t) = \nabla.\left(D_l(\boldsymbol{r})\nabla M_l(\boldsymbol{r}, t)\right) - \frac{1}{T_{2l}} M_l(\boldsymbol{r}, t) - i\gamma f(t)\boldsymbol{g}.\boldsymbol{r} M_l(\boldsymbol{r}, t)$$

Extended solver simulates IVIM MRI signals by solving the generalised Bloch-Torrey partial differential equation:

$$\frac{\partial}{\partial t} M_l(\boldsymbol{r}, t) = \nabla.\left(D_l(\boldsymbol{r})\nabla M_l(\boldsymbol{r}, t)\right) - \frac{1}{T_{2l}} M_l(\boldsymbol{r}, t) - i\gamma f(t)\boldsymbol{g}.\boldsymbol{r} M_l(\boldsymbol{r}, t) - \boldsymbol{v}(\boldsymbol{r}, t)\nabla. M_l(\boldsymbol{r}, t)$$

which includes a velocity term of $\boldsymbol{v}(\boldsymbol{r}, t)\nabla. M_l(\boldsymbol{r}, t)$. Unlike previous approaches, the extended version incorporates diffusion phenomena within the intravascular space by accounting for volumetric microvasculature during simulation, for the first-time.

# RESULTS

The objective of this experiment is to investigate whether alterations in intravascular diffusivity have an effect on the simulated IVIM imaging signal and the estimation of three parameters within the IVIM bi-exponential model. To achieve this, we employ three distinct intravascular diffusivity values: $10^{-7}\ \mu m^2/ms$, $1.26\ \mu m^2/ms$ and $3\ \mu m^2/ms$—The value of 1.26 $\mu m^2/ms$ is the experimental coefficient of blood diffusivity reported in [9]. Figure 2 shows the simulated signals in which the variations in colors denote different values of intravascular diffusivity.

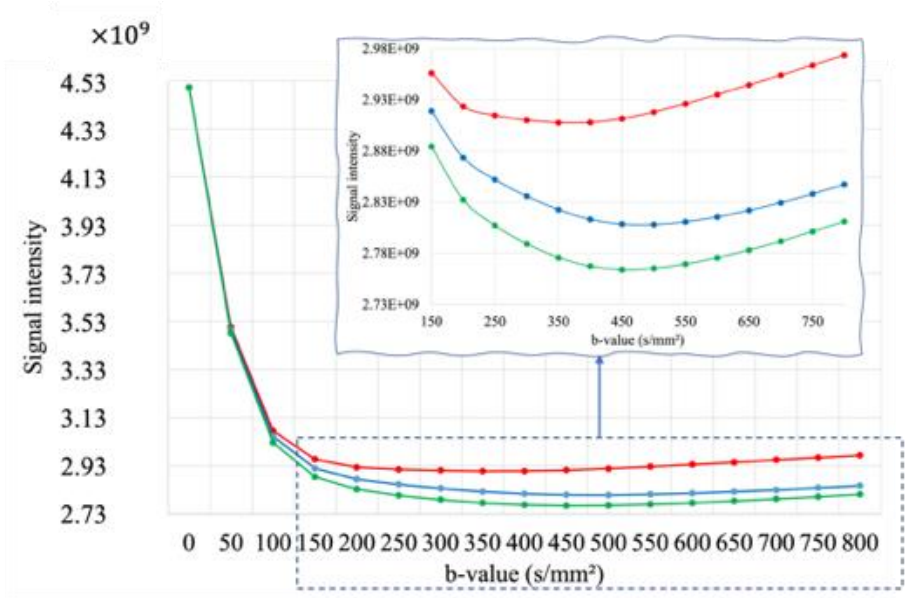

*a*

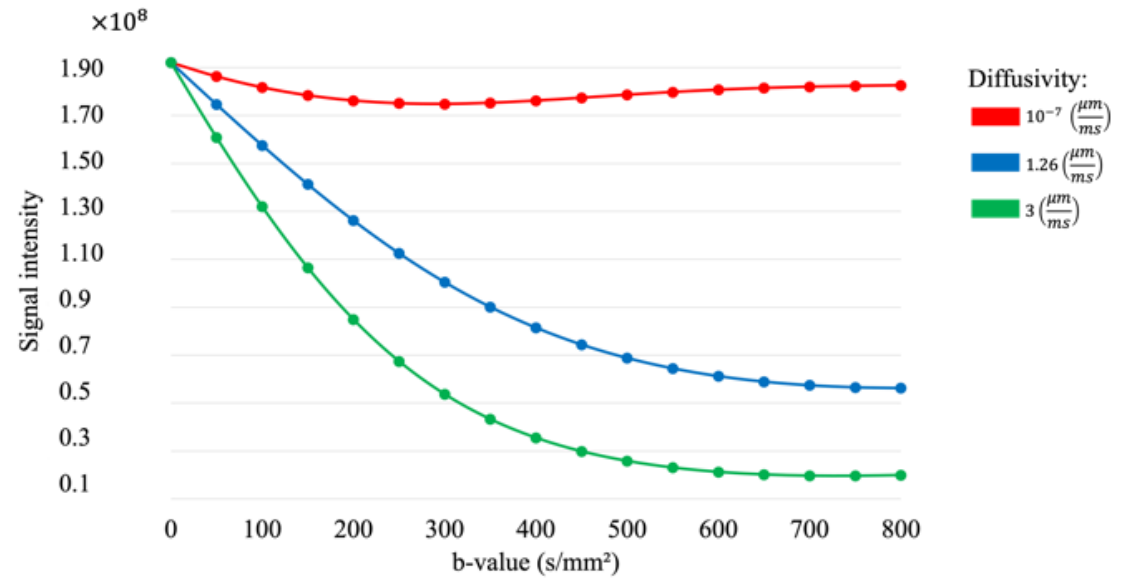

*b*

*Figure 2. Effect of changing the blood diffusivity on IVIM imaging signal originated from both intravascular and extravascular spaces at the velocity of 0.01 m/s b) IVIM imaging signal originated from intravascular spaces at the velocity of 0.01 m/s*

Moreover, in this study, three models are fitted to the simulated IVIM MRI signal with intravascular diffusivities of $10^{-7}$ $\mu m^2/ms$ and 1.26 $\mu m^2/ms$. These models include segmented model [10], Bayesian model [11], and Adaptive model [12]. Figures 3.a, 3.b, and 3.c compare changes in the estimated values of $D^*$, $D$, and $f$ for different values intravascular diffusivities. As it can be seen in Figure 3 for different blood flow velocities and for each fitting method, increasing intravascular diffusivity increases the estimated $D$, decreases $D^*$, and increases $f$.

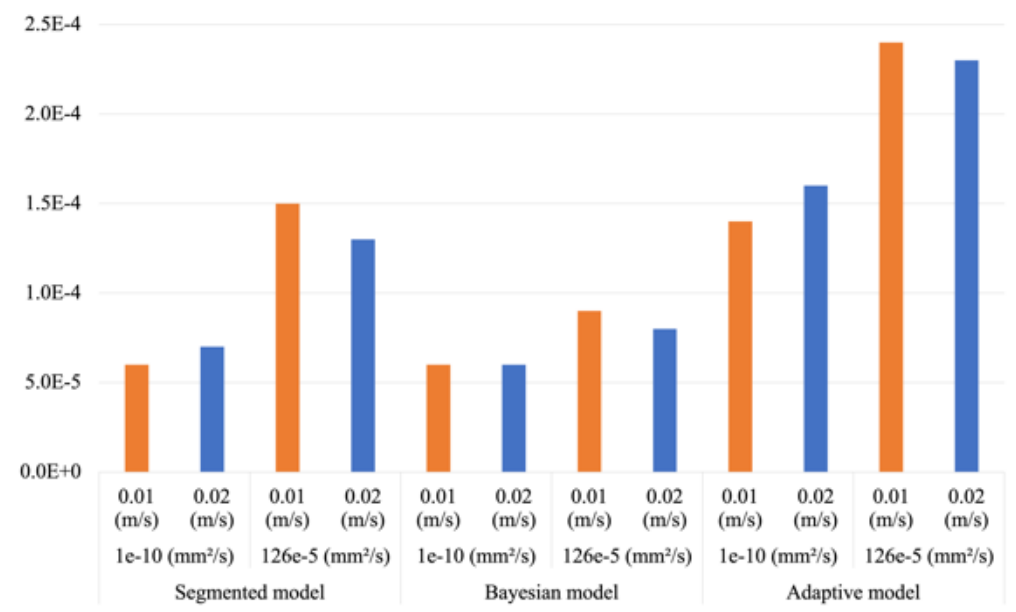

*a*

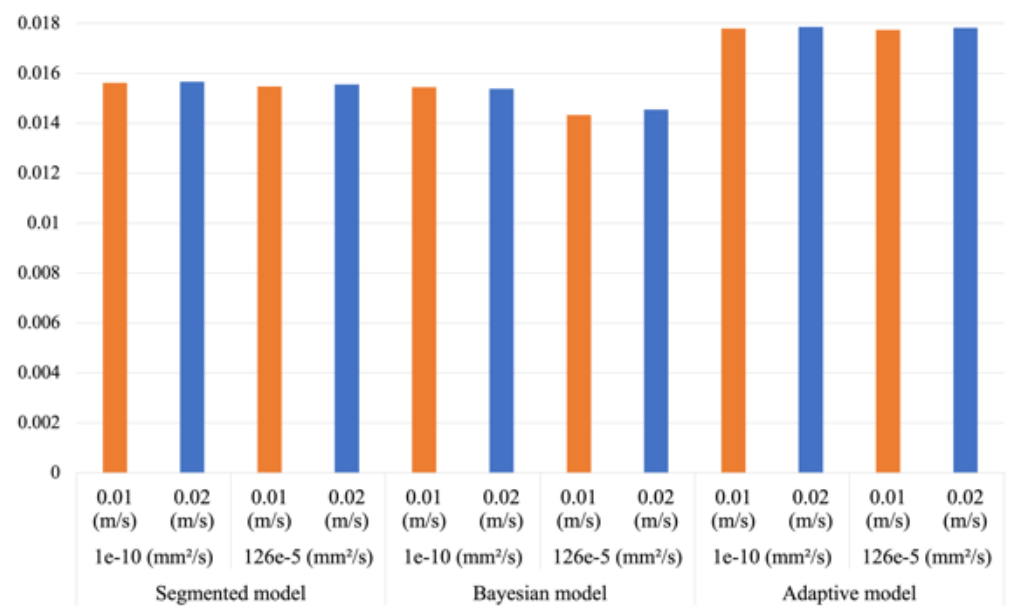

*b*

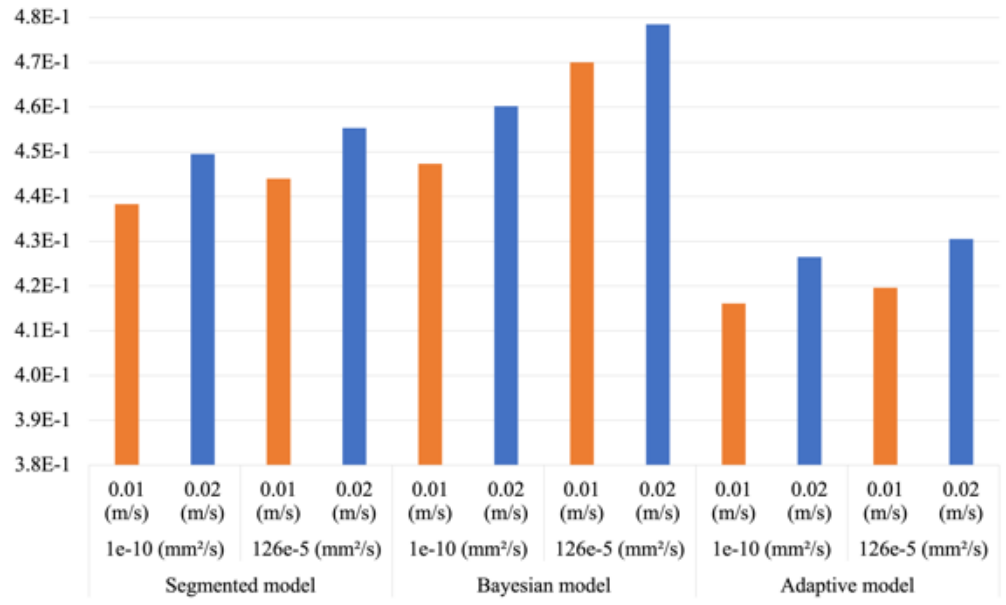

*c*

*Figure 3. Effect of changing the blood diffusivity on the estimated parameters of IVIM imaging a) D; b) D*; c) f; for the velocity of 0.01 m/s (orange) and 0.01 m/s (blue).*

# DISCUSSION

It has been suggested by [5] [9] that intravascular diffusion attenuates signals differently than blood velocity, but flow-compensated diffusion gradients contradict this theory. Since it was impossible to do an experiment in real world in which intravascular diffusivity is dismissed, a new virtual MRI simulator is developed to resolve this long-standing conflict. Using this new simulator, the theory suggested in [5] is confirmed and we gain a new insights into the effects of blood movement on IVIM MRI that may change the way we understood blood movement effects on IVIM MRI.

# CONCLUSION

Virtual imaging trial demonstrates the substantial impact of intravascular diffusivity on the IVIM MRI signal, which agrees with previous research [5] [9]. Therefore, future numerical studies of IVIM MRI should consider the inclusion of intravascular diffusivity in their study.

# Accelerated Container Applications for the Development and Integration of Virtual Imaging Platforms

Bruno Barufaldi[1], Miguel A. Lago[2], Ehsan Abadi[3], and Andrew D.A. Maidment[1]

[1]Department of Radiology, University of Pennsylvania, Philadelphia, PA, USA
[2]Division of Imaging, Diagnostics and Software Reliability, OSEL, CDRH, U.S. Food and Drug Administration, Silver Spring, MD, USA
[3]Department of Radiology, Duke University School of Medicine, Durham, USA

## BACKGROUND AND PURPOSE

Virtual imaging trials (VIT) have advanced significantly in medical imaging through the development of human anatomy models, simulation of scanner-specific devices, and interpretation of images by virtual observers.[1] VITs can be conducted with accessible ground truth and complete control over specific parameters, offering a faster, efficient, and objective approach to developing new imaging technologies.[2–4] However, because of the complexity involved in conducting these trials, VIT utilization can be challenging and requires efforts to enhance their widespread use.

The American Association of Physicists in Medicine (AAPM) Task Group No. 387 (TG387) has been created to establish and provide consensus guidelines for the reliable use of VITs, and to ensure their wide adoption in the medical imaging research community. In this task group, we define and discuss advanced VIT methods to establish robust frameworks and integrated VIT platforms to promote their broad adoption in medical imaging research.

In this work, members of the AAPM TG387 aimed to develop an accelerated and containerized environment to enhance collaborative research and encourage broader use of VITs.

## METHODS

### 1. Virtual Imaging Platforms

One initiative of the AAPM TG387 involves developing strategies to ensure reproducibility and communication across multiple VIT platforms. Every VIT platform encompasses specific documentation that must cover the setup of the virtual environment, description of input and output data, procedures for data manipulation, among others. Establishing a "universal language" for facilitating communication across platforms poses a significant challenge.

Two VIT pipelines renowned for breast imaging simulations were used in this work: VICTRE[2,5] and OpenVCT.[6] These pipelines facilitate the creation of *in-silico* breast radiographic images for evaluating digital mammography (DM) and digital breast tomosynthesis (DBT) technologies. They encompass software tools to generate anthropomorphic breast phantoms and lesions, simulate phantoms under mammographic compression, simulate x-ray projections data, and collect data through reader models for evaluating lesion detection and search tasks.

Despite significant similarities, the algorithms and methods underlying the simulations differ markedly. Considering these differences, we have described a method to integrate and manipulate data across these platforms.

## 2. Creation of a Containerized Environment and Package Installation

Docker (version 4.27, Docker, Inc., San Francisco, CA) was used for OS-level virtualization and for delivering the VIT platforms in packages called "containers." Using Windows Subsystem for Linux (WSL) 2 backend OS, Docker automates the deployment of applications in lightweight containers so that applications can work efficiently in different environments in isolation. The NVIDIA containerized environment (HPC SDK 24.3, NVIDIA, Ubuntu22.04+CUDA 12.3) was selected for the containerized environment for both tested VIT platforms. Importantly, Docker containers can be used in any OS, such as Linux, Windows, and Mac.

The required installation packages of each VIT platform were installed in each OS virtualized images using two commands: "docker build -t *tag_name .*" and "docker run -it --ipc=host -v *host_path*:/*container_path*/ --gpus all *tag_name* sh *config_File.sh*". In this process, all required libraries, packages, and dependencies are installed; users do not need to be preoccupied in the installation of package versions, virtual environment and variables, and any other potential specific requirement. The docker files describe and create the working environment for the users. After setting up the OS virtual environment, users can explore the containerized image using the standard Docker commands such as run, exec, cp, etc.

## 3. Data Management and Communication via Wrappers

Wrappers were created to simplify the setup and execution of VIT pipeline software in the container-based environments. The wrappers streamline the data formatting and pipeline execution, ensuring compatibility and interoperability between diverse software components. Using wrappers, the specific VIT methods can be executed through a command-line interface.

In this work, the language selected for creating the wrappers was Python (version 3, PSF). The wrappers must be simple and intuitive to facilitate the execution. Examples of wrappers were provided and published at the VICTRE GitHub page.[2] The commands below show an example of process of creation of an anthropomorphic phantom using the VICTRE pipeline:

```
from Victre import Pipeline

#Initialization
pipeline = Pipeline() #add optional arguments and constants

#Run
pipeline.generate_phantom() #add optional methods below for compression, lesion insertion, etc.
```

By calling this Python script using the command "docker run", both input and output data will be saved directly on your local host machine. Importantly, data and files can be transferred between containers using various methods. The most straightforward approach involves creating an environment to simulate and store all data in the local host machine (--ipc=host); however, users can also extract data from containers using the command line instruction "docker cp *source_file destination_file*". This command allows for file transfers either to the host machine or within the virtualized containers (specified as "*container_name*:/*file*"). Alternatively, file transfer and data share can be conducted over a network by establishing a Docker network (designating an IP address and port) and setting up servers for synchronized file transfer. For this option, users must allow containers with root access which can be a constraint that must be considered in the network. The command to initiate a network is "docker network create *network*".

The use of containers and wrappers enables communication between independent VIT frameworks (Fig. 1). Wrappers and containers were tested using a hybrid environment with two VIT pipelines for simulating breast imaging.

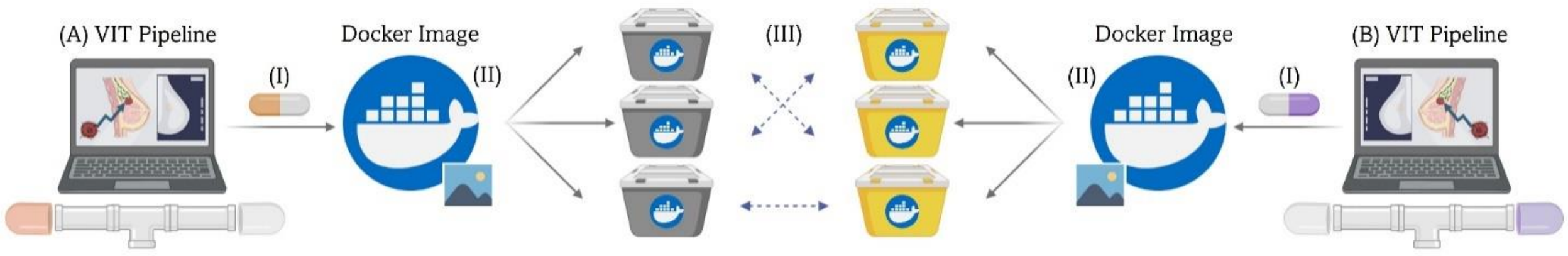

*Figure 1 Flowchart illustrating the development process of an integrated cross-platform VIT framework within a containerized virtual environment. (I) Utilizing wrappers to encapsulate essential packages for VIT pipeline software installation and Docker image creation. (II) Instantiating Docker containers for the execution of specific VIT methods. (III) Employing wrappers to facilitate the streamlined execution of VIT methods, data management within containers, and communication between containers.*

We used wrappers to streamline the data exchange process between the VICTRE and OpenVCT containers, ensuring continuous input and output operations across platforms. These wrappers are designed to convert data into the required format for each respective pipeline, enabling users to effortlessly access, modify, and leverage the full range of resources offered by both platforms. As an example, we used anthropomorphic breast phantoms generated with a VICTRE container and transformed the output phantoms via interface wrappers. Finally, using the OpenVCT container, the phantoms were modified using Simplex methods[7] and x-ray projections were simulated using a ray tracing algorithm.[8] Importantly, the inverse path for breast phantom and image simulations is also possible via these interface wrappers, characterizing a hybrid and containerized cross-platform environment.

## RESULTS

This approach significantly accelerated the creation of the software environment by simplifying package installation and code compilation. The packages required to set up specific virtual environments were embedded in Docker images ensuring that a functioning software environment was readily available for users.

The containerized environments support the simulation of anthropomorphic breast phantoms and x-ray images using different methods, such as Monte Carlo (Fig. 2, left) and ray tracing (Fig. 2, right). The environments facilitated data sharing among containers, ensuring data reproducibility of phantoms and images. The images simulated using VICTRE and OpenVCT methods were evaluated and found to be visually comparable. Furthermore, we adapted the original tissue structures simulated with the VICTRE pipeline by integrating parenchymal complexity simulated using the OpenVCT pipeline.

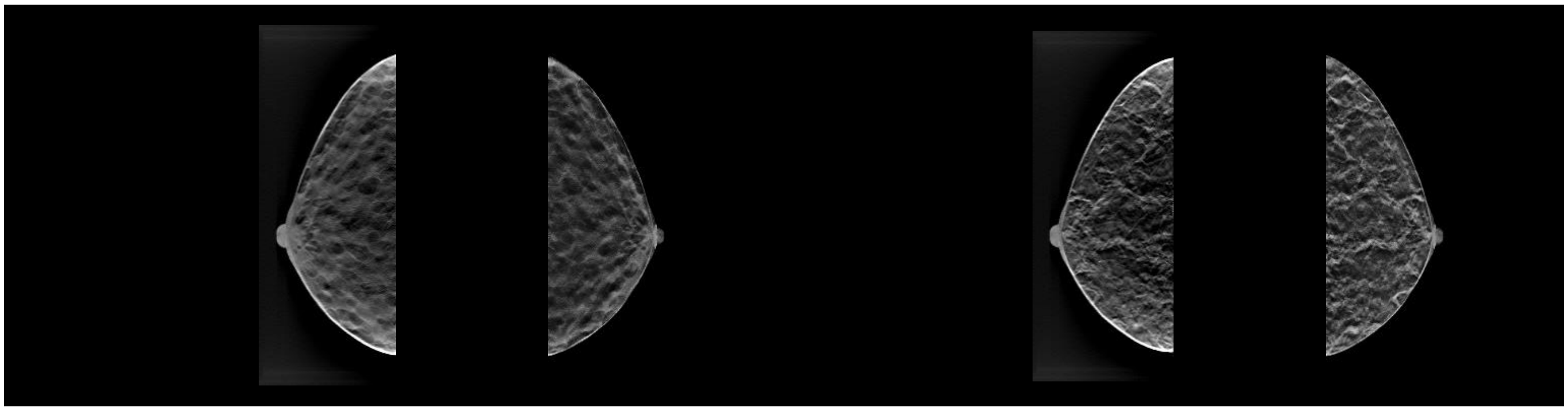

(A) (B)

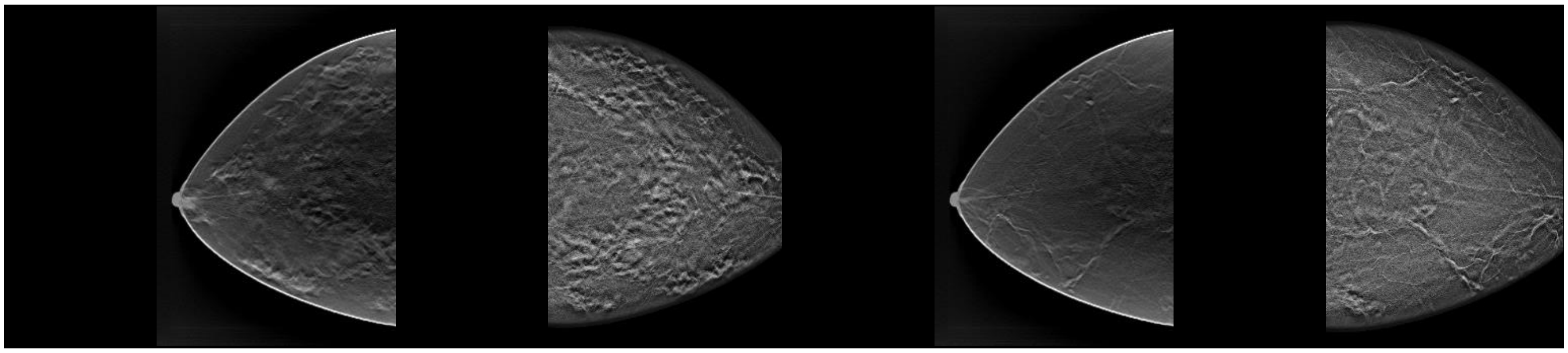

(C) (D)

*Figure 2 Examples of VICTRE phantoms simulated using containerized (left) VICTRE pipeline and (right) hybrid approach with containerized (right) OpenVCT pipeline. Central slices of DBT reconstructions slices are shown: (A) dense phantom, (B) dense phantom with Simplex-based parenchyma, (C) Fatty phantom, and (D) Fatty phantom with Simplex-based parenchyma.*

## DISCUSSION

The adoption of containerized environments for VIT simulations enhances data exchange and interoperability across different platforms, helping users and researchers to select and work on multiple components of different VCT pipelines, without the requirement to adhere to a uniform standard for data manipulation. However, converting data formats, integrating, and synchronizing the communication between VIT platforms are significantly challenging. Interface wrappers depend on a thorough understanding the workflow of multiple VIT platforms for seamless operation. To avoid the need for data translation between platforms, it is essential to develop standardized strategies. Such strategies have been explored and discussed in the AAPM TG387.

For future work, we will explore the use Dockers in a cloud computing environment (e.g., AWS, Google Collab, Azure, etc.) and also the use of Kubernetes[9] for automating deployment, scaling, and operations of VIT containers. Kubernetes can orchestrate Docker containers, managing their lifecycle across a cluster of machines. They automate the deployment, scaling, and management of containerized applications, ensuring they run efficiently and are available to meet user demand. If traffic to a container is high, Kubernetes will balance and distributing the network traffic.

## B. CONCLUSION

To make VITs widely accessible, the development of strategies to streamline coding, manage data effectively, standardize I/O formats, and develop user-friendly interfaces are crucial. Container applications, such as Docker, provide a solution that addresses these requirements, supporting cross-platform integration and facilitating the widespread use of VITs.

## C. ACKNOWLEDGEMENTS AND CONFLICT OF INTEREST

This research was funded by the Terri Brodeur Breast Cancer Foundation; the Susan G. Komen Foundation, Career Catalyst Research Grant; and the American Association of Physicists in Medicine, 2020 Research Seed Funding grant.

The authors would like to express their gratitude to Real Time Tomography, LCC (Villanova, PA) for reconstructing the DBT projections simulated through ray tracing (OpenVCT pipeline).

## D. REFERENCES

# Towards personalized imaging: a pilot virtual imaging evaluation of interpatient variability in chest CT

Rajagopal, Aboulbanine, Abadi, Segars, Kapadia

## Purpose

A personalized approach to imaging requires adding patient specific flexibility to existing generalized protocols. Such developments will be driven by an understanding of the interaction between interpatient variability and clinical task-specific imaging features. In this study, a pilot virtual imaging trial (VIT) was run with a small cohort of virtual patients and a generalized protocol to assess interpatient variability in chest CT.

## Methods

A validated CT simulation platform (DukeSim) was used to simulate a generic single-source CT scanner (64x1.0 mm collimation). Ten adult computational chest phantoms (XCAT, BMI range: 16.8-41.19), six containing visible lesions, were imaged with a generalized chest/lung protocol. Images were acquired at 120 kV, 1.0 pitch, and 0.5 s rotation time. Moderate tube current modulation was applied with a quality reference mAs of 100 mAs. Images were reconstructed using an open-source reconstruction package (MCR Toolkit) using iterative reconstruction, 40 cm field-of-view, and 1.0 mm slice thickness. Ground truth masks were used to segment target organs in reconstructed images for analysis. CT number histograms across patient cases were evaluated for the heart and lungs using the first three statistical moments. A region-of-interest was extracted around each lesion and evaluated for local lesion contrast-to-noise ratio.

## Results

When averaged across all patient cases, histogram analysis of the lungs found a mean value of -700 HU, standard deviation of 32.7 HU, and a skew of 2.75. For the heart, the histogram analysis found a mean value of 33.2 HU, a standard deviation of 5.6 HU, and a skew of -1.76. Measured signal for lungs showed a wider range (49.9 HU) when compared to the heart (26.4 HU). Lesion local contrast-to-noise ratio ranged from 1.43-2.2 and showed a slight dependence on patient size.

## Conclusion

We successfully performed a small scale VIT to highlight the interpatient variability across a population of virtual patients for a chest/lung protocol.

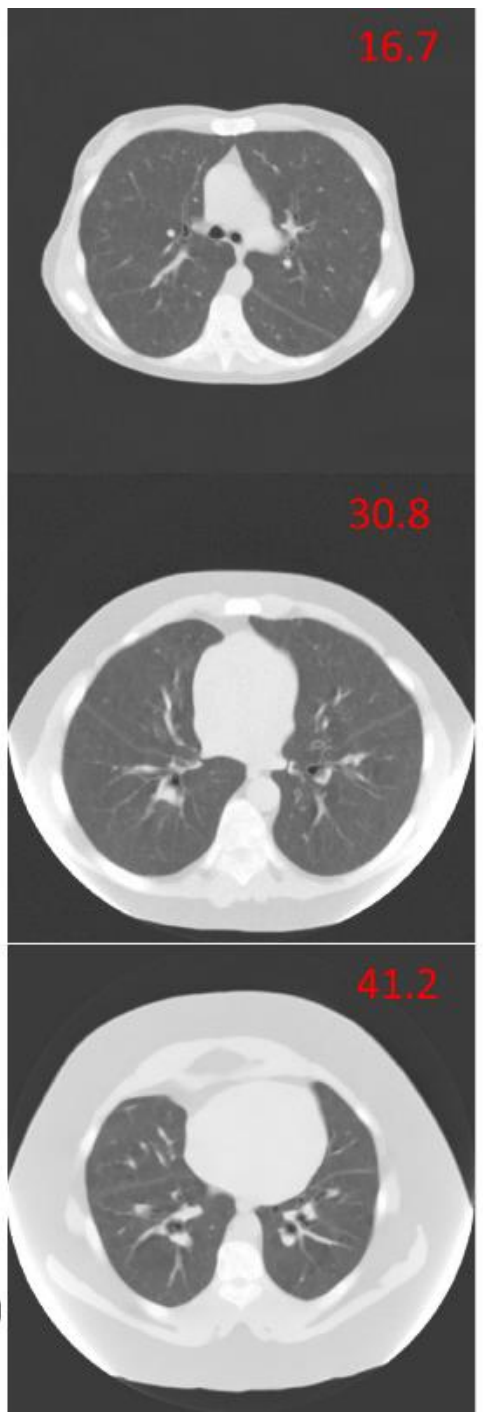

(a)

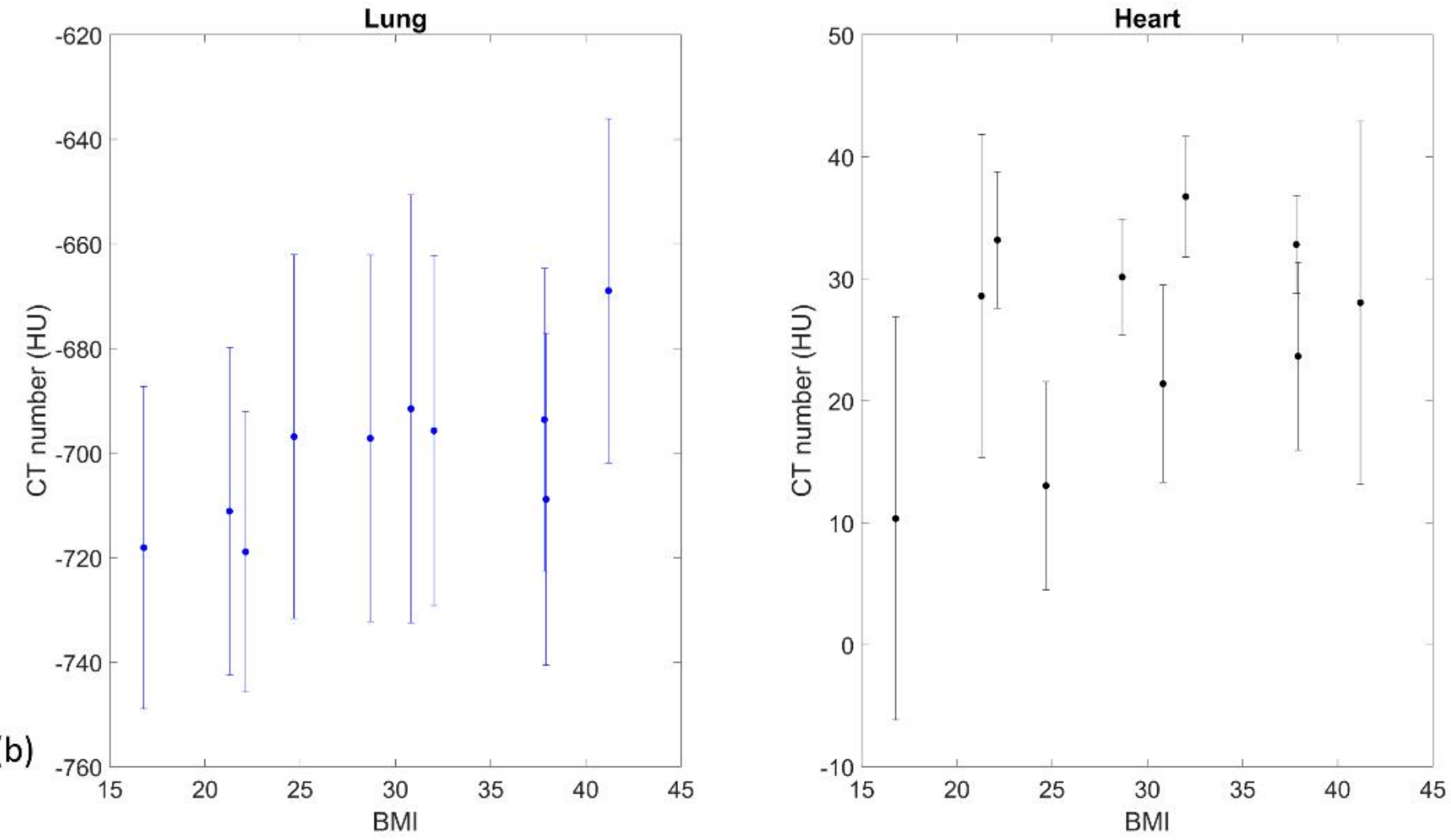

(b)

(a) Example of virtual patient cases used in this study at three different sizes. BMI values represented in red. Images presented with lung window (W/L – 1500/-600)

(b) Plots of intensity mean and standard deviation (HU, y-axis) for lung (left) and heart (right) across patient cases. Plotted against patient BMI (x-axis).

# Poster minute digest 1

Tuesday, April 23, 2024

12:10 am - 12:30 am

Chaired by Francesco Ria and Liesbeth Vancoillie

# Measurement of scapulohumeral rhythm in shoulder dynamic digital radiography: A virtual imaging trial

Yu Homareda[1], Rie Tanaka[1], Futa Goshima[1], Hiroyuki Okamoto[1], Shuhei Minami[2]
1. Graduate School of Medical Sciences, Kanazawa University, Ishikawa, Japan
2. Division of radiology, Kanazawa University Hospital, Ishikawa, Japan

## BACKGROUND AND PURPOSE

Dynamic digital radiography (DDR) is reportedly useful for measuring scapulohumeral rhythm (SHR), an important parameter in assessing shoulder kinematics. This study aimed to develop optimal imaging conditions for SHR assessment using DDR through a virtual imaging trial (VIT).

## METHODS

Three male virtual patients (4D extended cardiac-torso: XCAT phantom) were remodeled with the right arm abducted using the Rhinoceros modeling tool and projected using an X-ray simulator at four total entrance surface dose (ESD) levels (1/10, 1/5, 1/2, and 1 of the reference ESD) and three imaging rates (3, 6, and 15 frames per second (fps)) (Figure 1). Glenohumeral angle and scapulothoracic angle were manually measured by six radiological technologists, and the measured SHR was compared to the reference SHR based on standard error (SE), measurement time, and inter-frame variability; the Wilcoxon signed-rank test was performed for mean absolute error (MAE). The image quality at each imaging rate and dose was evaluated visually.

## RESULTS

SEs analysis indicated no trend among the three imaging rates. However, the measurement times for images with 6 and 15 fps took approximately 1.4 and 2.7 times longer than with a 3 fps image; the inter-frame variability of the measured angles was also greater. In addition, there was no significant difference in MAE between images at 1/2 and 1 ESD, whereas the MAE of images at 1/10 and 1/5 of ESD were significantly worse than those of images at 1 ESD (Figure 2). These results indicate that reasonable imaging conditions for SHR measurements with DDR are a total ESD of 1/2 of the reference ESD at an imaging rate of 3 fps.

## CONCLUSION

VIT allowed the development of optimal imaging rates and doses for SHR assessment with DDR, which would be useful in personalized imaging of shoulder DDR.

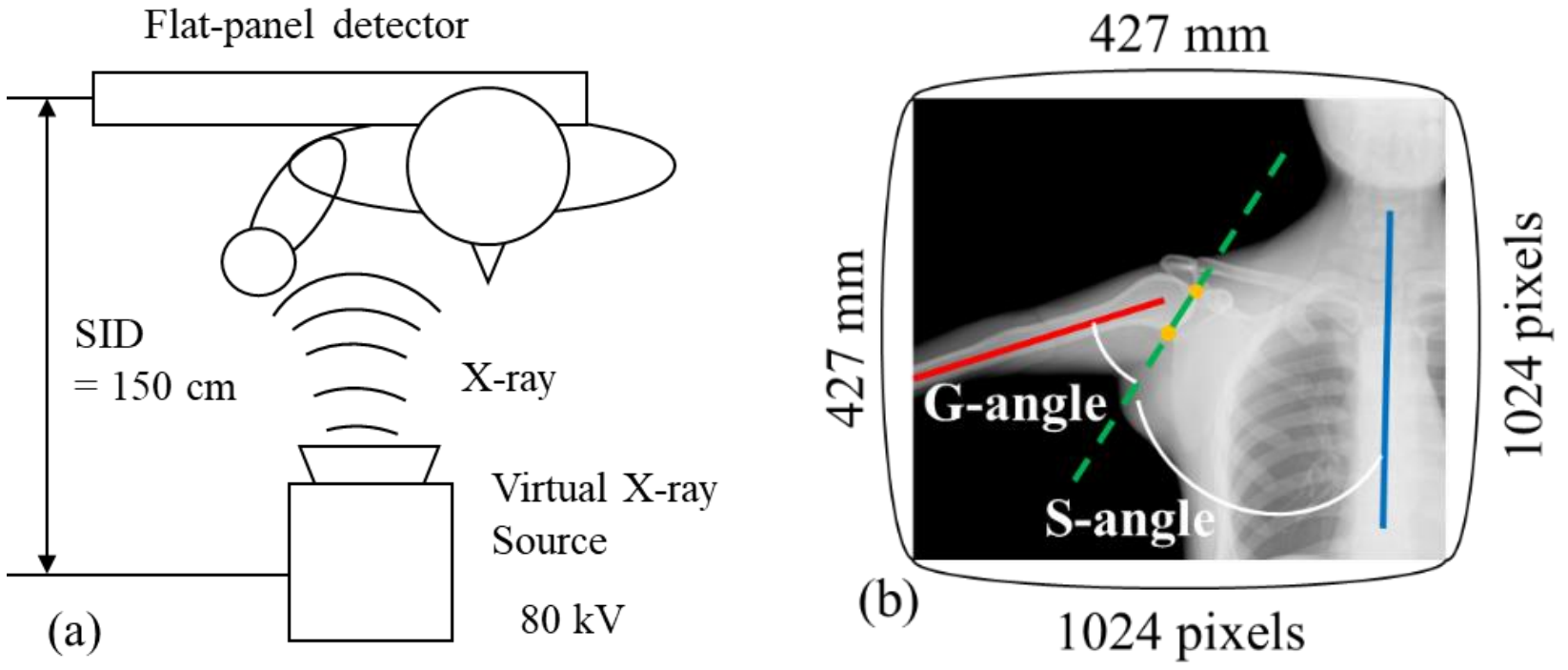

*Figure 1 Illustrations of (a) virtual image acquisition and (b) angle measurements in shoulder dynamic digital radiography. The red, green, and blue lines and yellow points represent the center line of the humerus, a line connecting the intersections of the humerus and scapula, the body axis, and tracking points, respectively. SID: source-to-image distance; G-angle: glenohumeral angle; S-angle: scapulothoracic angle*

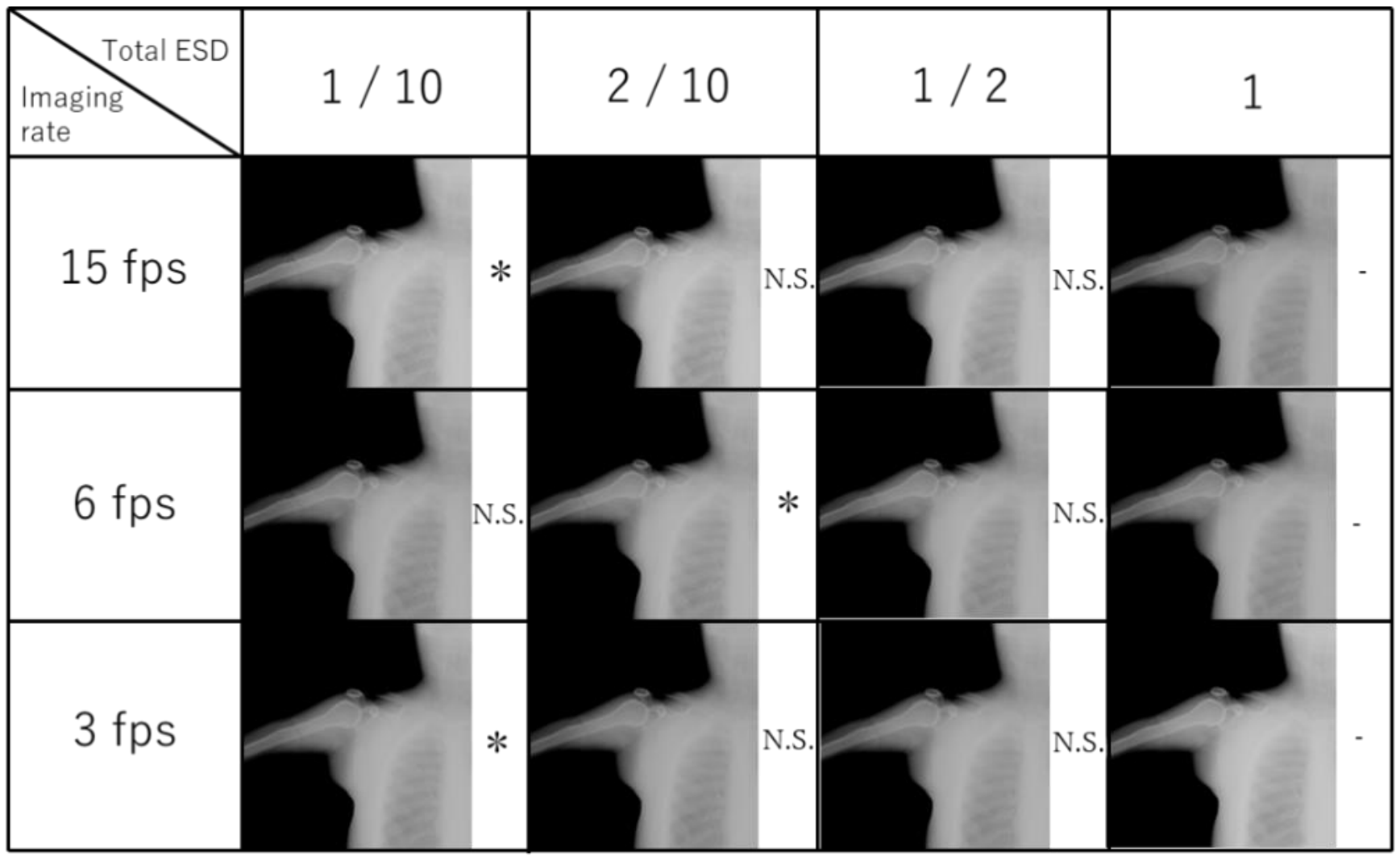

*Figure 2 Summary of the relationship between a total entrance surface dose (ESD) and imaging rate The ESD of conventional shoulder radiographs measured at Kanazawa University Hospital was used as the reference dose. The symbols (*, N.S., -) indicate a significant difference, no significant difference, or standard. As you can see, images obtained at one-tenth and two-tenths of reference ESD are not of sufficient image quality for informed consent.*

*fps: frames per second*

# In-silico flow diverter performance assessment in posterior communicating artery aneurysms

Michael MacRaild[1], Ali Sarrami-Foroushani[1], Shuang Song[1], Qiongyao Liu[1], Christopher Kelly[1], Nishant Ravikumar[1], Tufail Patankar[2], Toni Lassila[1], Zeike A. Taylor[1], Alejandro F. Frangi[1]

[1]Centre for Computational Imaging and Simulation Technologies in Biomedicine (CISTIB), [2]Leeds General Infirmary

E-mail: scmm@leeds.ac.uk

## ABSTRACT

Flow diversion is an established treatment method for intracranial aneurysms. The Pipeline Embolization Device (PED) is the most widely used flow diverter but is not FDA-approved for use in posterior communicating artery (PComA) aneurysms, which make up 25% of all aneurysms. PED use has extended to various off-label indications regardless, including PComA aneurysms. Retrospective analysis has shown that the PED is less effective in patients with fetal posterior circulation (FPC) but the reason for this is unclear. Clinical trials to date have not focused on PComA aneurysm flow diversion and it would be costly and unethical to perform a clinical trial in patients with FPC where the treatment may not be effective. In-silico trials based on computer modelling and simulation can overcome these concerns and provide a comprehensive understanding of the mechanisms underlying treatment failure.

We present an in-silico trial into PED flow diversion of PComA aneurysms, aiming to understand the relationship between PComA patency and aneurysm occlusion in patients with and without FPC. Our cohort is comprised of 64 patients processed from the AneuX and AneurIST databases. We apply flow-split outlet boundary conditions to model non-fetal and fetal configurations. We simulate transient haemodynamics pre- and post-treatment for each patient. The surrogate occlusion end-point for successful treatment is a reduction in aneurysm space-and-time-averaged velocity greater than 35%.

Preliminary results in a cohort of 17 patients with zero-pressure outlet boundary conditions demonstrate a significant relationship between aneurysm flow reduction and PComA outlet flow and between aneurysm flow reduction and PComA size. Aneurysm size, aspect ratio and inflow physiology (rest, hypertension and exercise) were not found to significantly affect aneurysm flow reduction. These results suggest that PED treatment of PComA aneurysms is complex. The full trial will help to understand this further.

## ACKNOWLEDGEMENTS AND CONFLICT OF INTEREST

We acknowledge support from ANSYS through an Academic Partnership agreement, and funding from EPSRC (EP/L014823/1) and the RAEng (CiET1819\19; LTRF2021\17115).

# Development of deep learning-based cardiac phase estimation in dynamic chest radiography using virtual patients

Saho Matsuo[1] Rie Tanaka[1] Ryuuichi Nagatani[1] Haruto Kikuno[1]
1. Graduate School of Medical Sciences, Kanazawa University, Ishikawa, Japan

## BACKGROUND AND PURPOSE

Dynamic chest radiography (DCR) is a flat-panel detector (FPD)-based functional X-ray imaging method used during respiration and cardiac pumping. As image-based functional analysis is affected by the target frame, determining the respiratory and cardiac phases in each frame is crucial. This study aimed to develop a deep-learning-based cardiac phase estimation method and to determine the estimation accuracy compared to the conventional motion analysis methods using virtual patients (4D extended cardiac-torso: XCAT phantom).

## METHODS

Thirty-one virtual patients (heart rate, 60 beats/min; breath holding; body mass index, 18.5~34.5 kg/m2) were generated by the XCAT phantom program and then projected by an X-ray simulator. A total of 1,395 DCR images were created, and the inter-frame subtraction images were used to train and test (21:4) a convolutional neural network (CNN) to estimate the cardiac phase on DCR images. The performance of CNN was compared with those of kernel correlation filter (KCF) tracking algorithm, template matching, and optical flow. The classification performance of each method was evaluated in terms of sensitivity, specificity, and accuracy, based on the cardiac phases derived from the log files of the XCAT phantom.

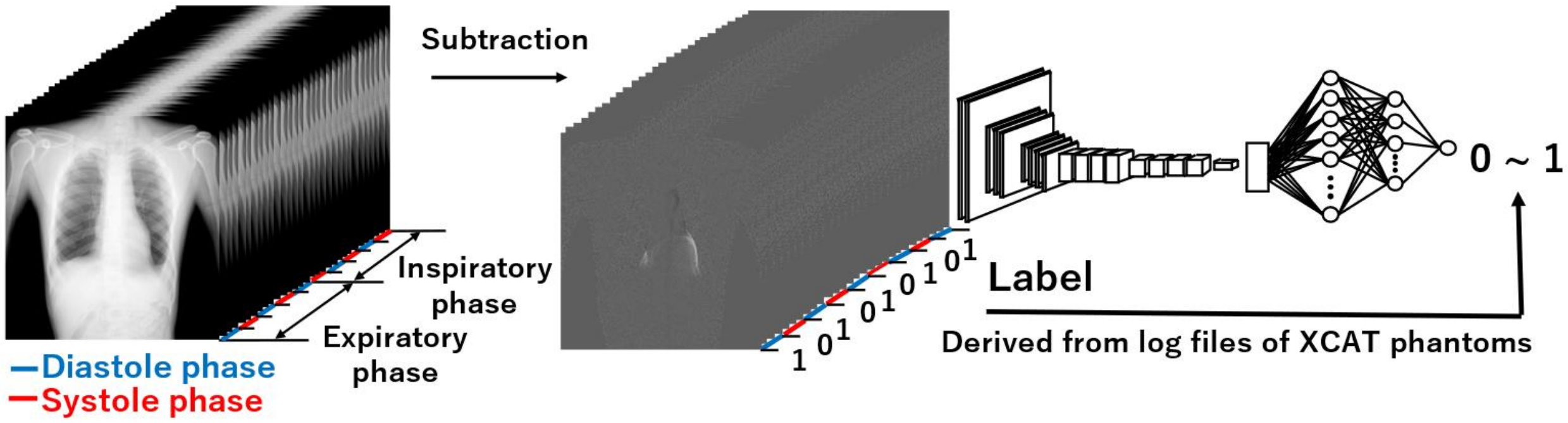

*Figure 1 Schematic illustration of the training session.*

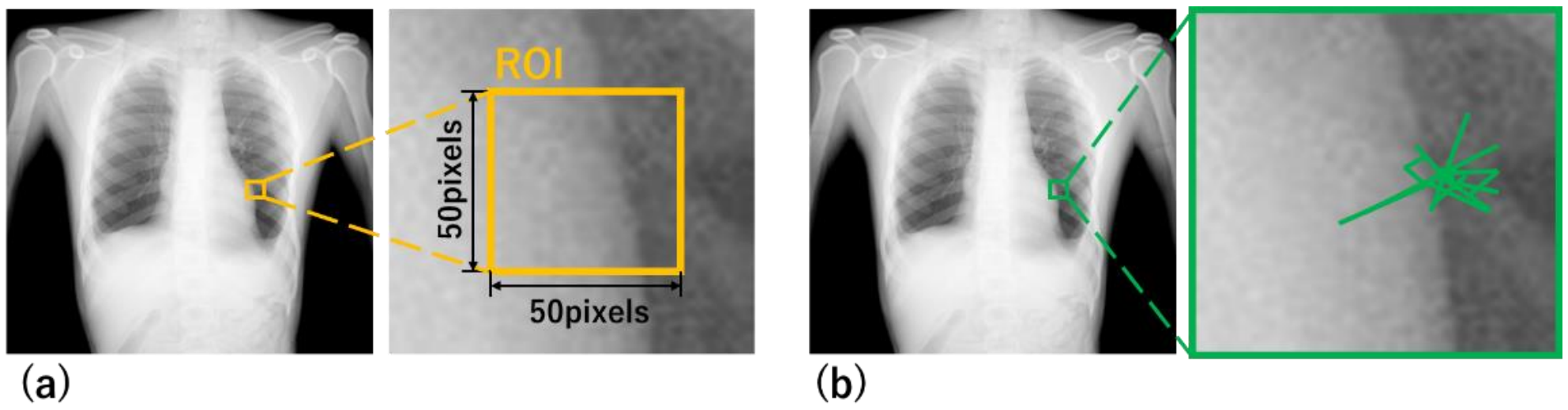

*Figure 2 Tracking area by (a) KCF tracker and template matching and (b) optical flow. The orange square represents the region of interest (ROI) for motion tracking by KCF tracker and template matching, and the green line represents the locus of optical flow.*

# RESULTS

The sensitivities of the CNN, KCF tracking algorithm, template matching, and optical flow were 86.9%, 69.7%, 71.4%, and 77.3%, respectively. The specificities were 93.4%, 92.5%, 84.9%, and 89.6%, and the accuracies were 90.0%, 80.4%, 77.8%, and 83.1%, respectively. The CNN showed the best performance and worked reasonably well in distinguishing between diastolic and systolic phases on DCR images.

*Table 1 Estimation performance of the cardiac phase in each method.*

| | Sensitivities [%] | Specificities [%] | Accuracies [%] |
|---|---|---|---|
| CNN | 95.6 | 99.2 | 97.1 |
| KCF tracking algorithm | 67.4 | 96.9 | 79.5 |
| Template matching | 76.8 | 99.2 | 86.0 |
| Optical flow | 82.3 | 97.6 | 88.6 |

*CNN: convolutional neural network; KCF: kernel correlation filter*

# CONCLUSION

Deep-learning-based cardiac phase estimation techniques have been successfully developed using virtual patients with known cardiac phases. The developed model is expected to realize stable functional analysis of DCR images, resulting in a better understanding of pulmonary and cardiac conditions using DCR.

# In-silico investigation of an innovative cone-beam CT configuration for quantitative imaging

Antonio Sarno[1,2,*], Paolo Cardarelli[3], Paolo Mauriello[1,2], Antonio Minopoli[1], Carlos Mollo[2], Silvio Pardi[2], Gianfranco Paternò[3], Mariagabriella Pugliese[1,2] and Riccardo de Asmundis[2]

[1]Dipartimento di Fisica"Ettore Pancini", Università di Napoli Federico II
[2]INFN sez. Napoli, Napoli, Italy
[3]INFN – Sez. Ferrara, Ferrara I-44122, Italy
*sarno@na.infn.it

## BACKGROUND AND PURPOSE

We performed in-silico investigations of an innovative scanning geometry in CBCT. Such a geometry (ExoCT) relies on beam collimation and an oscillating scanning trajectory for reducing detected scatter and cone beam artifacts with the aim of improving accuracy and precision in quantitative analysis. Such an oscillating geometry is obtained by substituting each of the projection of CBCT conventional protocol with several collimated projections over an oscillation period of the source.

## METHODS

We simulated the scanner geometry via a GPU Monte Carlo software base on the GGEMS platform (https://ggems.fr). The used software yielded computation times one order of magnitude shorter than codes on CPU architecture. We developed a digital phantom, which replicated the Defrise phantom for the evaluation of the image conspicuity over the FOV. It embodied 4 material inserts at different locations for evaluating accuracy, precision and reproducibility of attenuation coefficients estimates.

## RESULTS

Narrowing the beam aperture consistently reduces the scatter-to-primary ratio (SPR) (fig. 1). Replacing a CBCT projection with 10 collimated projections in ExoCT reduces the SPR of 86%. This implies the improvement of the accuracy in the estimations of attenuation coefficients of the materials included in the FOV (tab. I). As example, an ExoCT scan with 4 projections per oscillation, determined an increase in the accuracy of the attenuation coefficients estimates of test inserts between 2.6% for PC and 13.1% for PVC (Tab. I). Additionally, the oscillation trajectory of the ExoCT determines an enhanced reproducibility of the quantitative evaluations over the FOV, as opposed to what observed in CBCT (Tab. II).

## CONCLUSION

We investigate the ExoCT scanning geometry via a GPU Monte Carlo software and a customized digital phantom. The ExoCT configuration was shown to increase the accuracy and reproducibility of quantitative evaluations, when compared to conventional CBCT.

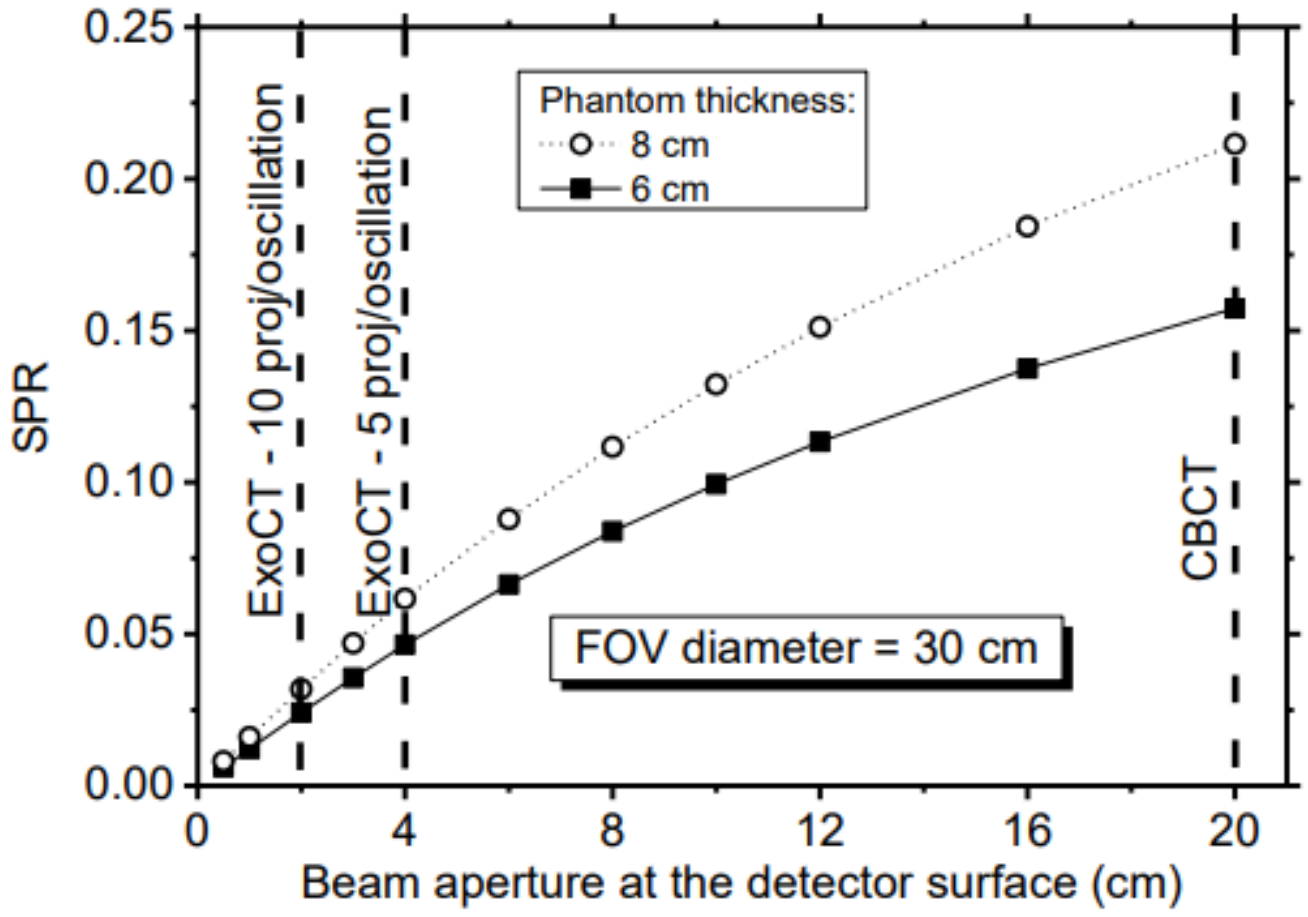

***Figure 1.*** *Scatter-to-primary SPR as function of the beam aperture at the detector surface; the irradiated sample was a 10 cm × 10 cm box made of soft tissue whose thickness was either 8 cm or 6 cm. Source-to-isocenter distance = 500 mm; source-to-detector distance = 750 mm; W/Al 80 kV spectrum.*

| **Attenuation coefficients estimates over simulated images to the theorical one, percent difference** | | |
|---|---|---|
| **Material** | **CBCT** | **ExoCT** |
| PMMA | 18.7 % | 8.5 % |
| PVC | 29.9 % | 16.8 % |
| PE | 36.0 % | 31.2 % |
| PC | 17.4 % | 14.8 % |
| Water | 23.0 % | 14.7 % |

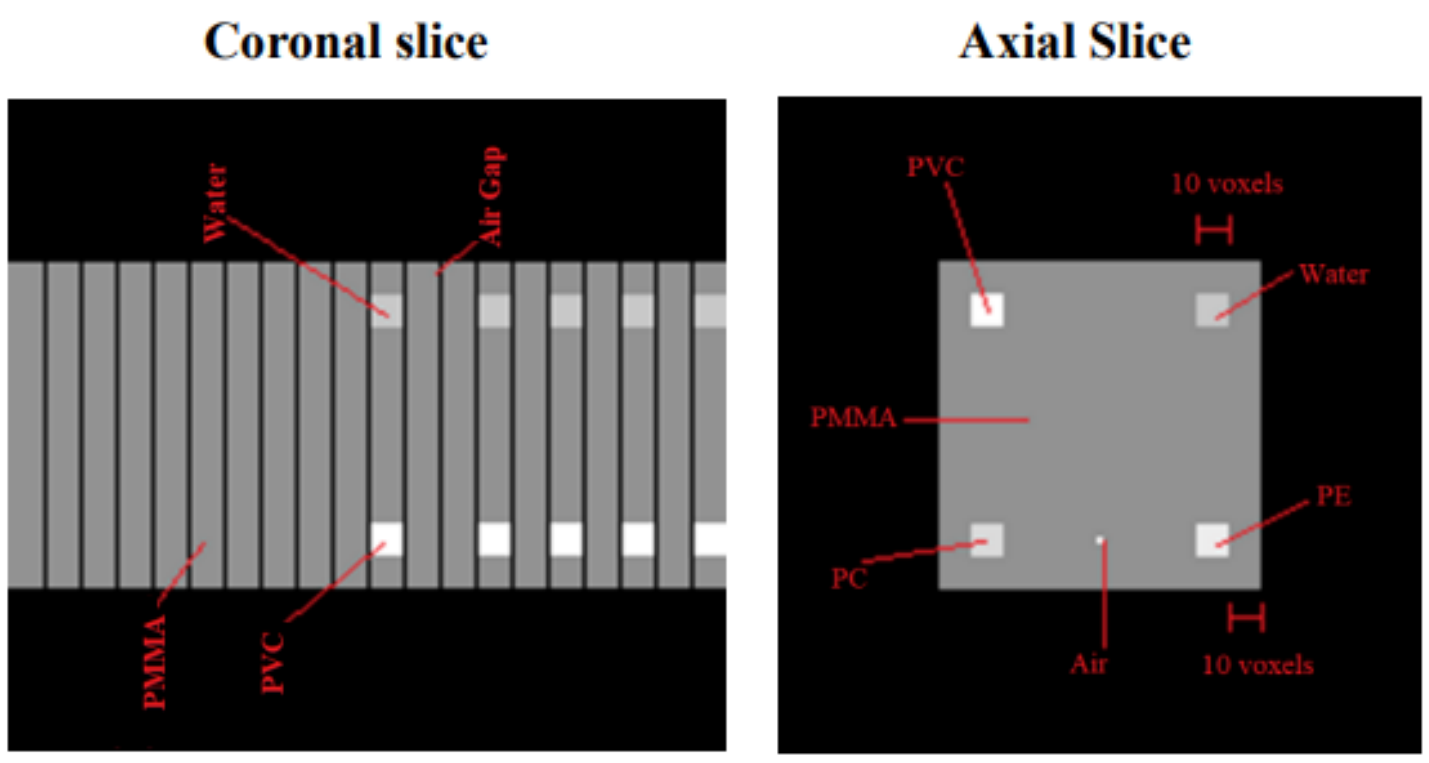

***Table I**. Percent difference between attenuation coefficients estimated in-silico and the theoretical ones at the central axial plane of the FOV. In the ExoCT, 4 projections per oscillations were adopted for 300 oscillations; in CBCT 300 projections were adopted. On the right, a schematic of the used digital phantom.*

| **Attenuation coefficients estimates over simulated images to the theorical one, percent difference** | | |
|---|---|---|
| **Distance from the central axial plane** | **CBCT** | **ExoCT** |
| 5mm | 29.8 % | 16.8 % |
| 55mm | 34.3 % | 18.2 % |
| 75mm | 36.7 % | 21.0 % |
| 95mm | 80.8 % | 25.2 % |

***Table II**. Percent difference between attenuation coefficients estimated in-silico and the theoretical ones as function of the distance from the central axial plane of the FOV.*

# The regulatory landscape of virtual tools in diagnostic imaging

Francesco Ria[1], Kelly Lindblom[2], Kyle Myers[3], Ehsan Samei[1]

[1] Carl E. Ravin Advanced Imaging Labs, Center for Virtual Imaging Trials, Department of Radiology, Duke University Health System

[2] Office of Regulatory Affairs and Quality, Duke University School of Medicine

[3] Puente Solutions LLC

## 1. Purpose

To assess a medical product's safety and effectiveness, clinical trials often require testing in real patients, which inevitably involve factors of cost, time-length, and patient risk. Such elements can be notably easier to manage by implementing computational models to virtually replicate realistic scenarios to simulate clinical trials. To increase the likelihood that regulatory bodies accept evidence from virtual clinical trials, designers of virtual tools should make use of standards and guidance provided by regulators. The purpose of this work is to delve into the virtual diagnostic imaging tools (VDITs) regulatory landscape to inform their design and implementation.

## 2. Methods

This study investigated guidance documents from the Food and Drug Administration Center for Devices and Radiological Health (FDA-CDRH), and standards from the American Society of Mechanical Engineers (ASME). In particular, the FDA-CDRH Qualification of Medical Device Development Tools guidance (1), the FDA-CDRH Assessing the Credibility of Computation Modeling and Simulation in Medical Device Submissions guidance (2), and the ASME-V&V40 standard to assess the credibility of computation modeling applied to medical devices (3) are reviewed.

## 3. Results

Based on current FDA guidance, VDITs can be characterized as medical device development tools (MDDTs) in the Non-clinical Assessment Model (NAM) category. NAM-MDDTs are non-clinical test models that measure or predict specific parameters and can be instrumental in the assessment of a device safety, effectiveness, or performance. The MDDT qualification process requires a 2-phase application for proposal and qualification (Figure 1). A qualified MDDT can be used to evaluate a medical device in a certain context of use without the need to reconfirm the tool suitability or utility. Computational models have been qualified through this program for use in the evaluation of the compatibility of medical implants for patients in a magnetic resonance imaging environment, eliminating the need for actual testing of patients with implants. Additionally, FDA is proposing a 9-step framework (ASME-V&V40) to assess the credibility of a computational model in a medical device regulatory submission.

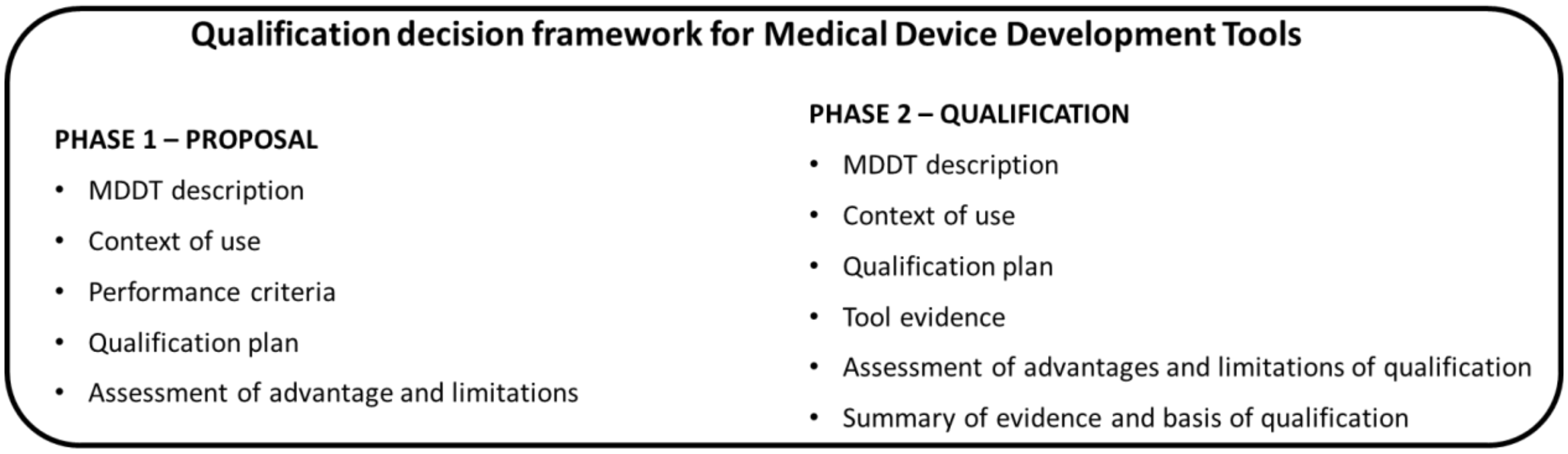

(a)

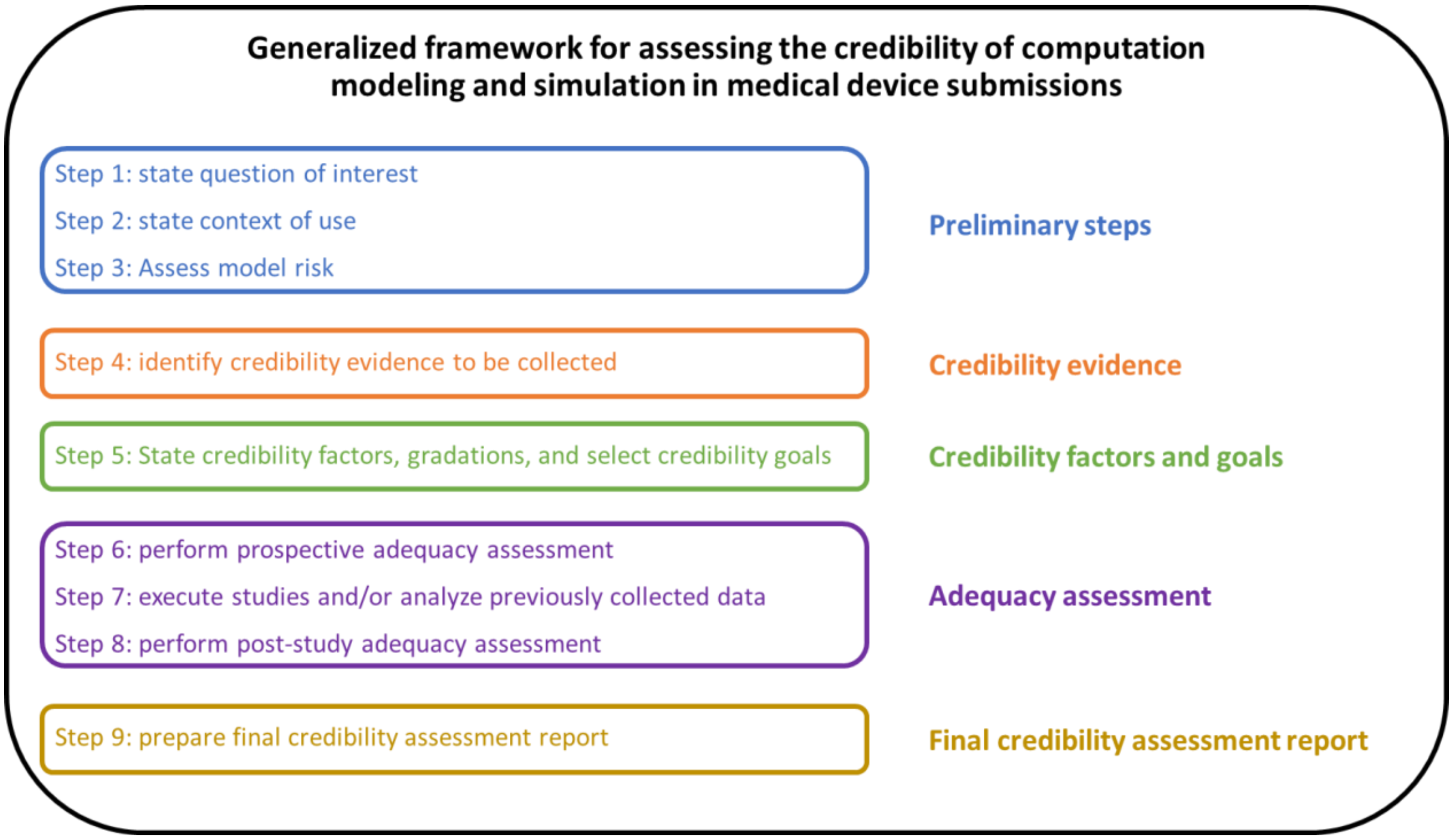

(b)

**Figure 1**. Decision framework for the qualification of Medical Device Development Tools (a); and overview of the generalized framework for assessing the credibility of computation modeling and simulation in medical device submissions (b).

## 4. Conclusion

Government agencies and standards development organizations are joining efforts to put forward guidance and standards that help ensure the safety and effectiveness of virtual tools in medical imaging and other medical devices. The systematic study of such resources can guide investigators in designing and implementing computational models that can support regulatory decision-making in the development of medical products.

# Optimization and validation of a mechanical compression model for realistic digital breast phantoms in mammography and digital breast tomosynthesis simulations

Gustavo Pacheco[a], Koen Michielsen[a], Eloy García[b], Oliver Díaz[b,c,d], Ioannis Sechopoulos[a,e,f]

[a]Dept. of Medical Imaging, Radboudumc, The Netherlands; [b]Institute of Computer Vision and Robotics (ViCOROB), University of Girona, Spain; [c]Facultat de Mahemàtiques i Informàtica, Universitat de Barcelona, Spain; [d]Computer Vision Center, Spain; [e]Dutch Reference Centre for Screening (LRCB), The Netherlands; [f]Technical Medicine Centre, University of Twente, The Netherlands

## BACKGROUND AND PURPOSE

To improve, optimize and validate a mechanical compression model to create realistic digital breast phantoms for mammography and digital breast tomosynthesis simulations.

## METHODS

Simulated finite-element breast compressions were performed to obtain phantoms from segmented patient breast CT scans (N=88). The compression model includes elastic properties of the breast tissues, boundary conditions, and newly added external forces simulating the compression paddle and radiographer handling during the examination. To validate and optimize the model parameters, a principal component analysis (PCA) model derived from 3D surface scans was used. The surface scans (N=236) were acquired during clinical breast compression in the cranio-caudal (CC) view. The comparison involved projecting the phantom data onto the patient PCA base and assessing the agreement between the distributions for each of the 15 principal components (PCs). The compression optimization aimed to maximize the agreement between these two distributions.

## RESULTS

The external forces, particularly those simulating the radiographer's handling, had a greater impact on phantom shape than elastic parameter changes. 10 out of 15 PCs showed significant disagreement before introducing external forces and optimizing all parameters. With the final parameter set, only 4 out of 15 PCs disagreed. Two are related to the overall breast volume, where the disagreement can be attributed to differences in scan coverage between the breast CT and optical surface scans. The other two describe subtler surface features, and although they still exhibited disagreement, their mean values were closer to the patient distributions.

## CONCLUSION

Using patient 3D surface scans, a compression model was optimized and validated for generating digital compressed breast phantoms in the CC view. Optimizing model parameters and introducing external forces simulating breast handling improved agreement between phantom shapes and patient scans.

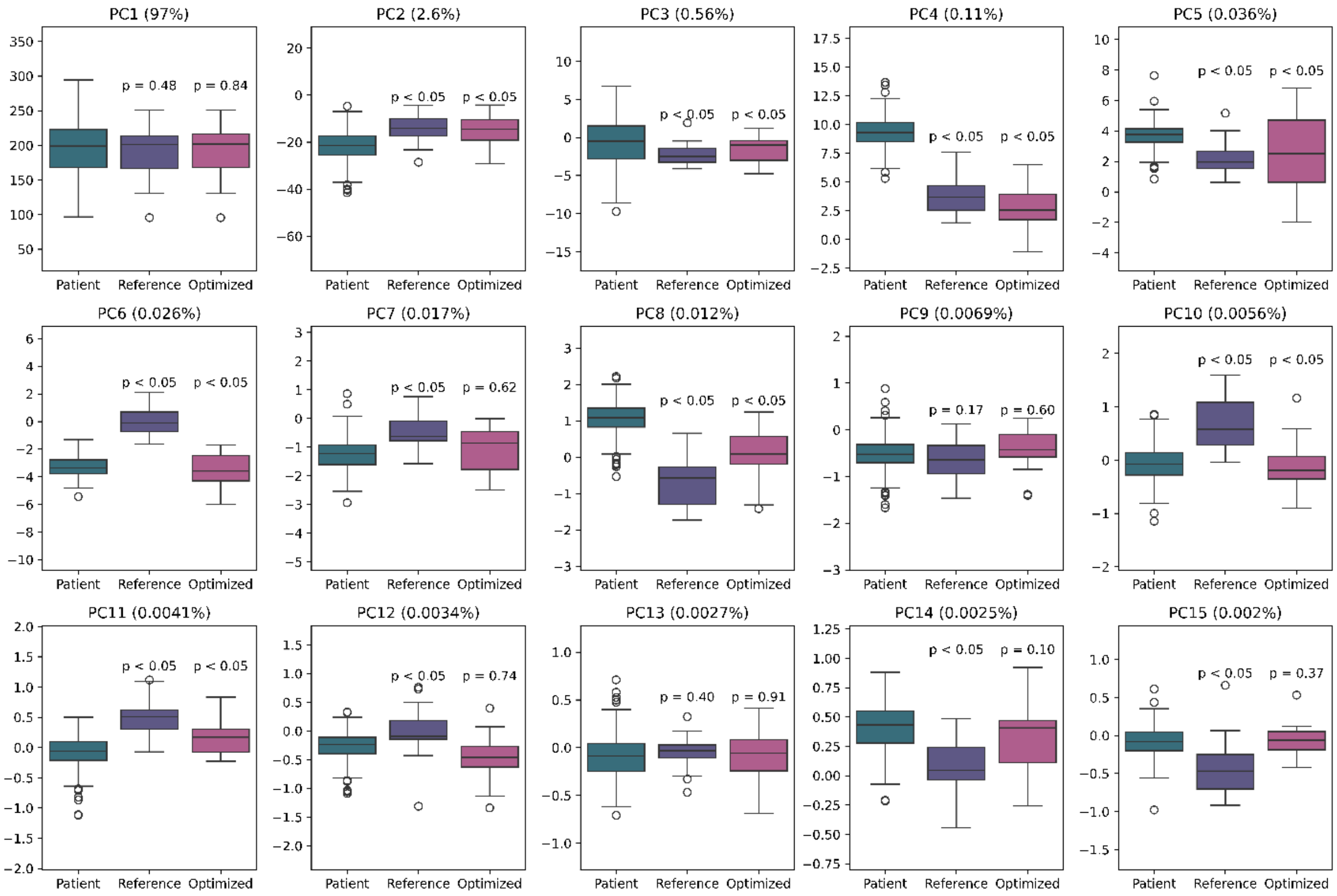

*Figure 1: Box-whisker plots for the 15 PC distributions. In each subplot, from left to right: patient scan distribution, reference compression model results, and optimized compression considering external forces. P-values < 0.05 denote statistically significant differences from the patient distribution. Subplot titles indicate the component and the percentage of explained variance (in parentheses).*

# Simulating ultrasound optical tomography – Distinguishing breast tumors from the surrounding healthy tissue

Adam Kinos[1,2,*], David Hill[2], Magnus Dustler[3,4], Sophia Zackrisson[4], Lars Rippe[2], Johannes Swartling[1], Stefan Kröll[2,†], Predrag R. Bakic[3,4,‡,]

[1] Deep Light Vision AB, Lund, Sweden
[2] Department of Physics, Lund University, Sweden
[3] Medical Radiation Physics, Department of Translational Medicine, Lund University, Sweden
[4] Diagnostic Radiology, Department of Translational Medicine, Lund University, Sweden

## BACKGROUND AND PURPOSE

Ultrasound Optical Tomography (UOT) is an innovative imaging technique that merges the capabilities of ultrasound and near-infrared optical imaging to provide a comprehensive and non-invasive approach to characterize biological tissue [1, 2]. By analyzing only the light that has been frequency shifted by the ultrasound, it provides the optical contrast of light with the spatial resolution of the ultrasound, see Fig. 1.

To analyze the technique's viability for the characterization of breast tumors, Monte Carlo simulations on a realistic breast tissue phantom are performed. The phantom consists of skin, adipose, and fibroglandular tissue compartments with a 1 cm$^3$ tumor region placed in one out of six different positions. The aim of this study is to investigate if the UOT signal from a tumor region can be distinguished from the surrounding tissue, despite the natural variations that occur in the UOT signal due to the varying tissue types.

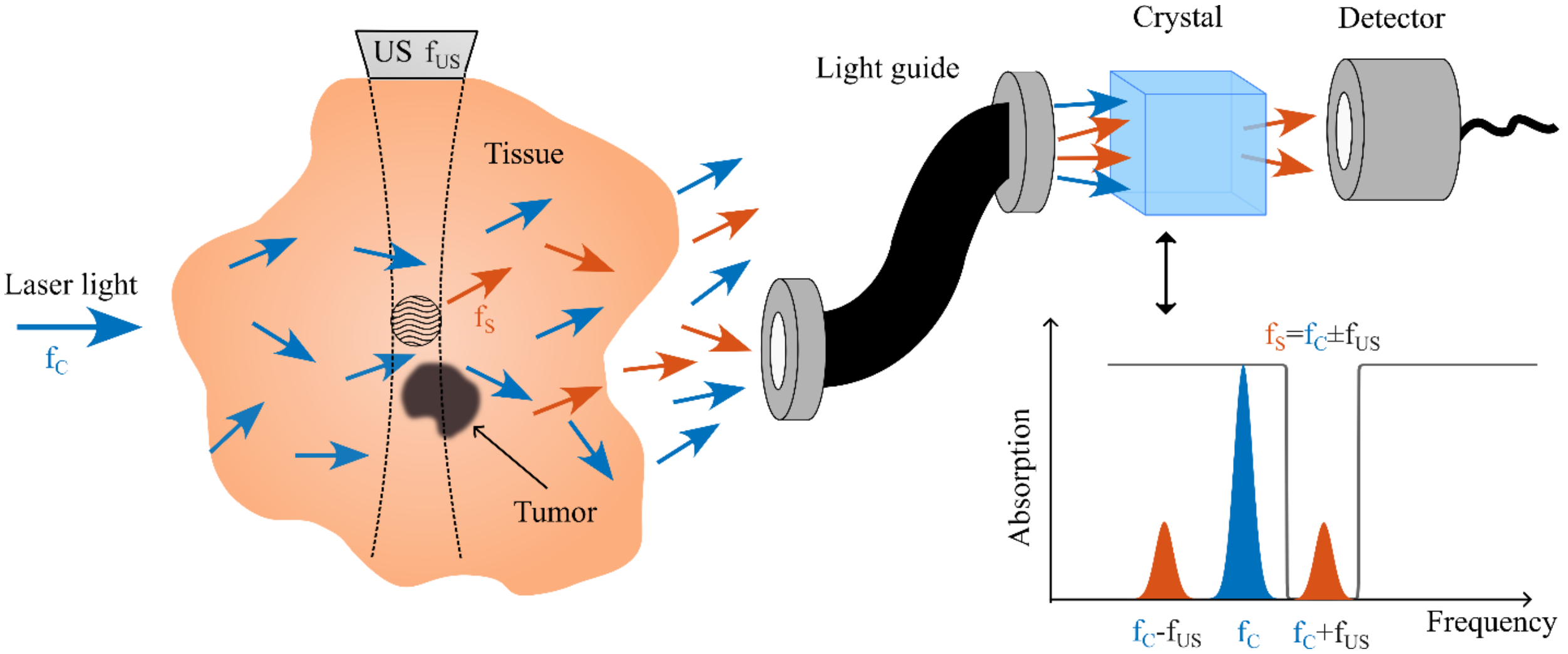

*Figure 1: In UOT, laser light with a carrier frequency $f_C$ (blue arrows) is scattered throughout the tissue simultaneously as an ultrasound (US) pulse with frequency $f_{US}$ propagates through it. At any instant, light can interact with the US and via the acousto-optical effect be frequency shifted to $f_S = f_C \pm f_{US}$ (red arrows). A portion of the light exiting the tissue is collected with a light guide that directs the light to a cryostat containing a rare-earth-ion-doped crystal. The crystal acts as a high étendue narrowband spectral filter capable of attenuating the carrier light (blue) while transmitting the frequency-shifted light (red). Since the frequency-shifted light, i.e., the UOT signal, has passed through the ultrasound pulse, detecting only this light provides spatially resolved information of the optical properties of the tissue. For example, if the ultrasound focus is located on a tumor with high optical absorption, the UOT signal is weaker since any frequency-shifted light comes from the tumor region where light is absorbed more.*

*ak@deeplightvision.com; †stefan.kroll@fysik.lth.se; ‡predrag.bakic@med.lu.se;

## METHODS

The computational breast tissue phantom was created with the OpenVCT platform and simulated using recursive partitioning and a Perlin noise-based tissue model [3]. The phantom, with a volume of 450 ml (B-cup), has a volume breast density of 10.9% and is compressed to 5 cm in the medio-lateral direction. The phantom voxel size was 200 µm, and in this study we focused on the four tissue types of skin, adipose, fibroglandular, or tumor, as shown in Fig. 2. The optical properties used in this study are presented in Table 1. The absorption coefficients are estimated based on the content of four chromophores: oxygenated and deoxygenated blood [4], water [5], and lipid, where the lipid content is based on an estimated linear relationship with the water content [6, 7, 8]. The absorption spectrum of each chromophore is taken from Ref. [9]. The tumor absorption coefficient is allowed to vary to mimic the increased blood volume or deoxygenation that might occur in tumors. The reduced scattering at 690 nm is estimated using $\mu_s'(\lambda) = A(\lambda/\lambda_0)^{-b}$, where $A$ at $\lambda_0 = 830$ nm is taken from Ref. [4] and $b$ is estimated using the linear relationship with the water content [6, 7].

A sequential Monte Carlo method [10, 11] is used to simulate how carrier photons are scattered throughout the tissue without any absorption, and the paths of the photons reaching the light guide are saved. Absorption is added afterwards, which significantly speeds up the process of studying variations in the absorption coefficients of the tissue. To estimate the UOT signal, we assume that carrier photons are frequency-shifted by the ultrasound with a probability that is linearly proportional to the distance the photon traveled in the ultrasound

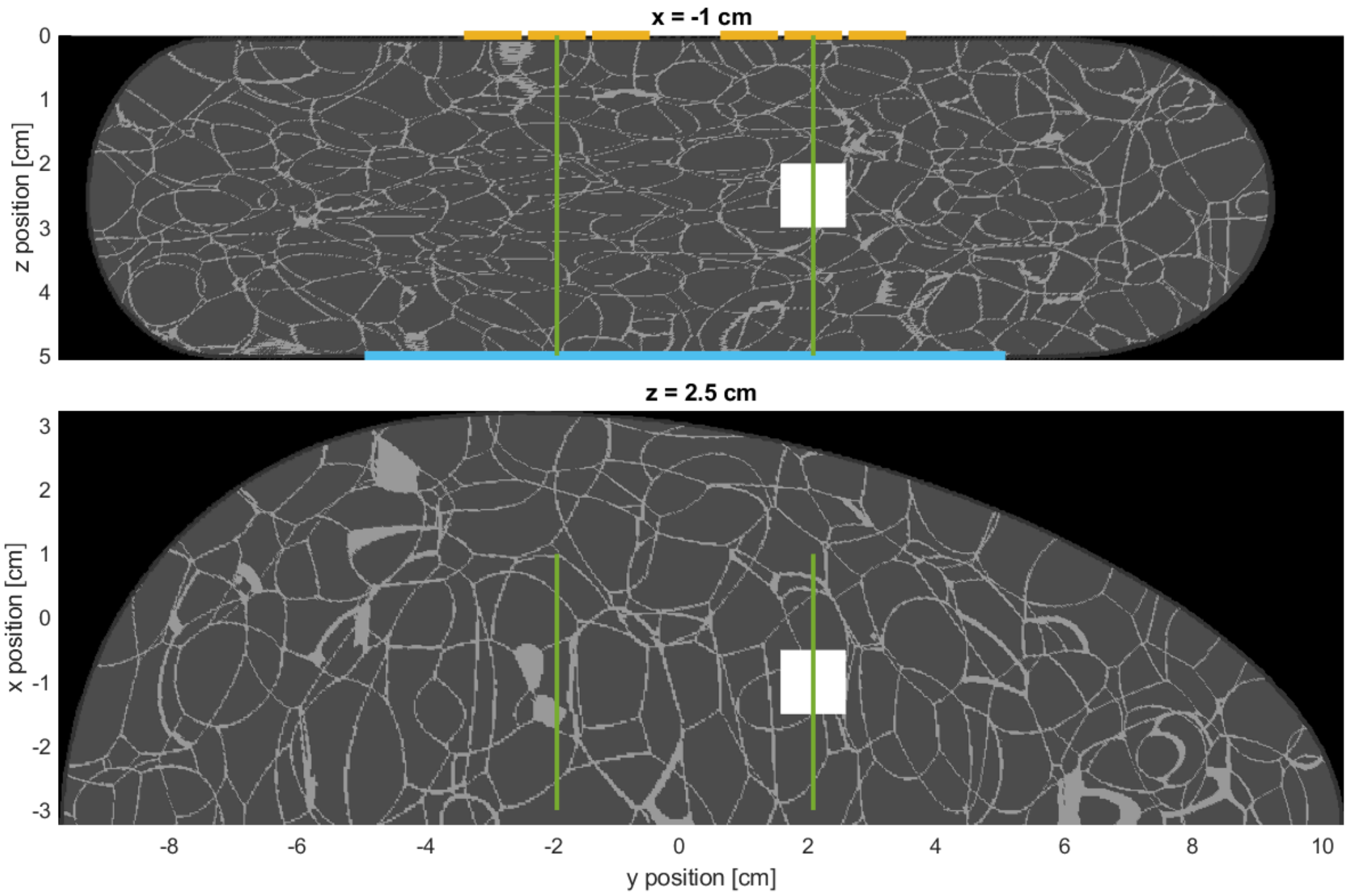

ents, $\mu_a$,

| Tumor |
|---|
| 11.1 |
| 46 → 0.092 |

*Figure 2: Cross sections through the computational breast tissue phantom showing adipose (dark gray), fibroglandular (light gray) and tumor (white). UOT simulations are performed in two locations (green lines), one over a healthy region and one over the tumor region. For each location, light is emitted from one of three 1 cm diameter sources (yellow lines in the top graph) and for each source, light is collected by a light guide centered right underneath the light source and with a 4 cm diameter (the blue line in the top graph covers the entire area used for detection).*

focus (2.5 mm size). Furthermore, we assume that the carrier light leaking through the spectral filter is negligible compared to the UOT signal, which is reasonable since the fraction of frequency-shifted light is usually in the order of 1% [12] and spectral filters can have attenuations much greater than 20 dB [13, 10]. Simulations of two-dimensional UOT images are performed at two locations, one over a healthy region and one over the tumor region, see Fig. 2, so that the difference in the two UOT signals can be studied.

# RESULTS

Typical UOT signals from healthy and tumor regions can be seen in Fig. 3(a)-(b). The largest variation in signal occur due to the difference in photon flux throughout the tissue, e.g., since all photons are emitted from the source that region has a larger UOT signal compared to regions in the center of the phantom where photons are less likely to visit a particular region of the same volume. However, by subtracting the healthy region signal from the tumor region signal one can detect the tumor due to its higher absorption, see Fig. 3(c). This can be compared to Fig. 3(d), where the difference between the same two regions is shown, except there is no tumor.

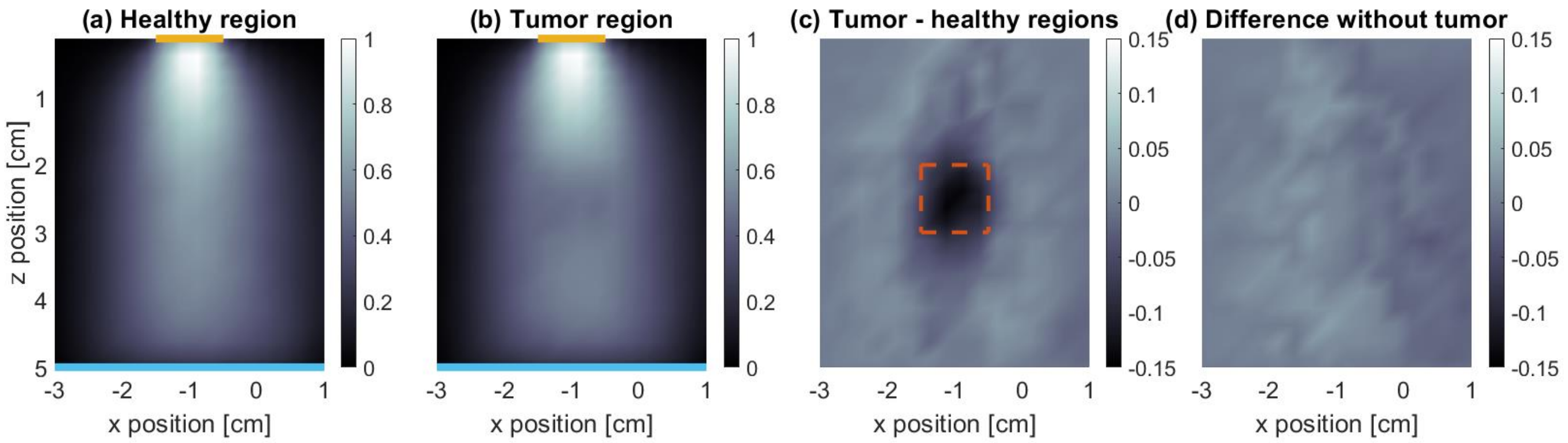

*Figure 3: UOT simulations performed over the (a) healthy and (b) tumor regions shown in Fig. 2. The positions of the sources (yellow lines) and detectors (blue lines) are also outlined. The signals have been normalized to a maximum value of 1. (c) The difference in signal with the tumor region marked (dashed red box). (d) The difference between the two regions when no tumor is present. The optical properties used are shown in Table 1, with a tumor absorption coefficient of 0.092 cm$^{-1}$.*

We define the contrast-to-noise ratio (CNR) as

$$\mathrm{CNR} = -\frac{\Delta S - \mu_{noise}}{\sigma_{noise}}$$

where $\Delta S$ is the average UOT signal difference in the 1x1 cm$^2$ tumor region, and $\mu_{noise}$ and $\sigma_{noise}$ are the mean and standard deviation of the average UOT signal difference in the 14 other non-overlapping 1x1 cm$^2$ regions covering $x = -2.5 \rightarrow 0.5$ cm and $z = 0 \rightarrow 5$ cm.

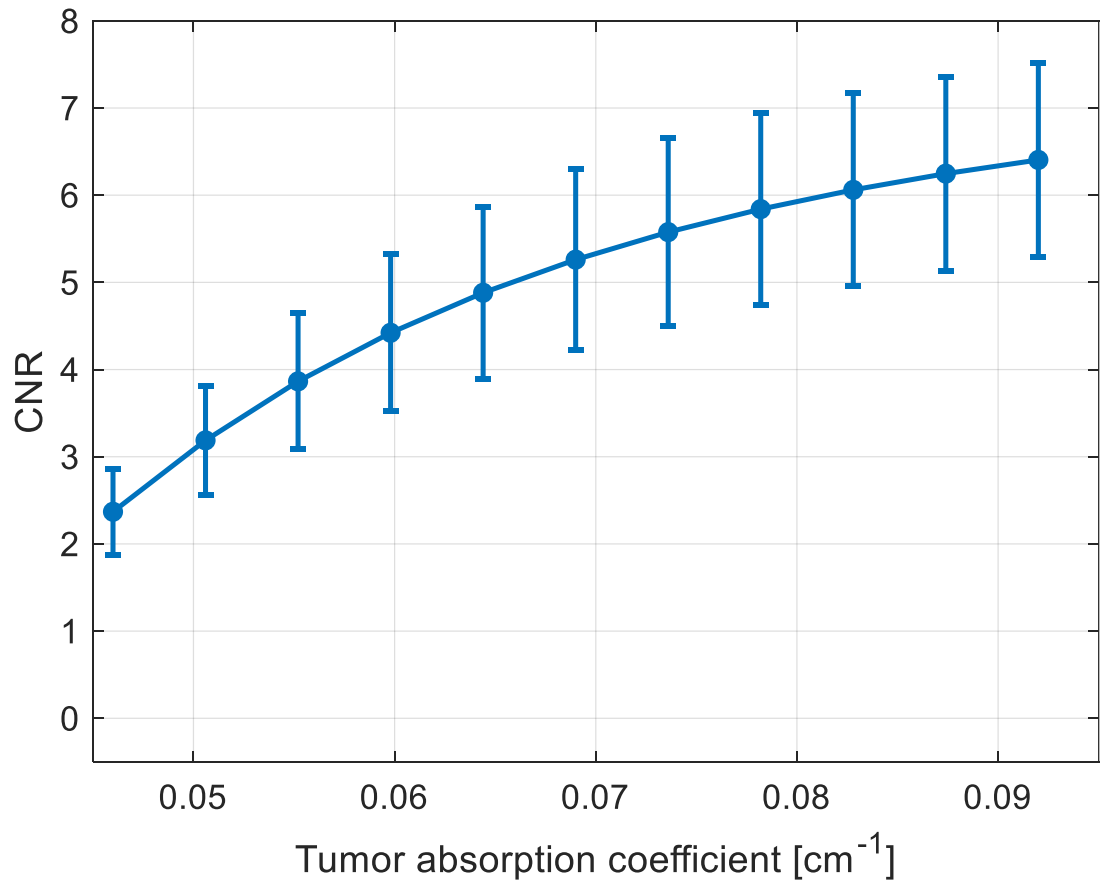

*Figure 4: Average CNR for the six tumor positions as a function of the tumor absorption coefficient. The error bars show one standard deviation.*

We perform six simulations where we vary the tumor position within the breast phantom shown in Fig. 2 and study the average CNR as a function of the tumor absorption coefficient, see Fig. 4.

## DISCUSSION

The present study is important as it sets a new standard for computationally investigating UOT by using realistic breast tissue phantoms. Furthermore, as can be seen from Figs. 3 and 4, it is clearly possible to distinguish the tumors from the surrounding tissue deep inside the breast, despite the natural variation of adipose and glandular tissue compartments. Even when tumors have the same absorption coefficient as glandular tissue (0.046 $cm^{-1}$) an average CNR above 2 is obtained. This is partly due to the higher scattering of tumors compared to the other tissue types, see Table 1, but the foremost reason is that both tumor and glandular tissues are more absorbing than adipose tissue, and a concentrated volume of either of those tissue types can therefore be distinguished from the surrounding tissue which is mostly adipose.

One open question is how well the photon flux compensation can be done in a clinical setting. For example, how reliably can one perform measurements at two different locations, as shown in Fig. 3(a)-(b), and how robust is their difference, shown in Fig. 3(c), with respect to potential clinical errors such as the positioning of the source, light guide, and ultrasound transducer, differences in breast thickness, or fluctuations in the laser power and the ultrasound pressure.

## CONCLUSION

This study improves the methodology for how to computationally investigate UOT by using realistic breast tissue phantoms. Furthermore, it indicates that UOT is capable of distinguishing tumors from the surrounding healthy tissue. In a similar follow-up work, accepted for the SPIE Photonics Europe 2024 conference, we have made further improvements by including more tissue phantoms with varying breast tissue composition and focused on the harder problem of distinguishing malignant tumors from benign glandular tissue, which has optical properties similar to that of tumors.

## ACKNOWLEDGEMENTS AND CONFLICT OF INTEREST

This research was supported by the Swedish Research Council (2019-04949, 2022-01140), the Knut and Alice Wallenberg Foundation (2016.0081), the Wallenberg Center for Quantum Technology (WACQT) funded by The Knut and Alice Wallenberg Foundation (2021.0009), the Wallenberg Launch Pad (WALP) the innovation program of Knut and Alice Wallenberg Foundation (2020.0294), Vinnova (2022-02853), and Mats Paulsson's Foundation (*Noninvasive optical biopsy for breast tumour characterization,* with PI Sophia Zackrisson).

# Modeling different body composition within a computational phantom library

Cornelio Salinas[1], Kirti Magudia MD, PhD[2], Ehsan Samei, PhD[1,2], and W. Paul Segars, PhD[1,2]
[1]Center for Virtual Imaging Trials, Duke University, [2]Department of Radiology, Duke University Medical Center

## BACKGROUND AND PURPOSE

Virtual anatomical phantoms have seen widespread use in medical imaging research and radiation dosimetry. These phantoms have the benefit of having a defined ground truth that can be used for comparison among the different imaging systems.[1] Due to their versatility, researchers have developed ways to improve the accuracy and detail of virtual phantoms. Current phantom designs include 3D clinically acquired patient scans, phantoms with organ motion, and modified standard body models based on references.[2] These phantom designs have shown promising results, yet there is still limited representation available for modeling the variability of muscle and fat content within the body, an increasingly studied facet of human anatomy known as body composition with wide-ranging clinical significance.[3,4]

Patient body composition greatly affects image quality as the interaction of ionizing and non-ionizing radiation is proportional with the amount and density of tissue they pass through before reaching the detector array of the imaging device.[5] Finding the exact parameters to offset this problem is challenging due to inter-patient and inter-device variability. Often, the approach is to tackle the problem by referring to the decision-making capabilities of experienced radiology technologists on an exam by exam basis.[6] Thus, a body composition-adjustable phantom could prove vital to improving current methodologies. Current methods to model muscle and fat variability can range from manual adding and scaling to predicting body mass adjusted surface layers.[7] Another method involves a deformation algorithm that translates each vertex point of the polygon models along the normal direction of the relevant face. The vertices stop once they are within a certain distance to the closest vertex. This method was used to develop phantoms of different weight distributions.[8] Though providing some variation, these approaches are suboptimal due to manipulating generic phantoms or using a solely geometry-based deformation approach.

Therefore, there is a clear unmet need for a more realistic way to model muscle and fat to represent the known spectrum of patient body composition[9] to more optimally leverage virtual phantoms. The purpose of this study is to model changes of the skeletal muscle, visceral fat, and subcutaneous fat within the XCAT patient phantom library using a physics-based approach to represent the range of body composition found in the overall population.

## METHODS

### Patient-Specific Phantoms

The initial phantoms used for this study were selected from the XCAT patient library. The phantoms were based on CT data obtained from the Duke University hospital, selected based on their good scan quality and representation of two body composition profiles: 1) relatively high skeletal muscle and low proportion of visceral and subcutaneous fat, and 2) relatively low skeletal muscle and high proportion of visceral and subcutaneous fat. These represent relative extremes of body composition encountered commonly in clinical practice by radiologists. A deep-learning segmentation algorithm was used to extract 117 total organs and structures[10]. These were saved into a label map as shown in Figure 1.

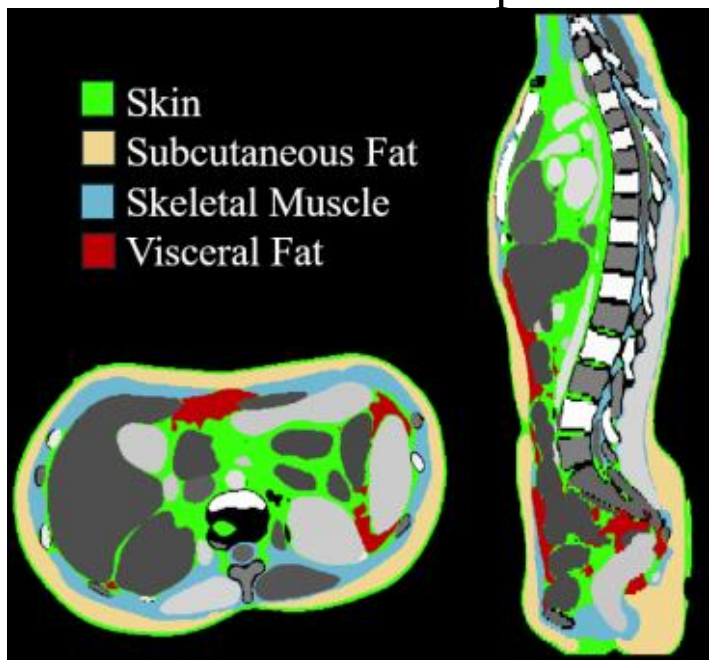

Figure 1. Label map volume with the four main structures colored and the rest in grayscale as viewed in 3DSlicer[11].

Polygon mesh surfaces were fit to each structure using a marching cubes algorithm (implemented through vtkDiscreteMarchingCubes[12]). The meshes were cleaned by removing defects and correcting face directions. The resulting meshes could have up to 1.8 million vertices. Owing to limited computational resources for the physics-based simulations, mesh size was reduced by 99% with the intent of modeling the displacement field and transforming the original resolution phantom.

**Physical dynamics using a live physics engine**

To modify the meshes in a realistic manner, the vertices and edges of the mesh faces need to be moved by changing their positions in space based on certain principles. One popular approach is using force-based dynamic simulations, where elements have force and velocity properties that are updated over time via integration methods until they reach certain conditions (e.g. sum of forces equal to zero). [13] Another approach that has seen more generalized and efficient results is the projective-based dynamic simulations, where instead of solving a different complex linear system in each time step, the calculations are decoupled into a local step (projecting an element onto a specific constraint) and a global step (combining the local steps and finding a compromise between them). For this study, the live physics engine Kangaroo2, which employs a projective-based approach, was chosen because of its capability to create intuitive and realistic deformations in an efficient manner.

**Kangaroo2**

Kangaroo2 is a physics plugin for the visual programming language Grasshopper that can be accessed through the digital modeling software, Rhinoceros3D, www.rhino3d.com. Kangaroo2 is primarily used as a form-finding tool for architectural design, such as in the development of tensile membranes, timber shells, and inflatable structures.[14] The engine's solver uses a modified form of the dynamic relaxation algorithm.[15] Initially, the structures along with the loads and constraints of the system termed as 'goals' by Piker are not in equilibrium due to having opposing rest states. The structures then start to oscillate, and the new position and forces are calculated to reduce the total energy, set as half of the squared distance from their rest states and multiplied by a weighting. This is an iterative process that stops if the energy is below a certain threshold or after several time steps.

**Simulation Parameters**

To set up the phantom modeling system, all 117 meshes were imported into Rhino3D. To optimize solver calculations, each mesh was converted into an isotropic triangular mesh. Structures were placed into 5 groups according to their unique interaction goals (Figure 2): subcutaneous, muscle, rigid, deformable, and visceral. To grow the specific group, a "pressure" goal is added wherein the vertices are subjected to a force in the normal direction that is proportional to the area of the mesh face multiplied by a weighting factor. A "stiffness" goal was then added that resists the mesh edges from stretching past their rest length when the vertices are moved by the forces. The weighting is changed depending on the group, with the rigid group having the highest weighting and the fat having the lowest. For collisions, a soft body collision goal was used wherein the mesh vertices are pulled away by a reaction force if they are checked to be in another mesh. For all goals, some velocity was conserved after every iteration. To keep calculation time low, 200 iterations were used, as this was found to produce satisfactory results.

Different parameters were setup for each phantom case to demonstrate increasing or decreasing the content of muscle or fat. To model visceral growth, muscle stiffness was reduced to allow fat to protrude outward instead of over compressing organs. For muscle growth, the goal for maintaining the subcutaneous fat volume was relaxed to let the fat form over the muscle. The result for each case was the deformed low resolution meshes of each phantom.

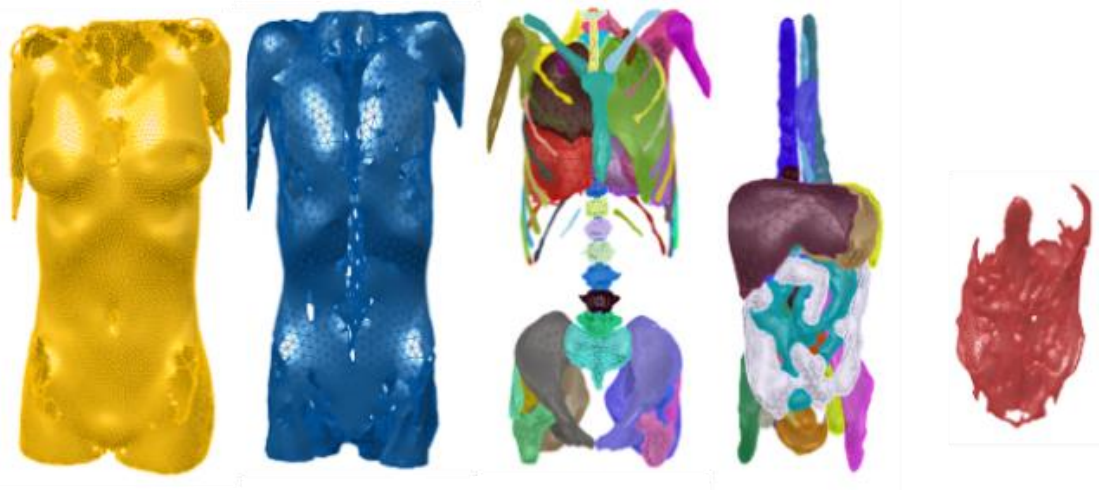

Figure 2. Patient specific phantom mesh groups (L-R: subcutaneous, muscle, rigid, deformable, and visceral). Female example is shown.

**Calculation of the deformation field**

To apply the changes to the original, high-resolution phantom, we used image registration to calculate the transform. The original low resolution meshes were voxelized into a pre-transform image while the deformed meshes were voxelized into a post-transform image. The displacement field transform was calculated by registering the pre-transform and post-transform images using a diffeomorphic registration algorithm.[16] The transform was applied on the original phantom image to model the relatively small deformations (organs and bones) and the large deformations (such as fat or skin) were added after. A priority queue function was used to ensure that important structures such as the digestive tract or the heart were maintained, and no overlapping occurred.

# RESULTS

The reduced meshes processed from the CT phantom are presented in Figure 2. Multiple holes are present in some meshes due to the thinness of the segmentation. Holes are also found in the muscles due to the segmentation algorithm not including the fascial strutures, thin layers of connective tissue that surround and hold muscle in place as well as separating the abdominal cavity into different anatomical spaces. For the physics simulations, soft organs like the heart chambers and lung lobes were placed in the rigid group to keep their original shape and relative location consistent.

Figure 3 highlights the workflow to calculate and apply the deformation field from the physics-based simulation. The displacement field consists of the changes to the moving image without the fat and muscle. Registration attempts that included the fat and muscle resulted in the structures overlapping near their boundaries. This defect would then propagate as the fat expands. To circumvent this, the transformed fat and muscle were instead combined as a mask and added to the original. This workflow takes an hour to complete, with most of the time spent on the simulations.

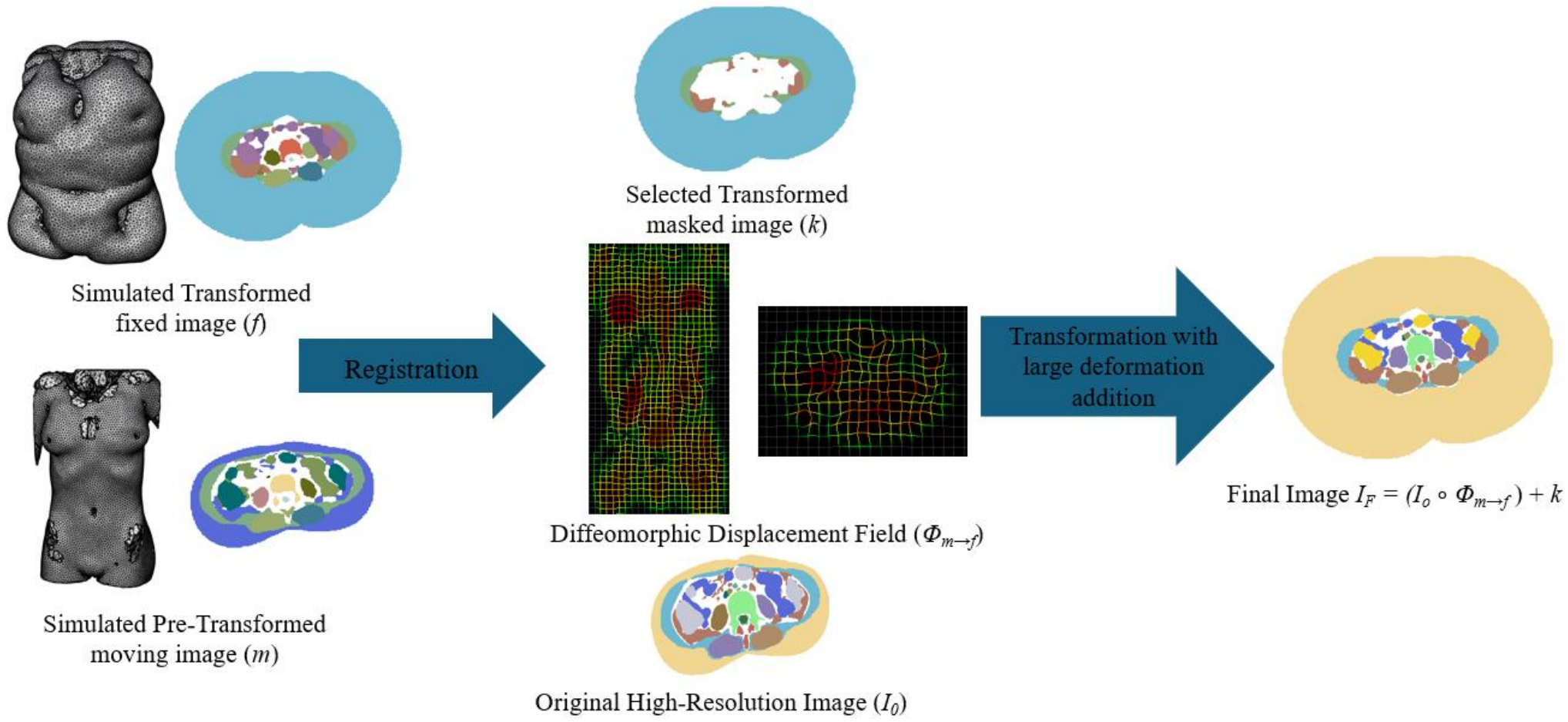

Figure 3 High resolution final image workflow.

Figure 4 highlights four different examples changing muscle and fat content using Kangaroo2. In each of the cases, organs local to the growth were deformed but were within ±2% of the original volume. The visceral growth example produced a phantom with a large protrusion in the abdomen, where most of the initial visceral fat accumulated. The subcutaneous growth example grew fat in all directions with emphasis on the back side, where considerable fat was already present. This 'obese' patient can be compared with the right image of an actual obese patient that was used for the fat reduction example. The reduction reduced both the subcutaneous and visceral fat of the phantom. The pressure applied was negative and 'pushed' the fat inward, collapsing, and thereby shrinking the mesh.

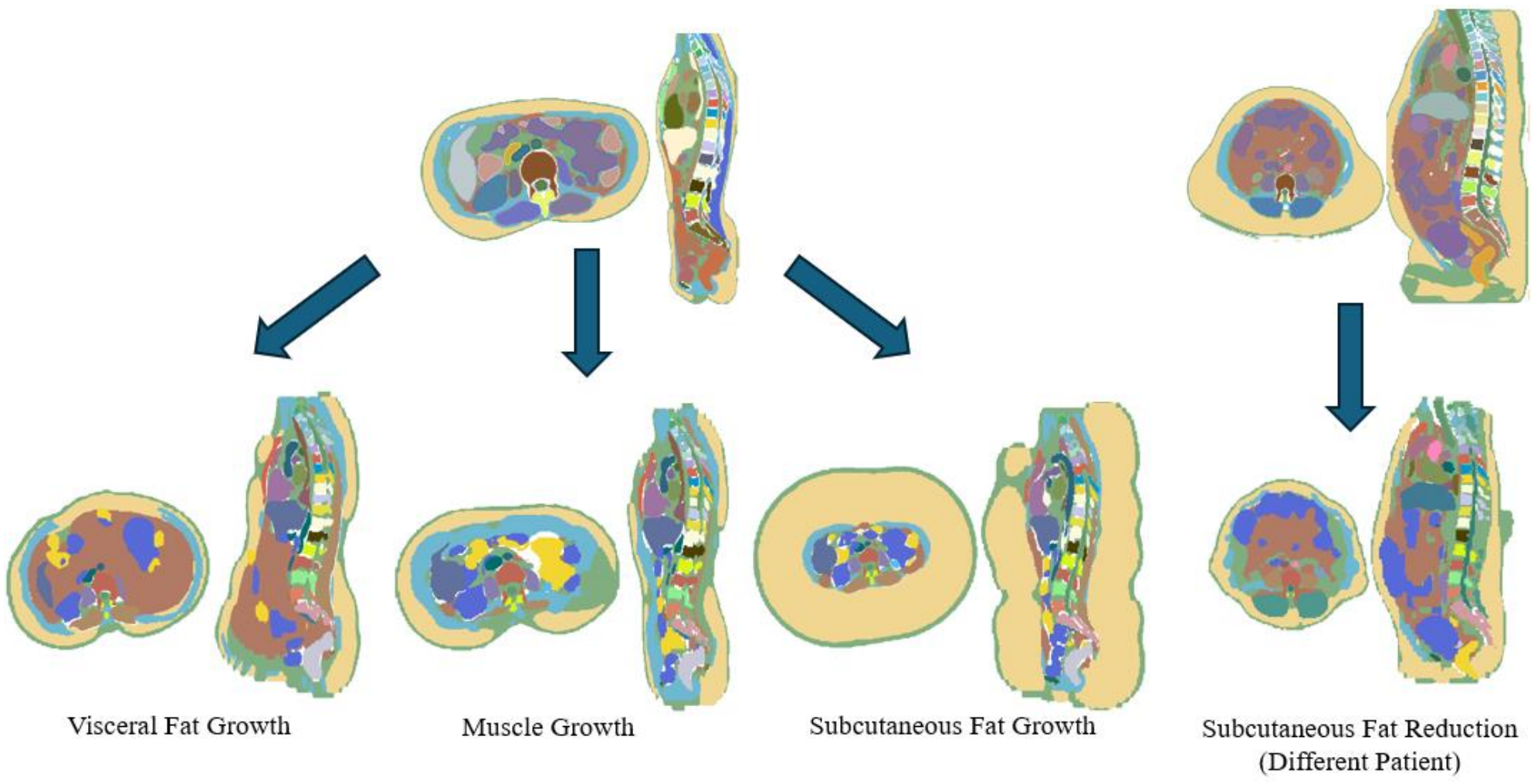

Figure 4. Phantom simulations with different growth goals.

## DISCUSSION

The techniques devised successfully changed the volumes of the different body composition components within the computational phantoms. The physics-based simulations along with the image conversion workflow offer a new way to modify phantoms to model variability in fat and muscle content. The focus of this work has been on the use of the modified phantoms in CT. However, the XCAT phantoms can be combined with imaging simulators modeling different modalities. As such, the modified phantoms can be applied to investigate the effect of body composition for modalities other than CT. The models constructed are also not limited to the extent of growth as shown in Figure 4 as the pressure can be changed easily. Prior to this study, only carefully pre-constructed mesh models were used as a basis for modifications.[2,8] Also, to produce realistic deformations where objects interact with another, one would either require skilled 3D artists or finite element modeling. Both would require a significant amount of resources and time. Using the modified dynamic relaxation method, models can now be constructed quicker while also creating an anatomically plausible phantom.

The simulations are highly dependent on the quality of the segmentation label map. The lack of segmented fascial layers may cause simulated parts to separate significantly, which reflects the significant role fascia plays in the anatomy of the human body. Skin was unusable for the simulation as it included all other parts that the segmentation did not assign a label to. This causes the computation time to increase dramatically. To overcome this, a shrink-wrap algorithm was used to recreate the skin after the simulation was completed.[17] The results of the simulation may produce self-intersecting and thin meshes and these must be corrected before moving onto the image workflow. Future areas of improvement are thus found in three key areas. Inclusion of more structures in the segmentation algorithm will create more realistic simulations. The growth mechanism can be expanded upon to include physiological processes such as cell growth instead of relying just on pressure. More robust registration methods are desirable to represent large deformations with sliding motions across object boundaries.

## CONCLUSION

A physics-based simulation set-up was developed to modify lower resolution phantom meshes. The deformed meshes were then used to calculate a displacement field to apply to the original high resolution meshes of the phantom. By successfully developing these techniques, one can expand on existing phantom libraries with phantoms that simulate the range of body composition known to exist in the overall population. The construction of modified patient specific models was made quicker and less demanding. Overall, these methods will provide a vital tool to expand phantom development and imaging research.

## ACKNOWLEDGEMENTS AND CONFLICT OF INTEREST

This work was supported in part by grants from the NIH (P41EB028744 and RO1EB001838).

# Harnessing the power of high-performance computing for virtual imaging trials

Authors: Rajagopal, Aboulbanine, Wang, Abadi, Segars, Kapadia

Purpose: Virtual imaging trials (VIT) require generating massive datasets that comprehensively cover the variations in clinical populations and image acquisition conditions. Such data generation requires extensive computational time and resources, which are often prohibitive in conventional computing settings. This challenge can be overcome using high-performance computing (HPC) techniques and advanced computational resources; however, doing so requires the problem to be optimized to use such resources. In this work, we used the HPC capabilities at Oak Ridge National Laboratories (ORNL) to perform a VIT on lung CT and evaluate the impact on data generation.

Methods: A validated CT simulation platform (DukeSim) was integrated with ORNL's HPC system. DukeSim uses a combination of ray tracing (RTCat) to estimate primary signal, and GPU-accelerated Monte Carlo (MC-GPU) for scatter and dose estimation. Simulations were performed on an adult computational XCAT chest phantom using a standardized lung scan protocol and run on two high-performance clusters: a standard compute server (CADES, 2 NVIDIA Tesla P100) and a dedicated GPU server (DGX2, 16 NVIDIA Tesla V100). A varied number of GPU resources were used to evaluate the benefit of additional parallelization on data generation.

Results: We benchmarked DukeSim on the standard and dedicated GPU compute systems as follows: MC-GPU: 19-45 mins (both single GPU). RTcat 121-240 min (single GPU) and 61-118 min (two GPUs); 18 min (8 GPUs). Projecting these benchmarks forward, we estimated that a virtual imaging trial consisting of 200 simulations would require approximately 143 computer hours with 8 GPUs. We are currently working on evaluating benchmarks for simulations on Frontier, the world's first exascale system. We found that parallelization of RTCat had a strong effect on reducing run time.

Conclusion: HPC methods at ORNL have the potential to boost VIT data generation by 5-10X using standalone GPU systems, and substantially higher on Frontier (Results TBD).

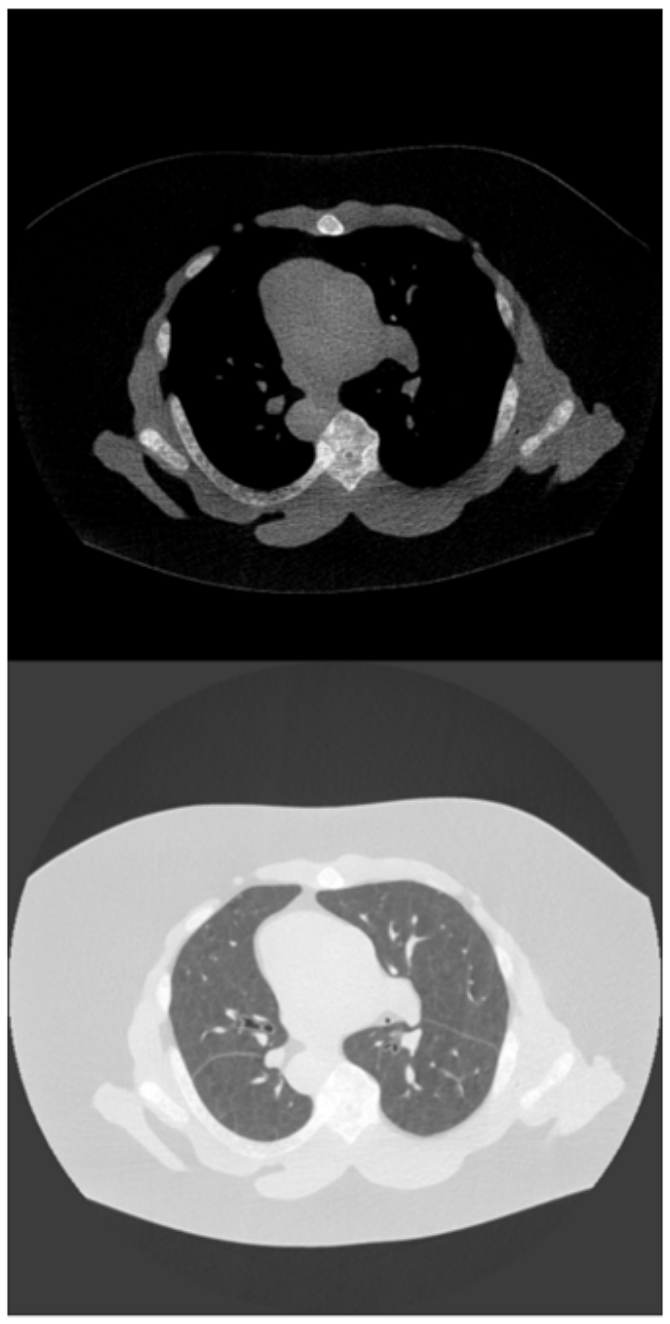

(a)

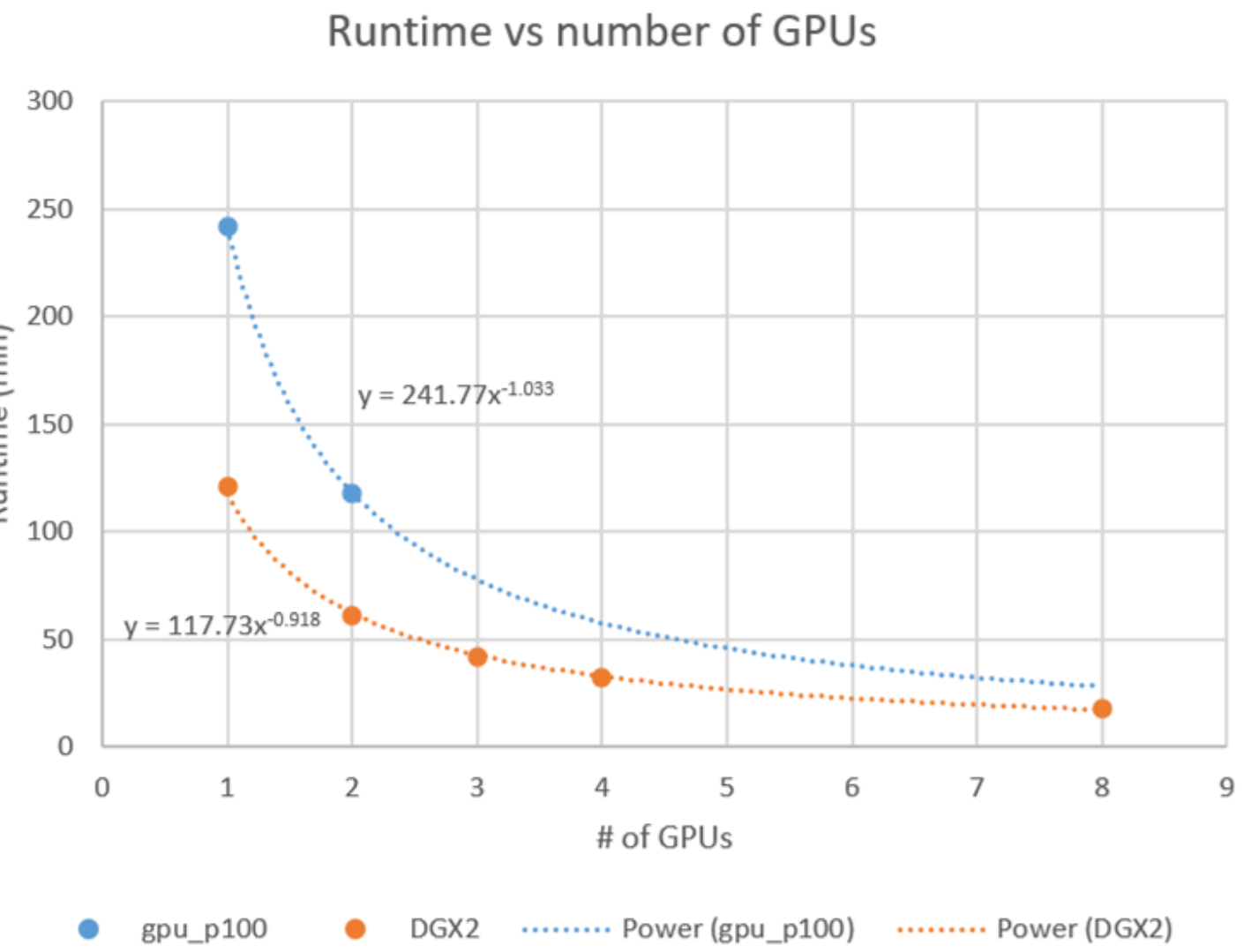

(b)

(a) Reconstructed example of the phantom simulated in this study. Images shown with mediastinum (Top, W/L – 350/50) and lung (Bottom, W/L – 1500/-600)
(b) Plot of runtime (min, y-axis) vs number of GPUs (x-axis) for both CADES (blue) and DGX2 (orange) clusters. Power regression estimates shown with dashed lines

# ISIT-DaT: An in silico imaging trial to objectively evaluate the performance of a scatter-window projection and deep learning-based transmission-less attenuation compensation method for dopamine transporter SPECT

Zitong Yu (Washington University in St. Louis), Md Ashequr Rahman (Washington University in St. Louis), Chunwei Ying (Washington University in St Louis), Hongyu An (Washington University, St. Louis), Abhinav Jha (Washington University in St. Louis)

## PURPOSE

Attenuation compensation (AC) is beneficial for the precise quantification in dopamine transporter single-photon emission computed tomography (DaT SPECT). Conventional AC methods typically rely on a separate computed tomography (CT) scan, leading to higher radiation exposure and costs, and the potential of quantification errors due to misalignment between the SPECT and CT scans (1-3). Furthermore, many SPECT systems do not have a CT capability (4). In this context, a range of transmission-less (Tx-less) AC methods have been proposed (5-8), including these based on deep learning (DL) (9-16). In this study, we conducted an in silico imaging trial, namely ISIT-DaT, to evaluate a scatter-window projection and DL-based Tx-less AC method for DaT SPECT (DLAC).

## METHODS

The primary objective of ISIT-DaT is to compare the DLAC method, a CT-based AC method (CTAC), and a uniform-attenuation-map-based AC method (UAC) on quantification of the DaT uptake within the caudate, putamen, and GP regions by calculating the intraclass correlation coefficient (ICC). The secondary objective is to assess the performance on fidelity-based figures of merit (FoMs). In ISIT-DaT, we calculated the sample size following the strategy proposed by Walter et al. (17) We that a sample size of N = 35 provides 80% power to detect a difference of 0.15 between the ICC under the baseline of 0.65 and an ICC under the proposed method of 0.85 at a significance level of 0.05. Considering a ~20% drop-out rate, we collected images from N = 45 patients who underwent 3T T1-weighted MRI and CT studies. From these images, we generated a clinically relevant trial population, as details shown in Fig. 1. We simulated a realistic imaging protocol with a GE Discovery 670 SPECT system and obtained the emission data in photopeak and scatter windows. For the primary objective, we calculated the ICC between the DaT uptake within the caudate, putamen, and GP regions estimated by DLAC and those estimated by the CTAC method. For the secondary objective, we evaluated the fidelity-based performance of DLAC using the normalized root mean-square-error (RMSE) and the structure similarity index measurement (SSIM). DLAC was pre-trained on N = 175 samples.

# RESULTS

The trial accrual in ISIT-DaT is shown in Table 1. In Fig. 2a, we observed that DLAC performed similarly to CTAC (ICC = 0.85, confidence intervals: [0.75, 0.86]) and significantly outperformed the UAC on regional DaT uptake quantification. In addition, we observed that the estimated attenuation and activity maps obtained by DLAC yielded a significantly lower normalized RMSE and a higher SSIM than those obtained by UAC (Fig. 2b). Fig. 3 shows some representative examples.

# CONCLUSION

Primary results from ISIT-DaT demonstrated that a scatter-window projection and DL-based Tx-less AC method yielded similar performance to a CT-based AC method and outperformed a uniform-attenuation-map-based AC method on the task of regional DaT uptake quantification.

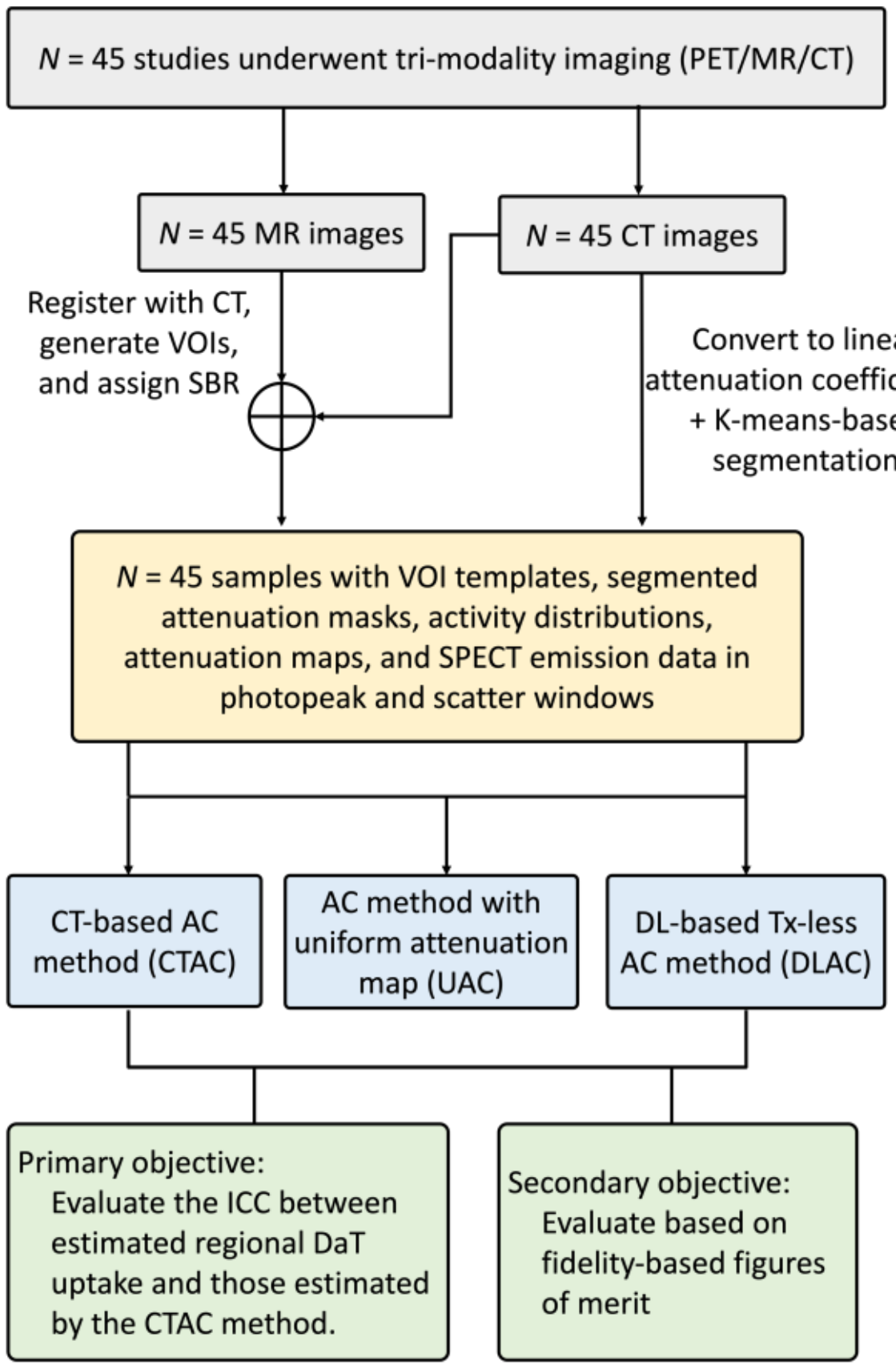

Figure 1. The workflow of trial population generation.

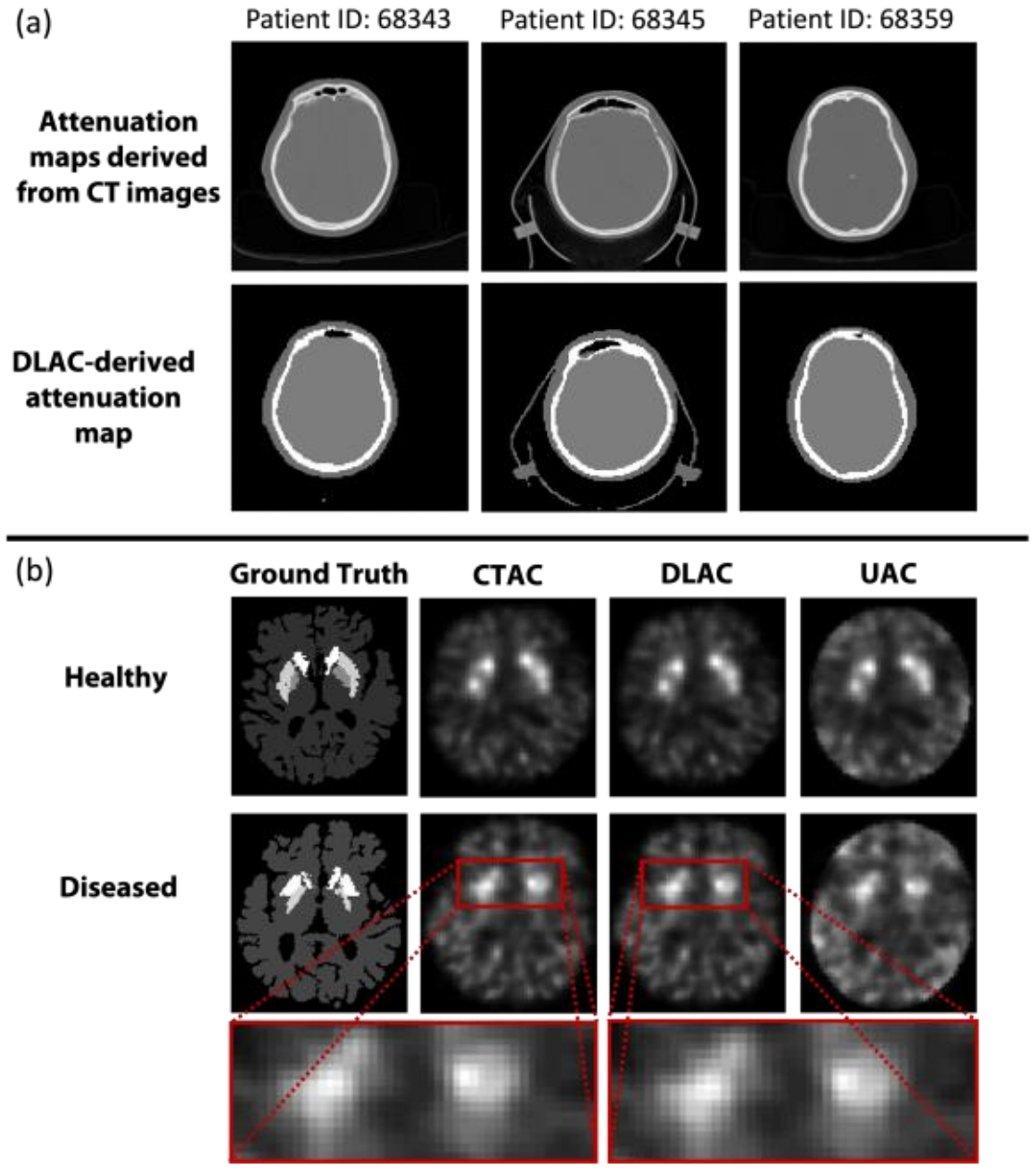

| Characteristics | Summary* |
|---|---|
| Number of patients | 45 |
| Number of MR images | 45 |
| Number of CT images | 45 |
| Age | 70 (64.25, 75) |
| Sex | 26 females (57.8%) |

*Count (%) for categorial characteristics and median (interquartile range) for age.

Table 1. The characteristics of the patients in ISIT-

(a)

| | ICC | 95% CI* lower limit | 95% CI upper limit |
|---|---|---|---|
| DLAC v.s. CTAC | 0.85 | 0.75 | 0.86 |
| UAC v.s. CTAC | 0.31 | 0.21 | 0.33 |

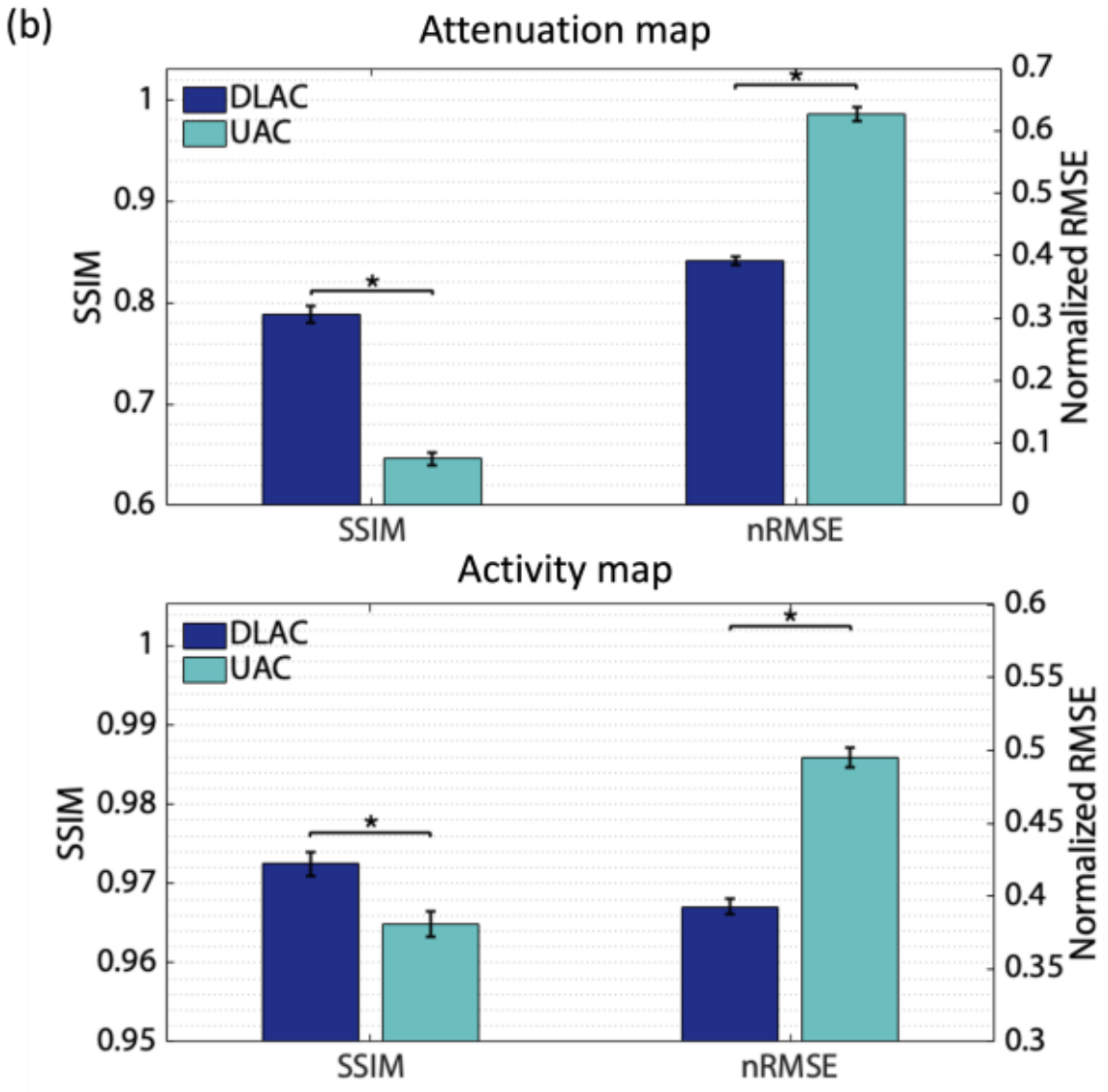

Figure 2. ISIT-DaT evaluation results. (a) The ICCs between the estimated regional DaT uptakes and those estimated by the CTAC method. (b) Fidelity-based performance evaluated on the normalized

Figure 3. Representative examples (a) Attenuation maps of three samples obtained by the DLAC method and derived from CT images. Note that the middle sample had a head holder, which was correctly predicted by DLAC. (b) Ground truth activity maps and activity maps obtained by the CTAC, DLAC, and UAC methods. The upper example was from a healthy sample, while the lower example was from a diseased sample. Zoomed version was given in the red-boxed regions.

# Generation of a representative synthetic phantom dataset for the training of neural networks in personalized CT dosimetry

Marie-Luise Kuhlmann[1], Stefan Pojtinger[1]

[1]Physikalisch-Technische Bundesanstalt Braunschweig, Germany

## BACKGROUND AND PURPOSE

Computed Tomography (CT) plays a crucial role in modern medicine. The currently used computer tomography dose index (CTDI) only provides a rough estimate of the effective dose[1]. Personalized CT dosimetry is therefore the next step in patient risk assessment but is based on time-consuming dose calculations for each individual patient. One appropriate method to increase calculation speed is using neural networks. The training of neural networks requires a huge amount of training data, linking patient geometries (phantoms) to the effective dose. This work investigates whether virtually created phantoms can be used in combination with Monte Carlo simulations to create synthetic training data for this purpose.

A dataset of 200 phantoms was generated based on a given statistic, addressing the need for realistic and representative data for the training of a neural network. As a proof-of-principle, three phantoms from the generated dataset were used to calculate effective dose values using Monte Carlo simulations.

This study showed, that XCAT[2] can be used to generate synthetic training data for the described purpose and that the resulting effective dose values show the expected differences compared to effective dose calculations based on CTDI and size-specific dose estimates (SSDE).

## METHODS

### Phantom generation

For the generation of the phantom data, the commercial software XCAT Version 2[2] was used. To obtain a balanced distribution in terms of body types, organ volume, and gender, the mean values from the ICRP Publication 145[3] were used. Based on these values, a uniform distribution was assumed with limits of ±10 % relative to the mean, and the phantom data was sampled from this distribution. To account for the range of body types, the thickness of the fat layer around the abdomen, thorax, and extremities was varied within ±10 % relative to the mean. 100 male and 100 female phantoms were generated with body heights between 1.5 m and 2.10 m.

For the simulations of effective dose, the three female phantoms (phantom A, B, and C) shown in Figure 1 were chosen from the dataset. The colored areas show the scan regions and the segmented organs. The chosen scan region from the seventh cervical vertebrae to the adrenal gland[4] is typical for a thorax scan.

## Radiation transport simulation

For the Monte Carlo simulation, a newly implemented particle source class was used. The particle source uses an aluminum equivalent bowtie filter representation of a typical clinical CT scanner and an aluminum attenuation-matched fluence spectrum. The used Monte Carlo Software was EGSnrc[5] in the 2020 version.

To use the phantoms in the Monte Carlo Software, the NIFTI files created with XCAT were converted to the egsphant format implemented in EGSnrc. For each tissue ID, the interaction cross sections and the appropriate density were assigned using a Python script.

For the CTDI phantom simulation, the phantom was defined as a PMMA cylinder with a height of 15 cm and a diameter of 32 cm. Four cylindrical holes were placed at the periphery and one in the center of the phantom.

## Calculation of dose values

For calculating the effective dose from the phantom simulations, the absorbed energy in the respective organs was scored ($E_{\text{dep,organ}}$). The organ volumes and media densities were used to calculate the organ mass ($m_{\text{organ}}$) and from this the organ doses. The effective dose was calculated ($E_{\text{eff,organ}}$) by multiplying the organ values with weighting factors $w_{\text{organ}}$ from ICRP Report 116[6].

$$E_{\text{eff,organ}} = \sum_{\text{organ}} {E_{\text{dep,organ}}}/{m_{\text{organ}}} \cdot w_{\text{organ}} \qquad (1)$$

To calculate the effective dose using the CTDI ($E_{\text{eff,calc}}$), the CTDI was multiplied by the scan length ($l$) and the conversion factor tabulated in the European Guideline for quality assurance[7] ($w$).

$$E_{\text{eff,calc}} = CTDI * l * w \qquad (2)$$

The size-specific dose estimate (SSDE) was calculated by multiplying the CTDI value by a patient size-dependent correction factor[8]. The effective dose based on the SSDE was then calculated in the same way as in equation (2), using the SSDE instead of the CTDI.

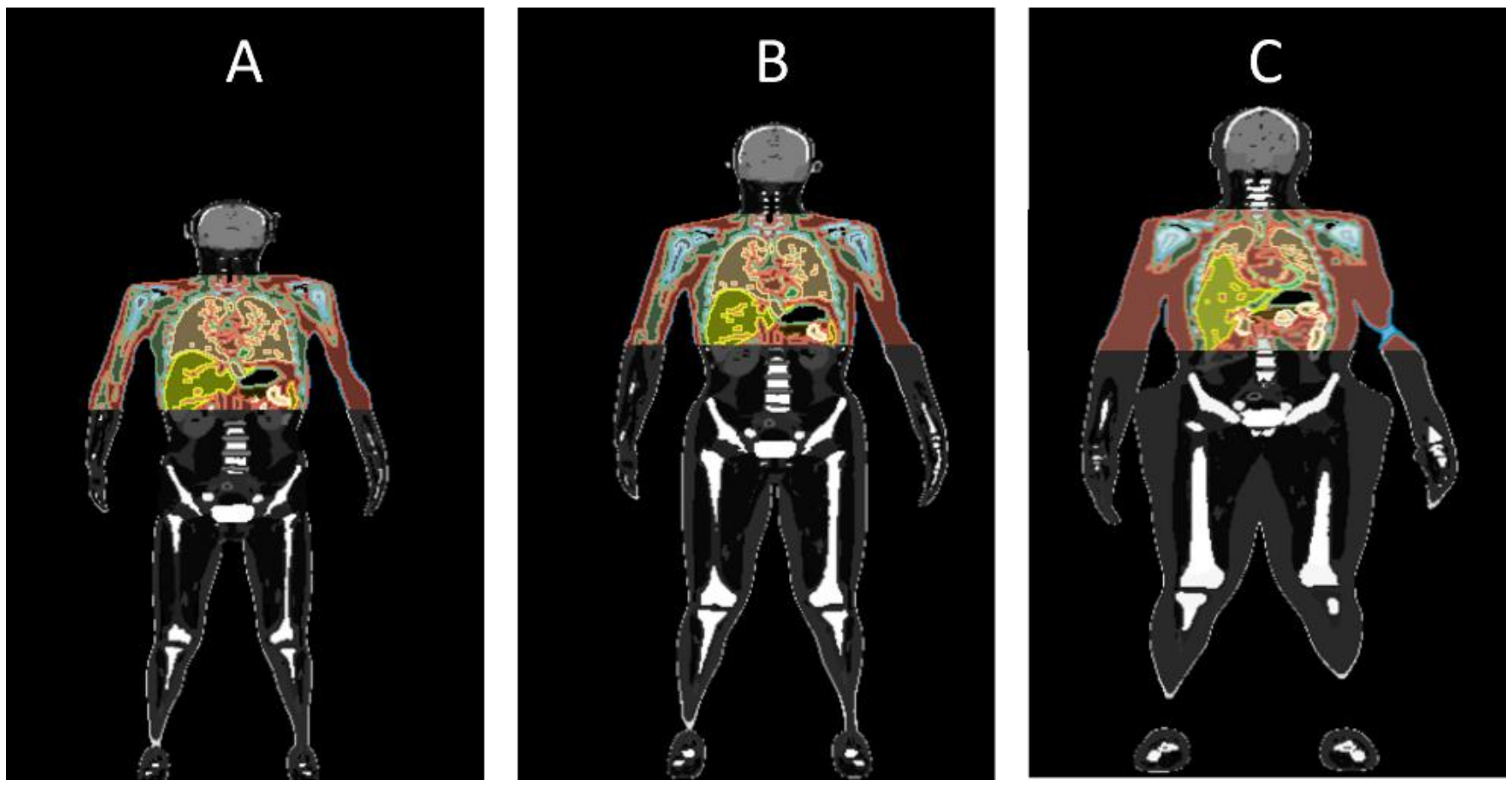

*Figure 1 Three phantoms from the dataset with three different body shapes. The colored areas show the simulated scan region with the segmented organs used to calculate the effective dose.*

## RESULTS

Figure 2 gives an overview of the generated data set.

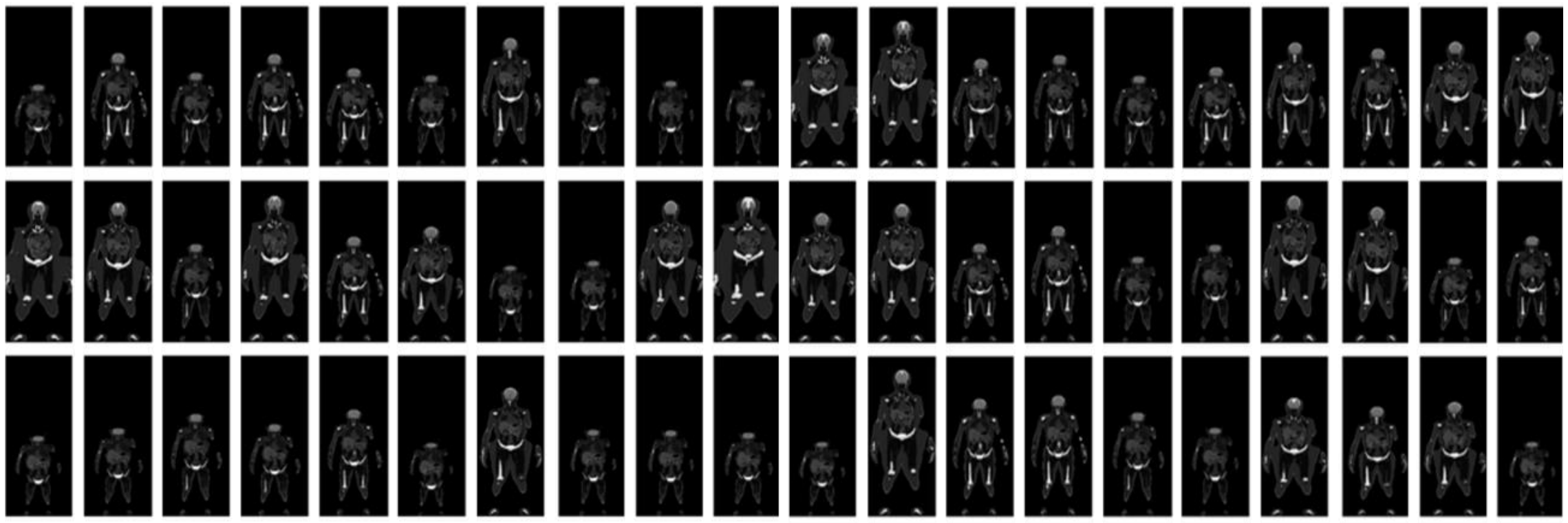

*Figure 2: The graphic shows a selection of phantoms from the generated dataset.*

Figure 3 presents the simulated effective dose values as the ratio ${E_{\mathrm{eff,calc}}}/{E_{\mathrm{eff,organ}}}$ in dependence of $E_{\mathrm{eff,organ}}$ for the three phantoms. The results show a dependence on the patient size. For the small phantom A, the effective dose calculated from CTDI underestimated $E_{\mathrm{eff,organ}}$ by a factor of two. The use of SSDE instead of CTDI results in a deviation of up to 30 %. For phantom B, the effective dose calculated from the CTDI shows a maximum deviation of 41 %. Here, the use of SSDE instead of CTDI leads to an agreement within 6 % with $E_{\mathrm{eff,organ}}$. For phantom C, the effective dose is slightly overestimated for CTDI as well as for SSDE. The largest deviation is seen for the smallest tube voltage.

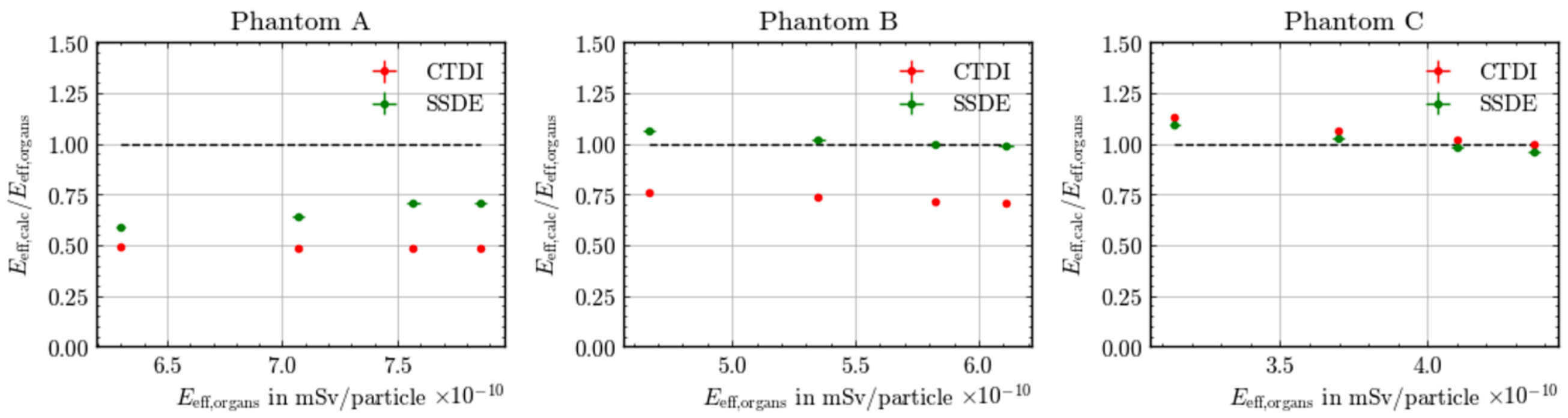

*Figure 3: The graphics show the ratio between $E_{eff,calc}$ from the CTDI and SSDE and the from the organ doses calculated effective dose $E_{eff,organ}$ for the three phantoms.*

# DISCUSSION

The simulated data showed a good agreement with the expected deviations between the SSDE and CTDI-based calculations compared to the calculations based on realistic patient phantoms, as described in the AAPM report 204[8]. For smaller patients, the CTDI-based calculation underestimated the effective dose by a factor of two. For adipose patients, the opposite effect was observed. The same trend can be observed for the simulated data shown in Figure 3.
The use of SSDE led to a better estimation.

# CONCLUSION

In conclusion, this study shows that the generated phantom data based on XCAT can be used for effective dose calculations in EGSnrc. The results show the expected deviations compared to calculations based on CTDI and SSDE. Therefore, this dataset can be used for the generation of synthetic data for the training of neural networks to be used for fast effective dose calculations in the context of personalized CT dosimetry.

# ACKNOWLEDGEMENTS AND CONFLICT OF INTEREST

The authors thank PTB Sec. Q.45 High Performance Computing (HPC) for computing resources.

The authors declare that they have no conflicts of interest.

# Pileup simulation for deep-silicon photon-counting CT

Erik Fredenberg,[a,b,c*] Daniel Collin,[b] Louis Carbonne,[b] Mingye Wu,[d] Fredrik Grönberg[b]

[a] Department of Physics, Royal Institute of Technology (KTH), Stockholm, Sweden
[b] GE HealthCare MICT, Stockholm, Sweden
[c] MedTechLabs, BioClinicum, Karolinska University Hospital, Solna, Sweden
[d] GE HealthCare Technology & Innovation Center, Niskayuna, NY, United States of America

* Corresponding author: fberg@kth.se

## BACKGROUND AND PURPOSE

The next major leap in detector technology for computed tomography (CT) is the introduction of photon-counting detectors. [1] As opposed to conventional integrating x-ray detectors, which detect the number of photons that have interacted in the detector during a period of time, photon-counting detectors register single photons along with their energies. Photon-counting detectors have the potential to significantly improve spectral and spatial resolution compared to current detectors, thereby enhancing, or enabling, a range of clinical applications.

Photon-counting detectors have been used in nuclear medicine for decades and were introduced for planar radiology (particularly mammography) 20 years ago, [2] while photon-counting CT (PCCT) is emerging now. One reason for the relatively slow evolution of photon-counting detectors in medical imaging is pulse pileup; the random staggering of two or more finite pulses, which leads to loss of counts and spectral distortion, and in the end image bias and reduced contrast-to-noise ratio.

Computer simulations and virtual clinical trials play an increasingly important role for evaluating imaging systems, [3] not least for the development of cutting-edge technologies, such as PCCT, where fast turn-around times are crucial. However, pileup is an intrinsic and non-negligible part of the detection process in any PCCT detector; even though the effect is reduced with the speed of the detector, it will never fully go away because of the randomness of photon interactions. [4]

To simulate PCCT over the full spectrum of clinical applications, it is therefore of essence to have an accurate pileup model. We have built a detailed Monte-Carlo-type pileup model for deep-silicon (multi-segmented) photon-counting detectors. The model simulates individual interactions based on the detector pulse shape and dead time. The purpose of this study is to evaluate the accuracy in simulating a real PCCT system with the model, for future use in system evaluation and virtual clinical trials. As a first application, the model is applied to investigate the effectiveness of a pileup correction algorithm.

## METHODS

### The deep-silicon photon-counting detector

GE HealthCare has developed a prototype wide-coverage PCCT system based on silicon detectors. [5] The advantages of silicon as a detector material include high-purity and high-quality elemental crystals with well-established manufacturing processes, high charge-collection efficiency, no K-fluorescence in the relevant energy range, and high intrinsic energy resolution. In addition, silicon is a non-toxic material with minimal impact on the environment. To reach a sufficient photon stopping power in the diagnostic energy range despite the relatively low atomic number of silicon, the silicon wafers are oriented in an edge-on geometry. [1] This sensor arrangement, several centimeters thick, is referred to as a deep-silicon sensor.

When a photon interacts in the detector material, charge is released, which drifts as electron-hole pairs and is collected by electrodes on the sides of the sensors. The drift distance corresponds to the thickness of the sensor (not to the depth) and is therefore short, which minimizes spreading of the charge cloud. The collected charge generates a pulse with a height that is proportional to the energy of the photon, and comparators, implemented in an application-specific integrated circuit (ASIC), sort the detected pulses into energy bins.

The relatively long interaction trajectories in the deep-silicon design enables splitting of the electrodes into depth segments. [1] Since each of these segment is connected to an independent ASIC channel, the count rate per channel, and hence the pileup effect, is substantially reduced. The deep-silicon design therefore alleviates the need to shrink the pixel size below optimal levels; for all photon-counting detectors, too small pixels result in increased pixel cross talk, which causes artefacts and reduced spectral resolution. [6] Nevertheless, pileup will always be present to some degree in photon-counting detectors because, by necessity, the pulse at each photon interaction has a finite length, whereas the photon interactions are Poisson distributed and can occur arbitrarily close in time. [4]

## The semi-paralyzable pileup model

The pileup behavior of real photon-counting detectors, and in particular the deep-silicon detector, has been shown to correspond well to the semi-paralyzable model, [7] which resides between the traditional paralyzable and non-paralyzable detector models. [4] In short, a photon interaction in the detector initiates a dead time during which only one photon can be registered. If two or more photons hit the detector during a deadtime their respective pulses are added, and a single photon is registered with an energy that is higher than the individual photons. However, even though the piled-up pulse is registered as a single event, depending on when during the deadtime the photons hit and the energy of the photons, the tail of the piled-up pulse may extend beyond the deadtime and trigger a new deadtime and a new registered photon with an energy that is generally lower than the individual photons.

The semi-paralyzable detector model offers a way to describe the deterministic and statistical effects of pileup on a general and non-spectral level. [7] The model was used for benchmarking in this study, to understand whether any discrepancies between measurements and simulations are reasonable.

## Deep silicon pileup simulation and correction

To simulate the pileup behavior in the deep-silicon detector, we have built a Monte-Carlo-type pileup model that emulates individual interactions based on the detector pulse shape and dead time. These detector and ASIC design parameters are generally known by the detector manufacturer. The pileup model has been integrated into CatSim, which is a detailed CT and x-ray simulation environment that has been developed at GE HealthCare over the last two decades. [8]

In addition to the pileup effect, an image chain for the deep-silicon system has been implemented in CatSim, including data-driven material-decomposition[9] and pileup-correction[10] algorithms. The material decomposition was calibrated at a reference count rate (determined by the mA and kV of the x-ray tube), and the effect of pileup on image quality was investigated by moving away from the calibration mA.

# RESULTS

Figure 2 shows simulated virtual 65-keV monoenergetic images of a Gammex-like phantom at ± 200 mA relative the calibration mA at which the material decomposition was calibrated. Nonlinearities caused by the pileup effect result in pixel-to-pixel bias, i.e., ring artefacts, and large-area bias (center row), compared to the simulation without pileup (top row). The pileup correction algorithm effectively removes the artefacts (bottom row).

Figure 3 plots observed count rate (the sum of all energy bins) in a detector segment with high count rate and in another segment with lower count rate, as a function of total incident rate per dead time ($\tau$), simulated and measured on a typical deep-silicon photon-counting detector. The high-count-rate segment shows clear signs of pileup, while the low-count-rate segment is virtually linear throughout the range of incident rates. Predictions by the semi-paralyzable detector model are also shown in the plots, where the model parameter $\beta$ has been chosen to best fit the data in a least-square sense.

Figure 4 shows pileup-induced large-area bias in terms of Hounsfield units (HU) measured in the simulated phantom in Figure 2. Severe bias is observed in air but is also evident in the center of the 20-cm-diameter water-equivalent phantom material. The pileup correction algorithm virtually eliminates the bias.

## DISCUSSION AND CONCLUSIONS

The importance of pileup correction for imaging with photon-counting detectors away from the calibration mA is illustrated in Figure 2 and Figure 4; pileup correction linearizes the signal so that material decomposition, which assumes a linear response to mA, works as expected and generates images that are free of artefacts and bias. As illustrated in Figure 3, the depth segments of the deep-silicon detector can be designed to see different count rate levels. Hence, some of the segments may exhibit a close-to linear response up to very high incident count rates and can be used as a guide to improve the pileup correction. [10]

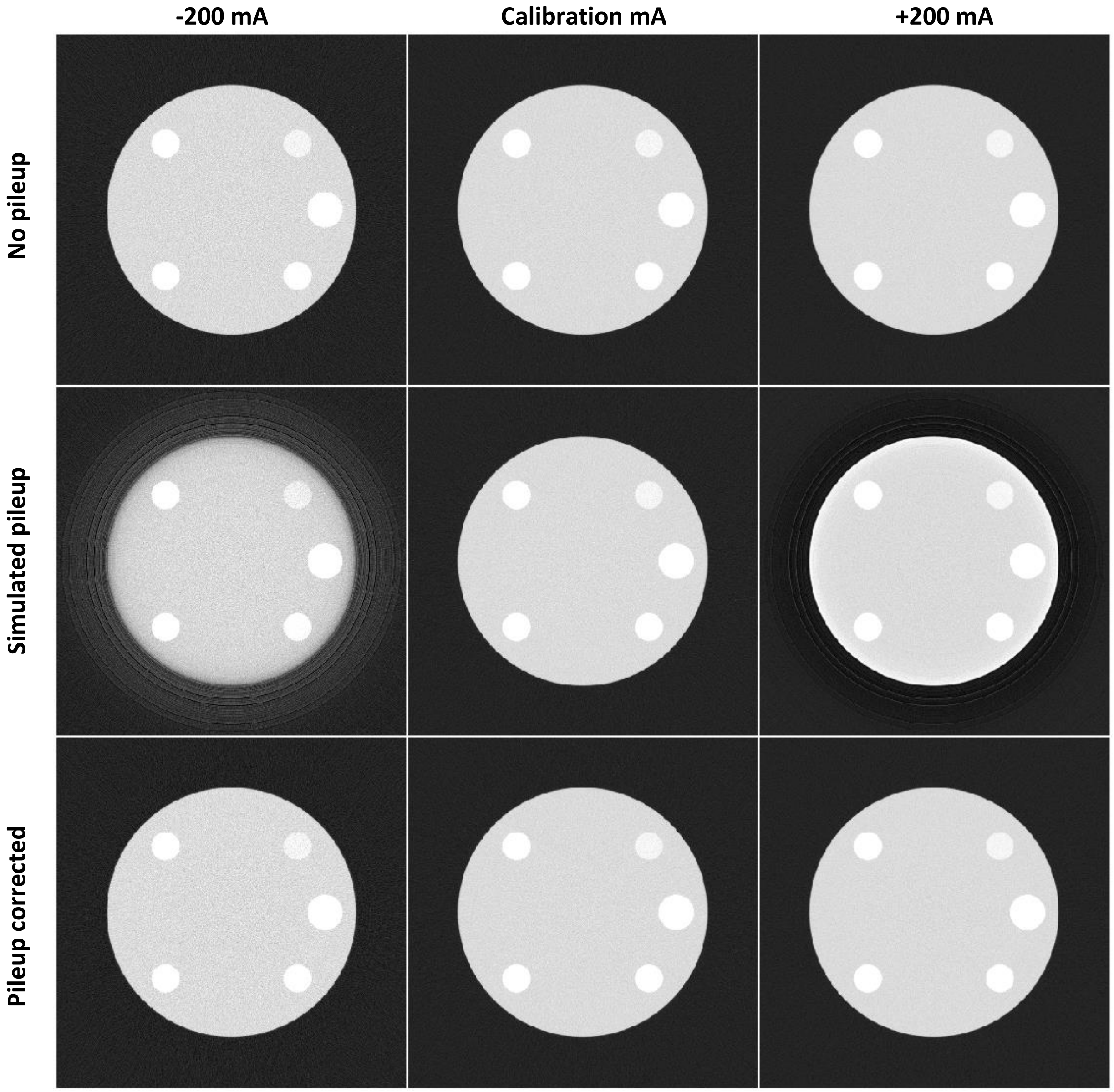

***Figure 2:*** *Simulated virtual 65-keV monoenergetic images of a Gammex-like phantom at ±200 mA relative the mA at which the material decomposition was calibrated. Window Level: -500 HU. Window Width: 1400 HU.*

The discrepancies between simulated and measured data that are seen in Figure 3 are largely captured by a difference in the parameter $\beta$ and can therefore be explained by unknown factors and non-idealities, such as uncertainties in the pulse shape, rather than errors in the model. Based on this, we conclude that the accuracy of the pileup model is high enough to simulate the behavior of typical deep-silicon photon-counting detectors and is suitable for future use in system evaluation and virtual clinical trials. One limitation is, however, that the validation was carried out on non-spectral data, i.e., the sum over energy bins. Future work will continue the validation on the level of energy bins and investigate spectral distortion.

The main advantage of simulating pileup on a per-interaction level is high accuracy and reliability. Simplifications in modelling generally come with constraints that need to be monitored, which adds complexity to the simulation and risk of false interpretations of the data. A major drawback is, however, high computational intensiveness, which leads to low simulation speed, especially at high simulated count rates and large data sets. A priority for future efforts will therefore be efficiency optimization.

Another upcoming focus area will be to expand the model to include other effects that will interact with pileup, such as electronic noise residuals and correlated noise between detector pixels. Due to the high generalizability of the interaction-based model, we expect this integration to be relatively straight forward.

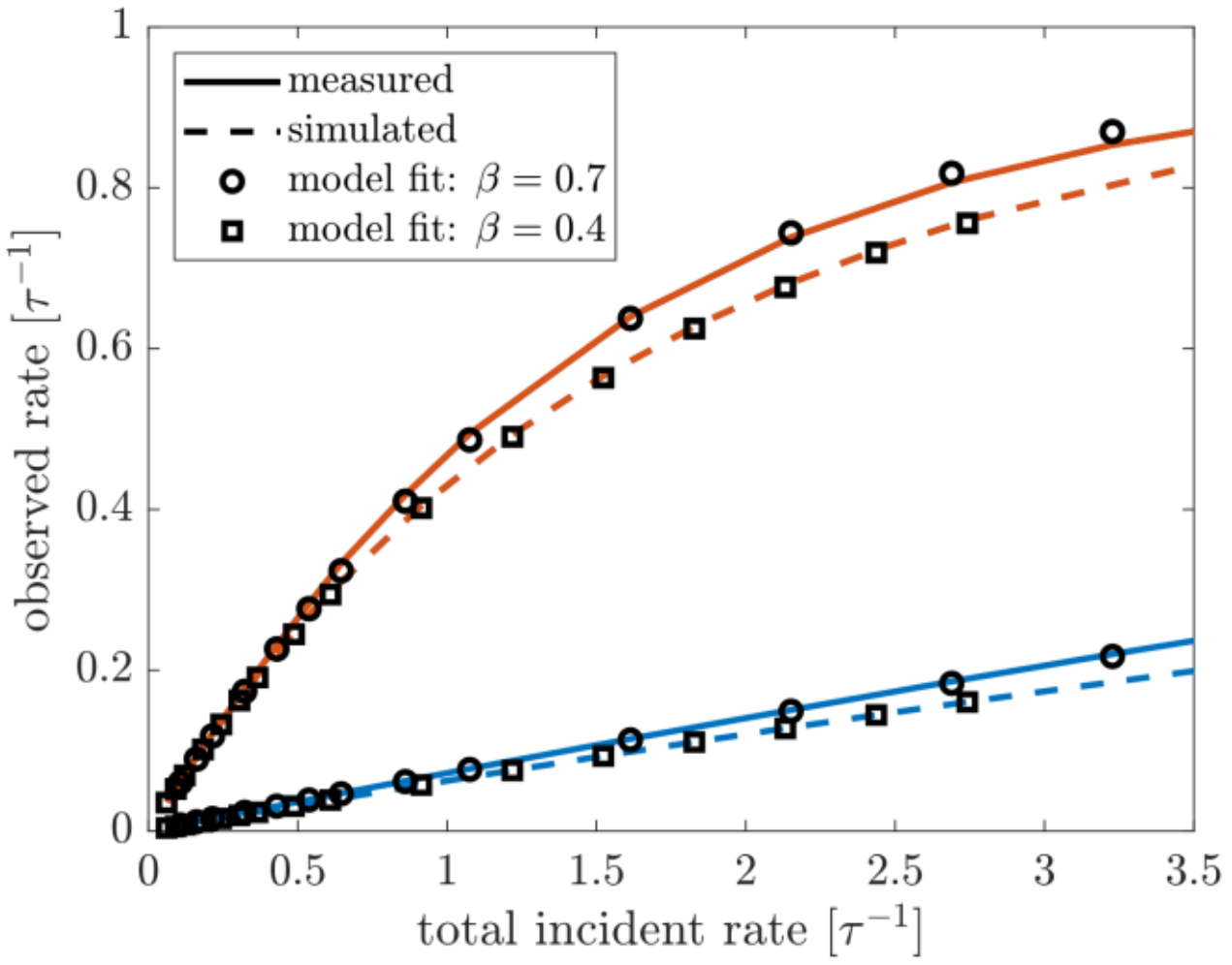

***Figure 3**: Observed count rate (the sum over energy bins) as a function of total incident rate per dead time (τ), simulated (dashed line) and measured (solid line), for two different detector segments (blue and red lines). Markers show results from the semi-paralyzable detector model for the values of β that best predict the measured and simulated data.*

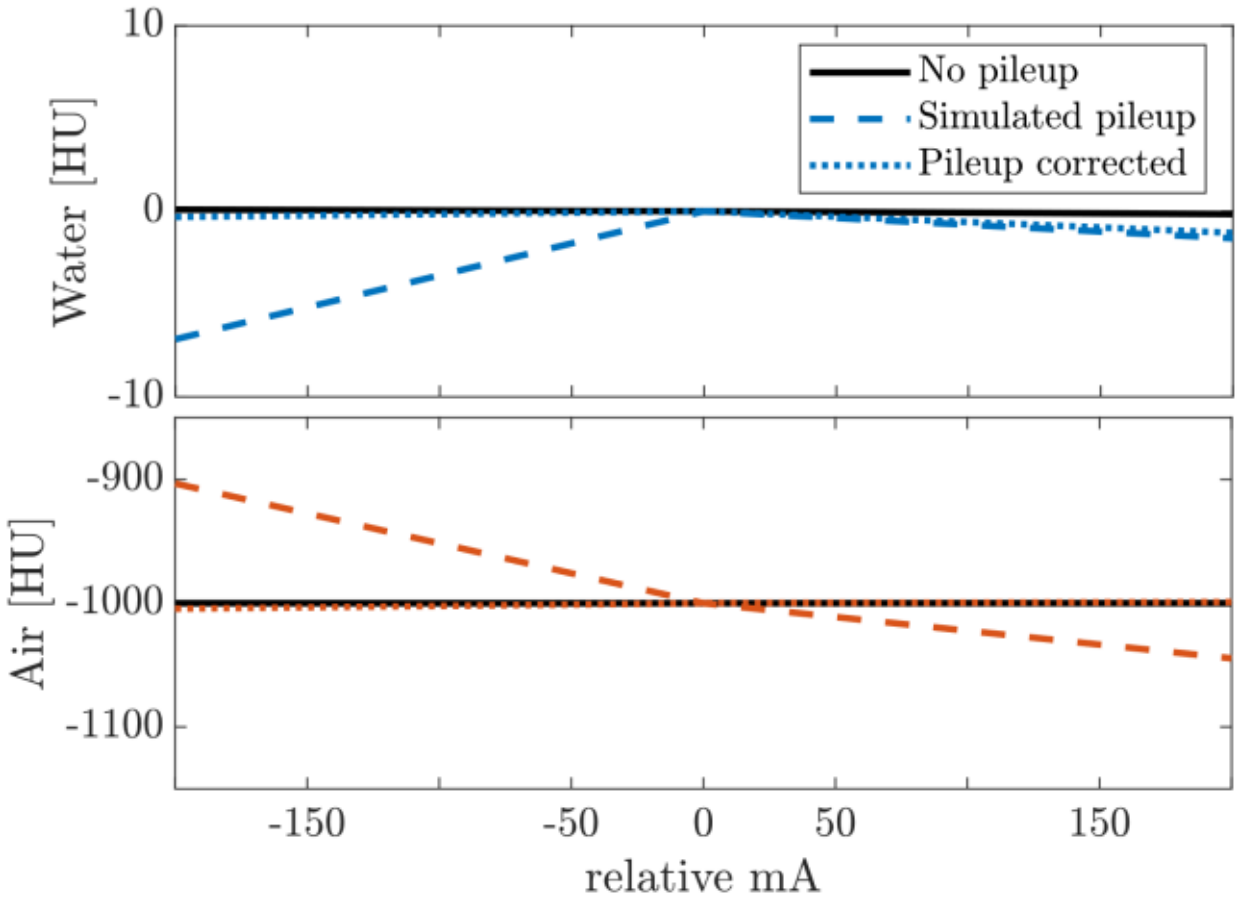

***Figure 4:** Pileup-induced bias in terms of HU measured inside (upper panel), and in the air surrounding (lower panel), the simulated phantom in Figure 2. The lines correspond to the rows in Figure 2, i.e., simulations without the pileup effect (solid), with pileup effect and no correction (dashed), and with pileup effect and pileup correction (dotted).*

# ACKNOWLEDGEMENTS AND CONFLICT OF INTEREST

The Authors wish to thank Bruno De Man at GE Healthcare for valuable discussions with respect to this work. All authors are employees of GE HealthCare. This work was funded by the Swedish Foundation for Strategic Research (grant number APR20-0012) and GE HealthCare.

# Comparison of lesion segmentation methodsusing simulated DBT images

Zhikai Yang[a*], Hanna Tomic[b], Victor Dahlblom[b], Sophia Zackrisson[b], Anders Tingberg[c]
Magnus Dustler[b,c], Örjan Smedby[a], Rodrigo Moreno[a] and Predrag Bakic[b,c]

[a]Division of Biomedical Imaging, Department of Biomedical Engineering and Health Systems, KTH Royal Instituteof Technology, Stockholm, Sweden;

[b]Diagnostic Radiology, Department of Translational Medicine, Faculty of Medicine, Lund University Hospital,Malmö, Sweden;

[c]Medical Radiation Physics, Department of Translational Medicine, Faculty of Medicine, Lund University, SkåneUniversity Hospital, Malmö, Sweden

*Corresponding author. E-mail: zhikai@kth.se

## BACKGROUND AND PURPOSE

To illustrate the use of simulated digital breast tomosynthesis (DBT) images, we compare the performance of two interactive segmentation methods: the deep learning-based method Segment Anything Model (SAM)[1] and the classic method Morphological Geodesic Active Contours (MGAC)[2].

## METHODS

We have used Perlin noise to generate 3D computational models of soft tissue breast lesions[3]. We simulated 14 breast lesions and inserted them in computational 3D breast phantoms that are based on our previous work of simulating breast tissue using Perlin noise[4-5]. We used OpenVCT software[6] to generate radiographic projections ofthe phantoms. The projections were reconstructed to DBT slices using an open-source method [7]. The simulated lesions consisted of a spherical 3D shape, within which the Perlin noise function is defined. To generate different lesion shapes, we modified the frequency and number of octaves of the Perlin noise. The corresponding reconstructed DBT images were used to compare SAM and MGAC methods.

SAM is a transformer-based method that uses manually delineated bounding boxes as prompts and can generate the segmentation result. The SAM pretrained weight is trained on a large variety of natural images. Several papers have already used it to validate medical images[8]. The MGAC is an energy-based method. It requires manually initializinga contour near the object; then the contour iteratively evolves to fit the object boundary when minimizing the energyfunction. We have computed the segmented tumor area within each DBT slice and compared it with the ground trutharea from the binary mask of the simulated tumor. The absolute and relative error in the segmented area is used as theperformance measure.

## RESULTS

The results suggested that both segmentation methods overestimate the tumor area near the central slice and underestimate the area in the peripheral slices. Figures 1 and 2 illustrate the segmentation performance of the two methods in the central and peripheral slices. The two methods have a better segmentation performance in the central slice and worse performance in the peripheral slice. Figure 3 illustrates the absolute error between the segmented areaand the ground truth area from the binary mask for each DBT slice. The average absolute error is 14.8 $mm^2$ and 21.7$mm^2$ in SAM and GMAC, respectively. Near the tumor center, we observed a relative error of 2.5 ± 1.8% and 7.6 ± 6.9% for SAM and GMAC segmentation, respectively. In the peripheral slices, we observed a relative error of 87.1 ±70.2% and 97.8 ± 51.2% for SAM and GMAC segmentation, respectively.

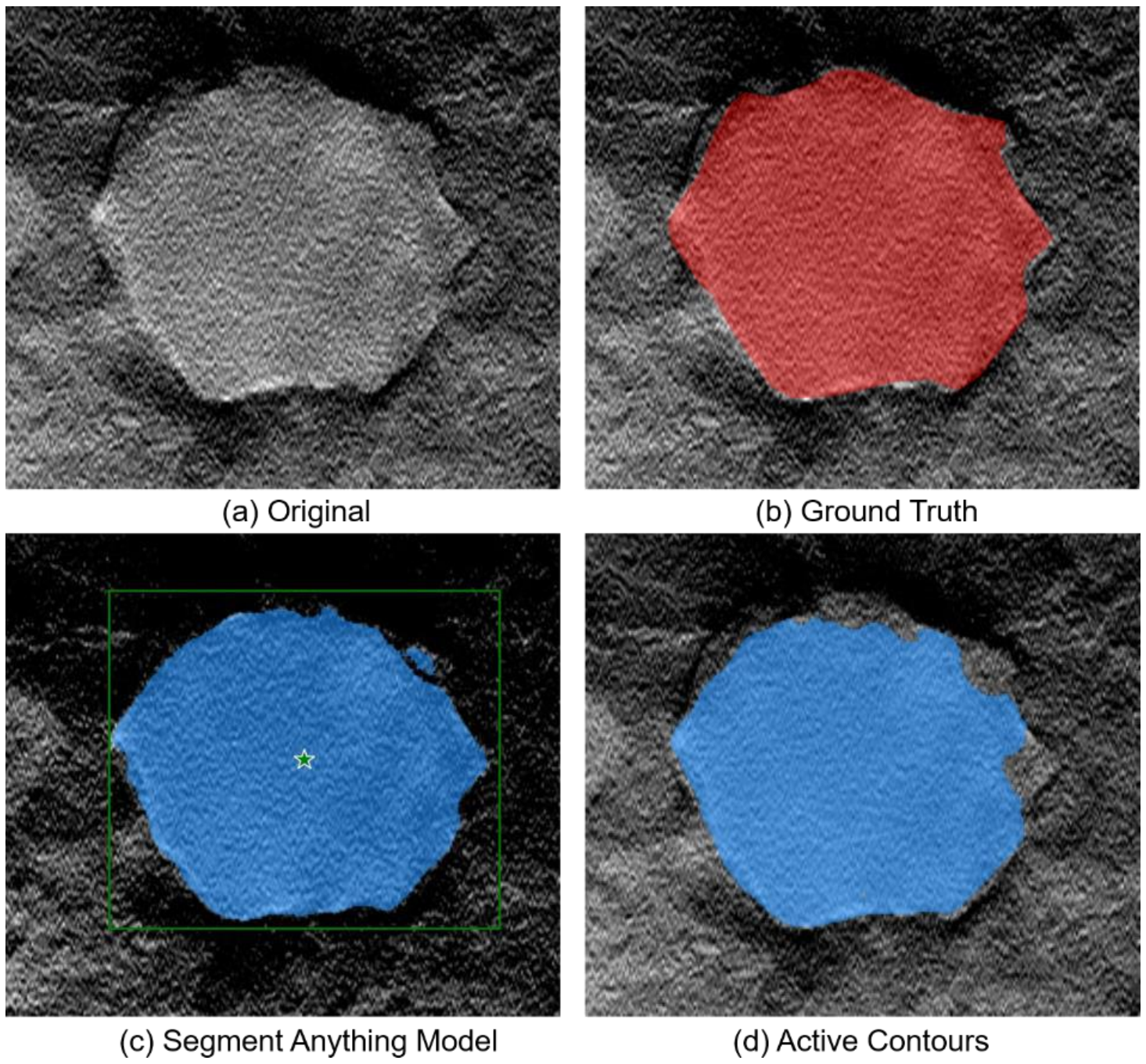

Figure 1. Visualization of segmentation results in central slice by SAM and Active Contours (GMAC) methods.

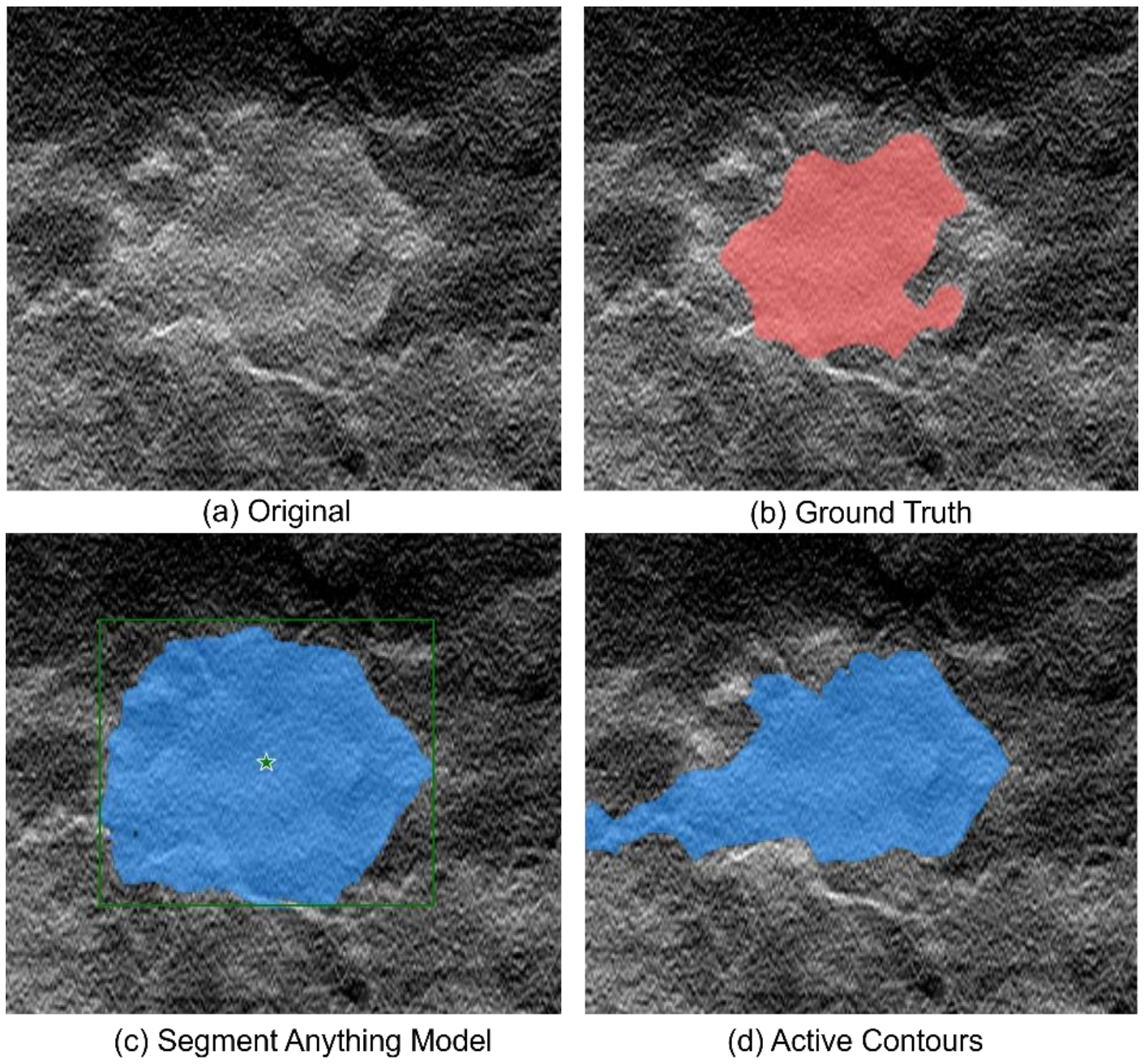

Figure 2. Visualization of segmentation results in a peripheral slice by SAM and Active Contour (GMAC) methods.

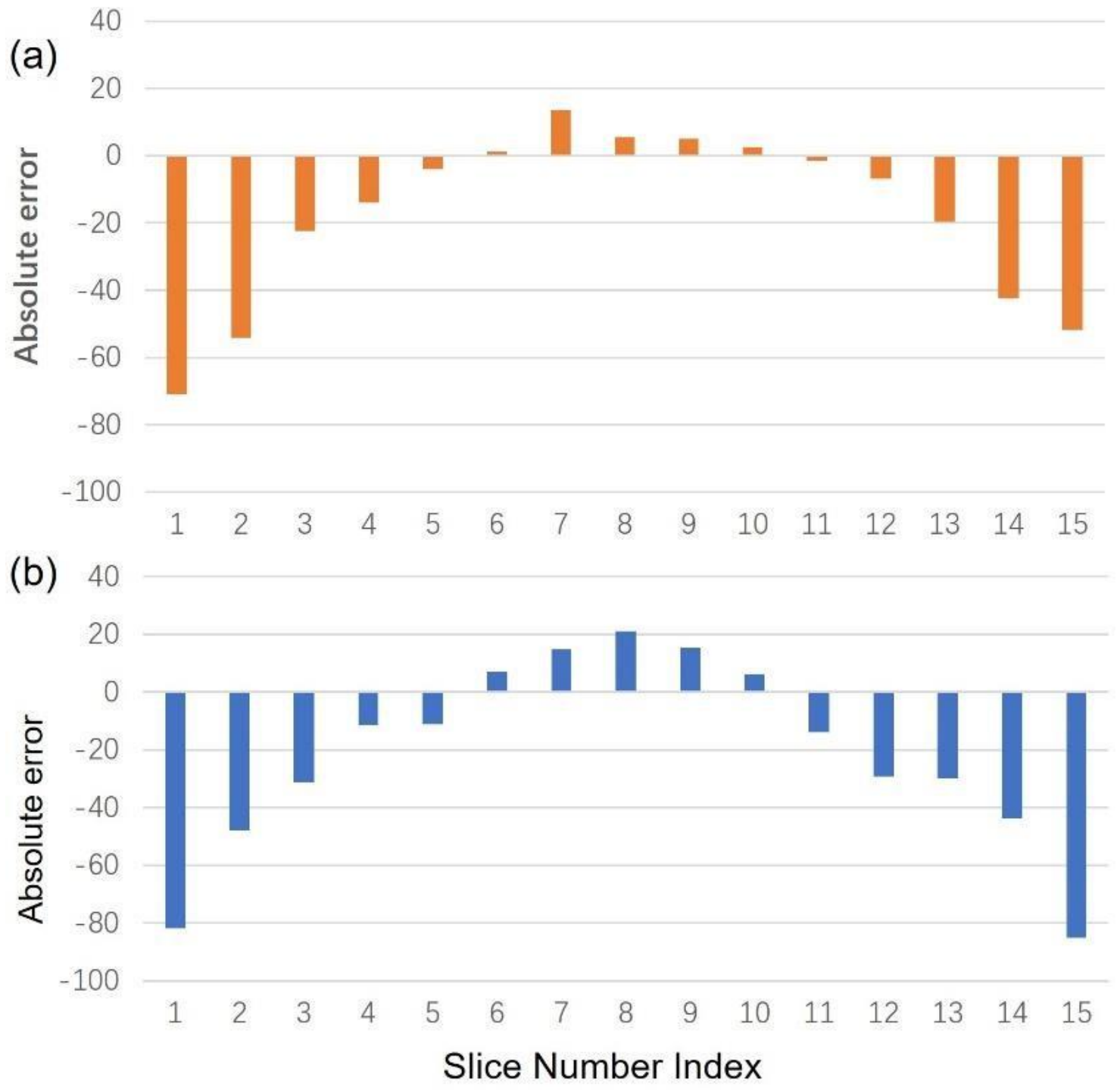

Figure 3. Comparison of the absolute error between the segmented area and the ground truth area at slice level, for the SAM and Active Contour (GMAC) methods. (a) shows the absolute error of the SAM segmentation for each slice number index, where slice index eight corresponds to the central DBT slice. (b) shows the absolute error of the GMAC segmentation method for each slice number index.

# DISCUSSION

In this study, we compared the performance of a deep learning-based (SAM) and a classic segmentation method (GMAC) without any data training. The results demonstrated that SAM performs better than MGAC when comparing the absolute relative errors between the segmented area and the ground truth area. Both methods over- estimate the true area in the central slice and slightly underestimate the area in the peripheral slice. Due to the limited angular range of the DBT imaging system, objects appear as out-of-plane artifacts in the peripheral slices. Consequently, due to this limitation, the tumor appears larger in peripheral reconstructed slices. To further improvethe segmentation performance, especially in the peripheral slices, one solution could be to use supervised learning on a large DBT dataset.

# CONCLUSION

We conducted preliminary experiments using the pre-trained SAM and MGAC to segment computer-simulated soft tissue breast lesions. The qualitative and quantitative results demonstrate that SAM has superior performance compared with the MGAC but still underestimates lesions in peripheral slices. In the future, we will conduct experiments compared with radiologists' delineation and various automatic segmentation methods.

# ACKNOWLEDGEMENTS AND CONFLICT OF INTEREST

This work was supported by grants from the European Commission H2020 Marie Skłodowska Curie Actions Fellowship (IF 846540), Cancerfonden (Grants 22-2389 Pj), the Swedish Breast Cancer Association, Region Skåne (Regionalt forskningsstöd), and Stiftelsen för Cancerforskning vid Onkologiska kliniken vid Universitetssjukhuset MAS. This work was supported by grants from Marie Skłodowska-Curie Doctoral Networks Actions (HORIZON- MSCA-2021-DN-01-01) with a grant number of 101073222.

# Tube current modulation effects on dose and image quality in computed tomography: a Monte Carlo simulation study

Z. Aboulbanine, J. Rajagopal, E. Abadi, P. Segars, and A. Kapadia

## Purpose

This work aims to estimate the effect of tube current modulation (TCM) on dose and image quality using a virtual imaging trial (VIT).

## Methods

A simulation of a chest CT-scan was performed using DukeSim and an XCAT phantom with a standard lung protocol at 120kV. Eleven TCM conditions were modeled for the study ranging from 0-1.0 in increments of 0.1. For each TCM setting, dose distributions were calculated at a spatial resolution of 1x1x1 $mm^3$ and evaluated using dose volume histograms (DVH), and image quality was assessed in terms of SNR within a region of interest (ROI) in the respective organ. Dose were reported for five organs in the chest: lung, heart, sternum, esophagus, and the skin, as well as the remaining tissue categorized as "body".

## Results

Increases in TCM strength were found to enhance the dose in the lungs, heart and sternum by a factor up to 1.12, 1.48 and 2.10 respectively. DVH analysis indicates that the sternum received the highest voxel dose.

## Conclusion

TCM increased the dose of superficial structures and organs, for instance the skin and the sternum. The dose outcome at the organ and voxel levels depended on TCM strength. We are currently evaluating the impact on SNR and also extending the study to include a larger number of XCAT phantoms.

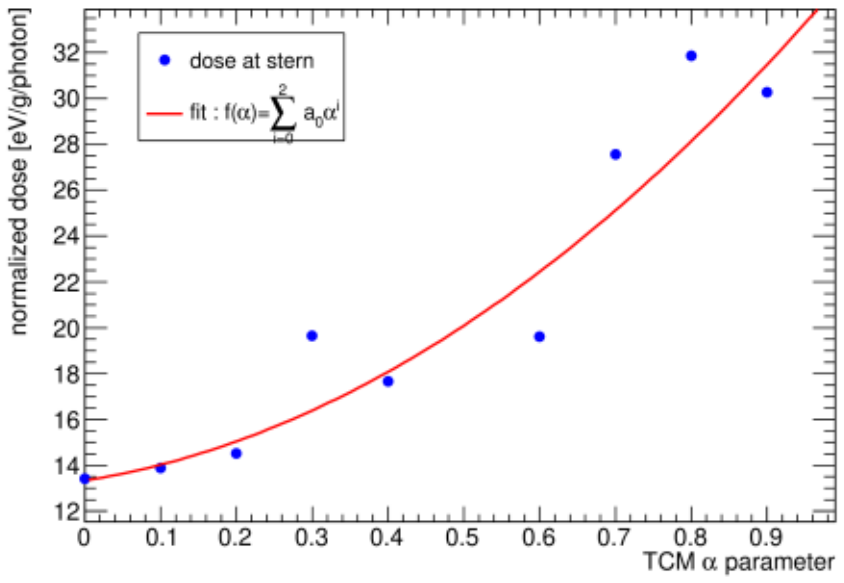

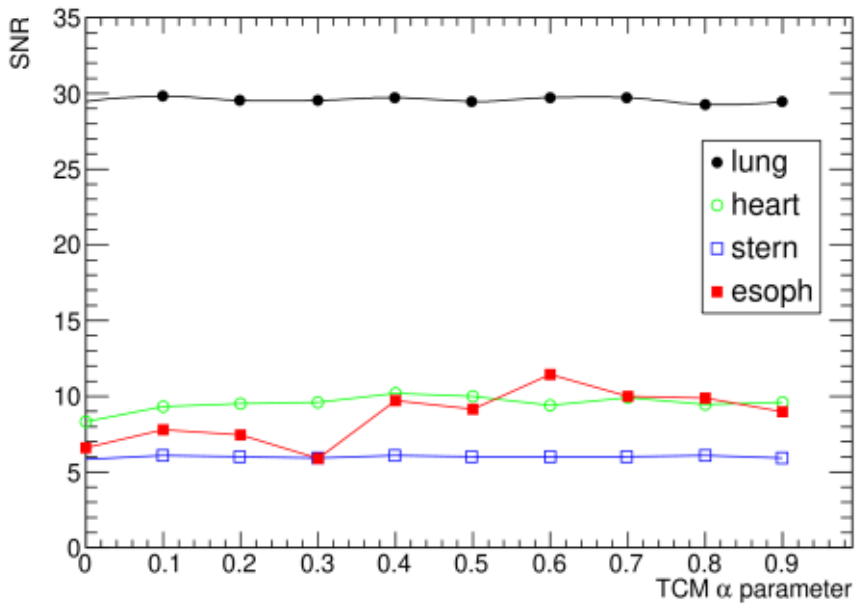

(a)

(b)

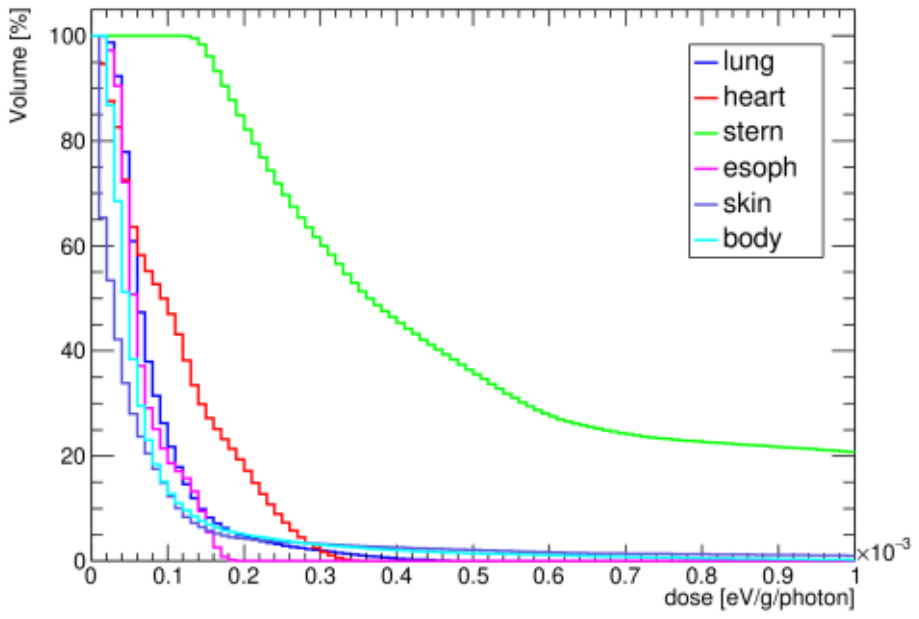

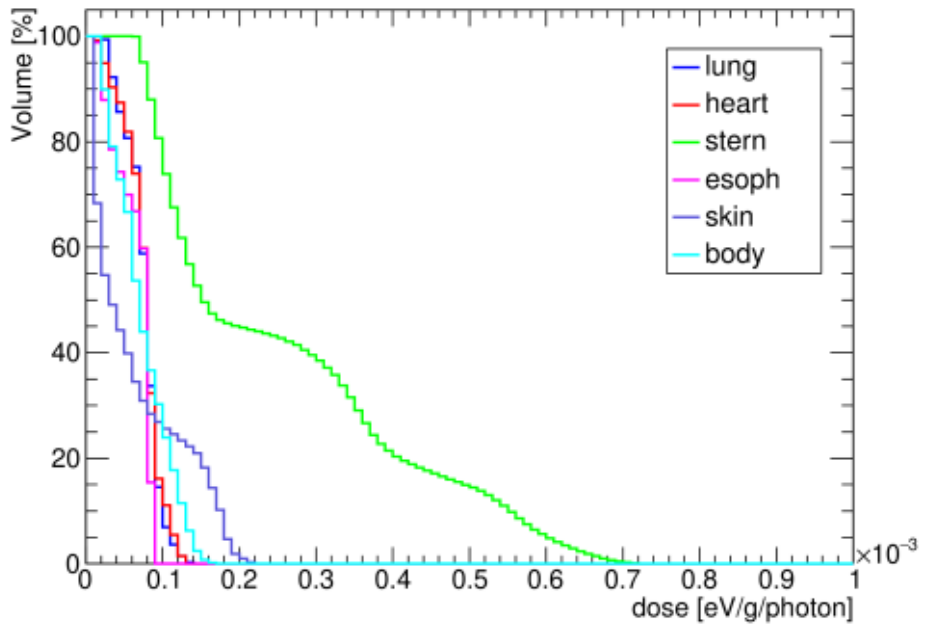

TCM on (α=0.9)

TCM off (α=0.0)

TCM effect on dose and reconstructed image SNR. (a) TCM strength vs normalized dose at the sternum; (b) : TCM vs SNR; DVH analysis for two different TCM strength values (c) 0.9 and (d) 0.

(c)

(d)

# How common is spontaneous thrombus formation in intracranial aneurysms? An in-silico observational study calibrated against clinical prevalence data

Qiongyao Liu[1], Ali Sarrami-Foroushani[1], Yongxing Wang[1], Michael MacRaild[1], Christopher Kelly[1], Fengming Lin[1], Yan Xia[1], Shuang Song[1], Nishant Ravikumar[1], Tufail Patankar[2], Zeike A. Taylor[1], Toni Lassila[1], Alejandro F. Frangi[1]

[1]Centre for Computational Imaging and Simulation Technologies in Biomedicine (CISTIB), [2]Leeds General Infirmary

E-mail: mnqli@leeds.ac.uk

## ABSTRACT

How prevalent is spontaneous thrombosis in a population containing all sizes of intracranial aneurysms? How can we calibrate computational models of thrombosis based on published data? How does spontaneous thrombosis differ in normo- and hypertensive subjects? We address the first question through a thorough analysis of published datasets that provide spontaneous thrombosis rates across different aneurysm characteristics. This analysis provides data for a subgroup of the general population of aneurysms, namely, those of large and giant size (>10 mm). Based on these observed spontaneous thrombosis rates, our computational modeling platform enables the first in silico observational study of spontaneous thrombosis prevalence across a broader set of aneurysm phenotypes. We generate 109 virtual patients and use a novel approach to calibrate two trigger thresholds: residence time and shear rate, thus addressing the second question. We then address the third question by utilizing this calibrated model to provide new insight into the effects of hypertension on spontaneous thrombosis. We demonstrate how a mechanistic thrombosis model calibrated on an intracranial aneurysm cohort can help estimate spontaneous thrombosis prevalence in a broader aneurysm population. This study is enabled through a fully automatic multi-scale modeling pipeline. We use the clinical spontaneous thrombosis data as an indirect population-level validation of a complex computational modeling framework. Furthermore, our framework allows exploration of the influence of hypertension in spontaneous thrombosis. This lays the foundation for in silico clinical trials of cerebrovascular devices in high-risk populations, e.g., assessing the performance of flow diverters in aneurysms for hypertensive patients.

# Beyond Detection: Bridging the Gap Between Virtual Imaging Trials and Clinical Impact

Fakrul Islam Tushar1,2, Liesbeth Vancoillie2, Dhrubajyoti Ghosh2, Kyle J. Lafata1,2, Joseph Y. Lo1,2

1Department of Electrical and Computer Engineering, Duke University, Durham, NC

2Center for Virtual Imaging Trails, Duke University, Durham, NC

## BACKGROUND

Virtual imaging trials (VIT) usually focus on the task of lesion detection. For true clinical relevance, however, VITs need to demonstrate they can act like real clinical trials by assessing patient-level endpoints such as cancer diagnosis. Objective: Using lung cancer screening as a clinical example, we introduce statistical labeling of lung nodules as benign vs. malignant based on Lung-RADS findings including nodule size. We performed virtual reader studies based on a cancer classification model to assess the robustness of lesion- and patient-level ROC performance.

## METHODS

Currently, our methodology harnesses real clinical data from the Lung Image Database Consortium and Image Database Resource Initiative (LIDC-IDRI). This data was combined with actual TCIA diagnosis data to maintain the authenticity of our simulations. A control prevalence mechanism was devised to ensure our dataset's balance. We constructed a logistic regression model, with the virtual nodule's maximum predicted size as a pivotal feature. This model's capacity to stratify virtual nodules based on their malignancy risk was meticulously evaluated. While our current focus is on real clinical data, we are in the advanced planning stages to adapt this approach to replicate the outcomes of the National Lung Screening Trial (NLST) in a VIT setting.

## RESULTS

Preliminary outcomes from the VIT underscore the potential of our approach in correlating nodule size with malignancy probability in a simulated milieu. The model exhibited nuanced performance across different thresholds in the virtual environment. A comparison between Diagnosis AUC and Statistical Labeling AUC displayed consistent performance up to a 0.8 threshold, with increased variability thereafter. A subsequent VIT analysis juxtaposed against virtual diagnostic data revealed the method's promise in emulating real-world diagnostic paradigms.

## Conclusion

Our statistical modeling technique provides a groundbreaking path for the evaluation of VITs using clinically relevant endpoints such as cancer diagnosis. This endeavor amplifies the clinical relevance of VITs, laying the groundwork for real-world diagnostic and therapeutic protocols.

# Poster minute digest 2

Tuesday, April 23, 2024

12:10 am - 12:30 am

Chaired by Francesco Ria and Liesbeth Vancoillie

# Hybrid (real and virtual) imaging trials in dynamic chest radiography: A feasibility study

Rie Tanaka[1], Nozomi Ishihara[1], Futa Goshima[1], Syunya Yamaguchi[1], Ryuichi Nagatani[1], Yu Homareda[1], Riku Yokoyama[1], Ryuunosuke Gotou[1], Saho Matsuo[1], Yuuna Yamawaki[1], Haruto Kikuno[1], Noriyuki Ohkura[2], Isao Matsumoto[3]

[1]College of Medical, Pharmaceutical & Health Sciences, Kanazawa University, Ishikawa, Japan
[2]Department of Thoracic Surgery, Kanazawa University, Ishikawa, Japan
[3]Department of Respiratory Medicine, Kanazawa University Hospital, Ishikawa, Japan

## BACKGROUND AND PURPOSE

Dynamic chest radiography (DCR) is a newly developed imaging technique to assess pulmonary function based on dynamic findings. Although its clinical usefulness has been reported, determining the optimal imaging conditions and diagnostic performance, and providing dedicated image-processing technology, is challenging. This study aimed to accelerate clinical research and technological development for DCR by hybrid imaging trials (Fig. 1).

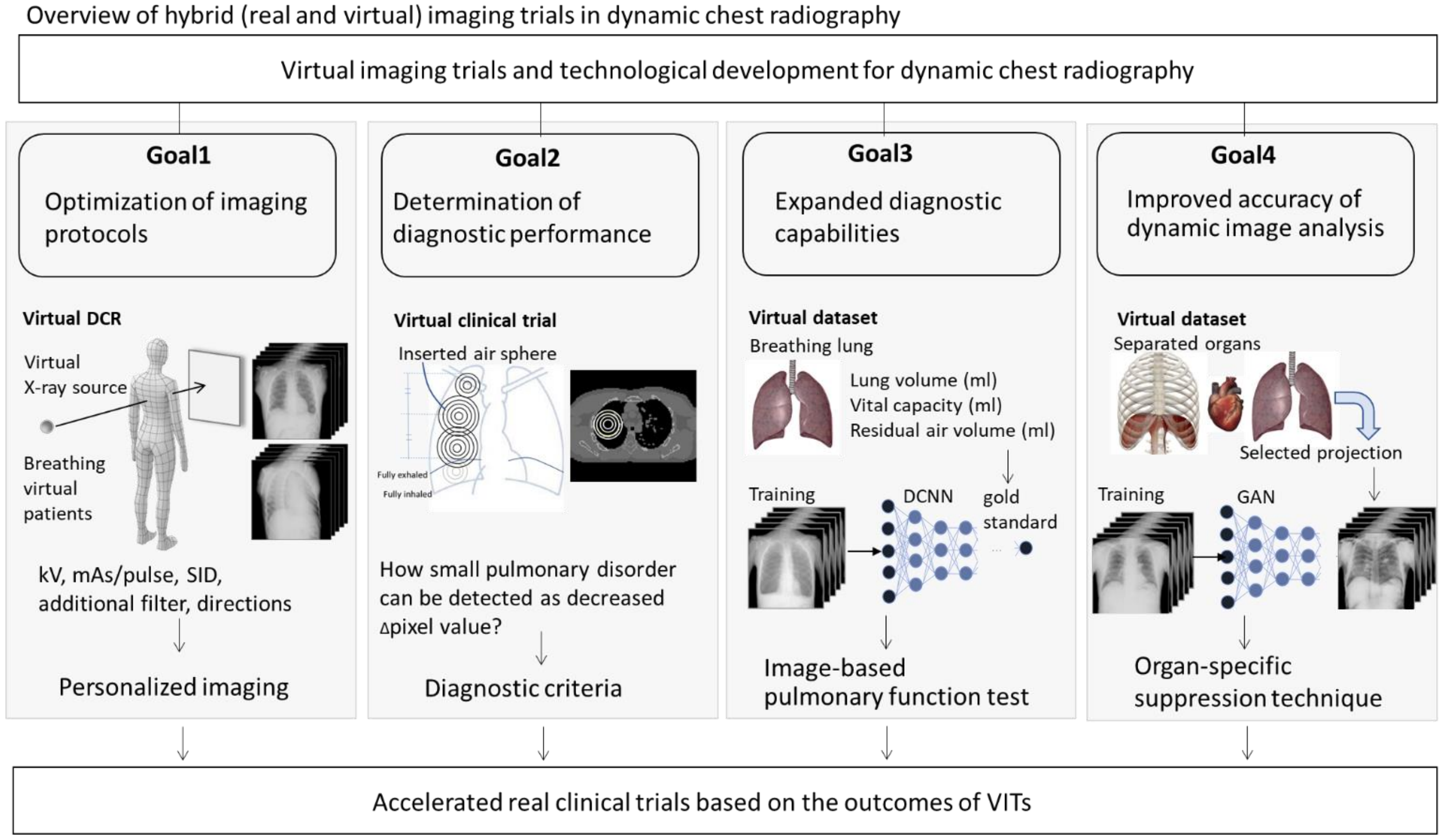

*Figure 1 Overview of dynamic chest radiography.*

## METHODS

Virtual imaging trials (VITs) were conducted in parallel with a clinical trial. Virtual, breathing patients with heartbeats were generated using the XCAT program and projected by an X-ray simulator at various tube voltage and mAs combinations. Optimal imaging conditions were determined based on the accuracy of functional analysis by DCR and visual evaluation of image quality. The diagnostic performance of the DCR-based pulmonary function assessment was evaluated in virtual patients with pulmonary impairment. Additionally, a deep learning-based artificial intelligence (AI) model to estimate lung volume from DCR images was trained on DCR-simulated images and ground truth derived from the log files of XCAT phantoms. AI-based multi-organ-specific suppression models were also trained on a pair of DCR simulation images with and without the target organs (rib, heart, breast, and diaphragm).

## RESULTS

Optimal imaging conditions and diagnostic performance of DCR were determined using VITs; real imaging trials based on the outcomes of VITs were effectively performed. Furthermore, AI-based lung volume estimation and multi-organ-specific suppression models for DCR were successfully developed using simulation images of virtual patients and worked effectively on real patients.

## CONCLUSION

This feasibility study highlights the success of hybrid imaging trials for emerging imaging devices and the potential role of VITs in facilitating the development of AI-based technologies and their translation into clinical applications.

# Development of deep learning-based left-right lung separation technique for lateral dynamic chest radiography using virtual patients

Riku Yokoyama[1], Rie Tanaka[1], Yu Homareda[1], Noriyuki Ohkura[2], Isao Matsumoto[3]
1. College of Medical, Pharmaceutical & Health Sciences, Kanazawa University, Ishikawa, Japan
2. Department of Respiratory Medicine, Kanazawa University Hospital, Ishikawa, Japan
3. Department of Thoracic Surgery, Kanazawa University, Ishikawa, Japan

## BACKGROUND AND PURPOSE

Dynamic chest radiography (DCR) is a recently developed technique for pulmonary functional imaging. However, lateral images are used only for visual assessment of thoracic motion because of the overlap between left and right lungs. This study aimed to develop a deep learning-based left-right lung separation technique for lateral DCR using a 4D extended cardiac-torso phantom (XCAT phantom).

## METHODS

Twenty-eight XCAT phantoms in 150 phases (12 respirations/min) and those with only left and right lungs extracted were pseudo-projected, respectively, and 4200 pairs of original and labeled (lung) images were prepared as a training dataset. Using the network structure of U-net, we constructed a left-right lung separation model for lateral DCR images using seven-fold cross-validation with training to validation ratio of 6:1 (Fig. 1). The separation accuracies of the left and right lungs in the output images were quantitatively evaluated based on Dice and Jaccard coefficients and by visual assessment.

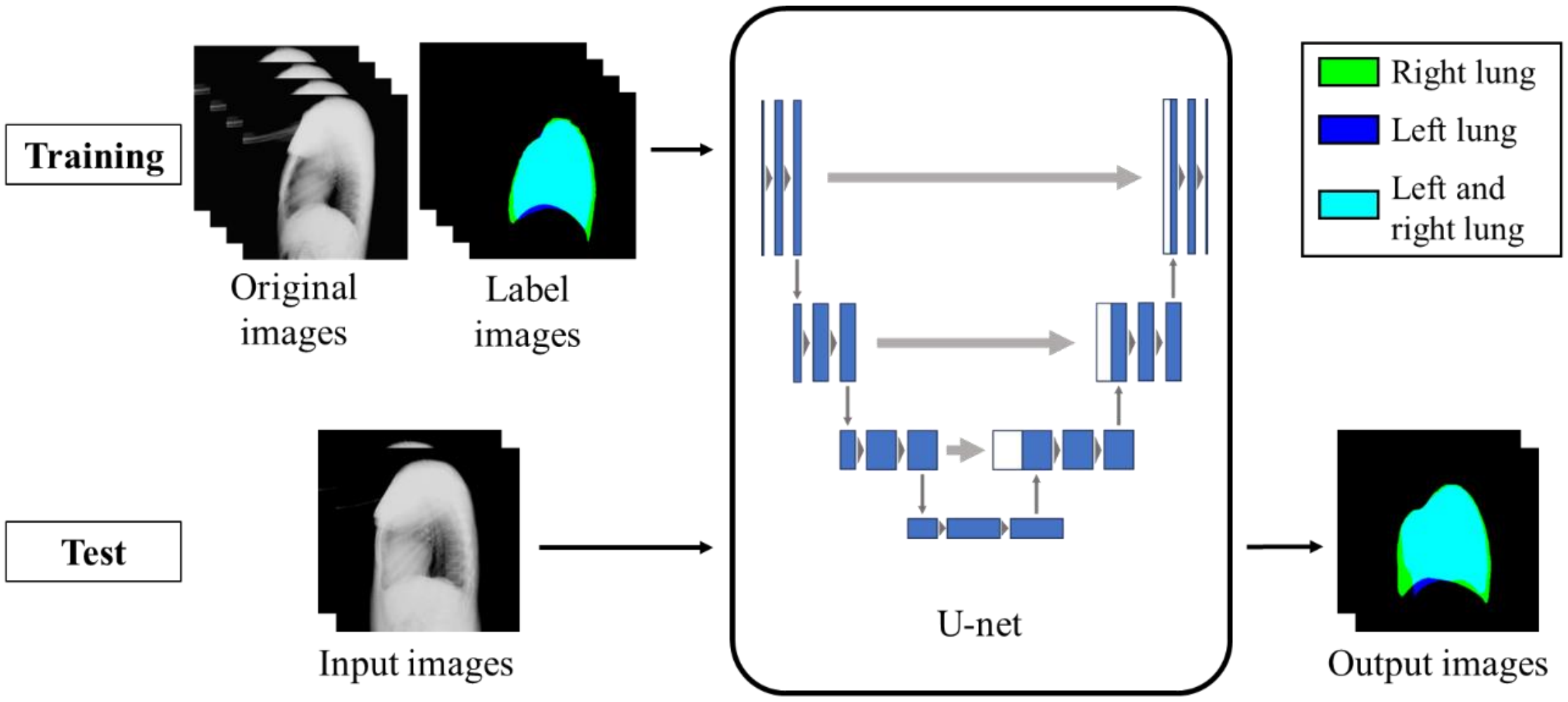

*Figure 1 Overview of the left–right lung separation model for lateral DCR images using U-net. DCR, dynamic chest radiography*

# RESULTS

The mean Dice and Jaccard coefficients were 0.948 (range, 0.834–0.974) and 0.903 (range, 0.716–0.950), respectively. In visual evaluations, misidentification of the lung fields and/or defects in the lung apex and base contours were observed in some cases, especially with large area under the diaphragm and/or with increased or decreased lung density in the lung area (Figs. 2(a)–(d)). These results suggest the need for additional training datasets for various body sizes.

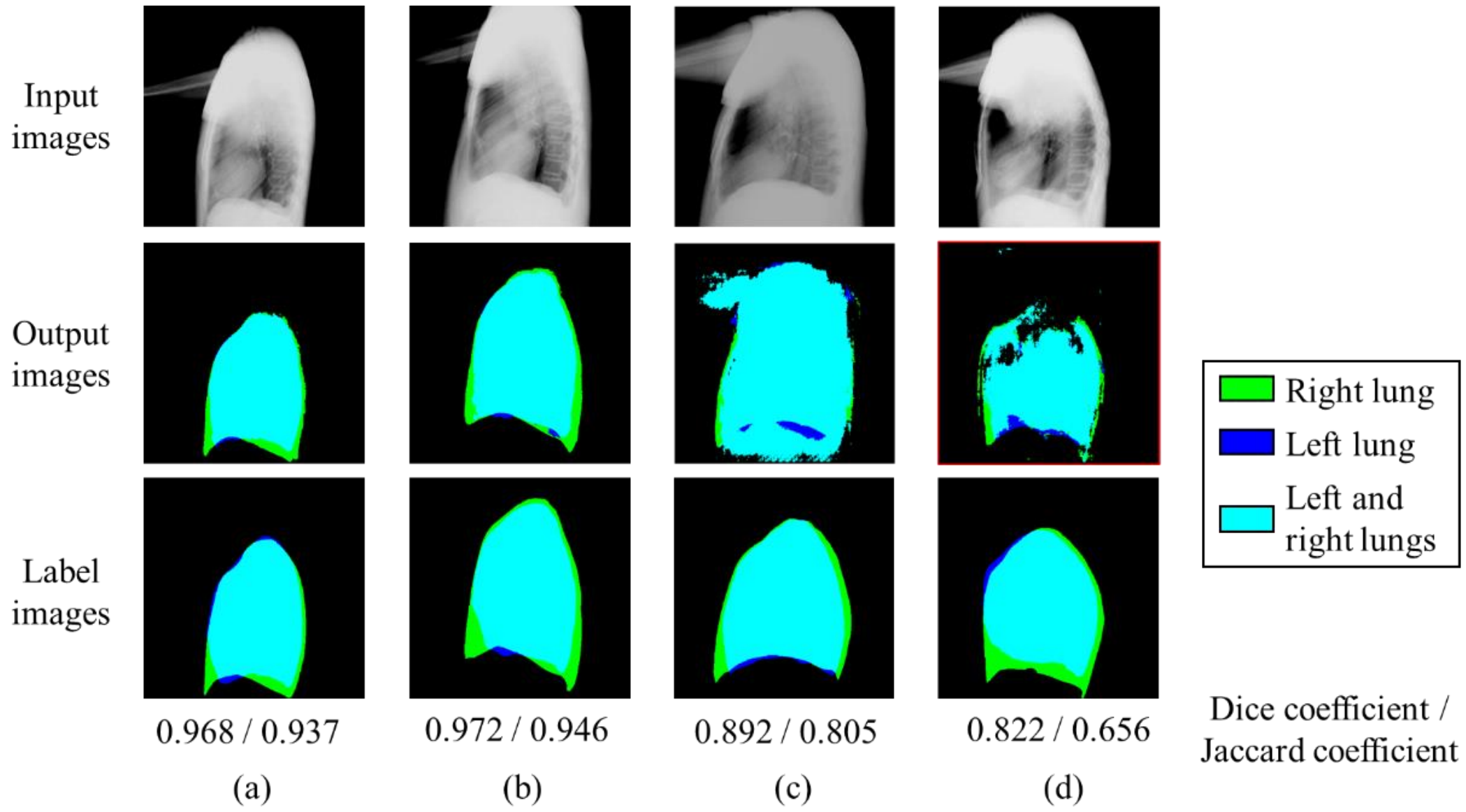

*Figure 2 Examples of the output images of the XCAT phantoms in the five-fold cross-validation from the left–right lung separation model for lateral DCR images: (a, b) reasonably segmented cases (male and male, BMI = 21.89 kg/m$^2$ and 26.88 kg/m$^2$); (c) incorrectly segmented case with misidentifications (male, BMI = 23.10 kg/m$^2$); (d) incorrectly segmented case with defection (male, BMI = 31.22 kg/m$^2$). BMI, body mass index*

# CONCLUSION

A deep-learning-based left–right lung separation technique for lateral DCR was successfully developed using virtual patients. The proposed method would be useful for a better understanding of pulmonary function with lateral DCR, such as quantitative evaluation of thoracic and diaphragmatic movements, as well as lung volume changes.

# Development of breast suppression techniques for dynamic chest radiography using virtual patient datasets

Authors: Yuuna Yamawaki[1], Rie Tanaka[1], Syunya Yamaguchi[1], Futa Goshima[1], Haruto Kikuno[1], Noriyuki Ohkura[2], Isao Matsumoto[3]

1. Graduate School of Medical Sciences, Kanazawa University, Ishikawa, Japan
2. Department of Respiratory Medicine, Kanazawa University hospital, Ishikawa, Japan
3. Department of Thoracic surgery, Kanazawa University, Ishikawa, Japan

## BACKGROUND AND PURPOSE

Dynamic chest radiography (DCR) is a flat-panel detector-based functional X-ray imaging technique. Pulmonary impairments can be detected as decreased changes in lung density due to respiration. However, breast shadows occasionally produce false-positive results in female patients. This study aimed to develop a deep learning–based breast suppression technique to improve the DCR-based evaluation of pulmonary function using four-dimensional (4-D) extended cardiac-torso (XCAT) phantoms.

## METHODS

Fourteen female XCAT phantoms with and without breasts were generated in ten respiratory phases with bone suppression using the XCAT program, and then projected using an X-ray simulator in the same settings in clinical practice. A total of 2,100 images were obtained for the training and testing datasets (6:1). Pix2pix, a type of conditional generative adversarial network, was trained to map the breast-suppressed images from original images through leave-one-out cross-validation. (Fig. 1) The resulting images were evaluated based on two metrics: peak signal-to-noise ratio (PSNR) and structural similarity index measure (SSIM). The trained model was applied to clinical images, and the effect of breast suppression was visually assessed. The color-mapped images depict changes in lung density, with higher color intensities representing higher X-ray translucency (increased air).

## RESULTS

The resulting images of XCAT phantoms had PSNR of 23.6 dB and SSIM of 0.963. Visual evaluation confirmed that the breast shadows are successfully reduced in output images of both the XCAT phantom and real patients (Fig. 2). In addition, the lung area behind the breast shadows was properly assessed using the color-mapped images after breast suppression processing (Fig. 3).

## CONCLUSION

The deep learning–based breast suppression model trained on the XCAT phantom was effective in contributing to improved pulmonary function assessment using DCR.

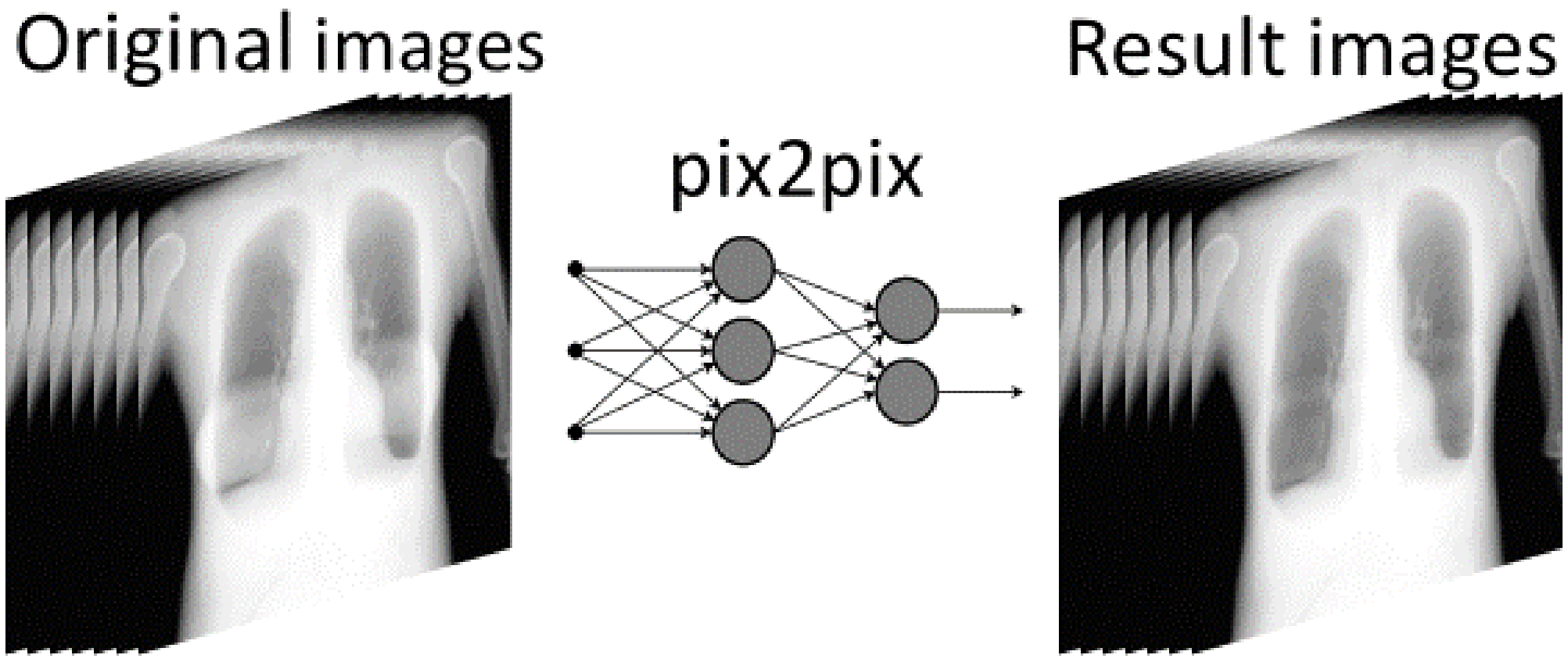

Figure 1 Overall scheme of proposed breast suppression technique.

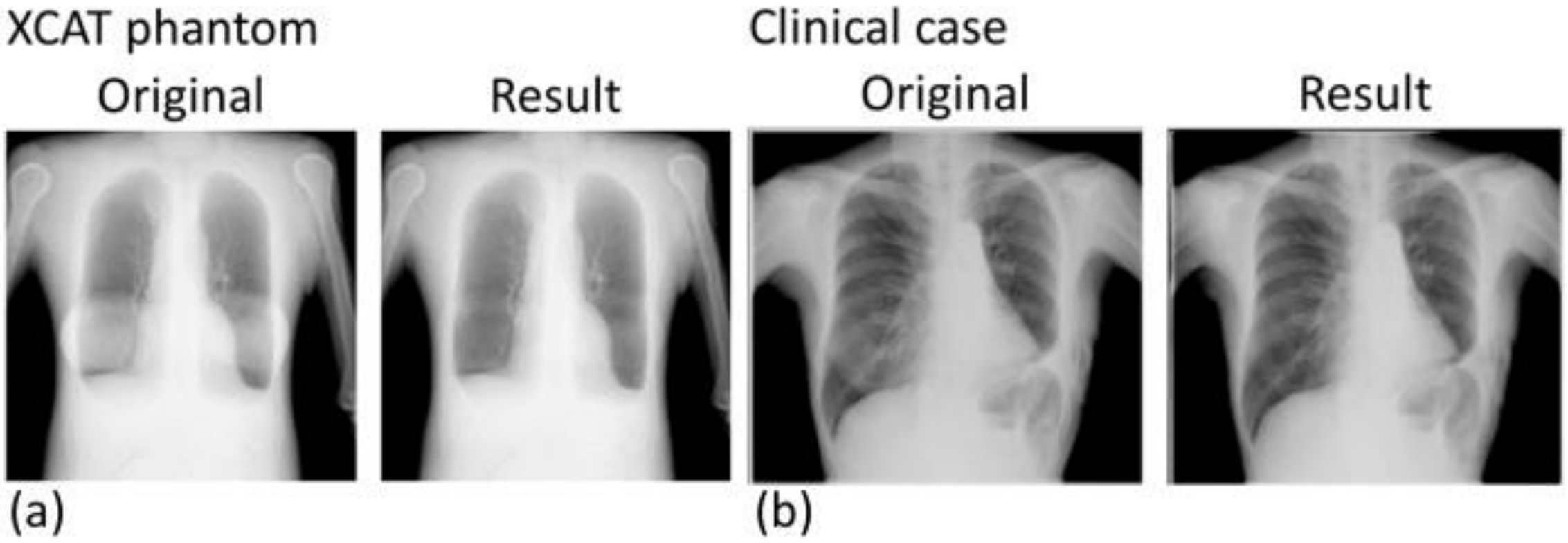

Figure 2 Representative images from (a) XCAT phantom (BMI 23.73 $kg/m^2$) (b) clinical cases (a 68-year-old female with postoperative left-lung cancer)

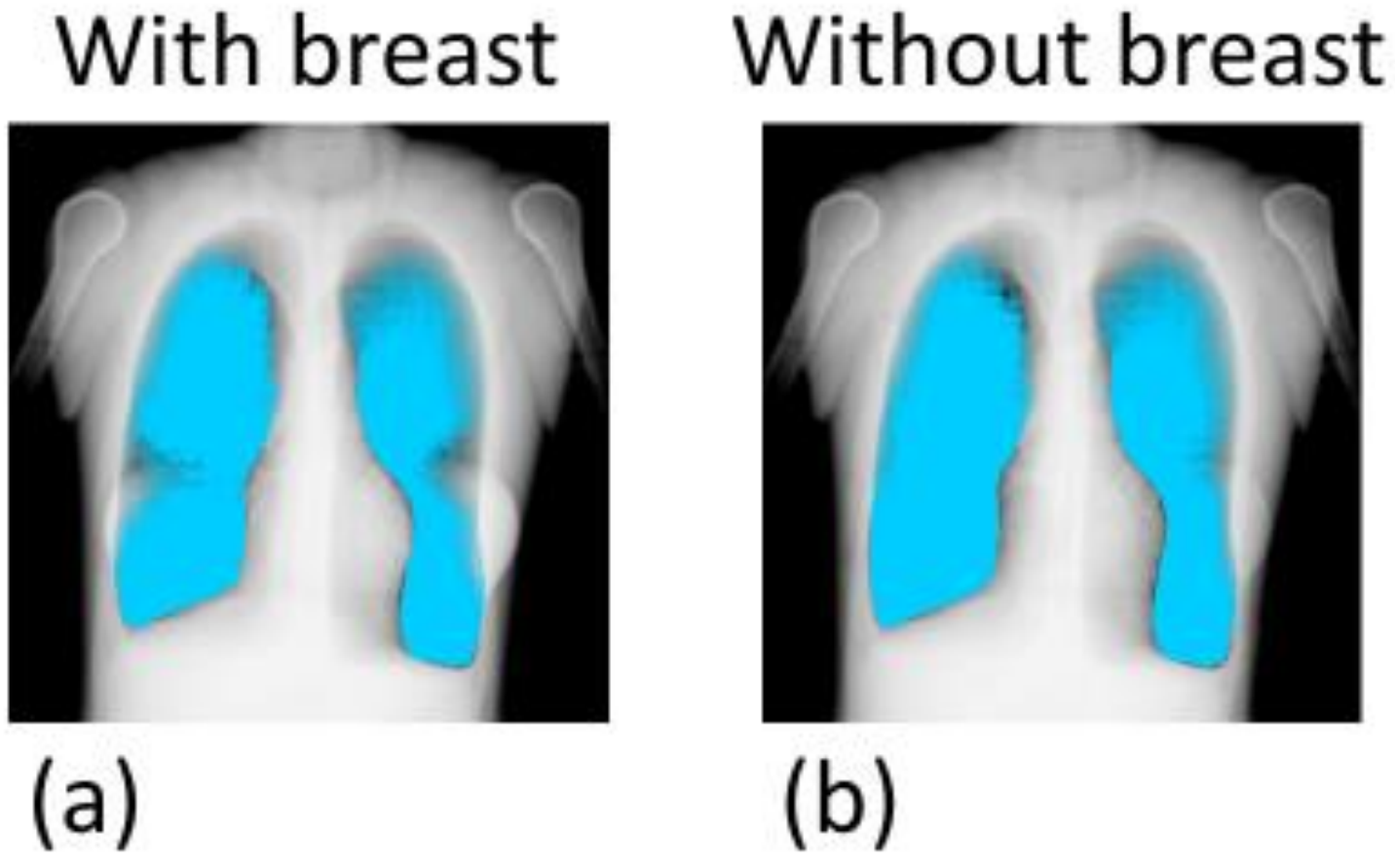

Figure 3 Results displayed as color-mapped images (a)with and (b) without breast shadows.

These images depict changes in lung density, with higher color intensities representing higher X-ray translucency (increased air).

# Generation of Synthetic brain vasculature using shape/anatomy guided latent diffusion models

Yash Deo (University of Leeds), Toni Lassila (University of Leeds), Haoran Dou (University of Leeds), Nishant Ravikumar (University of Leeds), Alejandro Frangi (University of Manchester), Fengming Lin (University of Leeds), Nina Cheng (University of Leeds)

This study addresses the significant anatomical variations within the Circle of Willis (CoW), a crucial cerebral artery network. CoW's diverse structures play a pivotal role in brain blood supply, reducing the risk of ischemic events by providing alternative routes when major vessels are blocked. However, existing classifications predominantly focus on common CoW configurations, leaving less prevalent variants understudied due to limited data. The research's core objective is to develop a generative model capable of synthesizing CoW segmentations while considering these anatomical variations. This model enables the creation of extensive virtual cohorts of brain vasculature, including less common CoW configurations. Such virtual cohorts can be invaluable for training deep learning algorithms, conducting in-silico trials, and furthering our understanding of cerebrovascular diseases. This study explores a conditional latent diffusion model that learns latent representations of brain vasculature, allowing for controlled generation of CoW configurations based on specific phenotypes. Evaluation metrics such as multiscale structural similarity index (MS-SSIM) and Fréchet inception distance (FID) are used to assess its performance, with comparisons against alternative generative models like 3D GANs and 3D variational auto-encoders (VAEs) to establish its effectiveness in replicating CoW variations.

# Evaluating SPECT Motion Correction: The Role of Realistic Phantoms

Maximilian P. Reymann[1, 2], Francesc Massanes[3], Wesley Gohn[3], A. Hans Vija[3], Andreas K. Maier[2]

[1]Siemens Healthineers AG, Forchheim, GERMANY
[2]Pattern Recognition Lab, Friedrich-Alexander-Universität Erlangen-Nürnberg, GERMANY
[3]Siemens Medical Solutions Inc., Hoffman Estates, IL, UNITED STATES OF AMERICA.

## BACKGROUND AND PURPOSE

Respiratory motion poses a significant challenge in slow rotating large Field-of-view SPECT imaging with multichannel collimator image formation. For instance, in myocardial perfusion imaging (MPI) it can impact diagnosis mis-interpreting from artifacts [1]. We have shown that simulations are useful to evaluate the effects of acquisition, motion, and geometry parameters on the accuracy of respiratory correction methods in SPECT using a set of water cylinder phantoms with different motions[2]. A simulation setup that is closer to clinical practice and could be used in Virtual Imaging Trials can be achieved using the Extended Cardiac-Torso Phantom (XCAT)[3]. While the XCAT phantom platform offers a highly versatile and realistic patient anatomy, the default respiratory motion cycle represents an idealized and averaged motion trace with a constant frequency and amplitude that might vary greatly from motions seen in clinical practice. We have already demonstrated that it is possible to use recorded patient motion traces as input to the XCAT phantom, thus achieving a more realistic, temporally varying respiratory motion affected SPECT simulation[4]. The purpose of this work is to investigate the effect of using realistic, patient-specific respiratory motion traces versus the default sinusoidal respiratory motion using the example of Data Driven Gating (DDG) for SPECT Myocardial Perfusion Imaging (MPI).

## METHODS

To simulate realistic respiratory motion using the XCAT phantom, we use the respiratory signal of 15 patients recorded with an Anzai respiration belt of patients undergoing SPECT MPI[5] and use the signal as input for the superior-inferior (SI) respiratory trajectory parameter of the phantom. For every phantom, we resampled a snippet of 15s of motion recording $f(t)$ by resampling it to 5Hz and adjusting the SI signal's amplitude to $f'(t) = \frac{1}{2*\sigma_f}$, where $\sigma_f$ is the standard deviation of $f(t)$, following the normalization procedure presented in[5]. This adjustment accounts for the respiratory motion belt's measurement of relative chest diameter changes and scales the signal to a realistic amplitude range in the order of 2 cm. Additionally, we set the amplitude of the anterior-posterior (AP) motion to 70% of the SI motion's amplitude, as the AP motion is typically smaller than the SI motion[6].

According to the approach presented in [4] we create 15 different realizations of the XCAT phantom with randomized anatomy and activity distribution parameters for each time step, resulting in 75 pairs of activity and attenuation maps. We also created a control phantom with default sinusoidal motion keeping the default respiratory motion cycle length of 5s, sampled at 5Hz for a total duration of 15s. Activity maps were created to represent SPECT MPI with 75-1400 MBq Tc99m of total injected activity and four sample projection views from right anterior oblique (RAO) to left posterior oblique (LPO) were simulated using the SIMIND Monte Carlo simulation package[7]. Subsequently, we used the center of light (COL) method and the non-linear DDG method based on Laplacian Eigenmaps (LE) introduced by Sanders et al.[5] to extract a respiratory surrogate signal for each phantom. We compute Pearson's correlation coefficient between the extracted respiratory surrogate signal and the input signal.

# RESULTS

In Figure 2 we show a comparison of the simulated respiratory motion trace from a real patient versus the default XCAT motion, showing irregularities in the patient data.

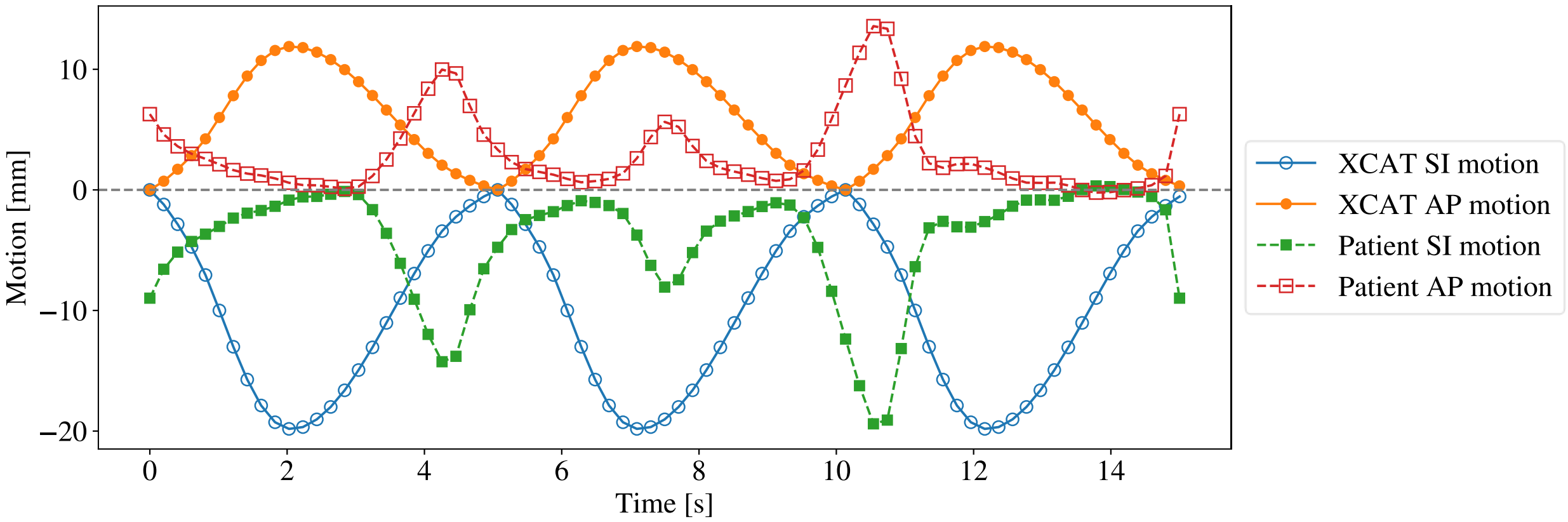

*Figure 2: Comparison of SI and AP motion between the XCAT phantom's default motion and a patient-derived signal used as XCAT input.*

We found significant (p< 0.01) differences between DDG results obtained on the phantoms with the default sinusoidal motion versus patient derived motion, irrespective of the DDG method used. Averaged over all simulations, the COL method showed a correlation of 0.90 ± 0.18 for simulated motion scenarios, versus 0.74 ± 0.26 for real motion. Meanwhile, the LE method demonstrated a near-perfect correlation of 1.00 ± 0.0034 with simulated motion, compared to 0.92 ± 0.086 for real motion, substantiating the superiority of the LE method across the board. We plot the results against the view angle and depending on the total simulated activity in Figure 3.

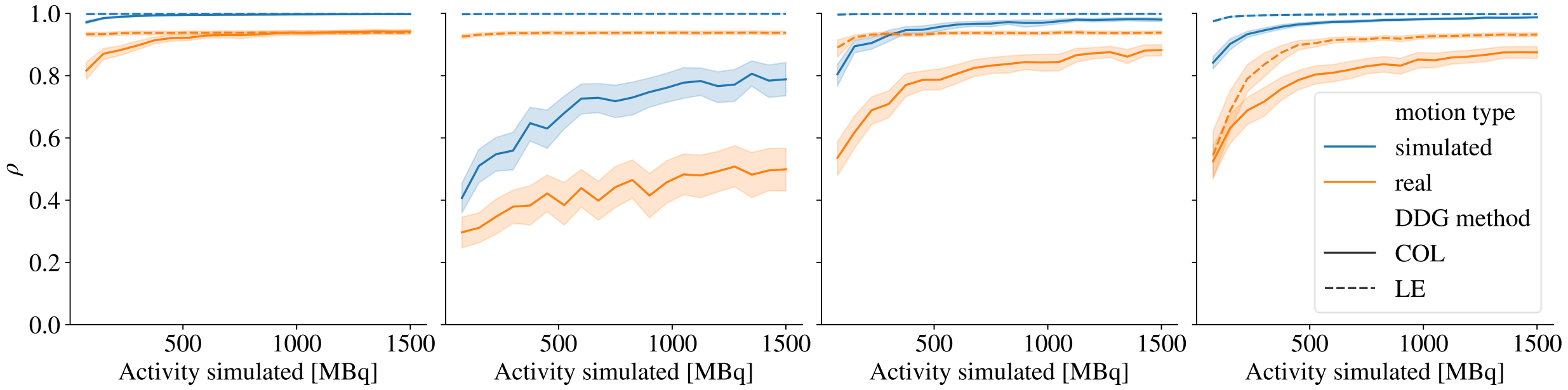

*Figure 3: Visualization of the mean and standard deviation (as envelope) for the COL (solid line) and LE (dashed line) DDG methods and motion types simulated (blue shows simulated motion, orange real motion). The columns show the 4 simulated views from RAO to LPO from left to right and each x-axis shows the relative count level.*

# DISCUSSION

We used DDG as an example to underscores the critical role of using patient-derived motion traces to increase the realism of simulated SPECT data. Figure 2 shows variations in frequency and amplitude that are often encountered in recordings of real patient data. Figure 3 illustrates the variation in DDG outcomes based on the detectability of respiratory motion across different view angles and the algorithms applied. It also highlights a consistent difference in results, suggesting that employing the XCAT phantom's default sinusoidal respiratory motion might inflate outcomes, though it is important to recognize the limitations of our study. The total duration of our simulated respiratory data was limited, the patient derived signal traces were not based on quantitative SI and AP displacements as the Anzai belt's output is non-quantitative, and our simulated phantoms lacked diversity. We manually selected 15s subsets of the patient recordings that represented diverse motion scenarios. These factors may restrict the study's ability to fully capture the variability of respiratory motion encountered in clinical practice, but nonetheless clearly demonstrate the benefits of incorporating real motion data over simplified simulations. Future work could include using measurements from different motion tracking systems to obtain quantitative inputs for the XCAT phantom, simulating longer motion signals and more diverse anatomies.

# CONCLUSION

Our findings highlight the significant advantage of incorporating real patient respiratory data over the default respiratory motion trace contained in the XCAT phantom library when creating Monte Carlo Simulations for SPECT MPI.

# ACKNOWLEDGEMENTS AND CONFLICT OF INTEREST

We thank Paul Segars for providing the XCAT licenses.
The authors declare no conflict of interest related to this study.

# Phantom for training in breast imaging techniques

Nikolay Dukov, Kristina Bliznakova, Zhivko Bliznakov

## A. BACKGROUND AND PURPOSE

The educational activities performed during practical exercises of future radiographers are closely linked to the utilization of physical phantoms. These phantoms offer the advantage of repeated use with X-rays, thereby minimizing unnecessary exposure of patients to radiation while simultaneously enhancing the trainees' proficiency in radiographic techniques. The advancements in 3D printing technology have paved the way for the creation of anthropomorphic phantoms, as well as phantoms tailored for educational purposes, surgical planning and simulation, medical device development, 3D biopritnting, quality control and specialized tasks such as refining imaging procedures[1-6]. 3D printing technology in creating custom phantoms for use in the quality control of artificial intelligence (AI) models in medical imaging has also been reported; however, this field has yet to be explored[7,8].

In today's era, access to high-precision 3D printers has become widespread, making them suitable for fabricating physical phantoms tailored specifically for educational and training purposes. For instance, the utilization of polylactic acid (PLA) in 3D printing has revolutionized the production of realistic and cost-effective prostate phantoms, which find extensive application in training programs for ultrasound-guided interstitial prostate brachytherapy[9]. Moreover, the integration of 3D-printed patient-specific skull and heart models has emerged as an indispensable tool in medical education. These models not only enhance the teaching of anatomy but also aid in surgical planning, providing a versatile platform for practicing many of medical procedures and manipulations[1,10].

In the field of breast imaging, 3D-printed breast phantoms are extensively employed in research activities aimed at developing novel techniques and refining existing ones[11,12]. However, their utilization in training programs focused on breast imaging remains relatively limited. The ELPIDA team is involved in the development of anthropomorphic breast phantoms for both, research and educational purposes. While phantoms intended for scientific research and quality control often utilize rigid plastic materials lacking elasticity, those employed in training students in mammography techniques necessitate elastic properties. These materials must exhibit the ability to rebound to their original state after compression, akin to the behavior of real breast tissue. Furthermore, it is advantageous for these materials to closely emulate the radiological properties of authentic breast tissues, thereby enhancing the fidelity of training simulations.

The ELPIDA group has previously engineered a phantom tailored for training radiographers on mammography simulators. This innovative design incorporated a flexible resin for printing the external breast shape, complemented by the use of water-absorbing polymer beads[13,14]. While it is well suited for the training of the compression with the mammography equipment, this phantom presented an unrealistic mammography image.

The aim of this study is to develop and initially assess an accessible and practical breast phantom designed to enhance the preparation of future radiographers and radiologists in the acquisition of mammography images and skills in breast biopsies.

## B. METHODS

### 1. Digital model

We developed a computational model of an uncompressed breast phantom (see Figure 1) using a dedicated software application called *BreastSimulator*[15]. The glandular structures of the breast were also considered, which in our case served also the purpose of the supporting structures necessary for the successful printing of the model. Additionally, a mathematically modeled tumor was generated separately from the breast model. Multiple spheres with varying diameters, including sizes of 1.6 mm, 3.2 mm, 6.4 mm, 9.6 mm, 12.8 mm, and 16 mm were separately modelled. These diameters were chosen based on our previous investigations[16,17]. Both, the tumor entity model and sphere models, are shown in Figure 2. The spheres were specifically designed to occupy the volume of the breast model, contributing to the creation of a texture resembling adipose tissue within the breast. The resulting computational models were converted to STL format using Fiji software[18].

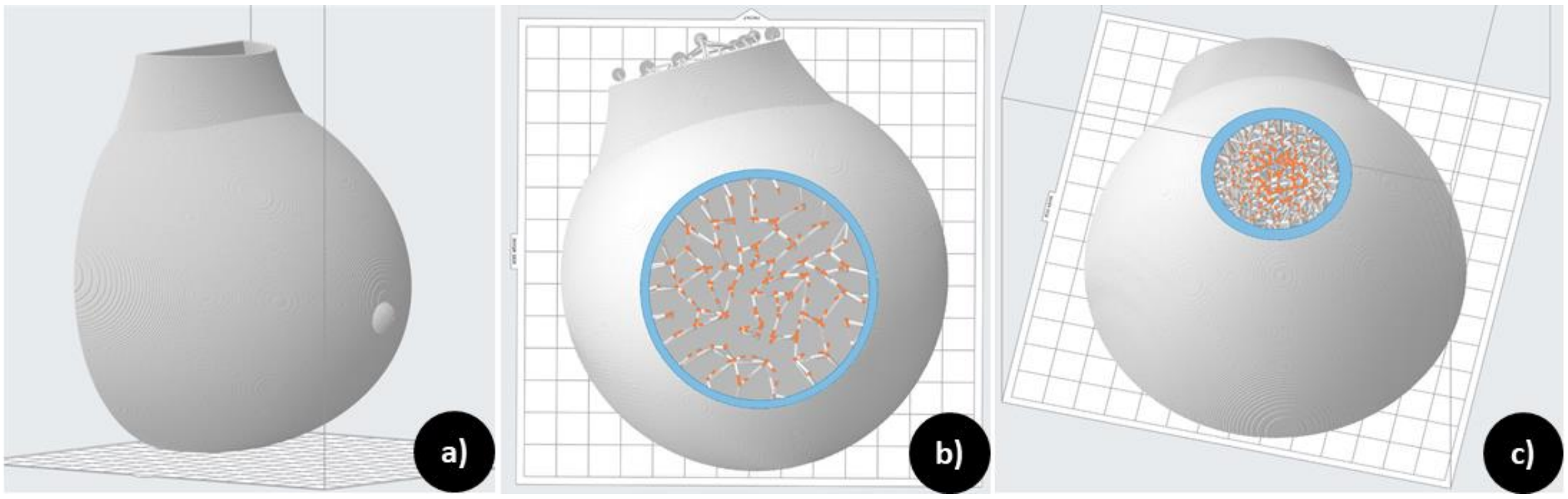

*Figure 4. Computational model of (a) the uncompressed breast phantom, (b-c) cross section of the breast model prepared for printing with a slicing software, the supporting structures added inside the model are visible, which were used also to mimic the glandular structures in the breast.*

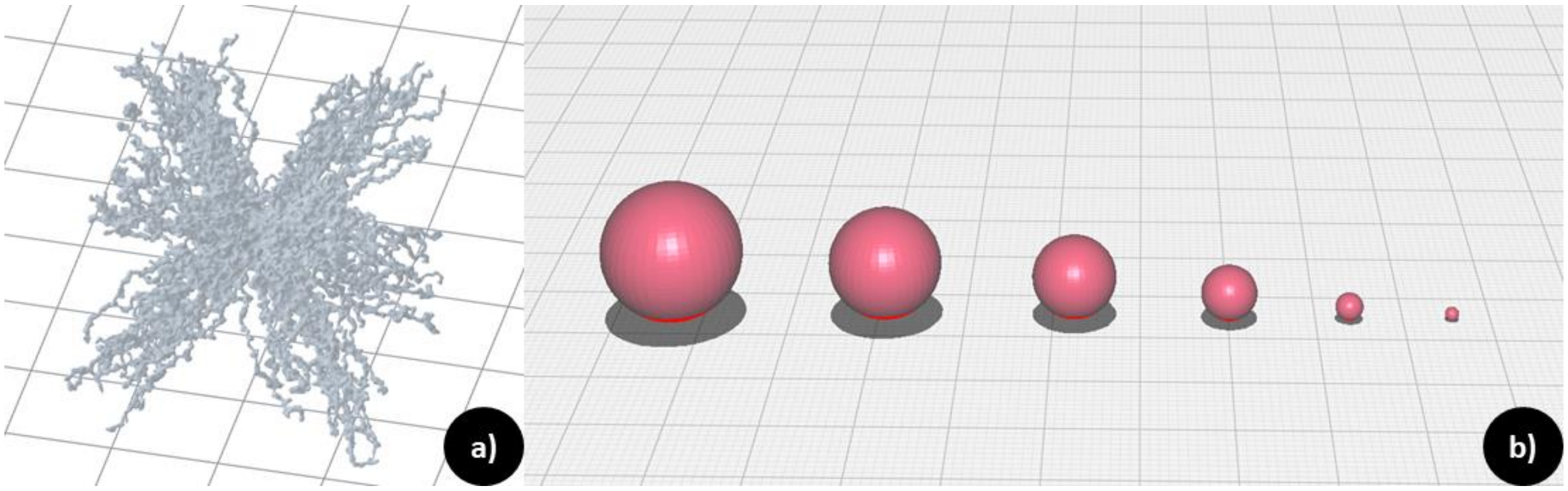

*Figure 5. Computational models of (a) the tumor entity and (b) the six types of sphere diameters used to fill the breast volume.*

### 2. Physical model

To physically create the phantom, we utilized a stereolithography (SLA) 3D printer for the shape (skin) of the breast model and the glandular tissue, as well as the tumor entity, while the spheres were printed with fused filament fabrication (FFF) printer. The printing material used with the SLA printer was resin with a shore rating of 50, suitable for flexible soft parts. The height of the printing layer was set to 100 µm. The spheres were printed with flexible filament, a layer height of 200 µm and 40% infill. For their printing no outer walls, as well as no top and bottom layers were used. This settings result with the only part of the sphere being printed, being the infill, which in the case was set to cross 3D pattern.

Upon integrating the 3D printed spheres into the breast phantom, any remaining air within the breast model was displaced by supplementing the residual volume with olive oil. This integration of 3D printed spheres and olive oil aims to replicate the characteristics of adipose tissue, potentially simulating its properties. Subsequently, the physical breast phantom was imaged using a clinical mammography unit operating in automatic mode, following standard protocol for mammary gland imaging.

## C. RESULTS

The resulting anthropomorphic breast phantom was semi-transparent, enhancing the overall visual experience. The components of the phantom are shown in Figure 3. Fabricated at a cost of under 100 euros in materials, it was utilized in radiographer training for mammography examination. The X-ray projections are depicted in Figure 3d-e, where Figure 3d is the raw image from the mammography unit, while the X-ray image in Figure 3e is the resulting processed image.

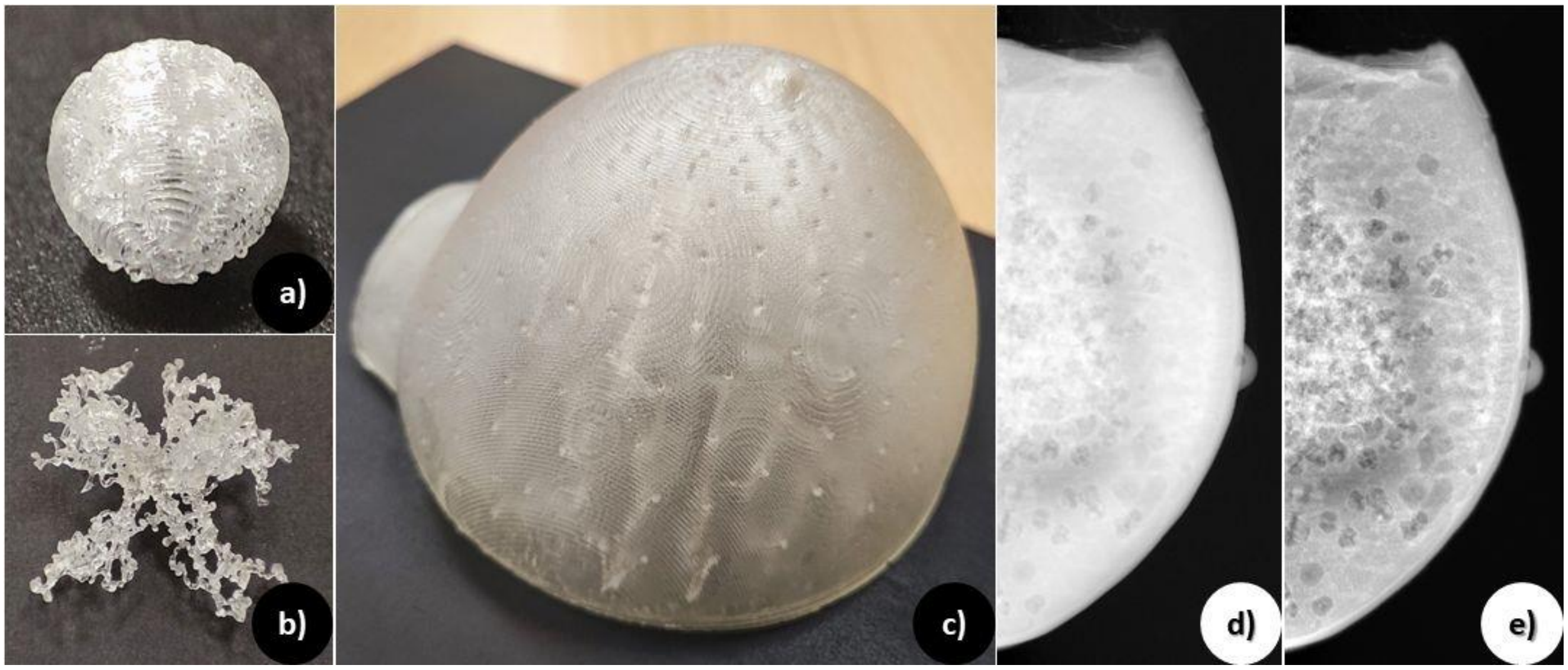

*Figure 6. Printed physical models of the (a) spheres, (b) tumor entity and (c) shape, as well as the (d) unprocessed and (e) processed X-ray mammography projections of the phantom.*

## D. DISCUSSION

The phantom was successfully used in mammography training of future radiographers in terms of training both compression and imaging skills. Furthermore, breast biopsy training was also initiated with the phantom, and while the suitability of the used materials and their durability is still under evaluation, there has been a positive initial feedback from the involved radiologists.

The resulting X-ray images show promise in the undertaken approach in utilizing the support structures needed for printing the model as glandular tissue. The printed spheres used for filling the volume of the breast model are soft and compressible in all directions, while the olive oil fills the remaining volume and gaps in the spheres. However, as depicted in Figure 3e-d, it is evident that not all of the spheres have effectively absorbed the olive oil, leaving some residual air. This leads to deviations from the characteristics of human breast adipose tissue.

Ongoing tests and evaluation of the phantom showed that it can withstand multiple compressions and irradiations, although the exposure to X-rays results in a noticeable visual change in the color of the used material for the skin to yellow. Further, due to the multiple compressions, some tearing is becoming evident in the upper part of the base of the phantom. Moreover, as a consequence of repeated compressions, there is a noticeable tearing emerging in the upper section of the phantom's base. Nevertheless, despite this issue, the phantom remains functional. Over twenty students and professionals have conducted compressions on the phantom thus far. The tearing is attributed to tension concentrated in that specific area, which will be addressed in subsequent iterations of the computer model's design.

When used for breast biopsy scenarios the inflicted damage from the needle on the phantom significantly diminishes its operational life to just a few manipulations. As a future iteration of the physical breast phantom, we are looking into silicone resins as a material for the skin.

Future work is related to the computer modelling and simulations using this phantom, with the objective of enhancing the imaging appearance of simulated breast structures and understanding the impact of compression force used in breast imaging on material elasticity. The ultimate goal is to optimize both the materials utilized for printing the external shape and those mimicking the internal breast structures.

## E. CONCLUSION

An anthropomorphic breast phantom was developed for mammography studies. The model was designed to be compressed and to quickly restore to its original form. Furthermore, although the X-ray imaging results are not identical to human breast anatomy, they show promise and will be further improved. An extensive feasibility evaluation with radiographers and radiologists is currently in progress.

## F. ACKNOWLEDGEMENTS AND CONFLICT OF INTEREST

This study is financed by the European Union-NextGenerationEU, through the National Recovery and Resilience Plan of the Republic of Bulgaria, project № BG-RRP-2.004-0009-C02. This study is delivered by the ELPIDA group.

The authors declare that they have no conflict of interest.

## G. REFERENCES

# Advancing in the physical fabrication of anthropomorphic breast phantoms

Kristina Bliznakova, Nikolay Dukov, Minko Milev, Zhivko Bliznakov

## BACKGROUND AND PURPOSE

Physical anthropomorphic breast phantoms are a key instrument for performing clinical and research work carried out in radiological departments[1-3]. Major challenge and requirement in the field of X-ray breast imaging is the use of materials exhibiting similar X-ray absorption coefficients to those of breast tissue, especially of these to be used in a wide energy kilovoltage (kV) diagnostic range. There are several methods for assessing the appropriateness of a particular material. One such method involves fabricating sample materials and analyzing their attenuation characteristics. This analysis typically involves calculating linear attenuation coefficients[4,5] and determining Hounsfield Units (HU) values through Computed Tomography (CT)[6-9] experiments.

Previously, we reported on two methods for manufacturing of physical anthropomorphic breast phantoms utilizing fused filament fabrication technology. In both cases, modified 3D printers and a custom-developed software were used[7,9]. The latter plays an important role in linking a series of breast patient CT scan images directly to the 3D printing process. One method involved the use of a single filament, a polylactic acid (PLA) one, with a constant filament extrusion rate for each voxel. This ensured that a precise amount of melted filament was extruded to every voxel corresponding to specific tissues, alongside a perimetric pattern to replicate smaller entities with more irregular shapes, such as glandular and tumor tissues[7].

The second approach utilized two filaments to replicate the breast tissues, a PLA filament and a polypropylene (PP) filament. Each layer of the phantom was printed twice, once with PLA and once with PP. The extrusion rate for each material was meticulously controlled voxel-by-voxel, determined by the HU of the imported CT images.

Extensive preparatory work preceded the final printing of the breast phantoms, including the production of cubes with varying percentage filling. These cubes were subsequently scanned at CT facilities to assess the HU of each printed cube. Following the proposed methodologies, the phantoms were successfully produced and evaluated. However, the phantoms manufactured with these methodologies are specifically tailored for use with a particular energy of the X-ray beam. Thus, for each X-ray beam, a new breast phantom would need to be printed.

An ideal breast tissue-mimicking phantom should be created from materials demonstrating HU values close to these of the human breast tissue at typical diagnostic CT energies, especially when printed at 100% infill density. This pursuit extends to design of breast phantoms, necessitating the identification of materials that faithfully mimic tissue characteristics. Our recent investigation demonstrates that for creating an anthropomorphic breast phantom with distinct filaments representing different tissue components (such as adipose and glandular), an effective solution involved utilizing Acrylonitrile Butadiene Styrene (ABS) for breast adipose tissue and Acrylonitrile Styrene Acrylate (ASA) for glandular breast tissue. This study presents the results for a new physical anthropomorphic breast phantom, developed with these materials used with FFF 3D printers.

# METHODS

In our recent investigation[10], we experimentally studied the radiological properties of twenty-two 3D printing materials, used with FFF printing technology. We precisely determined the HU values of these materials by scanning test samples at a clinical CT within the range of incident x-ray beam with anode voltages of 80 kV and 120 kV. The CT numbers for each material were analyzed, and sets of filaments were identified as potential representatives of breast tissues: Thermoplastic copolyester (TPC, https://formfutura.com/c/filaments/tpc-filaments/) and ASA as potential representatives for glandular tissue and High Impact Polystyrene (HIPS) and ABS as representatives of the adipose tissue. In this study we chose to manufacture a breast phantom with ASA and ABS materials.

The computational breast model in this study is derived by applying segmentation techniques to a patient MR image[7]. The resulting digital phantom contains two distinct types of tissues: adipose and gland. The glandular tissue was assigned to the skin tissue to simplify the production model. Subsequently, one pair of identified suitable materials (ASA, ABS) was used to replace the corresponding breast tissues, as illustrated in Figure 1.

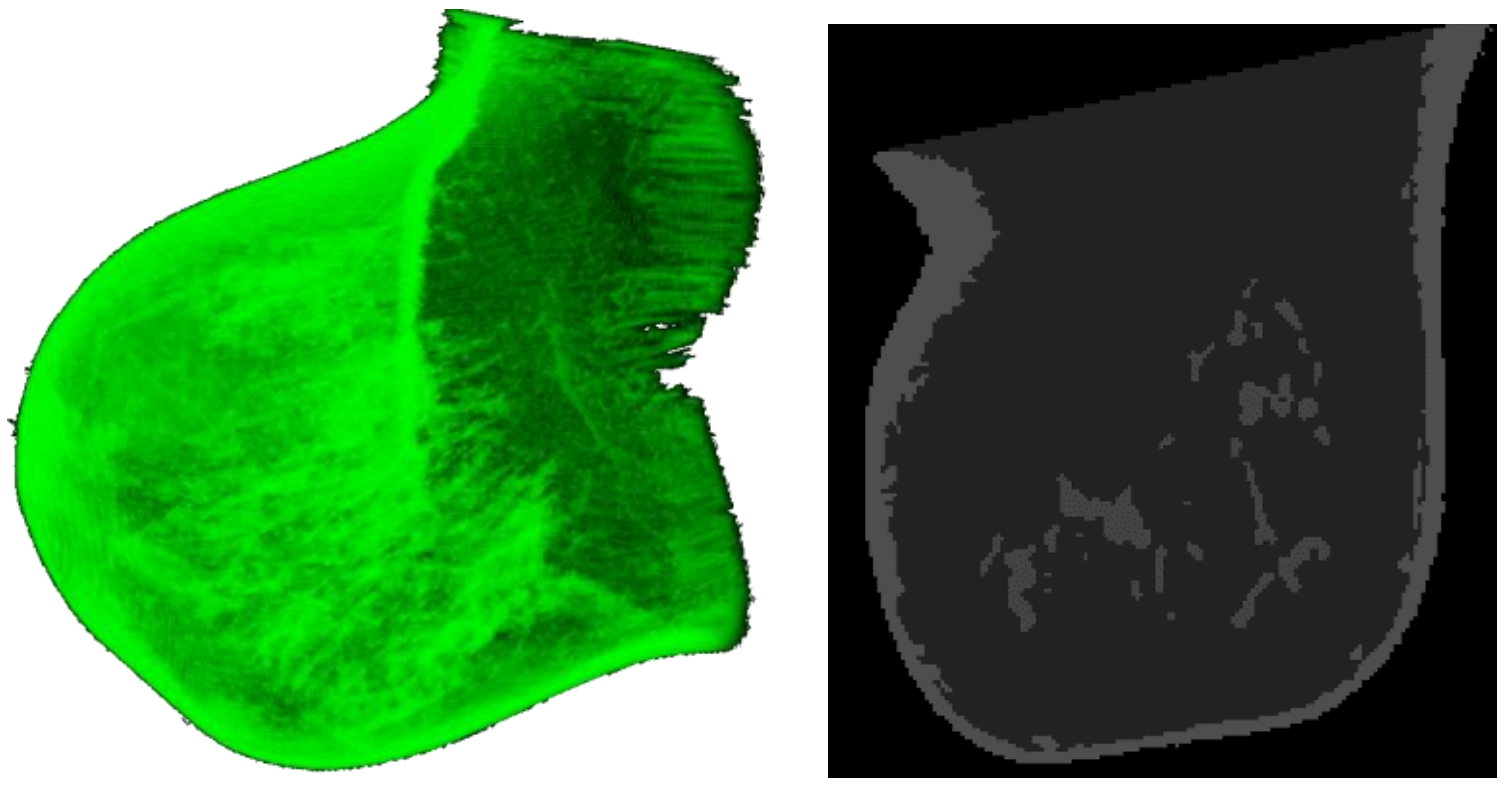

*Figure 7. Computational breast phantom (left) and a slice from the segmented breast (right).*

In this study, an FFF dual extrusion 3D printer is utilized. The phantom was printed with X1-Carbon Bambu Lab printer. The selection of ASA and ABS, offered a distinct advantage due to their closely matched printing temperatures of 240°C[10].

The assessment was conducted at a clinical CT unit, the Siemens Somatom Force, under three distinct tube voltages: 70 kVp, 120 kVp, and 150 kVp. A slice thickness of 0.5 mm was maintained throughout the scans, following the Thorax standard protocol. The distances between the source and the patient, as well as between the source and the detector, were 595 mm and 1085.6 mm, respectively. Image slices were sized at 523 pixels x 512 pixels, with each pixel measuring 0.285 mm by 0.285 mm, resulting in a width of 149 mm and a height of 146 mm.

# RESULTS

Figure 2 displays the printed anthropomorphic breast phantom. The white material of the breast phantom corresponds to ASA, while the green material corresponds to ABS. The printing process required a total of 4 days to complete. As a preparatory measure, preliminary slices were printed in advance and subsequently evaluated using a radiography system, as shown in Figure 2. This approach aimed to visually ascertain whether the selected materials accurately replicated the radiological appearance of breast tissues.

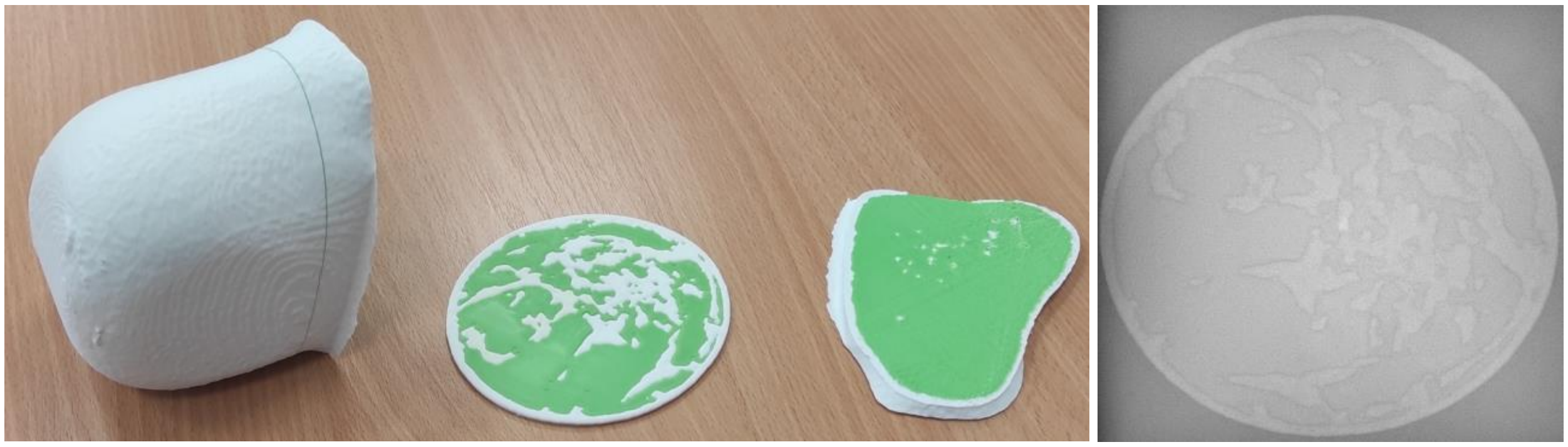

*Figure 8. Printed breast phantom and two printed slices. The radiography image on the right is the x-ray image of the slice, shown in the middle.*

CT images, reconstructed from the 70 kV set of the scanned breast are shown in Figure 3. The values of the HU of the glandular representation were assessed across 10 slices within Regions of Interest (ROIs) located at the skin level. Each ROI measured 8 x 8 pixels. Meanwhile, the HU of the adipose simulation was measured through a single ROI, spanning 100 x 100 pixels, across 15 consecutive slices for each imaging set. Further on, to evaluate the average breast tissue, dedicated software designed for feature extraction from X-ray images was utilized[11] as illustrated in Figure 4. Results are summarized in Figure 5 for the three different kV imaging sets.

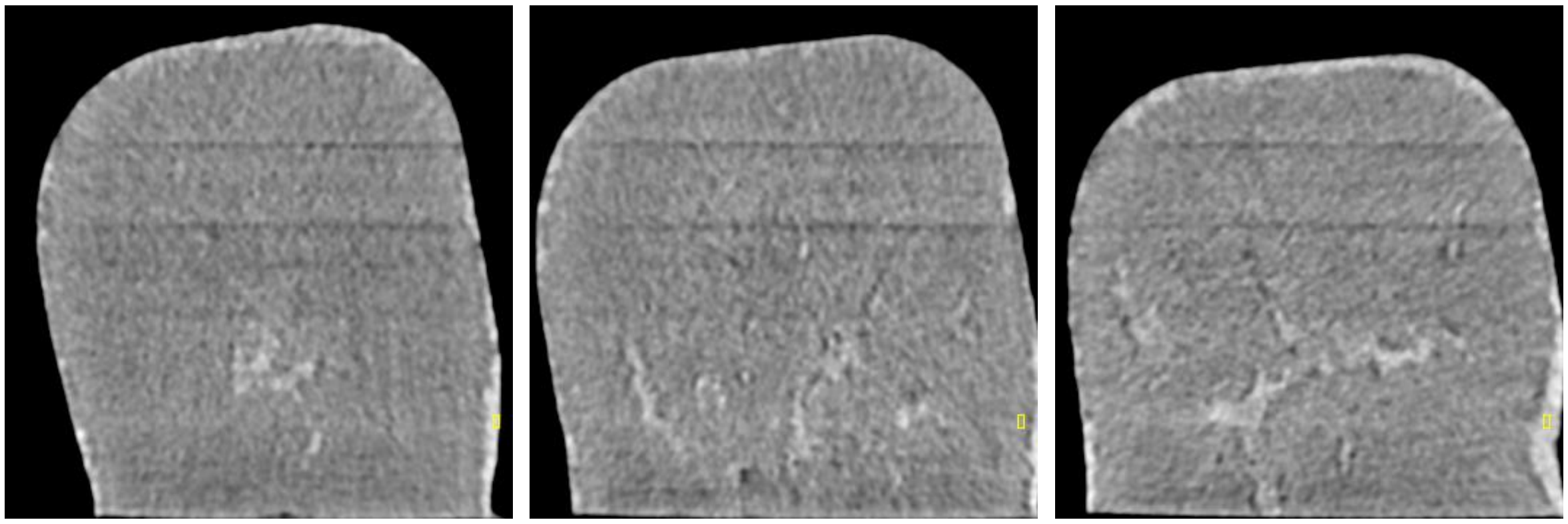

*Figure 9. CT images, reconstructed from projections acquired at 70 kV.*

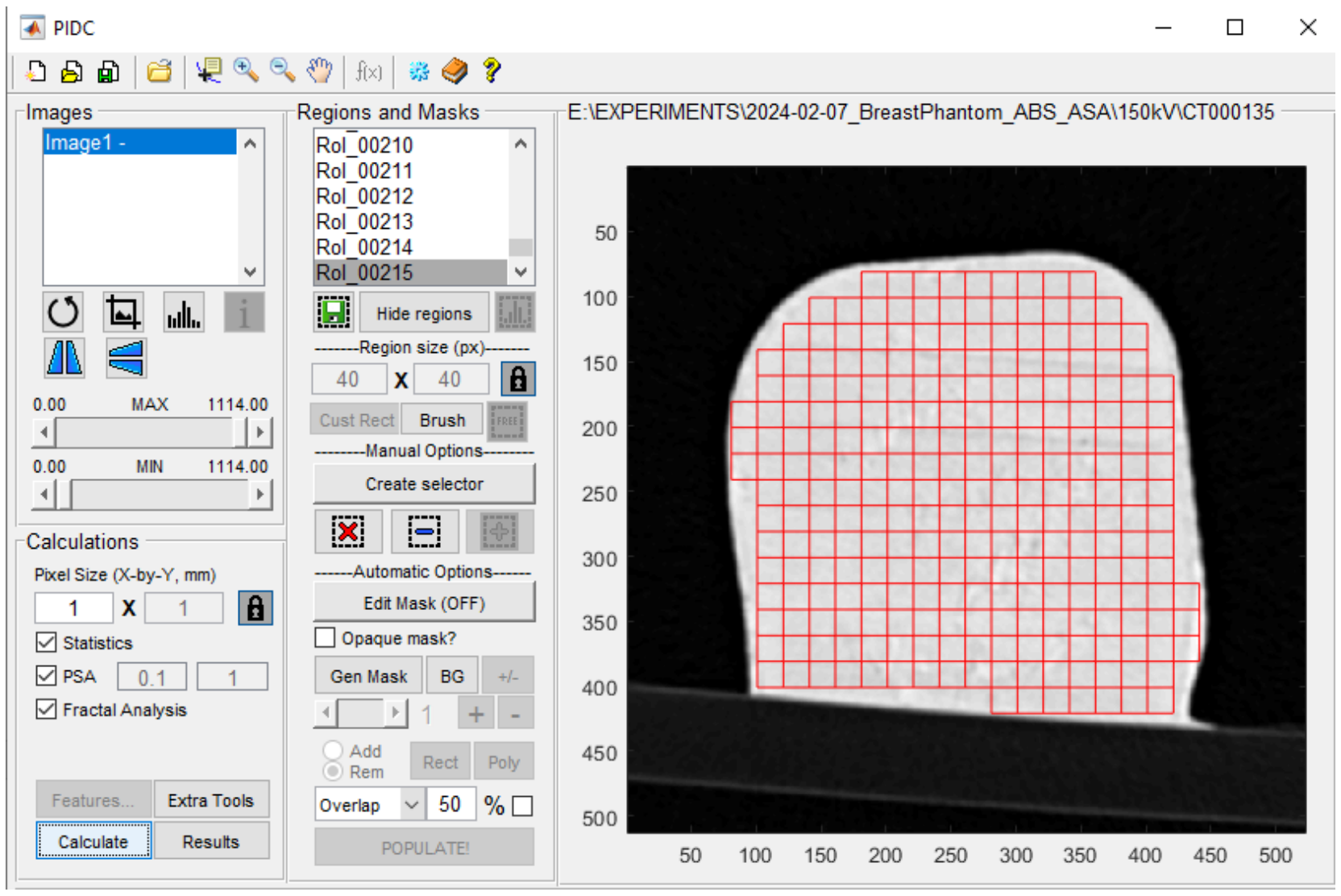

*Figure 10. Use of a dedicated software to measure the HU of the materials approximated by ASA and ABS.*

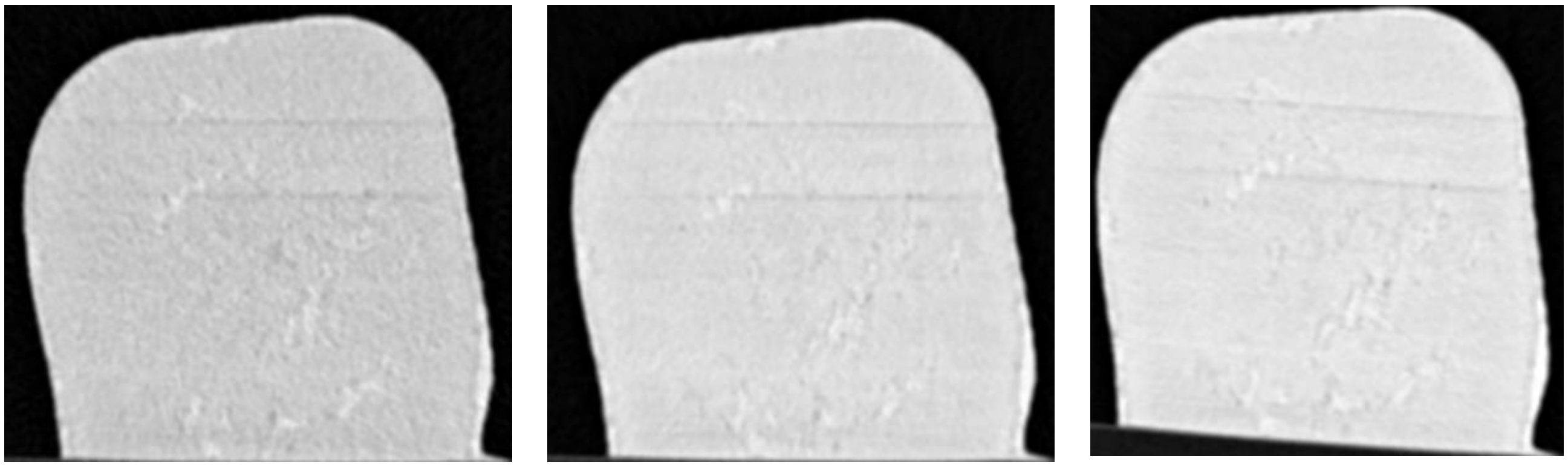

| | | |
|---|---|---|
| Mean HU = -114.86 ± 7.03 | Mean HU = -72.76 ± 5.62 | Mean HU = -63.51 ± 6.70 |
| Adipose HU = -116.142 ± 1.15 | Adipose HU = -75.89 ± 0.22 | Adipose HU = -60.82 ± 0.33 |
| Glandular HU = 45.80 ± 10.419 | Glandular HU = 57.94± 10.34 | Glandular HU = 65.87 ± 6.72 |
| **70 kV** | **120 kV** | **150 kV** |

*Figure 11. Measured HU values, based on CT images of the breast phantom for the three different kV imaging sets.*

# DISCUSSION

The range of measured CT numbers for the replicated adipose breast tissue is between -116 HU and -60 HU for the incident beams of 70 kV-150 kV. The observed trend of increasing the HU with increasing the kV well compares with the one reported by Ma *et al.* [6], where the authors modelled the HU of the adipose tissue in kV as a sum of the HU, measured at 120kV and a kV dependent difference function Δ. For the representation of glandular tissue, the measured values within the mentioned kV range ranged from 46 HU to 66 HU. Further experimental studies are necessary to validate the observed trend of increasing HU values of glandular tissue with higher kV settings. Ma *et al.* [6] reported no energy dependence of the CT values with the kV for porcine leg, bovine tail and bovine ribs soft tissue samples. Other studies have shown that HU values for adipose tissue range from approximately -138 (at 80 kV) [7] to -100 (at 120 kV)[12], while fibroglandular breast tissue is reported to be around 40 HU[7,12] for incident beams ranging from 80kV to 120kV. However, data for glandular tissue are limited, hindering comprehensive comparison.

Currently, ongoing work is focused on producing three additional breast phantoms based on a combination of TPC and ASA with HIPS and ABS, and their comparison with the two methods proposed by the PHENOMENO team.

## CONCLUSION

Utilizing ASA and ABS materials has proven effective in manufacturing an anthropomorphic breast phantom capable of reproducing radiological properties of the breast tissue across a broad spectrum of X-ray photon energies. Ongoing research activities will further clarify the optimal techniques and materials with the FFF printing technology for fabricating physical breast phantoms.

## ACKNOWLEDGEMENTS AND CONFLICT OF INTEREST

The PHENOMENO project "Physical breast anthropomorphic models and technology for their production" has received funding from the European Union's Horizon 2020 research and innovation programme under the Marie Skłodowska-Curie grant agreement No 101008020.

The authors declare that they have no conflict of interest.

# Creating a realistic virtual cohort from historical data

Authors: W. Gohn[1], W. McDougald[1], M. Reymann[2,3], F. Massanes Basi[1], A.H. Vija[1]

[1]Siemens Medical Solutions USA, Inc, Hoffman Estates, IL, USA

[2]Siemens Healthineers AG, Forchheim, Germany

[3]Friedrich - Alexander Universität Erlangen – Nürnberg, Germany

## BACKGROUND AND PURPOSE

For virtual imaging trials (VIT) it is important to create a realistic virtual cohort. Often cohorts are populated with some statistical distribution to represent patient demographic data such as height, weight, and other variables. We improve on this method by instead sampling from our large database of historical SPECT data. By pulling real demographic data to feed to the simulation, we have a more realistic cross-section of the population that might participate in a clinical trial.

## METHODS

Data has been extracted from the DICOM headers of more than 3500 studies and used to populate a NoSQL database. We randomly sample from the database to create sets of metadata for each patient in the VIT. For each patient sampled in the database, a virtual patient is constructed with the same demographic information (height, weight, age, and sex). This method creates a random and yet realistic virtual cohort.

## RESULTS

By sampling patient metadata, we create a virtual cohort that maintains correlations between demographic data, and yet is completely realistic. It allows us to specify certain groups for a cohort based on age or location,

## CONCLUSION

The utilization of historical data to create a virtual cohort allows for the ability to perform a more realistic VIT. Data for each virtual patient is random and yet properly correlated. VITs can thus be used to test an application on certain demographic segments of the population.

# Reconstruction kernel matching for end-to-end virtual imaging trials in X-ray CT

Darin P. Clark
Quantitative Imaging and Analysis Lab
Center for Virtual Imaging Trials, Carl E. Ravin Advanced Imaging Laboratories
Department of Radiology, Duke University, Durham, NC, USA

Cristian T. Badea
Quantitative Imaging and Analysis Lab, Dept. of Radiology, Duke University, Durham, NC, USA

## BACKGROUND AND PURPOSE

Recently, the digital XCAT human phantom,[1] the DukeSim CT scanner simulator[2] (cvit.duke.edu/resources/), and the open-source Multi-Channel CT Reconstruction (MCR) Toolkit[3] (gitlab.oit.duke.edu/dpc18/mcr-toolkit-2) have been combined for state-of-the-art virtual CT imaging trials.[4] The MCR Toolkit provides robust reconstruction options for 3rd generation, multi-detector row clinical CT, but discrepancies remain between Toolkit and commercial CT reconstructions. In this work, we propose two strategies for reconstruction kernel fitting to reproduce spatial resolution and noise properties of commercial CT reconstructions, yielding more realistic digital surrogates of clinical CT. With the first strategy, we propose three parametric reconstruction kernel prototypes which govern the trade-offs between noise and spatial resolution in reconstructed images, and we show that one of these prototypes can be used to fit a clinical reconstruction kernel. With the second strategy, we propose the use of a surrogate, digital contrast and resolution phantom and a reconstruction kernel optimization procedure to reproduce clinical reconstruction kernels when real clinical projection data is unavailable.

## METHODS

In Section 1 we detail three parametric reconstruction kernel window prototypes included in the MCR Toolkit for kernel fitting. In Section 2 we detail an open-ended kernel fitting procedure which uses a surrogate digital phantom. Finally, in Section 3 we describe two experiments which illustrate the strengths and weaknesses of the proposed approaches.

### Approach 1: reconstruction kernel window prototypes

We propose three parametric reconstruction kernel windows. When multiplied element-wise with a standard ramp filter,[5] these window functions trade between spatial resolution and image noise in reconstructed images (Figure 1). The "Smooth" kernel window is a Gaussian function with a full width at half maximum of $2 \cdot \mathrm{w}$ and evaluated at spatial frequencies, f:

$$\sigma(\mathrm{w}) = \frac{2\mathrm{w}}{2\sqrt{2\log_e(2)}}, \quad Smooth(\mathrm{f},\mathrm{w}) = \exp\left(\frac{-\mathrm{f}^2}{2\sigma(\mathrm{w})^2}\right).$$

Smooth reconstruction kernels prioritize noise reduction over spatial resolution. Conversely, "Sharp" reconstruction kernels preserve spatial frequencies below a target value, m, and then transition smoothly toward zero to control noise at high spatial frequencies:

$$Sharp(\mathrm{f},\mathrm{m},\mathrm{w}) = \begin{cases} 1.0, & \mathrm{f} \leq \mathrm{m} \\ \exp\left(\frac{-(\mathrm{f}-\mathrm{m})^2}{2\sigma(\mathrm{w})^2}\right), & \mathrm{f} > m \end{cases}.$$

Finally, "Enhancing" reconstruction kernels emphasize task-relevant spatial frequencies. Frequencies are enhanced around spatial frequency, $\mathrm{m}_1$, with scaling factor, s. The window's modulation transitions to zero at spatial frequency, $\mathrm{m}_2$.

$$Enhancing(\mathrm{f}, \mathrm{s}, \mathrm{m}_1, \mathrm{m}_2, \mathrm{w}_1, \mathrm{w}_2) = \begin{cases} \mathrm{s} \cdot \exp\left(\frac{-(\mathrm{f}-\mathrm{m}_1)^2}{2\sigma(\mathrm{w}_1)^2}\right) - \exp\left(\frac{-(\mathrm{f}-\mathrm{m}_2)^2}{2\sigma(\mathrm{w}_2)^2}\right) + 1, & \mathrm{f} \leq \mathrm{m}_2 \\ 0, & \mathrm{f} > \mathrm{m}_2 \end{cases}.$$

Notably, enhancing kernels can skew reconstructed Hounsfield Unit (HU) values, making them inappropriate for imaging tasks involving quantitative measurements of HU values. This is illustrated in Figure 1, where the enhancing kernel reconstruction shifts the HU values measured in the left ventricle from 23.5 HU to 34.1 HU.

## Approach 2: surrogate digital phantom

The second kernel fitting procedure takes Modulation Transfer Function (MTF) and Noise Power Spectrum (NPS) measurements as inputs. In practice, these measurements would be taken in a physical quality assurance phantom scanned and reconstructed using the commercial system to be emulated. The MCR Toolkit forward projection and reconstruction parameters are adjusted to best match the commercial settings (similar or exact source-detector configuration, identical reconstruction slice thickness, etc.). A digital quality-assurance phantom can then be used to quantify MTF and NPS discrepancies between the commercial and Toolkit reconstructions.

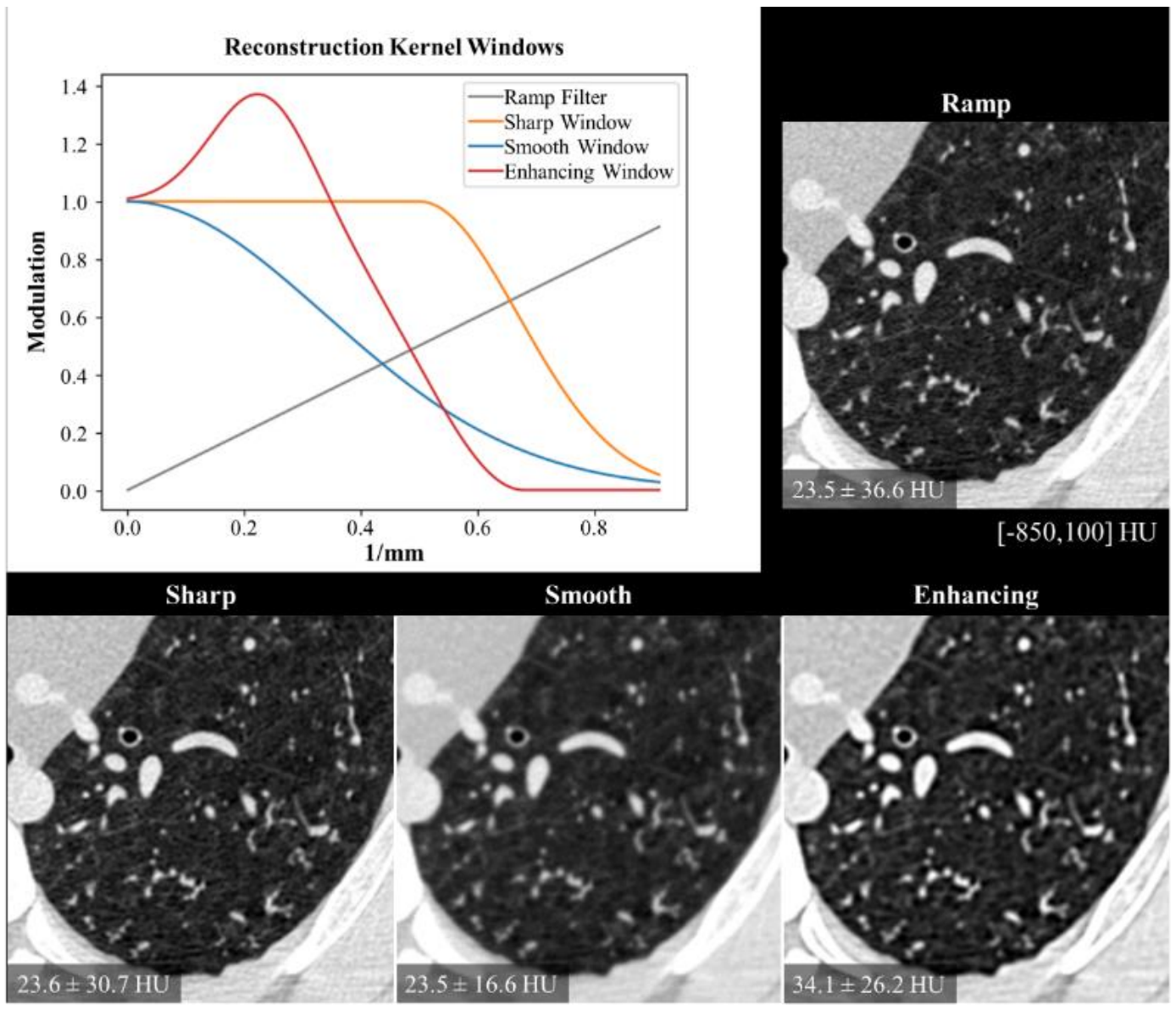

*Figure 1. Proposed reconstruction kernel windows. Sample reconstructions are shown for an axial region of interest in the left lung of an XCAT phantom modeling emphysema. Mean ± noise standard deviation measurements are reported for a region of interest in the left ventricle of the heart (reported at bottom left of each reconstruction; measurement region not shown). Kernel parameters shown (units of mm$^{-1}$): Sharp, w=0.2, m=0.5; Smooth, w=0.4; enhancing, s=0.4, $m_1$=0.23, $m_2$=0.68, $w_1$=0.1, $w_2$=0.2.*

In this work, we employ the Nelder-Mead optimization algorithm[6] to solve for the parameters of the reconstruction kernel window to best match the Toolkit reconstruction MTF and NPS measurements with those from a reference scan. The reconstruction kernel window is parameterized with 10 evenly spaced control points spanning spatial frequencies from 0 mm$^{-1}$ to the Nyquist frequency determined by the effective detector pixel width at the scanner's isocenter. The parameter at each control point corresponds with the modulation of the kernel window (Figure 1). The 10-parameter kernel window is then interpolated to intermediate spatial frequencies using the MATLAB software package (R2023a; The Mathworks, Inc.) and its implementation of modified Akima interpolation (avoids under, overshoots when slope changes rapidly).[7] To reproduce the target MTF and NPS measurements in the digital phantom, the following cost function is minimized:

$$cost(\mathbf{n}_\mathrm{d}, \mathbf{m}_\mathrm{d}, \mathbf{t}_\mathrm{d}, \mathbf{p}) = \frac{\|\mathbf{n}_\mathrm{d} - \mathbf{n}_\mathrm{r}\|_2}{\|\mathbf{n}_\mathrm{r}\|_2} + \frac{\|\mathbf{m}_\mathrm{d} - \mathbf{m}_\mathrm{r}\|_2}{\|\mathbf{m}_\mathrm{r}\|_2} + 10 \cdot \frac{\|\mathbf{t}_\mathrm{d} - \mathbf{t}_0\|_2}{\|\mathbf{t}_0\|_2} + 0.02 \cdot \sum|\mathbf{p}|.$$

The 'd', 'r', and '0' subscripts denote measurements from the digital phantom reconstructed with the parameterized window, the real phantom, and the digital phantom reconstructed with a ramp filter, respectively. The vector $\mathbf{n}$ is the reshaped 2D NPS after adding one and computing the natural log, the vector $\mathbf{m}$ contains MTF measurements, and the vector $\mathbf{t}$ contains the mean intensity value of each reconstructed axial slice. The vector $\mathbf{p}$ contains the 10 model parameters. The final cost term penalizes the magnitude of these parameters to encourage modulation values less than 1.

### Kernel fitting experiments

Two experiments were conducted to gauge the fitness of the proposed methods for reconstruction kernel matching. For the first experiment, full-dose CT projection data from a scan of the Gammex 464 phantom was used. The projection data set is available from The Cancer Imaging Archive.[8,9] The projection data was acquired with a Siemens SOMATOM Definition Flash scanner, and the provided reference reconstruction was performed with a B30 kernel, a 250 mm field of view, a 1 mm slice thickness, and a pixel size of 0.49 mm in the axial plane. The objective of the first experiment was to match the spatial resolution and noise power in the reference reconstruction of this phantom using the MCR Toolkit and the proposed kernel window prototypes (Figure 1). To accomplish this objective, the intensity modulation in high-contrast line pairs included in module 4 of the Gammex phantom (4-12 lp/cm) was measured and compared between the reference B30 reconstruction and the MCR Toolkit reconstruction using a ramp filter. The ratios of the B30 values over the Toolkit values were then used as modulation values to be fitted with the kernel prototypes.

The second experiment tested the surrogate digital phantom approach and kernel window fitting using the Nelder-Mead optimization method. Ultimately, this approach is intended for fitting clinical reconstruction kernels when projection data or exact knowledge of the system geometry are unavailable. In this work, however, we provide a proof of concept where the reconstruction to be matched is instead a Toolkit reconstruction performed with a ramp filter followed by convolution with a known edge enhancement kernel in the image domain. The objective is then to reproduce the spatial resolution of this convolved image with only a fitted kernel window applied during reconstruction. For added realism, the "Duke1" CT scanner geometry is used for creating projections and performing reconstruction.[4] Notably, the edge enhancement kernel used in this experiment is not arbitrary. It is used when performing regularized iterative reconstruction of CT data with the MCR Toolkit,[3] and it can be viewed as an image domain analogue of an equivalent reconstruction kernel window.

## RESULTS

Figure 2 summarizes the results of the reconstruction kernel window prototype fitting experiment. In row 1, a 1 mm axial slice through module 4 of the physical Gammex phantom is compared between the MCR Toolkit reconstruction using a ramp filter, the commercial reconstruction using a B30 kernel, and the MCR Toolkit reconstruction using a fitted sharp kernel window ($\mathrm{w} = 0.27\ \mathrm{mm}^{-1}$, $\mathrm{m} = 0.30\ \mathrm{mm}^{-1}$). Comparing the reference B30 kernel MTF values with the MCR Toolkit values (row 1, red text), the fitted sharp kernel window closely reproduces the spatial resolution measurements of the B30 kernel from 4 to 8 lp/cm. Row 2 shows 2D NPS measurements taken in another axial slice of the phantom containing only plastic with water equivalent attenuation. The Toolkit reconstruction with the fitted window visually resembles the reconstruction with the B30 kernel; however, there is a notable discrepancy in the total noise power: 1.2e6 $\mathrm{HU}^2\cdot\mathrm{mm}^2$ with the B30 kernel vs. 2.0e6 $\mathrm{HU}^2\cdot\mathrm{mm}^2$ in the Toolkit reconstruction.

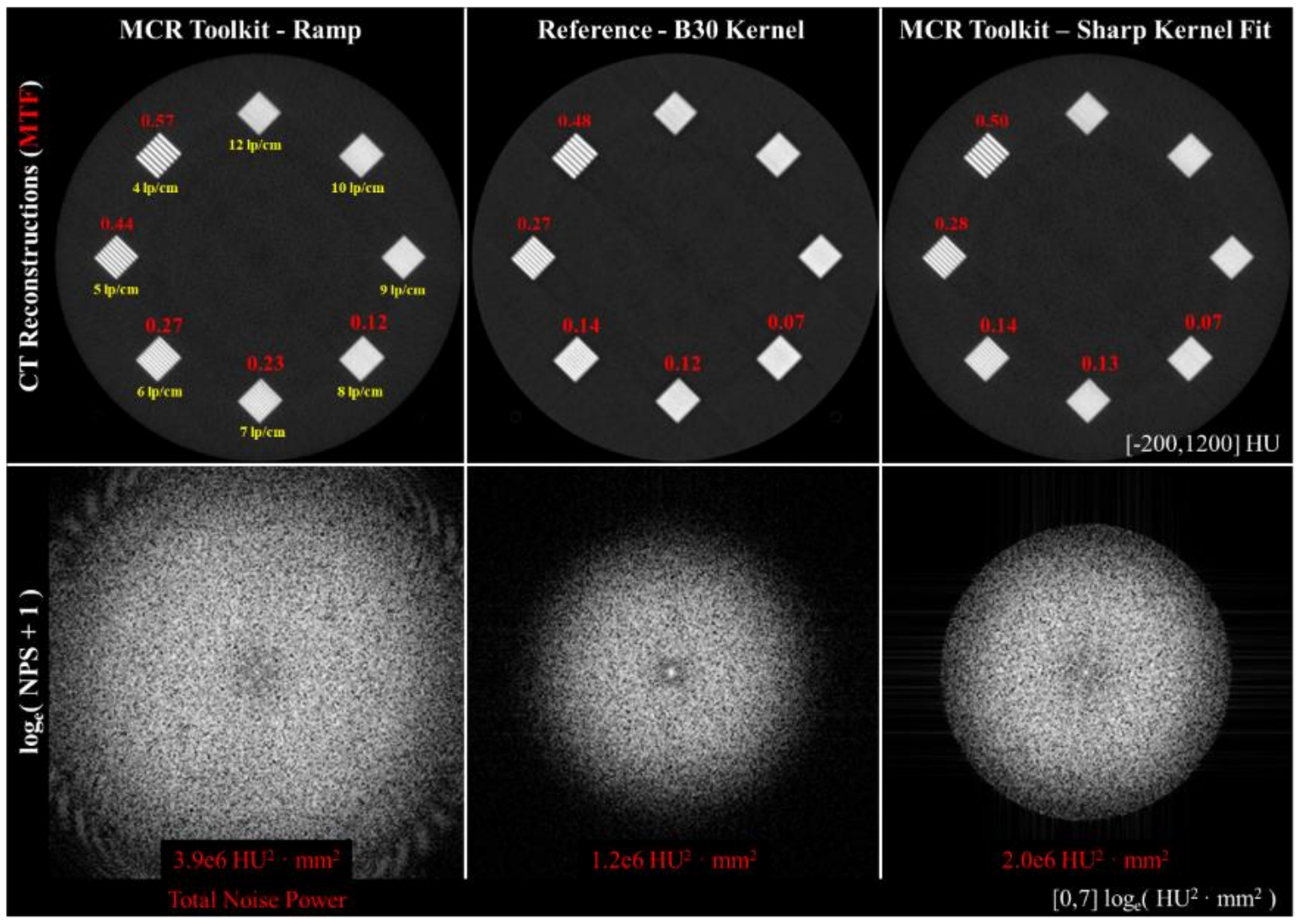

*Figure 2. CT reconstruction results in Module 4 of the Gammex ACR 464 phantom ("high contrast resolution" module). (Row 1) The phantom contains 8 sets of line pairs representing spatial frequencies from 4 to 12 lp/cm (line pairs / cm; yellow labels). Modulation transfer function (MTF) measurements (red text) are compared between an MCR Toolkit reconstruction with a ramp filter, a Siemens reconstruction with the B30 kernel, and an MCR Toolkit reconstruction with the sharp kernel window fitted to reproduce the MTF measurements of the B30 kernel (columns 1-3). (Row 2) Noise power spectrum (NPS) measurements taken in a uniform region in Module 3 of the phantom ("uniformity and noise") for each reconstruction.*

Figure 3 summarizes the results of the surrogate digital phantom experiment. Specifically, it shows the digital phantom used for MTF and NPS measurements (A) and an edge-enhancing 3D convolution kernel, applied in the image domain, which simulates an unknown reconstruction kernel (B, E). Reconstruction residuals are compared without (D, ramp only) and with (F) the fitted kernel window applied during reconstruction. Following kernel fitting, the L2 error norm cost function terms decrease in the MTF measurements (from 25.8% to 0.8%) and in the NPS measurements (4.5% to 0.8%; initial kernel window: inverse ramp). Panel (F) illustrates that the fitted kernel window reconciles the high spatial frequencies; however, some intensity bias (< 5 HU) is visible at the edges of the phantom.

## DISCUSSION

In this work we have discussed two strategies for reconstruction kernel matching to improve the accuracy of virtual imaging trials in X-ray CT. For the first approach, we assume projection data and geometry information are available for the system being modeled, and we propose a series of parametric kernel window prototypes to simplify kernel matching. This approach proved effective for reproducing the spatial resolution of the B30 kernel; however, discrepancies remain in noise power measurements (Figure 2). These discrepancies are likely due to differences in the handling of projection data during reconstruction such as the type of projection rebinning performed, the angular range of data used to reconstruct each axial slice (short scan vs. full scan), and the numerical method used to backproject line integrals.

For the second approach, we assumed a more general scenario where quality assurance phantom scans are available for a CT system, but projection data and exact system geometry may not be. To focus on the method and avoid decoupling of the MTF and NPS measurements, we used an image domain kernel convolution operation in lieu of real clinical data with an unknown kernel. The method was successful in reproducing the image domain kernel with a reconstruction kernel window. Reproducing a complex edge-enhancing kernel is non-trivial, but further work is needed to demonstrate the stability of this optimization procedure using real clinical data where the NPS and MTF may be decoupled, as in the first experiment. Notably, both approaches will likely struggle to reproduce commercial reconstructions with sharp kernels, as discrepancies between exact and assumed system geometries will tend to cause spatial blurring in 3$^{rd}$ party reconstructions.

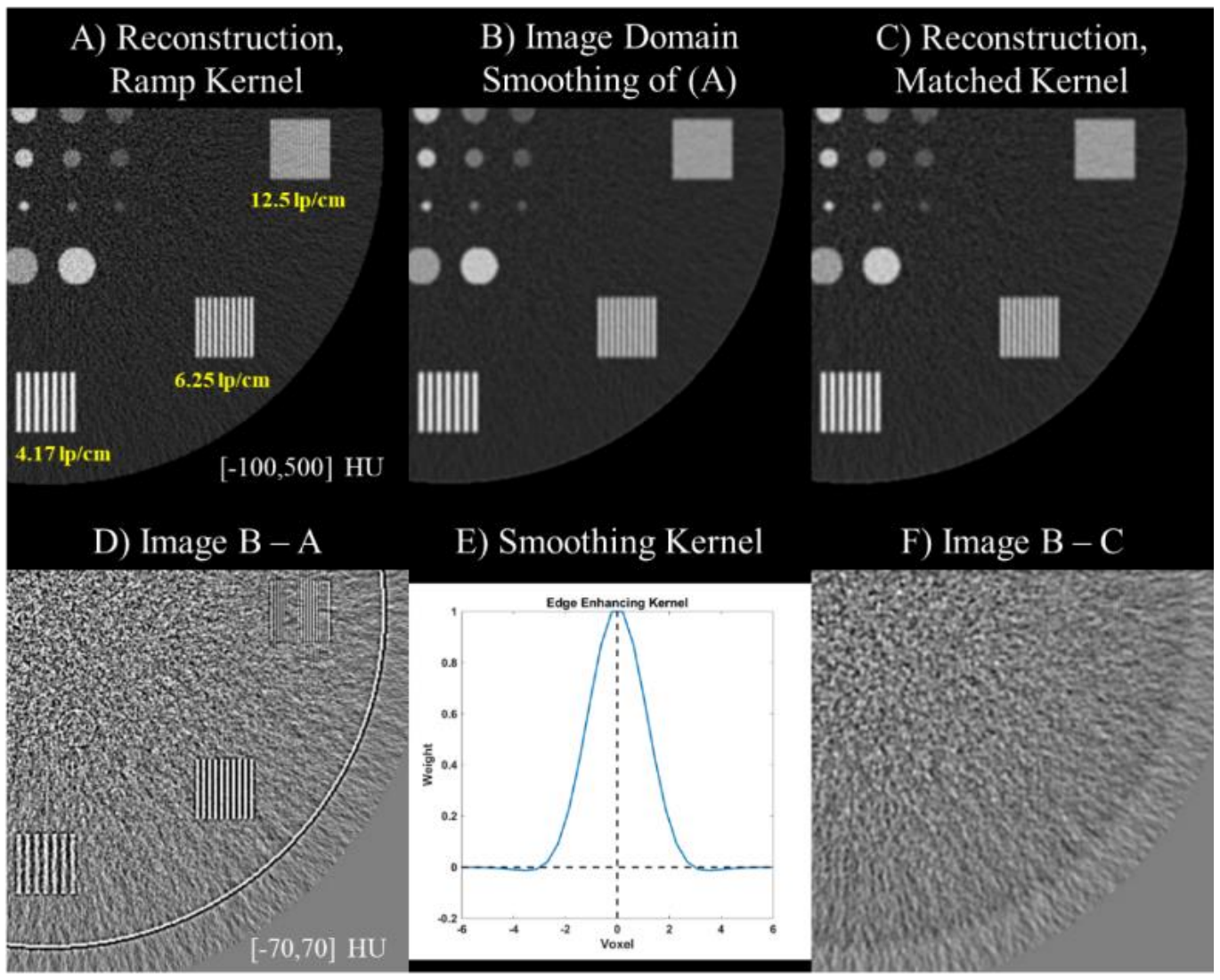

*Figure 3. Reproduction of an image domain resampling kernel with a matched analytical reconstruction window. (A) Reconstruction of a digital contrast-resolution phantom with a ramp kernel (18 cm diameter; sharpest possible reconstruction). (B) Image A after 3D image domain convolution with the edge enhancing kernel plotted in (E) (central line profile shown). (C) Image A reconstructed with a spline-interpolated reconstruction kernel window to best match the MTF and NPS measured in Image B. (D and F) Differences from Image B before and after reconstruction kernel matching.*

## CONCLUSION

The proposed kernel fitting approaches reduce discrepancies between physical and virtual scanners during reconstruction, improving the realism of virtual CT imaging trials.

## ACKNOWLEDGEMENTS AND CONFLICT OF INTEREST

Work at the Quantitative Imaging and Analysis Lab was supported by the NIH NIA (RF1AG070149-01). Work with the Center for Virtual Imaging Trials is funded by the NIH NIBIB (P41 EB028744, R01 EB001838) and NHLBI (R01HL155293). The authors have no conflicts of interest to report.

# The implementation of the DukeSim software in the medical physicist training curriculum

Allan Alves (Sao Paulo State University), Joel Mesa ( Sao Paulo State University ), Guilherme Soares ( Sao Paulo State University )

## Objective

Medical physics residencies in Brazil are essential in preparing the next generation of professionals. To ensure the competence of future medical physicists, it is essential to employ effective educational tools. This study investigates the prospective impact of DukeSim v1.2, a versatile software platform, on the quality of medical physics residency training, in the diagnostic imaging modality.

## Methods

DukeSim v1.2 will be implemented as a pilot program within the curriculum of the Bachelor's program in Medical Physics and the Medical Physics residencies at Sao Paulo State University. This software encompasses various modules, offering simulations of computed tomography projections, image reconstructions, and dose estimations. Over the coming months, medical physics residents and students will utilize DukeSim v1.2 to generate computed tomography projections that emulate the characteristics of CT scanners found in the clinical routine of residency hospitals. Data will be gathered regarding their engagement with the software and projected performance in simulated scenarios.

## Results

The integration of DukeSim v1.2 holds the potential to significantly enhance the educational experience of medical physics residents and students. DukeSim v1.2 is expected to serve as an interactive and invaluable tool for residents and students to practice, engage discussion, and acquire more knowledge regarding image quality parameters in computed tomography image formation.

## Additional Information

Brazil's vast territorial expanse presents challenges in ensuring equitable access to medical equipment and resources. The heterogeneity in the distribution of medical resources used in healthcare further emphasizes the need for advanced training tools. As the field of medical physics continues to evolve in Brazil, embracing innovative solutions becomes imperative for maintaining high educational standards. The future prospects of DukeSim v1.2 are promising in this context, offering a potential solution to these challenges.

# Sex-specific cardiac motion behavior analysis based on XCAT simulated X-ray angiography

Fariba Azizmohammadi1,Paul Segars2, Ehsan Samei2,Luc Duong1

1 Interventional Imaging Lab, Department of software and IT engineering, E´cole de technologie sup´erieure, 1100 Notre-Dame West, Montreal H3C 1K3, Canada

2 Department of Radiology, Carl E. Ravin Advanced Imaging Laboratories, Duke University Medical Center, Durham, NC, USA

## Abstract

Background: Sex-specific cardiac and vascular patterns are significant issues that need to be considered for diagnosing and treating cardiovascular diseases. This is especially important when designing patient-specific cardiac motion models.

Purpose: The goal of this study is to evaluate the bias in sex specific motion prediction models for X-ray angiography using the XCAT phantom.

Methods: We have evaluated a motion prediction model using a recurrent neural network architecture on two datasets, using motion pattern described in the literature. Using the XCAT simulator, we have simulated X-ray angiography sequences for 33 patients (17 females and 16 males) between 11 to 45 yr, including the contrasted left and right coronary arteries. This dataset was generated based on the default parameter setting for the heart cycle. The default parameter setting for a normal length of beating heart cycle in males and females and all ranges of ages is equal to 1 second. The second simulated dataset included only 20 patients (10 females and 10 males) over 50 years old with different heart curves for males and females. Then, we investigated the cardiac motion behavior represented by 2D affine transformation matrices resulting from the frame-by-frame sequence registration.

Results: Our results show that not only are these parameters different between male and female, but the prediction error for predicting the future frames of angiography sequences is slightly higher for female than male.

Conclusion: The integration of sex-specific information is important when designing motion prediction models and virtual simulators such as the XCAT phantom might help to better understand the bias due to sex in motion prediction models.

# Introduction

Cardiovascular disease (CVD) is the number one killer among women worldwide. Controver- sially, women have been under-represented in most clinical trials on cardiovascular disease1. In the past 20 years, the number of clinical and experimental research on investigating sex- related differences has grown significantly. Compared to men, there is a greater prevalence of heart failure with preserved ejection fraction for female patients, and this is why the differences between males and females are important for age-related cardiac remodeling.

Moreover, how coronary artery disease appears and develops typically or atypically over the life course mainly depends on sex variation. Hormonal and non-hormonal factors underlay the sex variance for cardiovascular aging2. Sex-based differences related to AF have been identified, the most concerning being that women with AF have an in- creased risk for car- diovascular events, including stroke. Previous studies have indicated that the annualized rate of stroke is 3% in women compared with 1.6% in men. This sex-based difference in patients with AF is concerning and not clearly understood3.

It has been proved that sex-specific patterns of cardiac and vascular aging are significant issues that need to be considered for the diagnosis and treatment of cardiovascular diseases2. In general, the female heart is smaller than the male heart (usually one-fourth of the men's heart). Accordingly, the number of heartbeats per minute (bpm) for adult men is between 70 to 72 bpm while for adult women is between 78 to 82 bpm. Hence, the heart cycle is shorter for women (average 0.75 sec) compared to men (average 0.85 sec). Apart from size differences, there are geometric differences such as heart mass, left and right ventricular mass, and wall thickness4. A general comprehension of the differences between males and females for cardiovascular diseases are still missing nowadays. With the recent popularity of artificial intelligence and machine learning approaches, the importance of considering the potential bias in the training datasets is paramount to ensure the development of clinical relevant models. Previously, we have introduced a patient-specific motion prediction approach for facilitating the navigation process of the cardiac intervention in coronary arteries 5.

The goal of this study is to evaluate the difference between male and female heart motion in motion prediction algorithm. The motivation of this study is to evaluate the potential bias with a general sex-specific motion prediction model.

# Materials and Methods

## II.A. Data simulation with XCAT

We have evaluated the prediction error due to the sex of patient in our LTSM-based motion prediction approach from X-ray angiography. The cardio-respiratory motion behavior in an X-ray image sequence was represented as a sequence of 2D affine transformation matrices, which provide the displacement information of contrasted coronary arteries in a sequence. The displacement information includes translation, rotation, shearing, and scaling in 2D. A many-to-many LSTM model was developed to predict 2D transformation parameters in matrix form for future frames based on previously generated images.

Using XCAT simulator6, we generated two datasets. First, we have simulated X-ray angiography sequences for 33 patients, 17 females and 16 males with age ranging from 11 to 45 yr, including the contrasted left and right coronary arteries. This dataset was generated based on the default parameter setting for the heart cycle. The default parameter setting for a normal length of beating heart cycle in males and females and all range of ages is equal to 1 second. A general blood pressure heart curve is defined by two heart geometries, end diastole (ED) and end systole (ES). Starting with the ED phase and 4 other parameters including ES, the heart curve is generated for motion simulation in all phantoms while the time duration to complete the heart curve is considered as 1 second. The second simulated dataset included only 20 patients (10 females and 10 males) over 50 years old with a different heart curve for males and females. All of the generated sequences had a length of 150 frames. Then, we investigated the cardiac motion behavior represented by 2D affine transformation matrices resulting from the frame-by-frame sequence registration. The respiratory motion was removed, and only the cardiac motion was simulated in X-ray sequences. For all the frames in each sequence, the arteries were segmented and registered in pairs using point cloud registration. The registration matrices including the affine transformation parameters were saved for each sequence. We generate sex-specific heart curves for adult males and females based on the differences in heart rates between men and women. The average heart rate for adult men is 75 bpm and for adult women is 80 bpm accordingly, the amount of time to complete a heart curve is 0.84 sec and 0.75 sec for males and females respectively. Fig. 1 shows the difference between the default heart curve for XCAT and new sex-specific heart curves.

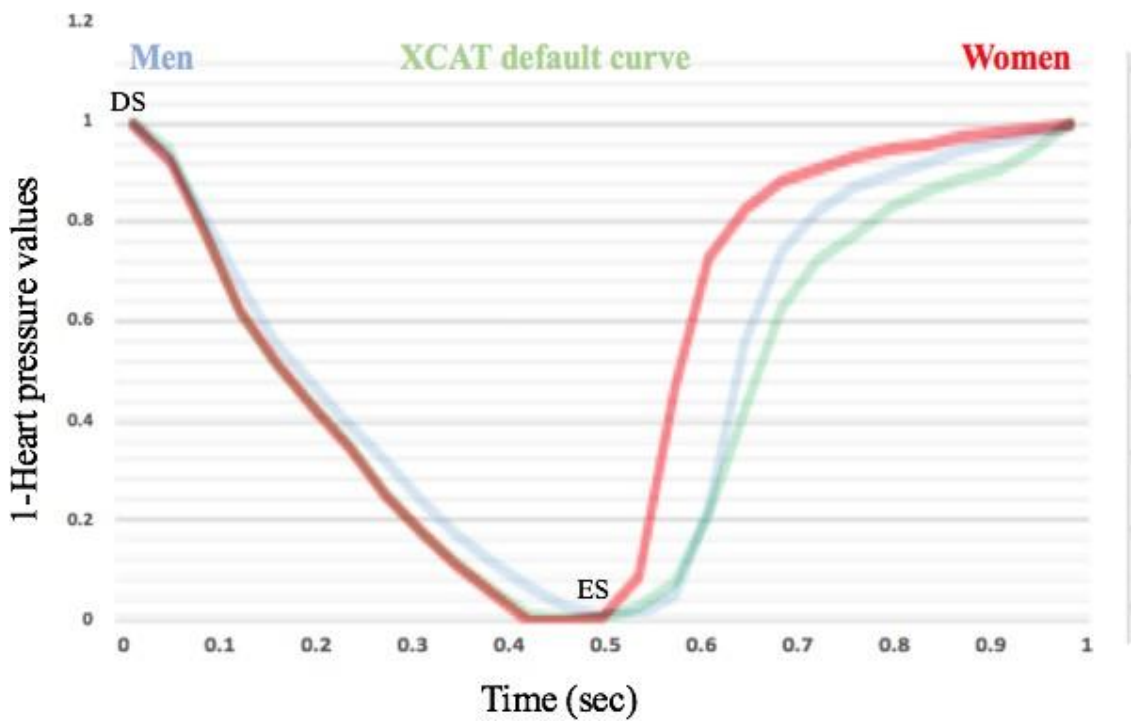

Figure 1: The default XCAT heart curve is illustrated in green. ED and ES are two main parameters for defining the heart curve. The women specific heart curve is outlined in red and the males' heart curve is outlined in blue.

## Results

For the first dataset, according to our previous work, the 2D affine transformation parameters of visible arteries in the X-ray images through a sequence represent cardiovascular motion. To obtain the arteries displacements from frame to frame in a sequence, we first segment the coronary arteries and extracted the center lines for each frame and applied Coherent Point Drift (CPD) algorithm to register the centerlines frame by frame through the sequence. The results of the registration with CPD are sequences of 2D affine transformation matrices. The transformation metrics contain information about translation in 2D (Tx and Ty), rotation, scaling and shearing (A00, A01, A10, A11). We compared the heart motion behavior of male and female patients using the 2D transformation matrices.

The range of motion for the male group is approximately 20 pixels in x direction and

68 pixels in y direction while for the female group, the range of motion in x direction is about 54 pixels and 60 pixels in y direction. Thus, according to these results, the range of motion for female patients in x is more than for male patients. For the other transformation parameters (A00, A01, A10, A11), the values for males and females are slightly different (approximately a pixel).

Figures 2 and 3 show the comparisons between the average values for transformation parameters over the male and female patients. Although there are outliers in both groups of females and males, each category has a specific pattern associated with each transformation parameter. Moreover, it can be seen that for female patients the irregularity for the motion parameters is seen more than men patients (A00 and A10) and for females the range of values is bigger compared to males. In terms of translation parameter differences between males and females, there is more visible diversity between different sexes in the x direction compared to the y direction.

The parameter predictions for affine transformation with our LSTM model worked well for patient-specific motion prediction. To come up with a more general model (sex-specific), we compared the parameters prediction errors in different groups (male and female). Based on the presented results, the transformation parameter prediction errors for females were slightly larger compared to the prediction error for men.

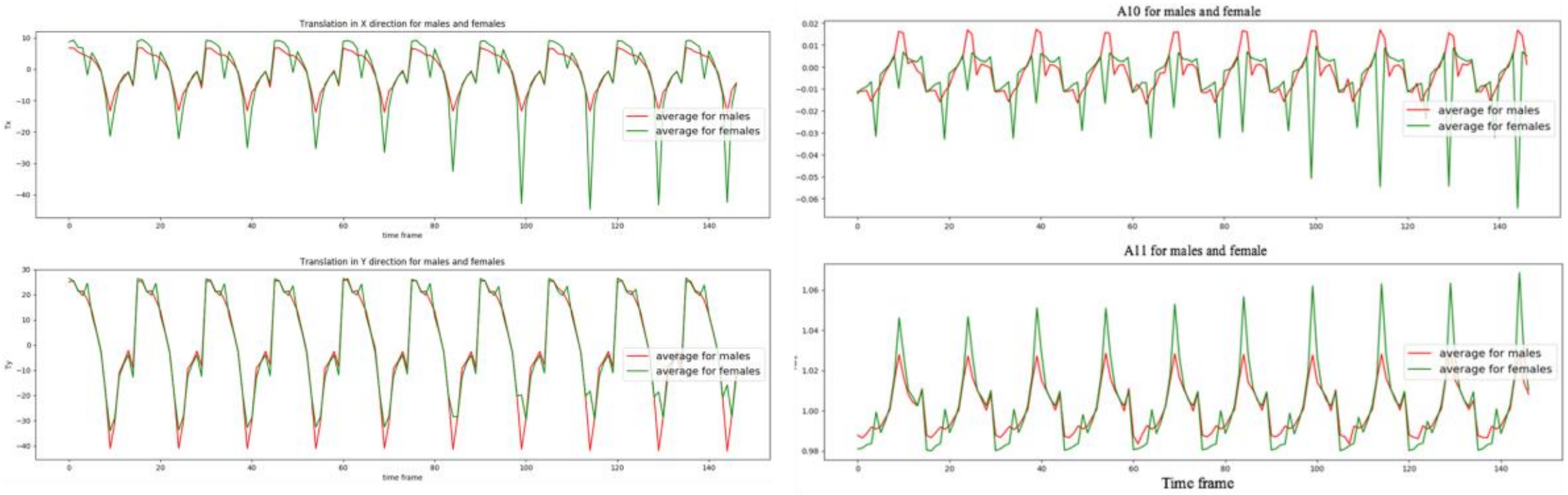

Figure 2: Average overall females and males for 2D translation parameters in pixel (left) and for 2D transformation parameter A00 and A01 in pixel

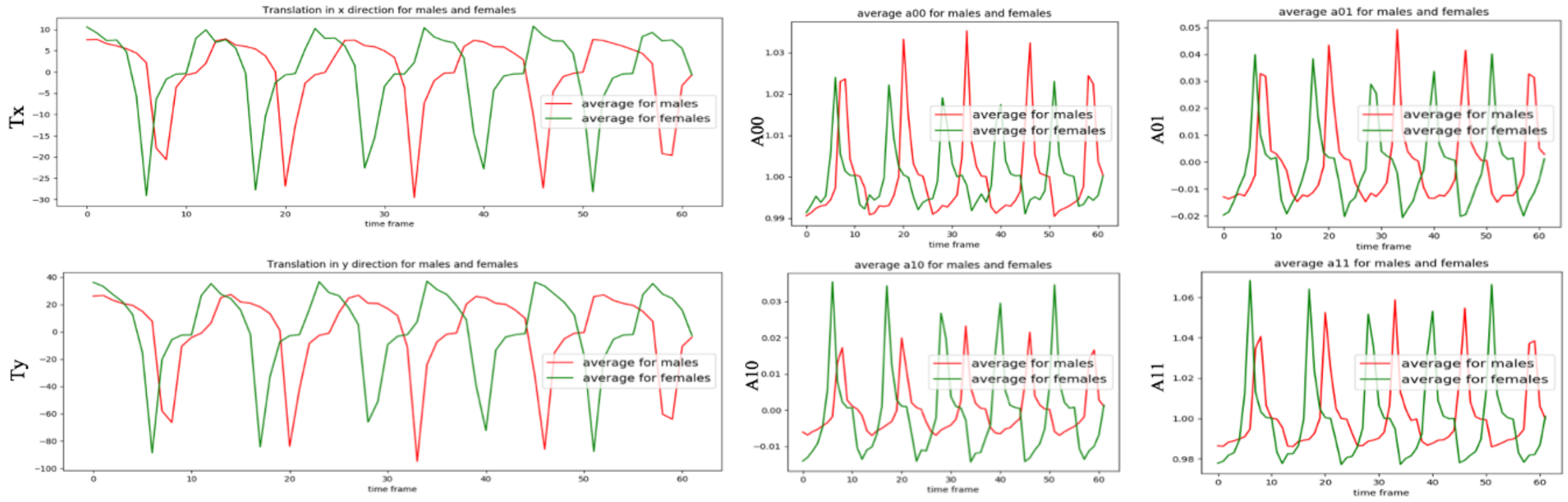

Figure 3: Average overall females and males over 50 yr for 2D translation parameters (Tx and Ty) in pixel and transformation parameters (A00, A01, A10 and A11) in pixel

## Conclusion

We investigated the differences between males and females in cardiac motion behavior based on two different datasets. Our results show differences between men and women, and higher prediction error for predicting the future frames of angiography sequences. This study out- lines the importance of reporting both male and female simulation results during imaging virtual trials, to ensure to evaluate any potential biases in the development of prediction models.

# Simulating PET Imaging with NURBS-based Objects

Darrin Byrd[1], Paul Segars[2], Ehsan Abadi[2], and Paul Kinahan[1]

1. Department of Radiology, University of Washington. Seattle, WA, USA.

2. Department of Radiology, Duke University. Durham, NC, USA.

## BACKGROUND AND PURPOSE

In PET imaging, simulations are commonly used to evaluate system design, reconstruction methods, and have also been used for a decade for virtual imaging trials of clinical task performance [1, 2]. Two general approaches are photon-tracking and ray-tracing. Photon-tracking software packages that generate and track PET annihilation photons have found wide use, and inherently include important physical effects within the imaging process, such as scatter and attenuation [3]. A drawback of these methods is the relatively high computational cost associated with accumulating signal from individual photons, which makes obtaining low-noise images or sets of statistically independent ensembles challenging. In contrast, ray-tracing simulations aim to approximate the true mean emission signal through geometric modeling. The resulting noise-free signal, although lacking random and scattered coincidences, can be used to rapidly generate independent noise instances that adequately represent the noise and bias in the imaging chain. A drawback of ray-tracing methods is their reliance on either simple analytic geometries or voxelization of test objects[4]. The present work evaluates a ray-tracing method using more complex surfaces that can be defined with non-uniform rational B-splines (NURBS) to represent emission and attenuation objects, which allows the simulation of detailed anatomy without the use of discretization.

## METHODS

To simulate PET data, we use the ASIM[5] package from the University of Washington. ASIM takes analytically defined activity and attenuation distributions and creates projection data by computing intersections of line segments without the use of discretization (i.e. no matrix or voxels).

*Object definitions.* We compare two methods for defining emission objects: elliptical and NURBS-based. Before recent modifications, analytic projection by ASIM could only be done using ellipses, which could be translated, rotated, and truncated by planes. The intersections in such cases were computed using the quadratic formula or linear equations in the case of planes. Figure 1 shows sample elliptical geometry and the corresponding system of equations, which can be solved precisely without resorting to discretization (quadratic projection).

| ASIM projection geometry | Equations for exact solution |
|---|---|
| 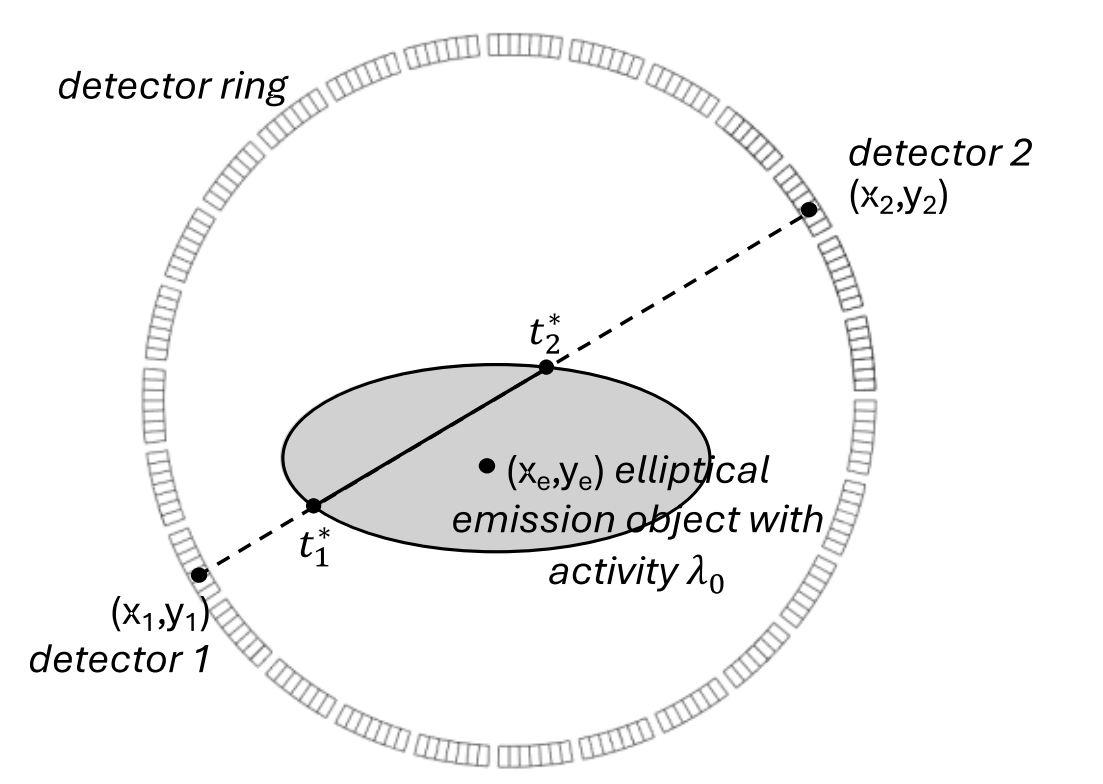 | parametric line segment<br>$x(t) = x_1 + t \times (x_2 - x_1)$<br>$y(t) = y_1 + t \times (y_2 - y_1)$ (1)<br>emission object boundary<br>$\frac{(x - x_e)^2}{a^2} + \frac{(y - y_e)^2}{b^2} = 1$ (2)<br>solutions to (1) and (2)<br>$t_1^*, t_2^*$<br>non-TOF signal for line of response<br>$\mathrm{S}_{\mathrm{LOR}} = \lambda_0 \times (t_2^* - t_1^*)$ |

*Figure 1: Quadratic, non-NURBS projection in ASIM. For detectors located at ($x_1$, $y_1$) and ($x_2$, $y_2$), the length of the intersection with an ellipse at ($x_e$, $y_e$) with axes a and b cab be found from the roots $t_1^*$ and $t_2^*$ of Equations (1) and (2). The product of $(t_2^* - t_1^*)$ with the object's activity $\lambda_0$ is the contribution to raw PET signal. The current ASIM detector model uses block gaps as shown, with curved block faces.*

In recently added code, for NURBS-based objects, the intersections were computed by approximating the curved spline surfaces as planes, for which the intersections can be analytically computed. The algorithm iteratively shrunk the planes around the point of intersection until the difference between the spline and its planar approximation were sufficiently small, then returned the intersection from the final iteration.

*Time of Flight.* We have also modified ASIM to generate time-of-flight (TOF) sinograms. For a given set of object boundaries, computed as described above, signal was distributed to TOF bins using integration, as shown in Equation 3, where $S_{ab}$ is the signal in a TOF bin having boundaries *a* and *b* along the line of response. The signal originates in an object of constant activity density $\lambda_0$ between $t_1^*$ and $t_2^*$ as shown in Figure 1. During detection, it is affected by timing uncertainty $\sigma$, modeled as the Gaussian convolution shown. Figure 2 shows the locations along the LOR for a particular object and TOF bin. The integral in (3) was computed using a power series expression of the integrand.

$$S_{\mathrm{ab}} = \lambda_0 \int_{\mathrm{a}}^{\mathrm{b}} \int_{t_1^*}^{t_2^*} e^{\frac{-(u-x)^2}{2\sigma^2}} \mathrm{d}u\mathrm{d}x \quad (3)$$

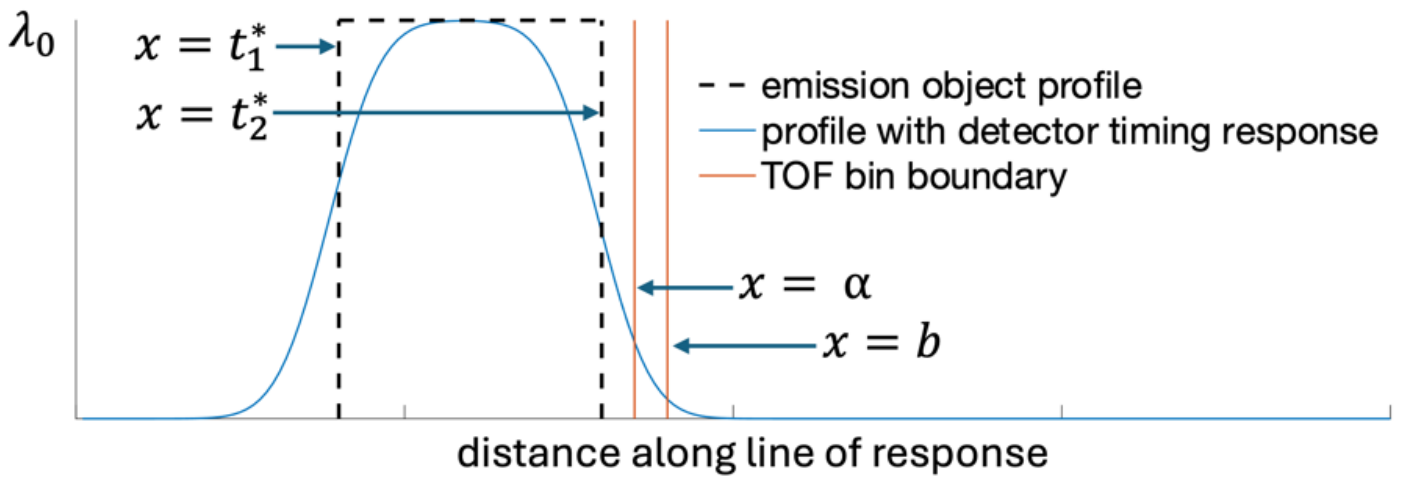

*Figure 2: The limits of integration used in Equation 3. The piecewise constant activity is convolved with the Gaussian timing uncertainty, and a second integration assigns signal to the TOF bin defined by the depicted boundaries.*

*Test objects.* We simulated two test objects. Test object 1 was defined with ellipses and NURBS surfaces, using matching coordinates. It was modeled after an Image Quality phantom and had six spheres (diameters: 10, 13 17, 22, 28, 37mm), with a round background region (diameter: 232mm). Attenuation in the phantom was equal to that of water. Figure 3 shows the voxelized rendering of the phantom. Test object 2 was the XCAT phantom[6], which was defined only as NURBS surfaces.

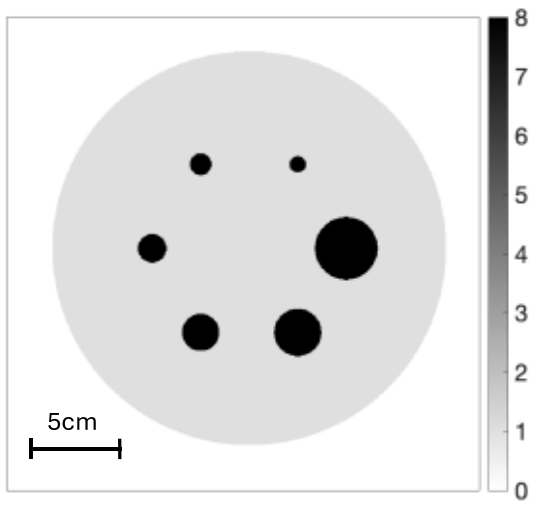

*Figure 3: Test object 1, which was used to test the agreement between NURBS-based and quadratic projection. The contrast of the spheres was 8 to 1, and the attenuation was equal to that of water.*

*Image Reconstruction.* Images were reconstructed in MATLAB using the Duetto package from GE HealthCare. The simulations were configured to match available settings in Duetto, and a clinical three-ring scanner (148mm axial field of view) with 386ps TOF resolution was selected, roughly modeling a Discovery MI (GE Healthcare, Waukesha, USA). Scattered coincidences were included by adding a blurred, noisy version of the emission sinogram. Random coincidences were added to each line or response. Both scattered and random events were perfectly corrected in the mean (i.e. they contributed noise but no bias to reconstructed images).

## RESULTS

The NURBS-based simulations showed good agreement with the analytic simulations. The iterative TOF reconstructions appeared realistic and had few artifacts. Figure 4 shows non-TOF signal from test object 1, created after the simulation was complete by summing the TOF bins. Figure 5 shows radial-time slices of the TOF sinogram, essentially showing the test object after being subjected to binning and the timing uncertainty. Both have the expected appearance, as well as good agreement between the quadratic and NURBS-based projection.

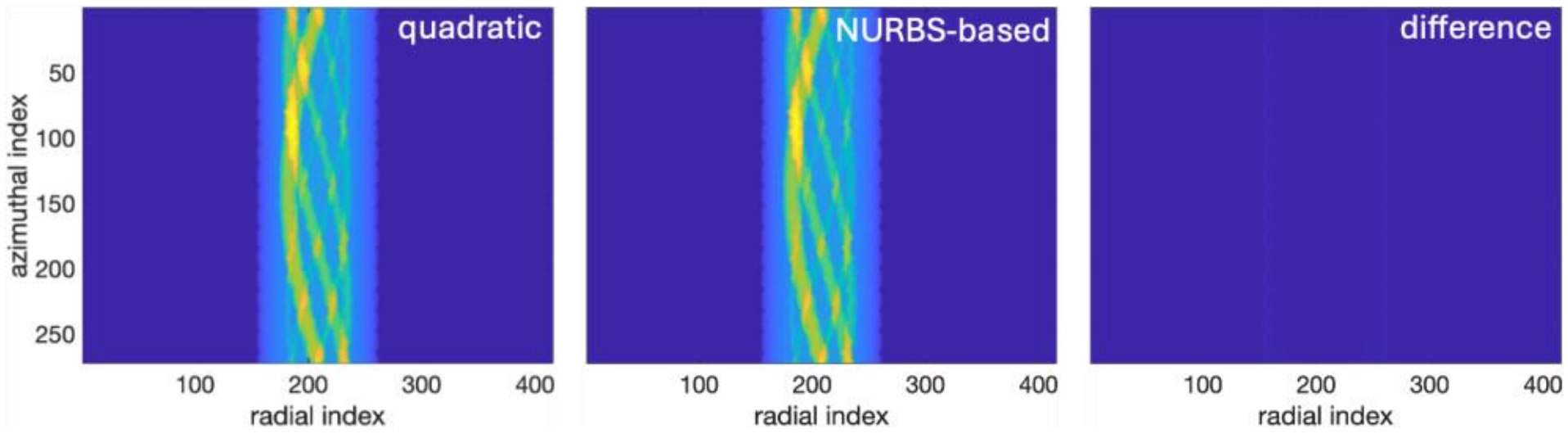

*Figure 4: Noise-free emission data for quadratic projection (left) and NURBS-based projection (center) for test object 1. The difference is shown on the right. Here, the TOF dimension has been summed. Color windows are matched in all panes.*

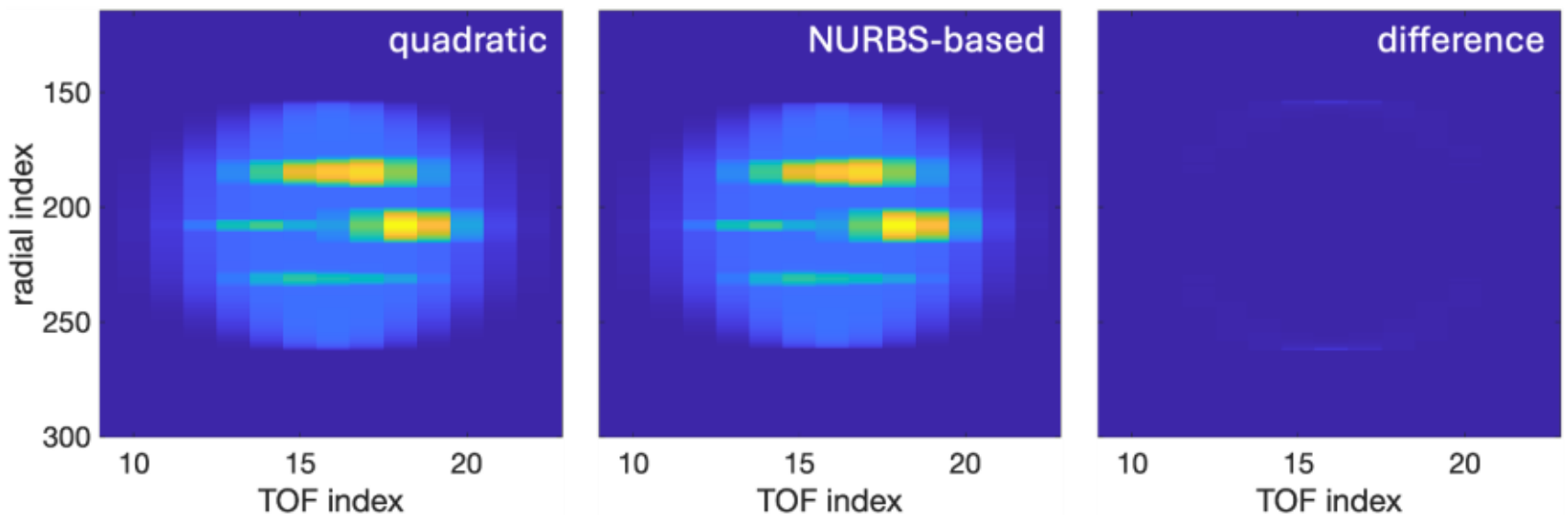

*Figure 5: Noise-free emission data in the time-radial dimensions for quadratic (left) and NURBS-based (center) projection data. The difference is shown on the right. Color windows are matched in all panes.*

Figures 6 and 7 show the reconstructed PET signal.

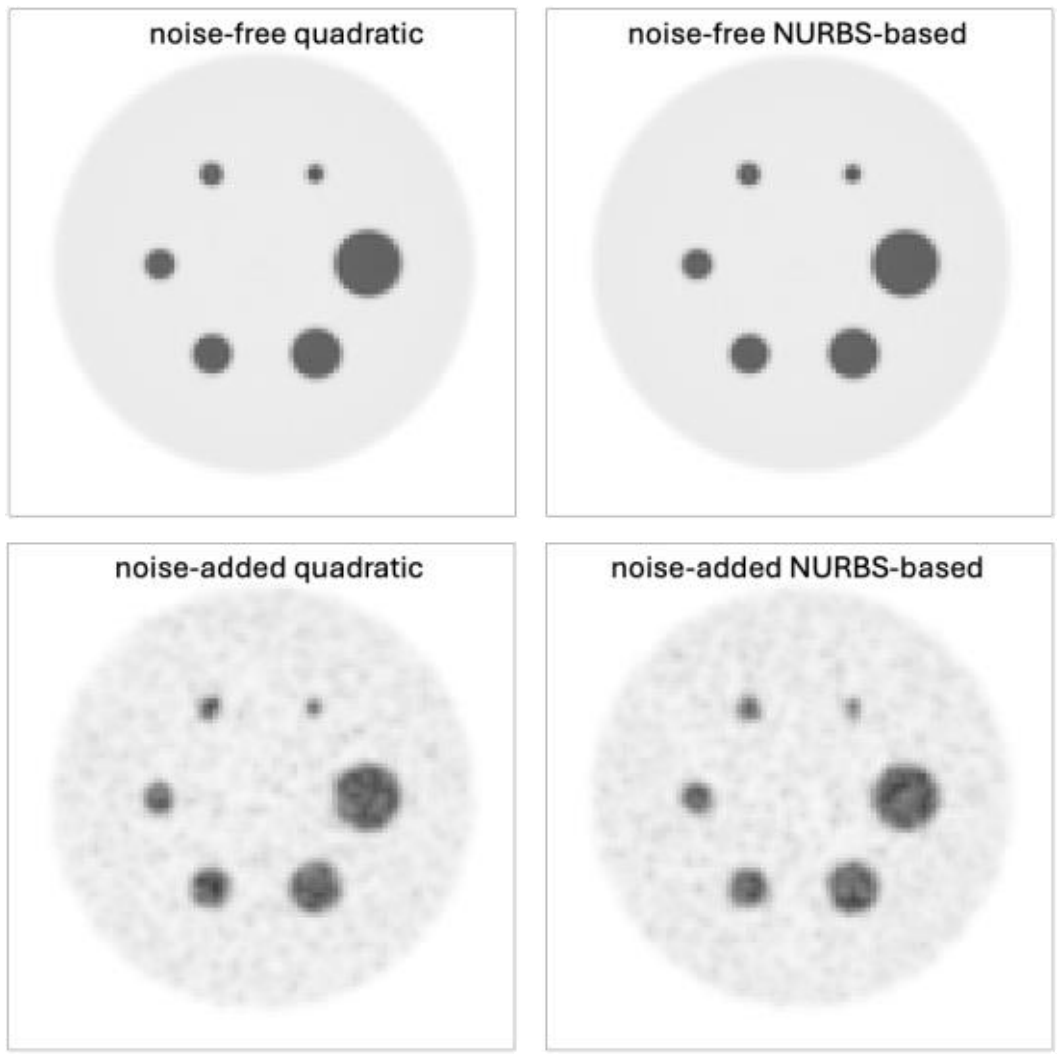

*Figure 6: Time-of-flight reconstructions of test object 1 made with Duetto in MATLAB using settings for a Dicovery MI 3-ring PET/CT scanner. Noisy coincidences in the final simulations were: 17.6M prompts, 4.0M randoms, 3.7M scatters.*

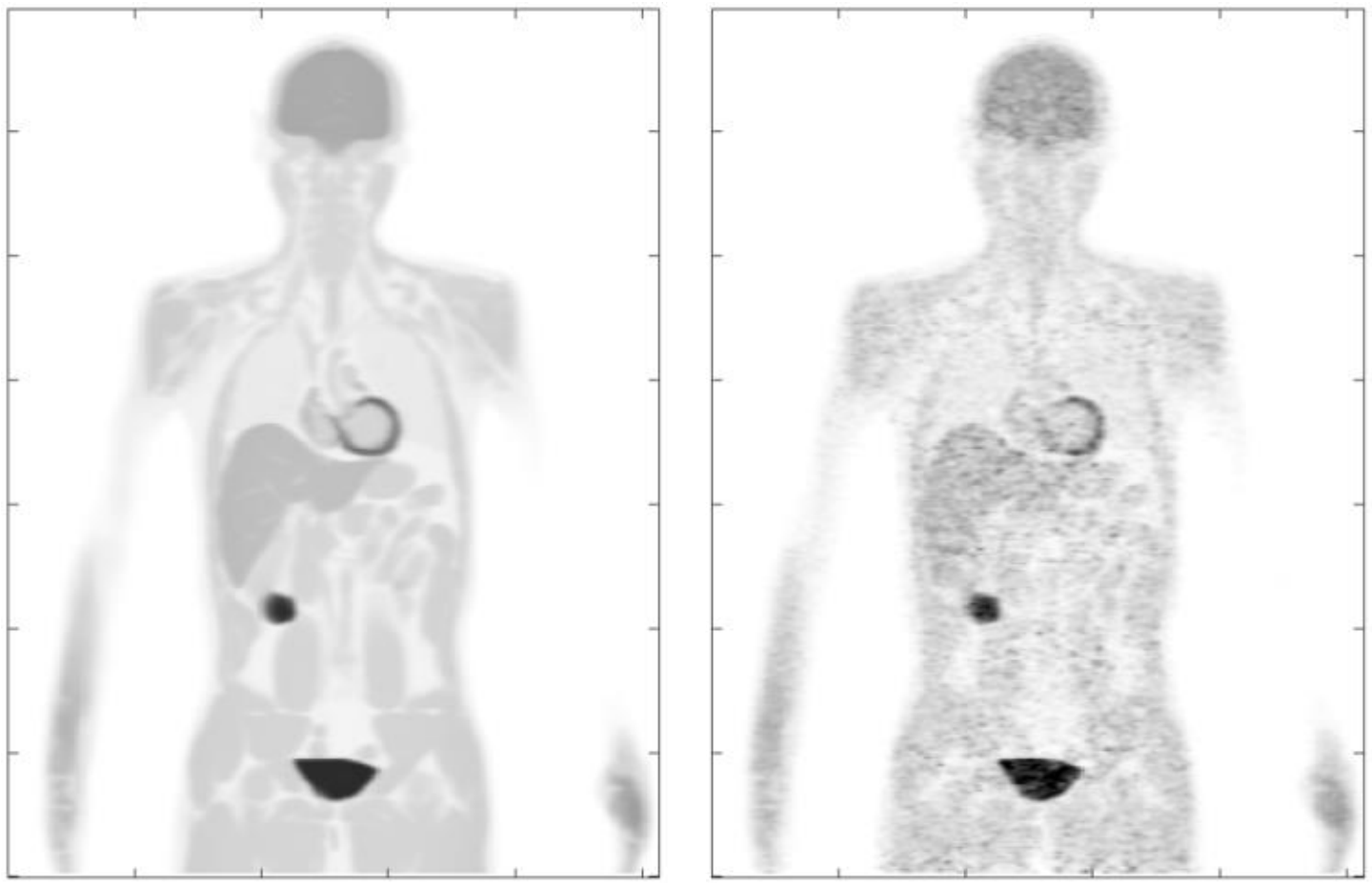

*Figure 7: For noise-free (left) and noise-added (right) simulations, time-of-flight reconstructions of the XCAT phantom made with Duetto in MATLAB using settings for a Discovery MI 3-ring PET/CT scanner.*

## DISCUSSION

The simulation of PET emission data with NURBS matches the precision of the previously validated quadratic projection for the simple test object. The XCAT phantom simulations work as expected.

## CONCLUSION

The extension of ASIM to use NURBS-based objects as inputs, as well as the introduction of analytically computed time-of-flight sinograms, appears adequate to create highly detailed simulated PET images with modern clinical scanner modeling.

## ACKNOWLEDGEMENTS AND CONFLICT OF INTEREST

1 R01 CA258298-01A1
P41EB028744
RO1EB001838

# Numerical x-ray image acquisition platform for 2D and 3D mammography: Impact of focal spot sampling number on system MTF accuracy

Ruben SANCHEZ DE LA ROSA* and Ann-Katherine CARTON*, Remy KLAUSZ, Xavier MANCARDI, Laurence VANCAMBERG

*Contributed equally as first authors

## BACKGROUND AND PURPOSE

Virtual Clinical Trials (VCT) play an important role in the design, evaluation and optimization of imaging devices and applications, as they constitute a method to forecast their potential clinical value. To design a VCT that effectively addresses a clinical question, it is essential to understand the specific design requirements, or 'level of realism' needed for each of the VCT components, which vary depending on the objectives of the VCT. Particularly in computational breast x-ray image acquisition platforms, accurately modeling spatial resolution is essential for replicating clinical scenarios that evaluate the detection and characterization performance of fine details such as microcalcifications or spiculated mass lesions. In addition to the x-ray transport process and blur originating from the image receptor, the shape and emission pattern of the focal spot impact spatial resolution, contribute to the loss of sharpness and contrast of fine details. Furthermore, those digital breast tomosynthesis (DBT) systems using a scanning protocol where the x-ray source moves during exposure, encounter additional geometric blurring caused by the apparent increase of the focal spot width. Thus, an accurate modeling of the focal spot and its motion is essential for evaluating their effect on detection and characterization performance of fine details in the breast. To create Full-Field Digital Mammography (FFDM) and DBT projection images from numerical phantoms, we developed a numerical model based on the CatSim SW platform [1] [2]. This computational model employs a cascaded workflow, raytracing techniques, and experimentally derived models [1] [2]. The model is intended for scintillator-based indirect energy-integrating detectors. In this work, we first explain the methodology used to model the finite focal spot and focal spot motion. To assess the accuracy of spatial resolution, the modulation transfer function (MTF) is used as metric. We evaluate the impact of the spatial sampling point density within the focal spot area and the pixel dimension of the oversampled image receptor matrix at different clinically relevant heights above the breast support table under both static and dynamic focal spot conditions.

## METHODS

### 1. Numerical Image Acquisition platform

**Figure 12** shows a 1D representation of the cascade of steps employed in the numerical image acquisition chain. Only the key elements that impact spatial resolution are shown. The focal spot of our 2D and 3D breast x-ray imaging device is represented by an N x N array of source points uniformly spaced over a rectangular area on the anode track. For the simulation experiments presented here, N was varied from 3 to 16. Each point source is assigned an identical weight to emulate a square emission pattern. This model provides a good first-order approximation of the real emission pattern. The 'x-ray path' originating from each source point and passing through a test object is traced towards a pixel of the numerical oversampled image receptor array, where each oversampled pixel measures $\Delta$µm x $\Delta$µm. For the simulation experiments presented here, $\Delta$ was varied from 5μm to 20μm. After modeling the processes of x-ray photon absorption and energy integration in the detection layer of the image receptor (details not shown), the blurring caused by optical light photons in the scintillator layer is modeled by applying an empirically determined filter to the oversampled detected image. Following this, the integration of the light photons over the sensitive areas of the photodiodes is simulated by filtering the over-sampled image in the Fourier domain with a 2D sinc-shaped kernel representative of the photodiode's aperture transfer function. Subsequently, the oversampled pixels are downscaled to the effective pixel size $\Lambda$ (100um pixel size) through a decimation step. Finally, all electronic noise sources are modeled as signal independent additive white noise (not shown).

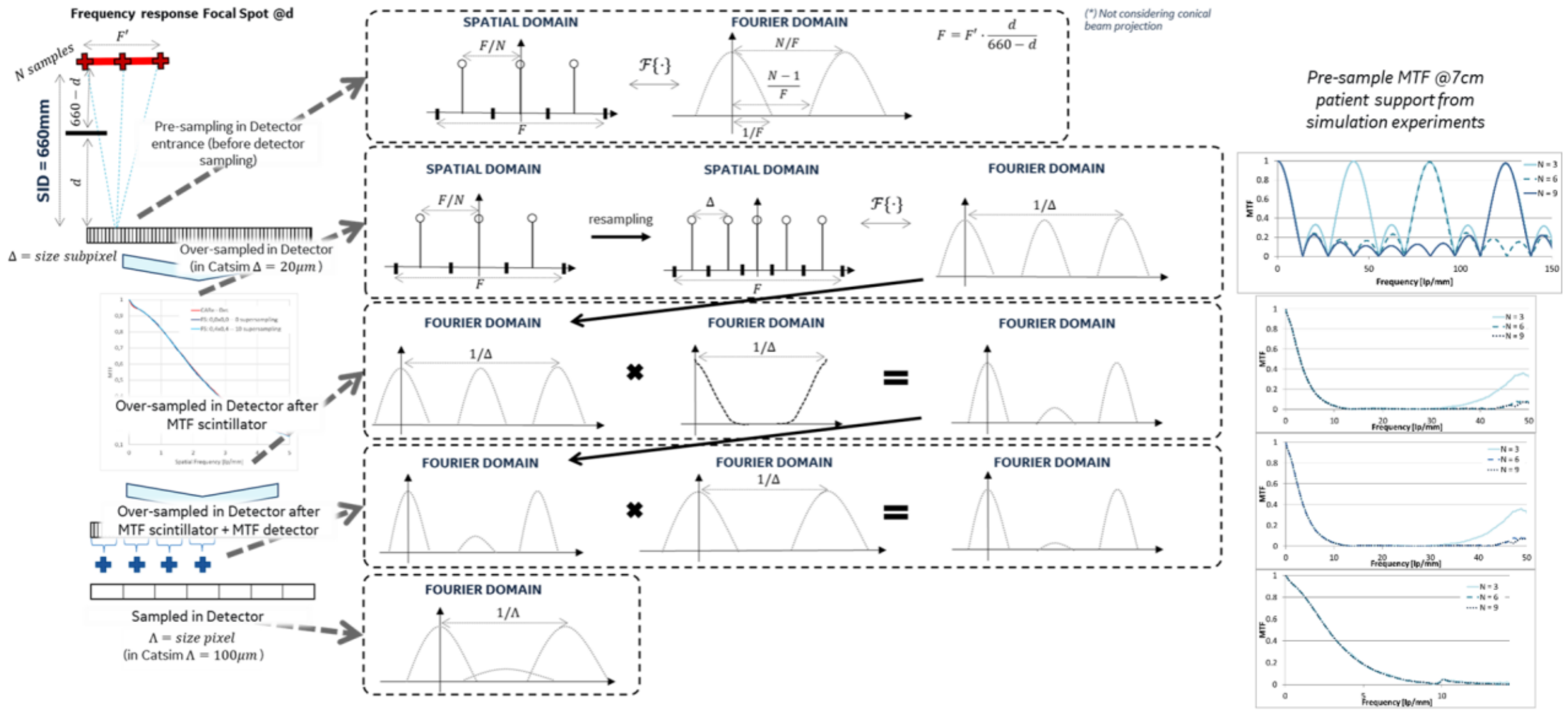

*Figure 12 1D representation of the numerical image acquisition chain in both the spatial domain and the Fourier domain highlighting only the modeling steps that impact the spatial resolution. Simulation results are shown on the RIGHT.*

**Figure 13** illustrates the methodology to model focal spot motion in 1D. At first, a total of n points is allocated along the direction of focal spot motion. The distribution of these sample points, for both the focal spot size (F') and the focal spot motion length (L), is determined by the ratio of the focal spot motion length to its width. More precisely, a number of sample points equal to ((n+1)/ratio+1), rounded up to the nearest integer, is allocated to the focal spot width. Similarly, an integer number of sample points equal to (ratio x (n+1)/ratio+1), rounded up to the nearest integer, is allocated to the focal spot motion length. The allocated sample points for both the focal spot size and its motion are evenly spaced and uniformly weighted. The emission pattern, resulting from the focal spot's motion, is obtained by convolving these two sample point distributions. Subsequent to the convolution, the resulting sample points are then resampled back to the initial total T, using linear interpolation for adjustment. Scan protocols without and with (0.6 to 1.0 mm) focal spot motion were designed.

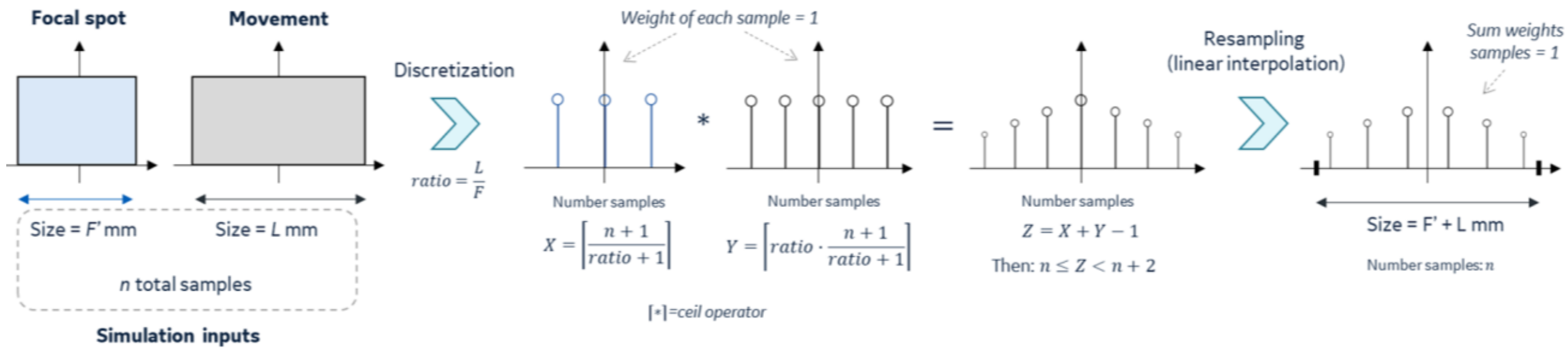

*Figure 13 1D representation of model used to simulate focal spot motion.*

## 1. Modulation Transfer Function

The Modulation Transfer Function (MTF) at image receptor level was used as a measure to evaluate the spatial resolution performance of the system. The MTFs for the computational models were assessed using the standard edge object method, and these results were then compared to a theoretical MTF analytically described by Marshall N. in 2022 **[3]**.

### Theory

The pre-sampled system MTF, $MTF_{pre\ system}\ (f)$, can be written as:

$$MTF_{pre\ system}\ (f) = MTF_{pre\ static\ FS}\ (f) \times MTF_{pre\ motion\ FS}(f) \times MTF_{pre\ detector}(f) \times MTF_{pre\ pixel\ aperture}\ (f)$$

Hereby, $MTF_{pre\ static\ FS}\ (f)$ , $MTF_{pre\ motion\ FS}(f)$, $MTF_{pre\ detector}(f)$, $MTF_{pre\ pixel\ aperture}\ (f)$ are the pre-sampled MTF contributions from the static focal spot, the focal spot motion, the blur from optical photons in the detector and the pixel aperture. The $MTF_{pre\ static\ FS}\ (f)$ corresponding to a focal spot with a rectangular shape and uniform emission pattern at a certain height above the image receptor can be mathematically represented by sinc($\pi \times D_{ImageReceptor}/D_{FS_Edge} \times FS_{optical}$). $D_{ImageReceptor}$ and $D_{FS_Edge}$ are the distances from the image receptor to the numerical edge object and from the focal spot to the numerical edge object, $FS_{optical}$ is the apparent optical focal spot size at 45 mm distance from the chest wall side and laterally centered with respect to the focal spot width. $MTF_{pre\ motion\ FS}(f)$ can be mathematically represented by sinc($\pi \times D_{ImageReceptor}/D_{FS_Edge} \times L$).

### Numerical image acquisition experiment

The numerical edge object was analytically defined and represented the physical edge MTF object according to the requirements of IEC 62220-1-2 for measuring the spatial resolution of mammography systems. This numerical edge object, designed with dimensions of 120 x 60 x 0.8 mm, was implemented to be composed of stainless steel. It was oriented at a 3° angle relative to the focal spot length and it was laterally centered with respect to the focal spot width. Image acquisition simulations of the edge object were performed placing the edge at various, clinically relevant heights (30, 45, 70 mm) above the numerical patient support. The source to image receptor distance was set to 660mm. Edge images were simulated using a 34 kVp Rh/Ag equivalent x-ray spectrum. The MTF were computed were computed using a method similar to that described by IEC62220-1-2 [4].

# RESULTS

Figure 14 presents the MTF assessments from the simulation experiment alongside the analytically determined 'ground truth' MTFs. The results are shown for an edge object positioned at various clinically relevant heights (3, 4.5, and 7 cm) above the patient support table, considering scan protocols both with and without focal spot motion. In scenarios without focal spot motion, the MTF error margin is smaller than 0.01 across all frequencies, irrespective of the focal spot point sampling density, at all edge object heights above the patient support. In scenarios with focal spot motion, the MTF error margin is smaller than 0.03 across all tested frequencies and edge object heights. For focal spot motions of 0.6mm, 0.8mm and 1.0mm, an MTF error < 0.01 is obtained at all frequencies and for all heights when using 16x16, 16x16 and 12x12 focal spot sample points, respectively. The pixel size of the oversampled numerical image receptor array has shown negligeable impact on MTF accuracy, with differences less than 0.7% observed within the 0 to 5 lp/mm range.

## DISCUSSION AND CONCLUSION

We presented an evaluation of the impact of focal spot modeling on MTF accuracy. For the investigated conditions, we observed that accurate system MTFs can be achieved in the presence and absence of focal spot motion, even when employing a sparse focal spot sampling point density (3x3). Of course, the number of focal spot sub-sampling points should be tailored according to the requirements of each VCT. We opted for a uniform emission pattern as a simple yet effective initial approach, although the actual emission patterns of focal spots in mammography tend to slightly diverge from this. For scenarios involving continuous focal spot motion, we modeled the movement as linear and constant. This simplification may not fully capture some DBT designs where the actual trajectory is described by an arc rather than a straight line. Our analysis focused on assessing MTF accuracy specifically in the direction of focal spot movement, without considering effects in the perpendicular direction.

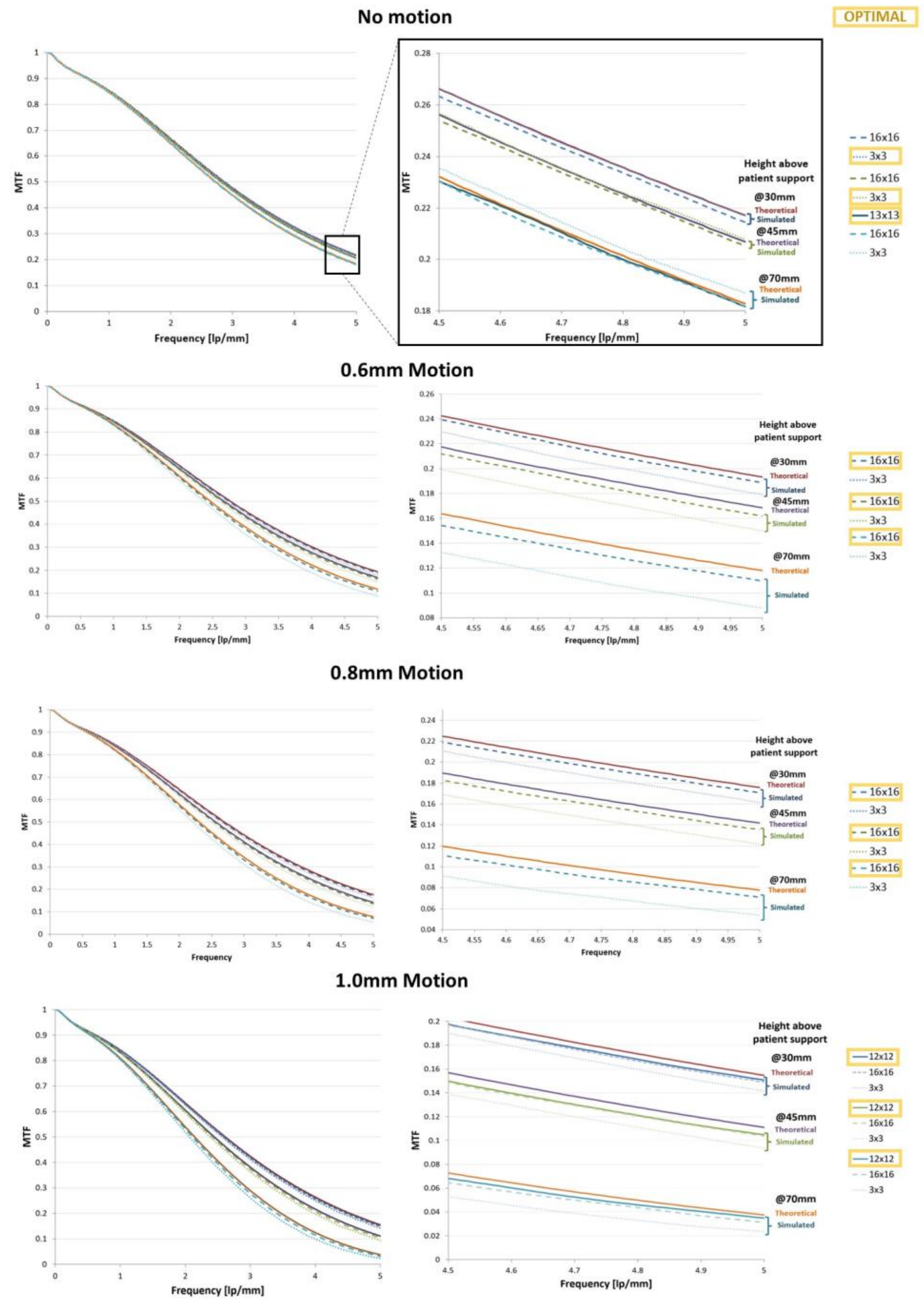

*Figure 14 MTFs assessed from simulated images of numerical slanted edge test objects at varying heights (30, 45, and 70 mm) above the patient support table, examined using scan protocols with and without focal spot (FS) motion. The results are compared against analytical 'ground truth' MTFs. Various focal spot sampling densities were utilized for the simulations.*

# Generalized Methodologies for Assessing the similarity and dissimilarity between two datasets

Dhrubajyoti Ghosh, Joseph Y. Lo, Liesbeth Vancoillie, Fakrul Islam Tushar, Lavsen Dahal, Ehsan Samei

**Purpose:** Virtual imaging trials are a crucial avenue for advancing the development and evaluation of imaging techniques in simulated environments. One fundamental aspect of replicating real-world data within virtual settings is ensuring that the covariates, or factors, within the virtual dataset closely resemble those found in the real world. This necessitates a thorough assessment of the similarity between these datasets, a process vital for the credibility and applicability of trial outcomes.

**Methods:** To assess the similarity of two distributions, various metrics can be employed. The Kolmogorov-Smirnov Distance (KS) measures the maximum vertical difference between cumulative distribution functions, making it valuable for basic goodness-of-fit testing. In contrast, the Wasserstein distance, also known as the Earth Mover's Distance, provides a more comprehensive measure. It considers both the magnitude of differences and their spatial arrangement within the entire distributions, offering a more detailed perspective. We applied these measurands of variability to data from the National Lung Screening Trial (NLST) dataset, which includes CT scans and Chest X-Ray from a large group of human subjects. NLST was assessed in terms of its similarity to a replica of the trial, named Virtual Lung Screening Trial (VLST). The similarity was assessed in terms of demographic information including age, education level, ethnicity, height, and weight. Both KS and Wasserstein distances were employed.

**Results:** The measures revealed that there is no significant evidence indicating distributional differences between the NLST and VLST datasets. In essence, the variability in these datasets is statistically similar, enabling one to be used as a training dataset and the other as a test dataset. Needless to say, such conclusion is dependent on the attributes that are used as a basis of any similarity assessment.

**Conclusion:** We are aiming to conduct similar analyses on virtual datasets, comparing them to the NLST database. This assessment aims to determine whether the demographics within the virtual dataset resemble those of the NLST dataset, facilitating the evaluation of detection methods in both datasets. By ensuring this similarity, virtual imaging trials can provide credible and insightful outcomes. Additionally, using these approaches, we can assess the similarity and dissimilarity between any two variables in a dataset. The methodology deployed here can be used both for the assessment as well as the design of similarity-aware trials.

# Oxygen Effect on Tumor Response to Irradiation: Insights from GATE Simulation Studies

Zakaria Ait Elcadi, Othmane Bouhali

## BACKGROUND AND PURPOSE

The Oxygen effect plays a pivotal role in how tumor cells respond to irradiation. Tumor hypoxia has been acknowledged as a limiting factor in treatment response for many years. Diverse approaches to enhance the rate of tumor control have been experimented with, aiming to amplify the radical cascade [1].

Cells with an oxygen partial pressure ($pO_2$) below 8-10 mmHg are generally categorized as hypoxic [2] and exhibit greater radio-resistance compared to cells with normal oxy- gen levels ($pO_2$ up to approximately 100 mmHg [2]). This phenomenon can be quantified through the Oxygen Enhancement Ratio (OER), representing the ratio between the dose at a specific oxygen pressure and the dose at normal oxy- gen pressure that produces an equivalent biological effect ICRU Report 30: Quantitative Concepts and Dosimetry in Radiobiology [3].

In this study, the dose distribution was assessed in a spherical tumor positioned at the center of the right lung. This tumor was generated using an XCAT phantom of an elderly woman, and various oxygen concentrations were considered. The evaluation was conducted using the GATE simulation code (V9.3) [4], employing the Monte Carlo method.

## METHODS

The investigation into the impact of oxygen on dose distribution was conducted using a female XCAT phantom, featuring a spherical tumor with a 1 cm diameter positioned at the center of the right lung (figure 1). This selection was influenced by the effect of low-density material on a spherical tumor.

with a density comparable to that of soft tissue. Various levels of oxygenation were applied to the spherical tumor to assess its response under different oxygen concentrations. The study utilized previously validated phase specifications of the Varian Clinac IX for a photon energy of 6MV [5]. The field size was fixed at 1.2x1.2 $cm^2$, employing removable jaws for a spherical tumor with a 1 cm radius, ensuring the tumor's central placement within the irradiation field. Oxygenation levels applied to the tumor ranged from 5 to 80 mmHg. The XCAT phantom was positioned at a source-surface distance of 100 cm. Results were analyzed based on deposited energy and OER values relative to $O_2$ pressure in mmHg, following a specific equation [6].

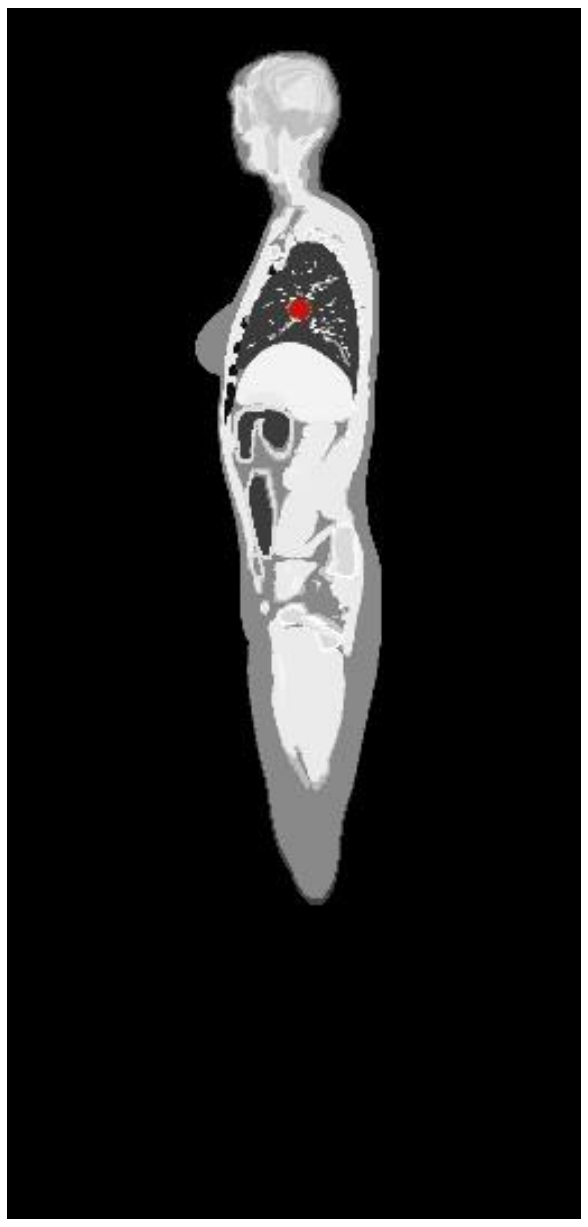

*Figure 15 Coronal view of the female XCAT phantom show a spherical tumor centered in the right lung with a diameter of 1 cm.*

## RESULTS

Nine simulations were conducted for the selected pressures in the initial phase of our study (figure 2). Figure (3) illustrates the deposited energy based on the $O_2$ pressure in the spherical tumor, while figure (4) displays OER values across $O_2$ pressures from 5 to 80 mmHg. The statistical error was a minimum of 2% for each simulation.

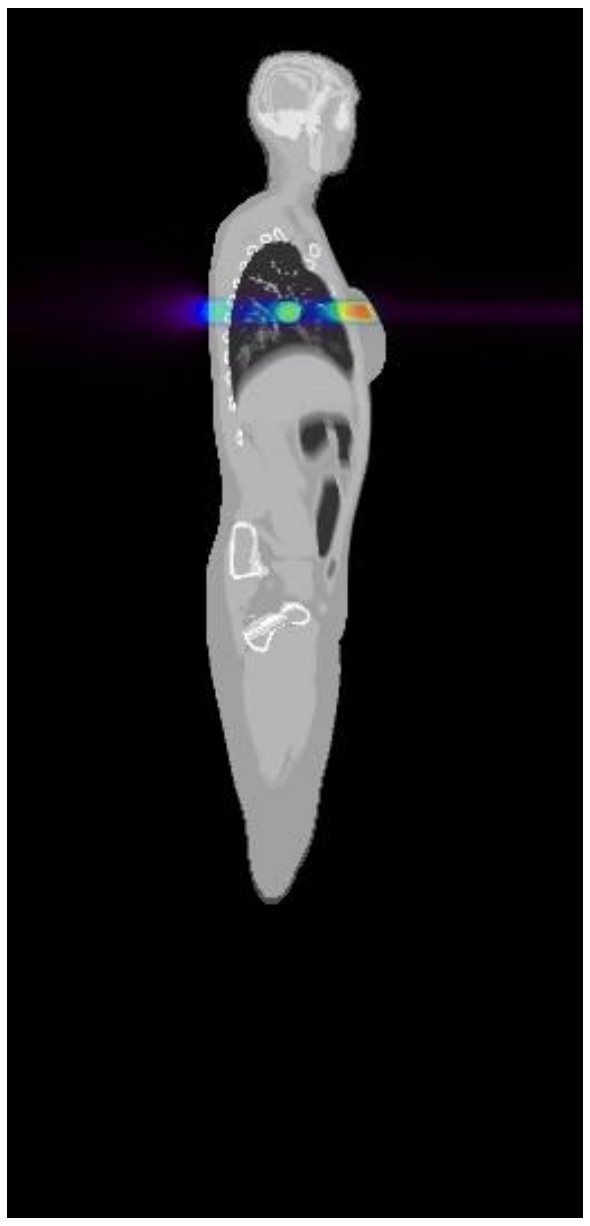

*Figure 2 Coronall view of the female XCAT phantom, simulated in GATE and targeting the spherical tumor in the right lung, exhibit a 1 cm diameter with a 1x1 cm² field size.*

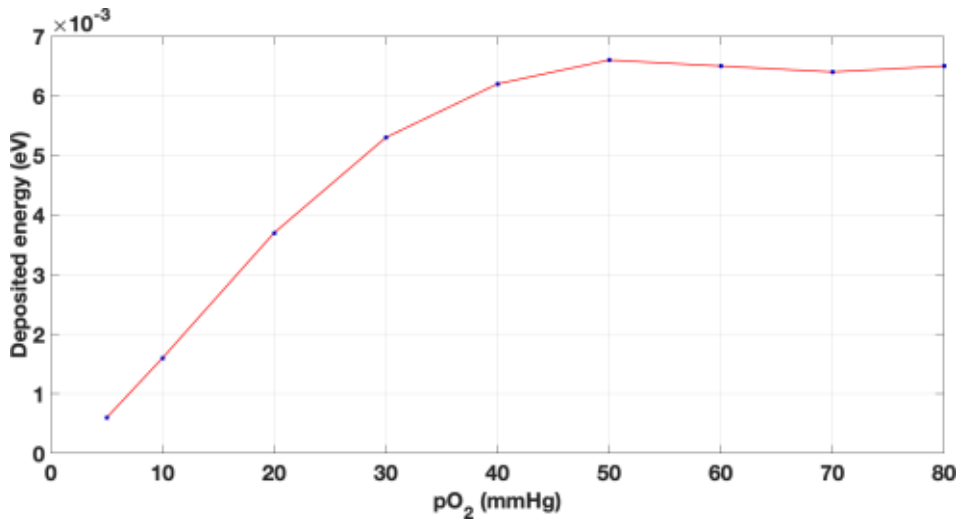

*Figure 3 Energy deposition for various $O_2$ pressures applied to the spherical tumor.*

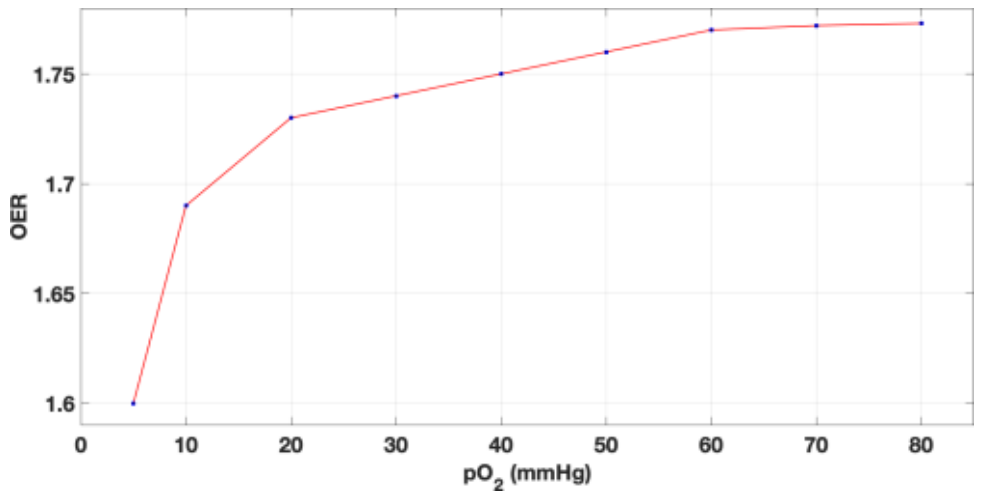

*Figure 4 OER values are associated with each $O_2$ pressure ap- plied to the spherical tumor.*

# DISCUSSION

The figure 3 indicate that the deposited energy increases as the $O_2$ pressure grows, reaching a plateau at 60 mmHg. Beyond this point, the ratio remains constant (>60 mmHg), as illustrated in the figure. The OER values relative to $O_2$ pressure exhibit a linear growth up to 60 mmHg, followed by stabilization (figure 4).

In summary, the figures lead to the conclusion that regions with low oxygen pressure necessitate characterization by low deposited energy, requiring a higher dose to achieve effects comparable to regions with normal oxygenation. This consideration extends to the same biological effects to be achieved and the specific environment of the tumor, influencing the choice of a region with low density for this study.

# CONCLUSION

The aim of this study was to explore the importance and necessity of oxygen ($O_2$) pressure in radiotherapeutic treatments. It was noted that the selected dosage for these treatments may differ, not dependent on the tumor and its location, but rather on the oxygen pressure within the tumor itself. This leads to the subsequent step of initiating a new study on physiological factors and biological effects that must be considered for tumors exhibiting hypoxia. Additionally, the study will examine the location of these tumors in relation to regions with varying densities, along with evaluating the dose received by neighboring organs.

# ACKNOWLEDGEMENTS AND CONFLICT OF INTEREST

This work was supported by an internal grant from Texas A\&M University at Qatar (RRSG2022). The simulation was conducted using the high-performance computing infrastructure at the same university.

The authors have no relevant financial or non-financial interests to disclose.

# Microcalcification cluster modelling for mammographic imaging: an automated toolbox with image-specific simulations

Astrid Van Camp*[a,b], Eva Punter[b], Katrien Houbrechts[b], Lesley Cockmartin[c], Nicholas W. Marshall[b,c], Henry C. Woodruff[a,d], Philippe Lambin[a,d], Hilde Bosmans[b,c]

[a] Department of Precision Medicine, GROW – Research Institute for Oncology and Reproduction, Maastricht University, Maastricht, The Netherlands
[b] KU Leuven, Department of Imaging and Pathology, Division of Medical Physics & Quality Assessment, Herestraat 49, 3000 Leuven, Belgium
[c] UZ Leuven, Department of Radiology, Herestraat 49, 3000 Leuven, Belgium
[d] Department of Radiology and Nuclear Medicine, GROW – Research Institute for Oncology and Reproduction, Maastricht University Medical Centre+, Maastricht, The Netherlands
*Corresponding author: Astrid Van Camp, a.vancamp@maastrichtuniversity.nl

## ABSTRACT

Accurate diagnosis and early detection of breast cancer rely on the correct interpretation of suspicious microcalcification clusters in mammographic images. This interpretation remains challenging due to the subtle nature and large variety of calcifications.
Simulated lesion models can be employed to optimise imaging techniques, improve deep learning models, and facilitate a better understanding of breast lesions. However, existing modelling techniques often lack the capacity to capture the diversity present in real lesions or to generate models customized to specific cases. In this work we present an automated, flexible toolbox to generate microcalcification clusters with specific characteristics and based on breast textures in mammographic images.
This toolbox provides a method to build three-dimensional (3D) cluster models utilizing geometrical shapes and morphological operations. Additionally, it includes a novel technique that employs a handcrafted radiomics analysis to create two-dimensional (2D) models for a specific 2D mammographic image. Considering the computed quantitative radiomics features, it determines a plausible location for insertion, and accounts for local breast textures in order to find possible individual calcifications and combine these into a cluster. Consequently, the simulated microcalcification cluster is relevant to the image, ensuring a seamless integration within the existing breast tissue.
The toolbox's interface guides the user through the parameter tuning process, allowing for either manual tuning, or the utilization of pre-determined settings for specific clinical calcification types. The ability to customise parameters related to the cluster characteristics enables the creation of models that range from regular to complex. The 2D generation method specifically guarantees the generation of a unique, realistic cluster for each image in a dataset. Such generated models are applicable for various tasks and are compatible with insertion in different types of mammographic images from multiple acquisition systems. Validation studies have confirmed the method's ability to simulate realistic clusters and capture clinical properties when tuned with appropriate parameter settings.

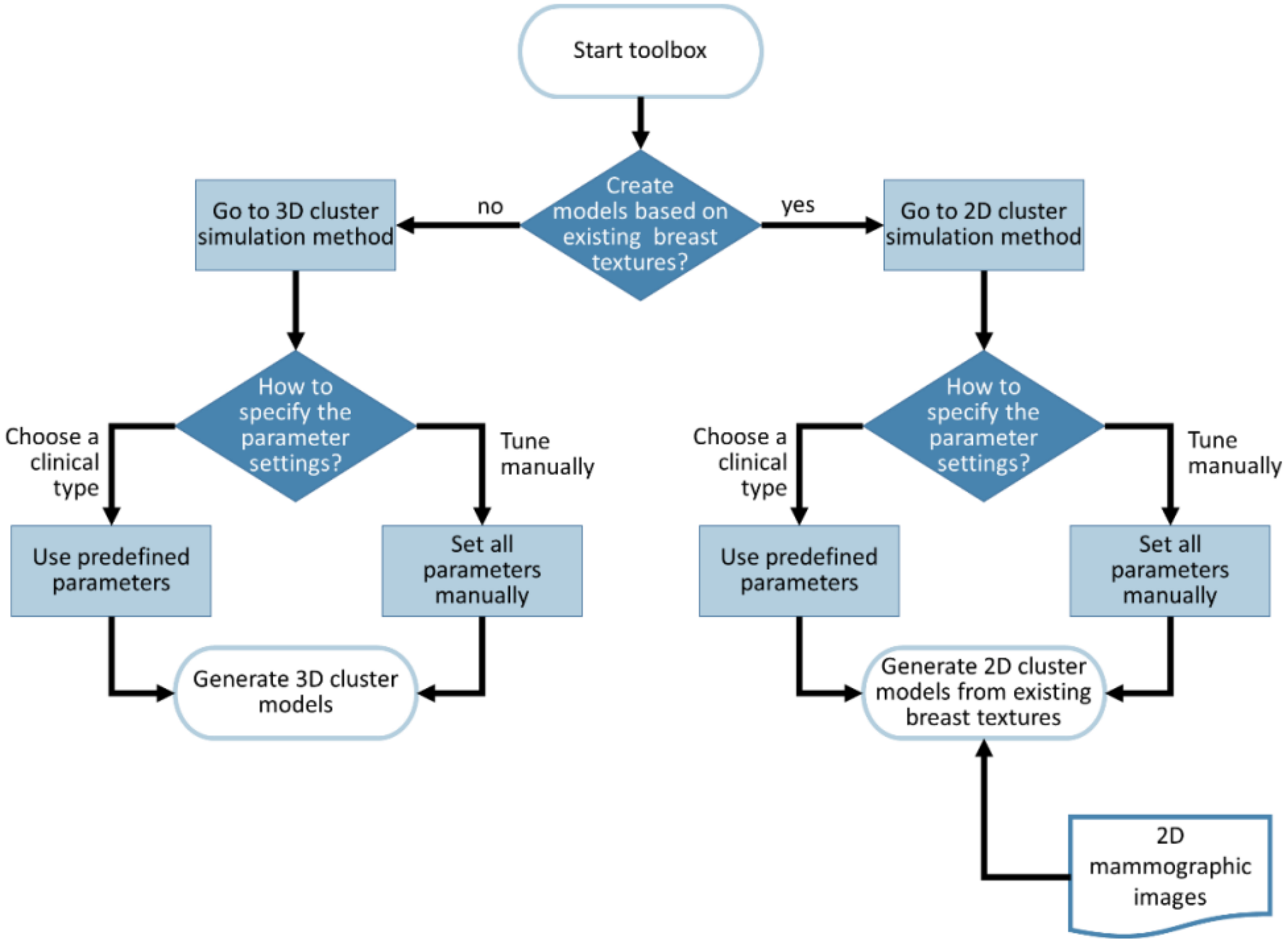

*Figure 1: Pipeline of the automated toolbox*

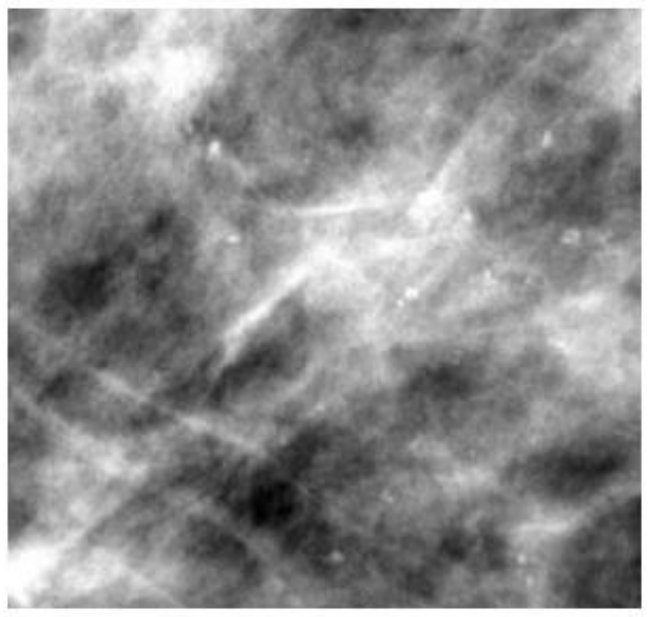

*Figure 2: A simulated 3D microcalcification cluster with coarse heterogeneous type calcifications inserted in a 2D lesion-free mammographic image*

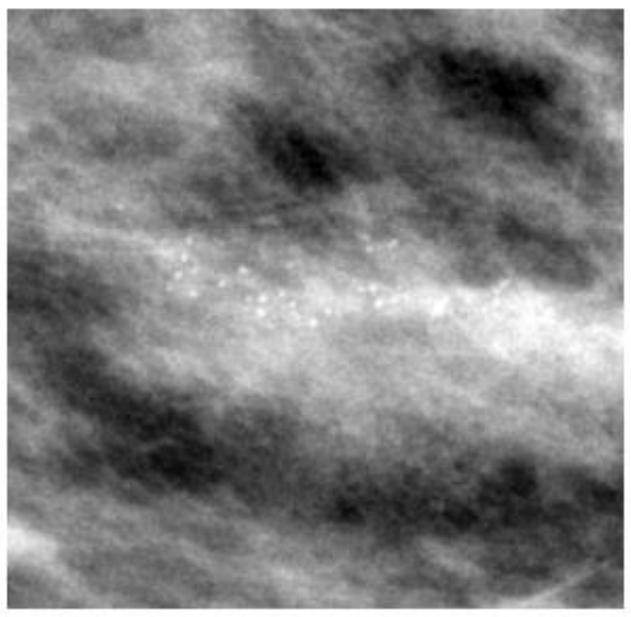

*Figure 3: A simulated 2D microcalcification cluster of suspicious morphology inserted in a 2D lesion-free mammographic image*